\pdfoutput=1 %arXiv admin added this line
\documentclass[twocolumn,amssymb,floatfix,aps,showpacs]{revtex4}

\usepackage{amsmath}
\usepackage{graphicx}% Include figure files
\usepackage{bm}% bold math
\usepackage{epsfig}
\usepackage[section]{placeins}
\newcommand{\bea}{\begin{eqnarray}}
\newcommand{\eea}{\end{eqnarray}}
\def\be{\begin{eqnarray}}
\def\ee{\end{eqnarray}}

\renewcommand{\c}[1]{\,{\rm cos}\,#1}
\newcommand{\s}[1]{\,{\rm sin}\,#1}
\renewcommand{\cos}[1]{\,{\rm cos}(#1)}
\renewcommand{\sin}[1]{\,{\rm sin}(#1)}
\def\pslash{\not{\hbox{\kern-2pt p}}}
\begin{document}
\preprint{CALT 68-2772}
\preprint{FERMILAB-PUB-10-019-T}
\title{
%% old 
%% new
Higgs look-alikes at the LHC} 
% Force line breaks with \\

%
\author{A. De R\'ujula${}^{a,b,c}$}
\author{Joseph Lykken${}^d$}
\author{Maurizio Pierini${}^c$}
\author{Christopher Rogan${}^e$}
\author{Maria Spiropulu$^{c,e}$}
\affiliation{  \vspace{3mm}
${}^a$ Instituto de F\'isica Te\'orica, Univ. Aut\'onoma de Madrid, Madrid, and 
CIEMAT, Madrid, Spain,\\
${}^b$ Physics Dept., Boston University, Boston, MA 02215,\\
${}^c$Physics Department, CERN, CH 1211 Geneva 23, Switzerland,\\
${}^d$Fermi National Accelerator Laboratory, P.O. Box 500, Batavia, IL 60510,\\
${}^e$Lauritsen Laboratory of Physics, California Institute of Technology, Pasadena, CA 91125}


\begin{abstract}
  The discovery of a Higgs particle is possible in a variety of search
  channels at the LHC. However, the true identity of any putative
  Higgs boson will, at first, remain ambiguous until one has
  experimentally excluded other possible assignments of quantum
  numbers and couplings.  We quantify the degree to which one can
  discriminate a Standard Model Higgs boson from ``look-alikes'' at,
  or close to, the moment of discovery at the LHC.  We focus on the
  fully-reconstructible ``golden'' decay mode to a pair of $Z$ bosons
  and a four-lepton final state.  Considering both on-shell and
  off-shell $Z$'s, we show how to utilize the full decay information
  from the events, including the distributions and correlations of the
  five relevant angular variables. We demonstrate how the finite phase
  space acceptance of any LHC detector sculpts the decay
  distributions, a feature neglected in previous studies. We use
  likelihood ratios to discriminate a Standard Model Higgs from
  look-alikes with other spins or nonstandard parity, $CP$, or form
  factors.  For a resonance mass of 200 GeV/c$^2$, we achieve a median
  discrimination significance of $3\,\sigma$ with as few as 19 events,
  and even better discrimination for the off-shell decays of a 145
  GeV/c$^2$ resonance.
\end{abstract}

\maketitle

%	\newcommand{\gsim}{\hbox{\rlap{$^>$}$_\sim$}}

%	\newcommand{\impostor}{\hbox{\rlap{$\, ^\star$}$_H$}}

%	$\gsim$ $\impostor$

%	$\hat H$ $\tilde H$ $H^\star$ $\star\atop H$

%%%%%         INTRO.TEX
%%%%%
\section{Introduction}
\label{sec:intro}
The CDF and D$\O$ experiments \cite{Collaboration:2009je} at the
Fermilab Tevatron are continuously improving their Higgs mass limits,
and the ATLAS and CMS detectors at the CERN LHC are designed to
discover \cite{Aad:2009wy,Ball:2007zza} the standard Higgs in all of
the unexplored mass range, up to the high masses at which its raison
d'\^etre is lost. While an un-discovery would be momentous, we focus
here on the possibility that evidence resembling the standard
expectation is found.

Because the idea is so venerable, one may have grown insensitive to
how special a Higgs boson would be.  Its quantum numbers must be those
of the vacuum, which its field permeates. Its couplings to the
electroweak gauge bosons $W^\pm$ and $Z$ are proportional to their
masses, as are its couplings to quarks and leptons.  Any deviation
from the predicted quantum numbers or couplings of a putative Higgs
boson would have deep ramifications for particle physics.  An
experimental program for Higgs physics must be focused on the rigorous
determination of these fundamental quantities.

A Higgs boson discovery at the LHC will arise from excesses 
observed in one or more final states. Since the couplings
and partial widths of a SM Higgs boson are predicted as a function of
its mass, the size of any excess, the width of a reconstructed
resonance, or a comparison of different channels may immediately give
clues as to whether the putative new particle is consistent with a SM
Higgs boson.  Nevertheless, the true identity of the new particle will
at first remain ambiguous, until one has experimentally excluded other
possible assignments of quantum numbers and couplings. We shall refer
to these other possibilities as Higgs look-alikes (HLLs).

The purpose of this paper is to quantify the degree to which one can
discriminate a Standard Model Higgs boson from HLLs at, or close to,
the moment of discovery at the LHC. There is a vast literature about determining
Higgs properties from signals in a variety of final states (for a
review, see \cite{Djouadi:2005gi}), but this research mostly addresses
only the related question of whether it is possible {\it at all} to
determine Higgs quantum numbers and couplings at a hadron collider.
The current situation in this respect is similar to the LHC
experimental program for supersymmetry, where only recently are there
quantitative studies of the potential to discriminate supersymmetry
look-alikes at the moment of discovery
\cite{Datta:2005vx}-\cite{Hallenbeck:2008hf}.

Our study focuses on the so-called ``golden channel" for Higgs
physics, namely the Higgs decay $H \to ZZ^{*} \to
\ell^+_1\ell^-_1\ell^+_2\ell^-_2$, where $\ell^\pm_{1,2}$ denotes an
electron or a muon, and $Z^{*}$ denotes that one of the $Z$s may be
strongly off-shell. This channel has the advantage that the kinematics
of the Higgs and its decay products are fully reconstructible from a
completely leptonic final state. Approximately half of the events will
be $\mu^+\mu^-e^+e^-$, where all four leptons are easily
distinguishable, and even in the $4\mu$ and $4e$ final states all four
leptons can be distinguished by the requirement that one or both $Z$
bosons are reconstructed within an on-shell mass window. A
well-measured, four-body, closed kinematic final state provides many
independent observables for determining properties of the observed
resonance; thus this channel provides more information than e.g.~the
Higgs decay into two photons, where the photon
polarizations are not measured.

The branching fraction for the golden mode is small; example values
for a SM Higgs $\to ZZ^{*} \to 4\ell$ are 0.0011 for $m_H$$=$$200$
GeV/c$^2$, 0.0014 for $m_H$$=$$350$ GeV/c$^2$, and 0.00036 
for $m_H$$=$$145$ GeV/c$^2$ \cite{Djouadi:1997yw}.  Even for favorable Higgs
masses, this branching fraction is two orders of magnitude smaller than that for
semileptonic $H \to W^+W^- \to \ell \nu jj$, a channel that, though
hampered by large backgrounds, is also fully reconstructible up to a
two-fold ambiguity in the determination of the longitudinal neutrino
momentum \cite{Gunion:1986cc,Dobrescu:2009zf}. The golden mode
branching fraction is also smaller than that for the fully leptonic SM
Higgs decay $H \to W^+W^- \to \ell^+ \nu \ell^-
\bar{\nu}$. Nevertheless, for a wide range of SM Higgs masses, this
mode is a promising discovery channel and would, in any event, be
populated at or around the time of a putative discovery in a different
channel.
 
We factorize the HLL problem into observables related to
production and observables related to decay. In this paper we perform
a systematic analysis including all of the information from the
putative Higgs decays, leaving the analysis of Higgs versus HLL
production to later work. While this factorization of production and
decay is not completely clean, we show that the resulting
model-dependent uncertainty introduced into the decay analysis is
small. A full analysis will include production information and could
produce stronger results than those presented here, since large cross
section differences are expected between SM Higgs production and the
production of many Higgs look-alikes.  However, including Higgs
and HLL production also introduces new theoretical and measurement
uncertainties involving associated hadronic jets and the parton
distribution functions that describe the initial state.

One advantage of focusing only on Higgs decay in the four-lepton final
state is that we can perform a realistic study without resorting to
full simulation of a detector. This is demonstrated in Section
\ref{sec:ana}, where we parametrize the relevant efficiencies,
resolutions and acceptances for an LHC detector.  Because both the
ATLAS and CMS detectors in general measure muons and electrons with
exquisite precision, the resolution and efficiency for detecting the
four leptons can be significantly degraded with no impact on our
results.

This is not to say that detector effects are not important. We will
show that the finite phase space acceptance of any LHC detector has
strong effects on the HLL analysis, causing a detector-induced
sculpting of the angular distributions used for HLL discrimination. We
demonstrate that these effects must be accounted for in order to avoid
serious biases in the characterization of a Higgs signal.

Our analysis depends on five distinct angles that describe the $H \to
ZZ^{*} \to 4\ell$ decay process. In the case where one of the $Z$
bosons is strongly off-shell, the SM Higgs versus HLL decays also
differ in their dependence on the reconstructed $Z^*$ invariant mass.
Because we are interested in HLL discrimination with small data
samples, at or near the moment of discovery, we need to use all of the
decay information in the events, including not just the distributions
but also the correlations between all five (or six) of the relevant
observables.

In the same spirit, we disentangle the Standard Model $ZZ$ background
from the putative Higgs signal using the $_sPlots$ technique
\cite{sPlots}.  This produces an effectively background-subtracted
data sample where, instead of making stringent requirements that
reduce the signal yield available for characterization, we reweight
the selected events according to how likely each event is considered
to be signal by the fit, keeping the normalization to the signal yield
found in the search.

Previous analyses of the Higgs golden mode decay properties have
examined the dependence on some of the relevant angular distributions
\cite{Dell'Aquila:1985ve}-\cite{Barger:1993wt} and have shown the
potential for LHC measurements to discriminate a SM Higgs from
look-alikes with different spin and parity assignments or $CP$
properties
\cite{Djouadi:2005gi},\cite{Soni:1993jc}-\cite{Cao:2009ah}.
However, none of these studies utilized all of the decay information
in the events, and all of them have ignored the effects of detector
phase space sculpting of the angular distributions.

In our analysis we compare a SM Higgs signal to a variety of Higgs
look-alikes.  We consider the most general Lorentz invariant
couplings of a massive, spinless
boson to $ZZ$ or $ZZ^*$; this corresponds to gauge-invariant couplings
up to dimension six.  Some of the corresponding HLLs can be considered
as  modifications of the SM Higgs properties via $P$ or $CP$
violation or Higgs compositeness.  Another spin 0 HLL corresponds to a
new massive pseudoscalar, a particle occurring in models with extended
Higgs sectors such as supersymmetry.

Our HLL analysis also includes the most general couplings
of a massive neutral spin 1 boson to $ZZ$ or
$ZZ^*$.  The off-shell case has not been presented before, to our
knowledge.  A spin 1 HLL is a special case of what is usually denoted
as a $Z^{\prime}$ vector boson. The spin 1 part of our results is then
also part of a $Z^{\prime}$ look-alike analysis, which is interesting
in its own right \cite{Keung:2008ve}.

We also discuss as one of our HLLs a massive spin 2 resonance coupling
to the $ZZ$ energy-momentum tensor, not necessarily with the
universality of a graviton-like coupling. Although universally-coupled
massive gravitons are already experimentally excluded in the relevant
mass range \cite{Davoudiasl:2009cd}, general spin 2 HLLs are a natural
example of our study of spin discriminations.

In Section \ref{sec:golden} we define our notation for the observables
of the four-lepton final state.  Section \ref{sec:couplings} contains the
general gauge and Lorentz invariant couplings of an HLL to $ZZ$ or
$ZZ^*$, with a discussion of other symmetry properties.  We describe
in Section \ref{sec:ana} event generation, detector simulation, and
the construction of effectively background-subtracted samples using
$_sPlots$; here also we show the sculpting of the angular
distributions and correlations by the finite phase space acceptance of
the detector.  In Section \ref{sec:stat} we describe our statistical
approach to HLL discrimination using hypothesis testing with
likelihood ratios.  We demonstrate in Section \ref{sec:disc} the
consistency of our methods by applying them to the discrimination of
signal from SM $ZZ$ background.  In Section \ref{sec:hllresults} we
detail many examples quantifying our ability to discriminate a SM
Higgs from a variety of HLLs, showing in each case the expected
discrimination significance as a function of the number of signal
events; we use benchmark Higgs masses of 145, 200, and 350 GeV/$c^2$.
We summarize, in Section \ref{sec:LHCpotential}, our results and
outlook for further improvements. Here we explicitly quantify the
extent to which our expected discrimination significance would be degraded by
using a less complete or less rigorous analysis.

\section{The golden channel}\label{sec:golden}
\begin{figure}[htb!]
\begin{center}
\includegraphics[width=0.45\textwidth]{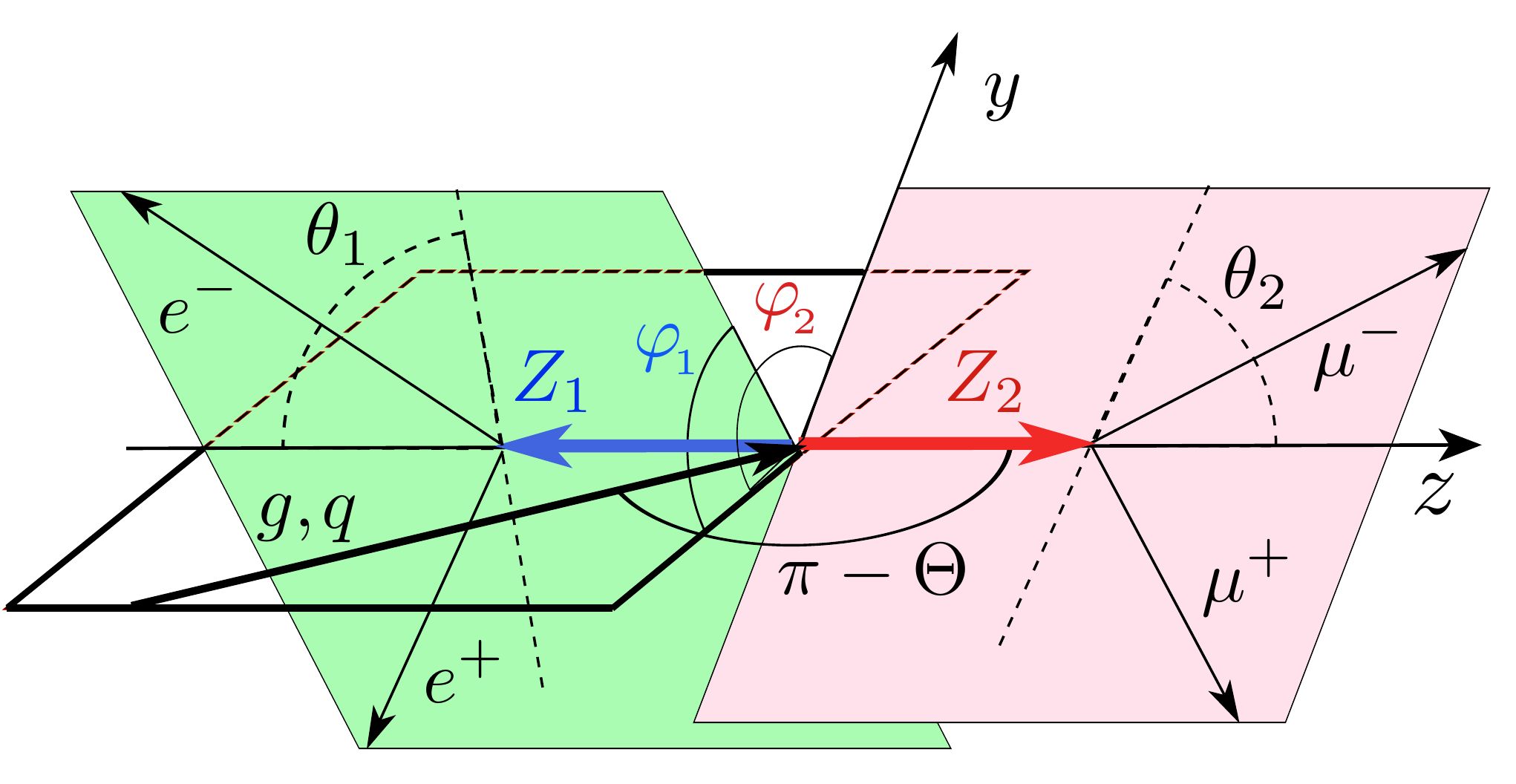}
\caption{The Cabibbo-Maksymowicz angles \cite{CMvariables} in the
  $H \to ZZ$ decays.\label{figangles}}
\end{center}
\end{figure}

We are interested in the case of a SM Higgs boson, or a Higgs
look-alike, decaying via $ZZ$ or $ZZ^*$ into a four-lepton final
state.  We will denote the putative Higgs and its mass by $H$ and
$m_H$, regardless of whether it is a SM Higgs or a look-alike. This
notation is also used to describe background events, where the
four-lepton object is treated as a Higgs or HLL in the sense that
$m_{H}$ stands for $m_{4\ell}$. Since the events are fully
reconstructible the lab frame kinematics of the candidate $H$
particles are known: their transverse momentum $p_T$, pseudorapidity
$\eta$, and azimuthal angle. These three variables define the
direction and boost from the lab frame to the $H$ rest frame. All
other observables can then be defined with respect to the $H$ rest
frame, as illustrated in Fig.~\ref{figangles}.

The $H$ azimuthal angle plays no physical role,
while the $p_T$ and $\eta$ distributions influence the way the detector
selects events,  sculpting the distributions of the final-state lepton's directions
and energies. Once an event is boosted back to the $4\,\ell$ rest-system
(the rest system of the two initial-state fusing partons),
the memory of $p_T$ and $\eta$ is lost, modulo these phase space
acceptance effects.

In the approximation that the final-state leptons are massless, 12
observables are measured per event. Since all 12 are well-measured
there is no experimental reason not to re-express these in terms of
whatever combinations most naturally capture the underlying
physics. Thus we choose four observables to be $m_H$ and the three
production observables just described that define the $H$ rest
frame. The remaining eight observables are taken to be the two
reconstructed masses of the $Z$ bosons together with six decay angles
defined with respect to the $H$ rest frame.

In the $H$ rest frame the reconstructed $Z$ bosons are back-to-back.
We label these bosons as $Z_1$, $Z_2$ and take the direction of $Z_2$
as defining the positive $z$-axis. Because of Bose symmetry, the
labeling is arbitrary; in the case of an $e^+e^-\mu^+\mu^-$ final
state we will follow the literature \cite{Buszello:2002uu} and choose
$Z_2$ to be the $Z$ boson that decayed to muons. We then adopt the
additional convention that the transverse direction of the $\mu^-$
lies along the positive $y$-axis; thus the $Z_2$ decay leptons lie in
the $y$-$z$ plane.

With the above choices, the reconstructed $Z$ boson masses $m_1$ and
$m_2$ also define the longitudinal boosts from the $H$ rest frame to
the rest frames of the decaying $Z_1$ and $Z_2$ bosons. The boost
parameters are given by \bea \gamma_1 &=& \frac{m_H}{2m_1}\left(
  1+\frac{m_1^2-m_2^2}{m_H^2}\right)
\;,\\
\gamma_2 &=& \frac{m_H}{2m_2}\left( 1-\frac{m_1^2-m_2^2}{m_H^2}\right)
\; .  \eea We let $\theta_1$, $\varphi_1$ denote the $\ell^-_1$ decay
angles in the $Z_1$ rest frame, while $\theta_2$, $\varphi_2$ denote
the $\ell^-_2$ decay angles in the $Z_2$ rest frame.

There are two additional angles $\Theta$, $\Phi$ defining the
direction of the initial state partons as reconstructed in the $H$
rest frame. For a gluon-gluon initial state these angles measure a
rotation from the $z$-axis defined above to the direction of the
initial state gluon with positive $z$-component of momentum. For
quark-antiquark ($q\bar{q}$) initiated production of an HLL we have
the problem that we do not know event-by-event which proton
contributed the antiquark; this is resolved by symmetrizing the
expected angular distributions under the replacement cos$\,\Theta \to
-$cos$\,\Theta$.

As expected, one combination of the three azimuthal angles $\Phi$,
$\varphi_1$ and $\varphi_2$ is physically redundant. We take advantage
of this fact to make the replacements $\varphi_1 \to \Phi + \phi$,
$\varphi_2 \to \Phi$. Thus $\phi$ then represents the azimuthal
rotation between the $Z_2$ and $Z_1$ decay planes.

In summary, the 4-momenta of the process $gg \to H \to Z_1Z_2 \to
\ell^-_1\ell^+_1\ell^-_2\ell^+_2$ are explicitly parametrized in the
$H$ rest frame as \bea p_{g_2} &=& \frac{m_H}{2}\; (\; 1,S\,{\rm
  cos}\,\Phi, S\,{\rm sin}\,\Phi,\; C)
\;,\nonumber\\
p_{g_1} &=& \frac{m_H}{2}\; (\; 1,-S\,{\rm cos}\,\Phi, -S\,{\rm
  sin}\,\Phi,\; -C)
\;,\nonumber\\
k &=& m_H\, (\;1,\;0,\;0,\;0)
\;,\nonumber\\
p_2 &=& m_2\, (\gamma_2,\; 0,\; 0, \beta_2\gamma_2 )
\; ,\nonumber\\
p_1 &=& m_1\, (\gamma_1,\; 0,\; 0, -\beta_1\gamma_1 )
\; ,\nonumber\\
p_{\ell^-_2} &=& \frac{m_2}{2}\,(\gamma_2(1+\beta_2 c_2),\; 0,\; s_2,
\gamma_2(\beta_2+c_2) )
\; ,\\
p_{\ell^+_2} &=& \frac{m_2}{2}\,(\gamma_2(1-\beta_2 c_2),\; 0,\; -s_2,
\gamma_2(\beta_2-c_2) )
\; ,\nonumber\\
p_{\ell^-_1} &=& \frac{m_1}{2}\,(\gamma_1(1+\beta_1 c_1),\; -s\;s_1,\;
-c\;s_1, -\gamma_1(\beta_1+c_1) )
\; ,\nonumber\\
p_{\ell^+_1} &=& \frac{m_1}{2}\,(\gamma_1(1-\beta_1 c_1),\; s\;s_1,\;
c\;s_1, -\gamma_1(\beta_1-c_1) ) \; .\nonumber \eea Here $k$ denotes
the 4-momentum of $H$, while $p_1$, $p_2$ are the 4-momenta of
$Z_1$, $Z_2$.  We used the condensed notation
$C,S$$=$cos$\,\Theta$, sin$\,\Theta$, $c,s$$=$cos$\,\phi$, sin$\,\phi$,
$c_1,s_1$$=$cos$\,\theta_1$, sin$\,\theta_1$, and
$c_2,s_2$$=$cos$\,\theta_2$, sin$\,\theta_2$.

Of the five relevant angles, $\Theta$ and $\Phi$ are $Z$-pair
production angles, while the remaining three are $4\ell$ production
angles. We will use the notation \bea
\vec{\Omega} &=& \{\Phi,\,{\rm cos}\, \Theta\}\; ,\nonumber\\
\vec{\omega} &=& \{\phi,\,{\rm cos}\,\theta_{1},\,{\rm cos}\,\theta_{2}\}\; .
 \label{eq:omegas} \eea For a SM
Higgs, the distributions in $\Theta$ and $\Phi$ are flat if we ignore
the phase space acceptance effects inherent in any experimental
analysis.  In previous studies these two angles have typically been
integrated over.

Although we have tried to conform to the literature in our
parametrization of the decay angles, we note that the literature
itself is divided over the choice of which decay plane orientation
corresponds to $\phi$$=$$0$ rather than $\phi$$=$$\pi$.  We conform to the
convention of Buszello et al.~\cite{Buszello:2002uu}, which is opposite
to that of Djouadi~\cite{Djouadi:2005gi} and 
Bredenstein et al.~\cite{Bredenstein:2006rh}.

The decay amplitudes defined in the next section depend on two
combinations of the boost parameters $\gamma_1$ and $\gamma_2$,
defined by \bea
\gamma_a &=& \gamma_1\gamma_2(1 + \beta_1\beta_2)\; ,\\
\gamma_b &=& \gamma_1\gamma_2(\beta_1 + \beta_2)\; , \eea which are in
fact just the cosh and sinh of the rapidity difference of $Z_2$ and
$Z_1$, such that \bea \gamma_a^2 - \gamma_b^2 = 1 \; .  \eea More
explicitly, we have \bea
\label{eqn:agamident}
\gamma_a = \frac{1}{2m_1m_2}\left( m_H^2 - (m_1^2+m_2^2) \right) \; .
\eea

%%%%%%%%%%%%%%%%%

\section{Couplings and angular distributions}
\label{sec:couplings}

\subsection{General couplings to $\mathbf{ZZ^*}$}
\label{generalcouplings}

The vertex Feynman rules for the most general coupling of a spinless particle to
the polarization vectors $\epsilon_1^\mu$ and $\epsilon_2^\alpha$ of two $Z$s 
of  four-momenta $p_1$ and $p_2$ are given by the expression:
  \begin{equation}
  L_{\mu \alpha }= 
  X\, g_{\mu \alpha } -(Y+i\, Z)\, {k_{\alpha } k_{\mu }\over M_Z^2}
  + (P+i \,Q) \,\epsilon _{\mu \alpha }{p_1p_2\over M_Z^2}\, ,
  \label{generalscalar}
  \end{equation}
where we have suppressed repeated indices in the contraction
of the four-index $\epsilon$ tensor,
$k$$=$$p_1+p_2$ and only Lorentz-invariance has been assumed. The 
dimensionless form factors
$X$ to $Q$ are functions of $k^2$ and $p_1\cdot p_2$ which, with no loss of generality,
can be taken to be real (but for their absorptive parts, expected to be perturbatively
small). The rescalings by $1/M_Z^2$ are just for definiteness, since the
true mass scale of the underlying operators is as yet unspecified.
In practice we also remove an overall factor of $igM_Z/$cos$\,\theta_W$, so
that $X$$=$$1$ corresponds to the tree level coupling of a SM Higgs boson.

Similarly, the most general vertex describing the coupling
of a spin $J$$=$$1$ particle to two Z-polarizations (indices $\mu$ and $\alpha$,
momenta $p_1$ and $p_2$, respectively) and to its own 
polarization (index $\rho$) is:
 \bea
 &&\hspace*{-27pt}
L^{\rho \mu \alpha } =
 X \left(g^{\rho
   \mu }\, p_1^{\alpha }\hspace*{-2pt}+\hspace*{-2pt}g^{\rho \alpha }\,
   p_2^{\mu }\right)+
       (P\hspace*{-2pt}+\hspace*{-2pt}i\, Q)\, \epsilon ^{\rho \mu \alpha}(p_1\hspace*{-2pt}-\hspace*{-2pt}p_2),
       \label{generalvector}
 \eea     
 again with $X$, $P$ and $Q$ real.

The most general parity-conserving vertex describing the coupling
of a $J$$=$$2^{+}$ particle of polarization tensor $\epsilon^{\rho\sigma}$
to our two vector bosons is:
 \begin{eqnarray}  
L^{\rho\sigma \mu \alpha } = &&
 X_0\,m_H^2\,g^{\mu\rho}\,g^{\alpha\sigma}\nonumber\\
&& 
 +(X_1+i\,Y_1)\left(p_1^\alpha\, p_2^\rho\,g^{\sigma\mu}
 +p_1^\rho \, p_2^\mu\,g^{\sigma\alpha}\right)\nonumber\\
 &&
+ (X_2+i\,Y_2)\; p_1^\rho\,p_2^\sigma\,g^{\mu\alpha}  ,
       \label{generaltensor}
 \end{eqnarray} 
where we have dropped contributions that have more than
two derivatives or are odd under parity, and    
again with all coefficients real. 
The special case of tree level graviton-like couplings corresponds to
\begin{equation}
X_0 = -\frac{1}{2}\kappa\;, \quad
X_1=\kappa \;, \quad
X_2 = -\kappa \;,
\end{equation}
with all other coefficients vanishing and
$\kappa$ an overall coupling strength.

These general couplings, with naive mass dimensions $d=3$, 4, and 5,
can arise from $SU(2)_L\times U(1)_Y$ invariant operators of dimension
5, 6, or higher. Since, for HLLs with non-vanishing weak charges, this parentage introduces model
dependence, we relegate it to a brief discussion in
Appendix $\ref{GIL}$.

%%%%%%%%%%%%%

\subsection{`Pure' cases of specified $J^{PC}$}

We specify in this section the results for four cases (scalar,
pseudoscalar, vector and axial vector) that would be `pure' in the
sense of having a single dominant term in their $HZZ$ couplings, which
we use to define their spin and parity. This allows one to illustrate
the mass and angular dependences of the predictions, setting the stage
for the later discussion of the impure cases for which $P$ and/or $CP$
are not symmetries of the theory, and to establish comparisons with
the existing literature (but for the $ZZ^*$ case for $J$$=$$1$, which we
have not found elsewhere).

The general expressions for the angular correlations in the $ZZ^*$
case (which includes $ZZ$ when the two $Z$ masses are fixed at $M_Z$)
are given in Appendices~\ref{generalformulae}
and~\ref{app:genspinone}, where 
\begin{equation}
  \eta\equiv {2\,c_v\,v_a\over (c_v^2+c_a^2)}\simeq 0.15,
  \label{eta}
\end{equation}
denotes the quantity arising from the SM couplings of the $Z$ bosons to the final state
leptons.

\subsubsection{The standard Higgs, $J^{PC}=0^{++}$}

The tree level SM coupling of the Higgs to two $Z$'s of polarisation
$\epsilon_1$ and $\epsilon_2$ is $\propto \epsilon_1\!\cdot\!
\epsilon_2$, see Eq.~(\ref{generalscalar}).  The angular distribution
of the leptons in $H\to ZZ\to 4\,l$ decay, for on or off-shell $Z$'s
of mass $m_1$ and $m_2$, is: \bea && {d\Gamma [0^+] \over dc_1\,dc_2
  \,d\phi} \propto m_1^2\,m_2^2 \,m_H^4 \bigl[ 1+c_1^2
c_2^2+(\gamma_b^2+c^2)s_1^2 s_2^2
\nonumber\\
&&\hspace*{10pt} + 2\gamma_a \,c \,s_1 s_2\, c_1 c_2 +2\eta^2 (c_1
c_2+\gamma_a\, c\, s_1 s_2) \bigr] \, .
\label{eq:standang}
\eea

\subsubsection{A pure pseudoscalar, $J^{PC}=0^{-+}$}

The coupling of a $J^{PC}$$=$$0^{-+}$ pseudoscalar to two $Z$'s of
polarisation $\epsilon_1$ and $\epsilon_2$ and four-momenta $p_1$ and
$p_2$ is proportional to $\epsilon[\epsilon_1, \epsilon_2,p_1,p_2]$, see
Eq.~(\ref{generalscalar}).  The angular distribution of the leptons in
its $ZZ\to 4\,l$ decay is:
   \begin{eqnarray}
 &&\hspace*{-20pt} 
 {d\Gamma [0^-] \over dc_1\,dc_2\,d\phi}
\propto m_1^4\,m_2^4\,\gamma _b^2 
\nonumber\\
&&   
\left( 1+ c_1^2 c_2^2
   -c^2 s_1^2 s_2^2 +2\, {\eta }^2\, c_1 c_2
    \right) \, .
   \label{purepseudoscalarP}
    \end{eqnarray}

\subsubsection{A pure vector, $J^{PC}=1^{--}$}

The coupling of a $J^{PC}$$=$$1^{--}$ vector particle of polarization
$\epsilon_H$ to two $Z$'s of polarisation $\epsilon_1$ and
$\epsilon_2$ and four-momenta $p_1$ and $p_2$ is $\propto \epsilon_H
\!\cdot\! \epsilon_1\;\epsilon_2\!\cdot\! p_1+ \epsilon_H \!\cdot\!
\epsilon_2\;\epsilon_1\!\cdot\! p_2$, see Eq.~(\ref{generalvector}).
Unlike for the scalar cases, the fully differential decay amplitude
depends nontrivially on the angles $\Theta$ and $\Phi$, representing
correlations between the helicities of the initial and final state
particles.  Assuming a quark-antiquark initial state this, in
principle, introduces two new parameters: the vector and axial
couplings of the (massless) quarks to the spin 1 HLL. However, once we
symmetrize over cos$\,\Theta \leftrightarrow -$cos$\,\Theta$,
reflecting our ignorance of which colliding proton contributes the
antiquark of the hard scattering, the dependence on these new
couplings disappears except for an overall factor. Performing this
symmetrization, we also introduce the notation
\be
m_d^2 \equiv m_1^2 - m_2^2
 \; ,
\ee
and find the angular distribution of the leptons in $H\to
ZZ^{*}\to 4\,l$ decay as follows:

 \begin{widetext}
% \onecolumngrid
 \bea
&&\hspace*{-15pt}
{d\Gamma [1^-] \over dC\, d c_1\,d c_2\, d\Phi \, d\phi} \propto
4 m_1^2 m_2^2  \gamma_b^2 \,
\Bigl[
S^2 s_1^2 s_2^2\, 
\bigl(
2 m_d^4 - m_H^2 
\bigr[
m_1^2  \cos{2(\Phi+\phi)}
+m_2^2  \cos{2\Phi}
\bigr]
\bigr)
\\
&&\hspace*{-30pt}
+m_H^2(1+ C^2)\bigl[
2m_2^2 s_1^2+2m_1^2 s_2^2-(m_1^2+m_2^2)s_1^2 s_2^2
\bigr]
+4   m_H m_d^2\, C\, S\, 
\bigl[
m_1 c_1\, s_1 s_2^2 \sin{\Phi+\phi} 
-m_2 c_2\, s_2 s_1^2  \s{\Phi}
\bigr]
\nonumber\\
&&\hspace*{20pt}
-2 m_H^2 m_1 m_2 s_1 s_2\, 
\bigl(
(1+C^2)  
(c_1 c_2 -\eta^2 )c
+S^2  (c_1 c_2 +\eta^2 )\cos{2\Phi+\phi}
\bigr)
\Bigr] \; .
\nonumber
 \eea
 \end{widetext}
%\twocolumngrid
 
\subsubsection{A pure axial vector, $J^{PC}=1^{++}$}

The coupling of a $J^{PC}=1^{++}$ axial vector particle of
polarization $\epsilon_H$ to two $Z$'s of polarisation $\epsilon_1$
and $\epsilon_2$ and four-momenta $p_1$ and $p_2$ is 
proportional to 
$\epsilon[\epsilon_H,\epsilon_1,\epsilon_2,p_1-p_2]$, see
Eq.~(\ref{generalvector}).  After the same symmetrization in
cos$\,\Theta$ described above, 
and introducing the notation
\bea
M_1^2 &\equiv& m_H^2 - 3m_1^2 -m_2^2 \; ,
\nonumber\\
M_2^2 &\equiv& m_H^2 - m_1^2 -3m_2^2 \; ,
\eea
the angular distribution of the final state leptons is
given by:

\begin{widetext}
\be
&&\hspace*{-15pt}
{d\Gamma [1^+] \over dC\, d c_1\,d c_2\, d\Phi \, d\phi} \propto
  m_H^2 S^2 s_1^2 s_2^2 \,
\bigl[
M_2^4 m_1^2\cos{2(\Phi+\phi)}+M_1^4 m_2^2 \cos{2\Phi}
\bigr]
%\nonumber\\
%&&\hspace*{0pt}
+8   m_1^2 m_2^2 m_d^4 S^2\,
\bigl[
\,c_1^2+c_2^2+s_1^2 s_2^2 s^2
+2\eta^2 c_1 c_2
\bigr]
\nonumber\\
&&\hspace*{60pt}
+m_H^2 (1+C^2) 
\bigl[
2M_1^4m_2^2s_1^2
+2M_2^4m_1^2s_2^2-(M_2^4 m_1^2+M_1^4 m_2^2)s_1^2 s_2^2
\bigr]
\\
&&\hspace*{60pt}
-8   m_H m_d^2m_1 m_2 C\, S\,
\bigl[
M_2^2 m_1 s_2\,  
\bigl(
 c_2 s_1^2 c \sin{\Phi+\phi}
+c_1( c_1 c_2+\eta^2 ) \s{\Phi}
\bigr)
\nonumber\\
&&\hspace*{150pt}
- M_1^2 m_2  s_1 \,
\bigl(
 c_1 s_2^2 c \s{\Phi}
+c_2( c_1 c_2+\eta^2 )\sin{\Phi+\phi}
\bigr)
\bigr]
\nonumber\\
&&\hspace*{60pt}
+2 m_H^2 M_1^2 M_2^2 m_1 m_2 s_1 s_2 
\bigl[
(1+C^2) ( c_1 c_2 
-\eta^2 ) 
c -S^2  ( c_1 c_2 
+\eta^2 )\cos{2\Phi+\phi}
\bigr]
 \; .
\nonumber
\ee
\end{widetext}

\subsubsection{A pure massive graviton, $J^{PC}=2^{++}$}

Since the general analysis of spin 2 coupling to
off-shell $Z$'s  is quite cumbersome, we
will only quote results for the example of
a positive parity spin 2 with graviton-like couplings produced
by gluon fusion and decaying
to two on-shell $Z$'s. Defining the on-shell ratio
$x \equiv m_H/M_Z$
%\be
%x \equiv \frac{m_H}{M_Z}
%\; ,
%\ee
and using the massive graviton formalism of \cite{Han:1998sg}, we obtain
the tree level angular distribution:
\begin{widetext}
\begin{eqnarray}
&&\hspace*{-45pt}
{d\Gamma [gg\to {\rm graviton}\to ZZ] \over dC\, d c_1\,d c_2\, d\Phi \, d\phi} \propto 
16  x^4 C^2+2 (x^4+16) S^4 
+s_1^2 s_2^2 [(x^4+16) S^4-4 x^2  (x^2+4) S^2+4x^4]
\nonumber\\
&&\hspace*{0pt}
+8x^2 S^2 \Bigl[ [2+S^2+(2-3S^2) c_2^2] s_1^2\, {\rm cos}(\Phi + \phi)^2
+ [2+S^2+(2-3S^2) c_1^2] s_2^2\, {\rm cos}^2\Phi \Bigr]
\nonumber\\
&&\hspace*{0pt}
+S^4 s_1^2 s_2^2 [x^4 \,{\rm cos}(2\Phi + \phi)^2+16\, c^2]
%\nonumber\\
%&&\hspace*{-30pt}
-(s_1^2+s_2^2)[(x^2+4)^2 C^4+2(3 x^4-16) C^2+(x^2-4)^2]
\nonumber\\
&&\hspace*{0pt}
%\nonumber\\
%&&\hspace*{-30pt}
+2 S^2 c_1 \, c_2\, s_1\, s_2\, \Bigl[ x^2 \,[2(x^2+4)-(x^2+12) S^2]{\rm cos}(2\Phi + \phi)
%\nonumber\\
%&&\hspace*{100pt}
+4\,[4 x^2-(3 x^2+4) S^2] c \Bigr]    
\; .
\end{eqnarray}
\end{widetext}
Note the cos$^4\,\Theta$ dependence characteristic of a spin 2 resonance.

%%%%%%%%%%%%%%%%%%

\subsection{Tests of symmetries}
\label{CPCB}

Now we discuss the behaviour of the $HZZ$ couplings under
various symmetries, including
$CP$ and Bose-Einstein statistics.
The discussion attempts to clarify
the literature on these issues.

Consider the $J$$=$$0$ case. The most general coupling of a
spinless particle to the polarization vectors $\epsilon_1$ and
$\epsilon_2$ of two $Z$'s is that of Eq.~(\ref{generalscalar}).  In
computing the ensuing $H\to ZZ^* \to 4\ell$ process one finds that the
$XP$ interference term is of the form:
\begin{eqnarray}
&&\hspace*{-10pt}
   {d\Gamma [0,{\rm {Todd}}] \over d c_1\,d c_2\,d\phi}
  \propto
  \nonumber\\ 
 &&\hspace*{-10pt}
  2 \,m_1^3\,m_2^3 \,m_H^2 \,\gamma_b\,s_1 \,s_2\,s 
  \left[ s_1
  \,s_2\,c+\gamma_a\,(c_1 c_2+ \eta^2)
  \right] \, ,
  \label{0Todd}
\end{eqnarray} 
where the term sin$\,\theta_1\,{\rm sin}\,\theta_2\,{\rm sin}\,\phi\propto \vec
p_{e^+}\cdot \vec p_{\mu^-} \times \vec p_{\mu^+}$.
By definition, this observable is $\,\tilde T$-odd: it changes sign as all
three-momentae are reversed (the tilde in ``$\,\tilde
T$-odd" emphasizes that past and future are not being
interchanged).

The Born approximation is, by definition, the result of squaring the
amplitude dictated by the Lagrangian to lowest order in its couplings:
a quadratic result, in our case, in any pair of the quantities $X$ to
$Q$ in Eq.~(\ref{generalscalar}).  To this order, a $\tilde T$-odd observable
must vanish if $CP$ is a symmetry, as shown in
\cite{Watson}. Thus, a non-vanishing $\tilde T$-odd observable such as that
of Eq.~(\ref{0Todd}) can only arise if $CP$-invariance is violated.

The $XQ$ interference term resulting from Eq.~(\ref{generalscalar})
is:
\begin{eqnarray}
  && {d\Gamma [0,{\rm Codd}] \over dc_1\,dc_2\,d\phi}
  \propto\nonumber\\ 
  &&\hspace*{-20pt}
   -2\, \eta\,
  m_1^3\,m_2^3 \,m_H^2\,\gamma_b\,[c_1+c_2]\, (1+c_1 c_2+\gamma_a\,s_1 s_2\, c)
  \, .
  \label{0Codd}
\end{eqnarray}
This term is $CP$ odd  and $\tilde T$-even, a combination not addressed
by the theorem quoted above. It is a
$C$-odd observable, in that it changes sign under the interchange of
$p_{e^+}\leftrightarrow p_{e^-}$ and $p_{\mu^+}\leftrightarrow
p_{\mu^-}$, tantamount to ${\rm cos}\, \theta_i \leftrightarrow -{\rm cos}\, \theta_i$
in our chosen notation. 

\subsubsection{Bose-Einstein statistics}

The general coupling, up to two derivatives, of a $J=1$ particle to two $Z$'s
is that of Eq.~(\ref{generalvector}). This is true whether or not the $Z$'s are
on-shell, which seems to be a point of confusion in the literature. Thus
for example \cite{Choi:2002jk}, whose authors were the first to emphasize
the importance of $M_{Z^*}$ as a discriminating variable,
contains extra
``off-shell" couplings, such as $g^{\mu\alpha}(p_1-p_2)^\rho$ and
$\epsilon^{\rho\mu\alpha}(p_1+p_2)$, that violate Bose symmetry and
vanish for two on-shell $Z$'s. However, Bose symmetry is a property manifest
at the Lagrangian level, and thus
independent of any on- or off-shell considerations.  The two $Z$'s
in an $H\to ZZ^*$ decay are described by the same bosonic $Z$ field,
whether or not they are on-shell, and they do not obey the laxer rules that
different particles ($Z\neq Z'$) would.

%%%%%%%%%%%%%%%%%

\subsection{Tests of compositeness }
\label{compo}

If the couplings of an HLL conserve $P$ and $CP$, but the object is not
point-like, there will be deviations from the standard $g_{\mu\nu}$
coupling to $Z$'s.  To lowest order in the dimensions of the
corresponding effective operators, these will be of two types. The
first is a non-vanishing $Y$ in Eq.~(\ref{generalscalar}), and the
second is a nontrivial form for $X$.  Barring large effects --quite
conceivable in a model with multiple SM Higgs-like fields-- deviations
in $X$ are much harder to limit or measure than a non-zero $Y/X$ which
is governed by the shapes of angular distributions. Contributions to
$Y$ can arise from gauge invariant operators of dimension 5 containing
a non SM-like spin 0 HLL (Appendix $\ref{GIL}$) or from higher
dimension operators containing the SM
Higgs~\cite{Hagiwara:1996kf}-\cite{Barbieri:1999tm}.

It is useful to introduce the notation $\tan\xi\equiv Y/X$.  In this
notation, the ``composite'' HLL angular distribution is of the form:
\begin{equation}
  d\Gamma_{C} = {\rm cos}^2\xi \,d\Gamma_{XX}
  +{\rm cos}\,\xi{\rm sin}\,\xi\,d\Gamma_{XY}
  +{\rm sin}^2\xi\,d\Gamma_{YY}\, ,
        \label{splusd}
    \end{equation}  
    where $d\Gamma_{XX}$ is the standard result of
    Eq.~(\ref{eq:standang}).  The interference term is:
\begin{eqnarray}
  &&\hspace*{-15pt}
  {d\Gamma_{XY} \over d c_1\,d c_2\,d\phi} \propto
  \nonumber\\ 
  &&\hspace*{-5pt}
   -2\, m_1^3\,m_2^3\, m_H^2\,\gamma_b^2\, s_1 \,s_2\, (c_1 c_2
  c+\gamma_a\, s_1 s_2 +\eta^2\,c) \, ,
  \label{interference}
\end{eqnarray} 
and the last term is:
\begin{equation}
  {d\Gamma_{YY} \over d c_1\,d c_2\,d\phi} \propto
  m_1^4\,m_2^4\,\gamma_b^4\, s_1^2 s_2^2 \; .
 \label{eq:YY}
\end{equation}   

Contrary to all of the other cases we study, the interference term in
this instance is between two operators whose $P$ and $C$ are
identical: the HLL is not point-like, but it is `pure' $0^{++}$.  As a
consequence, the angular distribution of the interference term is not
very different from that of the $XX$ and $YY$ terms and the
interference can, for certain values of $Y/X$, be very
destructive. This can be seen even at the level of the $H\to ZZ$
branching fraction, the integral of Eq.~(\ref{splusd}) over
cos$\,\theta_1$, cos$\,\theta_2$, and $\phi$:
\begin{equation}
\Gamma_{C}\propto m_1^2 m_2^2 \,[ 2 {\rm cos}^2\xi + \,(\gamma_a {\rm cos}\, \xi - m_1 m_2 \gamma_b^2  {\rm sin}\, \xi)^{2}]\, .
 \label{derivativeintegral}
\end{equation}

%\begin{eqnarray}
%&&\Gamma_{C}\propto
%m_1^2 m_2^2 \,[ 2 {\rm cos}^2\xi
%\\
%&& + \,(\gamma_a {\rm cos}\, \xi - m_1 m_2 \gamma_b^2  {\rm sin}\, \xi)^{2}]\, .
  % \nonumber
 %\label{derivativeintegral}
%\end{eqnarray}

If $\xi$ has a value close to the (mass-dependent) point of maximal
interference, the golden mode channel can be suppressed by a large
factor. For this to happen $X$ and $Y$ ought to be of the same order
of magnitude, signifying a low dynamical scale for a composite Higgs.

%%%%%%%%%%%%%%%%
%%%%%%
%%%%%%     ANA.TEX
%%%%%%
%%%%%%%%%%%%%%%%%%

\section{Analysis\label{sec:ana}}
In this section we describe the modeling of the detector effects and
the analysis strategy to extract an effectively pure sample of signal
events. We describe the Monte Carlo (MC) event generation and the
simulation of the detector response. We use parameterized
reconstruction resolutions and efficiencies based on the published CMS
performance results~\cite{CMStdr1}.  A similar study can be performed
with parameterizations based on the ATLAS detector.  We focus on the
four-muon ($4\mu$) final state, but the results can be generalized to
include final states with electrons. Since a four-lepton final state
is relatively ``clean'' in the LHC environment, we apply a loose event
selection and use a maximum likelihood (ML) fit technique to separate
the signal from the background. This maximizes the statistical power
and the possibility of characterizing the nature of the discovered
particle through the study of the multi-dimensional angular
distribution of the four leptons in the resonance rest frame.

\begin{figure}[htbp]
\begin{center}
\includegraphics[width=0.23\textwidth]{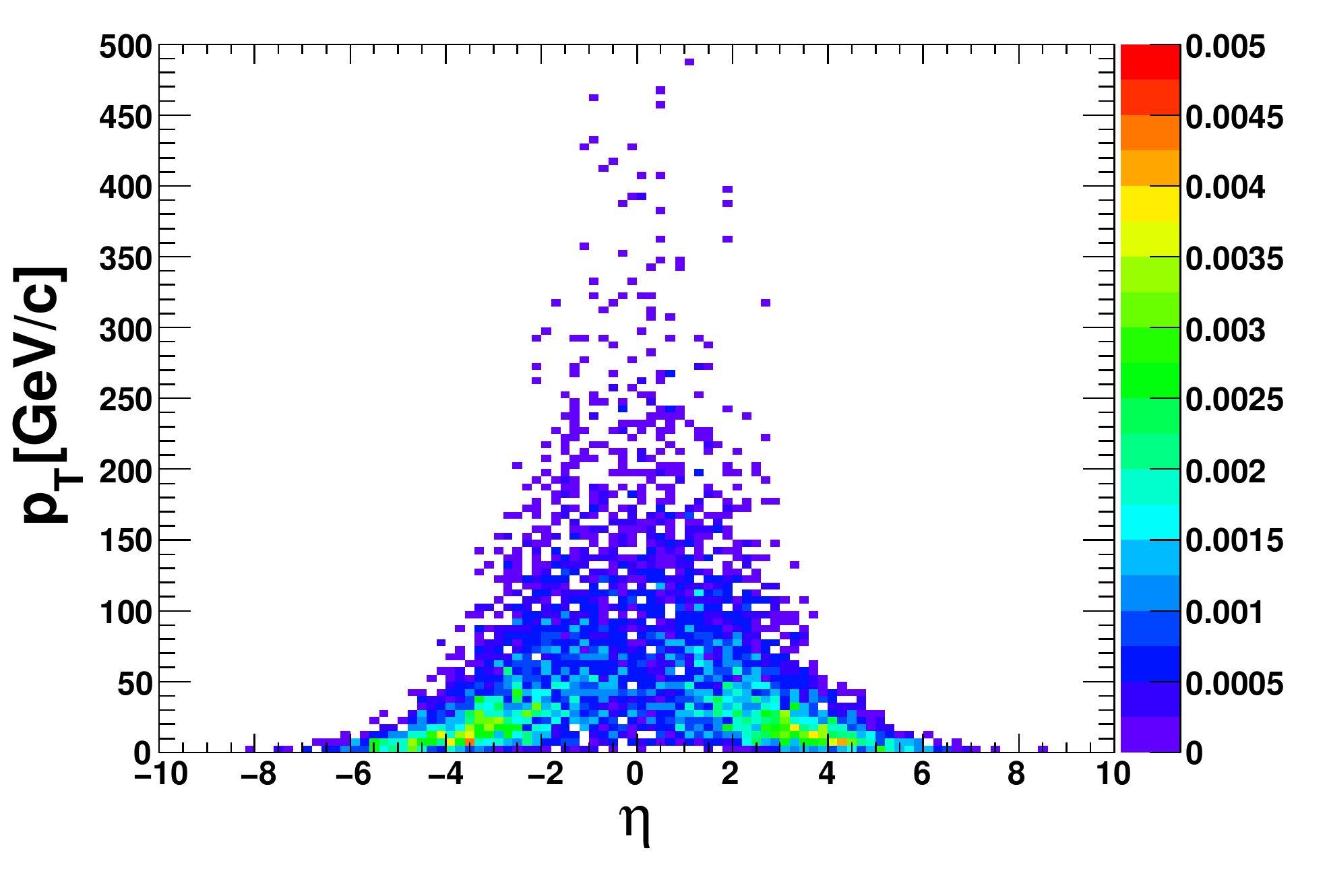}
\includegraphics[width=0.23\textwidth]{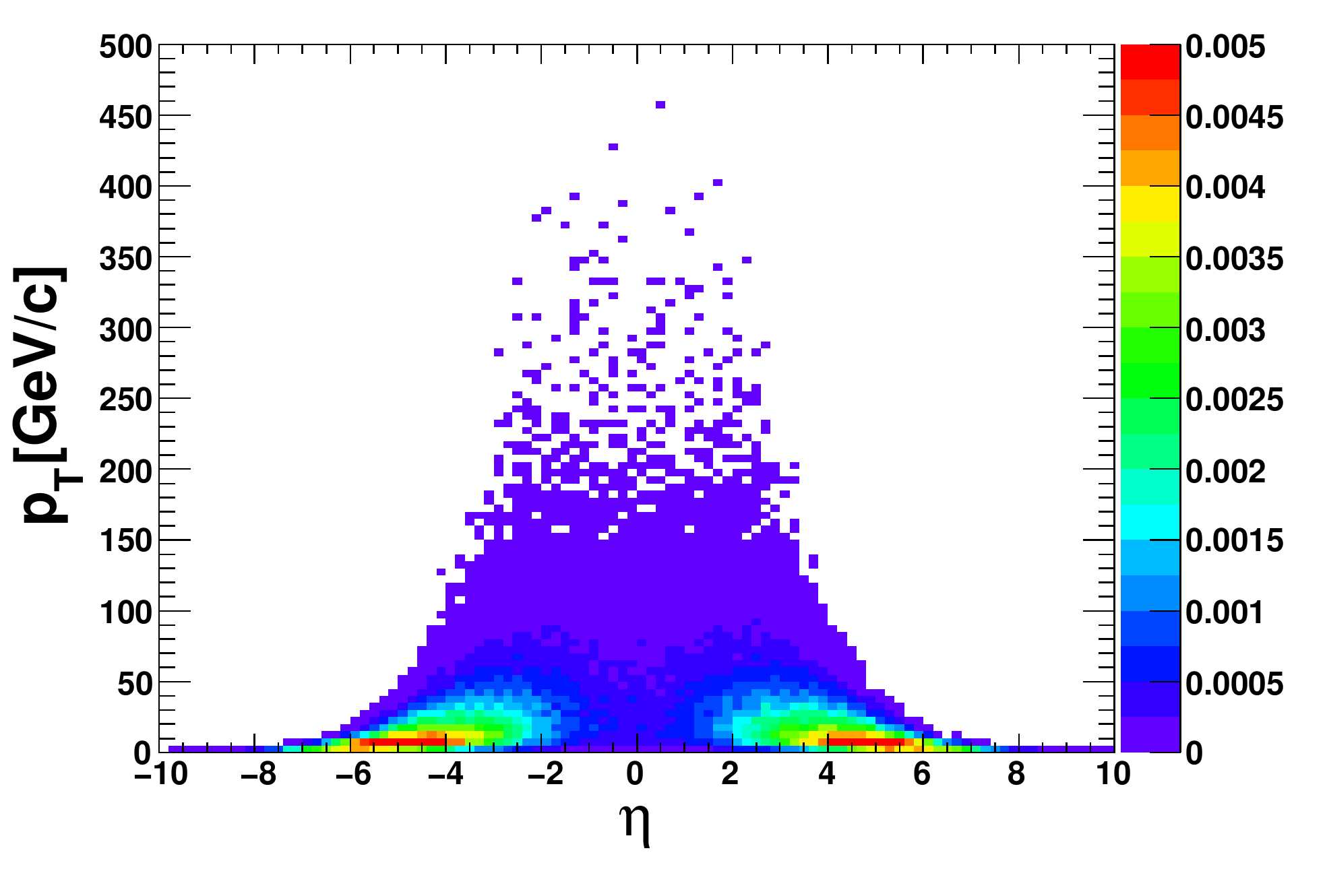} \\
\includegraphics[width=0.23\textwidth]{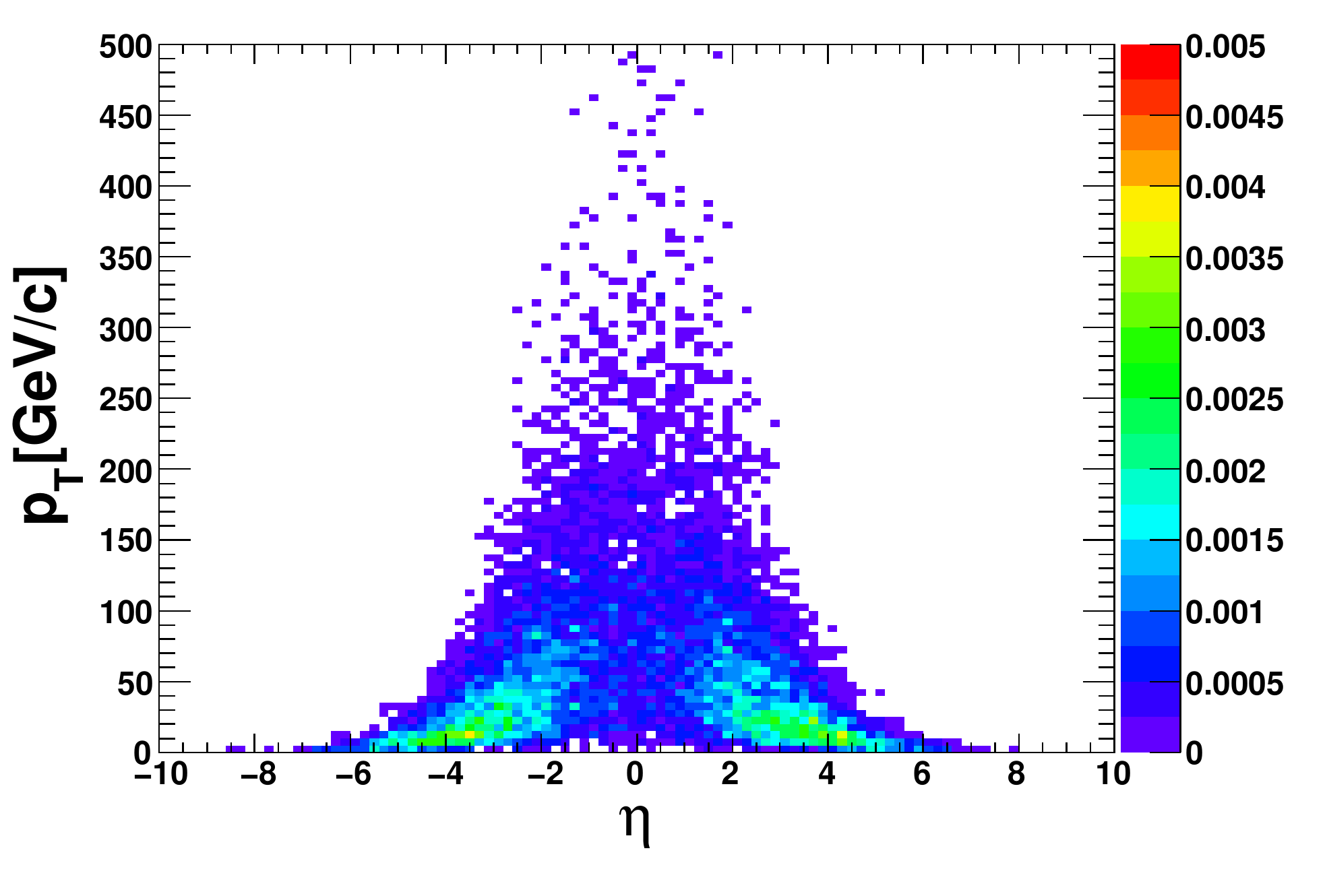}
\includegraphics[width=0.23\textwidth]{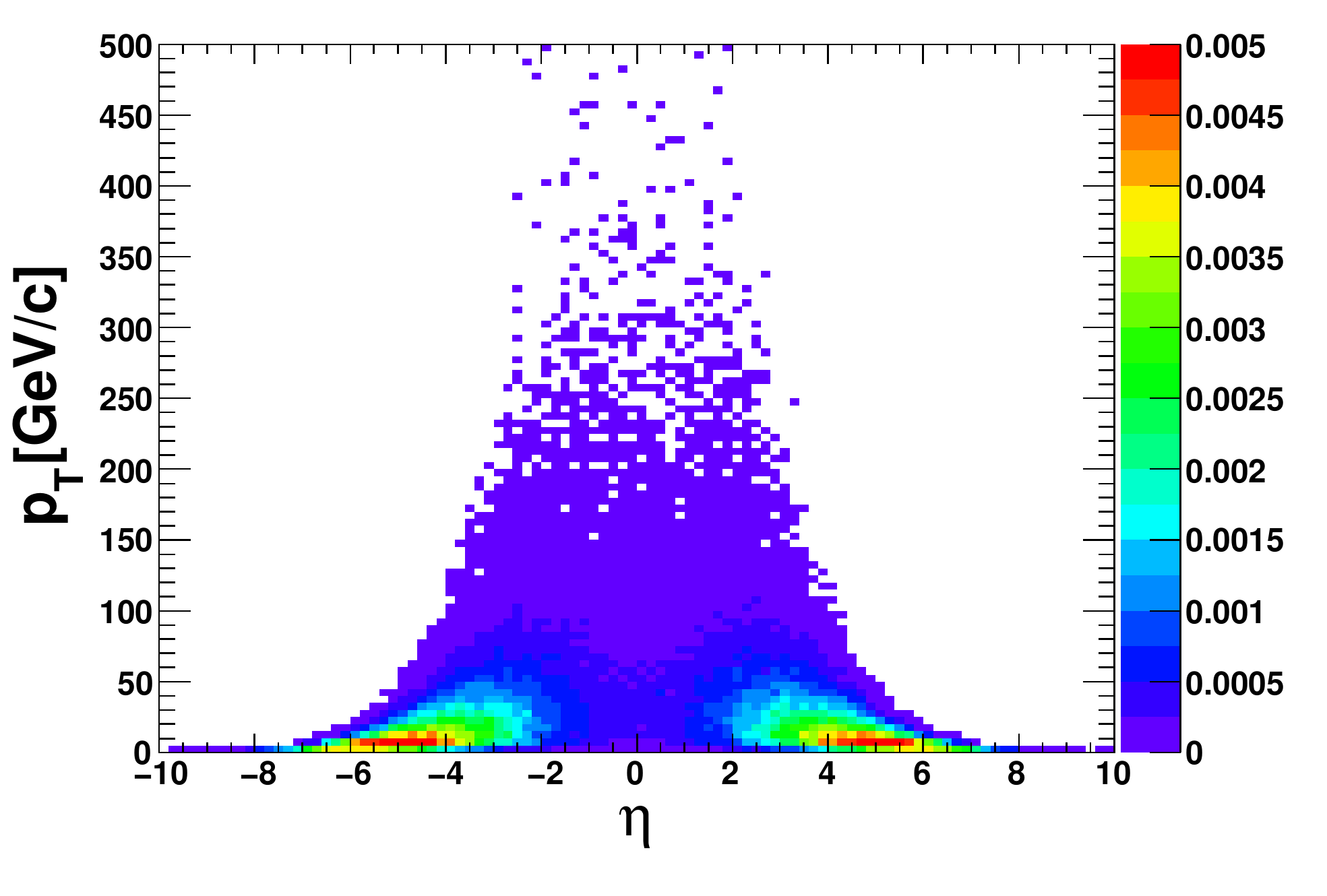} \\
\includegraphics[width=0.23\textwidth]{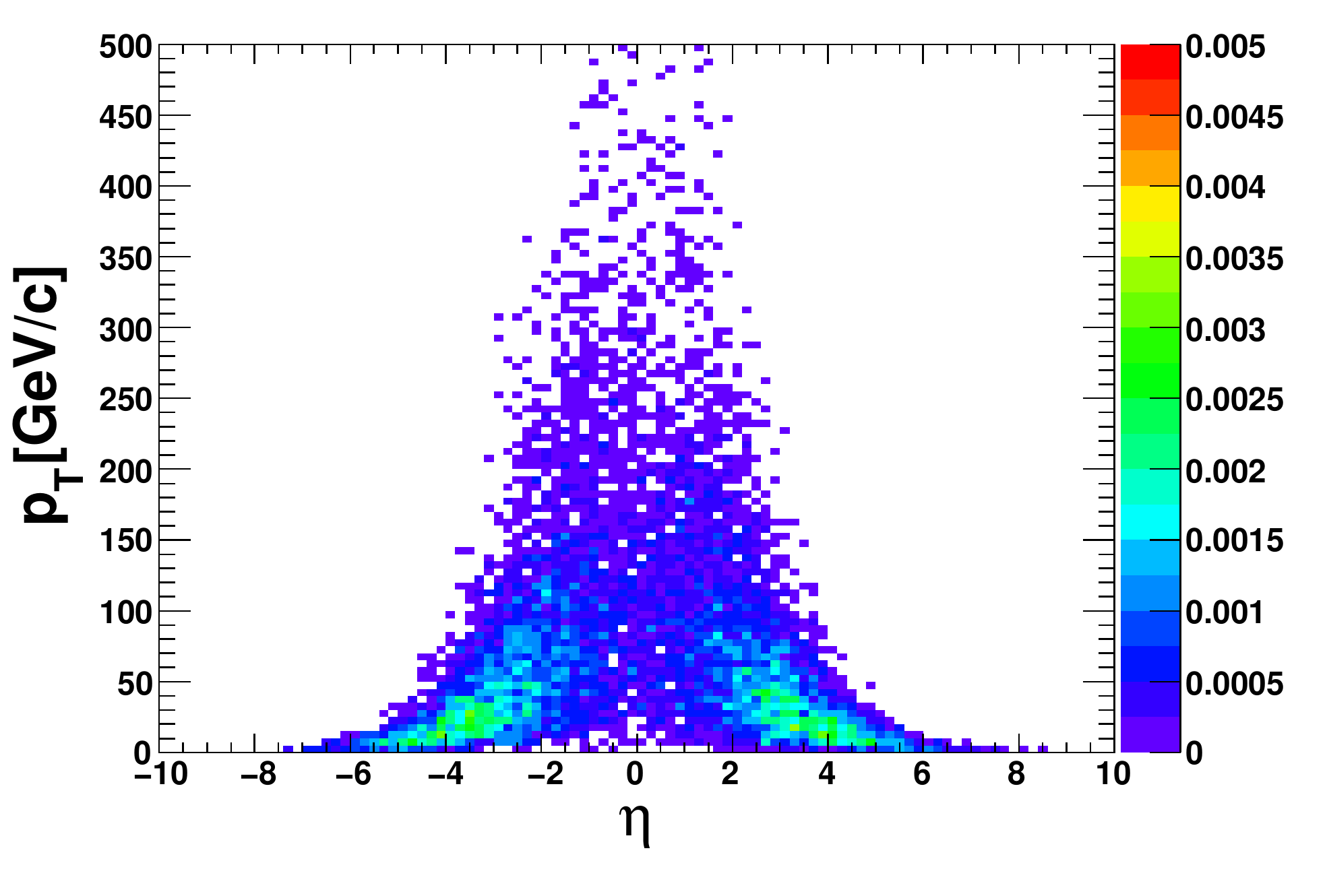}
\includegraphics[width=0.23\textwidth]{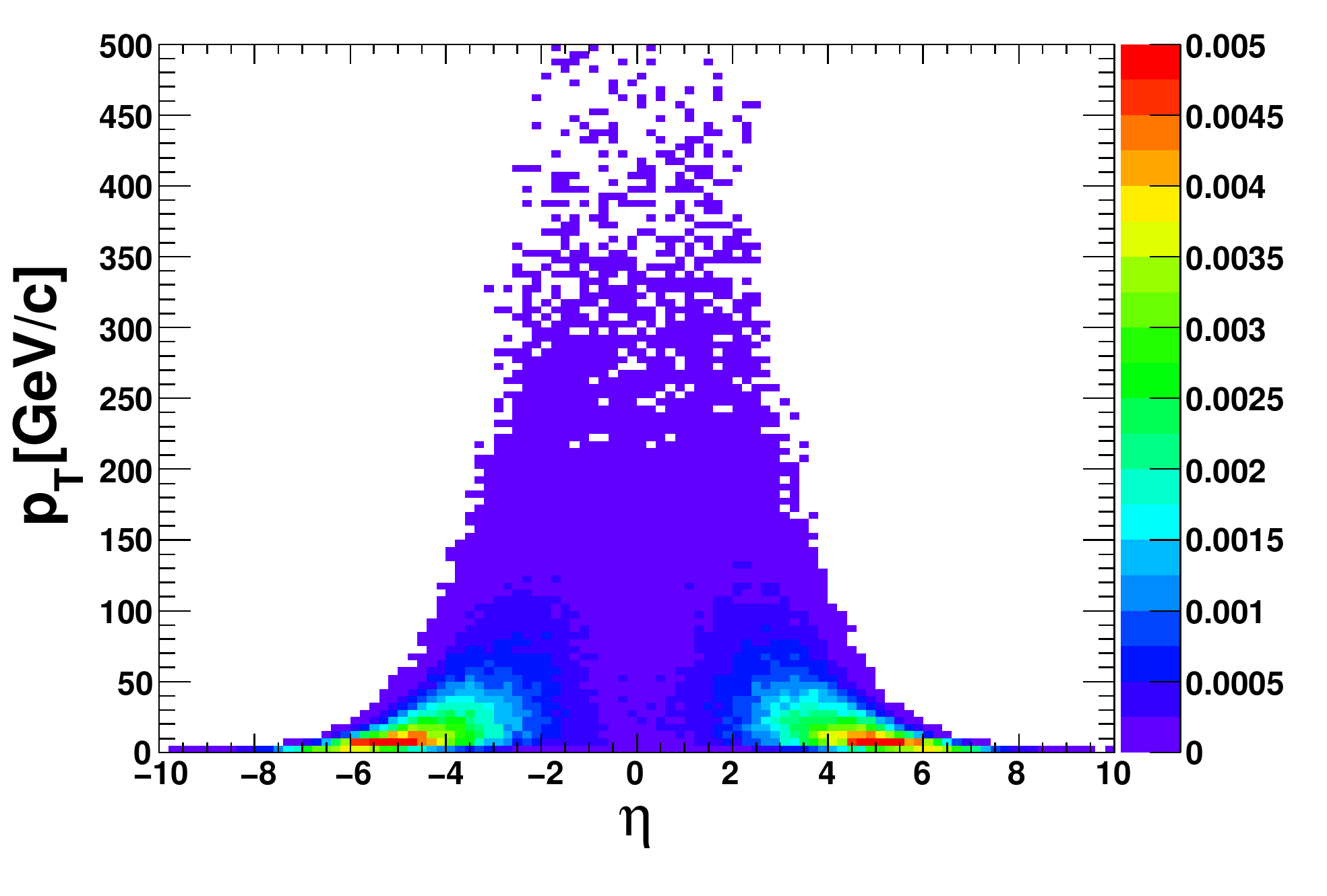}
\caption{2D $p_T$-$\eta$ {\it pdf} of a ${0^+}$ HLL resonance (left)
  and a $1^-$ one (right) for $\sqrt{s}$$=$$ 10$ TeV collisions, obtained
  using {\tt PYTHIA} and the CTEQL5 parton density functions and for
  $m_H$$=$145, 200, 350 GeV/c$^2$ (top, middle and bottom).
  \label{fig:etavsptScalarVector}}
\end{center}
\end{figure}

\subsection{Event generation}
The knowledge of the four-momenta of the leptons fully specifies the
information needed in this analysis. We generate the four-momenta of
the leptons from the five- or six-dimensional 
probability density functions ({\it pdfs}) of 
\bea 
\vec X &\equiv& \{\vec\omega, \vec\Omega\}\;\; \mathrm{for}\,ZZ\; , \nonumber\\
\vec X &\equiv& \{\vec \omega,\;\vec \Omega,\,M_{Z^*}\}\;\;{\rm for}\; ZZ^*\;,
\eea
where $\vec\Omega , \vec\omega $ are given in Eq.~(\ref{eq:omegas}).
The $\vec X$ quantities are generated in the rest frame of the
decaying resonance. The muons are then boosted to the laboratory frame,
and the detector effects (acceptance, efficiency and resolution) are
applied to the boosted momenta.  We use the azimuthal symmetry of the
LHC detectors to reduce the remaining kinematic degrees of freedom to
the knowledge of the $p_T$, $\eta$ and the invariant mass $m_{4\mu}$
of the $4\mu$ system. The $p_T$, $\eta$ for the signal is taken from a
two-dimensional {\it pdf} generated using {\tt MC@NLO}
~\cite{Frixione:2002ik}. We consider proton-proton collisions at
$\sqrt{s}$$=$$10$ TeV, and we model the parton distribution
functions (PDFs) using CTEQ5L~\cite{cteq}.

In this analysis we do not assume a specific signal production
mechanism and cross section, instead relying on the
discrimination provided by the angular distributions of the leptons in
the final state. Figure~\ref{fig:etavsptScalarVector} has the $p_T$
vs.~$\eta$ {\it pdfs} for a spin-0 and a spin-1 HLL.  As discussed in
Section~\ref{sec:intro}, for all the signal generation we use the
$p_T$ vs.~$\eta$ {\it pdfs} of the scalar.
For the SM $ZZ$ background the $p_T$, $\eta$ and $m_{4\mu}$ are taken
from a three-dimensional {\it pdf} generated using the {\tt
  PYTHIA}~\cite{PYTHIA} leading-order MC generator.  The momenta of
the four muons in the rest frame of the $ZZ^{(*)}$ system as a
function of $m_{4\mu}$ are generated according to the theoretical
distributions.

\begin{figure}[hbp]
\begin{center}
\includegraphics[width=0.42\textwidth]{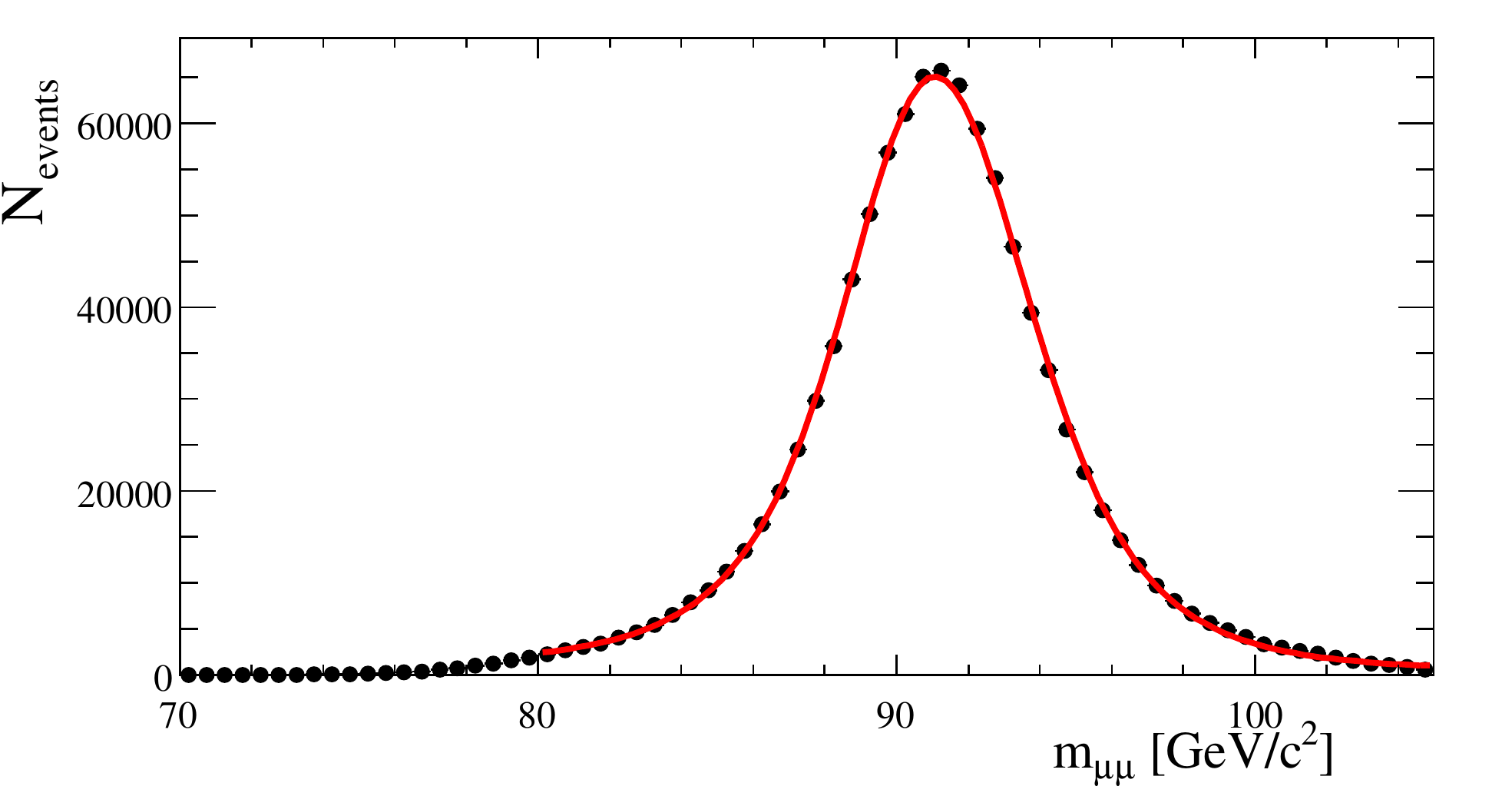}
\caption{ Distribution of the dimuon invariant mass for a
  sample of signal $H \to ZZ$ events, generated using our very-fast
  muon simulation. The parameters of the superimposed fit are
  extracted from~\cite{candle}. 
  \label{fig:Zpdf}}
\end{center}
\end{figure}

\subsection{Detector emulation and event selection}\label{detector}
Muon reconstruction efficiency and resolution are parameterized as a
function of the muon $p_T$ and $\eta$ according to~\cite{CMStdr1},
where the muon reconstruction efficiency is close to $100 \%$ for
muons with $p_{T}\geq 10$ GeV/$c$ and $|\eta|\leq 2.3$, corresponding
to the event selection in our analysis.  The reconstruction
efficiency is applied through a hit-or-miss technique.  For muon
candidates accepted by the efficiency filter, the reconstructed
momentum is determined by applying Gaussian smearing functions to the
true $p_T$, $\eta$ and $\phi$ with $p_T$- and $\eta$-dependent
resolutions. We verified the goodness of our {\it very-fast} muon
simulation by comparing the parameters of the fit of the $Z$ invariant
mass distribution obtained in our analysis, see
Fig.~\ref{fig:Zpdf}, with the corresponding ones from a published
full-simulation analysis~\cite{candle}.

\begin{figure}[htbp]
\begin{center}
\includegraphics[width=0.23\textwidth]{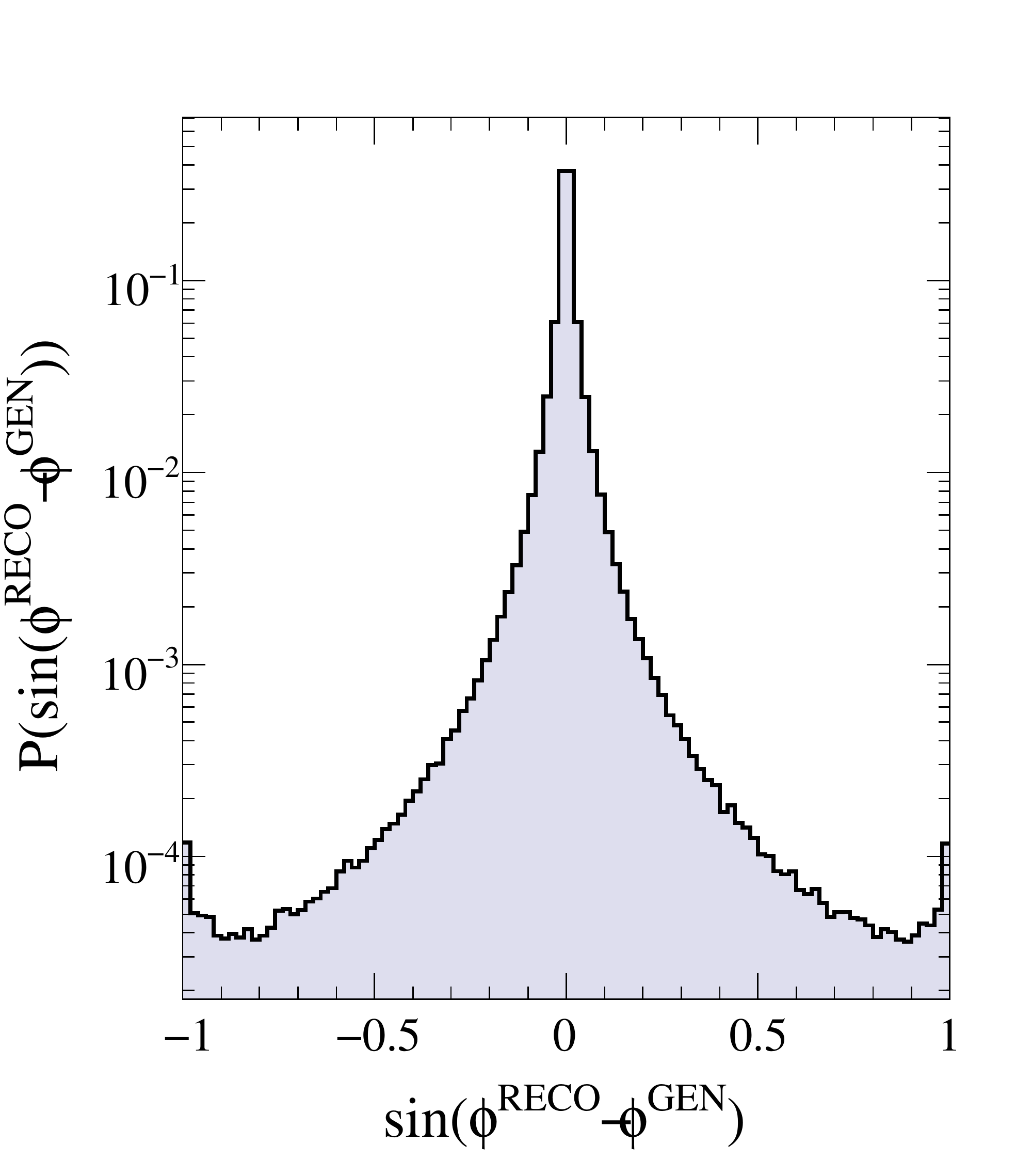}
\includegraphics[width=0.23\textwidth]{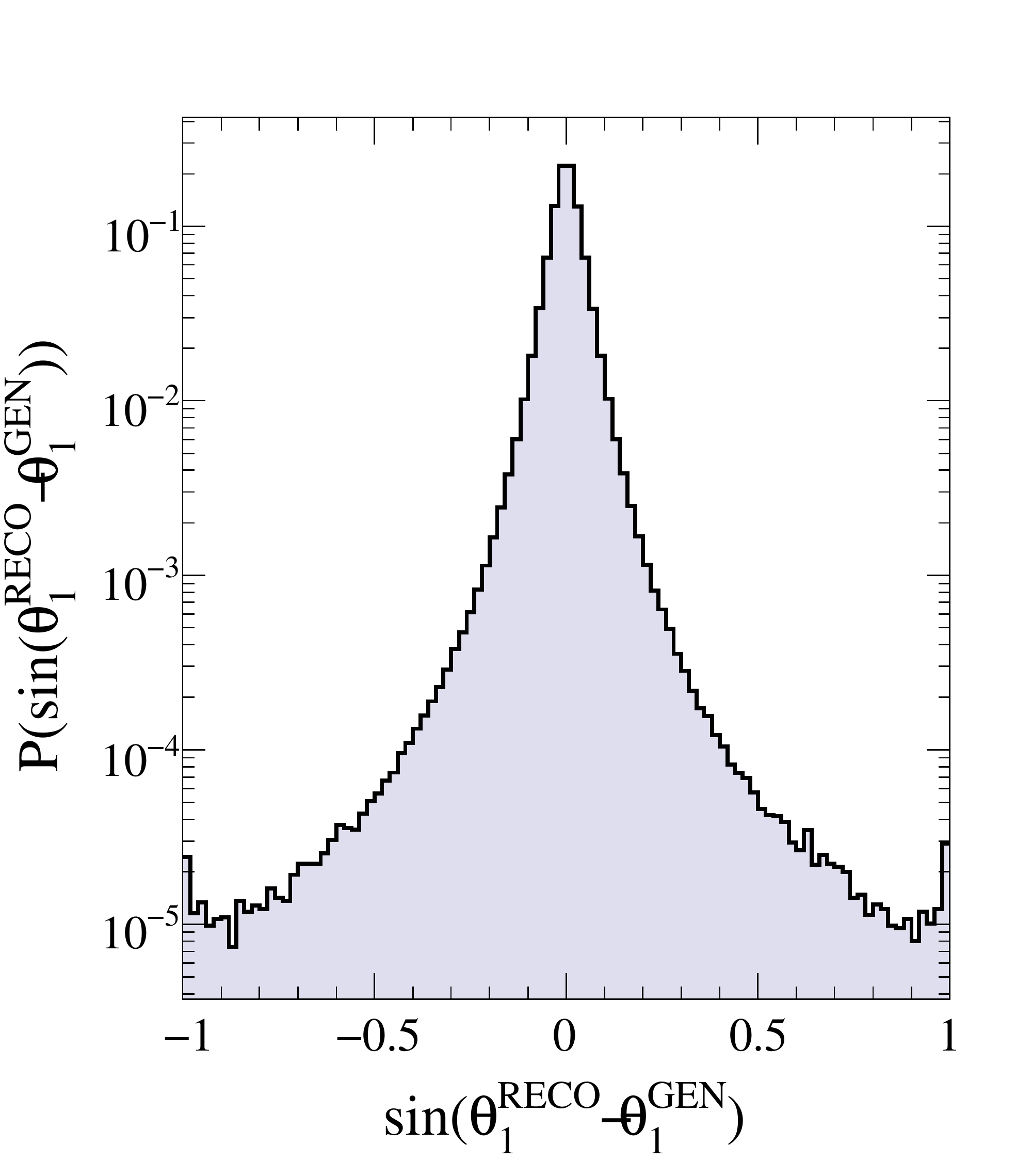}\\
\includegraphics[width=0.23\textwidth]{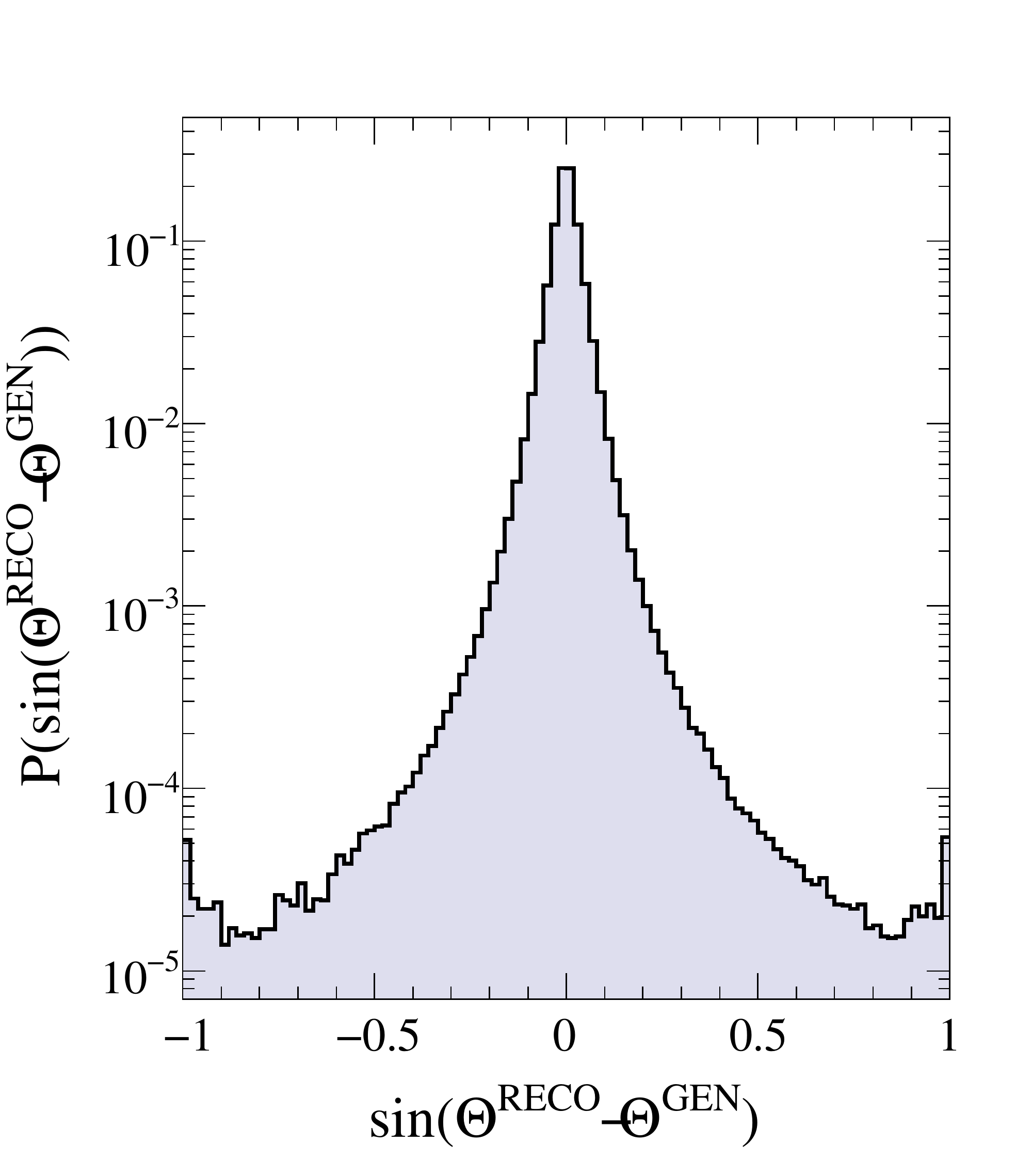}
\includegraphics[width=0.23\textwidth]{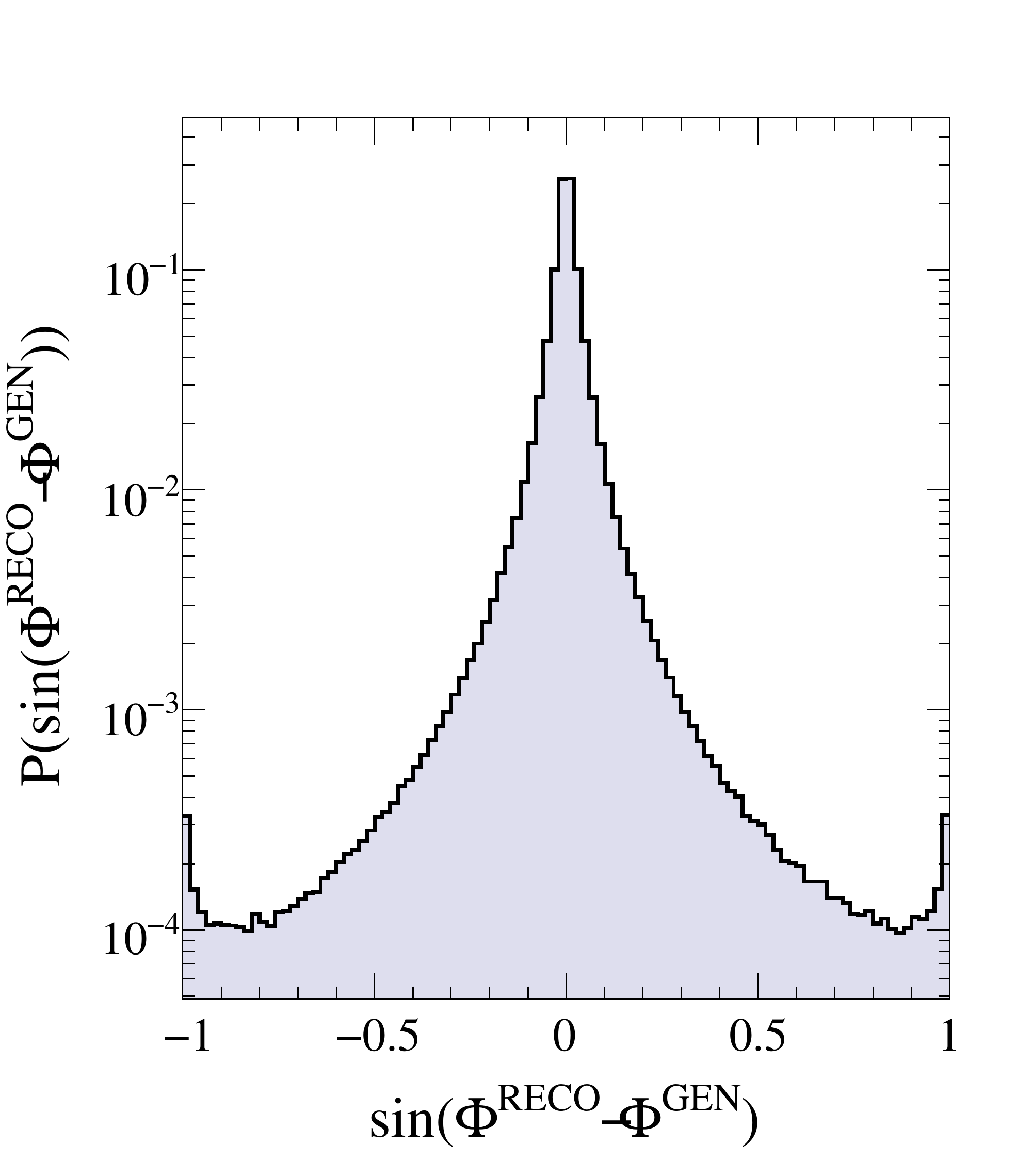}
\caption{Reconstruction resolution for the angular variables of
  $\vec{X}$ shown here for a resonance with mass 145 GeV/$c^{2}$. The
  cos$\,\theta_2$ and cos$\,\theta_1$ distributions are very similar
  in this case.  Only events surviving the signal selection are
  included. All distributions are normalized to unit
  integral. \label{fig:DET_res}}
\end{center}
\end{figure}

\begin{figure}[tbp]
\begin{center}
\includegraphics[width=0.23\textwidth]{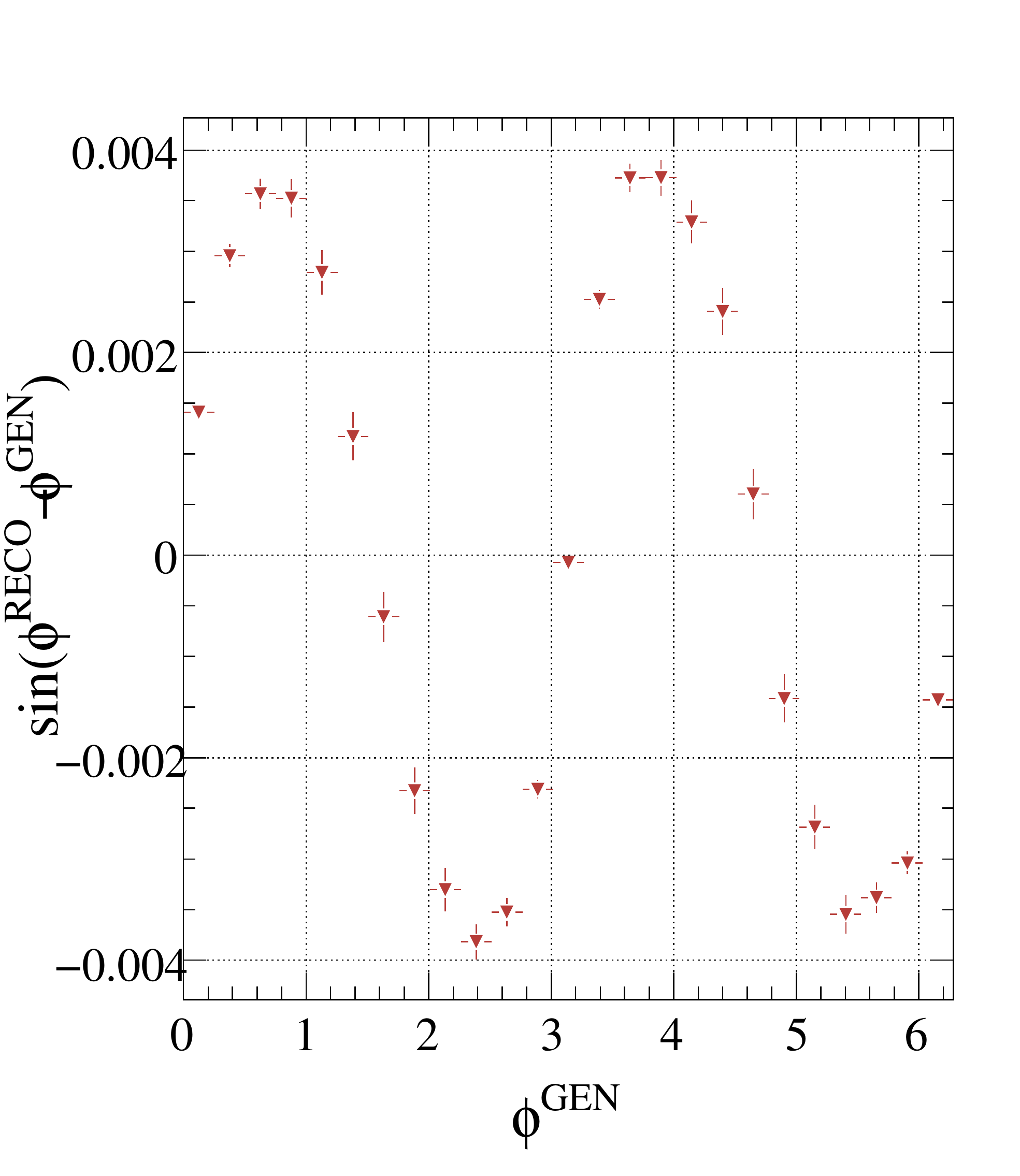}
\includegraphics[width=0.23\textwidth]{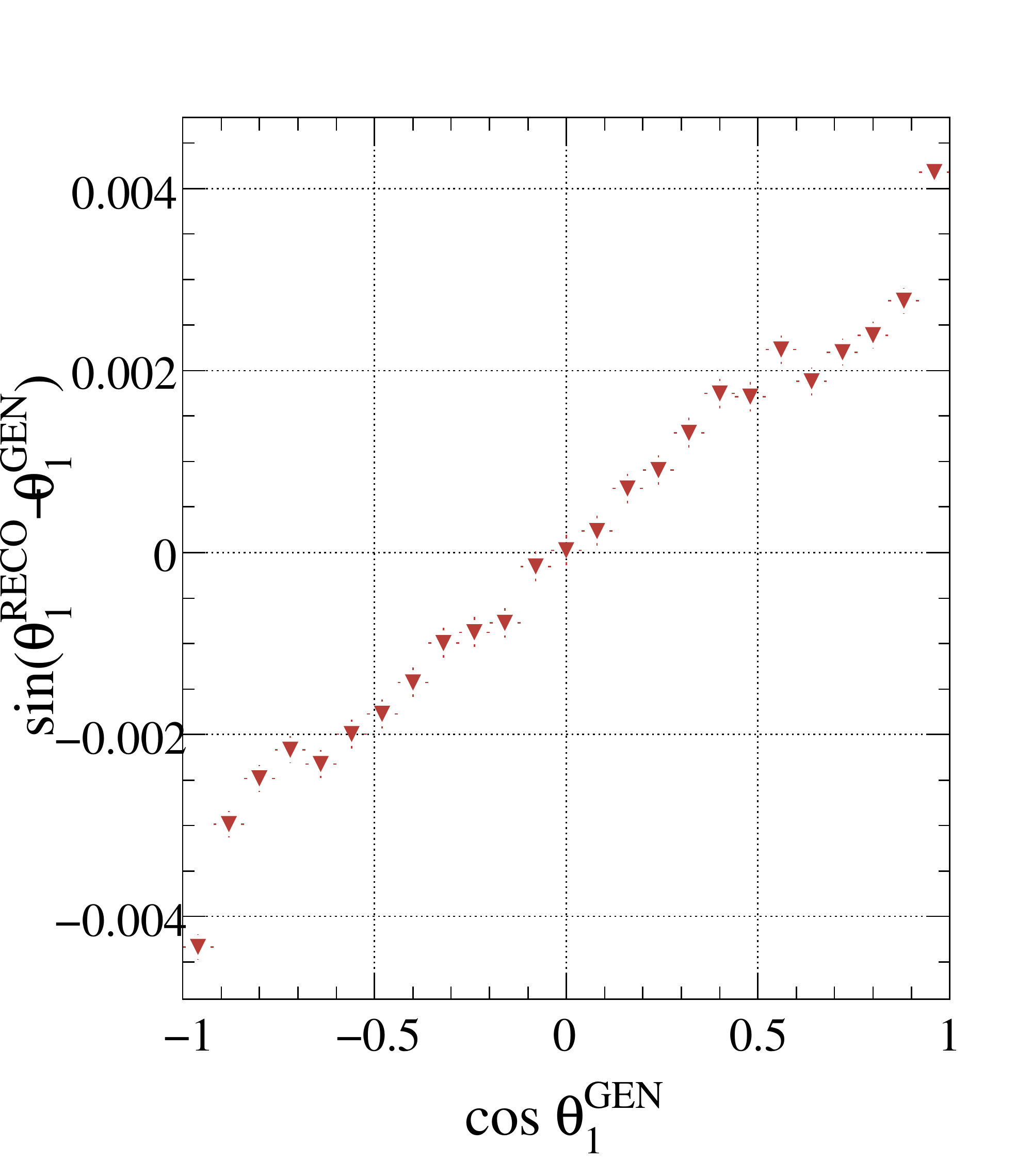}\\
\includegraphics[width=0.23\textwidth]{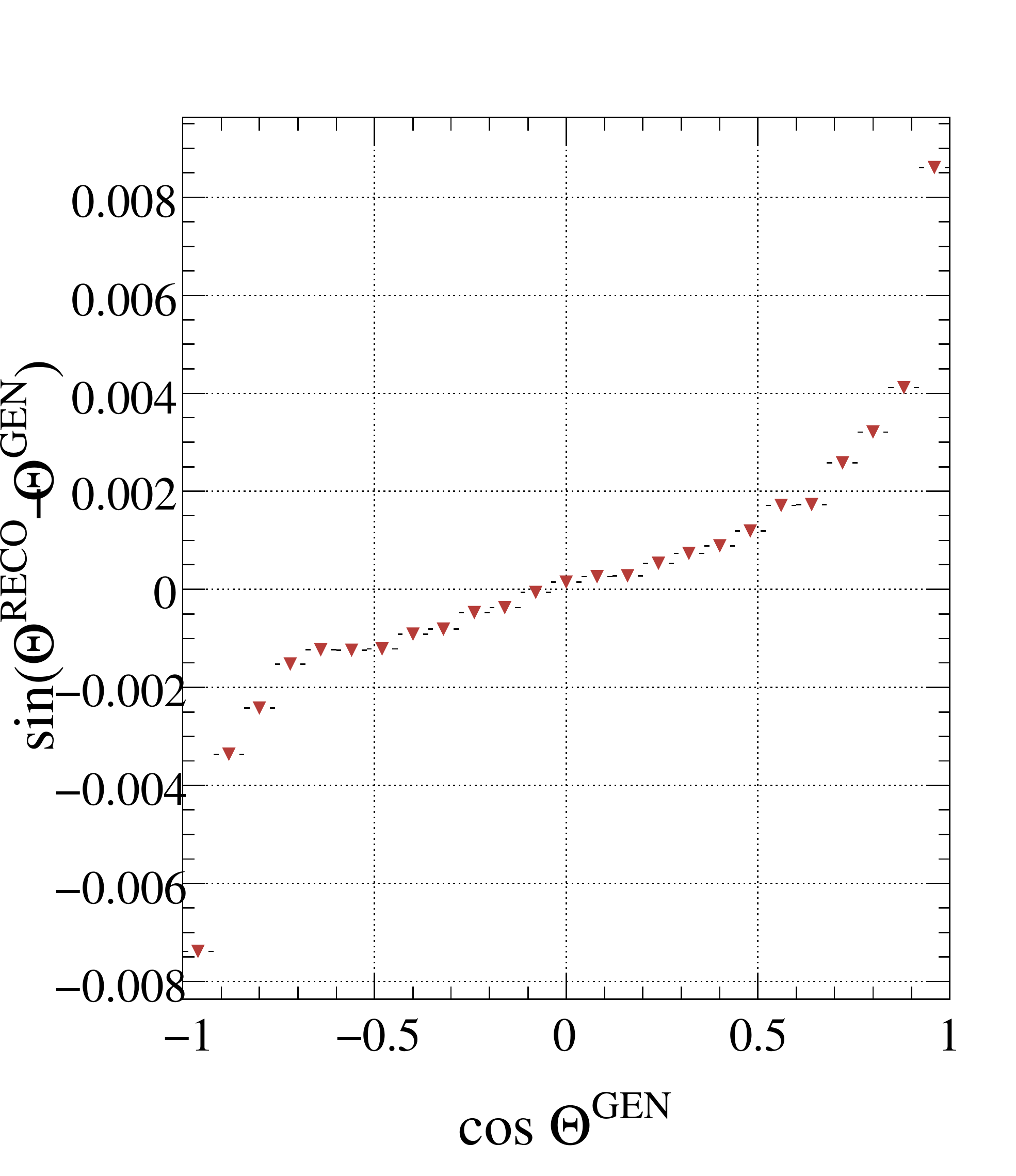}
\includegraphics[width=0.23\textwidth]{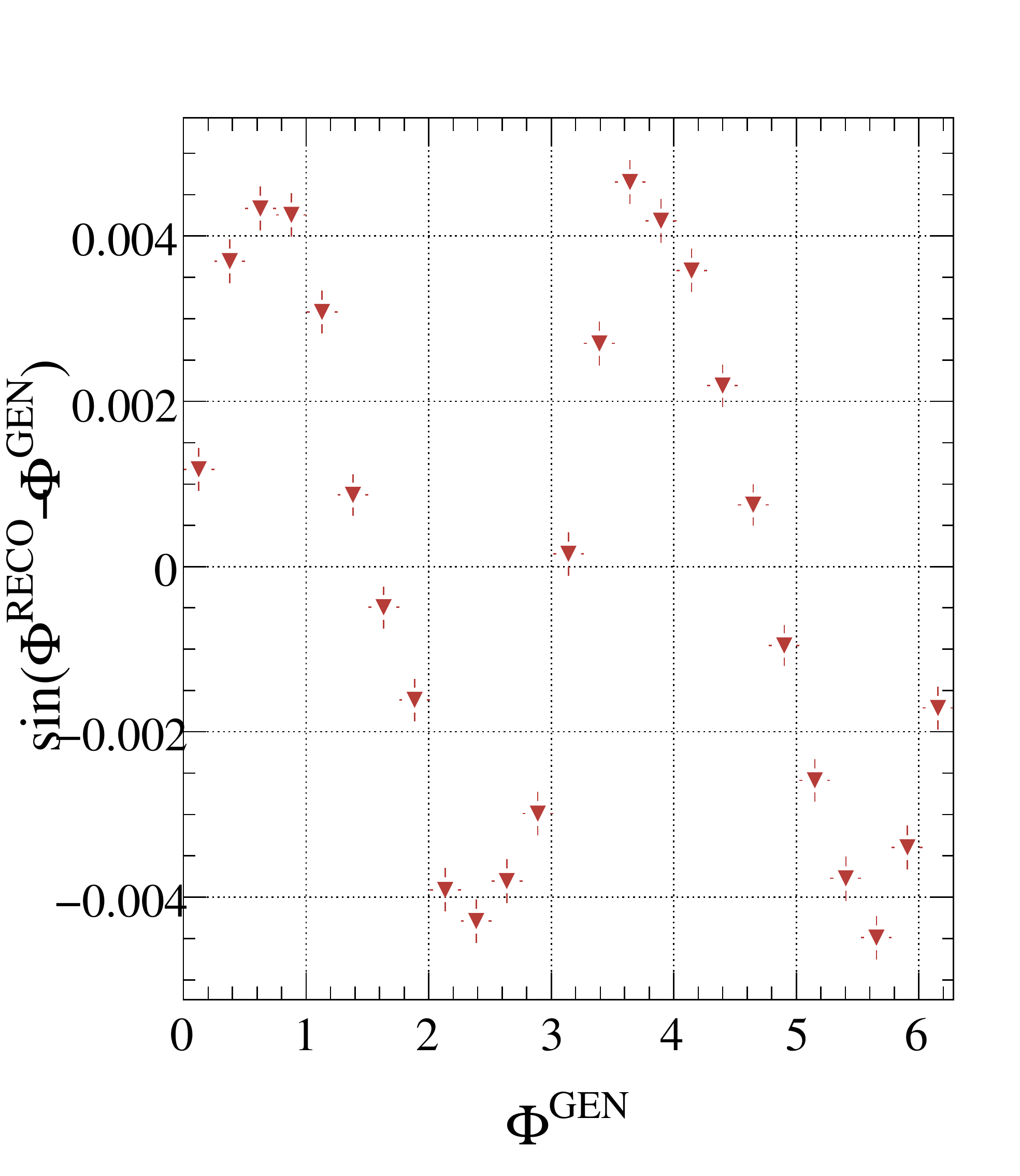}
\caption{Distributions showing systematic biases for a subset of the
  reconstructed variables $\vec{X}$ for a resonance with mass 145
  GeV/$c^{2}$.  Only events that survive the signal selection are
  included. All biases are negligible.\label{fig:DET_prof}}
\end{center}
\end{figure}

\begin{figure}[tbp]
\begin{center}
\includegraphics[width=0.23\textwidth]{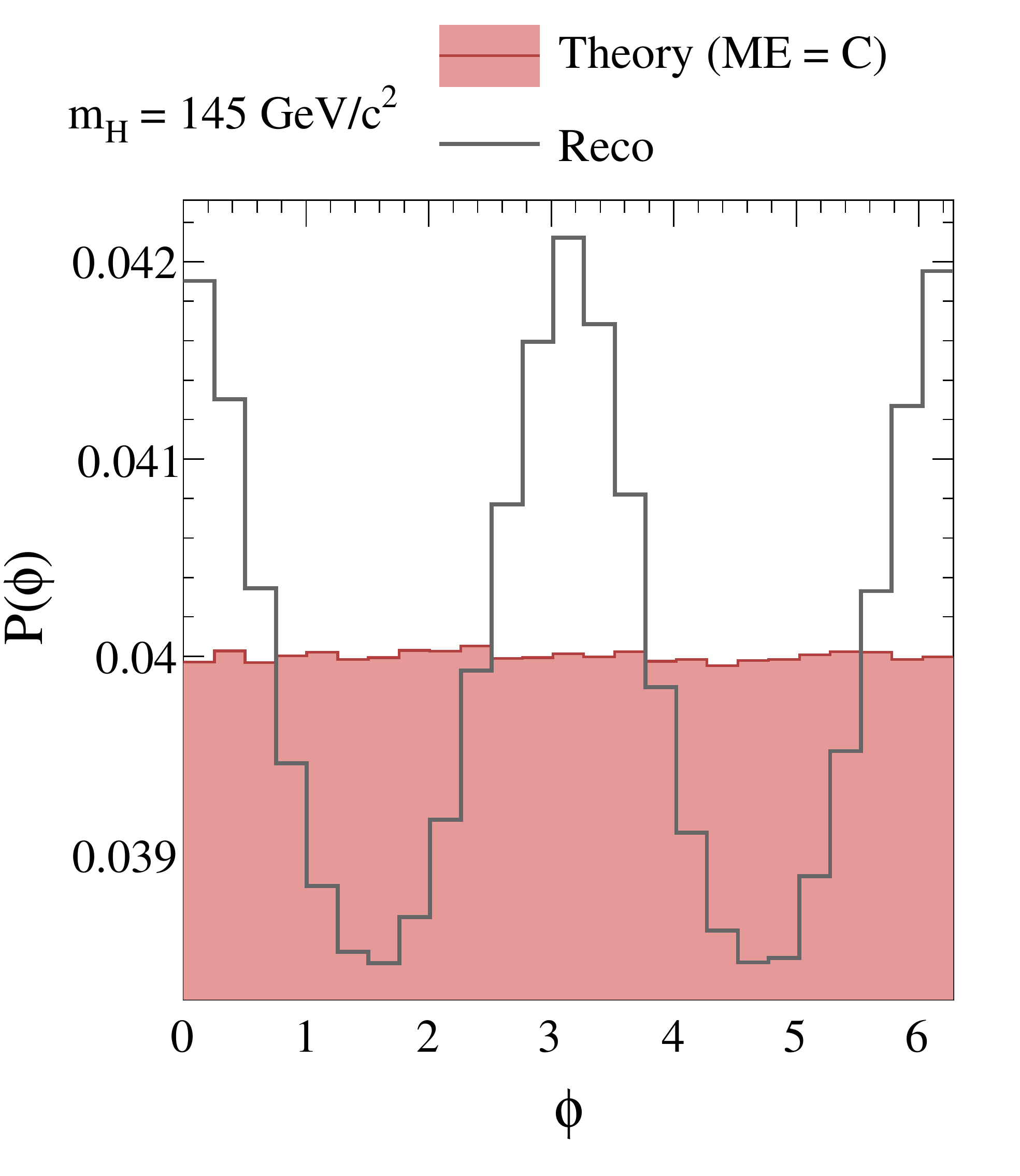}
\includegraphics[width=0.23\textwidth]{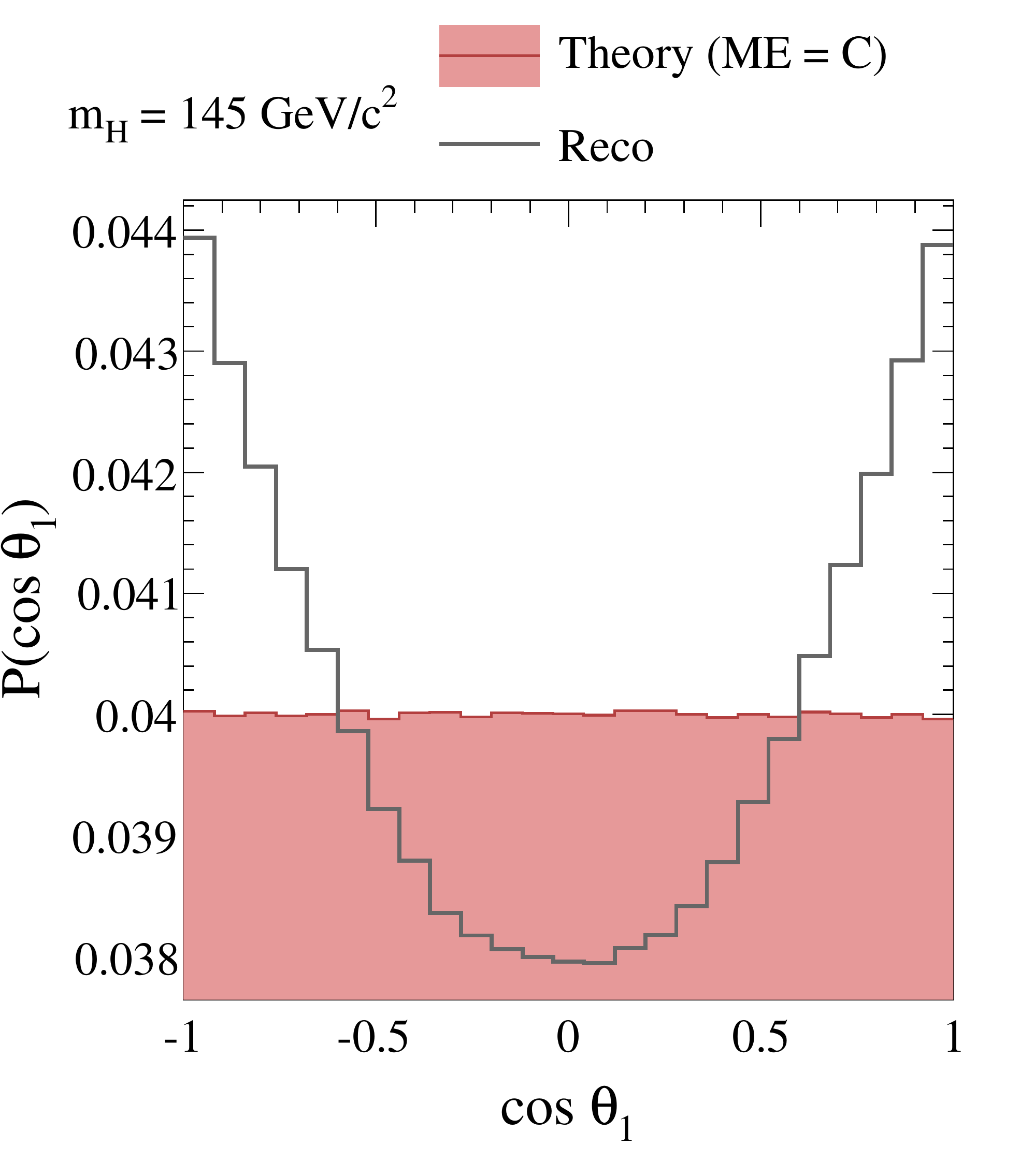}\\
\includegraphics[width=0.23\textwidth]{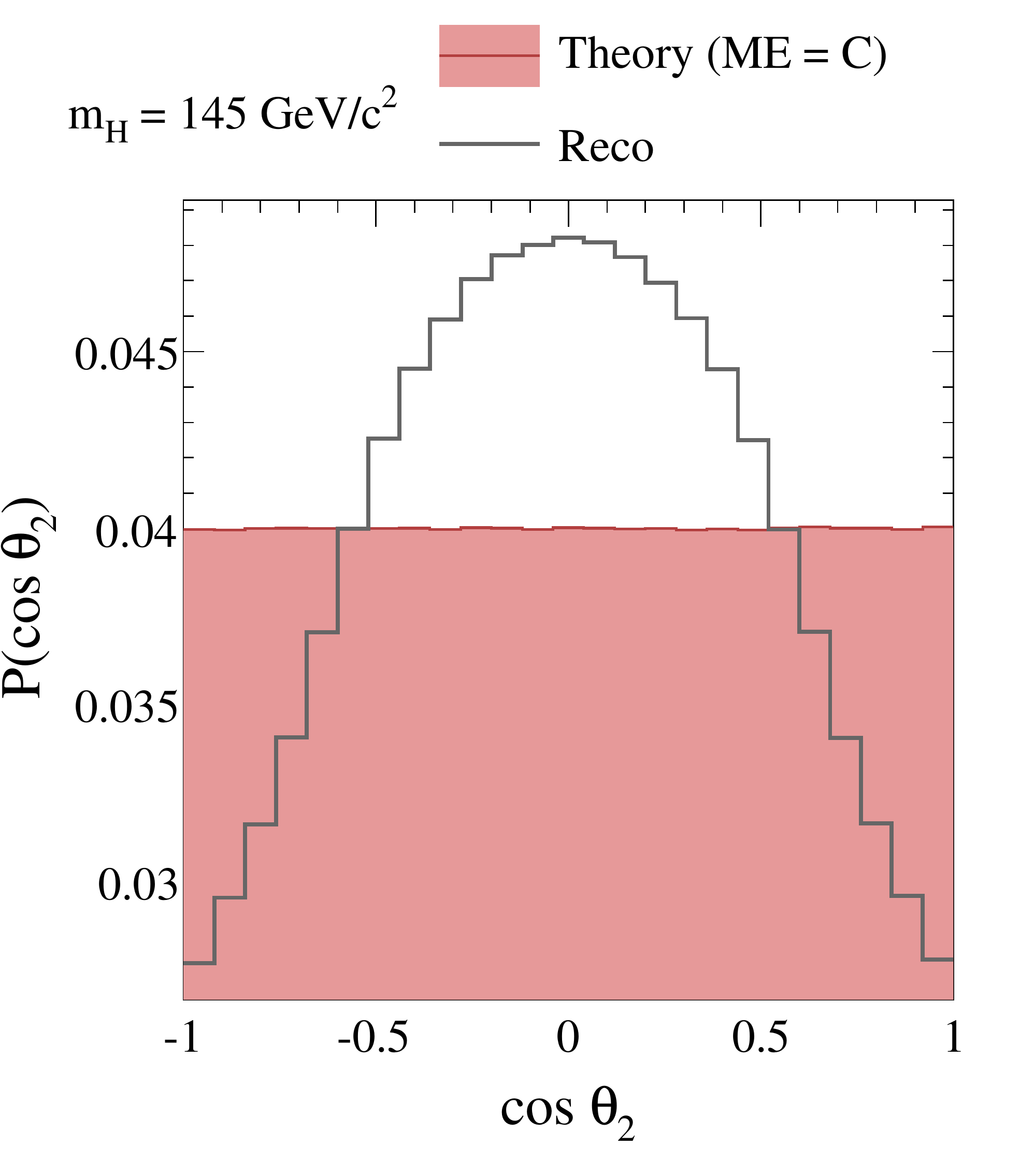}
\includegraphics[width=0.23\textwidth]{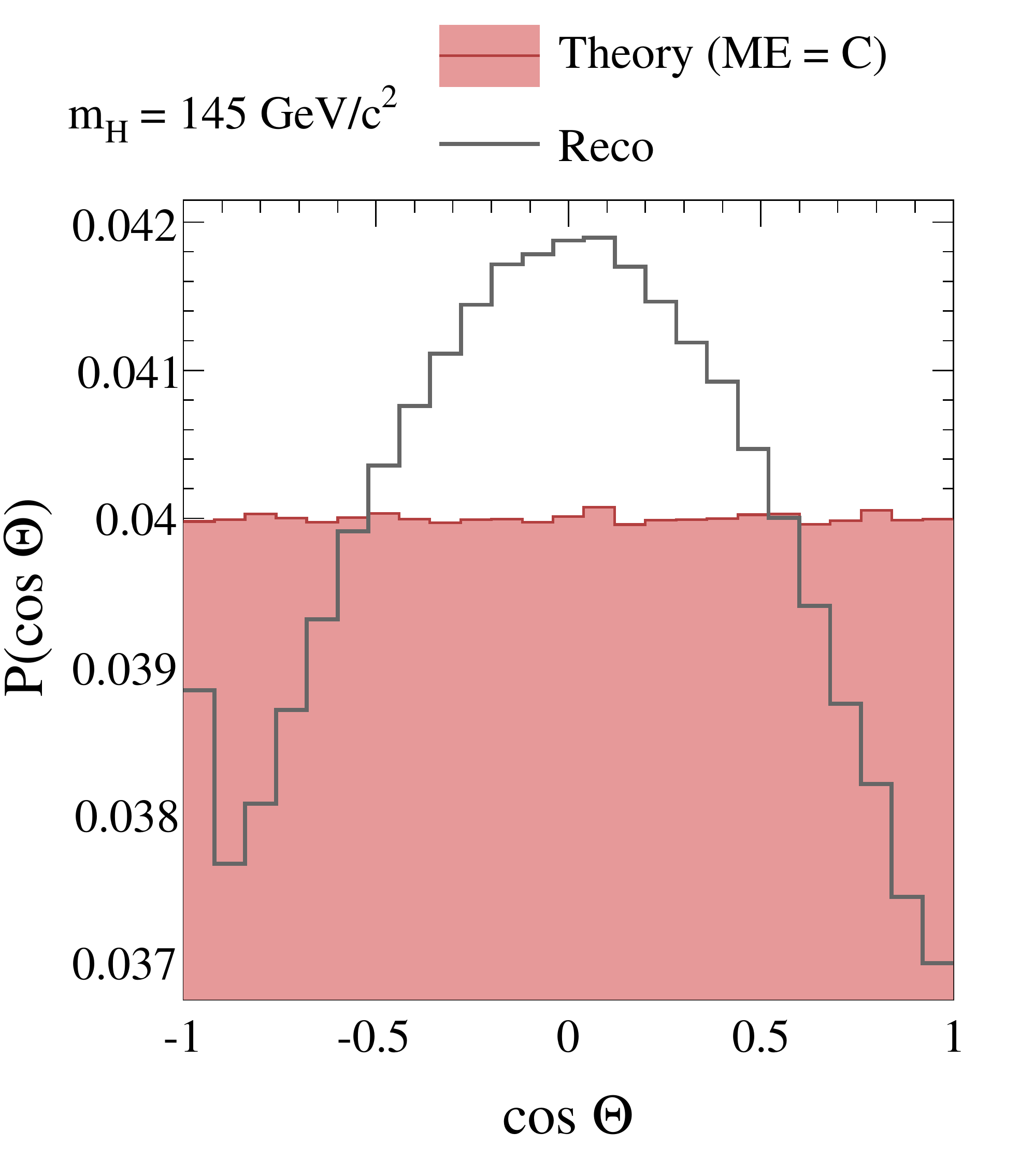}\\
\includegraphics[width=0.23\textwidth]{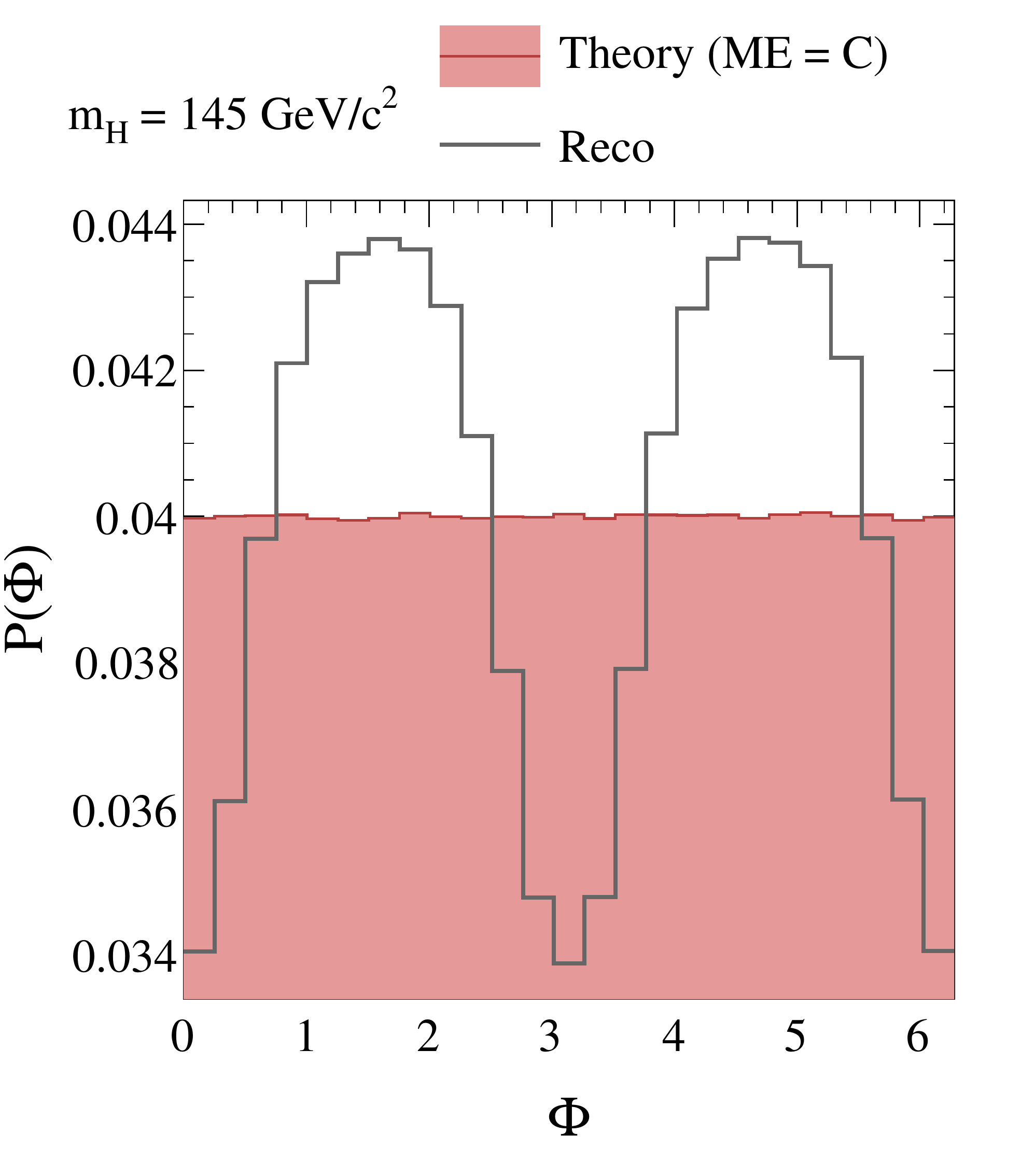}
\includegraphics[width=0.23\textwidth]{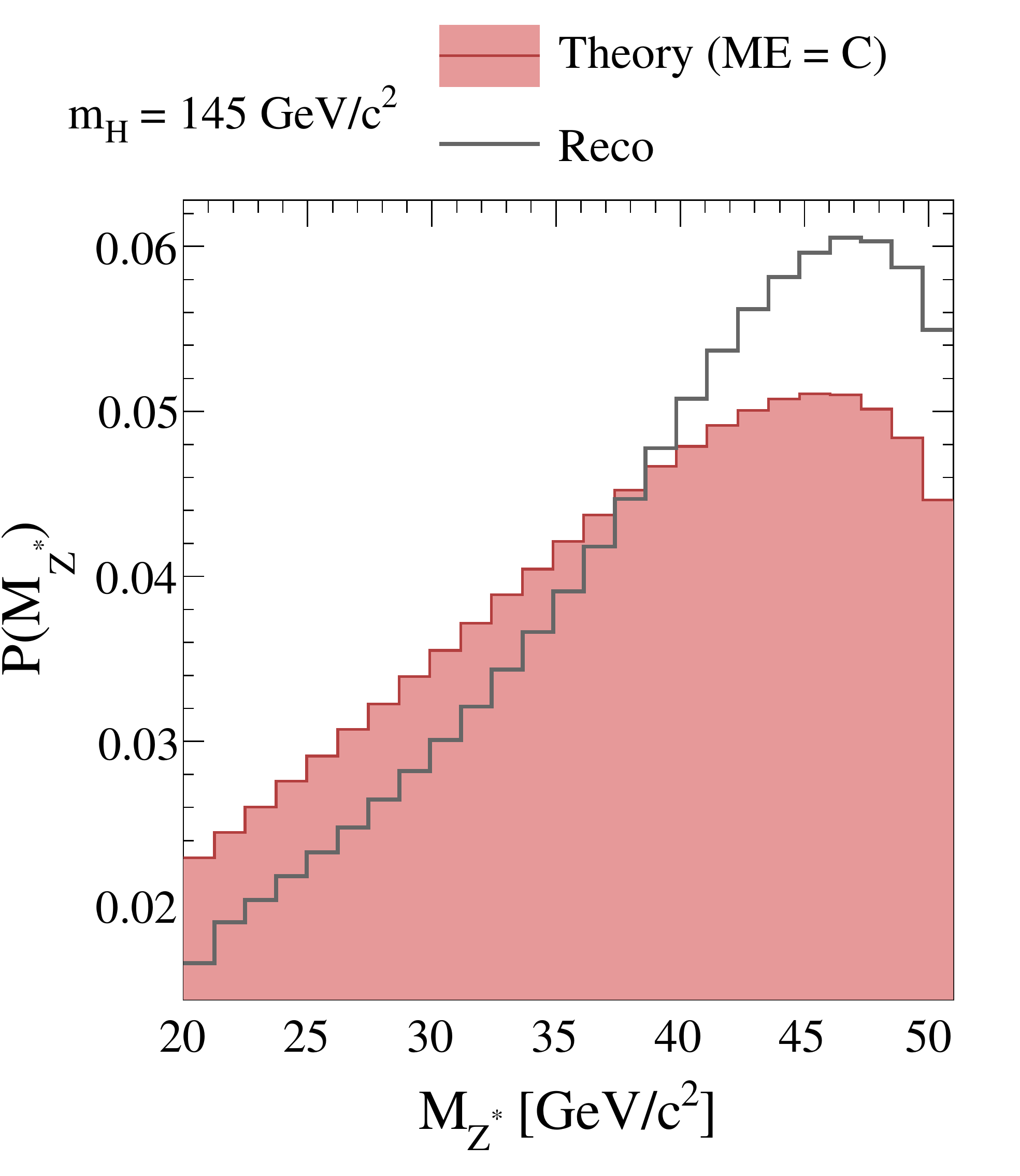}
\caption{The variables $\vec{X}$ used in this analysis for a 145
  GeV/$c^{2}$ resonance.  The off-shell $M_{Z^*}$ is required to lie a
  window between 20 and 50 GeV/c$^2$.  The shaded histograms are the
  1D distributions using a constant matrix element (i.e.~no
  angular correlations included).  The overlaid histograms show the
  same distributions for reconstructed events passing the $p_T$ and
  $\eta$ signal selection after the detector parameterization.  All
  distributions are normalized to unit integral. \label{fig:DET_SM}}
\end{center}
\end{figure}

\begin{figure}[tp]
\begin{center}
\includegraphics[width=0.42\textwidth]{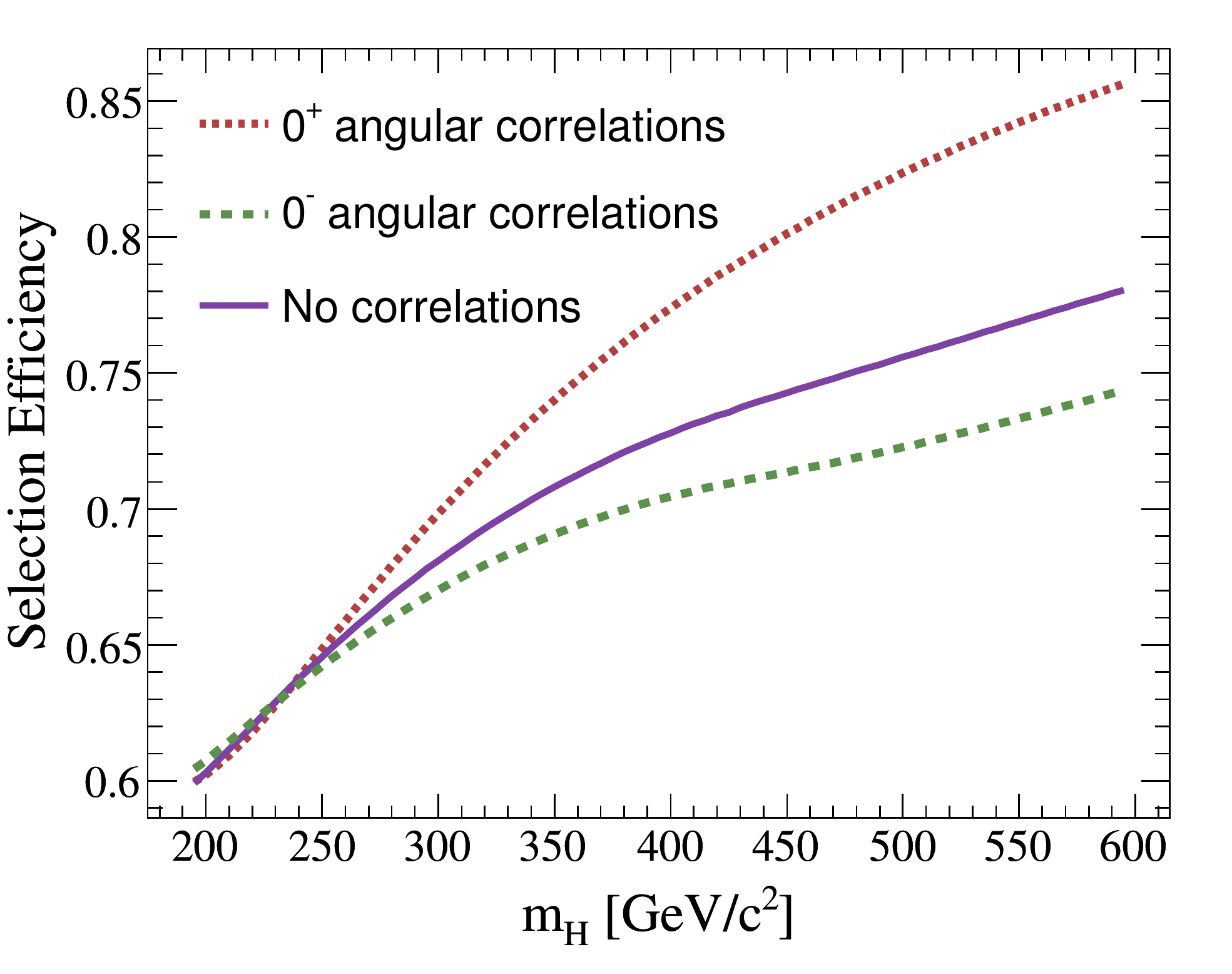}
\caption{The analysis efficiency for $0^+$, $0^-$  as a function of the resonance mass.
The case with no correlations is also shown for comparison.
\label{efficiency}}
\end{center}
\end{figure}

A number of detector related effects can modify the $\vec{X}$
observables' {\it pdfs}. The resolution of the observables used in the
analysis is shown in Fig.~\ref{fig:DET_res} and is found to be small
independent of the HLL resonance mass and quantum numbers.  The
systematic bias in the reconstruction of the same variables is shown
in Fig.~\ref{fig:DET_prof} and is found to be negligible. This
shows that the sculpting of the observables' {\it pdfs} is not a
result of reconstruction resolution or bias.
Rather, it depends on the simulated kinematics of the HLL resonance,
including its mass, and on the particular model considered ($0^{+}$,
$0^{-}$, etc). Specifically, the overall phase space acceptance,
implemented in the signal selection by means of the $p_T$ and $\eta$
requirements, produces the largest effects on the observables. This is
shown in Fig.~\ref{fig:DET_SM} for a resonance of mass 145 GeV/$c^2$
generated with no explicit angular correlations.
Adding the angular correlations can enhance or
reduce the overall selection 
%%%%%%%%%
%%%%%%%%%%%%%
efficiency depending on the details of
the multidimensional {\it pdf}.  Our selection is $60\%$ ($74\%$)
efficient for a $0^+$ 
resonance of mass 200 GeV/c$^2$ (350 GeV/$c^2$)
as shown in Fig.~\ref{efficiency}.  The same figure demonstrates that
the efficiency has a non-trivial dependence on the nature of the spin
correlations. Specifically, for a $0^{-}$ resonance of 200 GeV/c$^2$
(350 GeV/$c^2$) the efficiency is $60\%$ ($69\%$).  With an absence of
explicit spin correlations the efficiency for a 350 GeV/$c^2$
resonance is $71\%$.

We find that changes in the $\vec{X}$ distributions are strongly
correlated with the kinematics of the off-shell $Z$, e.g. for
cos$\,\theta_2$ the largest inefficiencies correspond to the
kinematic configurations where at least one of the muons is soft.
When the correlations between the variables $\vec{\omega}$ and
$\vec{\Omega}$ appear explicitly in the differential cross-sections, as
is the case for $J$$=$$1^{\pm}$, the phase space acceptance effects are
amplified. The consequences on model discrimination are discussed in
Sec.~\ref{sec:SM_v_1}.

\begin{figure}[htbp]
\begin{center}
\includegraphics[width=0.237\textwidth]{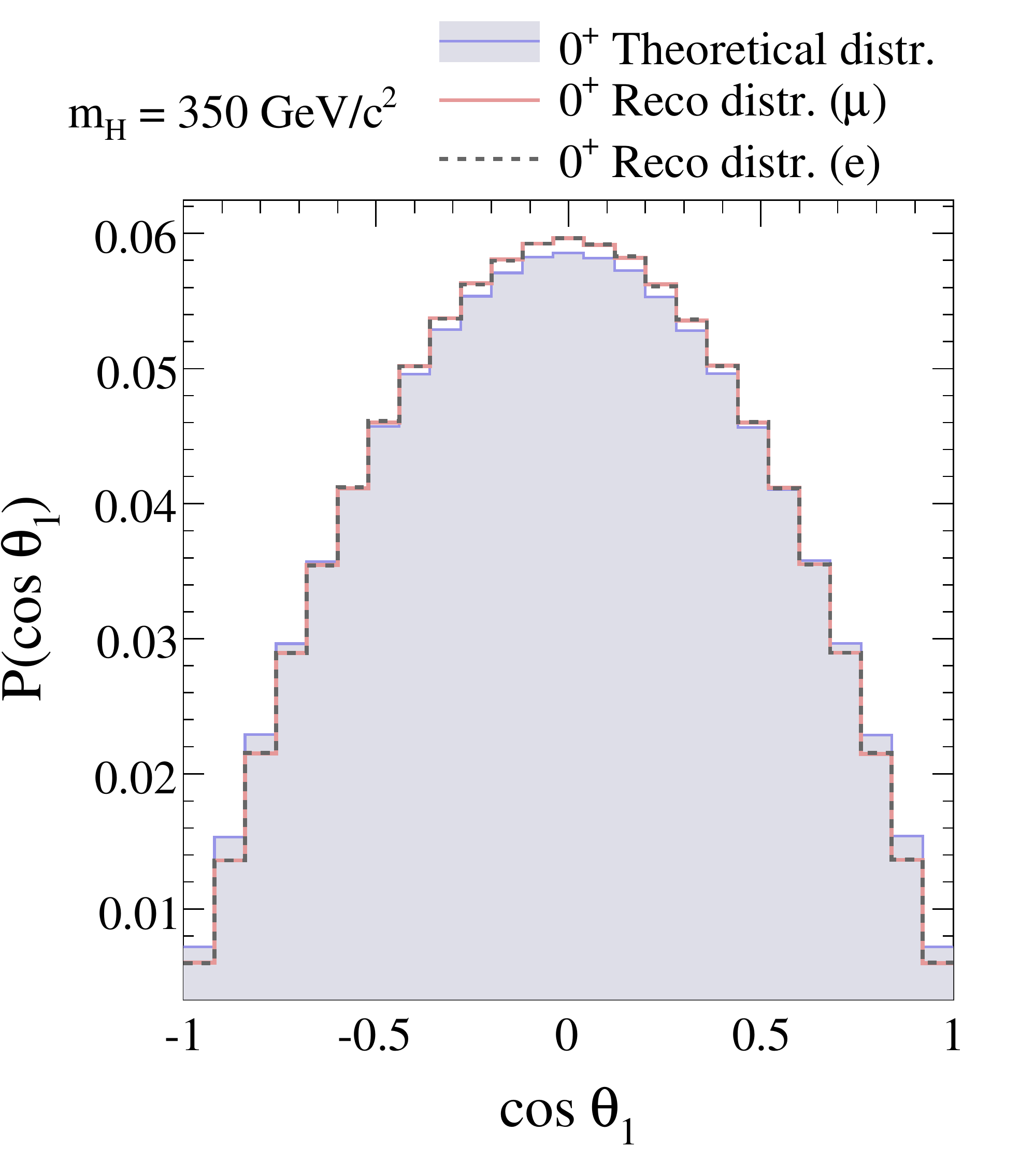}
\includegraphics[width=0.237\textwidth]{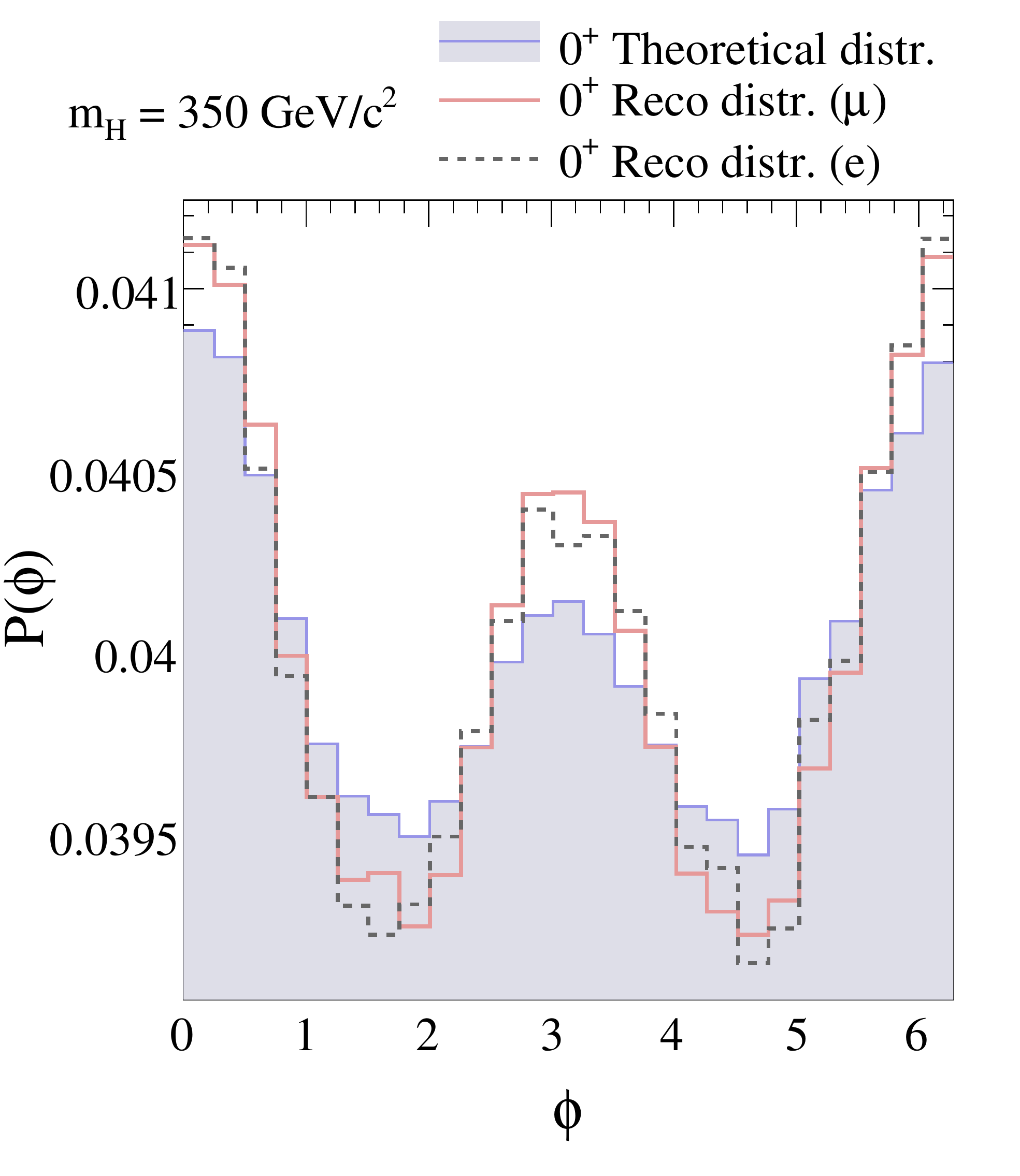}
\caption{Kinematic distributions for the variables cos$\,\theta_1$
  (left) and $\phi$ (right) for a $0^{+}$ resonance with mass 350
  GeV/$c^{2}$. The shaded histograms show the 1D projections of the
  variables as described by the analytic {\it pdfs}.  The overlaid
  histograms (blue, red) show the same 1D projections for
  reconstructed events passing the $p_T$ and $\eta$ signal selection
  after the detector parameterization for $4\mu$ and $4e$ final
  states.  All distributions are normalized to unit
  integral. \label{fig:DET_mu_v_e}}
\end{center}
\end{figure}

The shapes of the reconstructed $\vec{\omega}$ and $\vec{\Omega}$
distributions depend on the phase space acceptance both for electron
and muon final states ($H \to ZZ \to 2e2\mu$ or $4e$).  
Figure~\ref{fig:DET_mu_v_e} shows the relevant kinematic distributions.  All
the results concerning model discrimination, as a function of the
number of observed signal events, will be nearly identical when the
additional final states are included ($2e2\mu,~ 4e$), especially when
the off-shell $Z$ mass is not used as an observable. This is not
necessarily the case for results concerning the discovery of a
resonance in these final states with respect to the background-only
hypothesis, since different backgrounds need to be considered for
electron and muon final states.

%%%%%%%%%%%%%
%%%%%%%%%%%

\subsection{Fit definition and signal extraction}\label{ZZfit}
\begin{figure}[htbp]
\begin{center}
\includegraphics[width=0.23\textwidth]{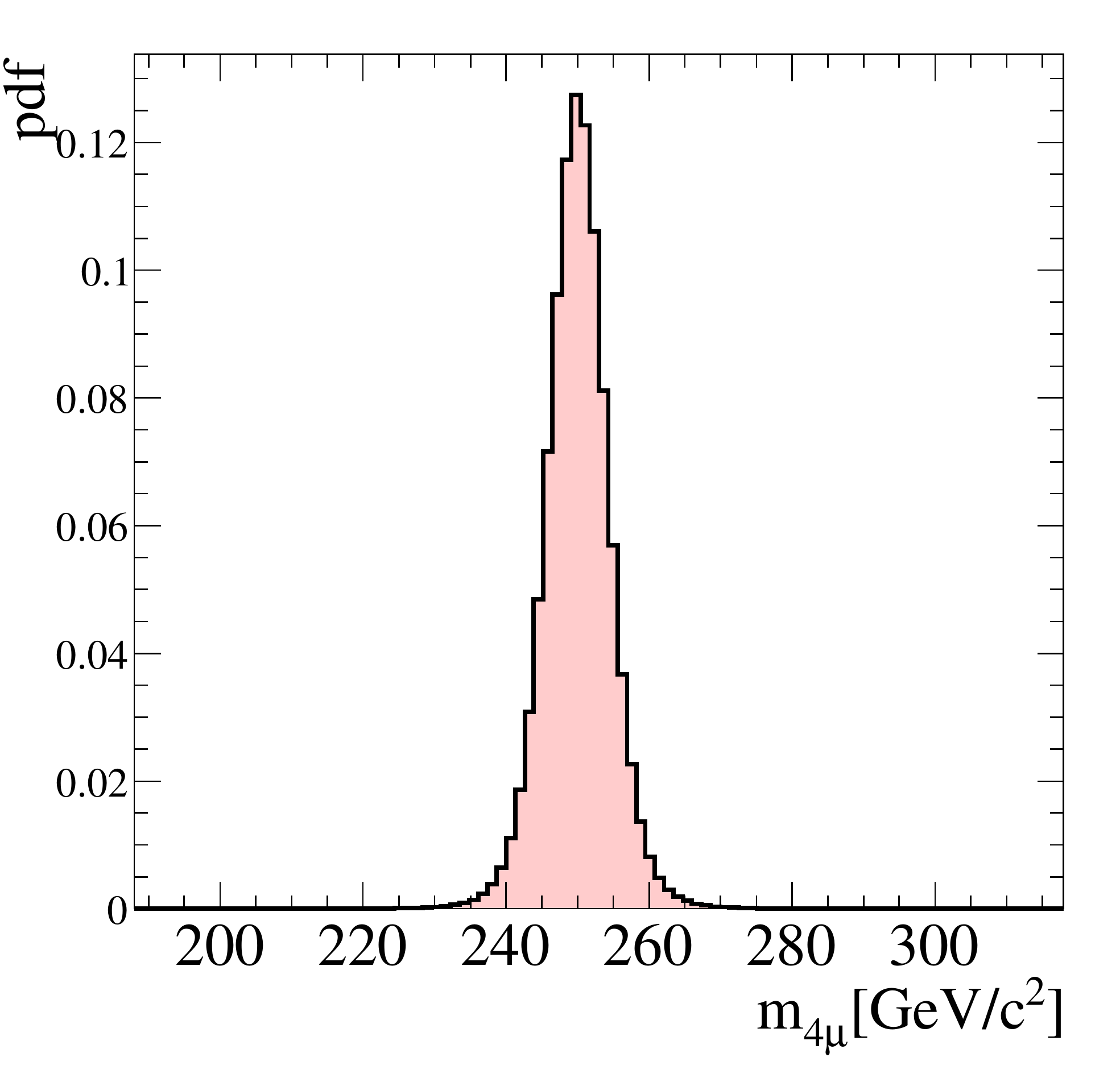}
\includegraphics[width=0.23\textwidth]{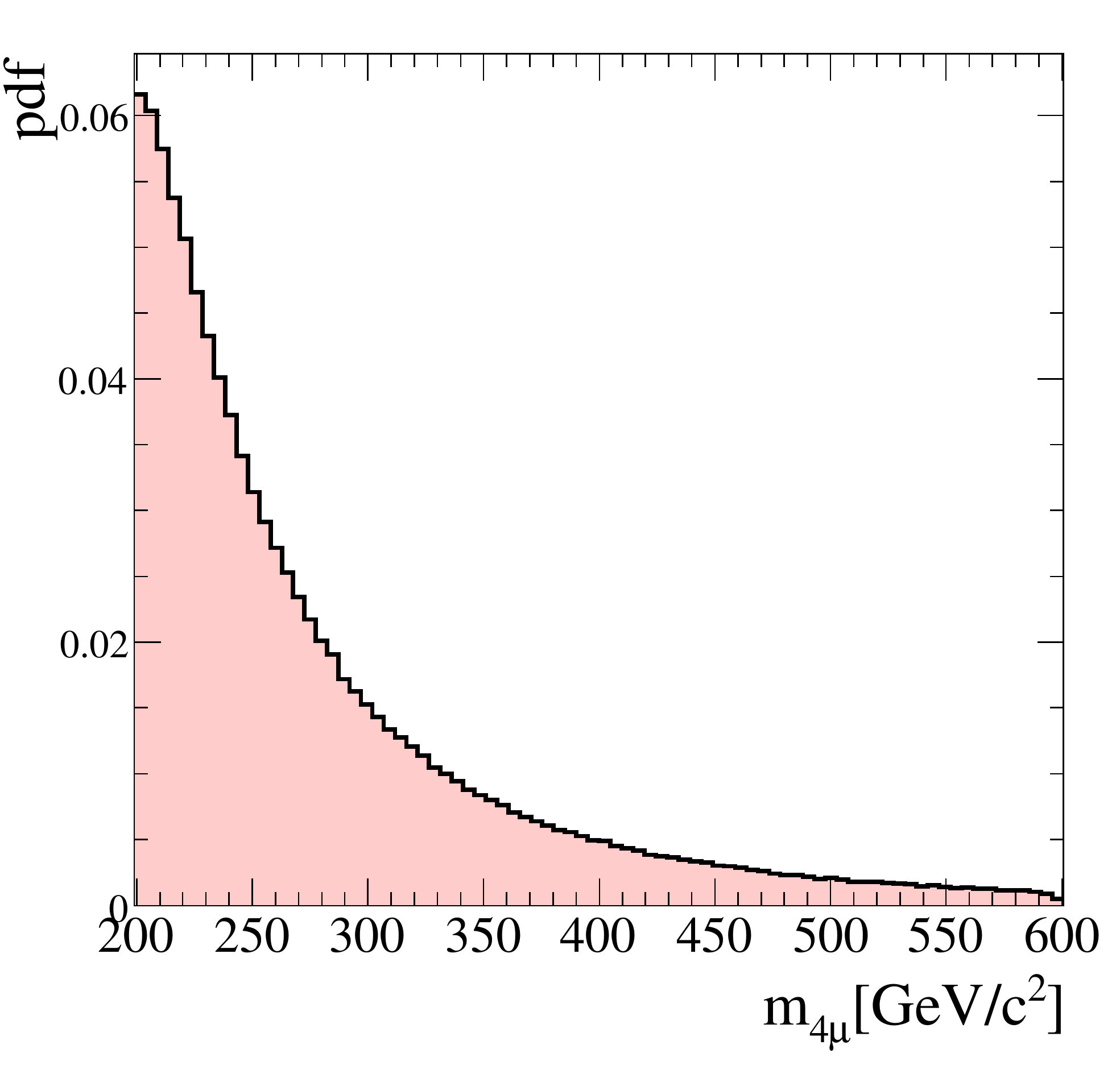}
\caption{Distribution of the $4\mu$ invariant mass for a sample of
  signal with $m_H$$=$$250$ GeV/c$^2$ (left), and background (right) $ZZ$ events.
\label{fig:simpdf}}
\end{center}
\end{figure}

The $H \to ZZ$ signal events can be discriminated from SM backgrounds
using an extended and unbinned ML fit. Since there is no resonant
$4\mu$ background in the SM, the fit can use as a discriminating
variable the $4\mu$ mass distribution. In the presence of a sizable
background due to fake $Z$ candidates (such as top decays) the $2\mu$
mass distributions can be included in the likelihood.  Since this is
not a conceptually different situation, we ignore this possibility and
assume for simplicity that the only relevant background is given by
events with two real $Z$ candidates. We write the likelihood function
as:
\begin{eqnarray}
&&\hspace*{-35pt}
  {\mathcal L} = \frac{1}{N!}\exp{
  \biggl(
  -\sum_{j}N_{j}
  \biggr)
  } \\
  &&\hspace*{10pt}
  \prod_{i=1}^N   \biggl( N_{S} P_{S}[m^i_{4\mu}] + N_{B}
  P_{B}[m^i_{4\mu}] \biggr) \; ,\nonumber
\label{likelihood}
\end{eqnarray}
where $N_j$ ($j$$=$$S, B$) represents the yield of each component,
$m^i_{4\mu}$ is the $4\mu$ candidate mass for the event $i$, and
$P_{S}[m]$ ($P_{B}[m]$) is the signal (background) distribution for
the variable $m$.
The {\it pdfs} for the signal and background components are described
using the template distributions from the simulation, as shown in
Fig.~\ref{fig:simpdf} for $m_H$$=$$250$ GeV/$c^2$.  This fit
configuration is appropriate
for the HLL characterization.

%%%%%%%%%%%%%
%%%%%%%%%%%%%%%

\subsection{Background subtraction\label{sec:splots}}
In order to establish if a newly-discovered resonance is indeed the
Higgs boson or not, a hypothesis test is performed (see
Sec.~\ref{sec:hllresults}).~In this context, a tool to disentangle
signal and background events from the selected dataset is an important
prerequisite.  We use the $_sWeight$~\cite{sPlots} technique and
reweight the selected dataset according to {\it how likely} each event
is considered to be signal by the fit.
\begin{figure}[htbp]
\begin{center}
\includegraphics[width=0.37\textwidth]{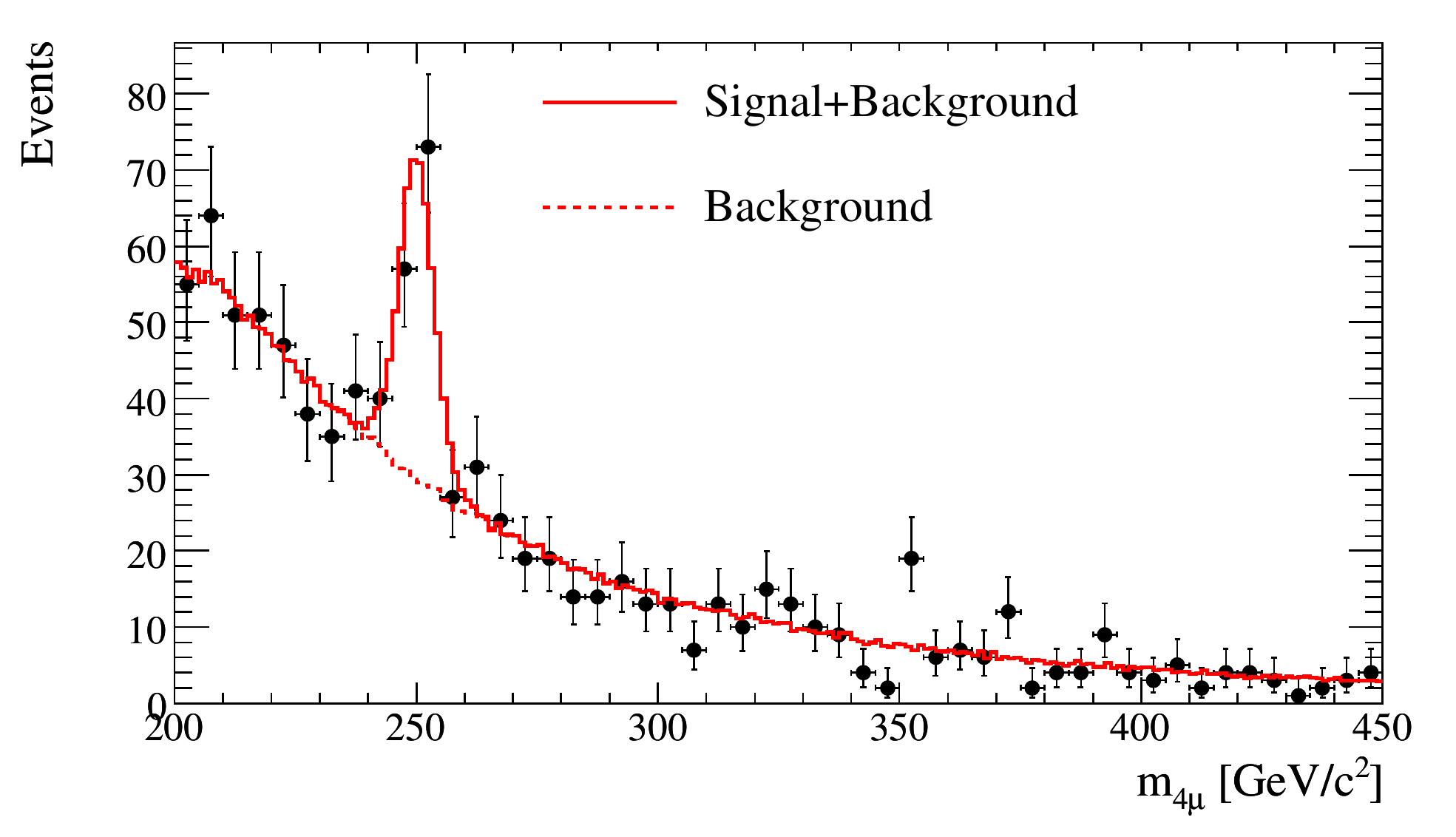}
\includegraphics[width=0.37\textwidth]{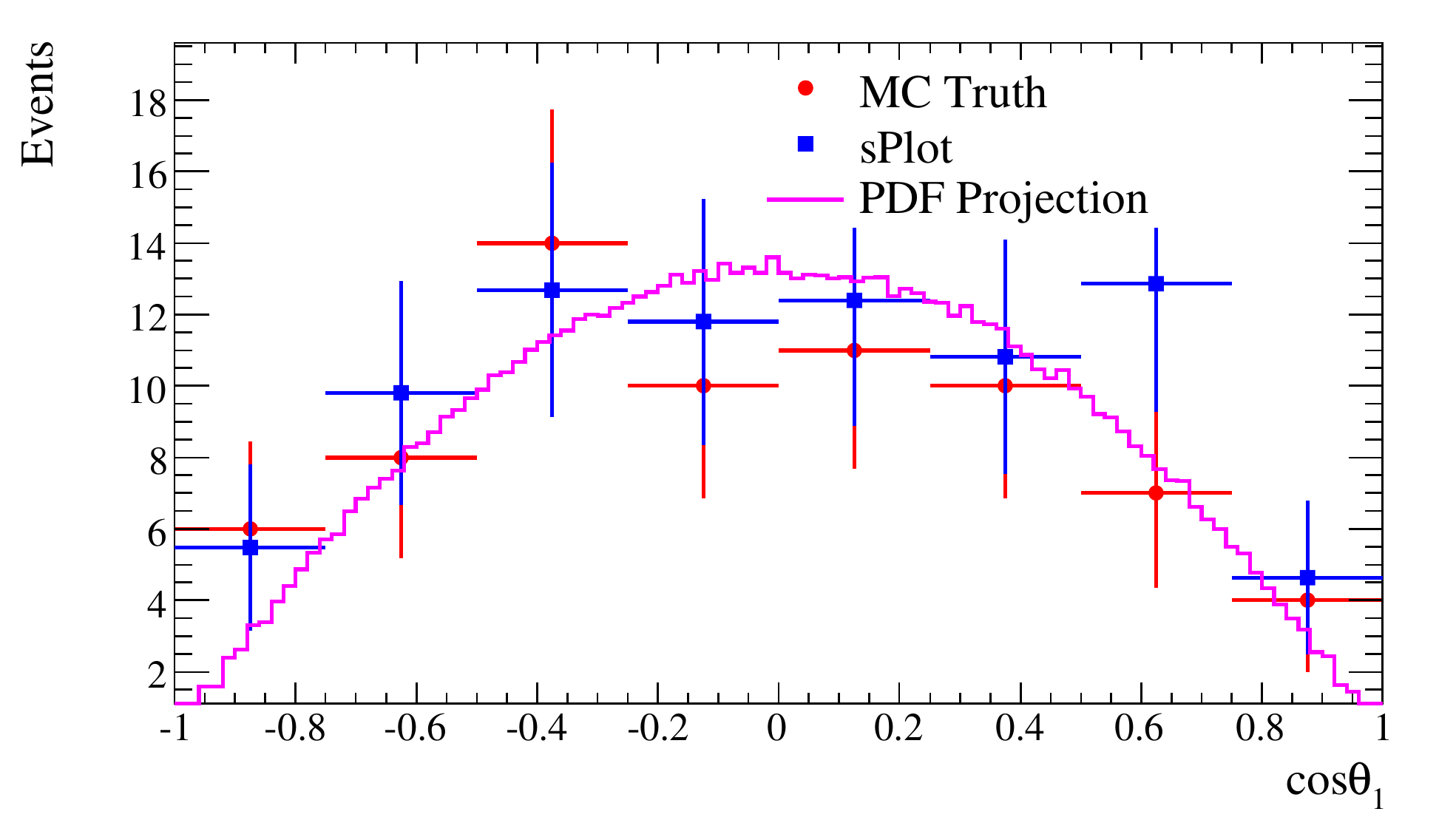}
\includegraphics[width=0.37\textwidth]{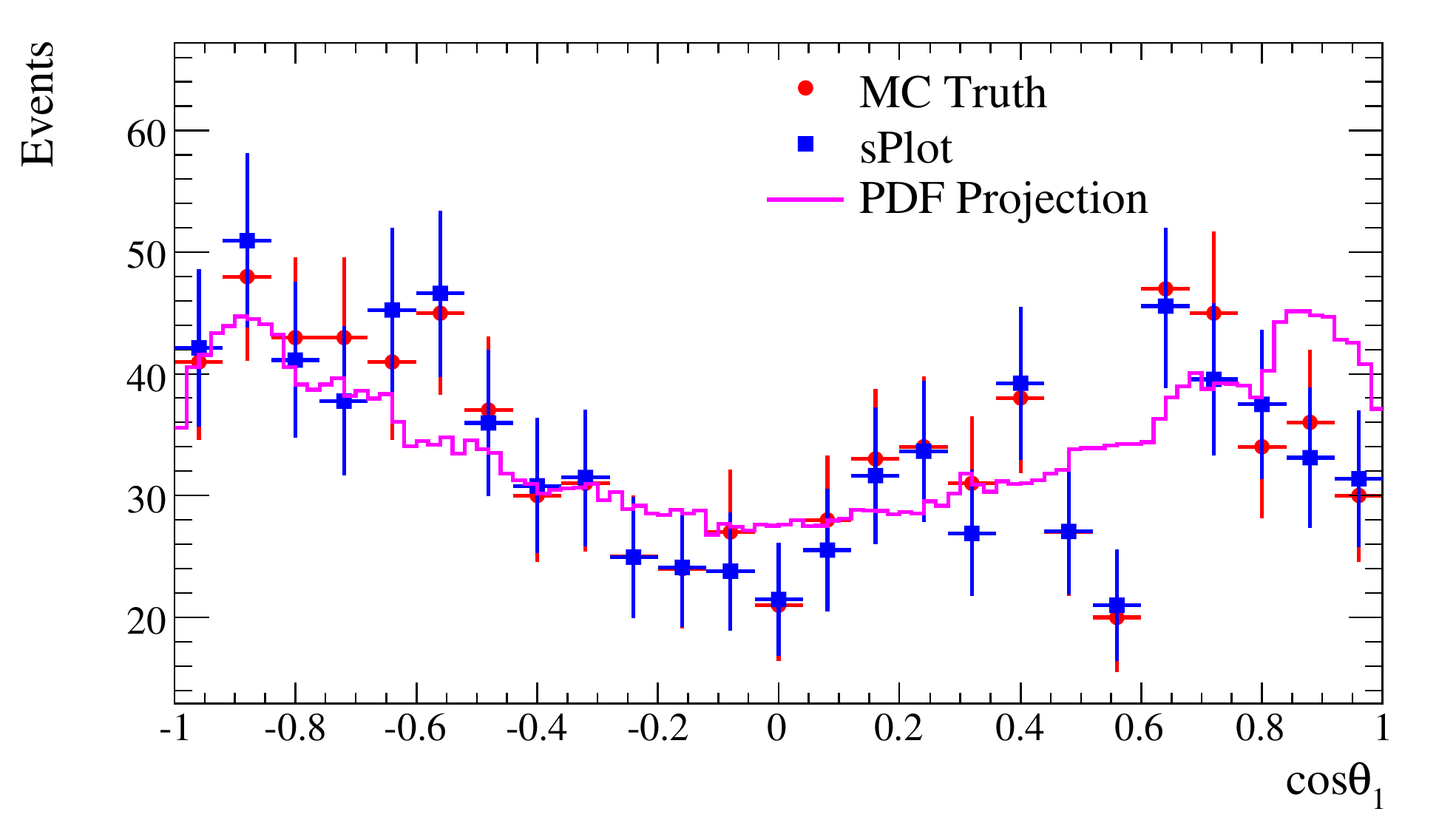}
\caption{The $4\mu$ invariant mass distribution for a sample of
  $N_{S}$$=$$70$ $H \to ZZ$ events with $m_H$$=$$250$ GeV/c$^2$ and
  $N_{B}$$=$$1000$ $ZZ$ background events. The superimposed curves
  represent the likelihood function returned by an ML fit, with
  $N_{S}$, $N_{B}$, and $m_{4\mu}$ as free parameters (top).  Comparison of
  the signal-only MC distribution of cos$\,\theta_{1}$, with the
  background-subtracted distribution obtained with the $_sWeight$
  technique (middle).  Comparison of the background-only MC
  distribution of cos$\,\theta_{1}$, with the signal-subtracted
  distribution obtained with the $_sWeight$
  technique (bottom).\label{fig:sweight}}
\end{center}
\end{figure}
The $_sWeight$ technique is statistically optimal when the
discriminating variable ($m_{4\mu}$ in our case) in the fit is
uncorrelated with the subsequently used variables ($\vec{X}$ in our
case).  On the upper plot of Fig.~\ref{fig:sweight}, the $4\mu$
invariant mass distribution is shown for a sample of $N_{S}$$=$$70$ $H \to
ZZ$ events (with $m_H$$=$$250$ GeV/c$^2$) on top of $N_{B}$$=$$1000$ continuum
$ZZ$ background events, corresponding to a $\simeq 5\sigma$ deviation
from the background-only hypothesis. The superimposed curves represent
the likelihood function returned by an ML fit (with $N_{S}$, $N_{B}$,
and $m_{4\mu}$ as free parameters). The middle plot shows the signal
$_sWeighted$ cos$\,\theta_{1}$ distribution.  Similarly, the bottom
plot shows the background $_sWeighted$ cos$\,\theta_{1}$ distribution.
The comparison of the two sets of points shows how the background
(signal) subtraction allows one to recover the signal (background)
distribution for the considered variable in the given sample, the
deviation from the expected {\it pdfs} being due to statistical
fluctuations already present at the MC level.

%%%%%%%%%%%%%%%%%%%%
%%%%%%
%%%%%          STATISTICS.TEX
%%%%%%
%%%%%%%%%%%%%%%%%%%%%%%%

\section{Statistical approach\label{sec:stat}}
In this section we discuss the statistical formulation we use to
address comparisons between different hypotheses as well as relevant
measurements for the characterization of an HLL resonance. We focus on
four statistical approaches:
\begin{itemize}
\item{(1)} Search analysis of a signal in the presence of backgrounds.
\item{(2a)} Comparisons between two ``pure'' spin-parity hypotheses
  (such as $0^+$ vs.~$1^-$).
\item{(2b)} Comparisons between two spin-parity hypotheses, with at
  least one of the two being an ``impure'' admixture of two pure HLL
  states (e.g.~$0^+$ vs.~a combination of $1^+$ and $1^-$). This case
  is similar to (2a), except for the presence of one or more nuisance
  parameters.
\item{(3)} The measurement of mixing parameters in the case of impure
  Higgs look-alikes.
\end{itemize} 
In case (1) we consider two hypotheses. $\mathbb{H}_1$ is the
``standard Higgs signal plus background'', and $\mathbb{H}_0$ is the
null, ``background only'' hypothesis.

Cases (2) and (3) involve attempting to establish the nature of a newly
discovered particle. Guided by our results on $_sPlots$, we contend
that it is a very good approximation to confront two different
``signal'' hypotheses in the absence of background -- the latter having been
statistically subtracted. This assumes that a resonance mass peak has
already been established.

The case (2) hypotheses refer to an $m_H$ peak with two different
$J^P$ interpretations. In the (2a) case the two hypotheses under
consideration are simple, i.e.~the corresponding likelihoods are
fully specified once the values $\vec{X}$ are fixed.  In the (2b) case
the unknown mixing angles for the impure hypothesis, referred to as $\vec{\xi}$ (and including
e.g.~various mixing angles $\xi$ and $\delta$ as discussed in
Sec.~\ref{sec:spinone}), are treated as
nuisance parameters. The analysis in case (3) is a traditional
parameter estimate, based on the ML fit, for which we obtain a
confidence interval by using the Feldman-Cousins
approach~\cite{FC}. We discuss the three cases starting from the last.

\subsection{Coupling admixtures}
\label{sec:case3}
Consider the example of a one-parameter mixture of two types of $HZZ$
coupling, such as the composite case discussed in Sec.~\ref{compo}.
For a fixed value of the resonance mass $m_H$ and the mixing angle
$\xi$, Eq.~(\ref{splusd}) is the theoretical
probability-distribution of the events as a function of the variables
$\vec{X}$ for $ZZ$ and $ZZ^*$ final states.  The experimental {\it pdf} 
  is a numerical representation of the result of sieving
--through a specific detector and its resolution, trigger and analysis
requirements-- a very large number of events, generated with the
theoretical {\it pdf} of Eq.~(\ref{splusd}). This experimental
 {\it  pdf}, referred to as $P$, is a function $P$$=$$P_{m_H}(\xi,\vec X)$ of
$m_H$, (which is kept fixed through this exercise), $\xi$, and
$\vec X$. The dependence on $\vec\Omega\equiv\{{\rm
  cos}\,\Theta,\Phi\}$ is, in this example, exclusively a phase space
acceptance effect.

Many experiments with a fixed number of events $N_S$ are simulated,
assuming the same detector response.  The probability of each event,
evaluated with the experimental {\it pdf}, is $P_i$.  The likelihood of
a given experiment is ${\cal L}(\xi)=\prod_{i=1}^{N_S}P_i$. The
experimentally measured value of the $\xi$ parameter,
$\hat{\xi}$ corresponds to the value that maximizes ${\cal L}(\xi)$.  
The simulation is repeated many times, as a function
of the true value of the mixing angle $\xi$. Running many
experiments one can derive the confidence interval, i.e.~the range
covering the true value of $\xi$ for some confidence level
and some measured value $\hat{\xi}$~\cite{FC}.

\begin{figure}[htbp]
  \begin{center}
    \includegraphics[width=0.38\textwidth]{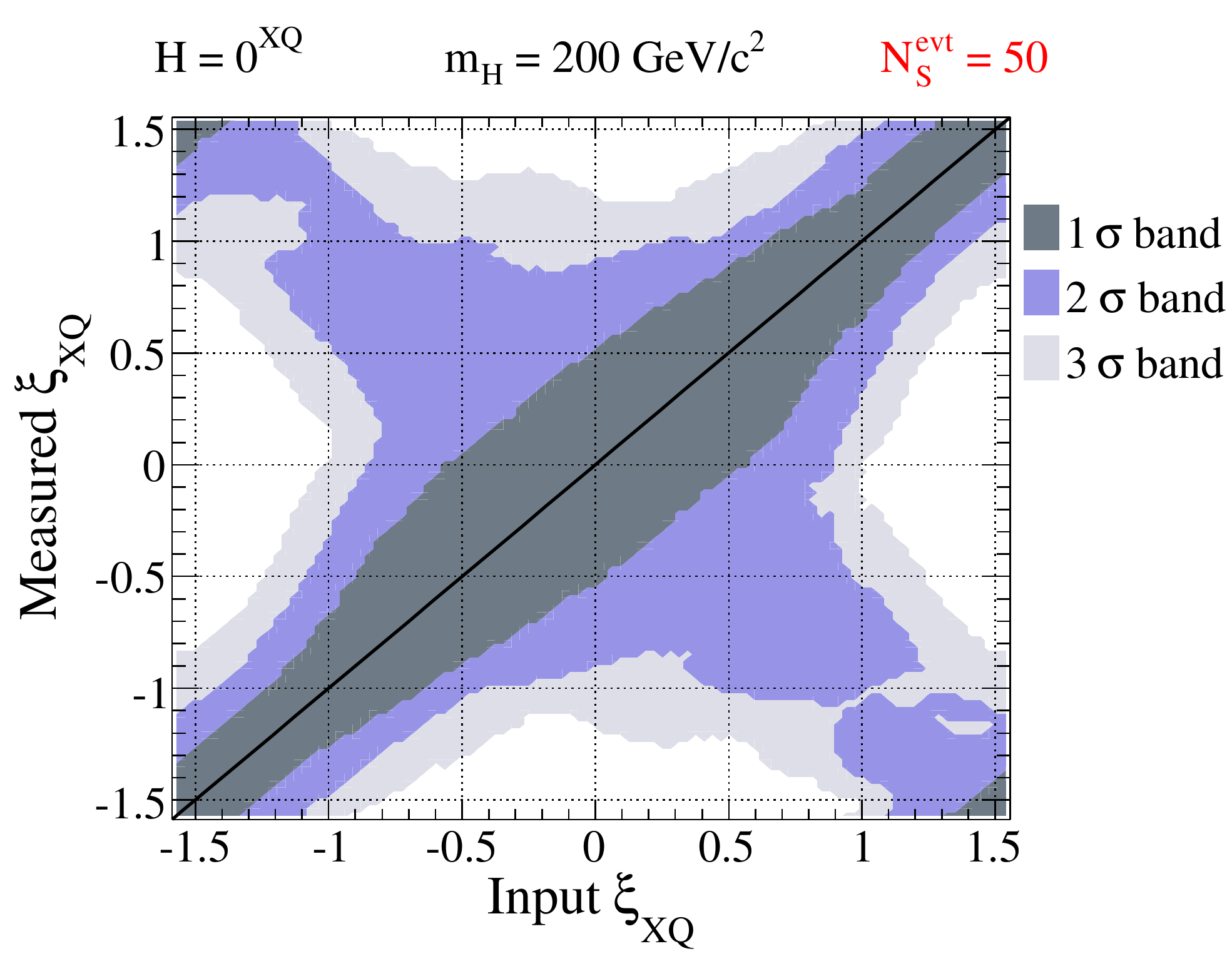}
    \caption{Confidence intervals for measured values of $\xi_{XQ}$
      for a $C$-violating $J$$=$$0$ resonance with a mass 200
      GeV/$c^{2}$.\label{fig:exampelFC}}
  \end{center}
\end{figure}

It is customary to estimate the error (or the number $n$ of standard
deviations $\sigma$) in the measured $\xi$ from the expression
${\cal L}(\xi_{max}\pm n\,\sigma)={\cal L}(\xi_{max})-n^2/2$.
While this method is accurate for large samples with Gaussian errors,
it is not the one used to draw the $\sigma$ contours in
Fig.~\ref{fig:exampelFC} (where $\xi$$=$$\xi_{XQ}$ 
as given in Eq.~(\ref{eq:XQdef}) and in the similar figures of
Sec.~\ref{sec:hllresults}). Instead, the confidence level (CL) is
evaluated measuring the frequency of a given result in the set of
generated pseudo-experiments.

\subsection{Confronting $J^P$ hypotheses}
\label{sec:case2}
\begin{figure}[b!]
\begin{center}
  \includegraphics*[width=0.50\textwidth]{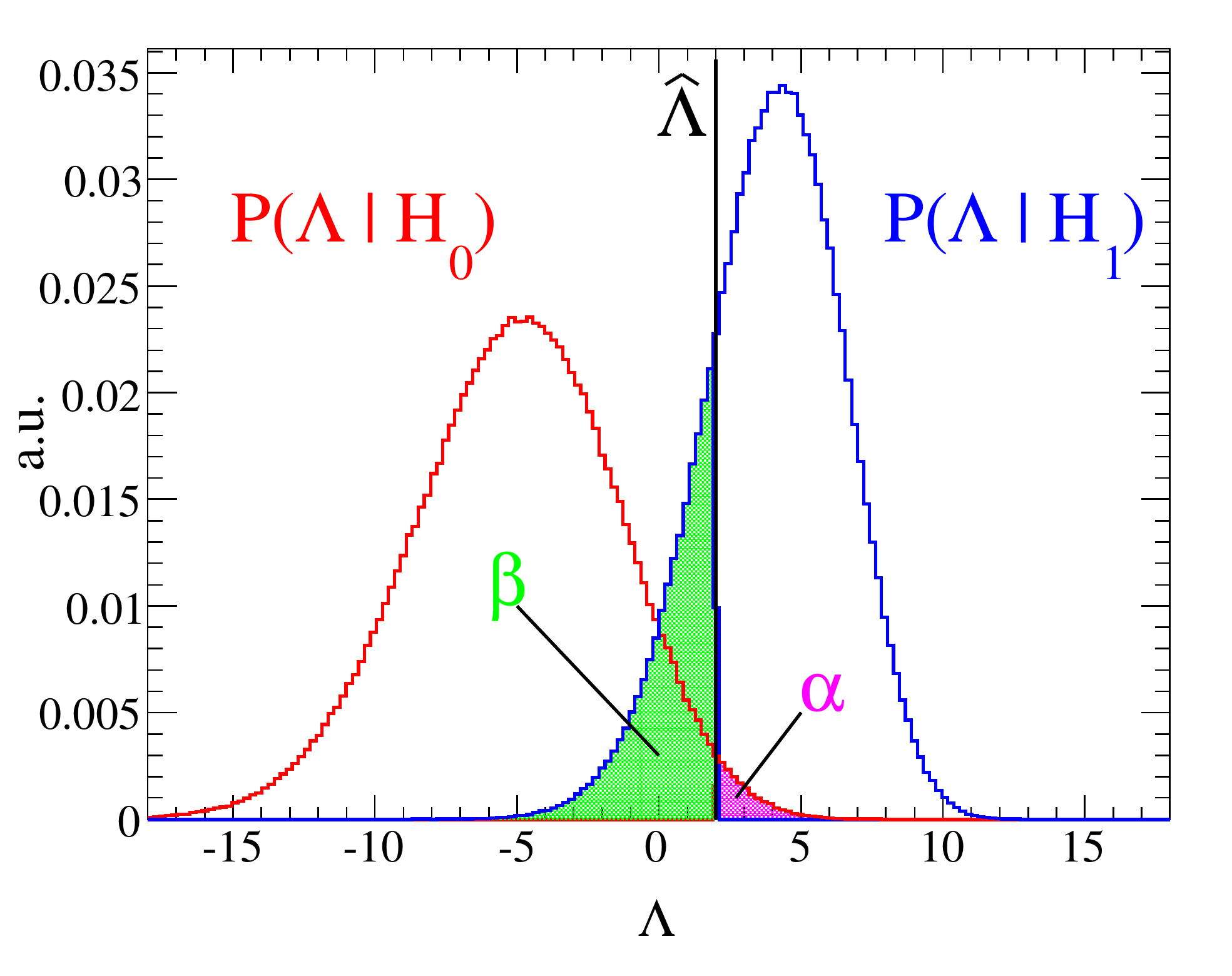}
  \caption{Distribution of $\Lambda$ for $m_H$$=$$200$ GeV/c$^2$ and
    $N_{S}$$=$$23$, constructed with $\sim10^{9}$ pseudo-experiments.  The
    hypotheses being confronted are $\mathbb{H}_{0}$$=$$0^+$ and
    $\mathbb{H}_{1}$$=$$0^-$.\label{fig:LambdasA}}
\end{center}
\end{figure}

Consider two hypotheses, $\mathbb{H}_{0,1}$, for the spin-parity
assignment of a signal candidate sample, detected via its $ZZ$ mass
peak and background-subtracted using the $_sPlot$ method. Large
numbers of events are generated assuming each hypothesis and used to
construct two unbinned experimental {\it pdfs}:
$P_{\mathbb{H}_{0,1}}\equiv P_{m_H}(\vec X\!\mid\!
\mathbb{H}_{0,1})$. 
For our pure spin-parity cases, the simple nature of the hypotheses
considered guarantees through the Neyman-Pearson (NePe)
lemma~\cite{NePe} that the hypothesis test is {\it universally most
  powerful}.
Next, we explicitly identify  one hypothesis as $\mathbb{H}_0$ and
the other as $\mathbb{H}_1$.
Additionally, we specify the test {\it statistic} $\Lambda$
which we define as the log-likelihood ratio $\log[{\mathcal
  L}(\mathbb{H}_{1})/{\mathcal L}(\mathbb{H}_{0})]$.
Finally, we must a priori choose the acceptable probability level
$\alpha$
of rejecting $\mathbb{H}_{0}$ in favor of $\mathbb{H}_{1}$, even
though $\mathbb{H}_{0}$ is true (Type I error).
We generate a series of pseudo-experiments with a fixed number of
events $N_S$ to construct the {\it pdf}
of $\Lambda$ for the two hypotheses.
A typical result is illustrated in Fig.~\ref{fig:LambdasA}.
We first generate pseudo-experiments considering  $\mathbb{H}_{0}$ as true.
For each experiment we construct two likelihoods ${\cal
  L}(\mathbb{H}_{0})\equiv\prod_{i=1}^{N_S}P_{\mathbb{H}_{0}}(\vec
X_i)$ for the correct interpretation of the true theory, and ${\cal
  L}(\mathbb{H}_{1})\equiv\prod_{i=1}^{N_S}P_{\mathbb{H}_{1}}(\vec
X_i)$ for its incorrect interpretation.  With the ensemble of
experiments one constructs the distribution $P(\Lambda\!\mid\!
\mathbb{H}_{0})$ with $\Lambda\equiv \log[{\mathcal
  L}(\mathbb{H}_{1})/{\mathcal L}(\mathbb{H}_{0})]$.
The result is the leftmost (red) curve in Fig.~\ref{fig:LambdasA}.
The exercise is repeated with the pseudo-experiments generated
considering $\mathbb{H}_{1}$ as true
and the result is the rightmost (blue) curve in the figure.
An a priori chosen value of $\alpha$
implicitly defines a value $\hat \Lambda(\alpha)$ via
\begin{equation}
  \alpha=\int_{\hat \Lambda(\alpha)}^{\infty}P(\Lambda\!\mid\! \mathbb{H}_{0})\,d\Lambda~.
\label{eq:alphadefA}
\end{equation}
This fixed value $\hat \Lambda(\alpha)$ implies that
\begin{equation}
  \beta(\alpha) = \int_{-\infty}^{\hat \Lambda(\alpha)}P(\Lambda\!\mid\! \mathbb{H}_{1})\,d\Lambda
\label{eq:betadefA}
\end{equation}
is the probability of accepting $\mathbb{H}_{0}$ even though
$\mathbb{H}_{1}$ is correct (Type II error).  The value $1-\beta$ is
called the {\it power} of the test.  When the real experiment is
performed, a specific value $\Lambda_{\rm exp}$, is obtained for
$\Lambda$.  The associated {\em p-value} $=\int_{\Lambda_{\rm
    exp}}^{\infty}P(\Lambda\!\mid\! \mathbb{H}_{0})\,d\Lambda~,$ is
compared to $\alpha$ to determine if the measurement favors one
hypothesis versus the other.

Instead of the $\alpha$ and $\beta$ values, the significance
$\sigma$ is commonly used.  To convert to an equivalent number of
$\sigma$'s using Fig.~\ref{fig:LambdasA} we calculate the same
$\alpha$-area in a Gaussian distribution centered at 0 with
$\sigma$$=$1.  The number $n$ of $\alpha$-equivalent standard deviations
is obtained by inverting
\begin{equation}
\alpha = \frac{1}{\sqrt{2\pi}}\int_{n}^{\infty}dx\, e^{-x^2/2} ~.
\label{nofalpha}
\end{equation}

The a priori (subjective) choice of $\alpha$ (and subsequently $\beta$
and corresponding significances) is heavily discussed in the
literature. {\it The Physical Review}, for example, requires a
$5\sigma$ ($3\sigma$) significance to claim discovery (evidence). The
caveat is, of course, that when one minimizes as much as possible the
probability of an error of Type I (wrongly claiming a discovery) one
risks making an error of Type II (and e.g.~delaying the claim of a
discovery to the next luminosity upgrade).

A pure vs.~impure HLL hypothesis test has an additional complication
due to the dependence of the likelihood function on the mixing angles
$\vec{\xi}$ in at least one of the two hypotheses. In this case, we
are testing the simple (i.e.~mixing angle independent) hypothesis
against a class of alternative hypotheses, connected by the variation
of a continuous unknown parameter(s).  The test is performed by
comparing the simple hypothesis to the impure hypothesis with values
of $\vec{\xi}$ that best fit the data.

The impure vs.~impure Higgs look-alike test is technically identical
to the pure vs.~impure. Here, we try to exclude some value of the
mixing angle parameter for one of the two composite hypotheses in
favor of the alternative impure hypothesis, where the mixing angles
are treated as nuisance parameters.  With fixed mixing angles, one
impure look-alike becomes a simple hypothesis (like a pure one) tested
against an impure hypothesis.

\subsection{Higgs searches}
\label{sec:case1}
When searching for a new particle two hypotheses are tested against
each other: the background-only, $\mathbb{H}_{0}$, and signal plus
background, $\mathbb{H}_{1}$.

Assuming that the event distributions for signal and background are
fully specified (an unrealistic situation in that the value of the
Higgs mass and width are not known a priori), one still has to
determine the signal and background yields. Hence, the likelihood
function has a parametric dependence on at least one nuisance
parameter. There is no guarantee that the Neyman-Pearson construction
is, in this case, the optimal hypothesis test one could perform.  We
are, however, not concerned with what the optimal statistical test is,
but rather on the physics content of the likelihood function.  Our aim
is to illustrate how different analyses that fully or partially
exploit the information in the theoretical {\it pdfs} compare with
each other. For this purpose it is sufficient to use a consistent
statistic among the various cases and discuss their relative
merits. We still perform a hypothesis test based on the likelihood
ratio. The dependence on the nuisance parameters is removed through a
maximization (profiling) of ${\cal L}(\mathbb{H}_{0,1})$ relative to
the nuisance parameter(s), prior to the construction of the likelihood
ratio, as done for the case of impure hypothesis-testing discussed in Sec.~\ref{sec:case2}.

Given a specific analysis setting (i.e.~a set of variables defining
the likelihood function) we evaluate its discovery power by computing
the significance (the number of $\sigma$'s) as a function of
the signal yield and for different values of the ratio of signal over background
yields. We define an expected value for the signal to background
ratio, $\langle N_S/N_B\rangle$, between the signal events
constituting the $m(ZZ)$ peak and the integral of the background
distribution in the same variable in the range 190 GeV/c$^2$ to 600
GeV/c$^2$. To address the uncertainties, we compare the two hypotheses
for various pre-selected values of $\langle N_S/N_B\rangle$, in a
large range including and bracketing the central current expectation.
The likelihood for $\mathbb{H}_{0}$ is then that of
Eq.~(\ref{likelihood}), expressed as a function of the angular
variables at fixed $m_H$, as opposed to a function of only
$m_H$. 

When adding the $\vec X$ variables to the likelihood, one should
consider the event-by-event dependence of their {\it pdf} on the value
of $m_H$. This can done using a different $\vec X$ {\it pdf} for each
bin of the template functions of Fig.~\ref{fig:simpdf}. This step is
straightforward when performing the real analysis, but CPU intensive
when performing hundreds of billions of pseudo-experiments.  The
resonance mass is narrow enough for the peak to be determined, in which
case the results are very close to the ones obtained with the full
mass-dependence of the $\vec X$ {\it pdf}.  For simplicity we did not
include the finite width of the resonance in the likelihood.

In our search results we compare the significance, as given by an
$m_H$-based peak search, with the corresponding quantity following
from the whole angular-distribution analysis.  In the case of a
discovery test, the $p$-value of any toy experiment is compared to the
equivalent of a $\ge 5\,\sigma$ significant $p$-value, in order to
establish if a discovery could be claimed for that experiment. By
repeating the exercise many times, we can associate a probability to
the discovery potential. The $5\, \sigma$ convention fixes the value
of $\alpha$ for the hypothesis test, as well as the value of $\beta$
for a given likelihood function.

%%%%%%%%%%%%%%%%%%%%
%%%%
%%%%      DISCOVERY_EXTRA.TEX
%%%
%%%%%%%%%%%%%%%%%%

\section{Signal significance using the angular
  information\label{sec:disc}}
As described in the two previous sections, discrimination of Higgs
look-alikes first requires an event sample following a putative Higgs
discovery.  
As noted already, the search analysis could be model-independent,
relying only on the reconstruction of a resonant excess over
non-resonant backgrounds. In this case a discovery is completely
factorized from its characterization.

Despite the natural factorization between discovery, HLL
discrimination based on production, and HLL discrimination based on
decay, it is important to check the consistency of the entire chain of
analysis.  This is especially true for the small datasets considered
here, where we demonstrate HLL discrimination with datasets
comparable to the original discovery sample.

A powerful check is to 
%fix the extracted (or assumed) values of $m_H$
%and $<N_{S}/N_{B}>$, and then 
compare the signal significance of two nominal analyses:
\begin{itemize}
\item An ``$m(ZZ)$ only'' fit, for which the discrimination between
  signal and background is given only by the $ZZ$ invariant-mass peak.
  This is an example of a model-independent discovery analysis
  (although not necessarily the actual discovery analysis used in the
  experiment).
\item An ``$m(ZZ)$$+$$\vec X$'' fit, in which the {\it pdf} for the
  angular variables $\vec X$ is also included.  Thus here we are
  using the angular information to improve the discrimination of the
  signal from the background, rather than discriminate SM Higgs from
  HLLs. The {\it pdf} of $\vec X$ corresponds to the 
  value of $m_H$ as extracted from the fit. 
\end{itemize} 
We compare the signal significance of the two analyses, corresponding
to different physics content for the likelihood function. A common
statistical framework is used, since we are interested to compare the
physics performance rather than determining the optimal statistical
approach.  The overall normalization is obtained by assuming
$\sqrt{s}$$=$$10$ TeV with a corresponding SM Higgs production cross
section \cite{CMStdr1}. 

A direct comparison of the two analyses in a common framework is a way
to quantify the price to pay in order to run a completely
model-independent search. At the same time, it is a consistency check
on the HLL discrimination analysis, since the background events are
themselves Higgs imposters. If, as we claim, HLL discrimination is
possible with datasets not much larger than, or identical to, the
original discovery sample, then we should also find that the
``$m(ZZ)$$+$$\vec X$'' fit offers comparable improvements in signal
significance over the ``$m(ZZ)$ only'' fit, for similarly small
datasets.

To make the likelihood comparison meaningful, a common fit setting is
used.  For the $ZZ$ invariant mass, we consider the range $190 < m_{H}
< 600$ GeV/c$^2$. The fit configuration is specified by the nominal expected
signal-over-background yield ratio $\langle N_{S}/N_{B} \rangle$ and
by the nominal number of signal events $N_S$.  We consider different
scenarios by fixing different values of $\langle N_{S}/N_{B} \rangle$
and perform the study as a function of $N_S$.

For each fit configuration we run a set of toy Monte Carlo
experiments. The actual number of background events are generated
according to a Poisson distribution around the nominal value, and the
event-by-event values of the variables used in the fit ($m_H$ and, if
used, $\vec X$) are randomly generated according to the signal and
background {\it pdfs}. The fit is then performed for each toy sample,
maximizing the likelihood as a function of the signal and background
yields and the value of $m_H$. The sets of fits provide a distribution
for the statistical significance obtained in a particular experiment.
%
% and in particular a $68\%$ probability range for the
%spread in this significance.

\begin{figure}[htbp]
\begin{center}
\includegraphics[width=0.23\textwidth]{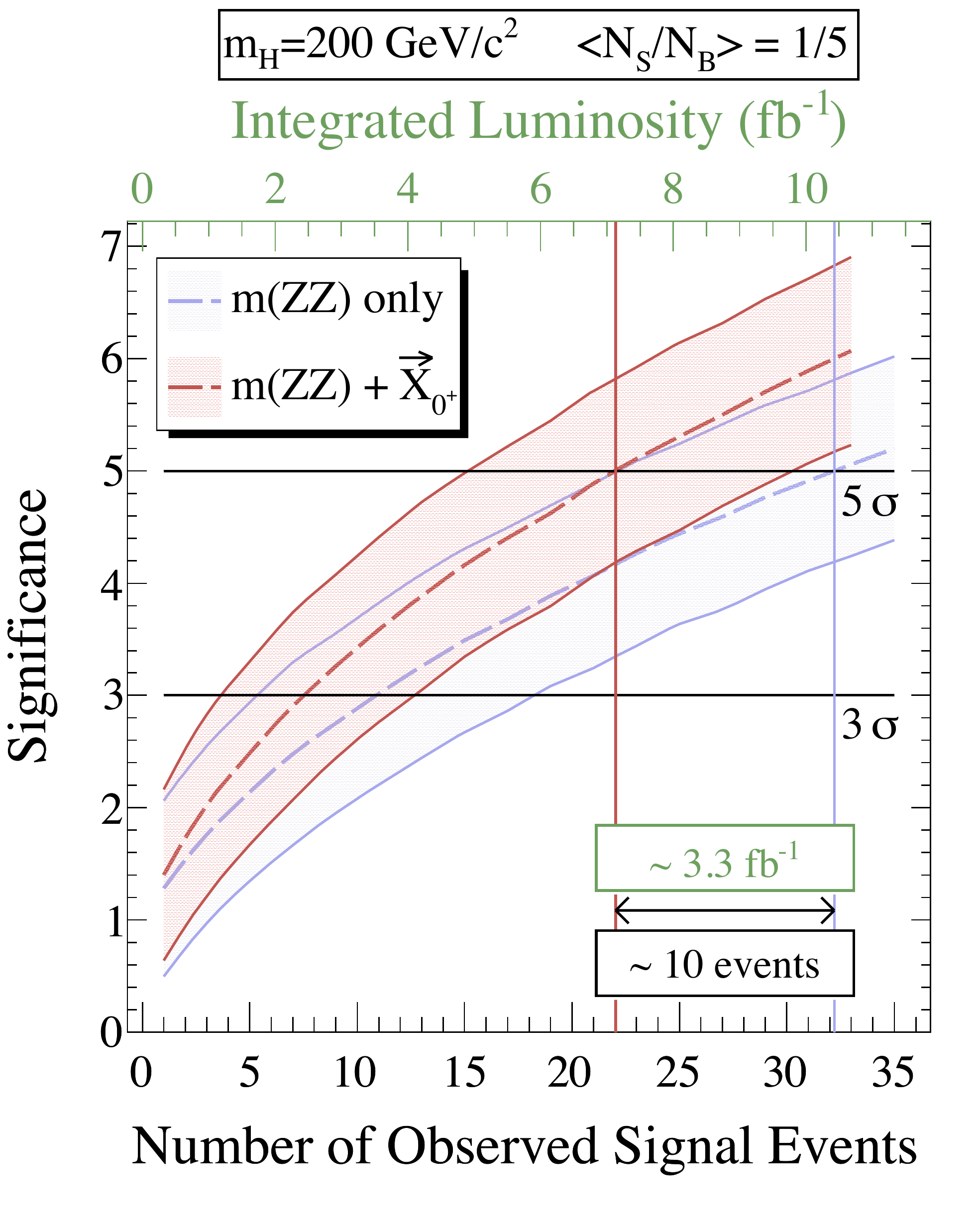}
\includegraphics[width=0.23\textwidth]{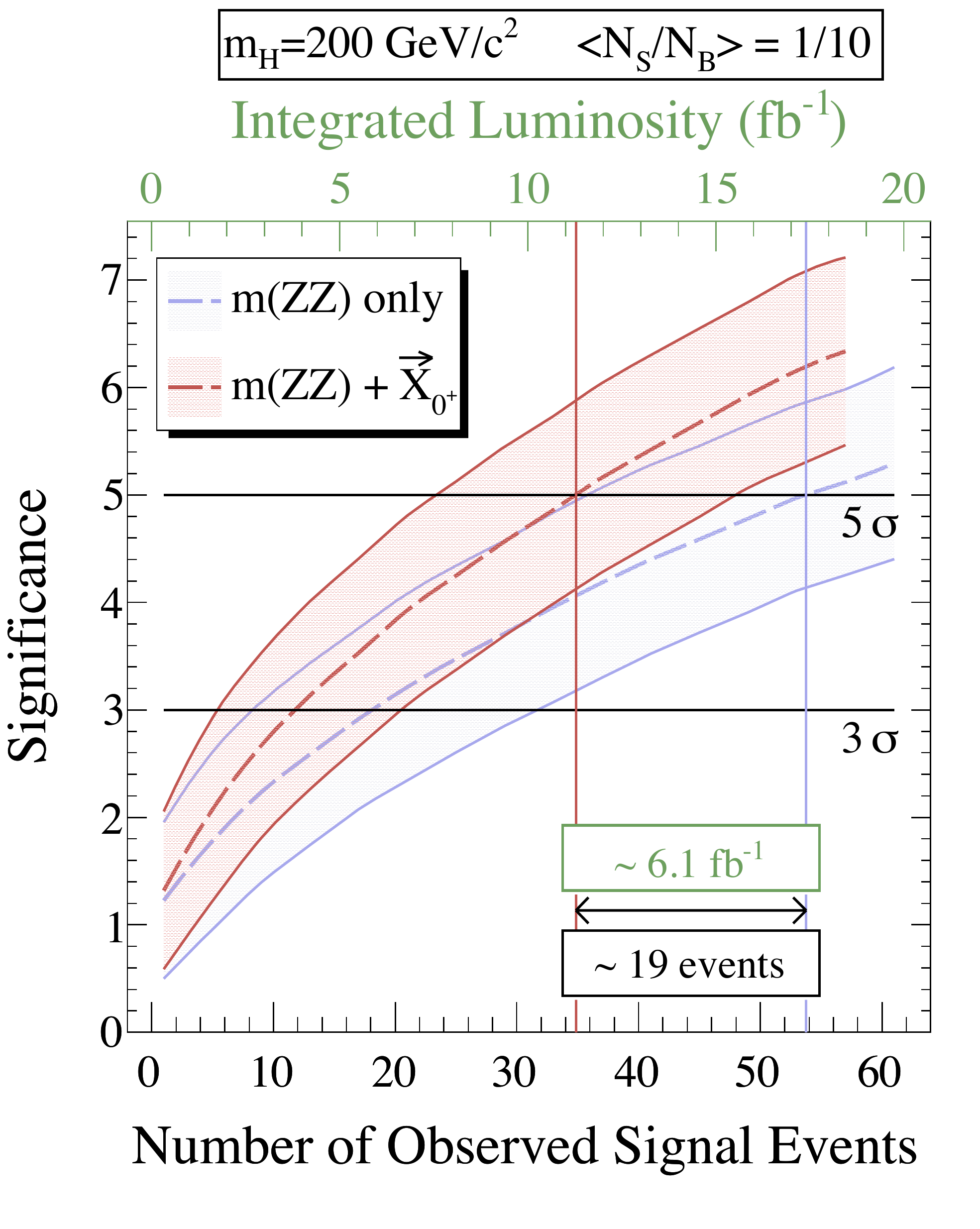}
\caption{ Distribution of signal significance for a 200 GeV/$c^{2}$ SM
  Higgs boson decaying in the $H \to ZZ \to 4\mu$ channel for $pp$
  collisions with $\sqrt{s}$$=$$10$ TeV.  The mean signal to background
  ratios used are $\langle N_{S}/N_{B} \rangle$$=$ 1/5 (left) and 1/10
  (right).\label{fig:DISC_200}}
\end{center}
\end{figure}

\begin{figure}[htbp]
\begin{center}
\includegraphics[width=0.23\textwidth]{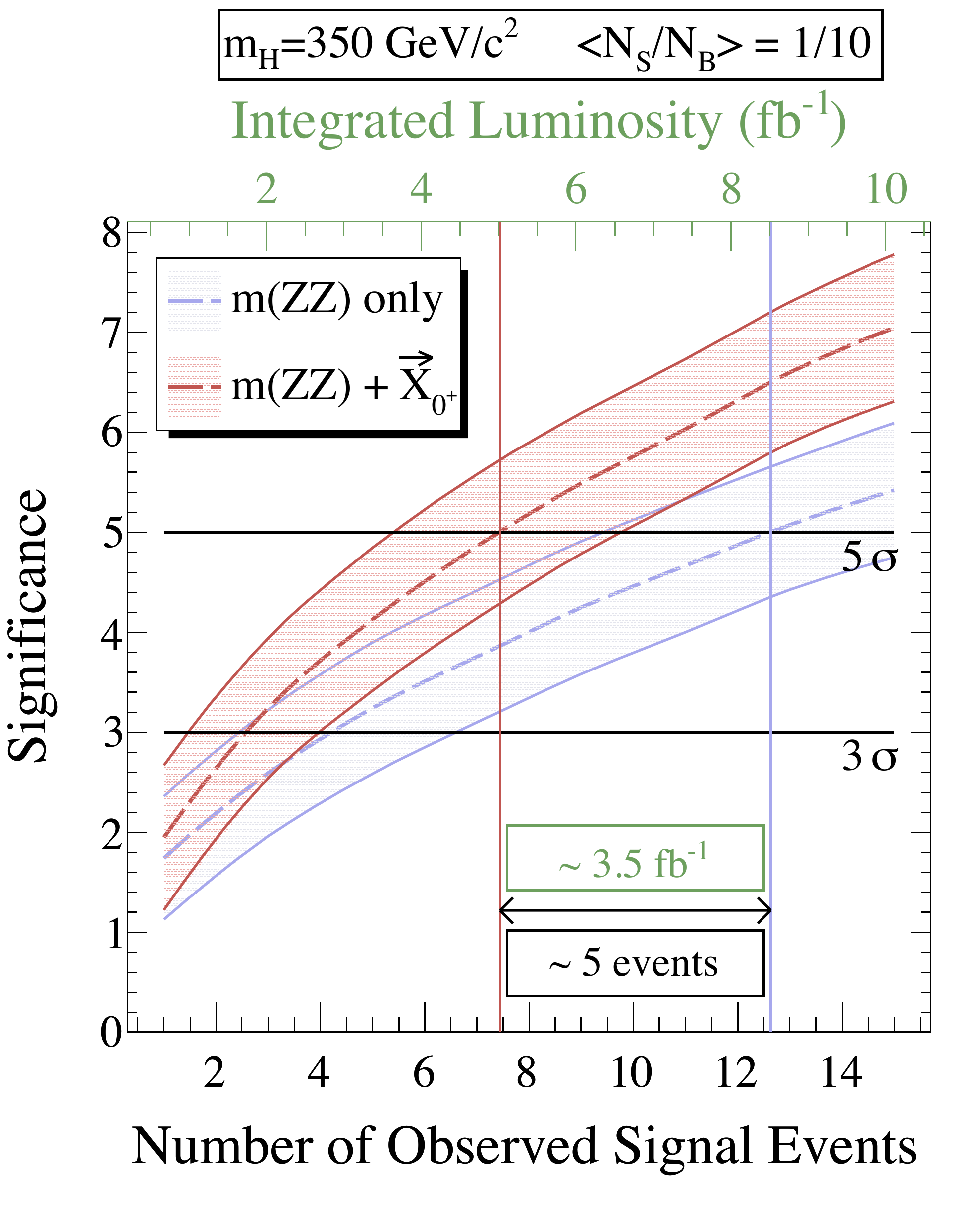}
\includegraphics[width=0.23\textwidth]{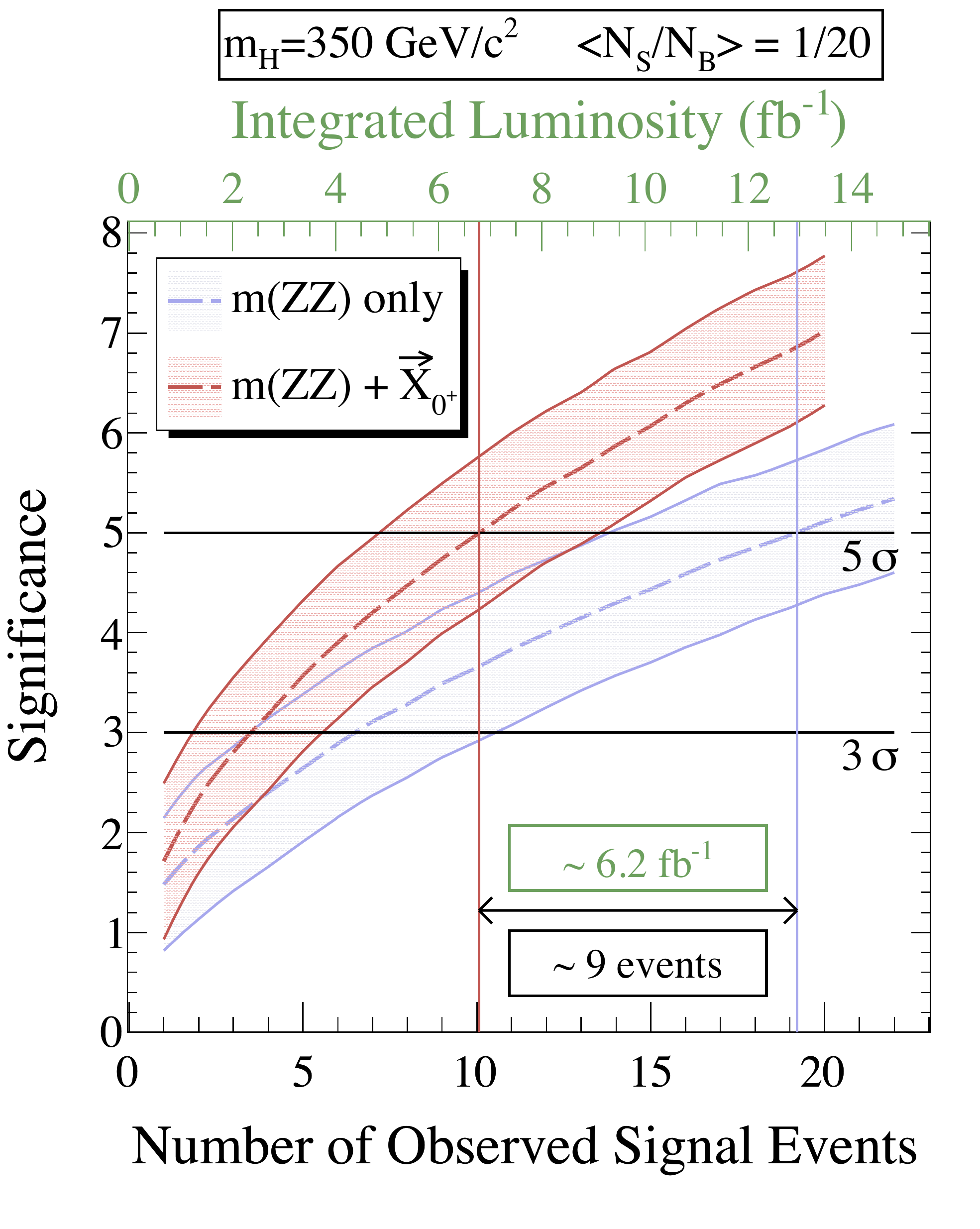}
\caption{Distribution of signal significance for a 350 GeV/$c^{2}$ SM
  Higgs boson decaying in the $H \to ZZ \to 4\mu$ channel for $pp$
  collisions with $\sqrt{s} $$=$$10$ TeV.  The mean signal to background
  ratios used are $\langle N_{S}/N_{B} \rangle$$=$ 1/10 (left) and 1/20
  (right).\label{fig:DISC_350}}
\end{center}
\end{figure}

\begin{figure}[htbp]
\begin{center}
\includegraphics[width=0.23\textwidth]{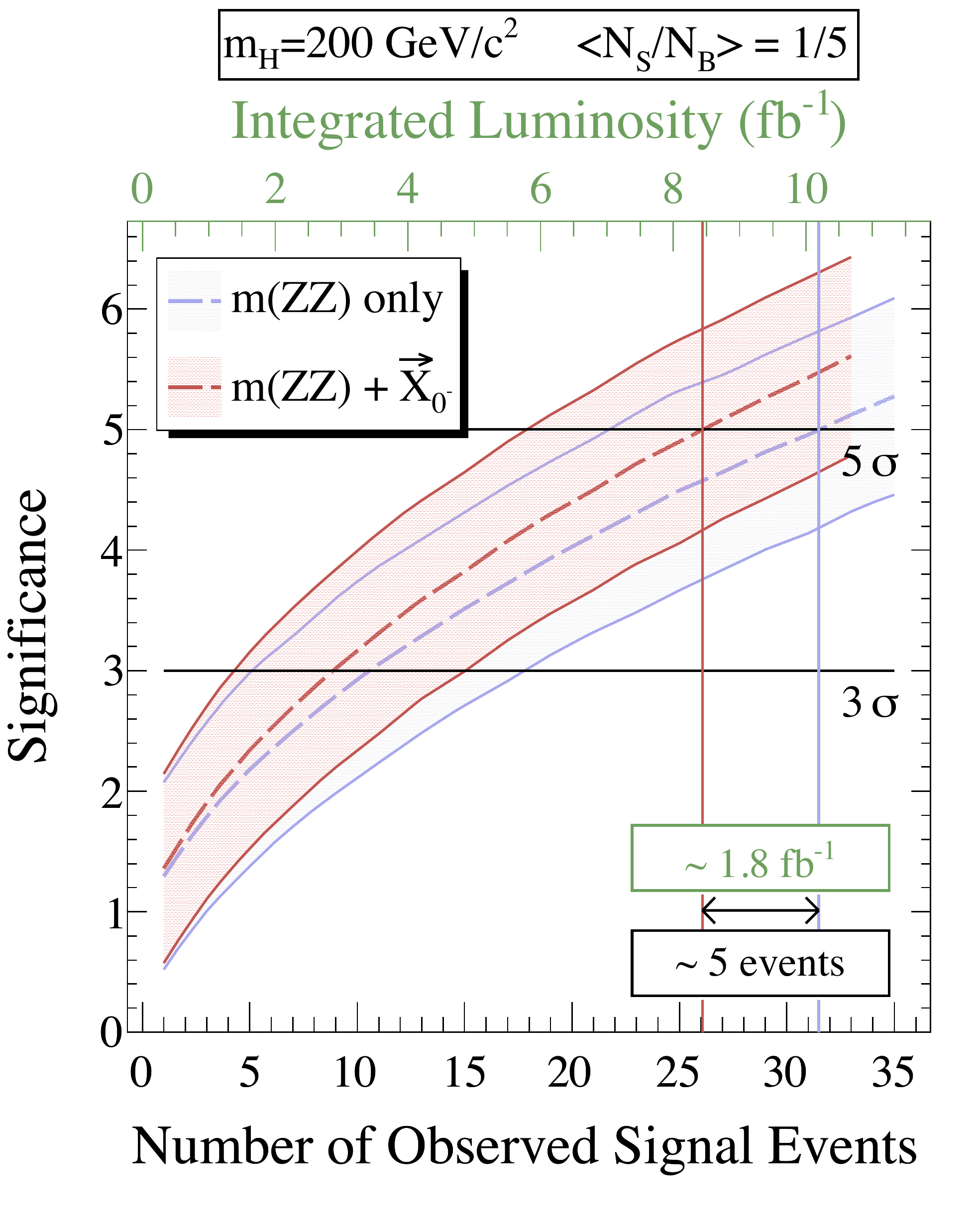}
\includegraphics[width=0.23\textwidth]{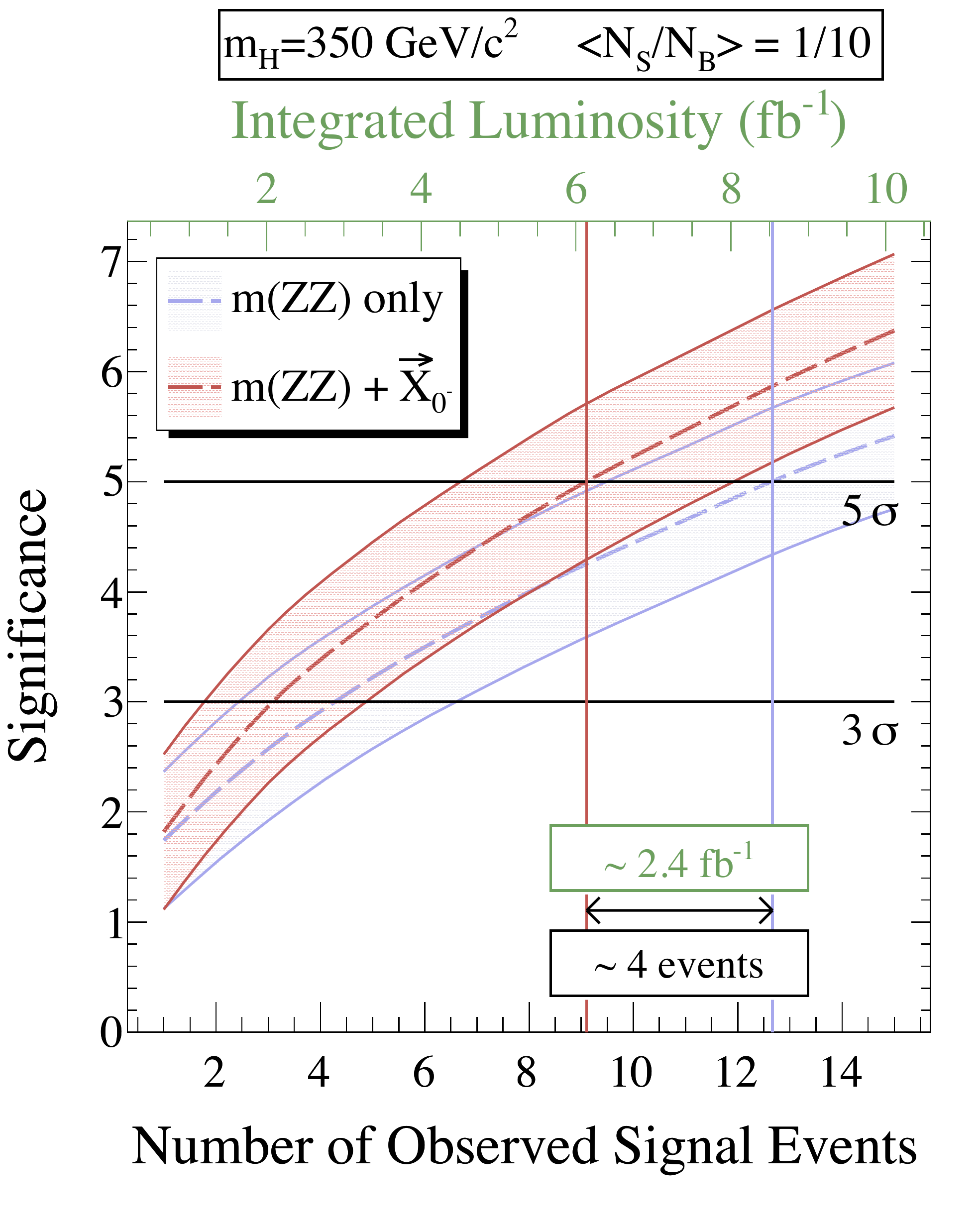}
\caption{Distribution of signal significance for a 200 (left)
  and 350 (right) GeV/c$^2$ pseudoscalar particle in the $0^{-} \to
  ZZ \to 4\mu$ channel for $pp$ collisions with $\sqrt{s}$$=$$10$ TeV.
  The mean signal to background ratios used are $\langle N_{S}/N_{B} \rangle$$=$ 1/5
  (left) and 1/10 (right), and assumes an NLO SM Higgs production cross
  section for the pseudoscalar.\label{fig:DISC_PS}}
\end{center}
\end{figure}

This is summarized in 
Figures \ref{fig:DISC_200}-\ref{fig:DISC_PS}.  
The two bands in the figures correspond to the
spread (at $68\%$ confidence level) for the signal significance
achieved in a single experiment, as a function of the signal yield
$N_S$, for the ``$m(ZZ)$ only'' fit (light band) and the
``$m(ZZ)$$+$$\vec X$'' fit (dark band).  The horizontal lines show the
$3\sigma$ and $5\sigma$ thresholds ({\it evidence} and {\it
  discovery}, in the usual convention). The intersection with each
band provides a corresponding range for the needed signal yield, the
spread being due to statistical fluctuations. For a correct
interpretation of the separation between the two bands, one should
consider that the statistical fluctuations in the two fits are
strongly correlated since they both depend on the invariant mass
observable, and background fluctuations for this mass distribution will
be the same for both.

Figure~\ref{fig:DISC_200} has the case of an $m_H$$=$$200$
GeV/c$^2$ SM Higgs boson, while Fig.~\ref{fig:DISC_350} illustrates
similar results for an $m_H$$=$$350$ GeV/c$^2$ SM Higgs boson. For each
mass, different values for $\langle N_{S}/N_{B} \rangle$ are
considered; we show here the results for $\langle N_{S}/N_{B}
\rangle$$=$$1/5$, $1/10$ for $m_H$$=$$200$ GeV/c$^2$ and $\langle
N_{S}/N_{B} \rangle$=$1/10$, $1/20$ for $m_H$$=$$350$ GeV/c$^2$.  We note
that better discrimination between the signal and background in the
higher mass case (compared to the lower mass) especially in the
invariant mass observable; despite the lower cross section this
results in higher significance for the higher mass case for the
same luminosity.
%%%%%%%%%%%%%%%%%%%%%%%%%%%%%%%%%%%%%%%%%
%%%%%%%%%%%%%%%%%%%%%%%%%%%%%%%%%%%%%%%%%
%%%%%%%%%%%%%%%%%%%%%%%%%%%%%%%%%%%%%%%%%

Similarly, Fig.~\ref{fig:DISC_PS} has the results for
$m_H$$=$$200$ GeV/c$^2$ and $m_H$$=$$350$ GeV/c$^2$ pseudoscalar HLLs. Here
the input parameters (such as the cross section) are assumed to be
those of a SM Higgs boson; only the shape of the {\it pdfs} defining
the likelihood (and in particular the correlations between the angles)
are different from the SM case.  The angular distributions and
correlations for a pseudoscalar resonance are similar to those of the
$ZZ$ background, resulting in a much smaller improvement in the signal
significance over the ``$m(ZZ)$ only'' fit, and thus a smaller
distance between the two bands in the plots.

%%%%%%%%%%%%%%%%%%%%%%%%%%%%%%%%%%%%%%%%%
%%%%%%%%%%%%%%%%%%%%%%%%%%%%%%%%%%%%%%%%%%%%%%%%%%%%%%%%%%%%%%%

\section{Results\label{sec:hllresults}}
We present results for three HLL masses: $m_H$$=$$145$, 200, and 350
GeV/c$^2$, using pseudo-experiments built with the full $\vec X$ {\it
  pdf}.

\begin{boldmath}
\subsection{$0^+$ vs. $0^-$}
\end{boldmath}

We consider here two different ``pure'' scalar hypotheses: $0^{+}$,
corresponding to a SM Higgs, and $0^{-}$, a pseudo-scalar. Neither of
these possibilities has an explicit dependence on the angles
$\vec{\Omega}$ in their differential cross-section, meaning that only
the variables $\vec{\omega}$ (and the off-shell $Z$ mass, 
$m_{2}$$=$$M_{Z^{*}}$, for $m_{H} < 2 M_{Z}$) are used to discriminate
between the two hypotheses.

\begin{figure}[t]
\begin{center}
\includegraphics[width=0.238\textwidth]{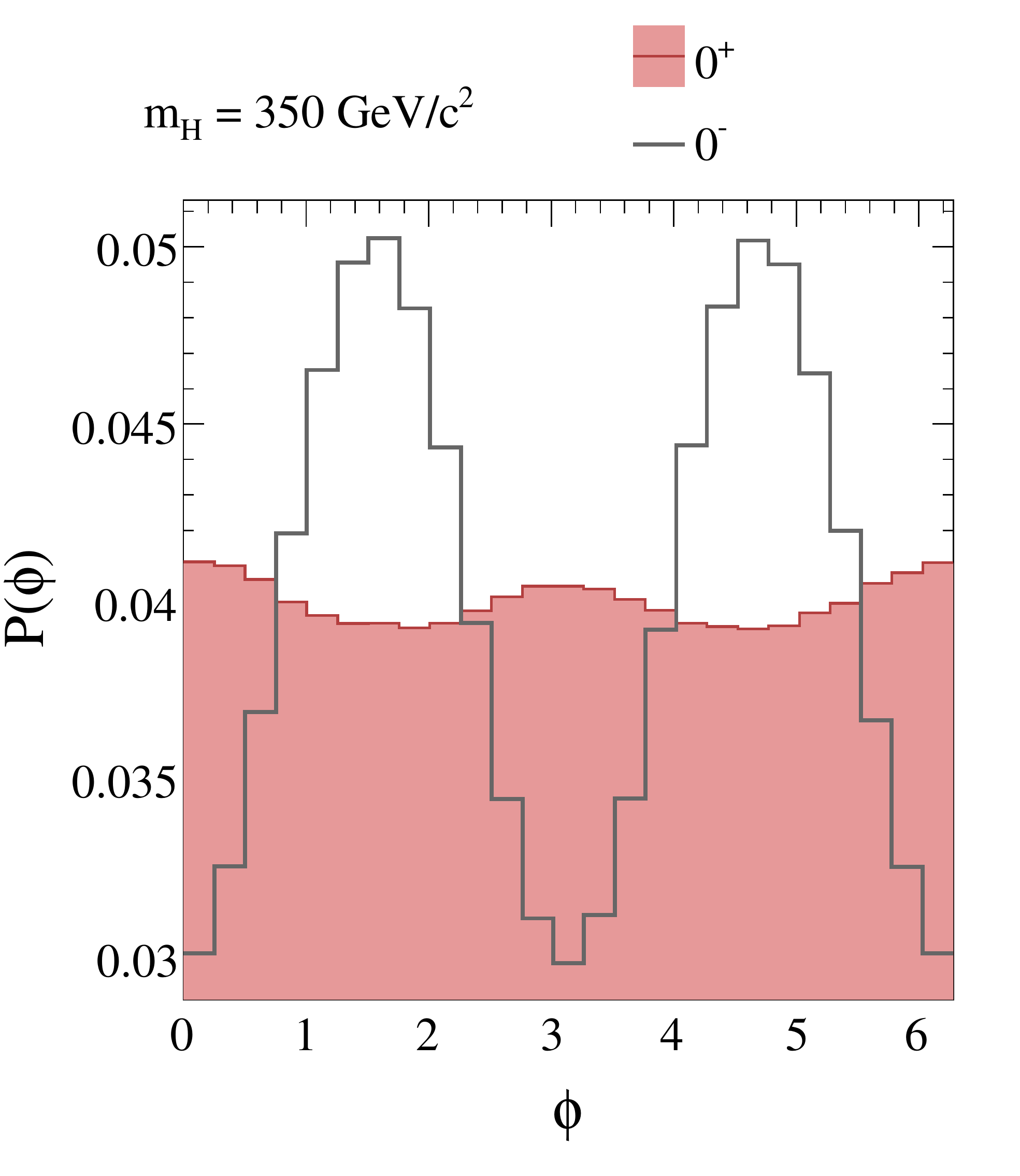}
\includegraphics[width=0.238\textwidth]{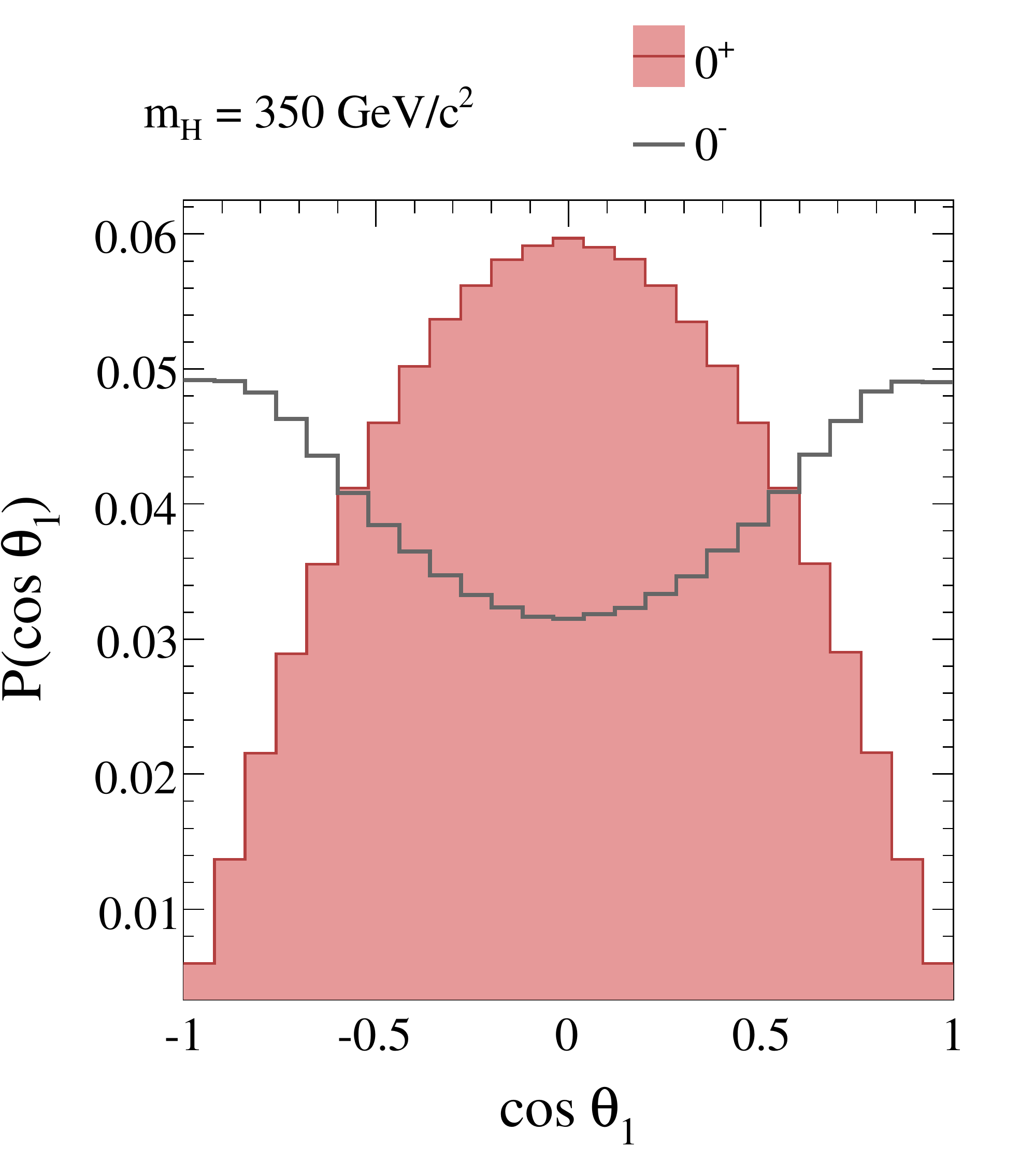}
\caption{Distributions of the variables $\phi$ (left) and
  cos$\,\theta_{1}$ (right) for $0^{+}$ and $0^{-}$ resonances with
  $m_H$$=$$ 350$ GeV/c$^{2}$. All distributions are normalized to a unit
  integral.
  \label{fig:KIN_SM_PS}}
\end{center}
\end{figure}

\begin{figure}[htbp]
\begin{center}
\includegraphics[width=0.238\textwidth]{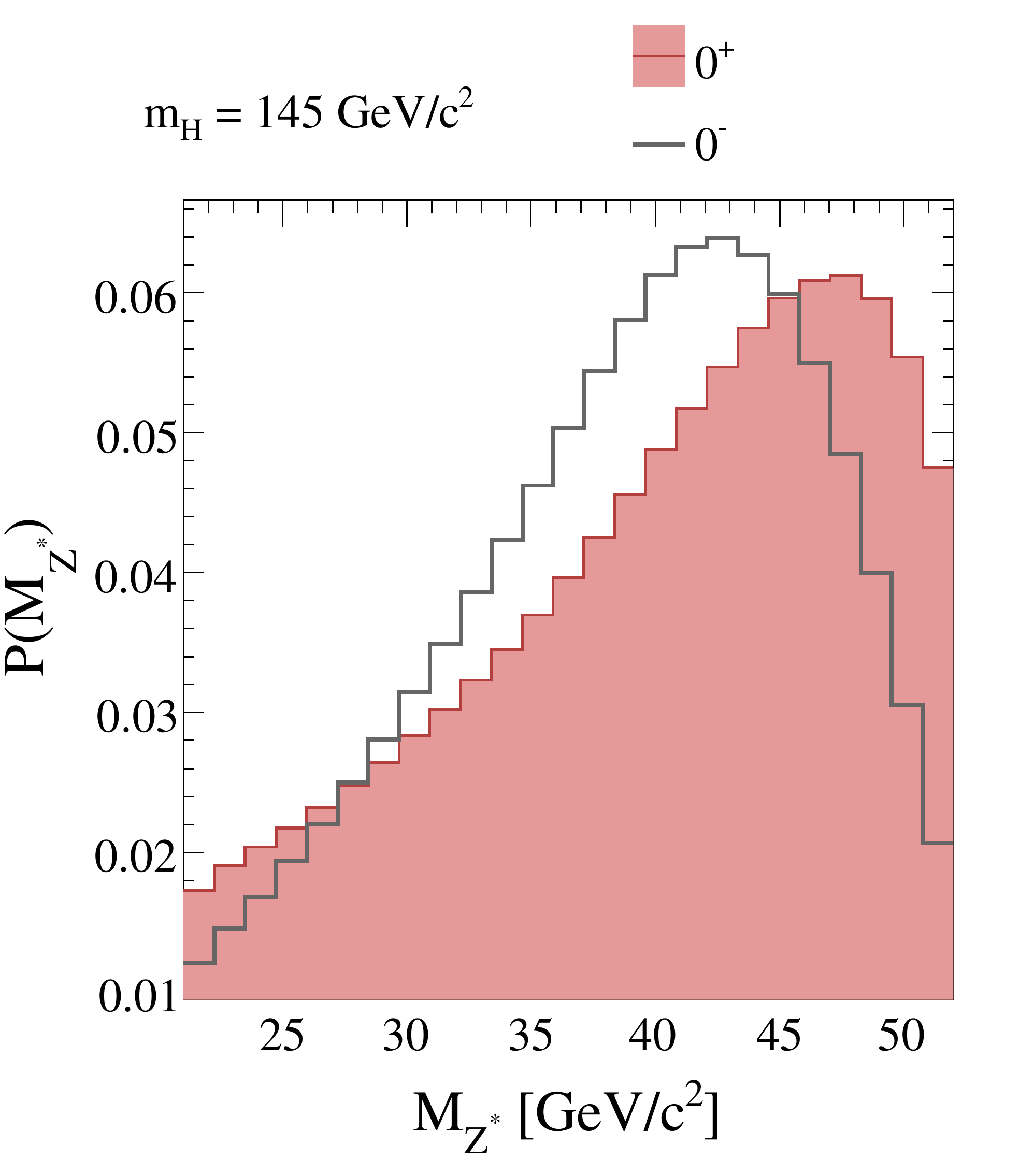}
\includegraphics[width=0.238\textwidth]{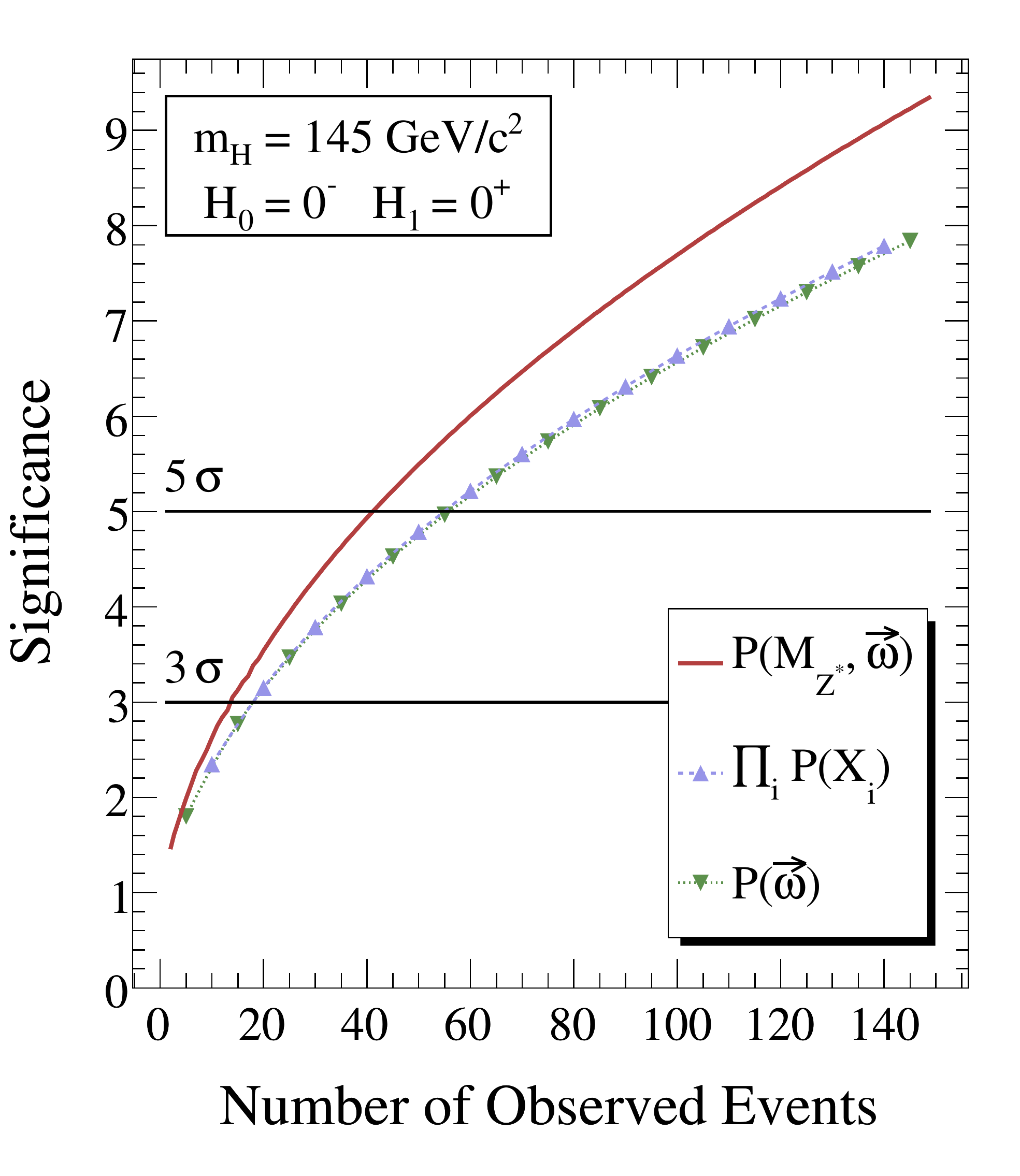}
\caption{Left: Distribution of $M_{Z^{*}}$ for $0^{+}$ and $0^{-}$ 
$H\to ZZ^*$ decays at $m_H$$=$$145$
  GeV/c$^{2}$, normalized to a unit integral.
  Right: Median significance for rejecting $0^{-}$
   in favor of $0^{+}$, assumed to be correct, as a function of $N_S$.
The different likelihood constructions are specified in the text.
\label{fig:SPEC_0_0}}
\end{center}
\end{figure}

In Fig.~\ref{fig:KIN_SM_PS} we show the distributions
in $\phi$ and cos$\,\theta_{1}$ at $m_{H}$$=$$350$ GeV/c$^{2}$ for
$J^P$$=$$0^+$ and $0^-$. These angular variables (along with
cos$\,\theta_{2}$, whose distribution is identical to that of
cos$\,\theta_{1}$ except when $Z_{2}$ is off-shell) provide the
discrimination between these two hypotheses at all masses $m_{H}$.
%%%%%%%%%%%%%%%%%%%%%%%%%%%%%%%%%%%%%%%%%%%%%%%%%%%%%%%%%%%%%%%%%%%
%%%%%%%%%%%%%%%%%%%%%%%%%%%%%%%%%%%%%%%%%%%%%%%%%%%%%%%%%%%%%%%%%%%
For masses $m_{H}$ below the $2\,M_{Z}$ threshold, the kinematic
factors in Eqs.~(\ref{eq:standang}),(\ref{purepseudoscalarP}) result in
the differential cross section dependences on the off-shell $Z$ mass
$M_{Z^{*}}$ that differ for the $0^{+}$ and $0^{-}$ cases. This is
illustated in Fig.~\ref{fig:SPEC_0_0} (left) for $m_{H}$$=$$145$
GeV/c$^{2}$. For all the discriminating variables we consider, the
ability to distinguish between two hypotheses is degraded when their
correlations are neglected.  This is shown in Fig.~\ref{fig:SPEC_0_0}
(right) where we present the results of the NePe hypothesis test
between $0^+$ and $0^-$ for likelihoods built using different subsets
of variables and correlations thereof.  Specifically
$P(M_{Z^{*}},\vec{\omega})$ denotes the use of the full set of
variables while in $P(\vec{\omega})$ the probability distribution of
$M_{Z^{*}}$ is ignored.  The product of all one-dimensional
probabilities, ignoring correlations, is $\prod_{i} P(X_{i})$.  As
expected, the likelihood including all discriminating variables and
their correlations is optimal.  The other two definitions give similar
results.  We note that, regardless of the results, the use of
$\prod_{i} P(X_{i})$ is an improper approximation, since the $X_{i}$
variables are far from being uncorrelated.
%%%%%%%%%%%%%%%%%%%%%%%%%%%%%%%%%%%%%%%%%%%%%%%%%%%%%%%%%%%%%%%%%%
%%%%%%%%%%%%%%%%%%%%%%%%%%%%%%%%%%%%%%%%%%%%%%%%%%%%%%%%%%%%%%%%%%%

\begin{figure}[htbp]
\begin{center}
\includegraphics[width=0.238\textwidth]{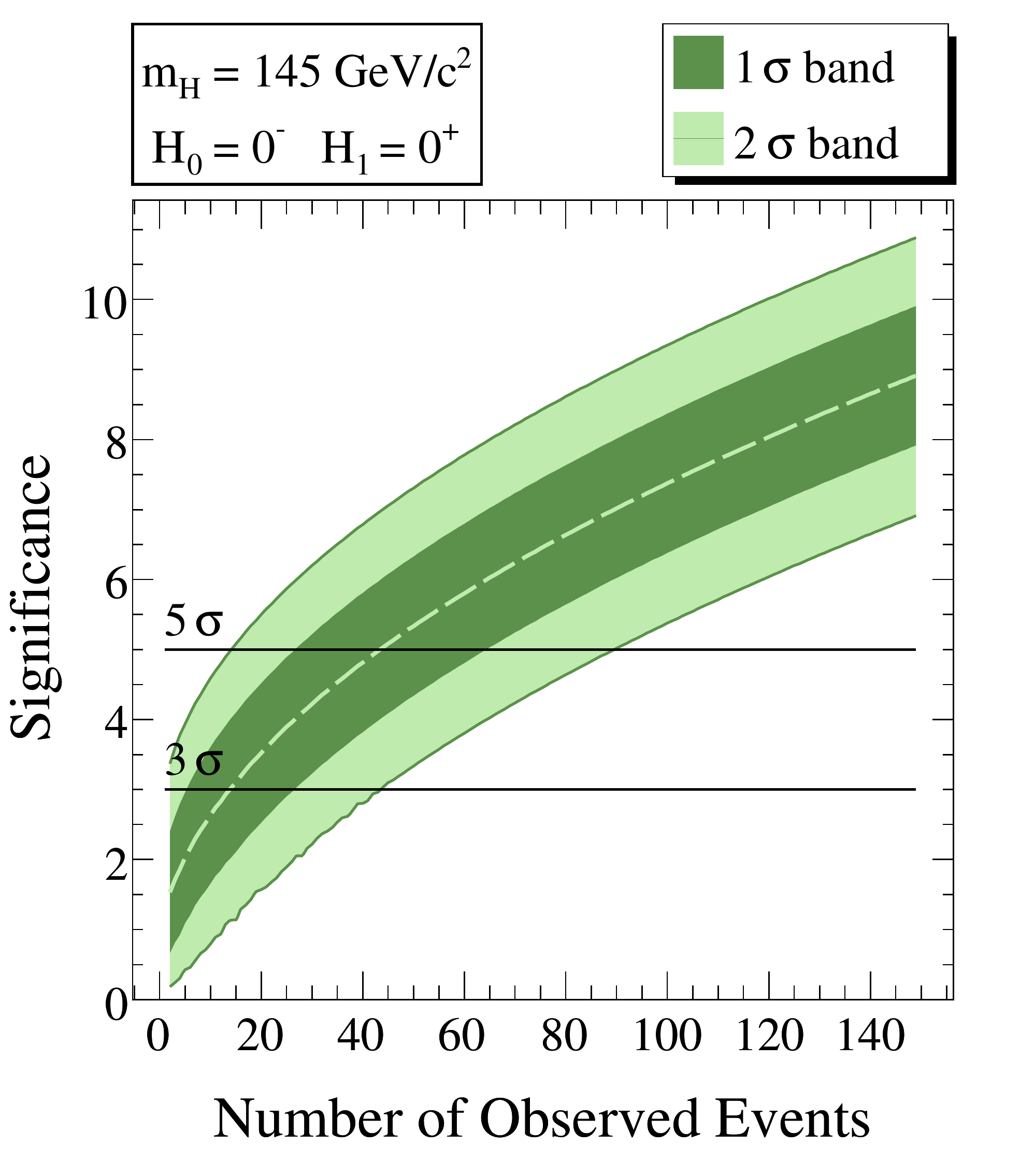}
\includegraphics[width=0.238\textwidth]{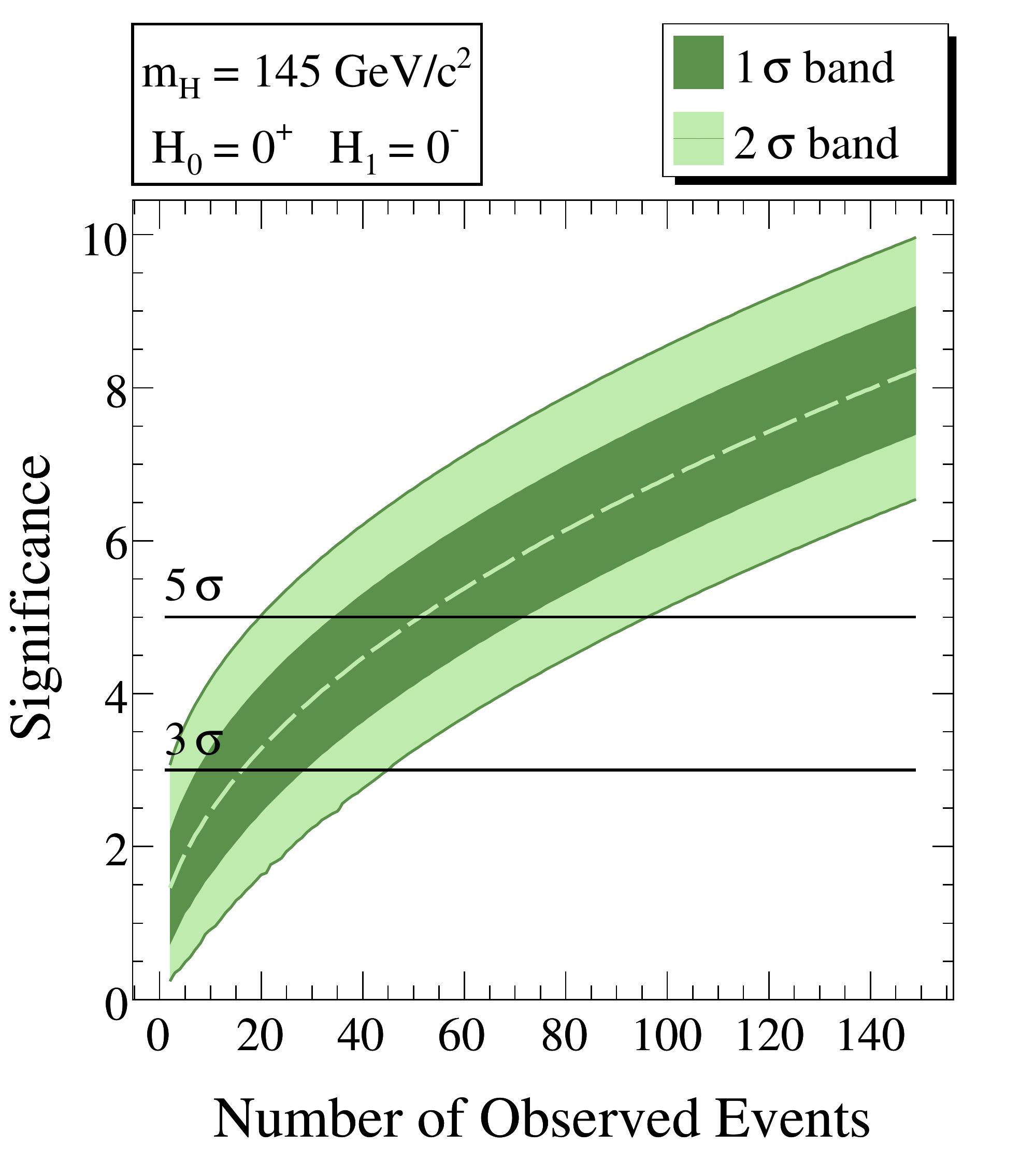}
\includegraphics[width=0.238\textwidth]{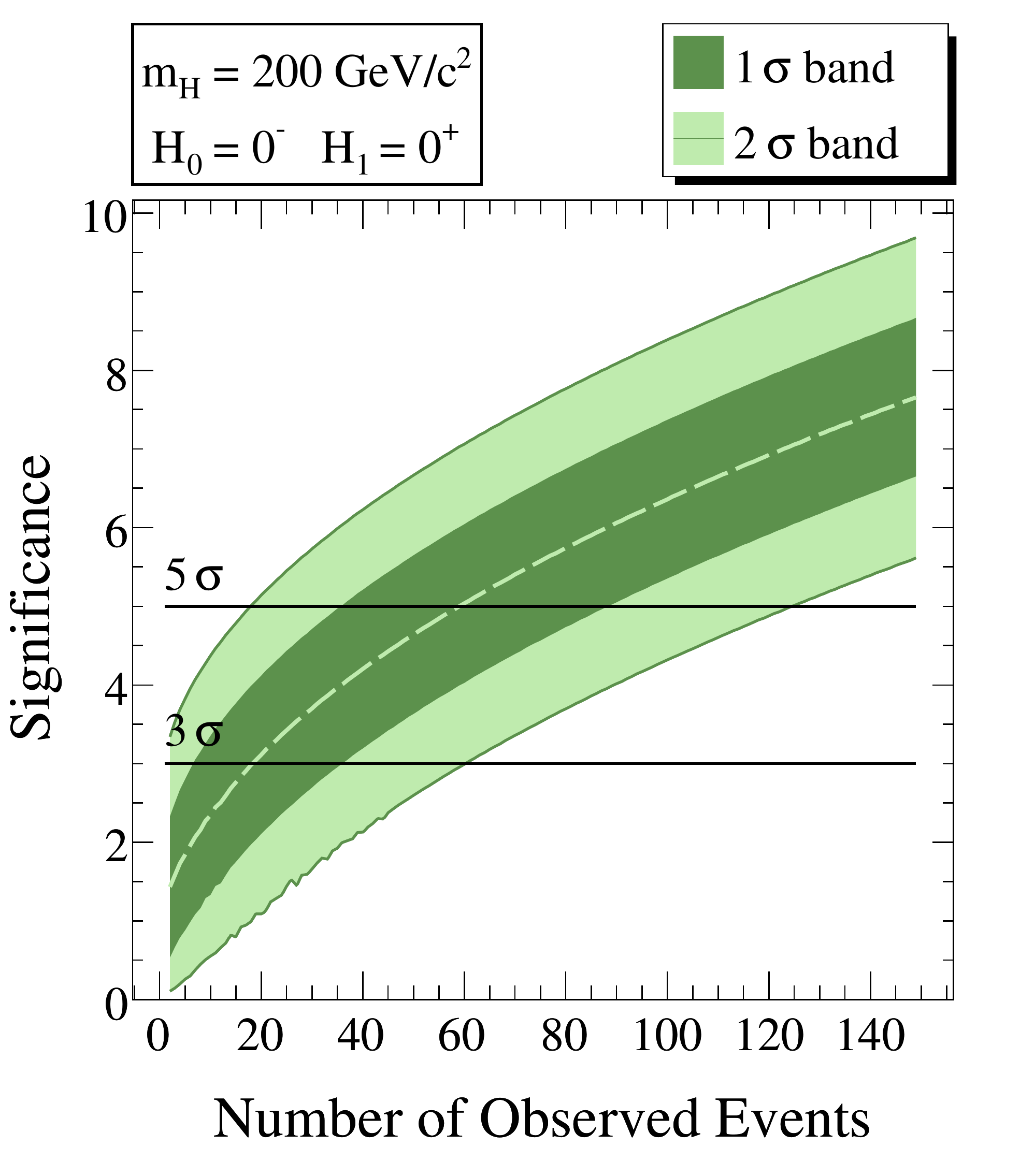}
\includegraphics[width=0.238\textwidth]{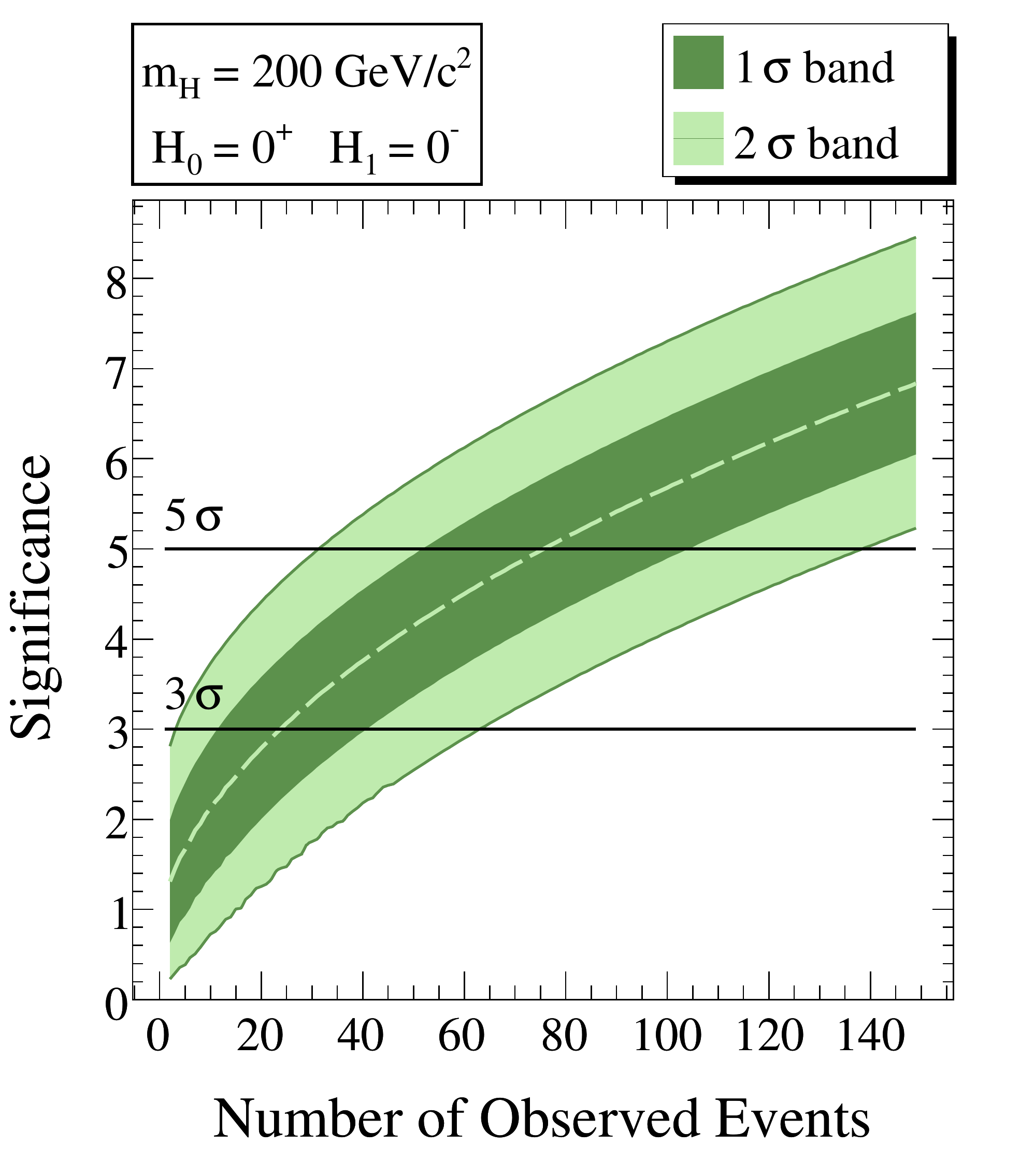}
\includegraphics[width=0.238\textwidth]{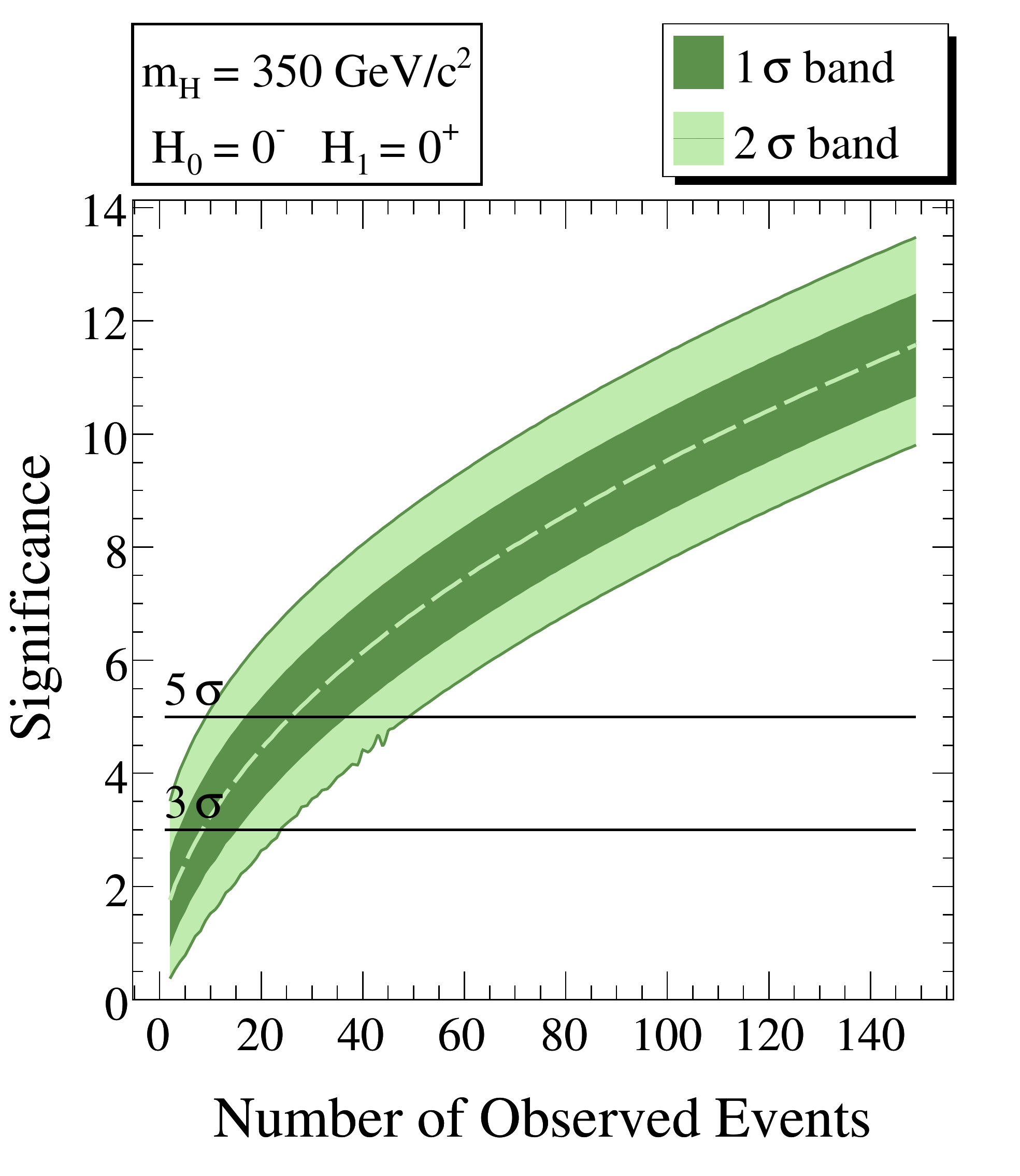}
\includegraphics[width=0.238\textwidth]{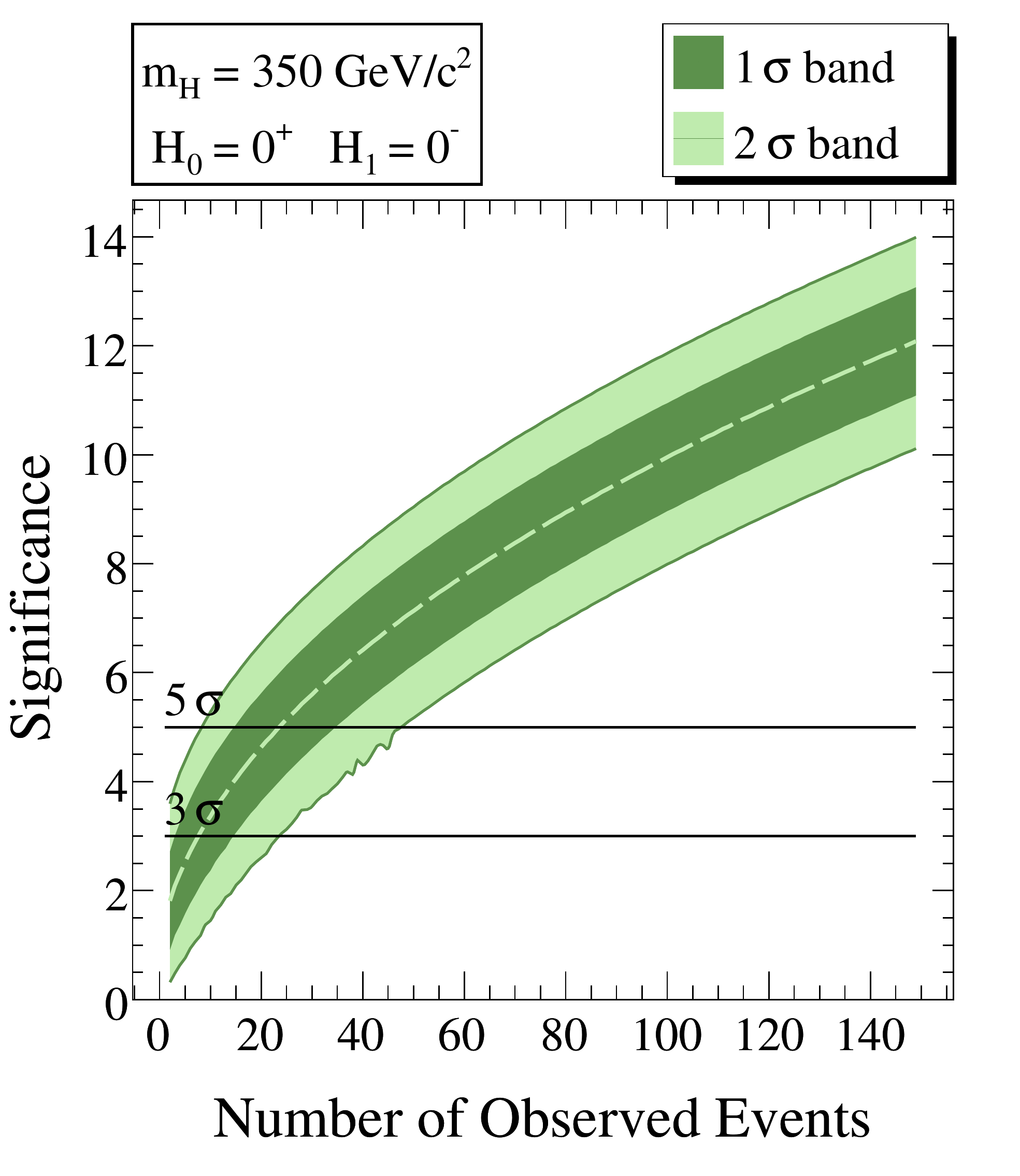}
\caption{Significance for rejecting $0^{-}$ in favor of
  $0^{+}$, assuming $0^{+}$ is true (left), and vice-versa, $0^+\!\leftrightarrow\! 0^-$
  (right), for $m_H$$=$$145$, 200 and 350 GeV/c$^{2}$ (top, middle and
  bottom). The dashed central line is the median significance. The 1
  and 2$\,\sigma$ bands correspond to 68\% and 95\% confidence
  intervals, centered on the median.
  \label{fig:COMP_SM_v_PS}}
\end{center}
\end{figure}

\begin{figure}[htbp]
\begin{center}
\includegraphics[width=0.238\textwidth]{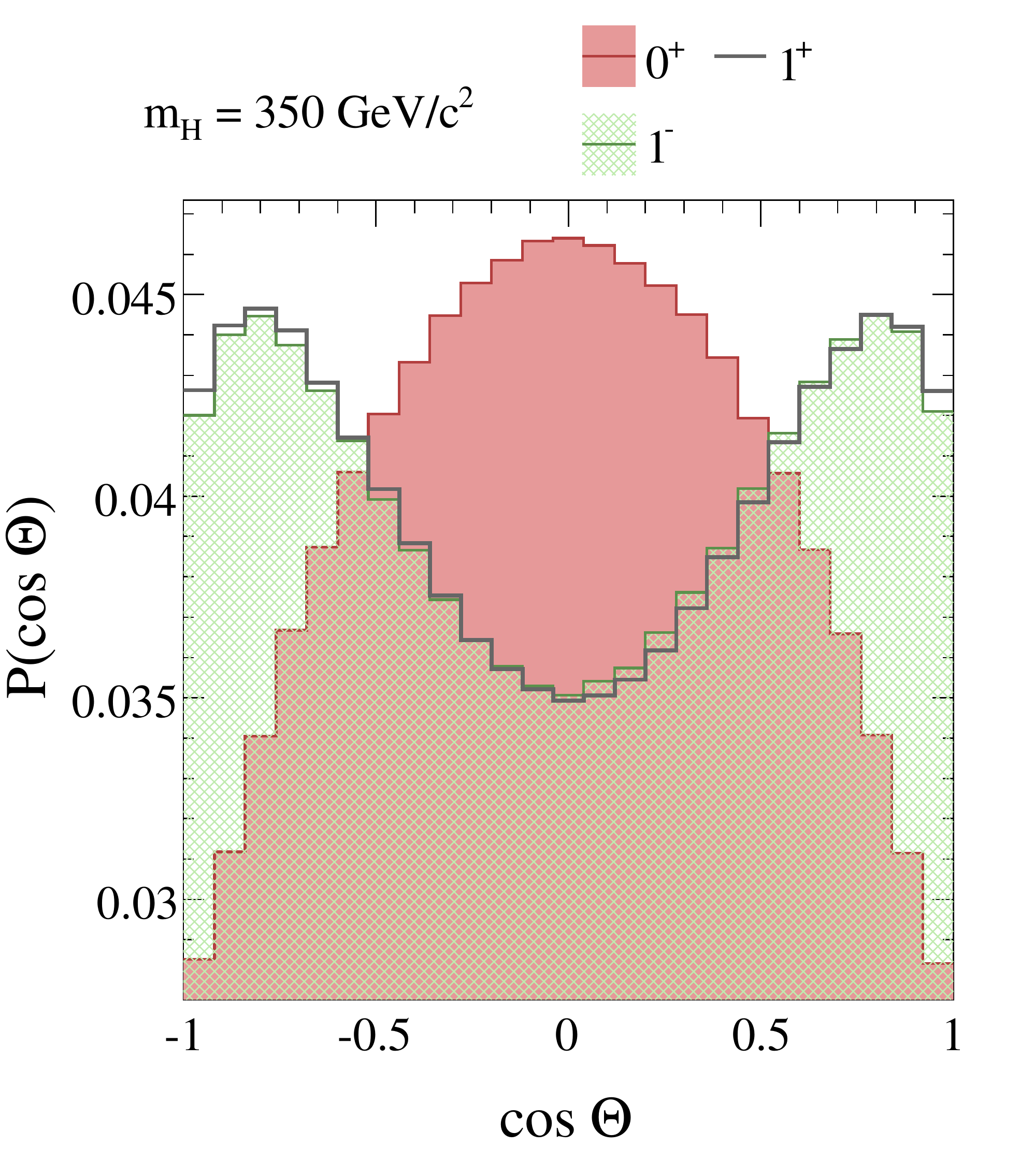}
\includegraphics[width=0.238\textwidth]{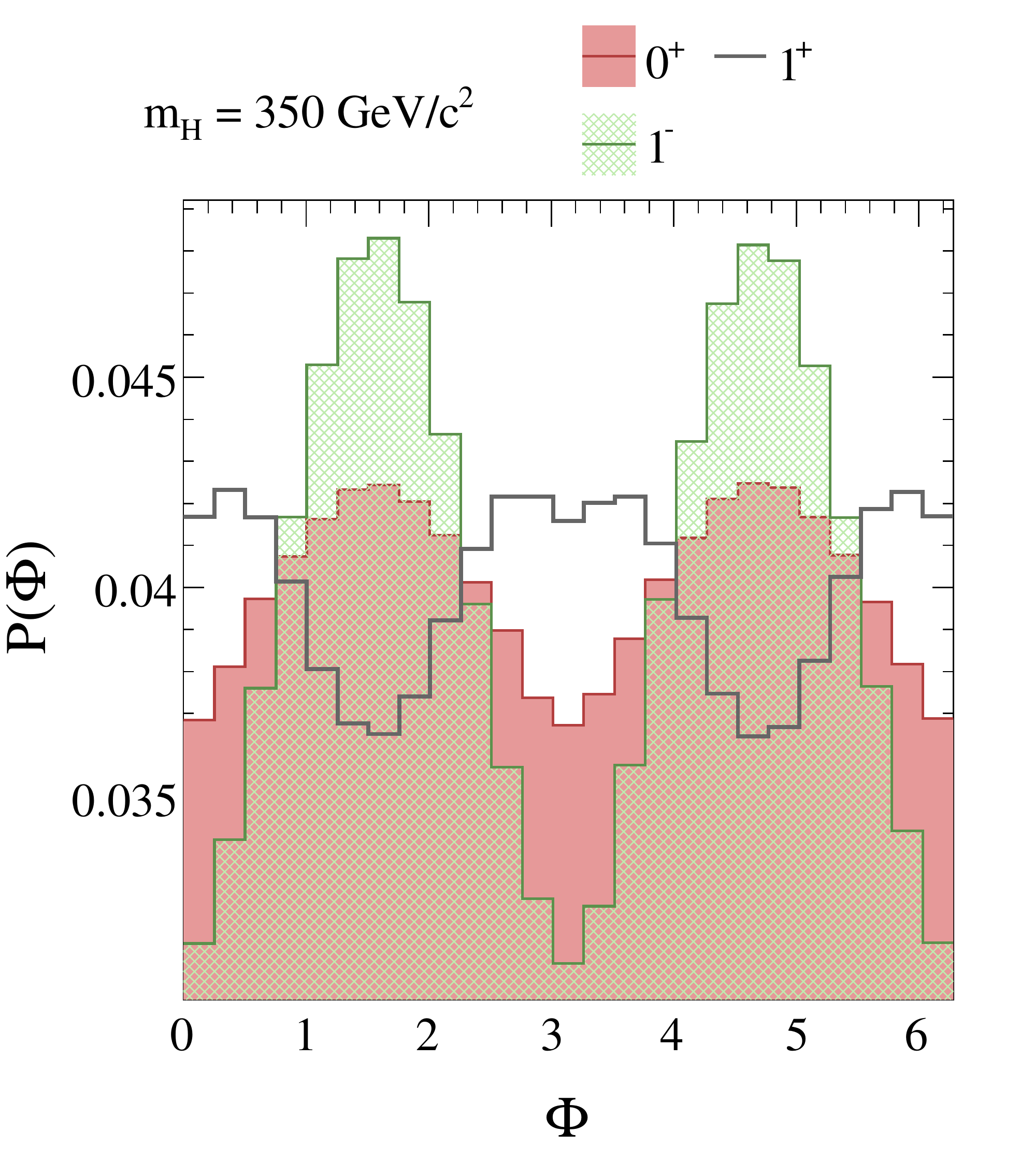}
\includegraphics[width=0.238\textwidth]{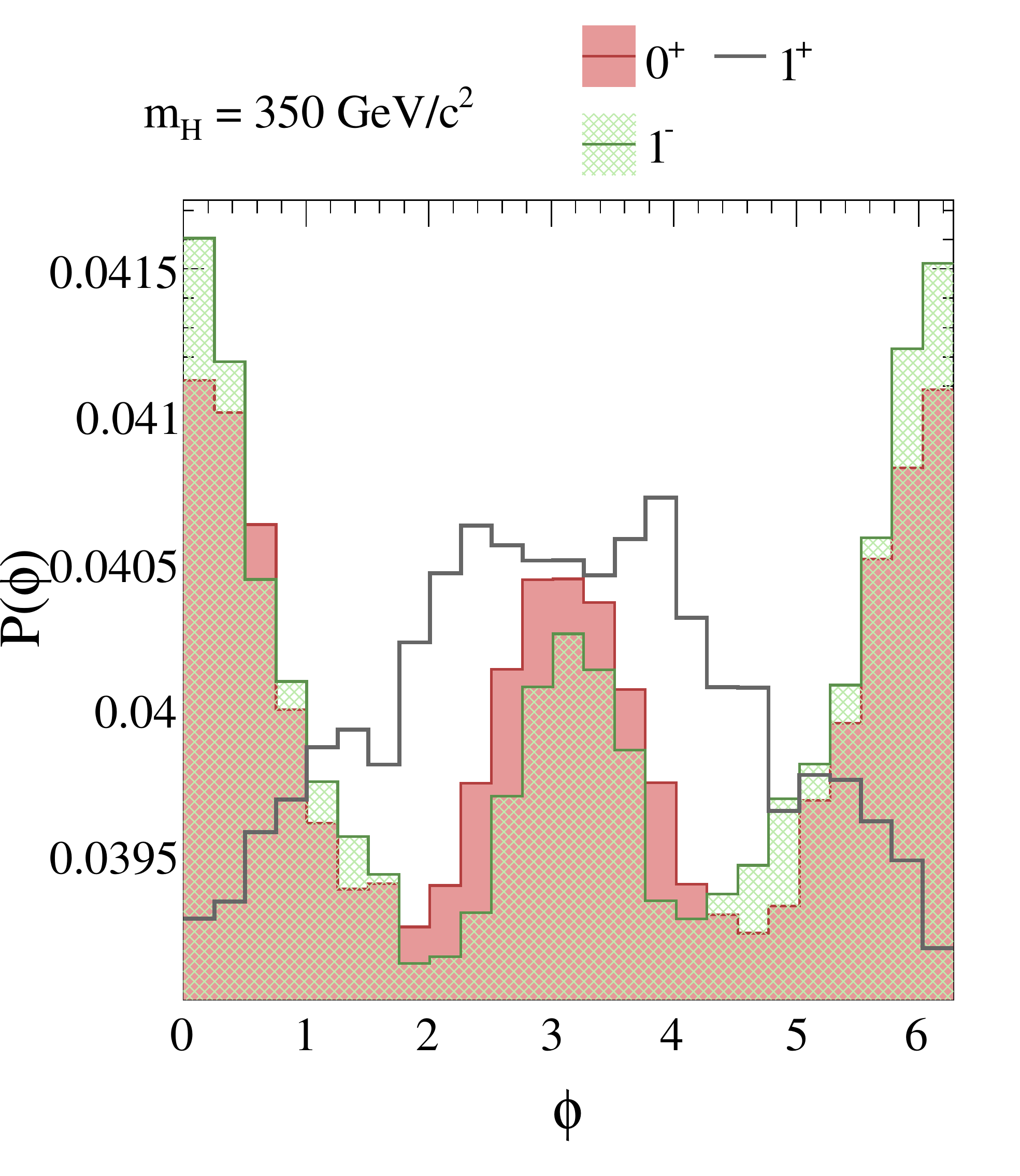}
\includegraphics[width=0.238\textwidth]{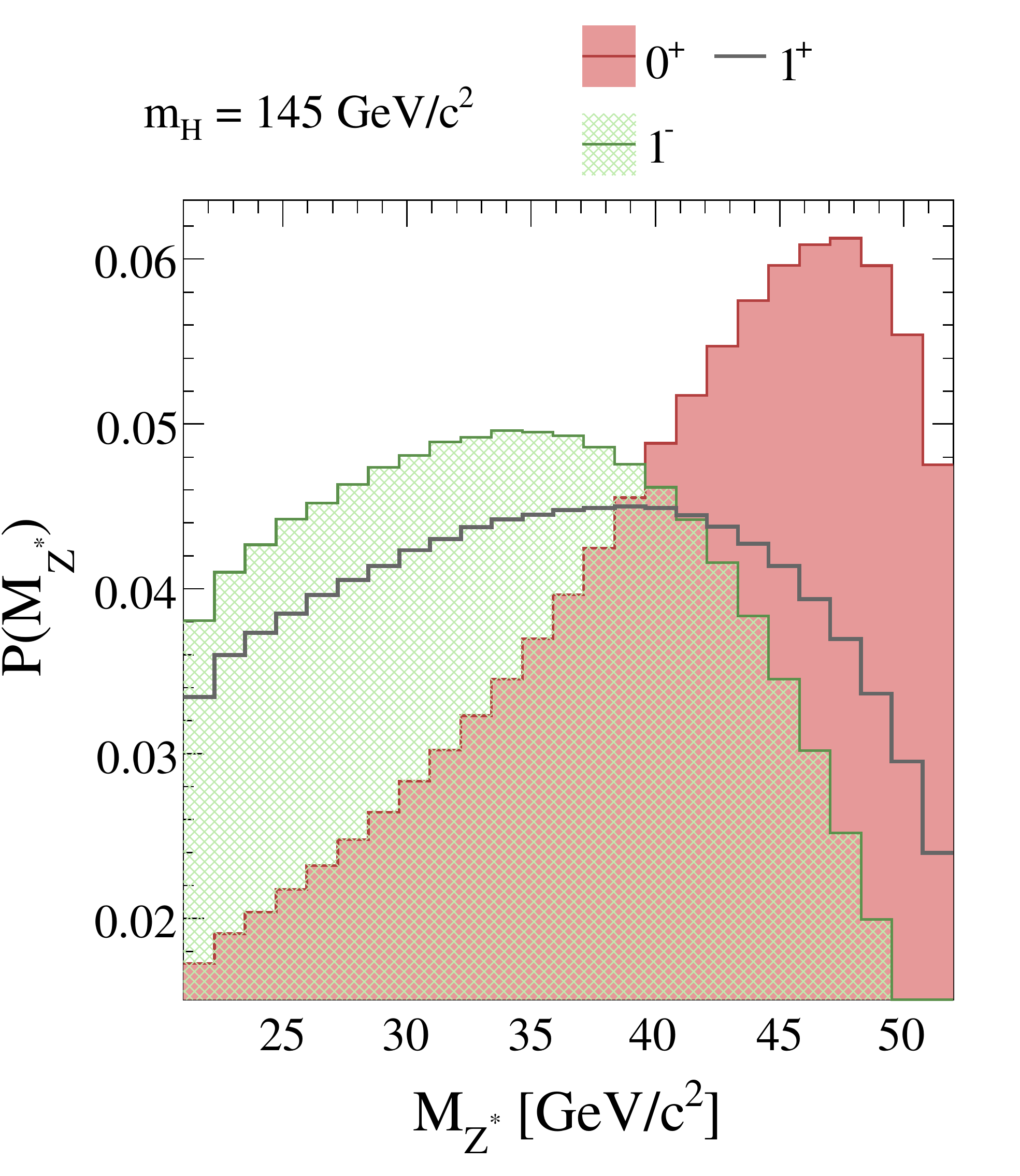}
\caption{Distributions of the variables cos$\,\Theta$ (upper left),
  $\Phi$ (upper right), $\phi$ (lower left) and $M_{Z^{*}}$ (lower
  right, $m_H$$=$$145$ GeV/c$^2$) for $0^{+}$, $1^{-}$ and $1^{+}$ resonances. All
  distributions are normalized to a unit integral.  The angular disctributions are
shown  for $m_H$$=$$350$ GeV/c$^2$. 
\label{fig:KIN_SM_V}}
\end{center}
\end{figure}

The significance for discriminating between the $0^{+}$ and $0^{-}$
hypotheses (assuming one or the other to be correct), as a function of
$N_S$, where $N_S$ is the number of observed signal events, is shown
in Fig.~\ref{fig:COMP_SM_v_PS} for $m_{H}$$=$$145$, 200, and 350
GeV/c$^{2}$. In all cases, results correspond to the case where
$\mathbb{H}_{1}$ is the true hypothesis (see Sec.~\ref{sec:stat}).
The model discrimination is based on a NePe test between these simple
hypotheses with test statistic $\log ({\mathcal L}[0^{+}]/{\mathcal
  L}[0^{-}])$. The variables $\vec{\omega}$ (and $M_{Z^{*}}$, when
applicable), along with their correlations, are used in the likelihood
construction.  The significance for rejecting one hypothesis in favor
of the other at the time of 5$\,\sigma$ excess (see
Sec.~\ref{sec:disc}) is better than 3$\,\sigma$ for $m_{H}$$=$$145$,
200, and 350 GeV/c$^{2}$ while a 5$\,\sigma$ discrimination can be
achieved with twice the observed signal events (less than $\sim$40
events in both mass cases presented here).
%%%%%%%%%%%%%%%%%%%%%%%%%%%%%%%%%%%%%%%%%%%%%%%%%%%%%%%%%%%%%%%%%%%
%%%%%%%%%%%%%%%%%%%%%%%%%%%%%%%%%%%%%%%%%%%%%%%%%%%%%%%%%%%%%%%%%%%
%%%%%%%%%%%%%%%%%%%%%%%%%%%%%%%%%%%%%%%%%%
% 0+ vs. spin 1 
%%%%%%%%%%%%%%%%%%%%%%%%%%%%%%%%%%%%%%%%%%
\begin{boldmath}
\subsection{$0^+$ vs. $1^-$ and $1^+$ 
\label{sec:SM_v_1}}
\end{boldmath}
We consider here two different ``pure" $J$$=$$1$ models specified by their
$HZZ$ couplings: ``vector" ($J$$=$$1^-$) and ``axial vector" ($J$$=$$1^+$).
Unlike in the $0^{+}$ case, the differential cross sections have
non-trivial dependences on the $Z$-production angles $\vec{\Omega}$
that provide additional discrimination between $0^{+}$ and $J$$=$$1$.  In
Fig.~\ref{fig:KIN_SM_V} we show the distributions for some of these
variables.
%%%%%%%%%%%%%%%%%%%%%%%%%%%%%%%%%%%%%%%%%%%%%%%%%%%%%%%%%%%%%%%%%%%

\begin{figure}[htbp]
\begin{center}
\includegraphics[width=0.238\textwidth]{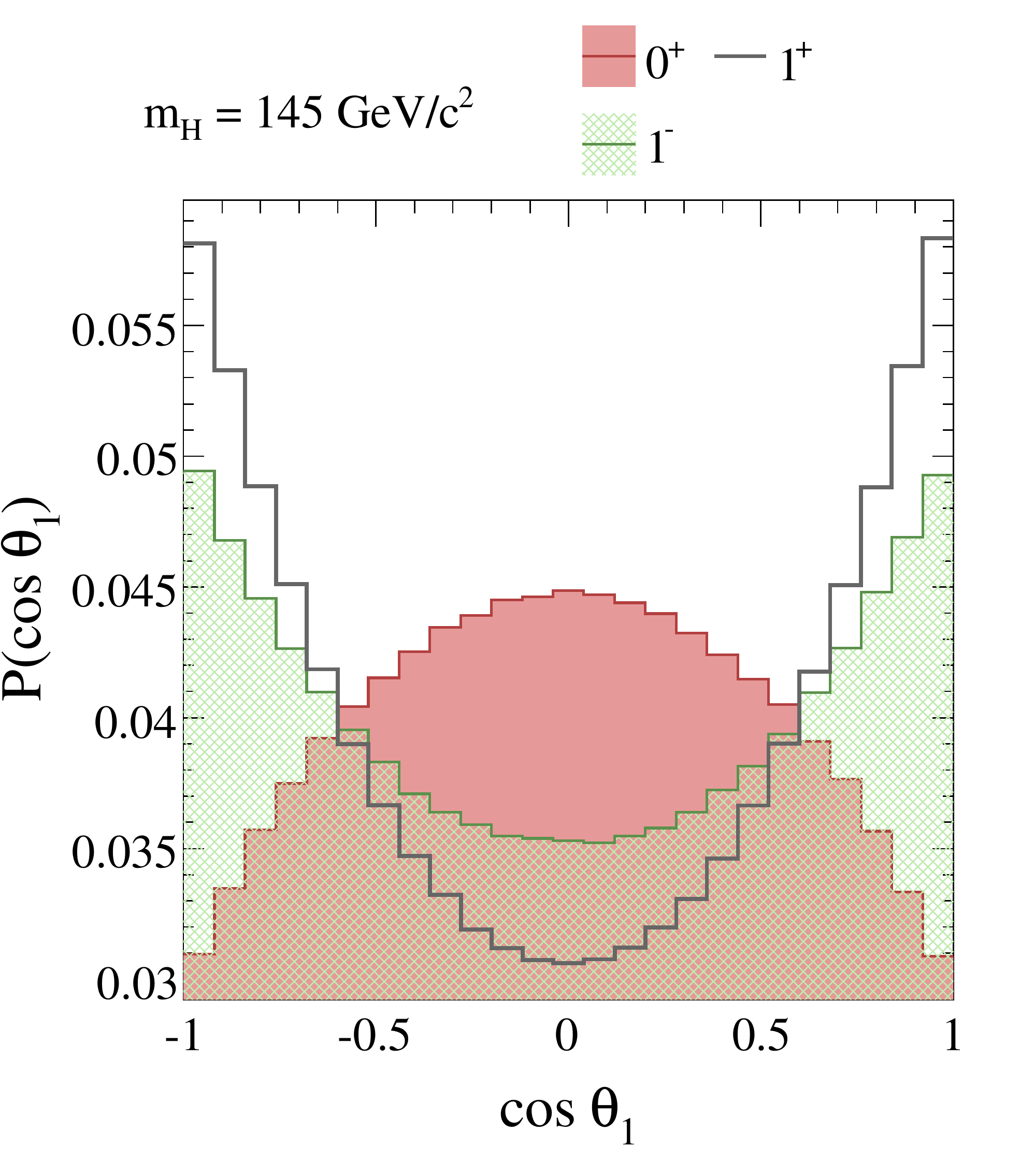}
\includegraphics[width=0.238\textwidth]{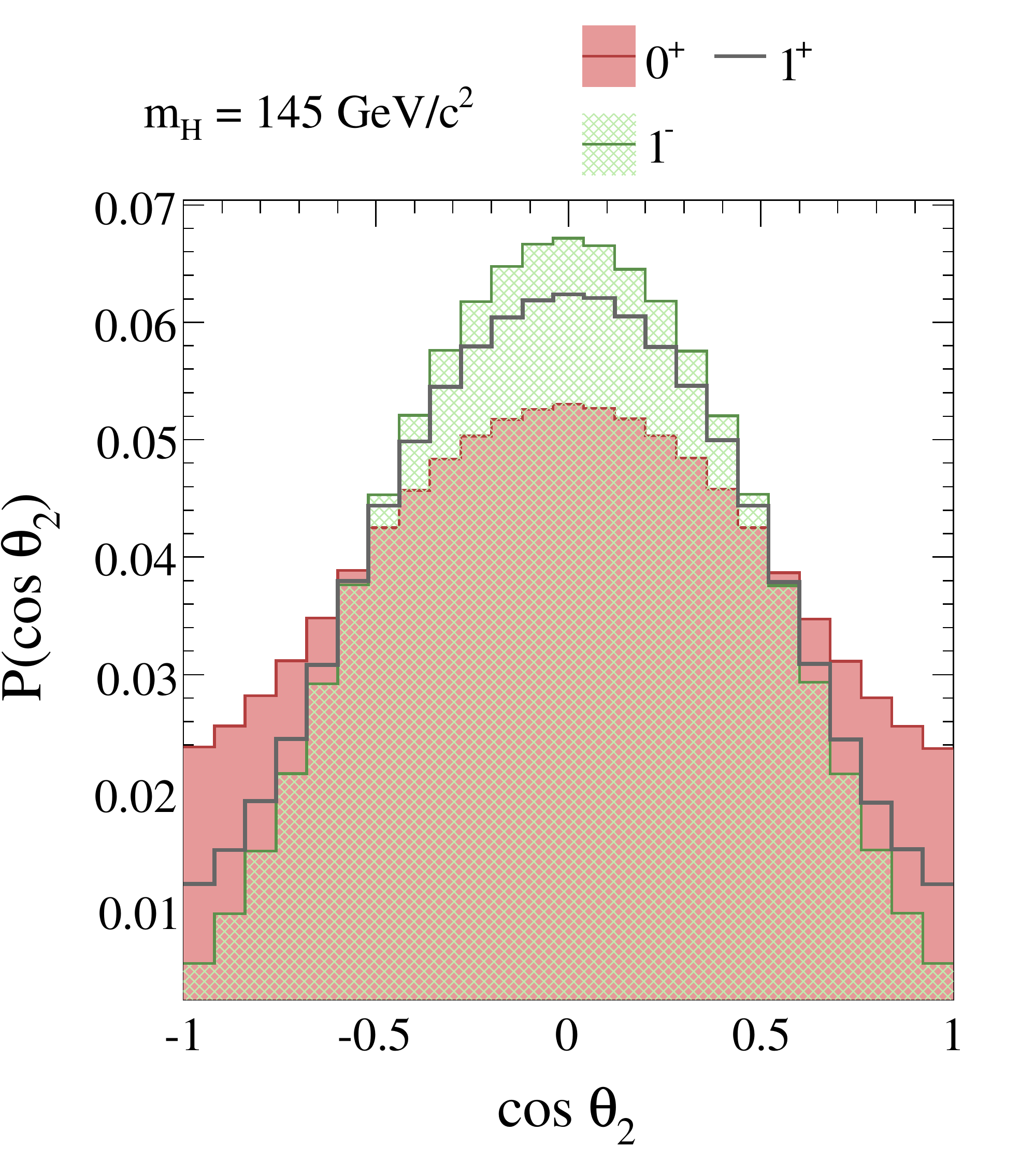}\\
\includegraphics[width=0.48\textwidth]{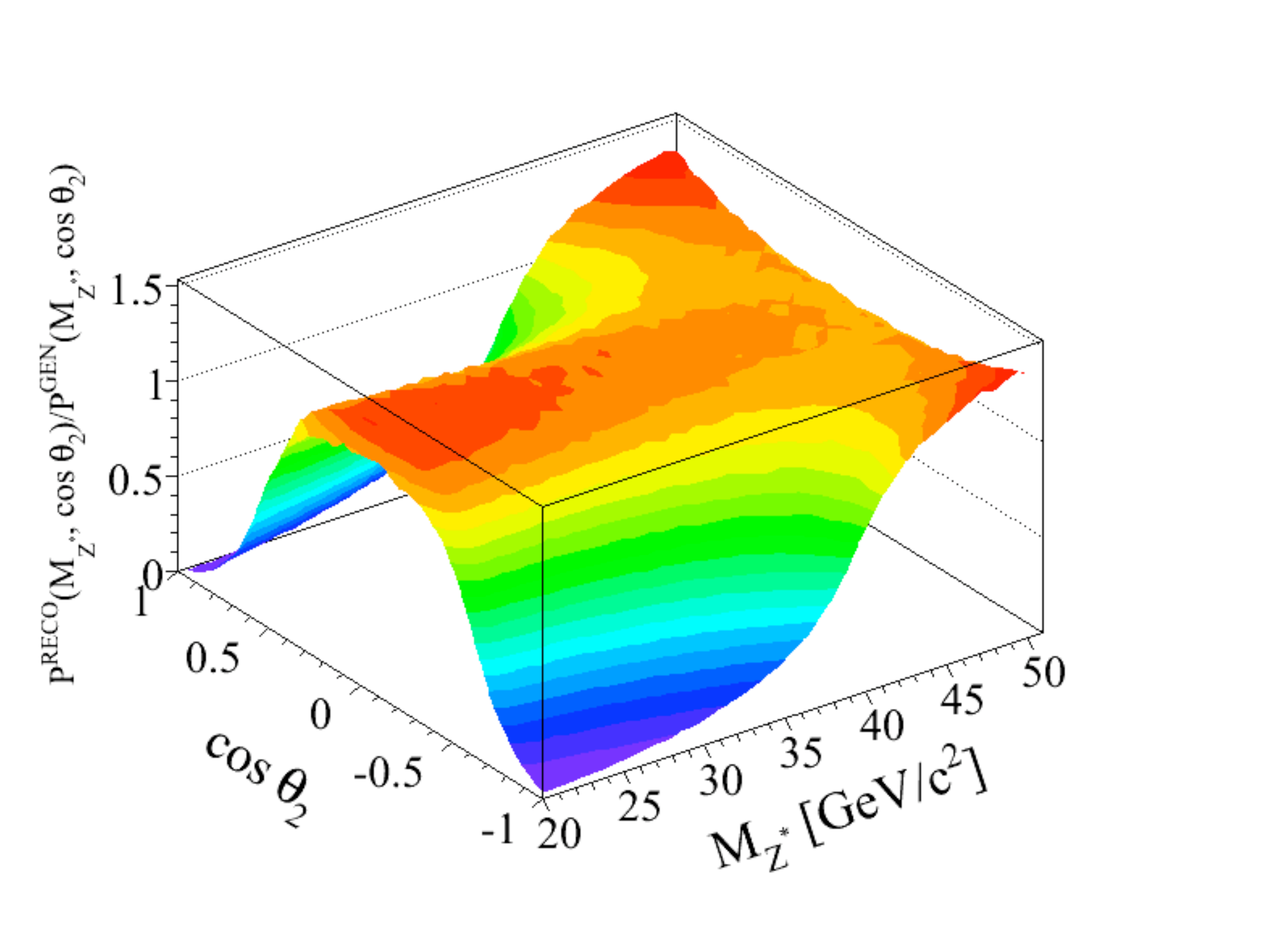}
\caption{Distributions in cos$\,\theta_{1}$ (left)  and cos$\,\theta_{2}$ (right), for  $0^{+}$, $1^{-}$ and $1^{+}$ resonances with mass 145 GeV/c$^{2}$, normalized to a unit integral. The very strong correlation between $M_{Z^{*}}$ and ${\rm cos}\,\theta_{2}$ is also shown (bottom) for the $J$$=$$1$ case.
  \label{fig:KIN_SM_V_flip}}
\end{center}
\end{figure}

We note that when a $J$$=$$1$ resonance decays in $ZZ^{*}$, the
distributions in
$c_1\equiv{\rm cos}\,\theta_{1}$ and $c_2\equiv{\rm
  cos}\,\theta_{2}$ are not any longer qualitatively similar, as
illustrated in Fig.~\ref{fig:KIN_SM_V_flip} (in striking contrast to
the $J$$=$$0$ cases).  Figure~\ref{fig:KIN_SM_V_flip} also shows the very
strong correlation between $M_{Z^{*}}$ and ${\rm cos}\,\theta_{2}$.
In the $J$$=$$1^-$ case, this asymmetric effect arises from the
configurations in which the object, in its rest system, is polarized
along the direction of motion of one of its $Z$-decay products. These
helicity configurations result in an addend proportional to
$m_{2}^{2}s_1^{2}c_2^{2}+ m_{1}^{2}s_2^{2}c_1^{2}$ in the {\it pdf},
which can be rewritten as $2 M_Z^2 (s_1^2 +s_2^2 - s_1^2 s_2^2) -
m_d^2 s_1^2 (2-s_2^2)$, with $m_d^2\equiv M_Z^2-m_2^2$. The second
term is 1$\leftrightarrow$2 asymmetric at fixed $m_d$ and induces the
difference between the $c_1$ and $c_2$ one-dimensional distributions.
In the $J$$=$$1^+$ case the asymmetric {\it pdf} term is, in the notation
of Appendix \ref{app:genspinone}, $2 M_{1}^{4} m_{2}^{2} s_1^{2} + 2
M_{2}^{4} m_{1}^{2}s_2^{2} - (M_{2}^{4}m_{1}^{2} +
M_{1}^{4}m_{2}^{2})s_1^{2} s_2^{2}$, and its origin is similar. These
asymmetric effects significantly enable the discrimination between
$J$$=$$1$ and $J$$=$$0$ models when $m_{H} < 2 M_{Z}$.
%%%%%%%%%%%%%%%%%%%%%%%%%%%%%%%%%%%%%%%%%%%%%%%%%%%%%%%%%%%%%%%%%%%
%%%%%%%%%%%%%%%%%%%%%%%%%%%%%%%%%%%%%%%%%%%%%%%%%%%%%%%%%%%%%%%%%%%

\begin{figure}[htbp]
\begin{center}
\includegraphics[width=0.42\textwidth]{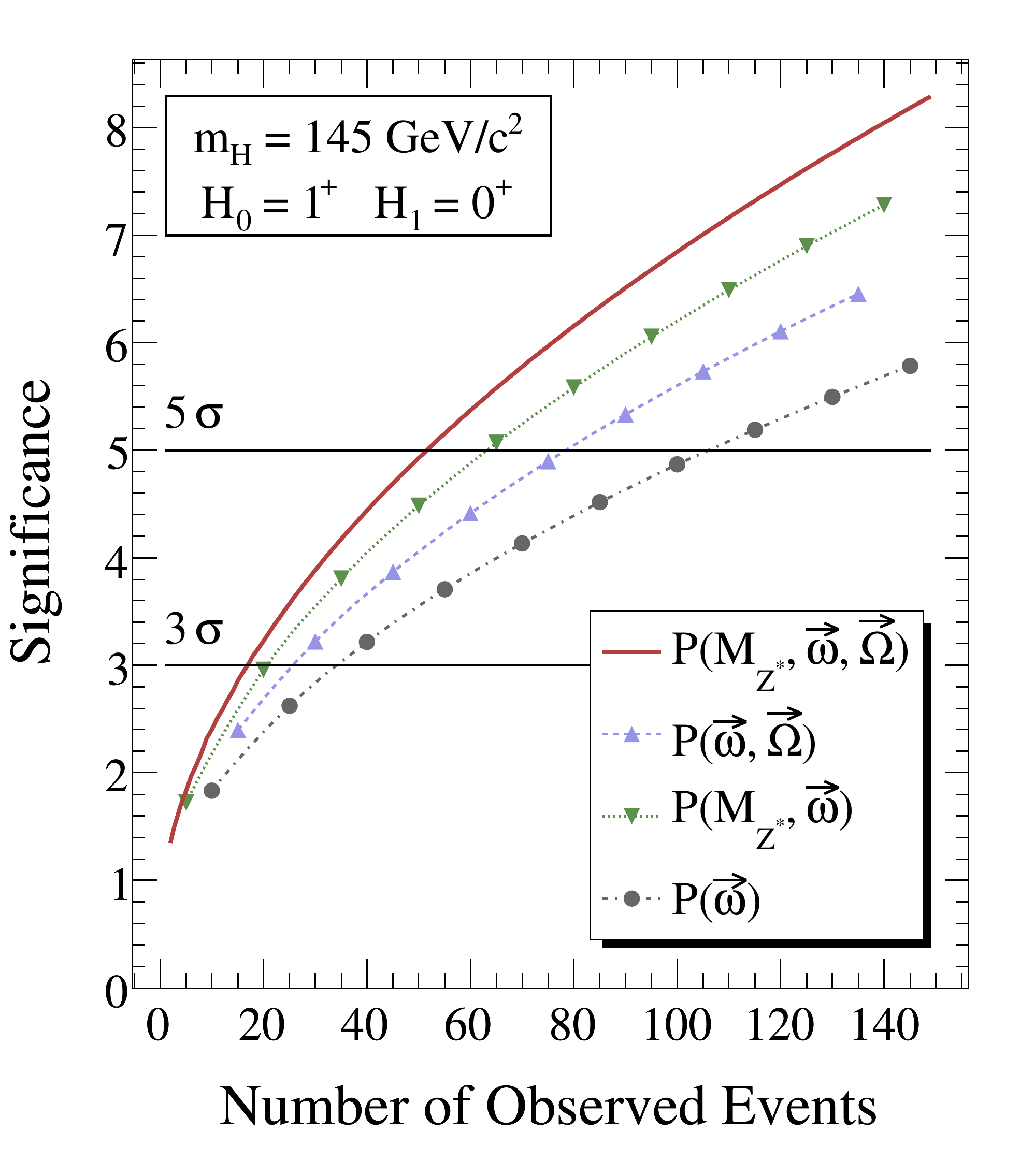}
\caption{Median significance for rejecting $1^{+}$
in favor of $0^{+}$ (assuming $0^{+}$ is true), for different likelihood constructions used in the
  likelihood ratio test statistic. \label{fig:SPEC_0_1}}
\end{center}
\end{figure}

In Fig.~\ref{fig:SPEC_0_1} we compare the discrimination between the
$0^{+}$ and $1^{+}$ hypotheses for likelihood definitions that exploit
different variables. The obvious qualitative conclusion is that
likelihoods defined in terms of {\it pdfs} containing the most
information are the most performant. The figure shows the relative
discriminating power of the different choices: $P(a_{1}, \cdots
,a_{N})$ denotes N-dimensional {\it pdfs} in the correlated variables
$\{a_{1},\cdots,a_{N}\}$. $\prod_{i} P(X_{i})$ is constructed from
one-dimensional {\it pdfs} for all variables, ignoring (erroneously)
their correlations. $P(\vec{\omega}\, |
\langle\vec{\Omega}\rangle_{\rm TH})$ are {\it pdfs} including the
variables $\vec{\omega}$ and their correlations, but with the
hypothesis $1^{+}$ represented by a {\it pdf} in which dependence on the variables
$\vec{\Omega}$$=$$\{\Phi, {\rm cos}\,\Theta\}$ has been integrated out of the analytic differential cross-section. The likelihood $P(\vec{\omega}\, | \langle\vec{\Omega}\rangle_{\rm
  TH})$ performs badly relative to $P(\vec{\omega})$, where the two differ only in
that the first construction implicitly assumes a uniform $4\,\pi$
coverage of the observed leptons, as if the muon $p_{T}$ and $\eta$
analysis requirements did not depend on the $\vec\Omega$ angular
variables. The primary reason for this difference is the strong
correlation between the variables $\Phi$ and $\phi$ in the $J$$=$$1$ {\it
  pdfs}, which causes phase space acceptance sculpting of the $\Phi$
distribution to alter the $\phi$ distribution, as discussed
in Sec.\ref{sec:ana}.
%%%%%%%%%%%%%%%%%%%%%%%%%%%%%%%%%%%%%%%%%%%%%%%%%%%%%%%%%%%%%%%%%%%
%%%%%%%%%%%%%%%%%%%%%%%%%%%%%%%%%%%%%%%%%%%%%%%%%%%%%%%%%%%%%%%%%%%

\begin{figure}[htbp]
\begin{center}
\includegraphics[width=0.238\textwidth]{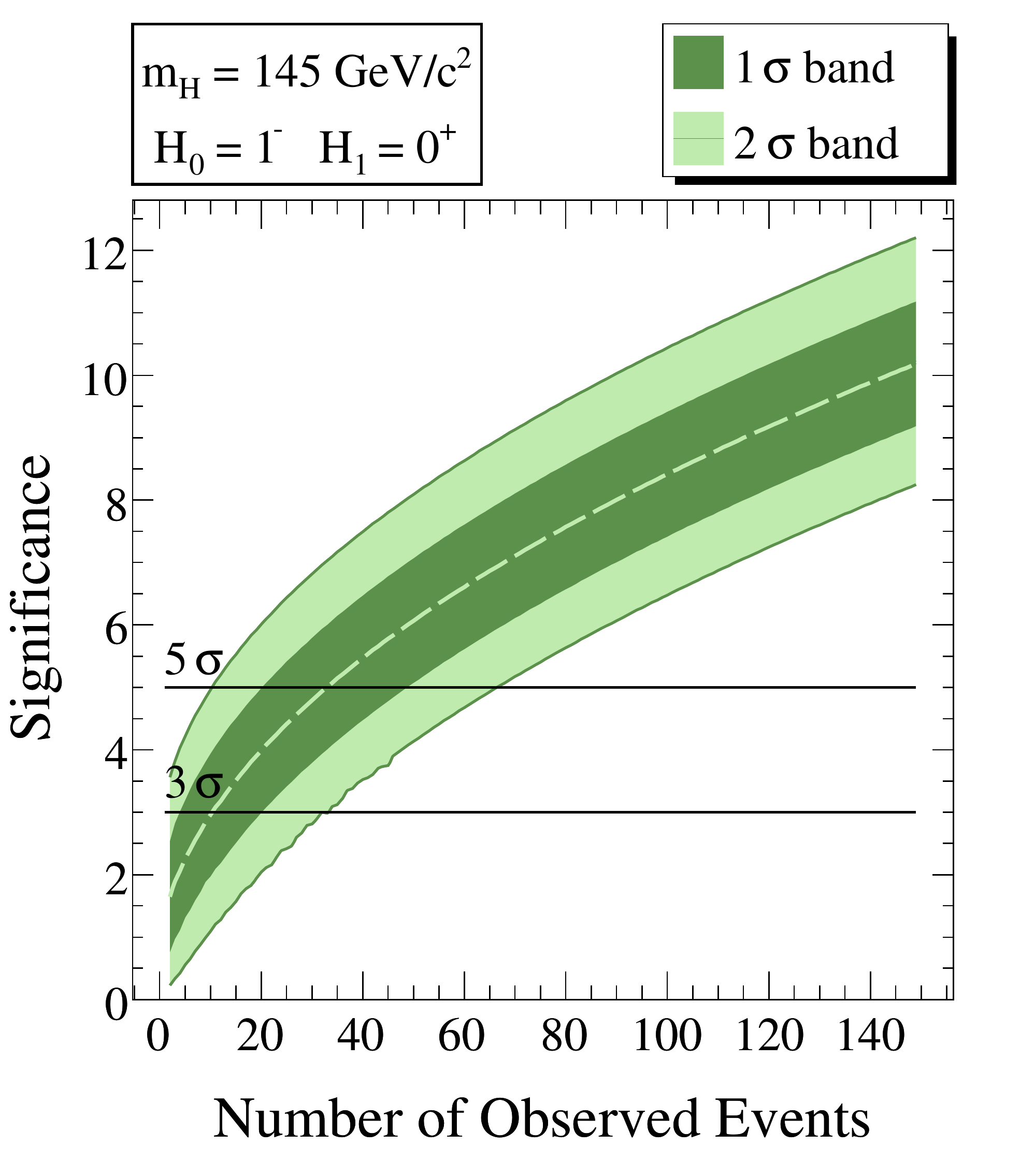}
\includegraphics[width=0.238\textwidth]{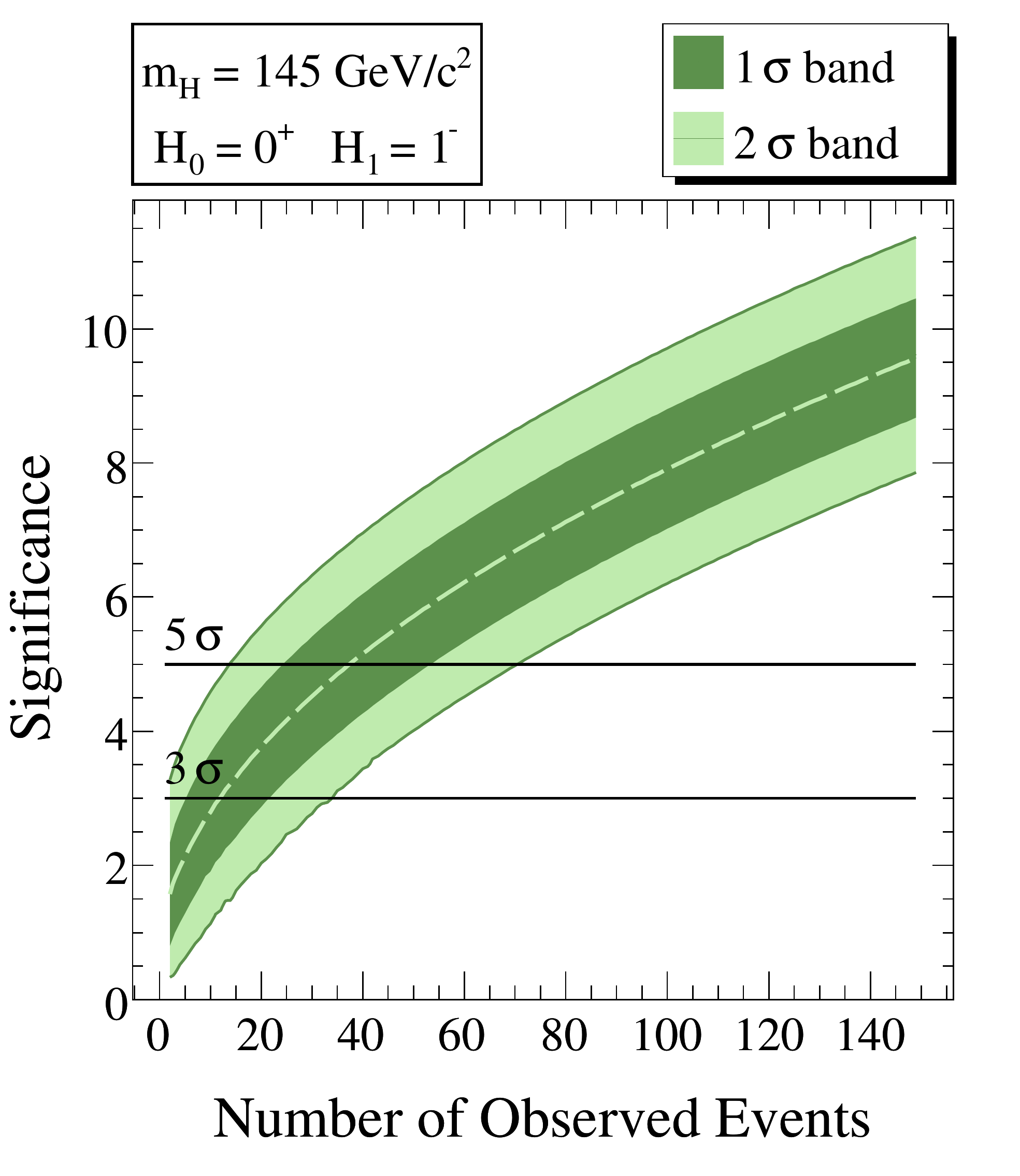}
\includegraphics[width=0.238\textwidth]{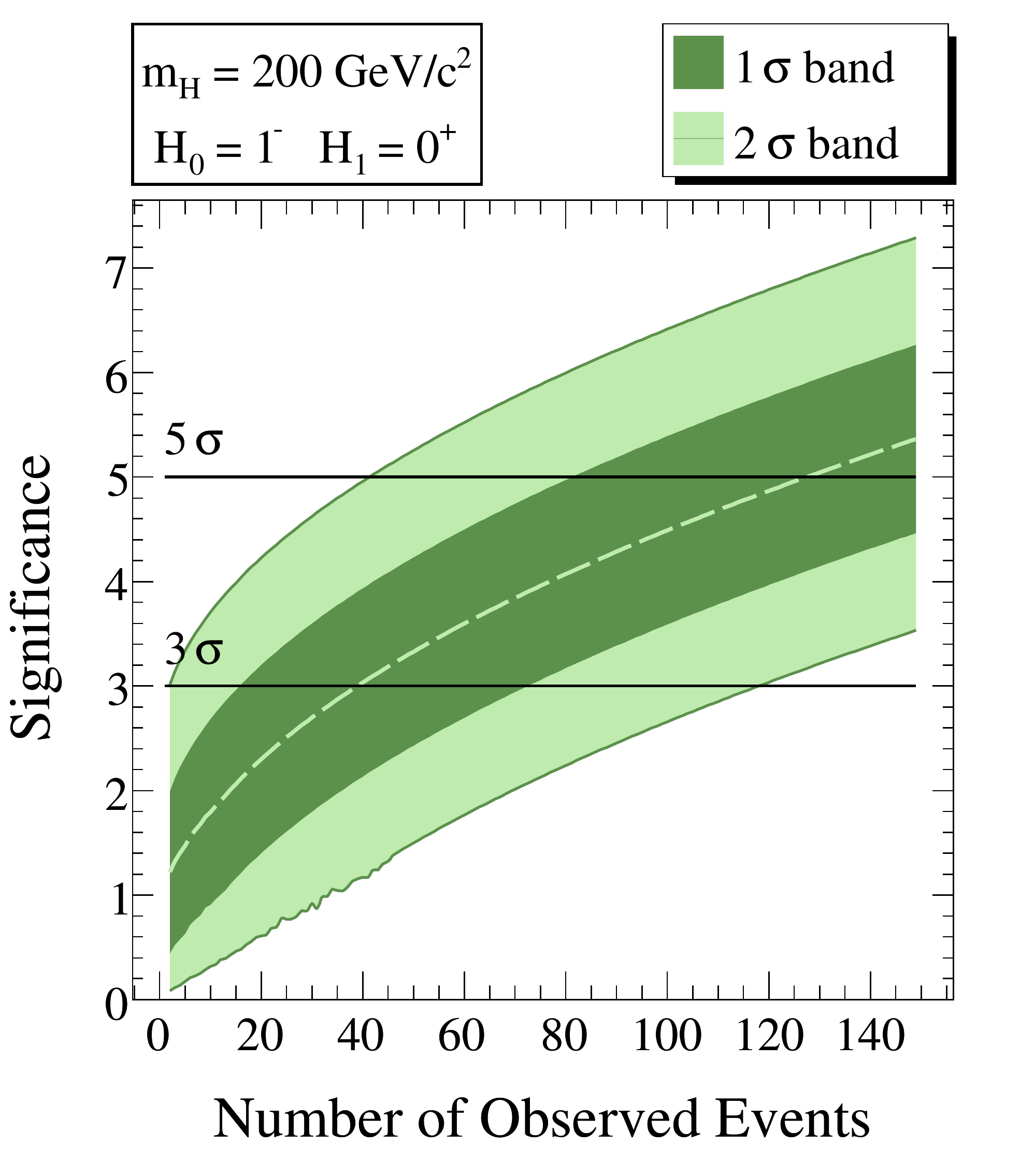}
\includegraphics[width=0.238\textwidth]{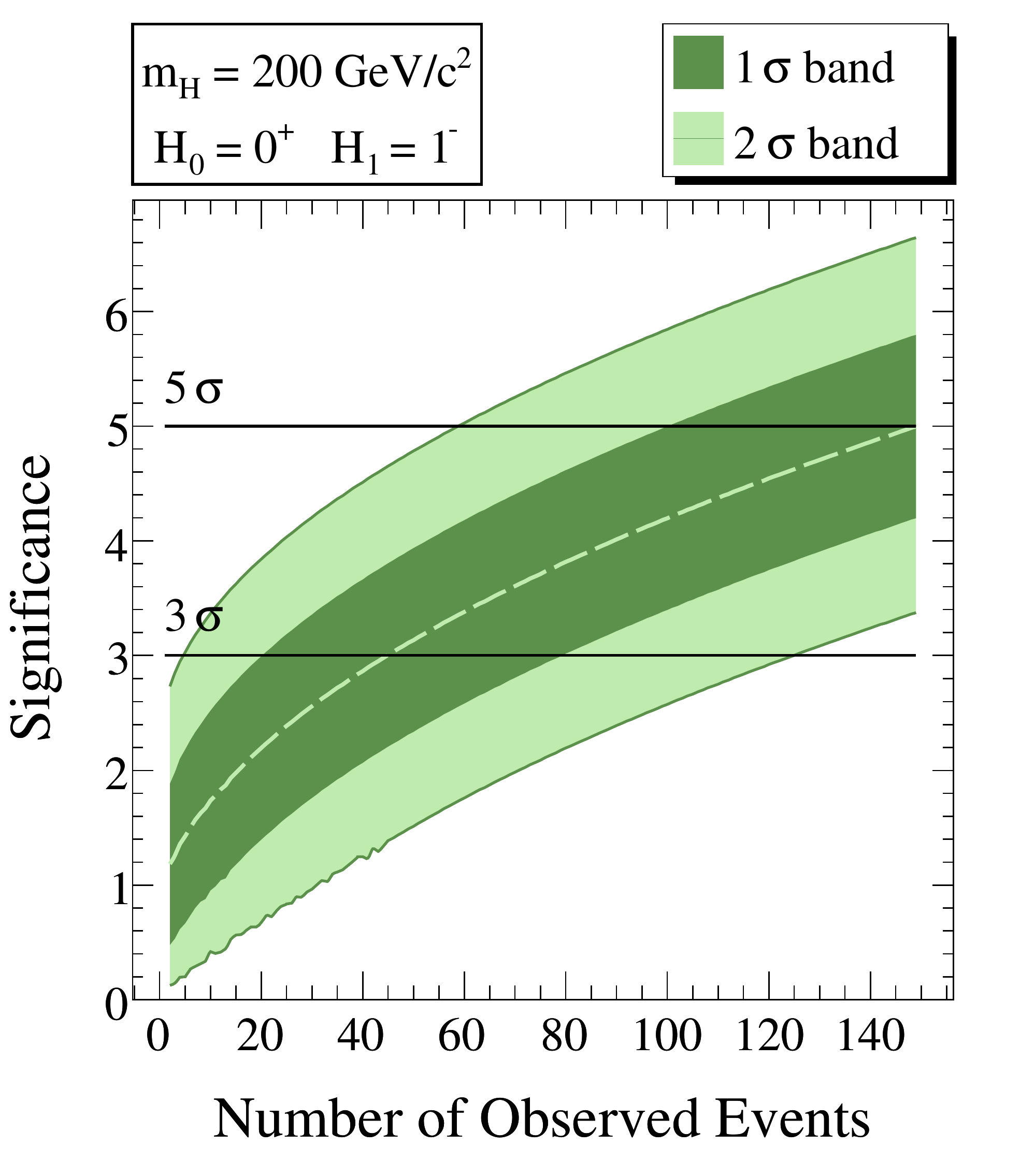}
\includegraphics[width=0.238\textwidth]{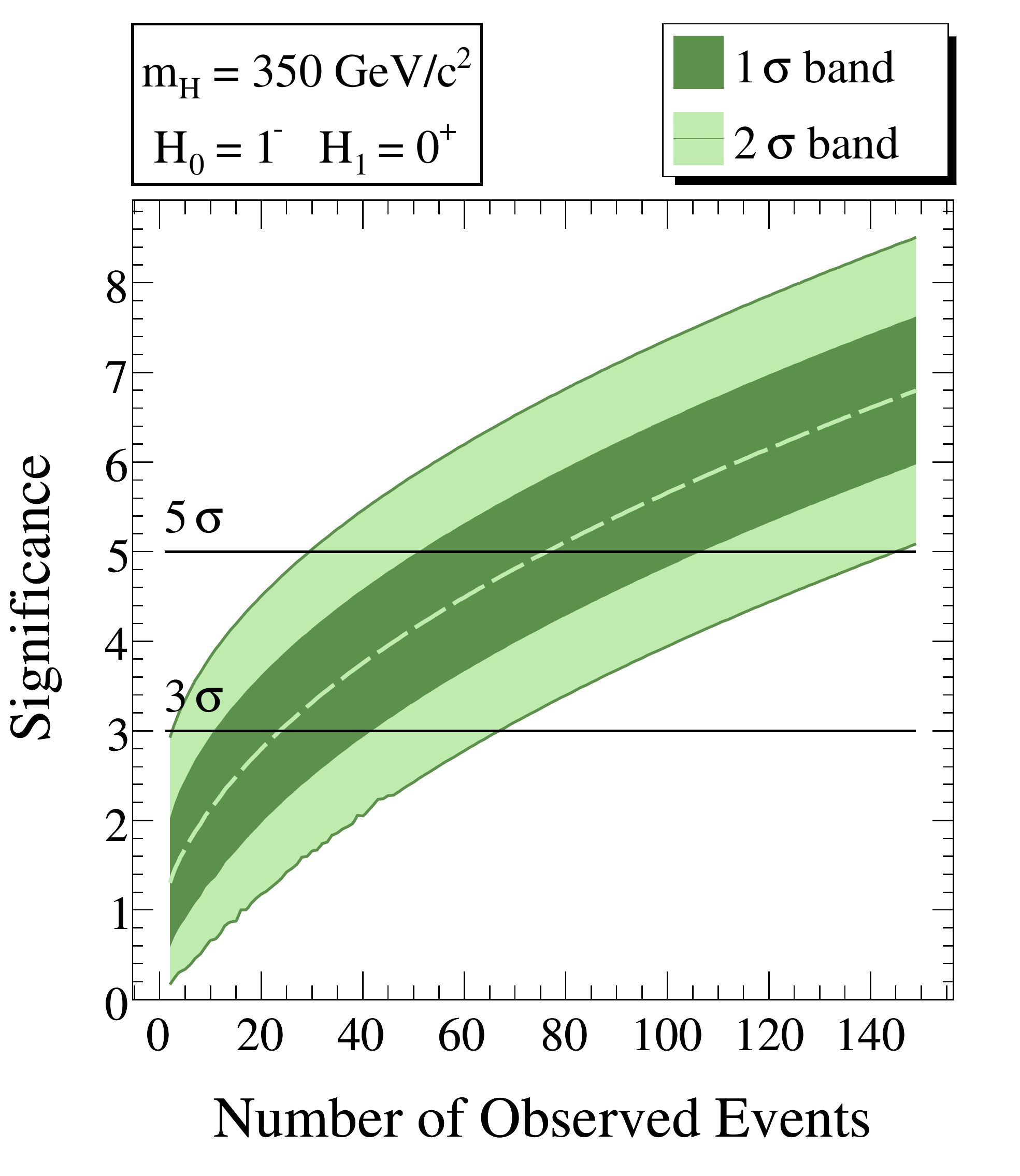}
\includegraphics[width=0.238\textwidth]{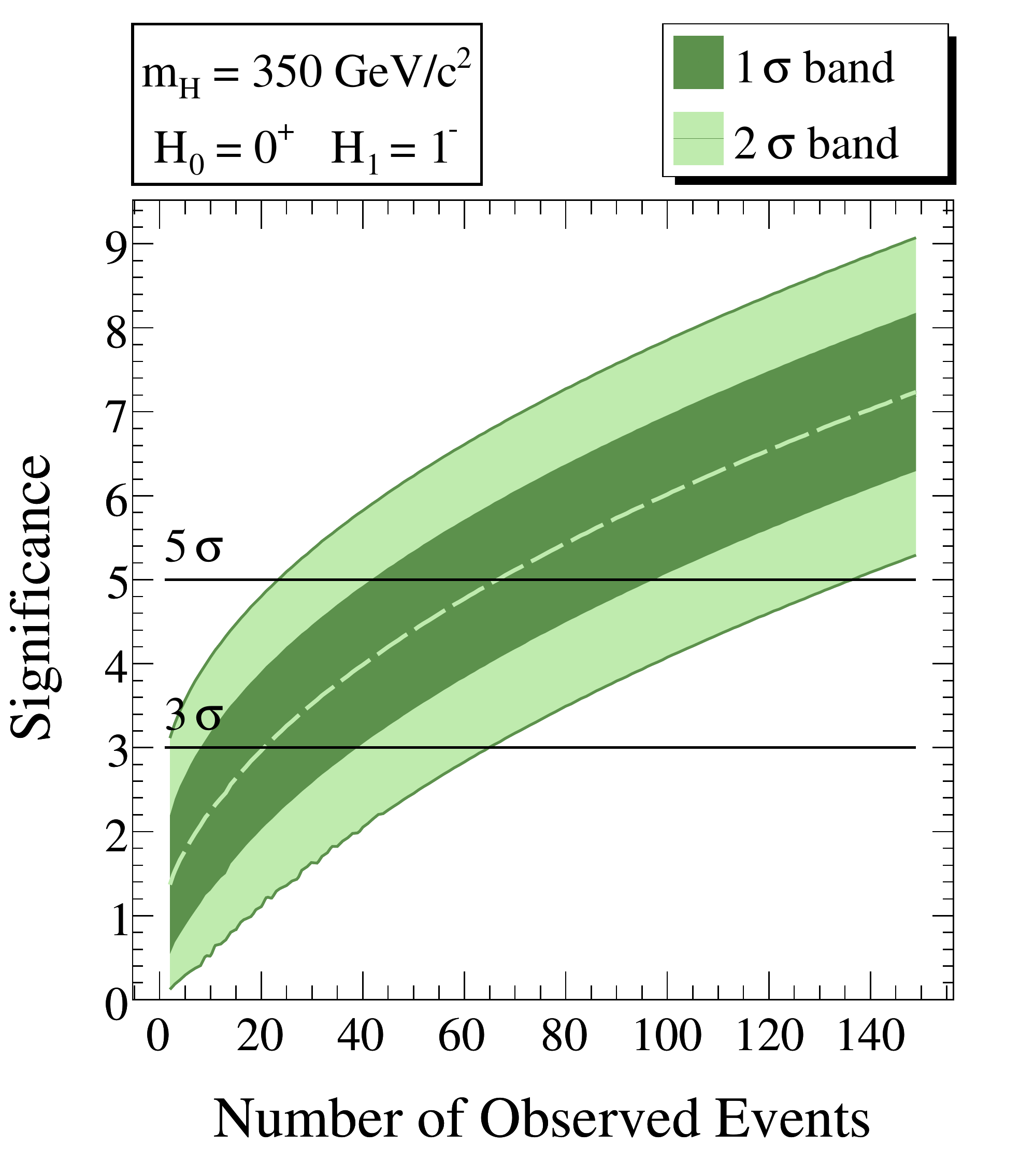}
\caption{Significance for rejecting $1^{-}$ in favor of
  $0^{+}$, assuming $0^{+}$ is true (left), or vice-versa ($0^+\!\leftrightarrow\! 1^-$, right),
  for $m_H$$=$$145$, 200 and 350 GeV/c$^{2}$ (top, middle and bottom).\label{fig:COMP_SM_v_PV}}
\end{center}
\end{figure}

\begin{figure}[htbp]
\begin{center}
\includegraphics[width=0.238\textwidth]{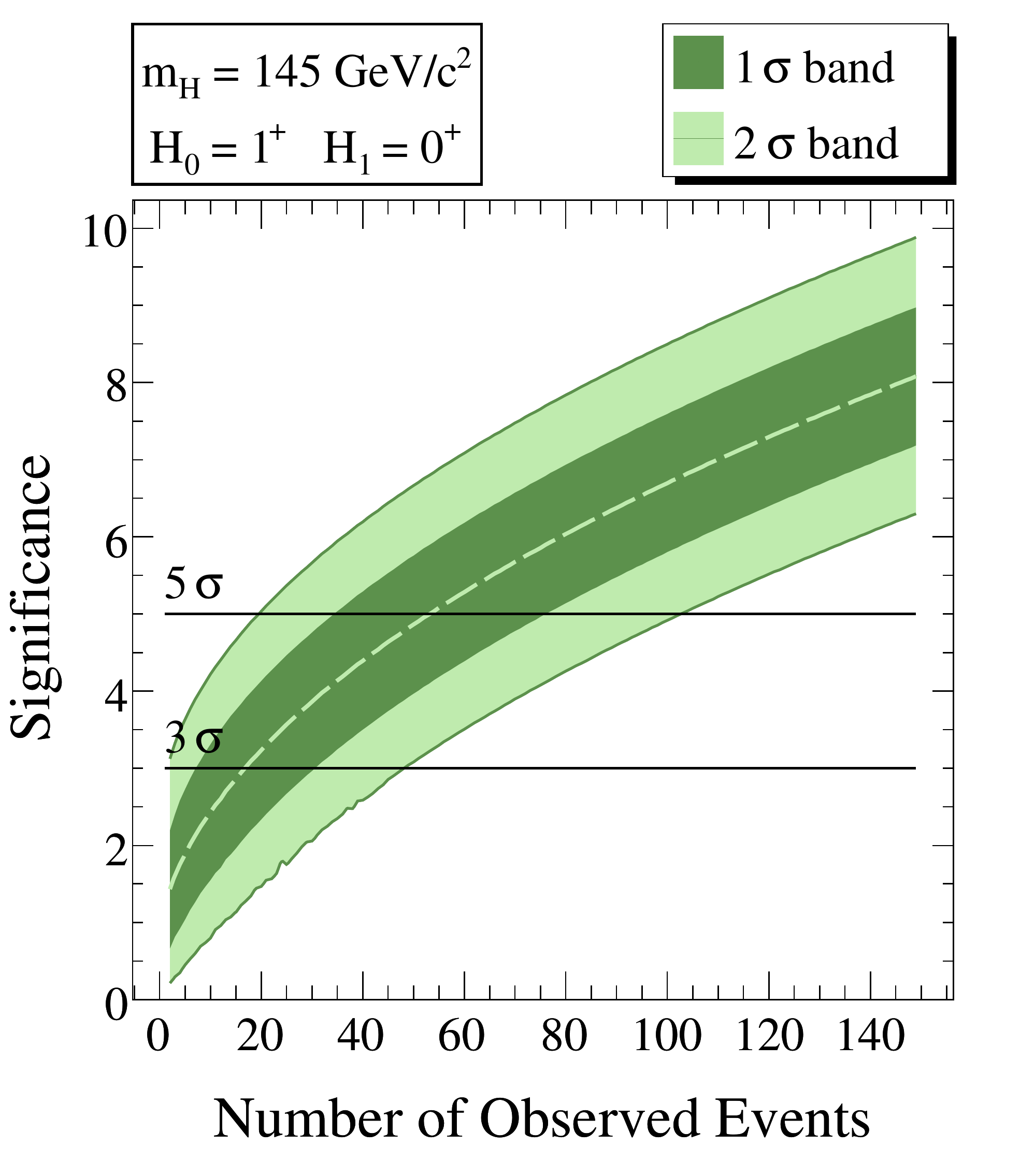}
\includegraphics[width=0.238\textwidth]{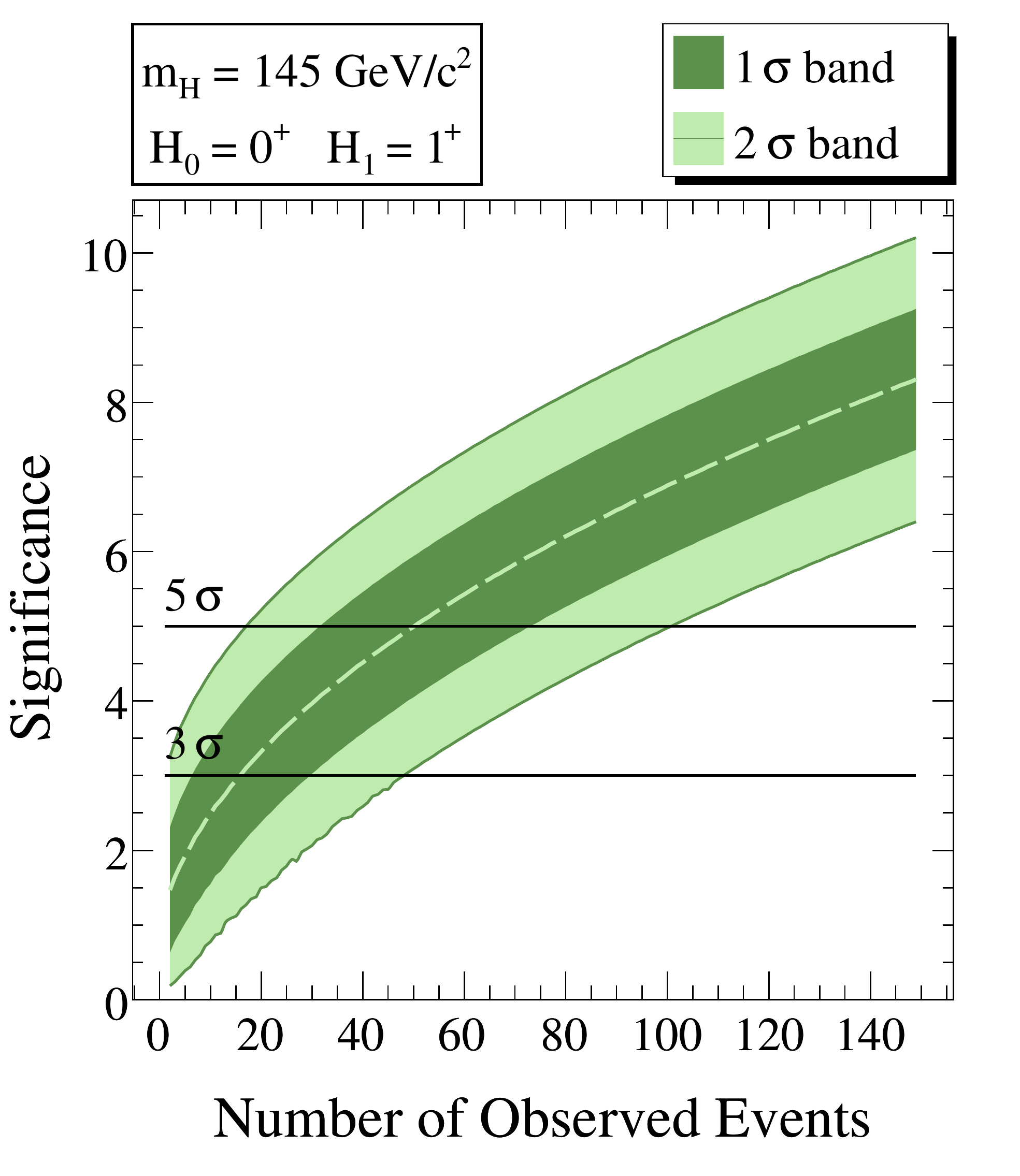}
\includegraphics[width=0.238\textwidth]{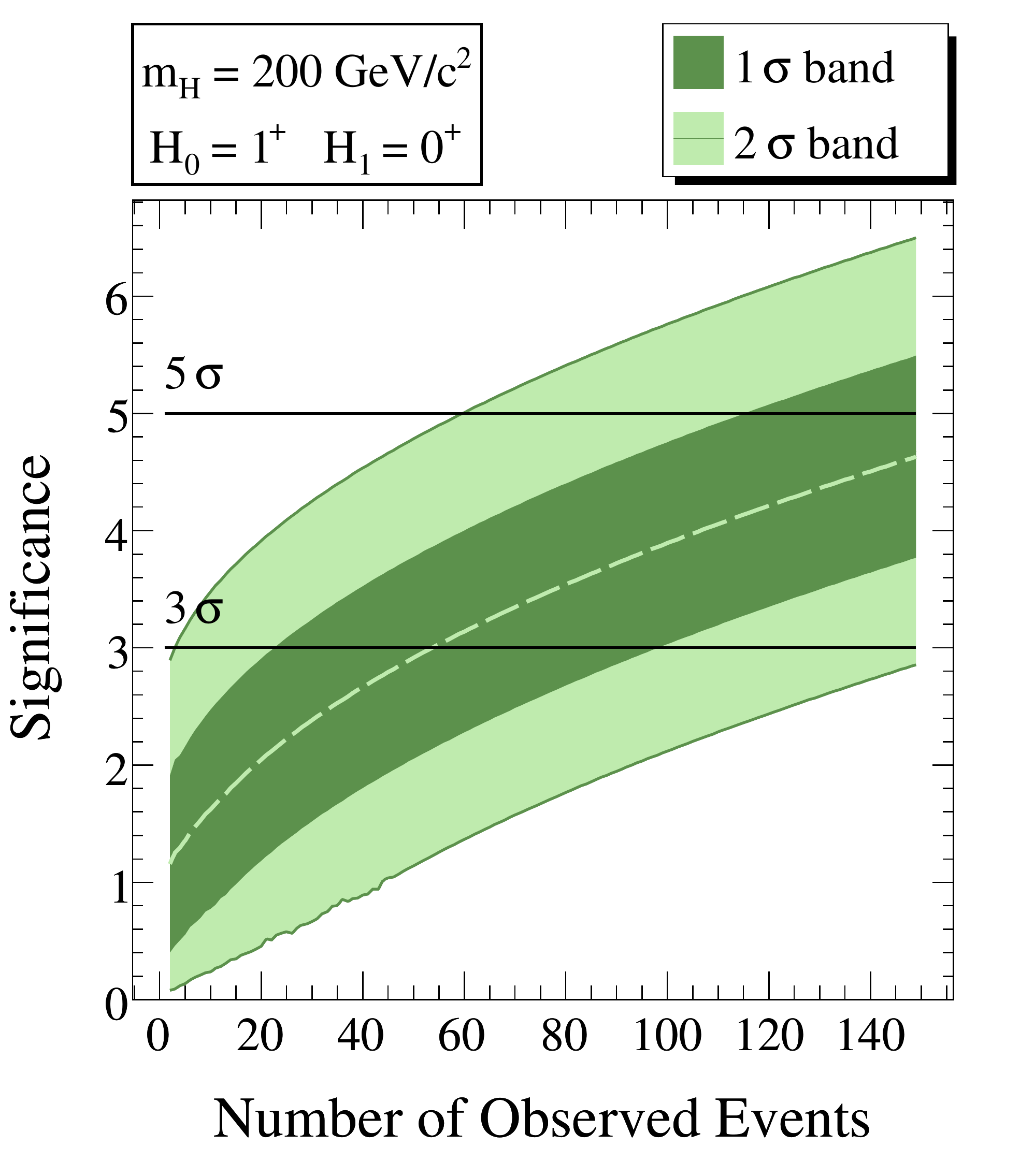}
\includegraphics[width=0.238\textwidth]{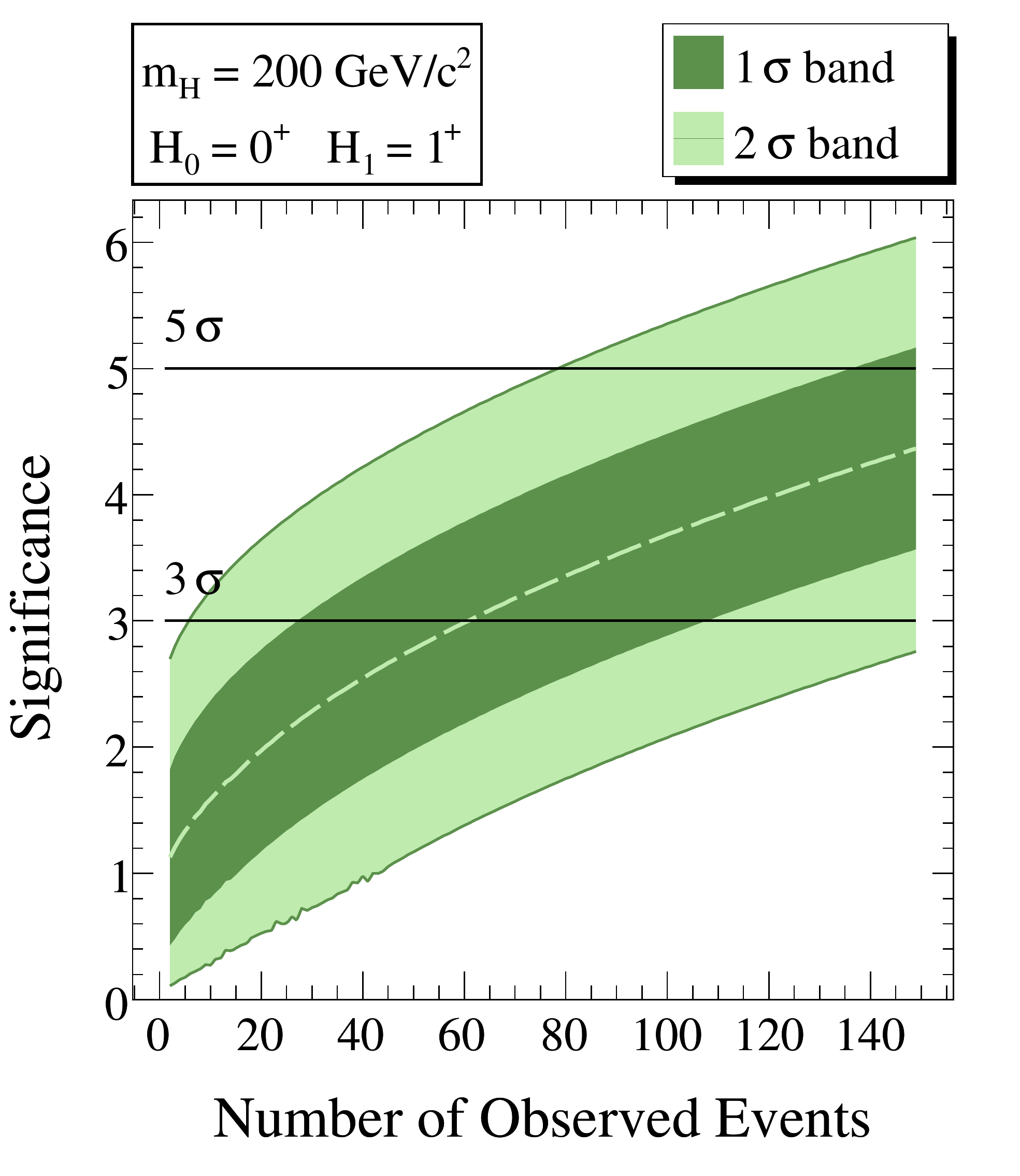}
\includegraphics[width=0.238\textwidth]{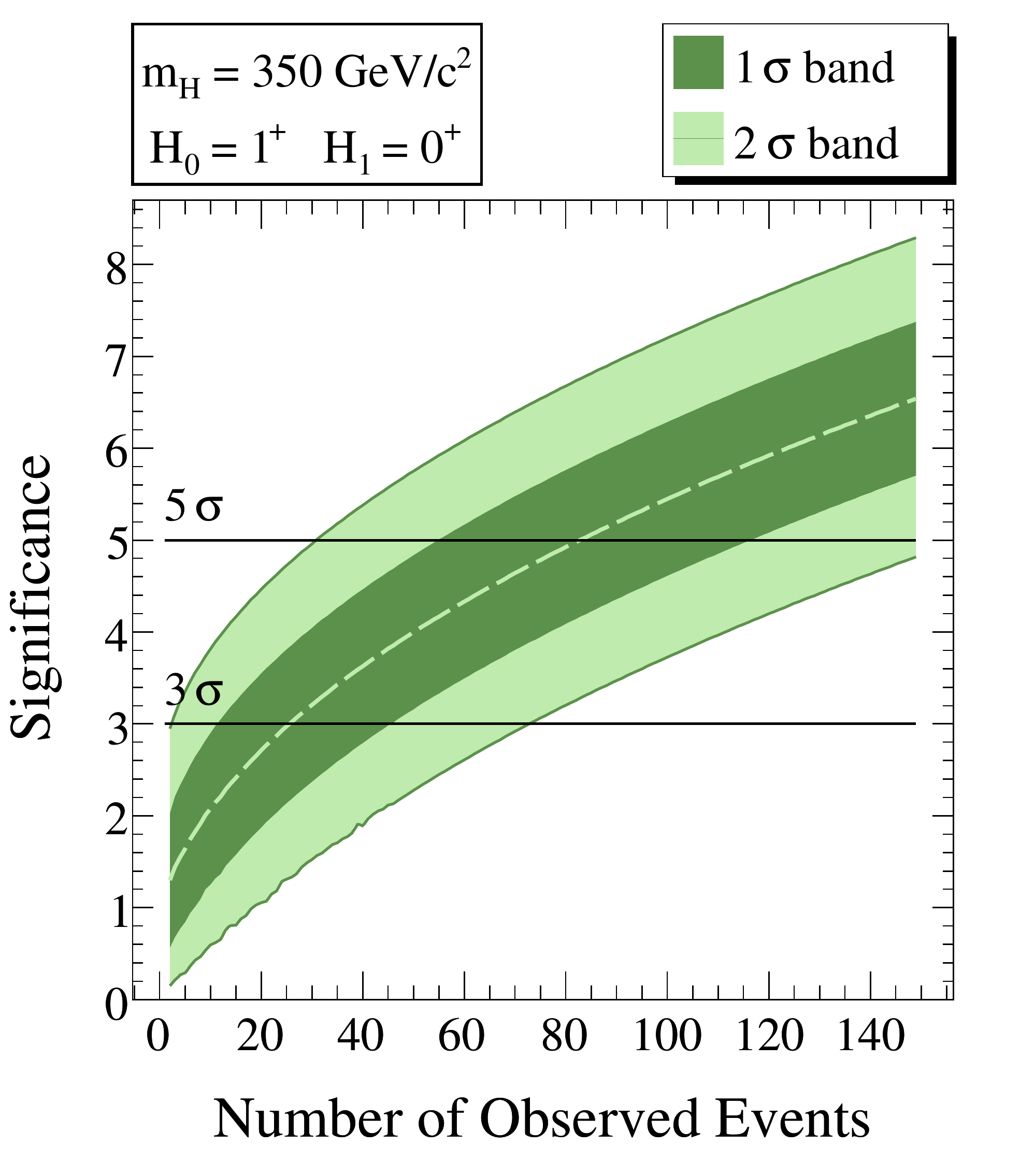}
\includegraphics[width=0.238\textwidth]{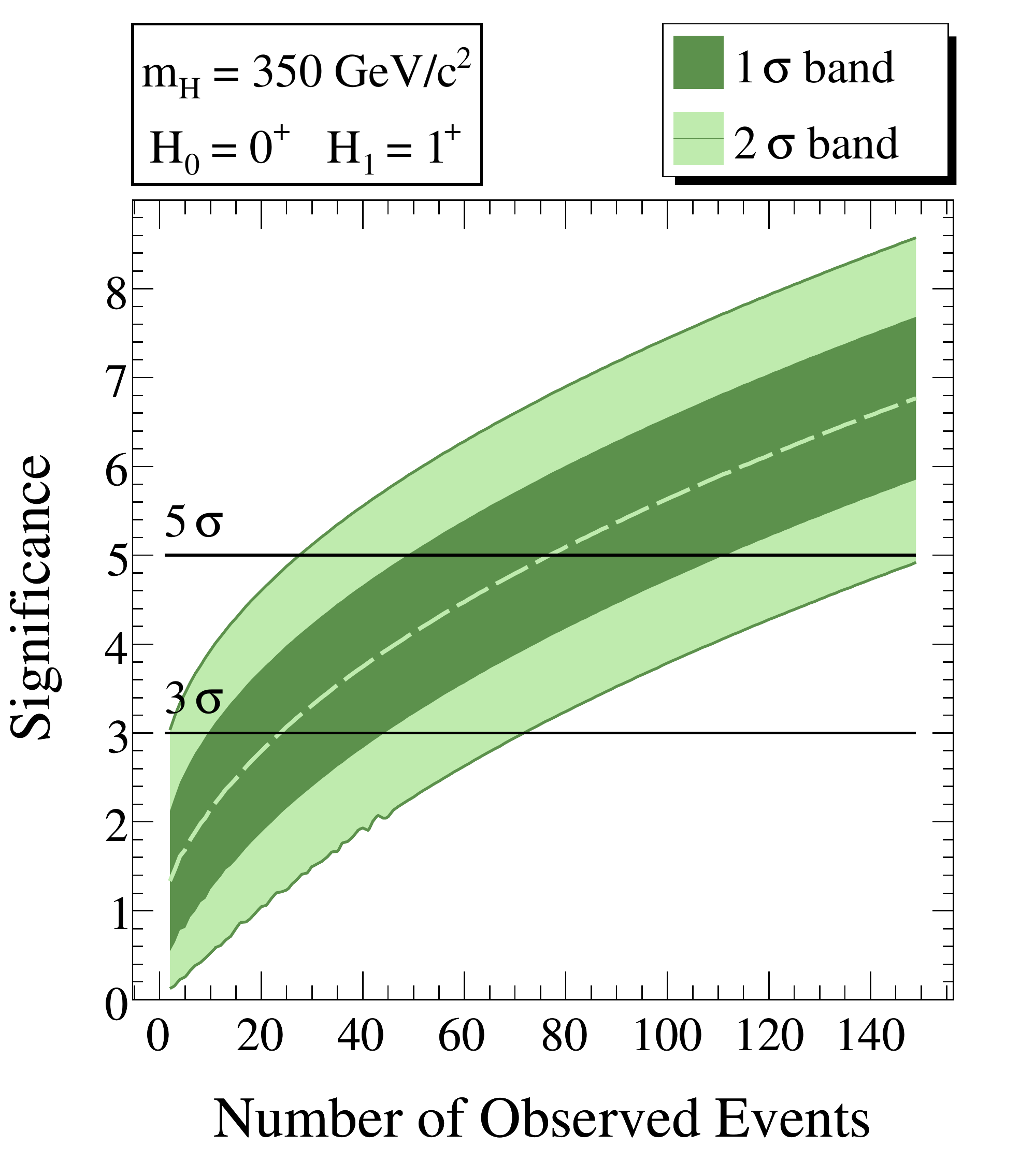}
\caption{Significance for rejecting  $1^{+}$  in
  favor of $0^{+}$, assuming $0^{+}$ is true  (left),
  or vice-versa ($0^+\!\leftrightarrow\! 1^+$, right),  
  for $m_H$$=$$145$, 200 and 350 GeV/c$^{2}$ (top, middle and bottom). \label{fig:COMP_SM_v_PA}}
\end{center}
\end{figure}

The significance for discriminating between the $0^{+}$ and
$1^{-}$ ($1^{+}$) hypotheses, as a function of $ N_S$, is summarized
in Fig.~\ref{fig:COMP_SM_v_PV} (Fig.~\ref{fig:COMP_SM_v_PA}). The full
correlated set of variables $\vec{\Omega}$, $\vec{\omega}$, and
$M_{Z^{*}}$ (when applicable) is used in the likelihood construction.
The discriminations are based on the NePe tests between simple
hypotheses with statistic $\log
(\mathcal{L}[0^{+}]/\mathcal{L}[1^{-}])$ ($\log (\mathcal{L}[0^{+}]/
\mathcal{L}[1^{+}])$). The discrimination between $0^{+}$ and $1^{-}$
or $1^{+}$ is similar.
%%%%%%%%%%%%%%%%%%%%%%%%%%%%%%%%%%%%%%%%%%%%%%%%%%%%%%%%%%%%%%%%%%%
%%%%%%%%%%%%%%%%%%%%%%%%%%%%%%%%%%%%%%%%%%%%%%%%%%%%%%%%%%%%%%%%%%%
%%%%%%%%%%%%%%%%%%%%%%%%%%%%%%%%%%%%%%%%%%%%%%%%%%%%%%%%%%%%%%%%%%%
%%%%%%%%%%%%%%%%%%%%%%%%%%%%%%%%%%%%%%%%%%%%%
% 0+ vs. spin two
%%%%%%%%%%%%%%%%%%%%%%%%%%%%%%%%%%%%%%%%%%%%%

\vspace*{10pt}
\subsection{$\mathbf{0^+}$ vs. $\mathbf{2^{+}}$}
\begin{figure}[htbp]
\begin{center}
\includegraphics[width=0.238\textwidth]{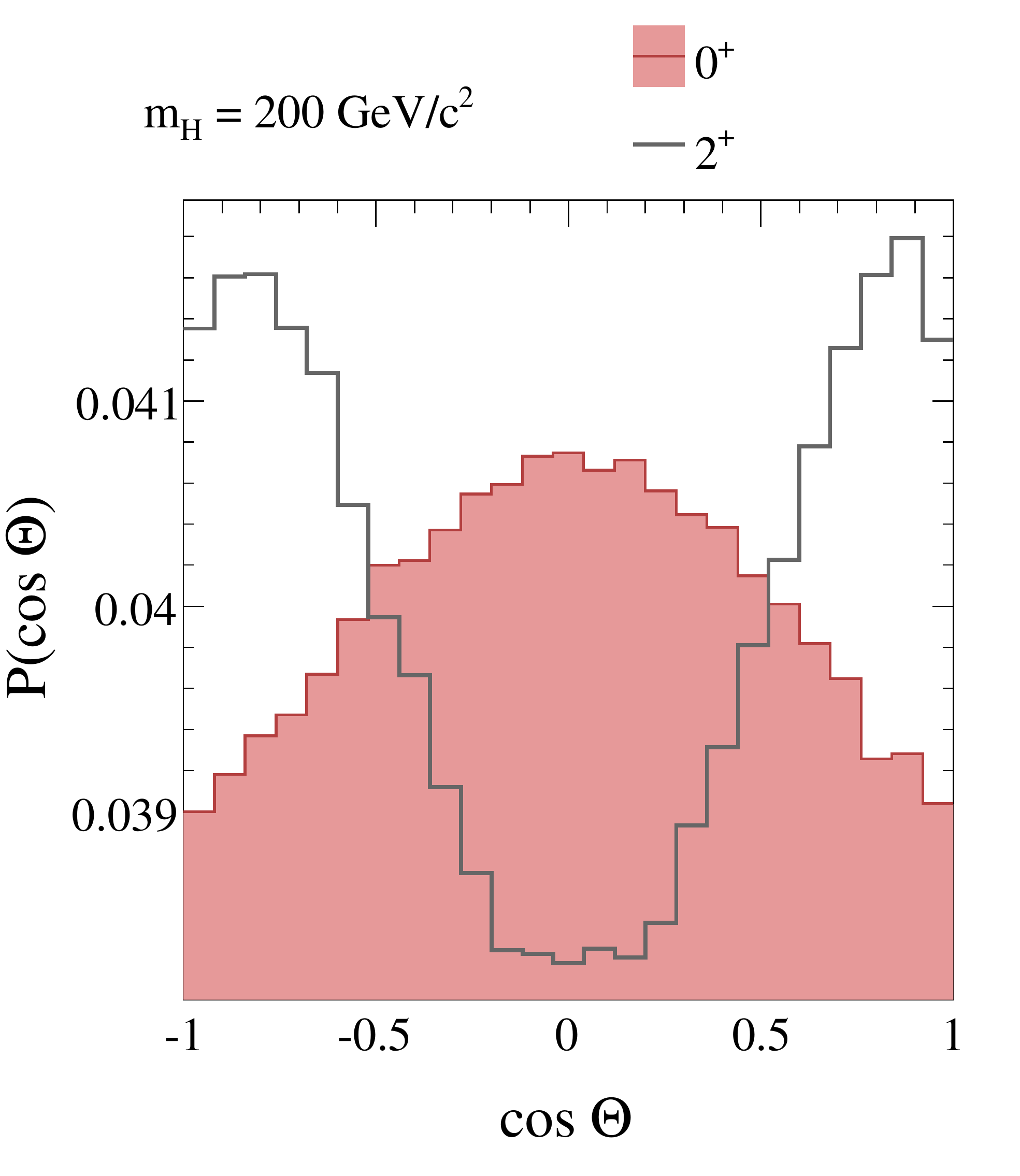}
\includegraphics[width=0.238\textwidth]{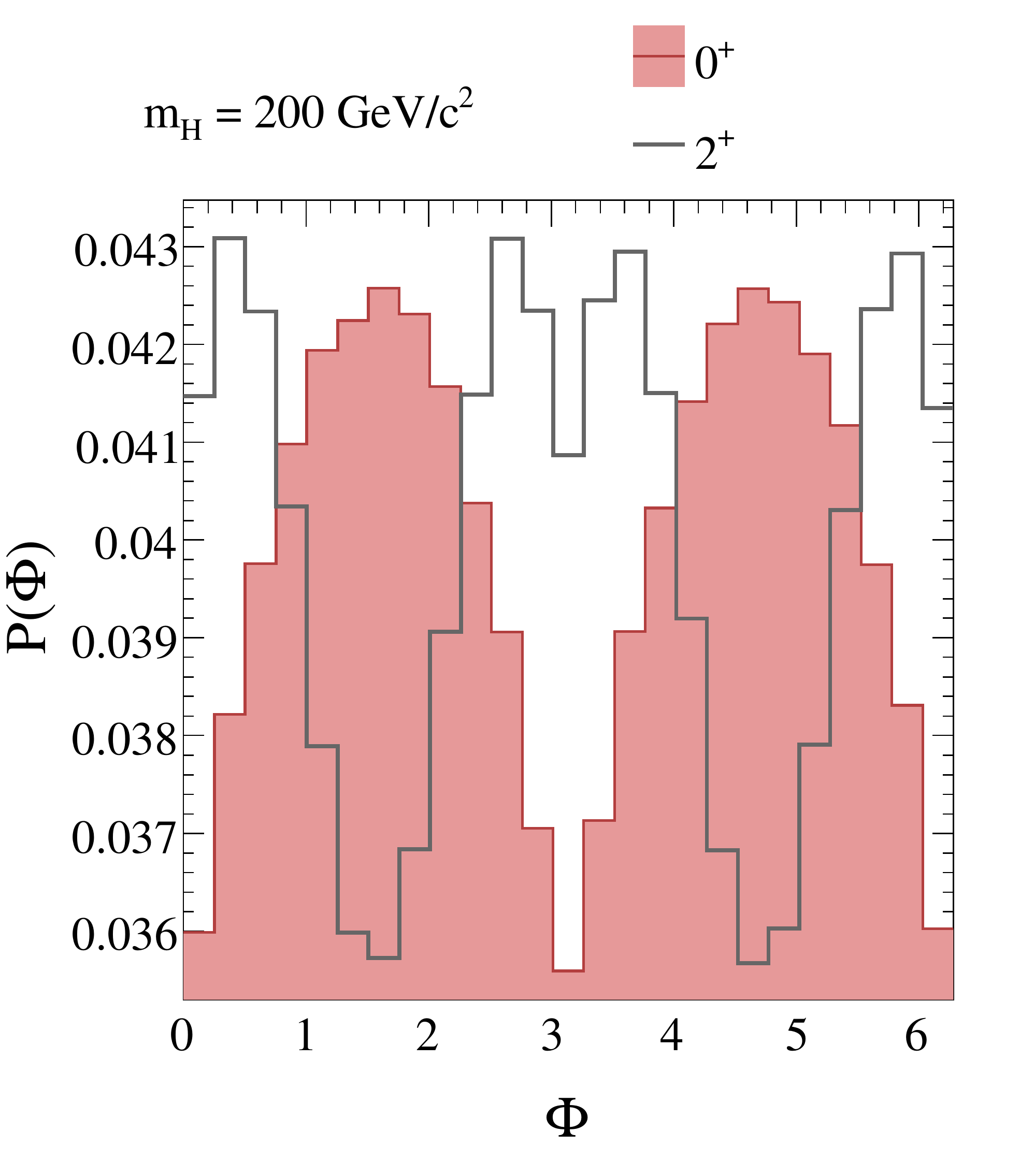}
\includegraphics[width=0.238\textwidth]{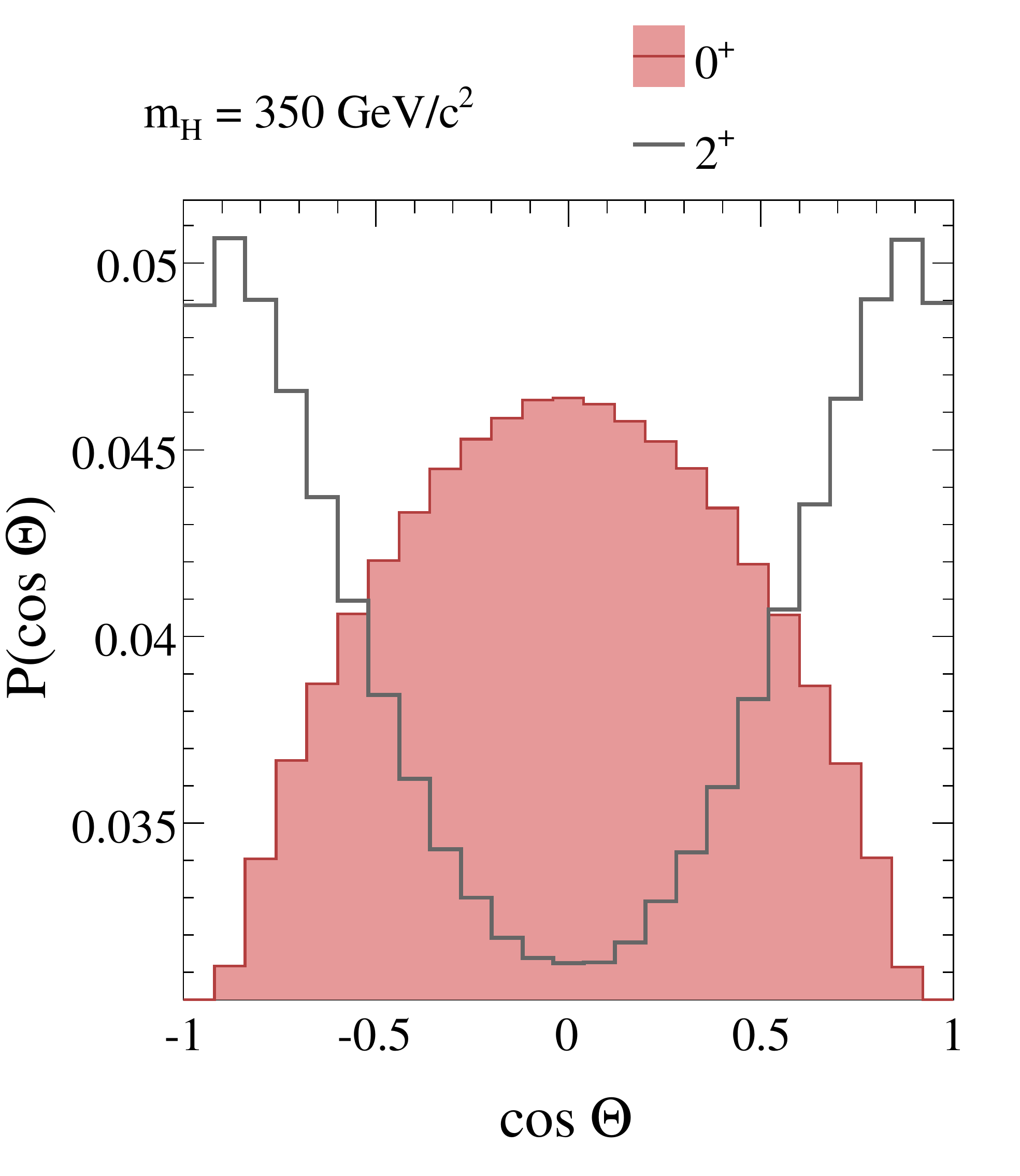}
\includegraphics[width=0.238\textwidth]{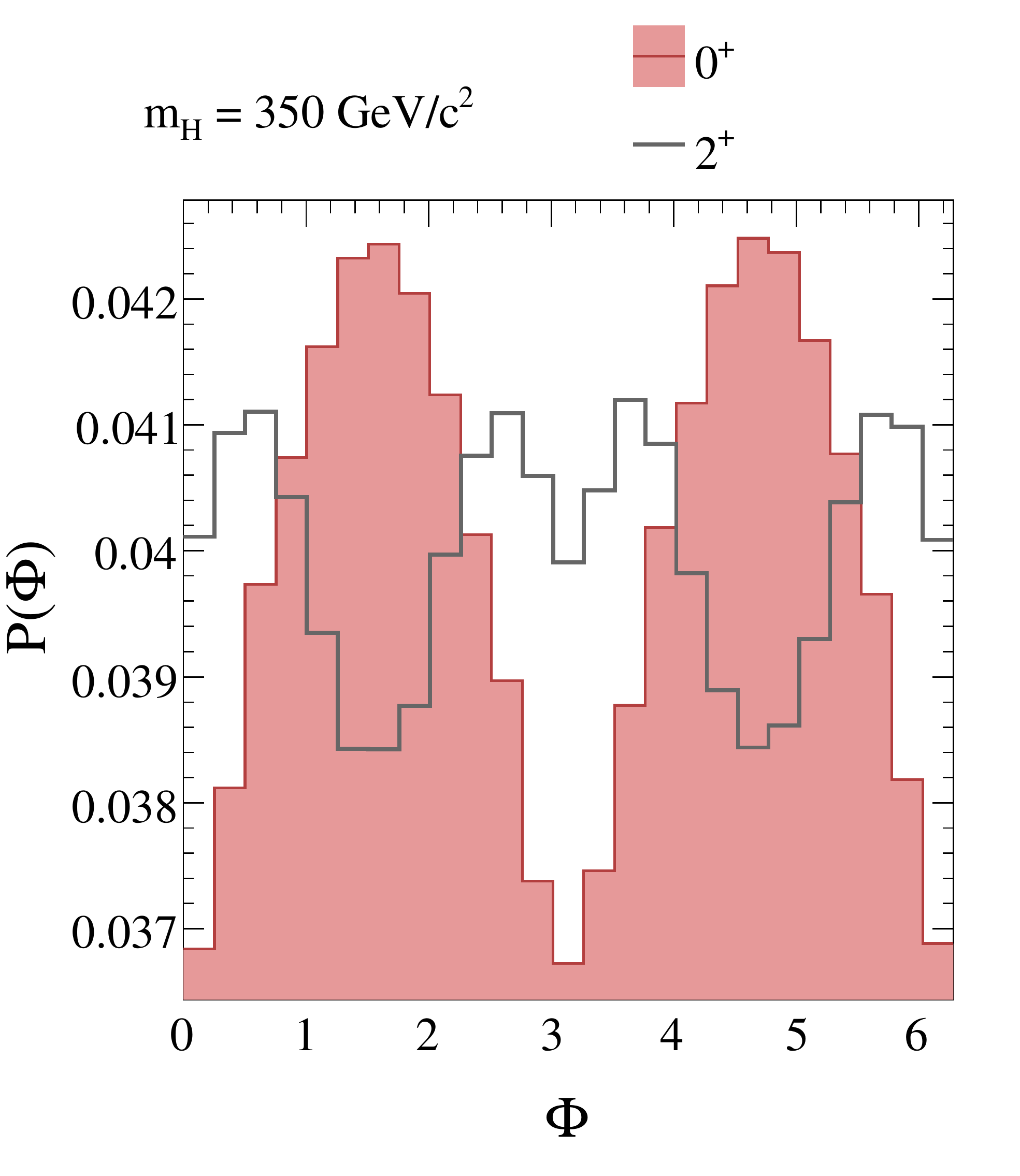}
\caption{Distributions  of the variables cos$\,\Theta$ (left) 
  and $\Phi$ (right) for $0^{+}$, $2^{+}$ 
  resonances with masses of 200 and 350 GeV/c$^{2}$ (top, bottom). 
  All distributions are normalized to a unit integral.
  \label{fig:KIN_SM_2_bomega}}
\end{center}
\end{figure}

\begin{figure}[htbp]
\begin{center}
\includegraphics[width=0.238\textwidth]{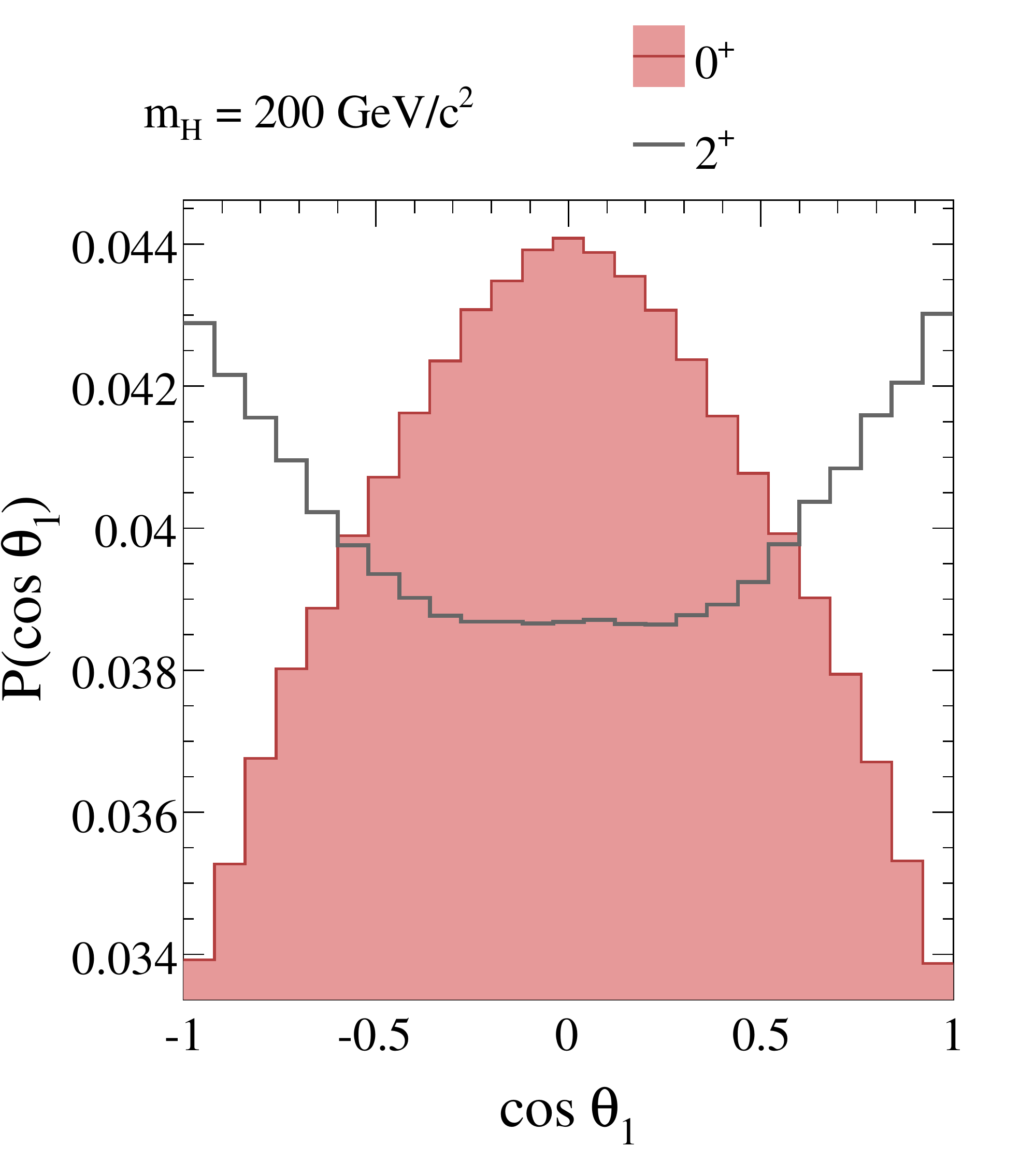}
\includegraphics[width=0.238\textwidth]{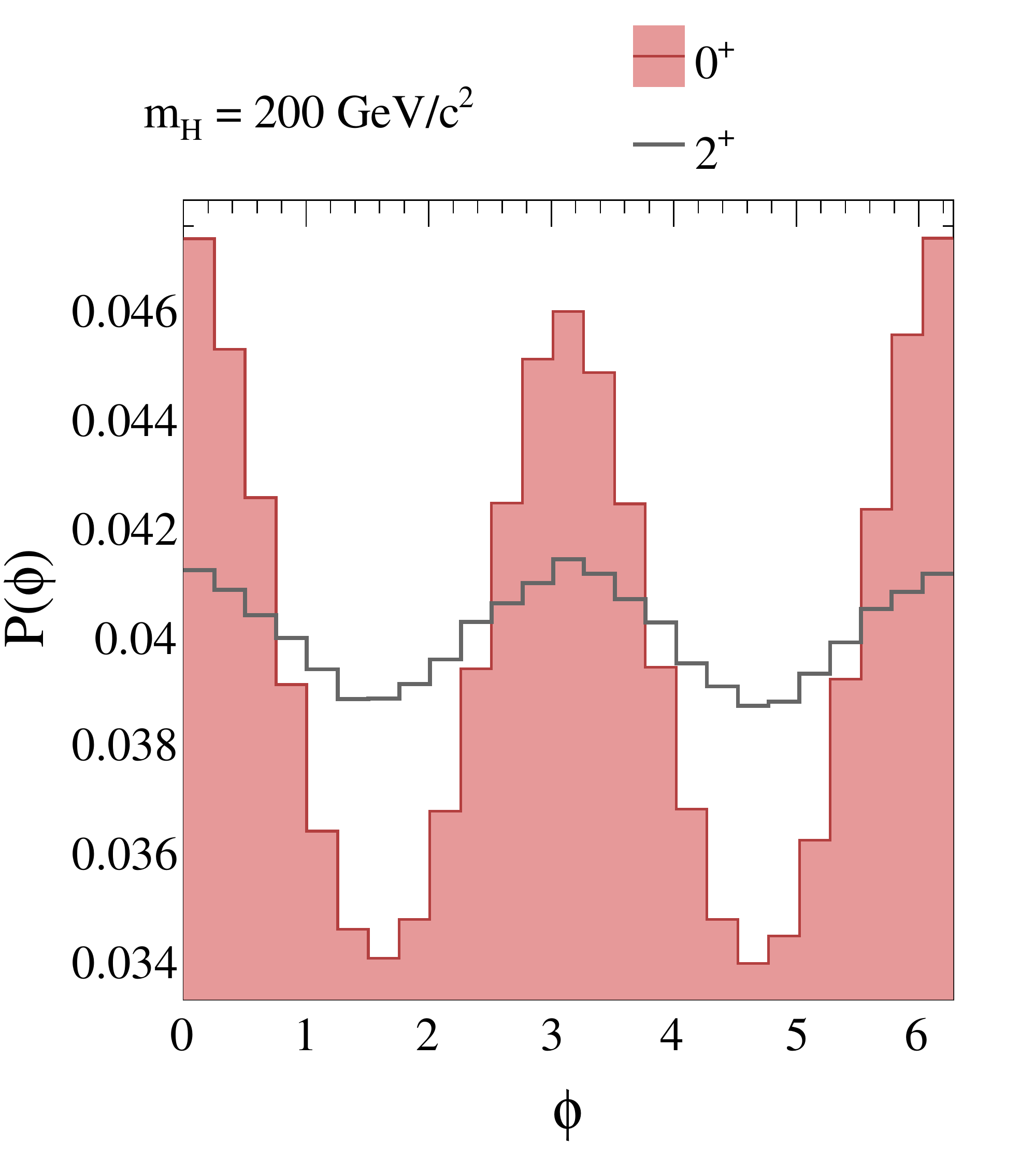}
\includegraphics[width=0.238\textwidth]{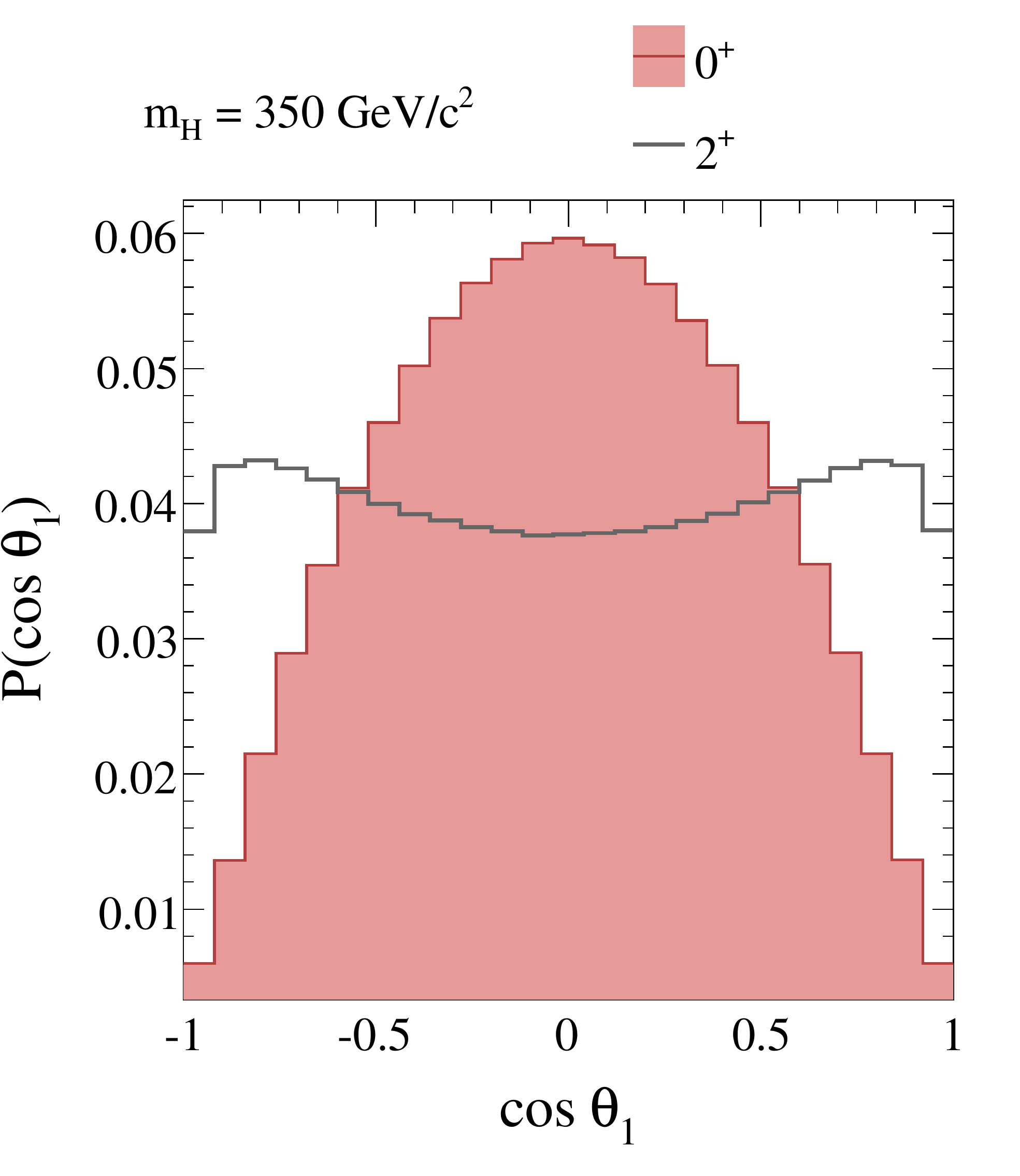}
\includegraphics[width=0.238\textwidth]{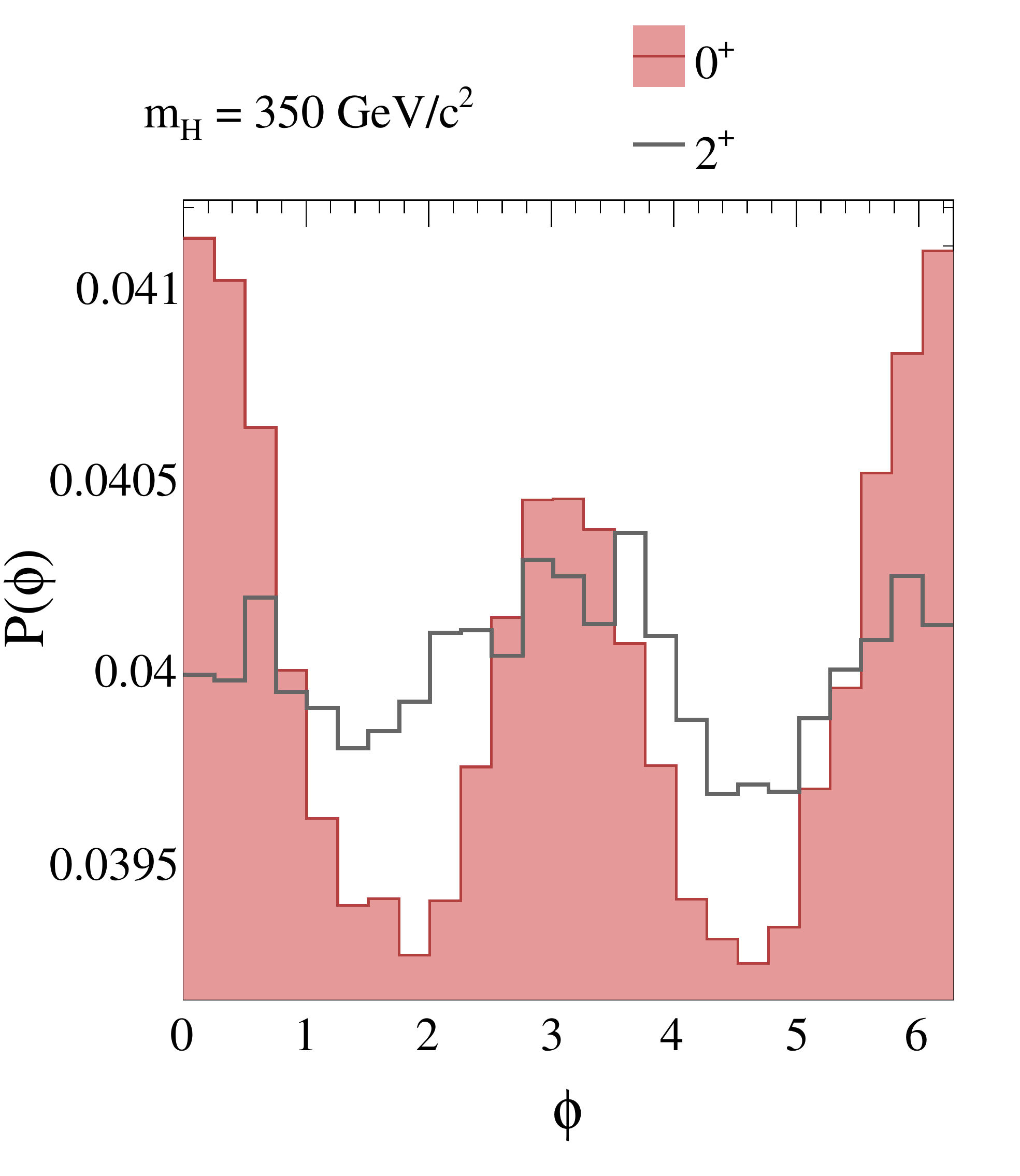}
\caption{Distributions of the variables cos$\,\theta_1$ (left) and
  $\phi$ (right) for $0^{+}$, $2^{+}$  resonances with
  masses of 200 and 350 GeV/c$^{2}$ (top, bottom).  All distributions
  are normalized a unit integral.
  \label{fig:KIN_SM_2_lomega}}
\end{center}
\end{figure}

\begin{figure}[htbp]
\begin{center}
\includegraphics[width=0.38\textwidth]{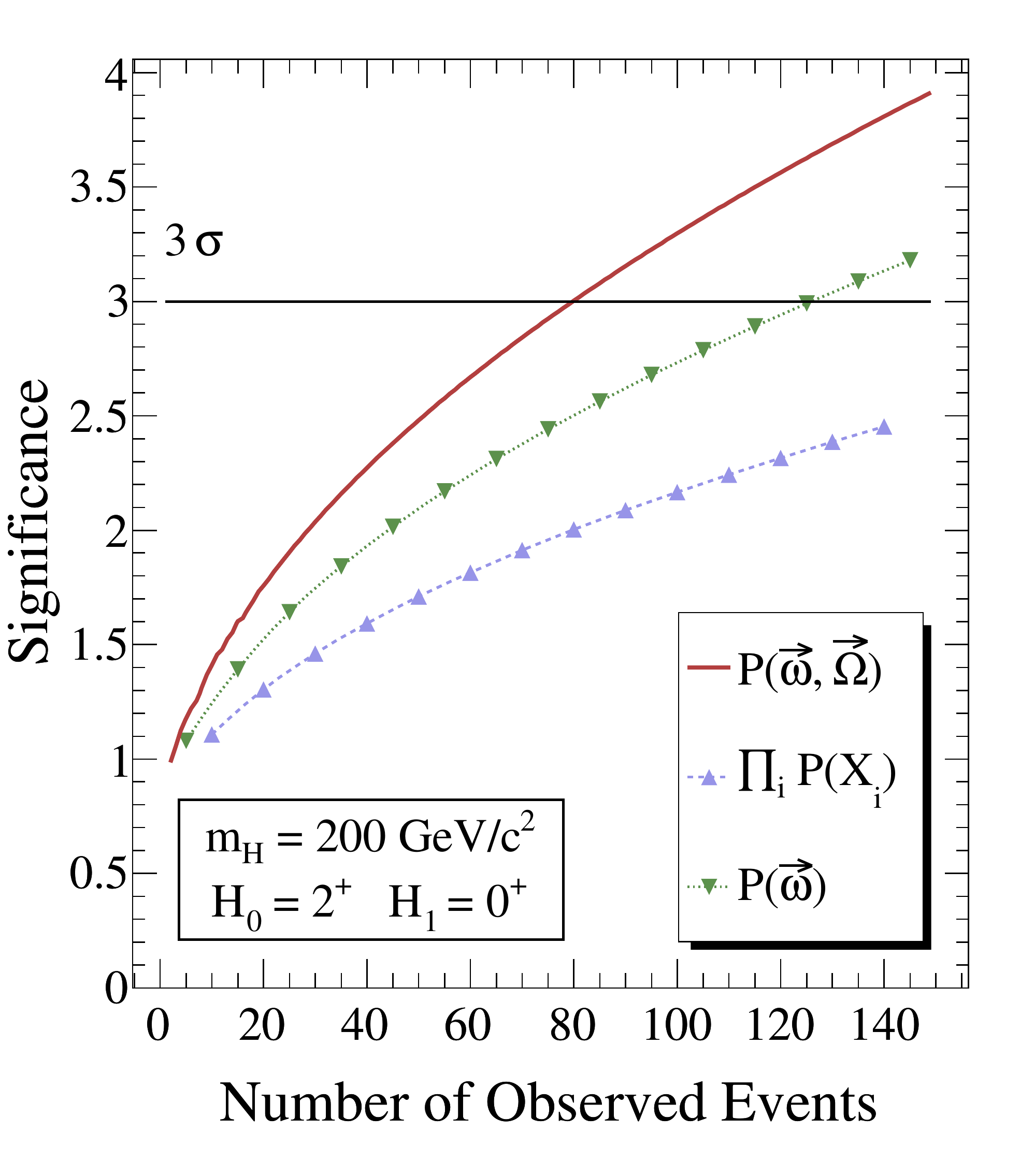}
\caption{Median significance for rejecting $2^{+}$
   in favor of $0^{+}$, assuming $0^{+}$ is true, for the different likelihood constructions discussed in the 
text. 
\label{fig:SPEC_0_2}}
\end{center}
\end{figure}

We consider one ``pure'' spin 2 model: a $J$$=$$2^+$ heavy graviton-like
resonance. A $J$$=$$2$ object has {\it pdfs} with non-trivial dependence
on the angles $\vec{\Omega}$ up to quartic order in cos$\,\Theta$.  In
Fig.~\ref{fig:KIN_SM_2_bomega} we show the corresponding distributions
in the $\vec{\Omega}$ variables for $m_H$$=$$200$ and $350$ GeV/c$^2$.
%%%%%%%%%%%%%%%%%%%%%%%%%%%%%%%%%%%%%%%%%%%%%%%%%%%%%%%%%%%%%%%%%%%
%%%%%%%%%%%%%%%%%%%%%%%%%%%%%%%%%%%%%%%%%%%%%%%%%%%%%%%%%%%%%%%%%%%
The ability to discriminate between the $0^{+}$ and $J$$=$$2$ hypotheses
improves with increasing resonance mass. Despite the presence of
quartic terms in cos$\,\Theta$ in the $2^{+}$ {\it pdf} and the
absence of this variable in the $0^{+}$ {\it pdf}, their corresponding
one-dimensional {\it pdfs} are similar for the $0^{+}$ and $2^{+}$
resonances for values of $m_{H}$ close to $2\,M_Z$, as shown in
Fig.~\ref{fig:KIN_SM_2_bomega}.  Similar behavior is observed in the
distributions of cos$\,\theta_{1}$ and cos$\,\theta_{2}$, as
illustrated in Fig.~\ref{fig:KIN_SM_2_lomega}.  Nevertheless, the
inclusion of all angular variables and their correlations improves
the discrimination power between these hypotheses as
shown in Fig.~\ref{fig:SPEC_0_2}.
%%%%%%%%%%%%%%%%%%%%%%%%%%%%%%%%%%%%%%%%%%%%%%%%%%%%%%%%%%%%%%%%%%%
%%%%%%%%%%%%%%%%%%%%%%%%%%%%%%%%%%%%%%%%%%%%%%%%%%%%%%%%%%%%%%%%%%%
%%%%%%%%%%%%%%%%%%%%%%%%%%%%%%%%%%%%%%%%%%%%%%%%%%%%%%%%%%%%%%%%%%%
%%%%%%%%%%%%%%%%%%%%%%%%%%%%%%%%%%%%%%%%%%%%%%%%%%%%%%%%%%%%%%%%%%%

\begin{figure}[htbp]
\begin{center}
\includegraphics[width=0.238\textwidth]{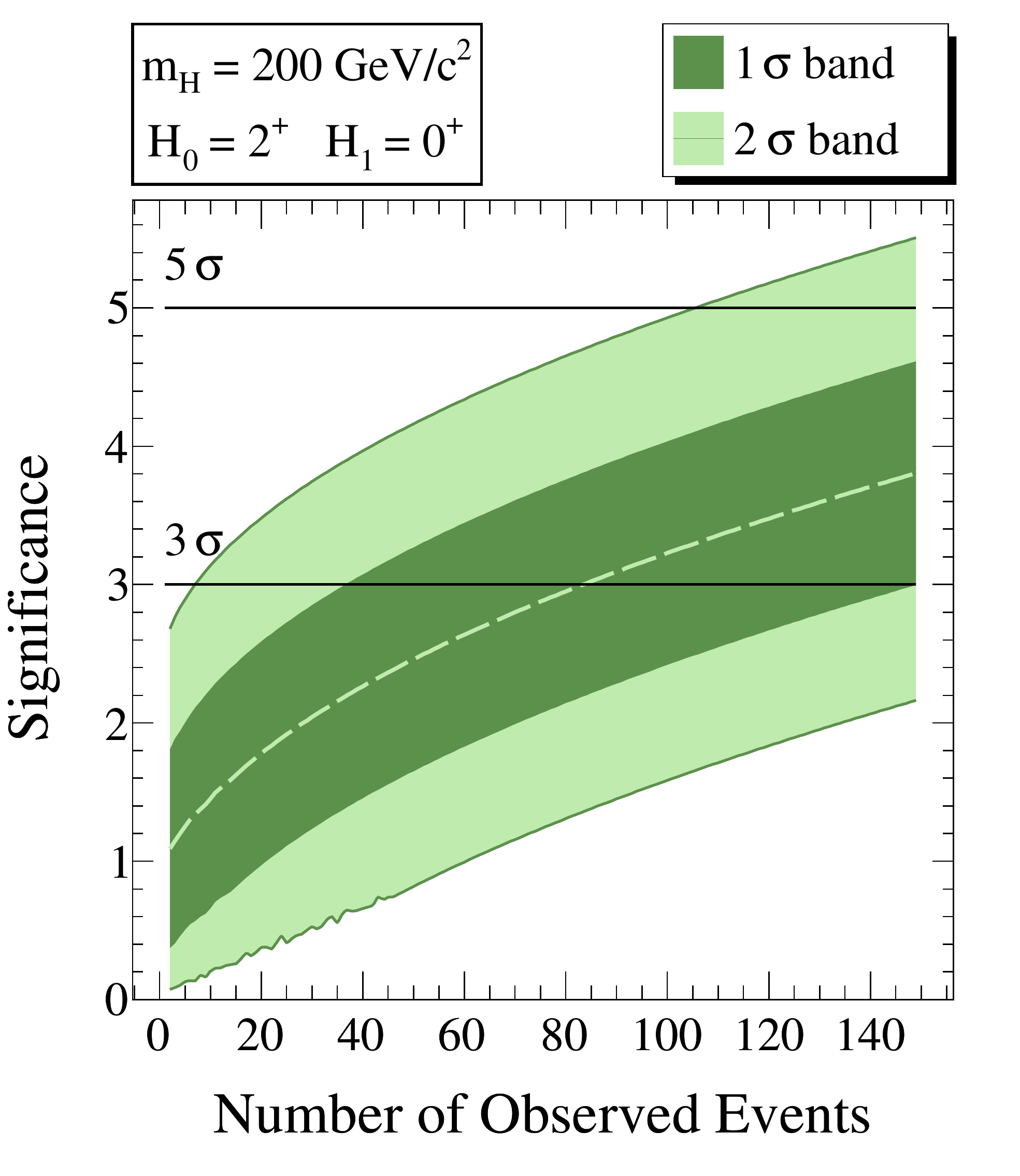}
\includegraphics[width=0.238\textwidth]{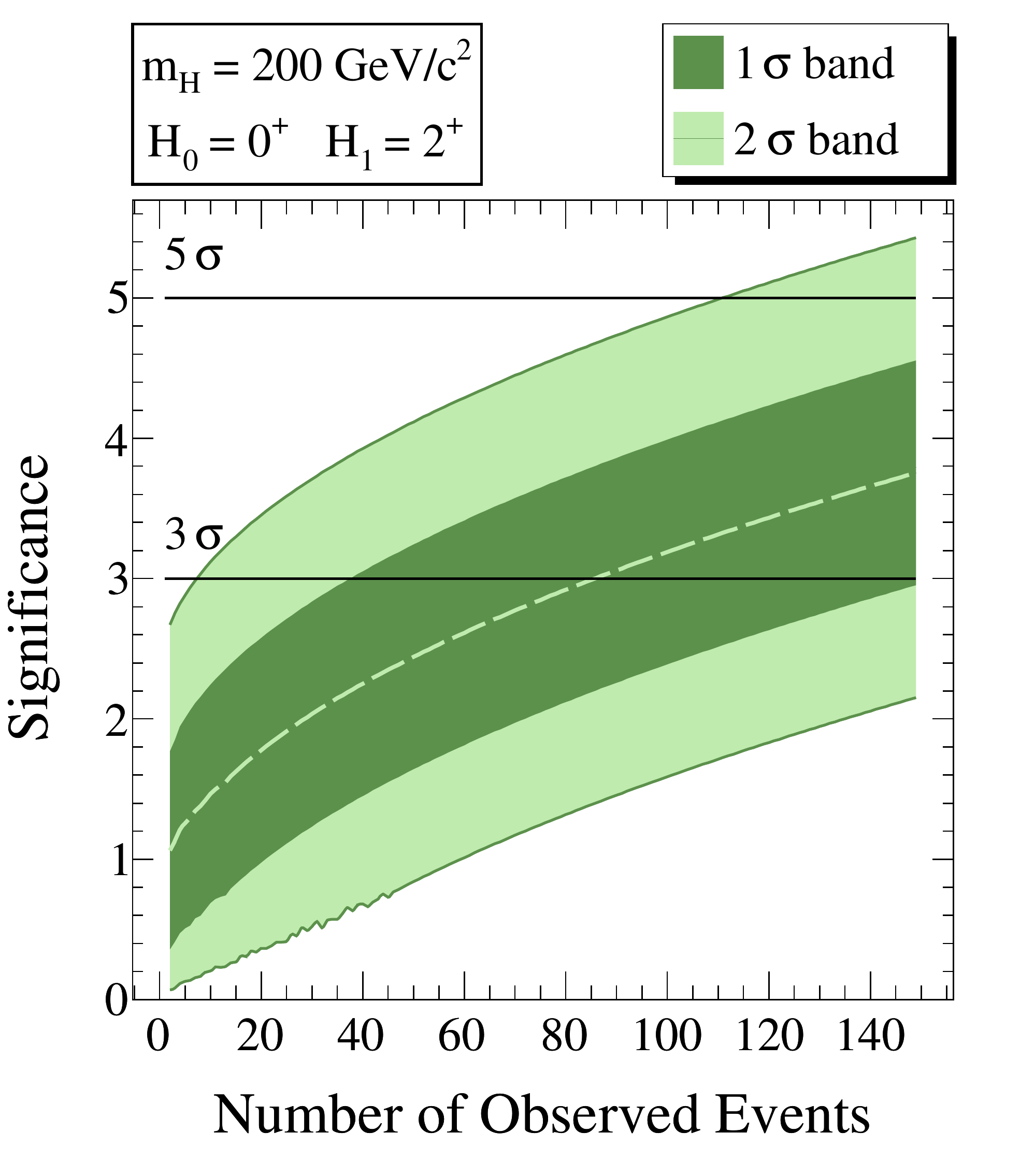}
\includegraphics[width=0.238\textwidth]{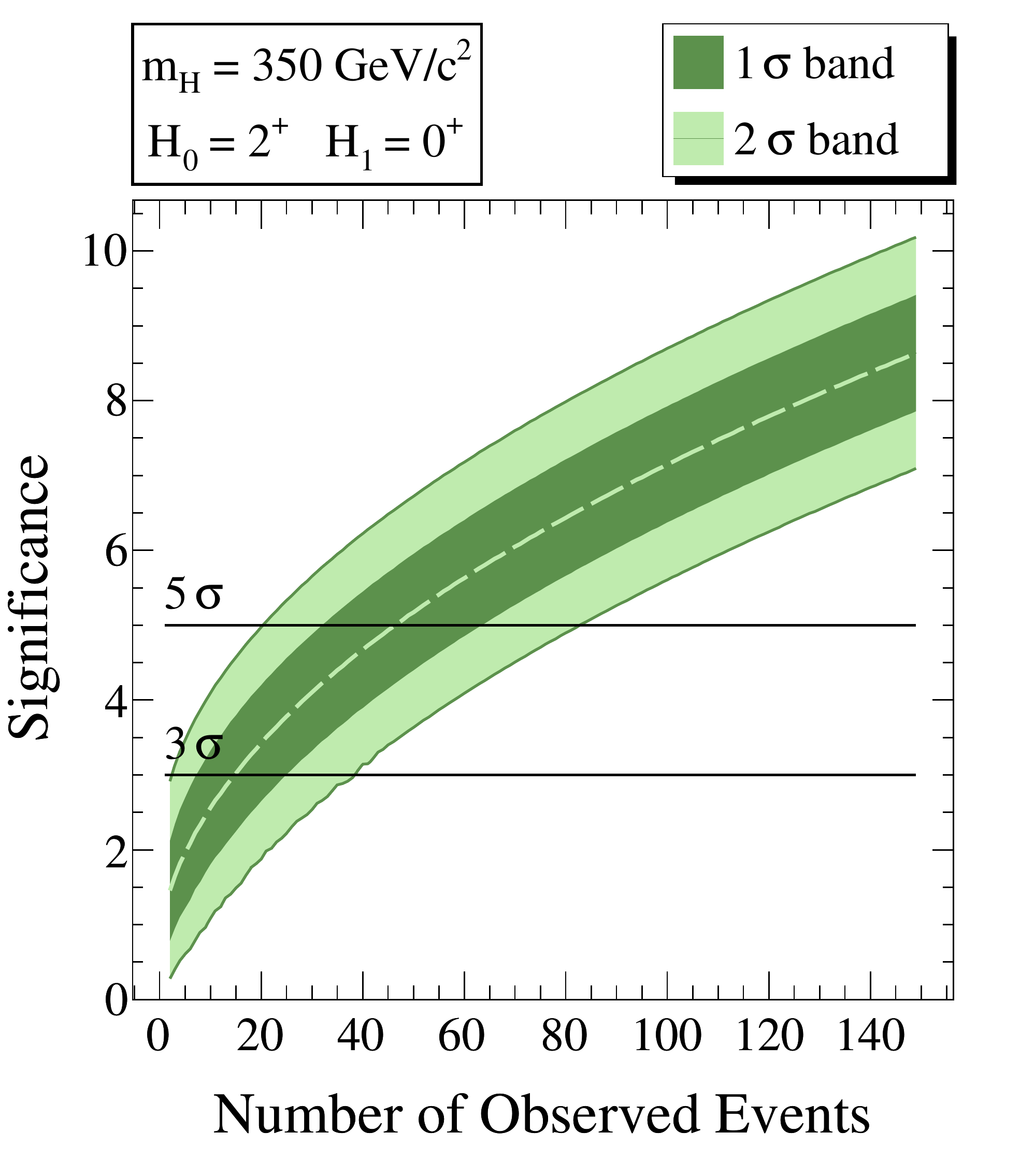}
\includegraphics[width=0.238\textwidth]{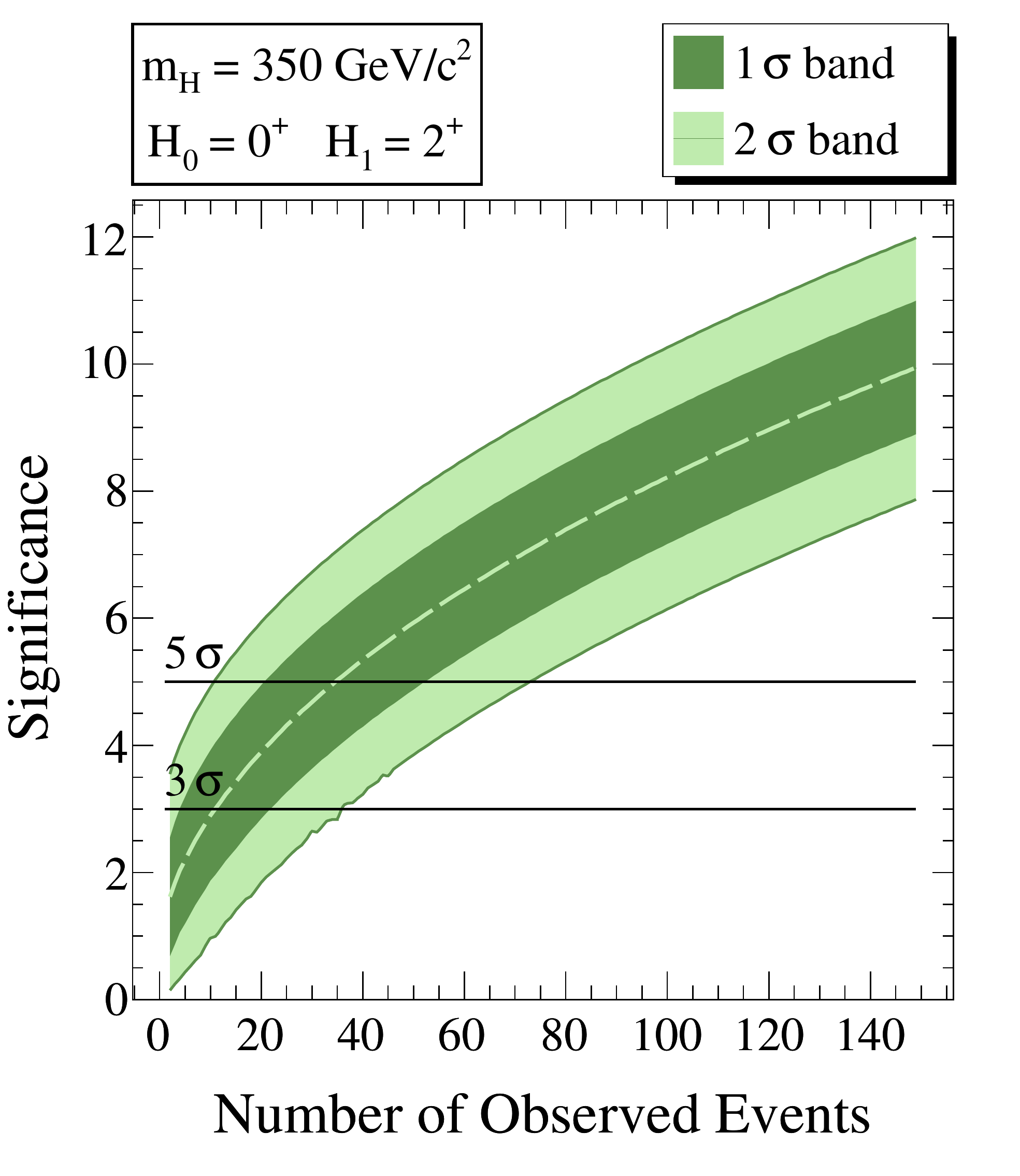}
\caption{Significance for rejecting $2^{+}$ in favor of
  $0^{+}$, assuming $0^{+}$ is true (left) or vice-versa ($0^+\!\leftrightarrow\! 2^+$, right),
  for $m_H$$=$$200$ and 350 GeV/c$^{2}$ (top, bottom). \label{fig:COMP_SM_v_KK}}
\end{center}
\end{figure}

The significance for discriminating between $0^{+}$ and
$2^{+}$ as a function of $N_S$, is summarized in
Fig.~\ref{fig:COMP_SM_v_KK} for $m_{H}$$=$$200$ and $350$ GeV/c$^{2}$. For
these tests the variables $\vec{\Omega}$ and $\vec{\omega}$ and their
correlations were used in the likelihood. Model discrimination is
based on the NePe test between simple hypotheses with test statistic
$\log ({\mathcal{L}}[0^{+}]/{\mathcal{L}}[2^{+}])$ and $\log
({\mathcal{L}}[0^{+}]/{\mathcal{L}}[2^{-}])$.
%%%%%%%%%%%%%%%%%%%%%%%%%%%%%%%%%%%%%%%%%%%%%%%%%%%%%%%%%%%%%%%%%%%
%%%%%%%%%%%%%%%%%%%%%%%%%%%%%%%%%%%%%%%%%%%%%%%%%%%%%%%%%%%%%%%%%%%
%%%%%%%%%%%%%%%%%%%%%%%%%%%%%%%%%
% other pure JPC resonances
%%%%%%%%%%%%%%%%%%%%%%%%%%%%%%%%%

\vspace*{2mm}
\subsection{Other pure $\mathbf{J^{PC}}$ comparisons \label{sec:other}}
If a resonance discovered in the $4\ell$ final state does not have the
quantum numbers of the SM Higgs boson, it is likely that $0^{+}$ will
be rejected in favor of other pure-$J^{PC}$ hypotheses.  The issue of
abandoning a particular $J^{PC}$ in favor of others becomes a
combinatoric exercise, where the compatibility of the data is assessed
against each possible pair of hypotheses in a simple NePe test, in view
of selecting the optimal assumption. In this section we present the
expected results for these comparison tests, as a function of the
observed number of events $N_S$. Following the results of the previous
section, we always use the full set of angular variables plus, when
appropriate, $M_{Z^*}$, corresponding to the optimal statistic
for model discrimination.
%%%%%%%%%%%%%%%%%%%%%%%%%%%%%%%%%%%%%%%%%%%%%%%%%%%%%%%%%%%%%%%%%%%
\begin{figure}[htbp]
\begin{center}
\includegraphics[width=0.238\textwidth]{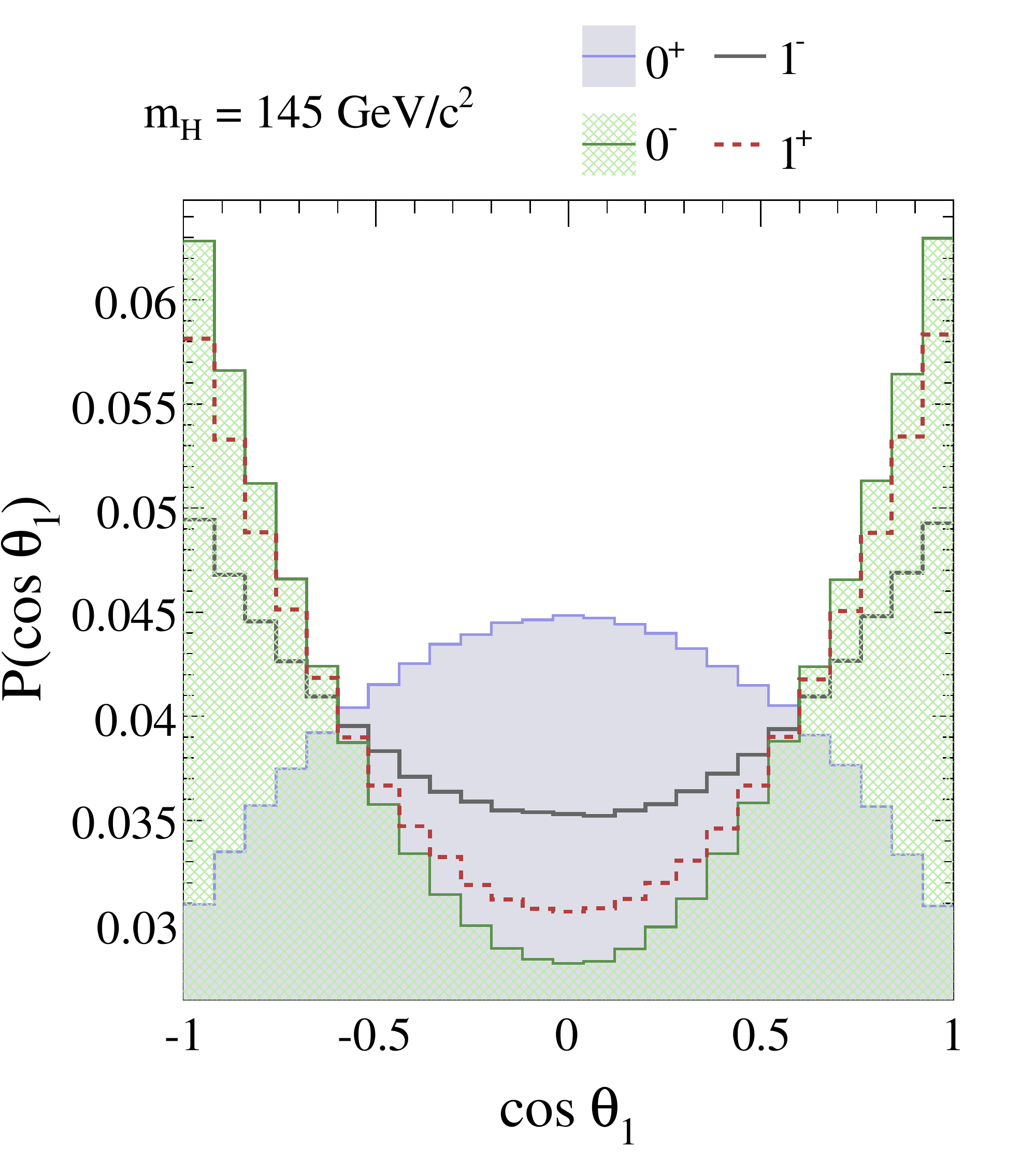}
\includegraphics[width=0.238\textwidth]{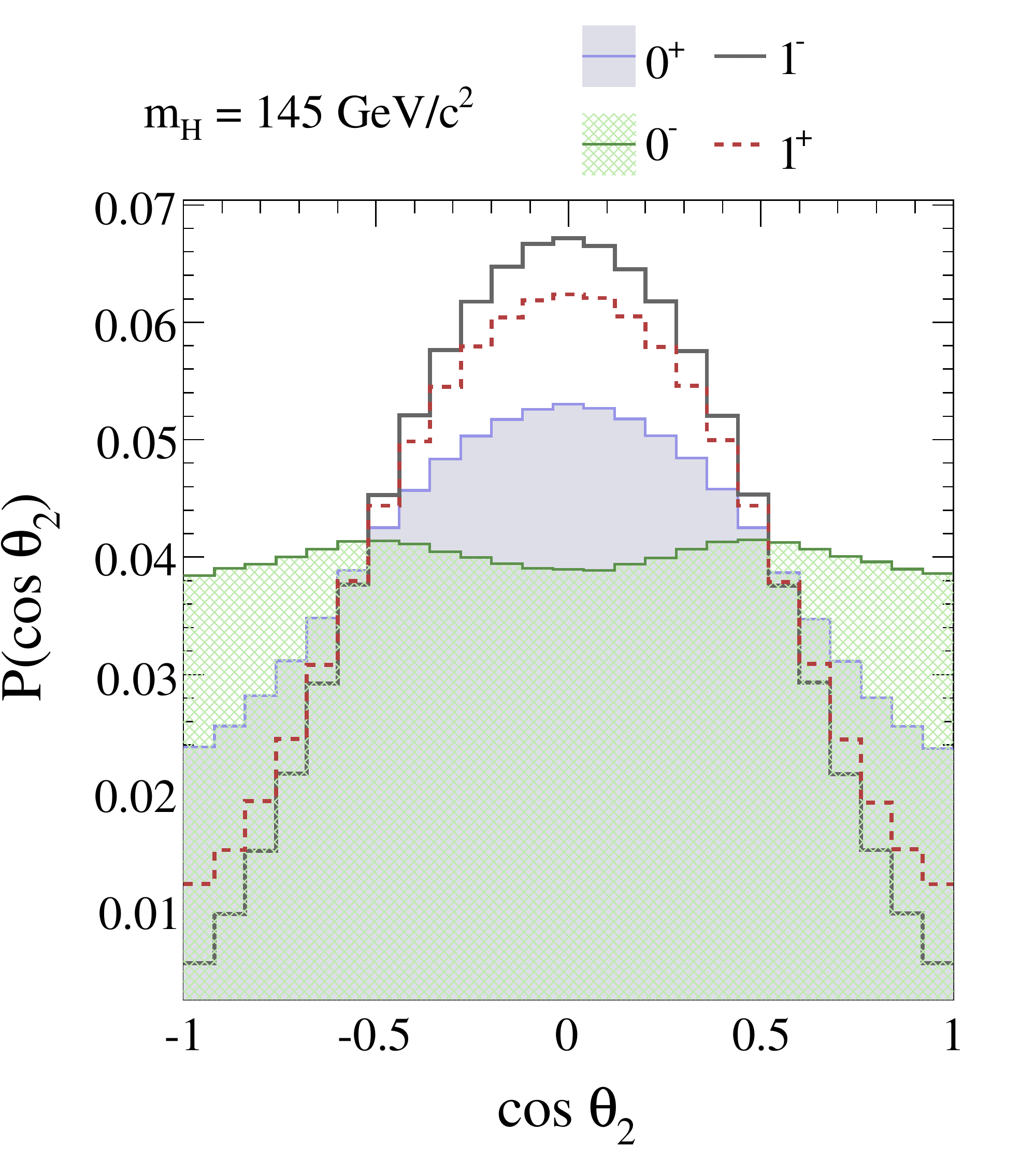}
\includegraphics[width=0.238\textwidth]{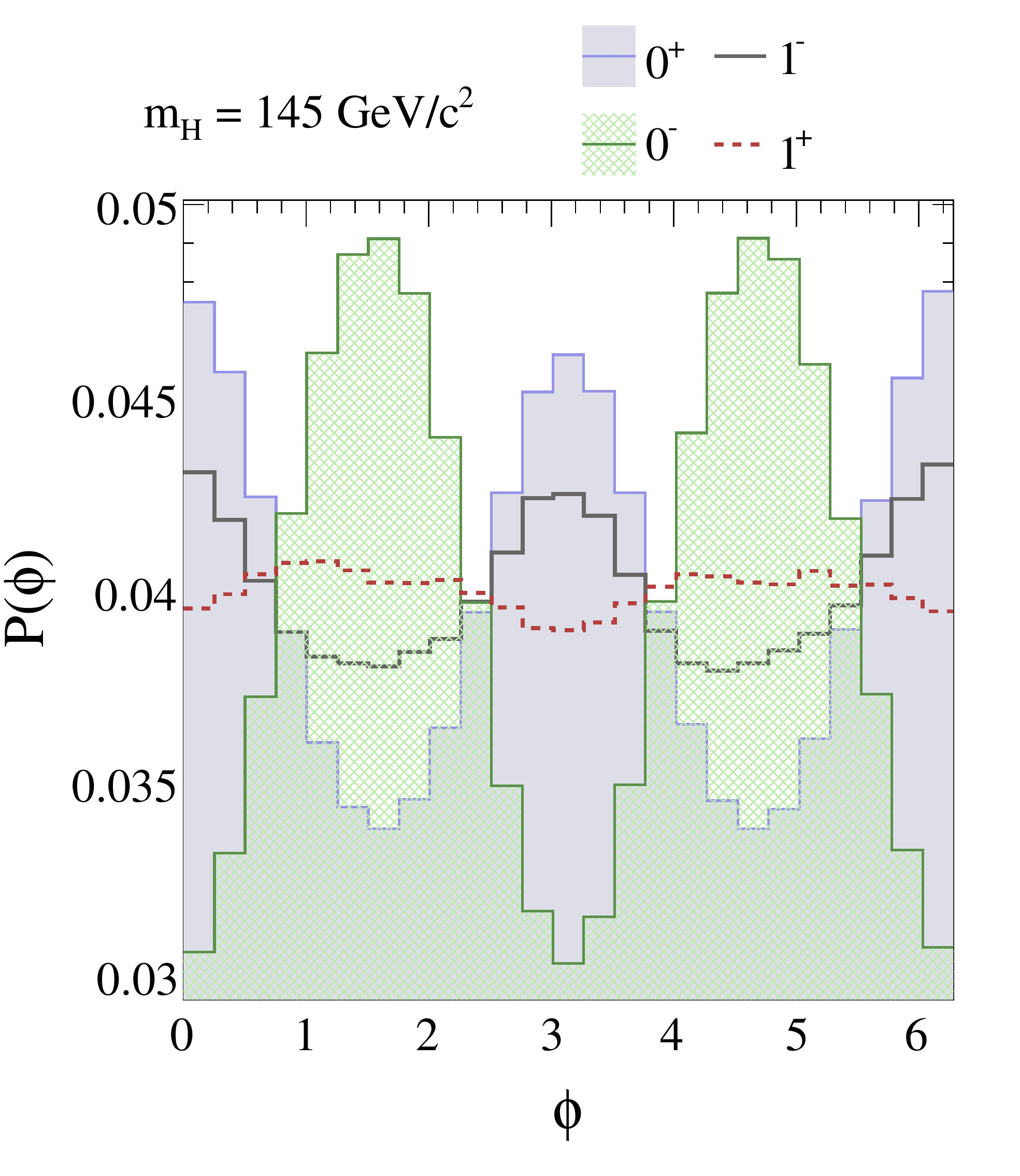}
\includegraphics[width=0.238\textwidth]{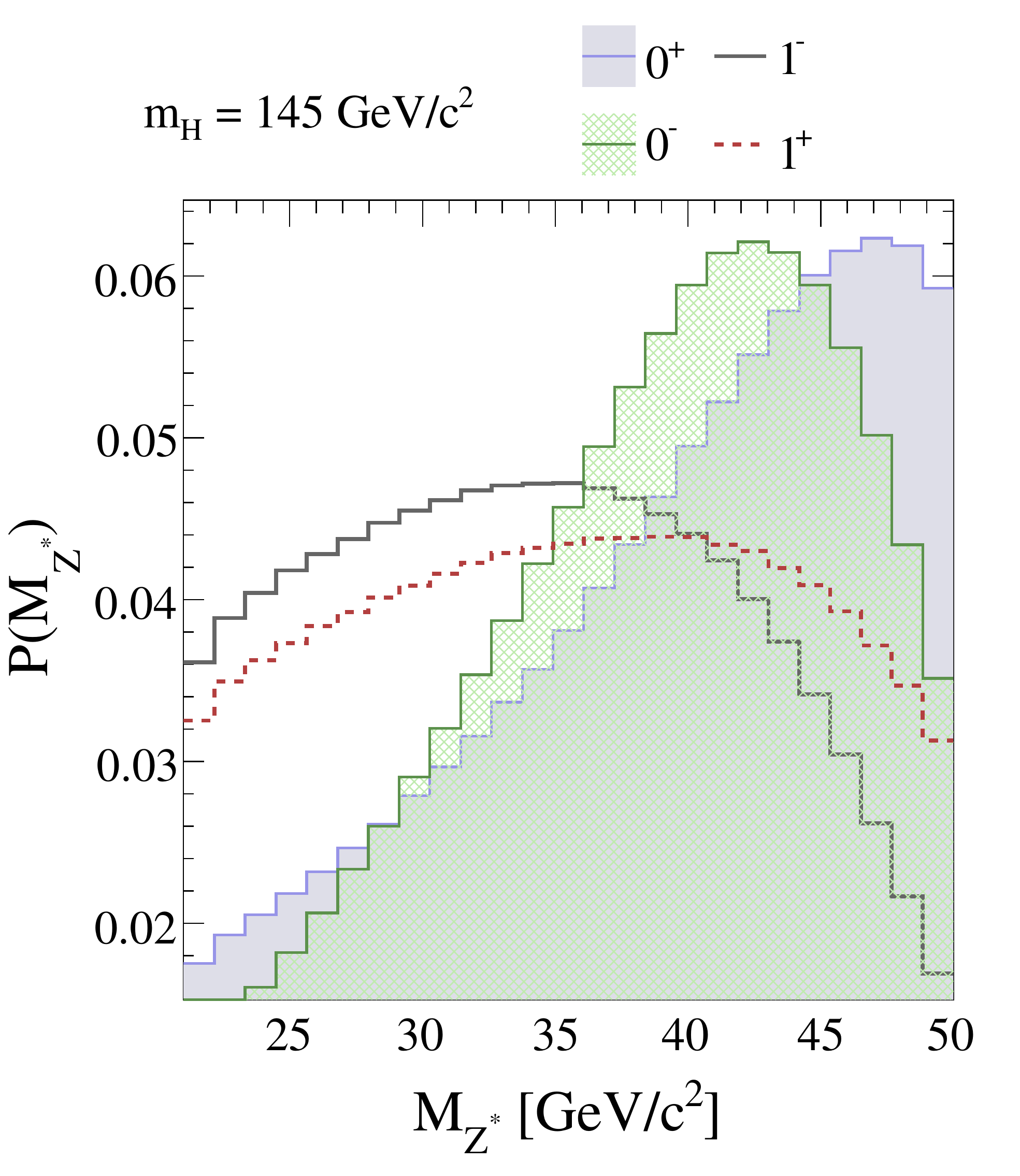}
\caption{Distributions, normalized to a unit integral, of the
  variables cos$\,\theta_1$ (top left), cos$\,\theta_2$ (top right),
  $\phi$ (bottom left) and $M_{Z^{*}}$ (bottom right) for $0^{+}$,
  $0^{-}$, $1^{+}$ and $1^{-}$ resonances with $m_{H}$$=$$145$
  GeV/c$^{2}$.
  \label{fig:KIN_145}}
\end{center}
\end{figure}
%%%%%%%%%%%%%%%%%%%%%%%%%%%%%%%%%%%%%%%%%%%%%%%%%%%%%%%%%%%%%%%%%%%

The discrimination between the $0^{-}$ hypothesis and the pure $J$$=$$1$
ones is very similar to the case of distinguishing the latter from
$0^{+}$, described in Sec.~\ref{sec:SM_v_1}. The {\it pdf} for $0^{-}$
has also no explicit dependence on the angles $\vec{\Omega}$.
Differences in the {\it pdfs} of these variables provide
discrimination between $0^{-}$ and $J$$=$$1$ states, as
Fig.~\ref{fig:KIN_145} illustrates.  The one-dimensional $M_{Z^{*}}$
{\it pdfs} are similar for $0^{-}$ and $0^{+}$, as well as for $1^{-}$
and $1^{+}$, while the differences between the two $J$-values are
maximal.  The cos$\,\theta_{1,2}$ distributions for $J$$=$$1$ have qualitatively different behavior when $m_{H} < 2\,M_{Z}$, as discussed in Sec.~\ref{sec:SM_v_1}. This results in the
$J$$=$$1$ cos$\,\theta_{1}$ (cos$\,\theta_{2}$) distribution being more
``$0^{-}$-like'' (``$0^{+}$-like''), resulting in similar levels of
discrimination between $J$$=$$0$ and $J$$=$$1$.
%%%%%%%%%%%%%%%%%%%%%%%%%%%%%%%%%%%%%%%%%%%%%%%%%%%%%%%%%%%%%%%%%%%
\begin{figure}[htbp]
\begin{center}
\includegraphics[width=0.210\textwidth]{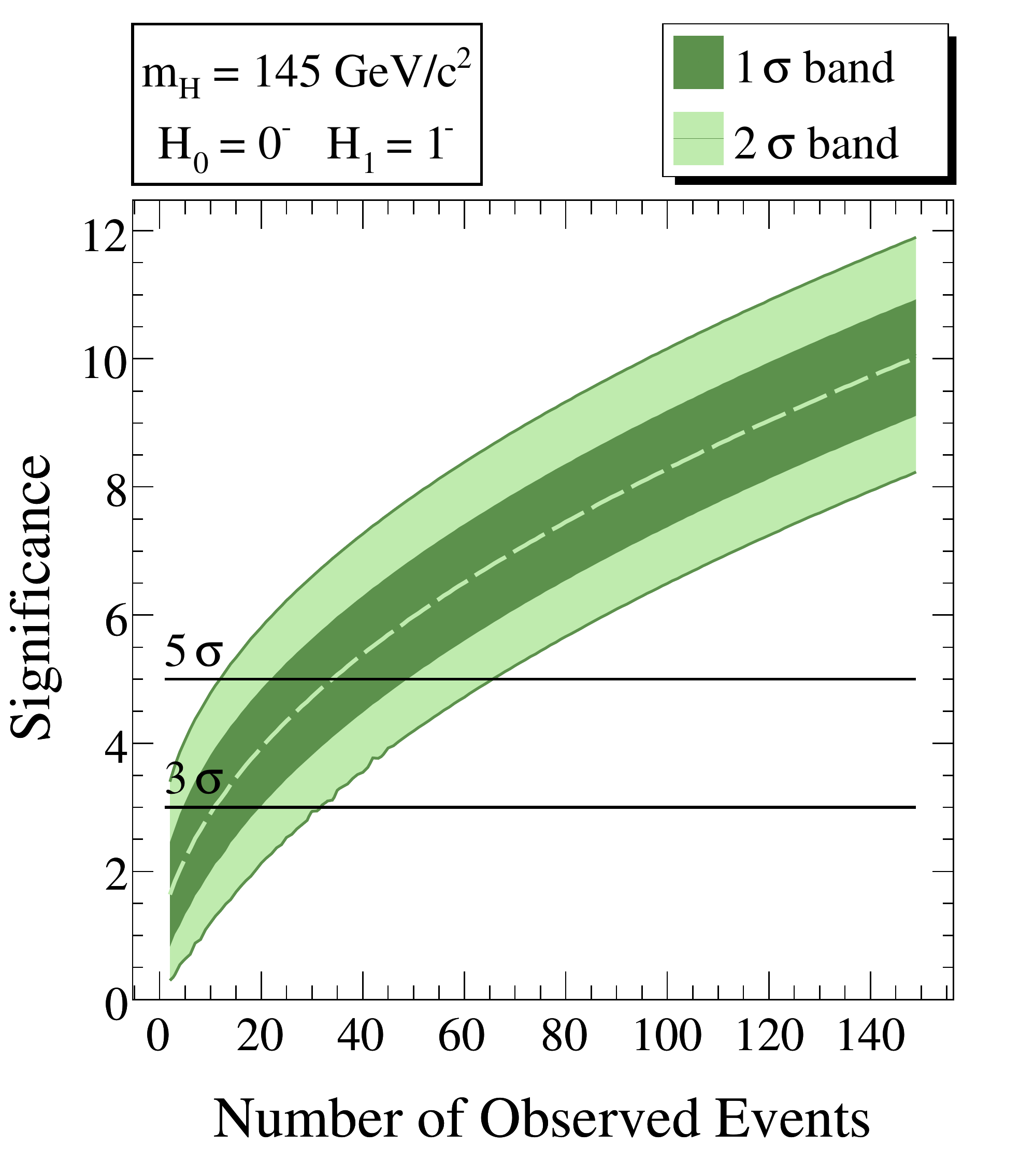}
\includegraphics[width=0.210\textwidth]{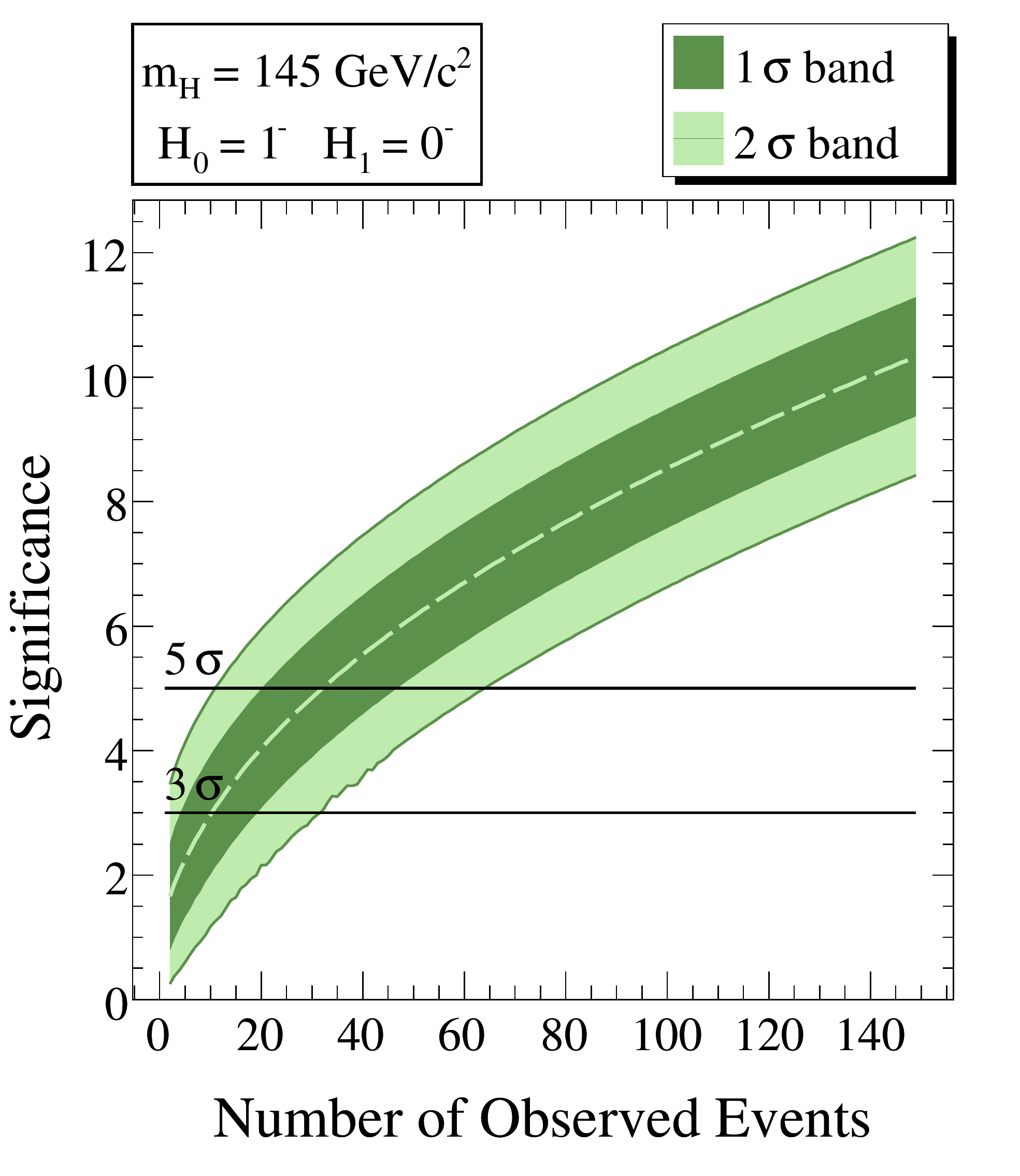}
\includegraphics[width=0.210\textwidth]{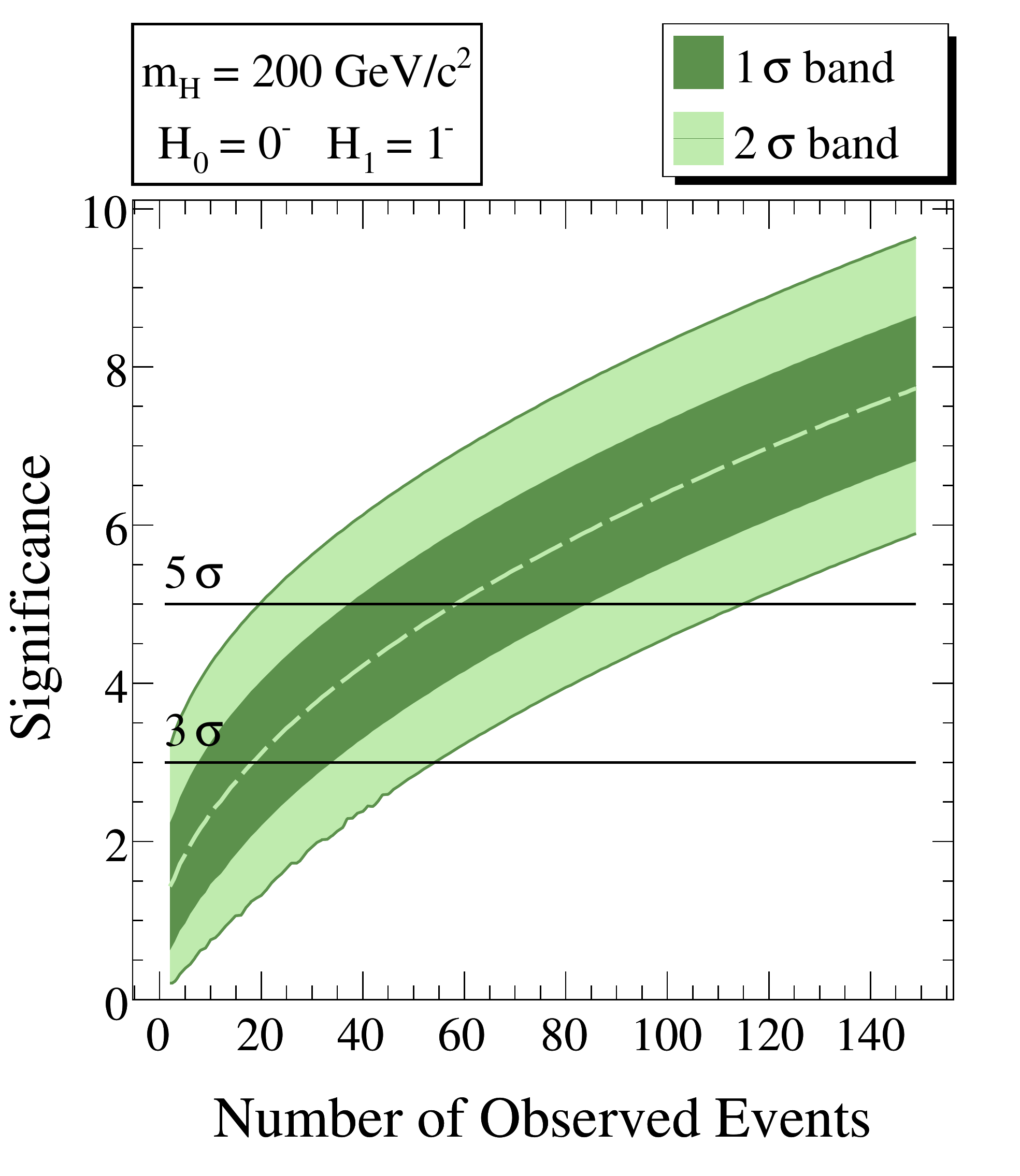}
\includegraphics[width=0.210\textwidth]{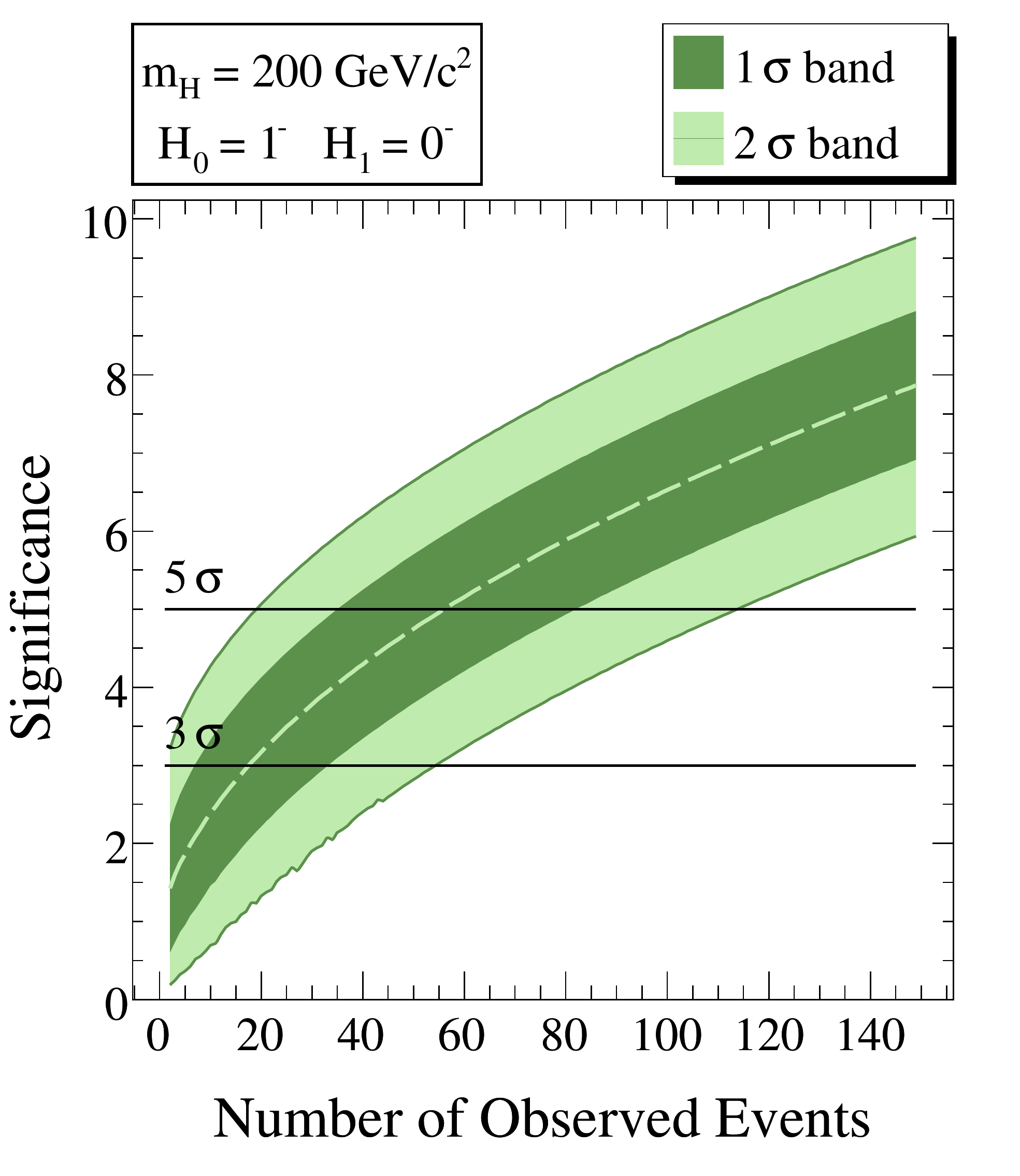}
\includegraphics[width=0.210\textwidth]{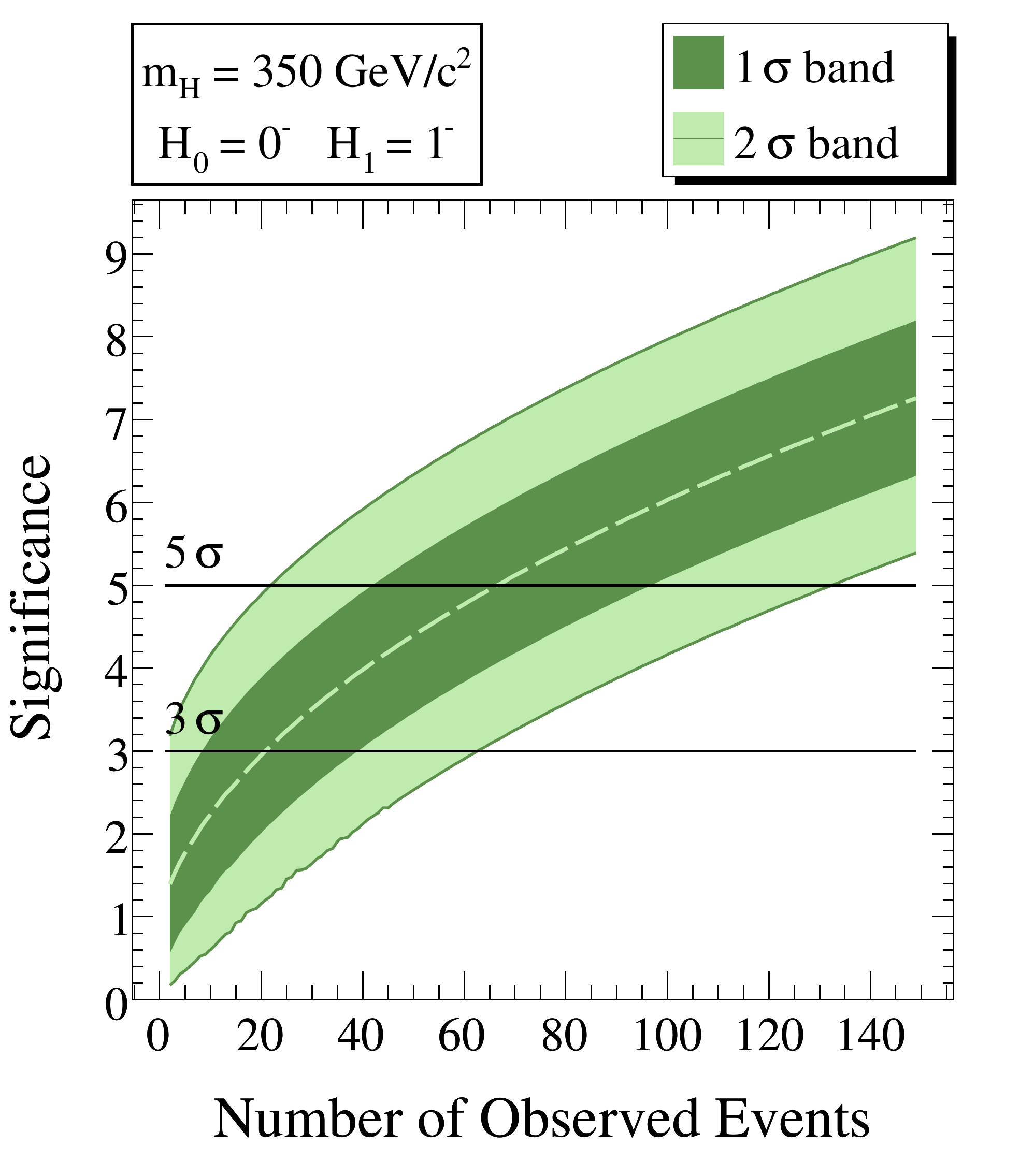}
\includegraphics[width=0.210\textwidth]{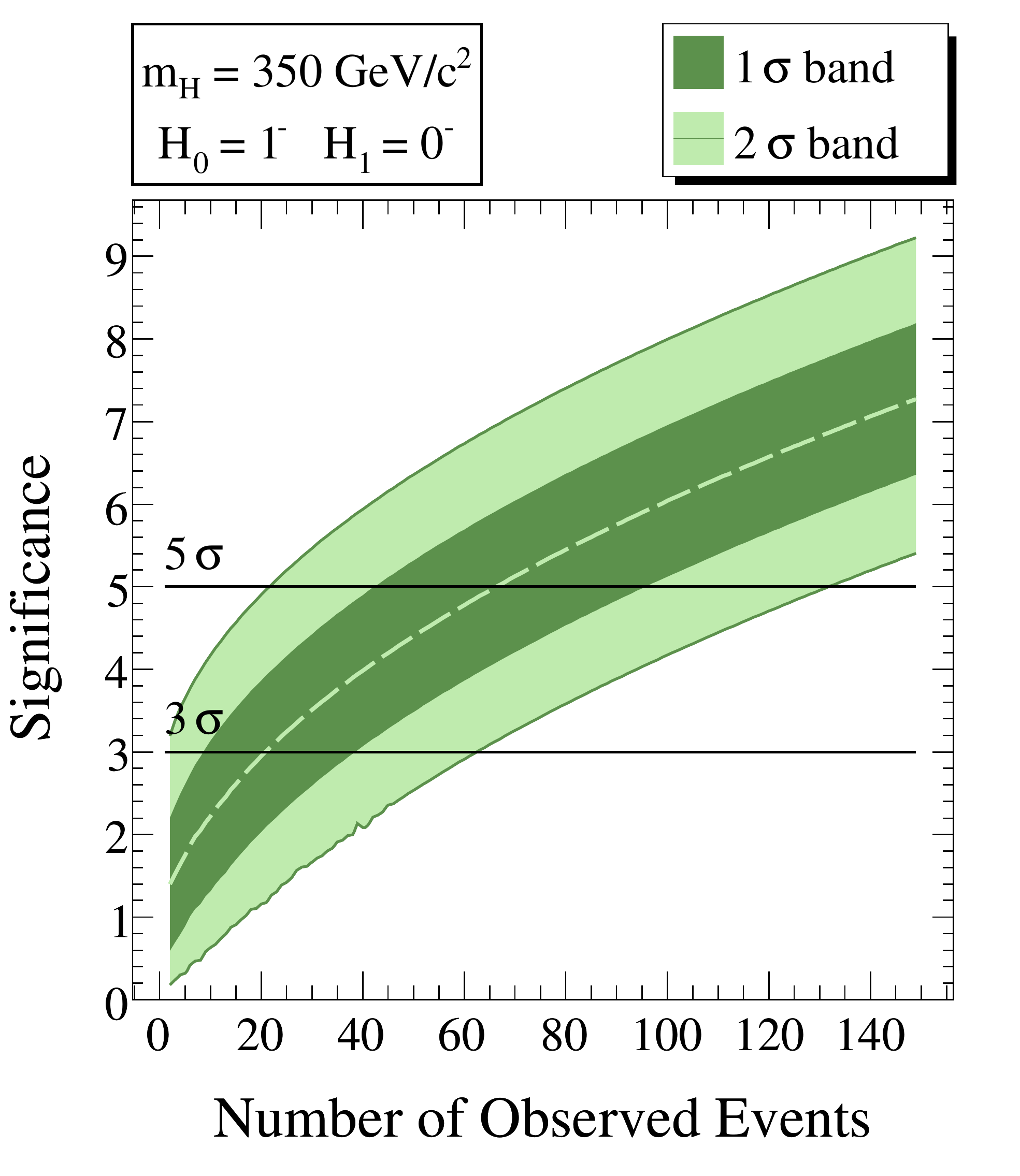}
\caption{Significance for rejecting $0^{-}$ in favor of
  $1^{-}$, assuming $1^{-}$ is true (left) or vice-versa ($0^-\!\leftrightarrow\! 1^-$, right)
  for $m_H$$=$$145$, 200 and 350 GeV/c$^{2}$ (top, middle and
  bottom).
  \label{fig:COMP_PS_v_PV}}
\end{center}
\end{figure}
%%%%%%%%%%%%%%%%%%%%%%%%%%%%%%%%%%%%%%%%%%%%%%%%%%%%%%%%%%%%%%%%%%%
\begin{figure}[htbp]
\begin{center}
\includegraphics[width=0.210\textwidth]{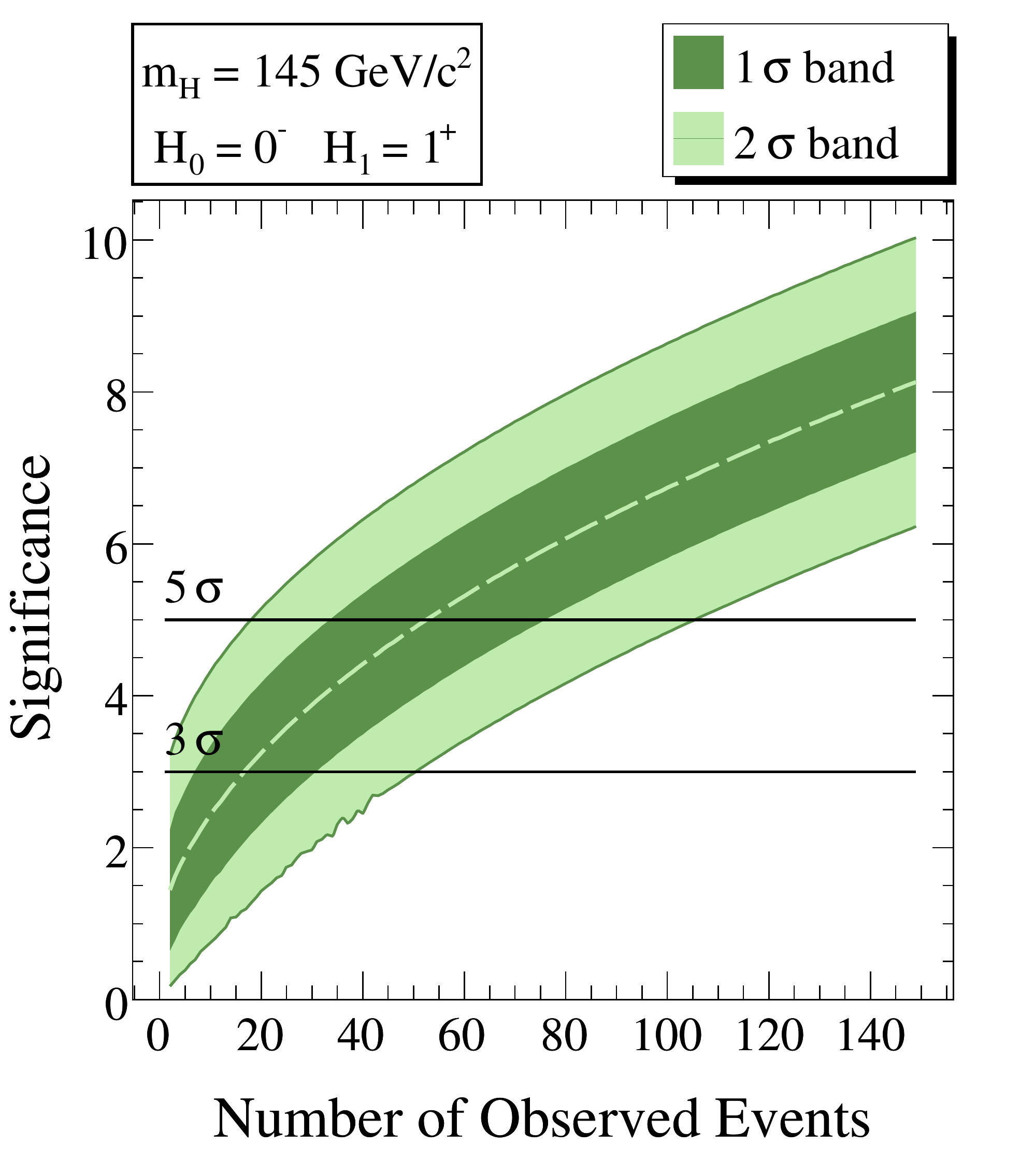}
\includegraphics[width=0.210\textwidth]{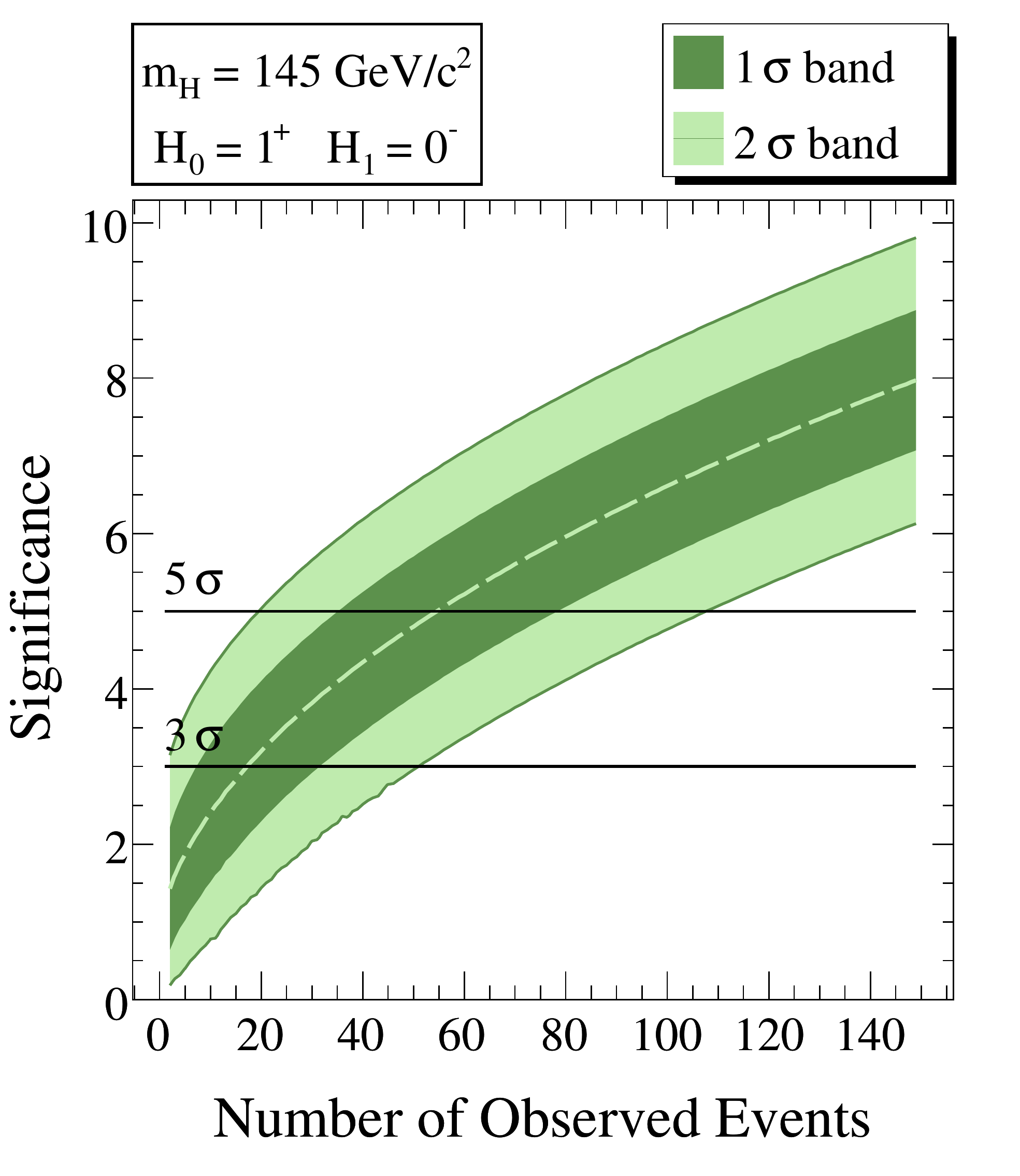}
\includegraphics[width=0.210\textwidth]{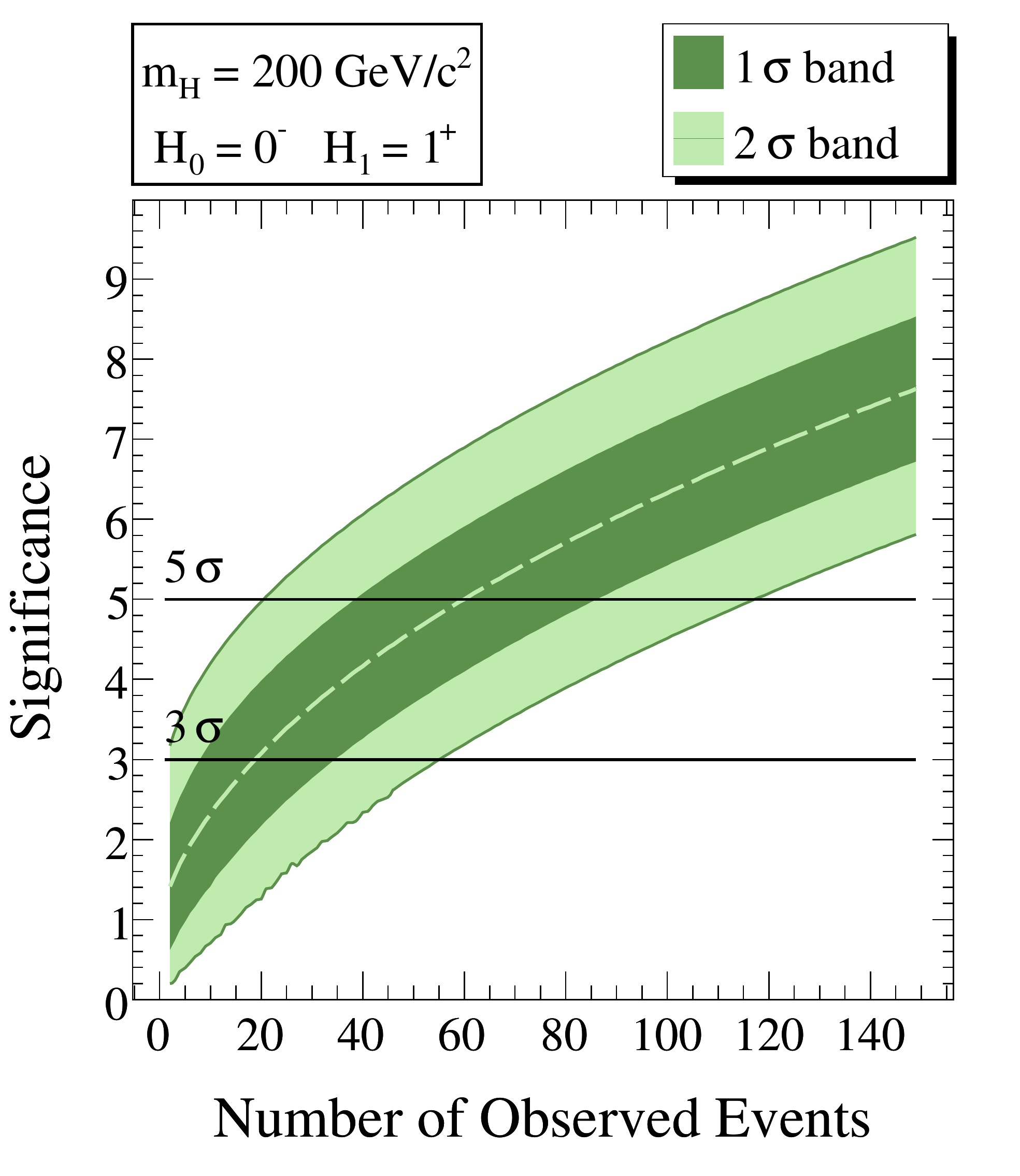}
\includegraphics[width=0.210\textwidth]{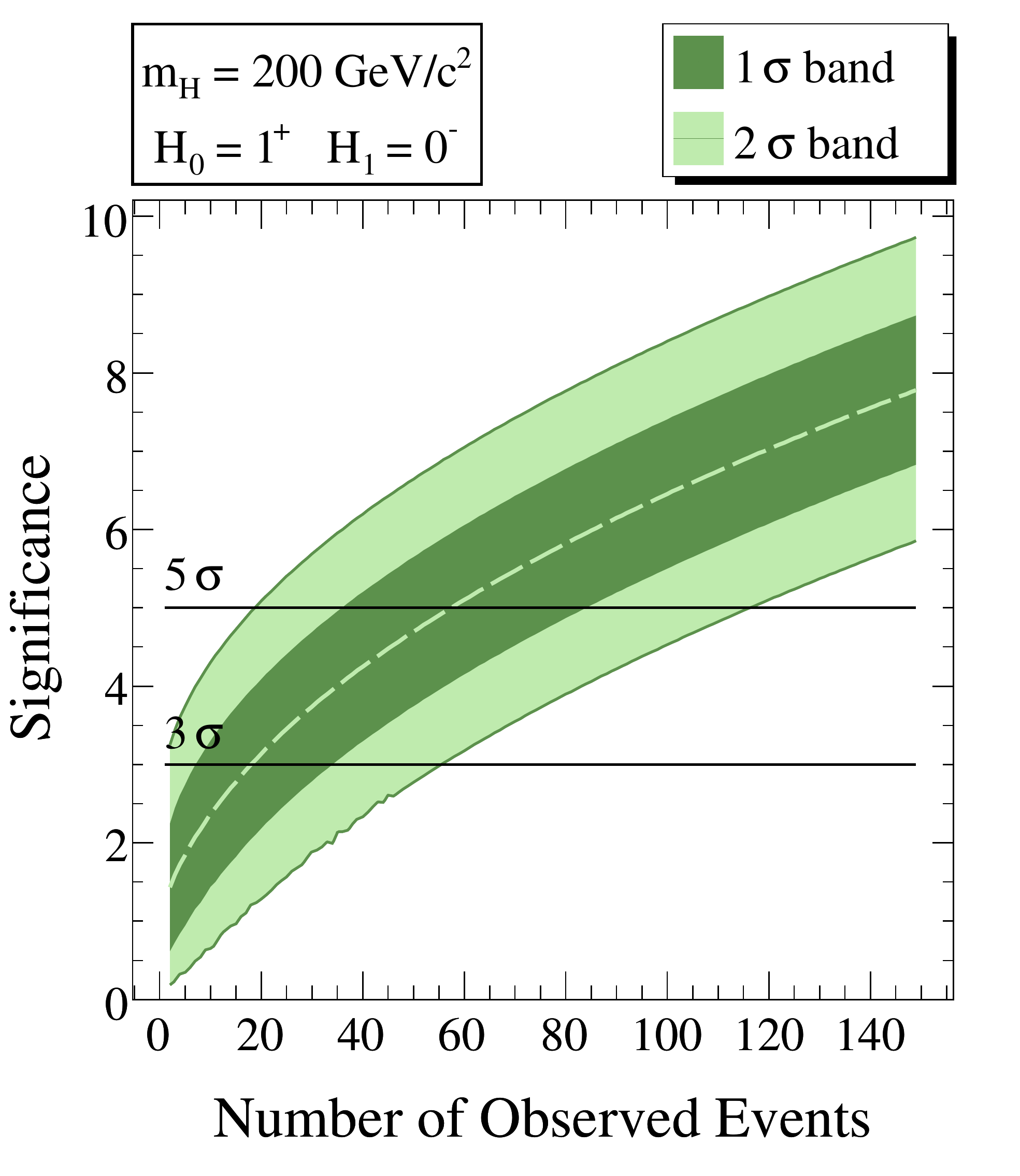}
\includegraphics[width=0.210\textwidth]{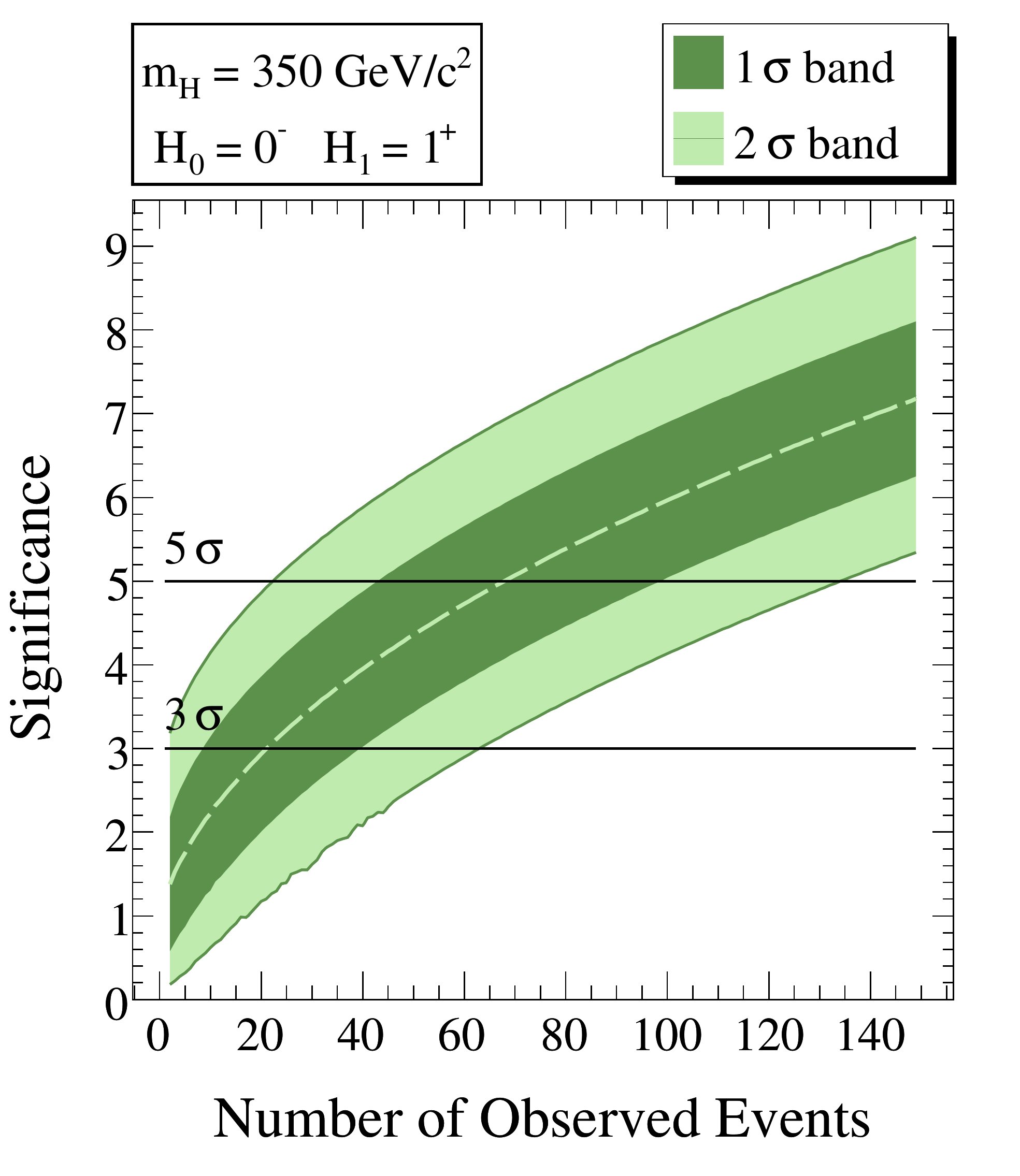}
\includegraphics[width=0.210\textwidth]{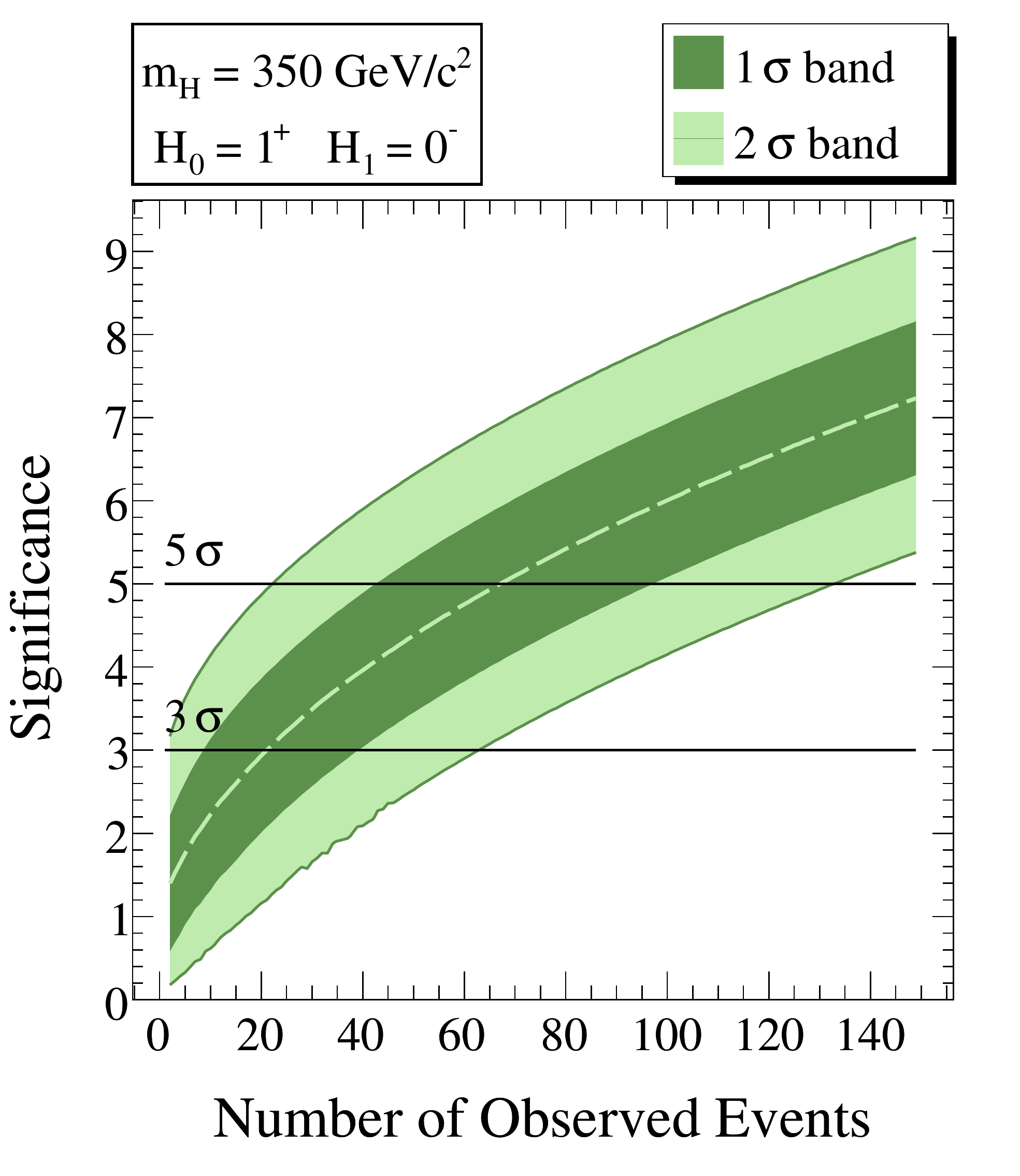}
\caption{Significance for rejecting $0^{-}$ in favor of
  $1^{+}$, assuming $1^{+}$ is true (left) or vice-versa
  ($0^-\!\leftrightarrow\! 1^+$, right) for $m_H$$=$$145$, 200 and 350
  GeV/c$^{2}$ (top, middle and bottom).   \label{fig:COMP_PS_v_PA}}
\end{center}
\end{figure}
%%%%%%%%%%%%%%%%%%%%%%%%%%%%%%%%%%%%%%%%%%%%%%%%%%%%%%%%%%%%%%%%%%%

The expected significance to distinguish the $0^{-}$ and $1^{-}$
($1^{+}$) hypotheses, as functions of $N_S$, is shown in
Fig.~\ref{fig:COMP_PS_v_PV} (\ref{fig:COMP_PS_v_PA}).  The $m_H$$=$$145$
GeV/c$^{2}$ results and the ones for $0^{+}$ vs.~$J$$=$$1$
(Figs.~\ref{fig:COMP_SM_v_PV} and \ref{fig:COMP_SM_v_PA}) are nearly
identical. A similar comparison of $0^{-}$ vs. $J$$=$$1$
(Figs.~\ref{fig:COMP_PS_v_PV} and \ref{fig:COMP_PS_v_PA}) with $0^{+}$
vs. $J$$=$$1$ (Figs.~\ref{fig:COMP_SM_v_PV} and \ref{fig:COMP_SM_v_PA})
for $m_H$$=$$200$ GeV/c$^{2}$ reveals that it is more difficult to
discriminate between $0^{+}$ and $J$$=$$1$ at this mass. This is
predominantly due to the {\it pdfs} for the angles cos$\,\theta_{1,2}$
(which are similar for $0^{+}$ and $J$$=$$1$ for $m_H > 2\,M_{Z}$).

\begin{figure}[htb!]
\begin{center}
\includegraphics*[width=0.238\textwidth]{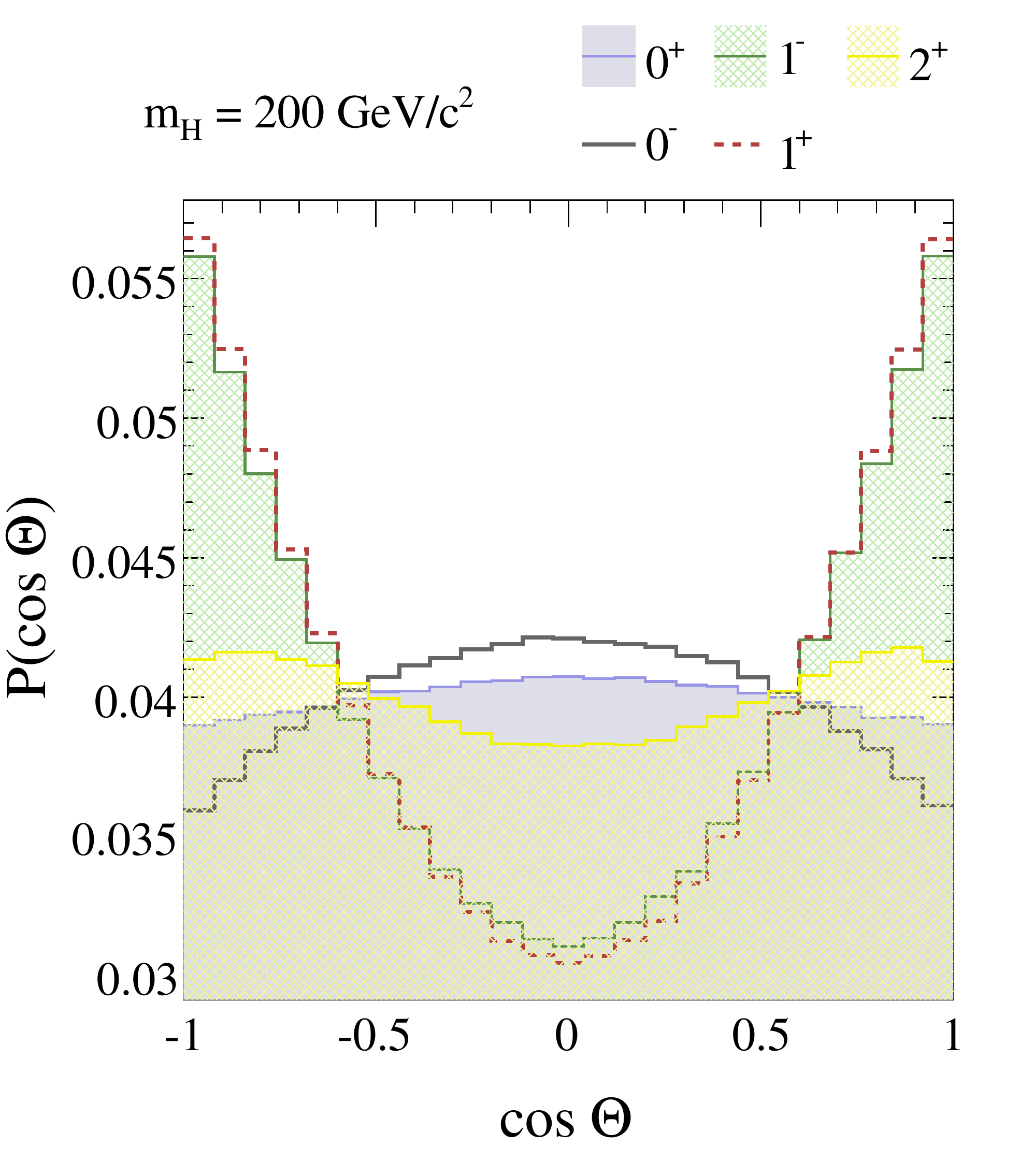}
\includegraphics*[width=0.238\textwidth]{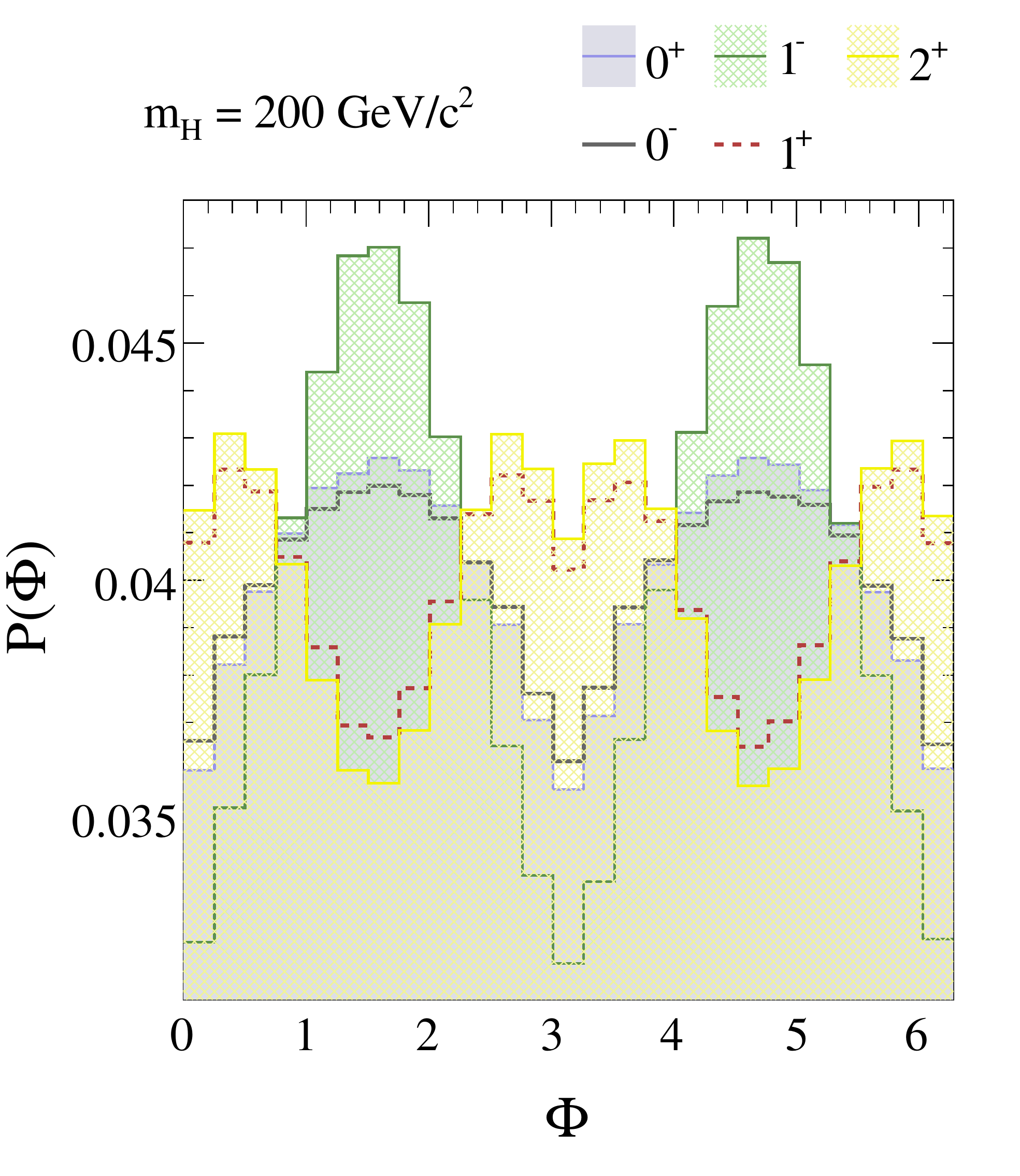}
\includegraphics*[width=0.238\textwidth]{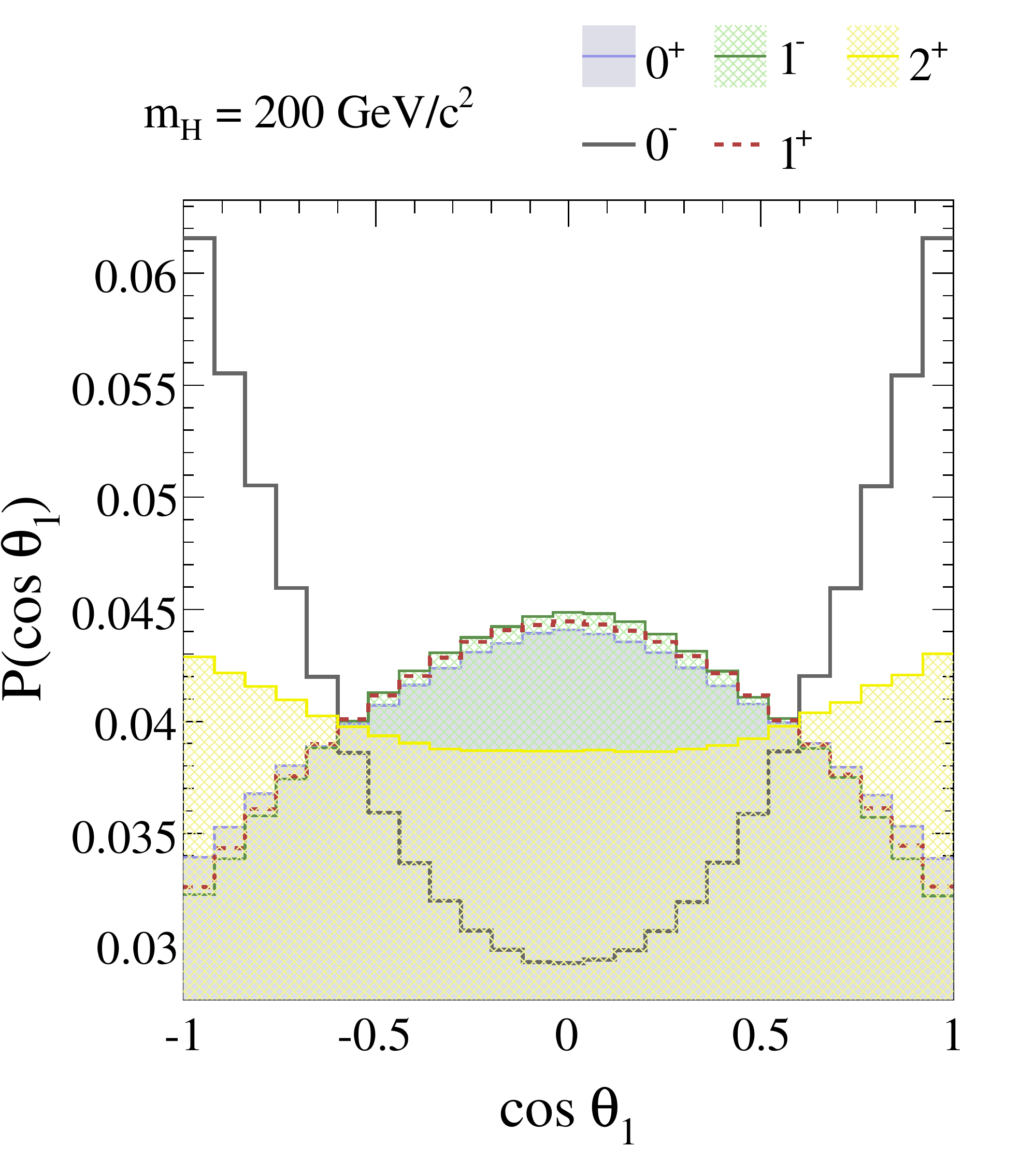}
\includegraphics*[width=0.238\textwidth]{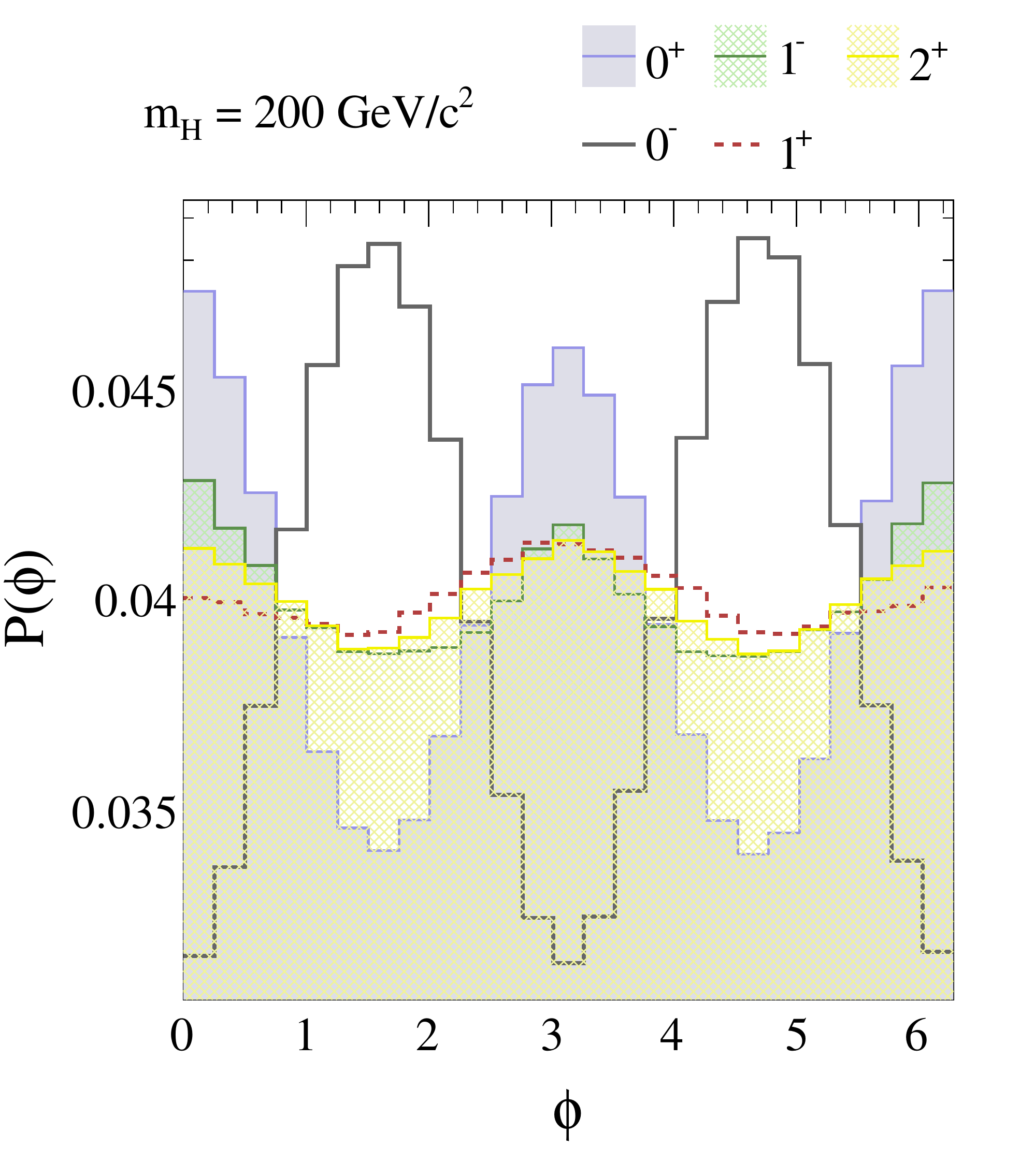}
\caption{Distributions of cos$\,\Theta$ (top left), $\Phi$ (top
  right), cos$\,\theta_{1}$ (bottom left) and $\phi$ (bottom right) for
  all the pure $J^{PC}$ choices we study, for $m_{H}$$=$$200$
  GeV/c$^{2}$. All distributions are normalized to a unit integral.
  \label{fig:KIN_200_all}}
\end{center}
\end{figure}

The distributions for the variables $\vec{\Omega}$ and $\vec{\omega}$
for all the pure $J^{PC}$ hypotheses considered in our analysis are
shown in Fig.~\ref{fig:KIN_200_all}, for $m_H$$=$$200$ GeV/c$^{2}$. The
$\vec{\Omega}$ distributions are nearly identical for the two $J$$=$$0$
cases, since they are only induced by detector limitations. 
%%%%%%%%%%%%%%%%%%%%%%%%%%%%%%%%%%%%%%%%%%%%%%%%%%%%%%%%%%%%%%%%%%%
%%%%%%%%%%%%%%%%%%%%%%%%%%%%%%%%%%%%%%%%%%%%%%%%%%%%%%%%%%%%%%%%%%%

\begin{figure}[htbp]
\begin{center}
\includegraphics[width=0.210\textwidth]{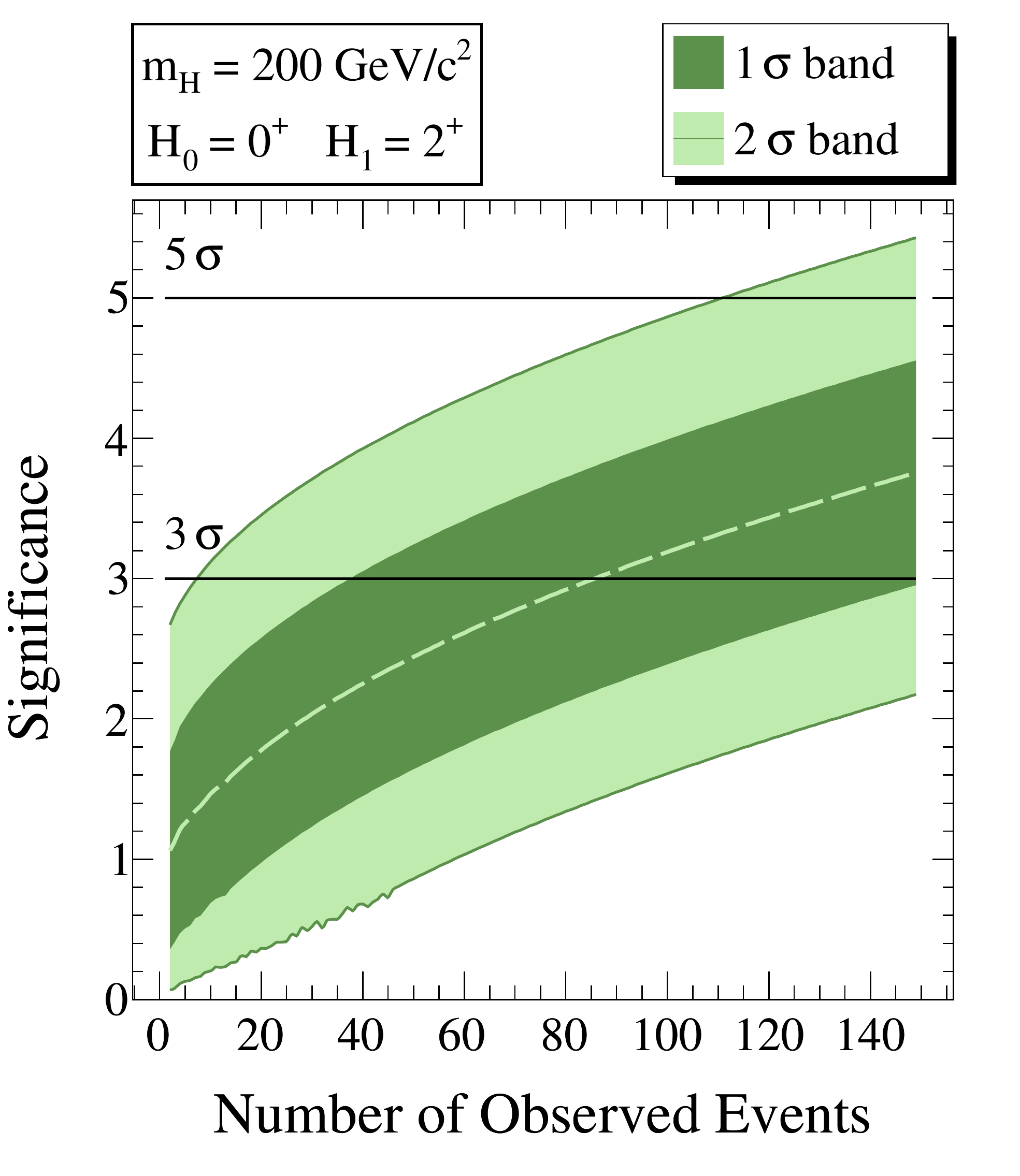}
\includegraphics[width=0.210\textwidth]{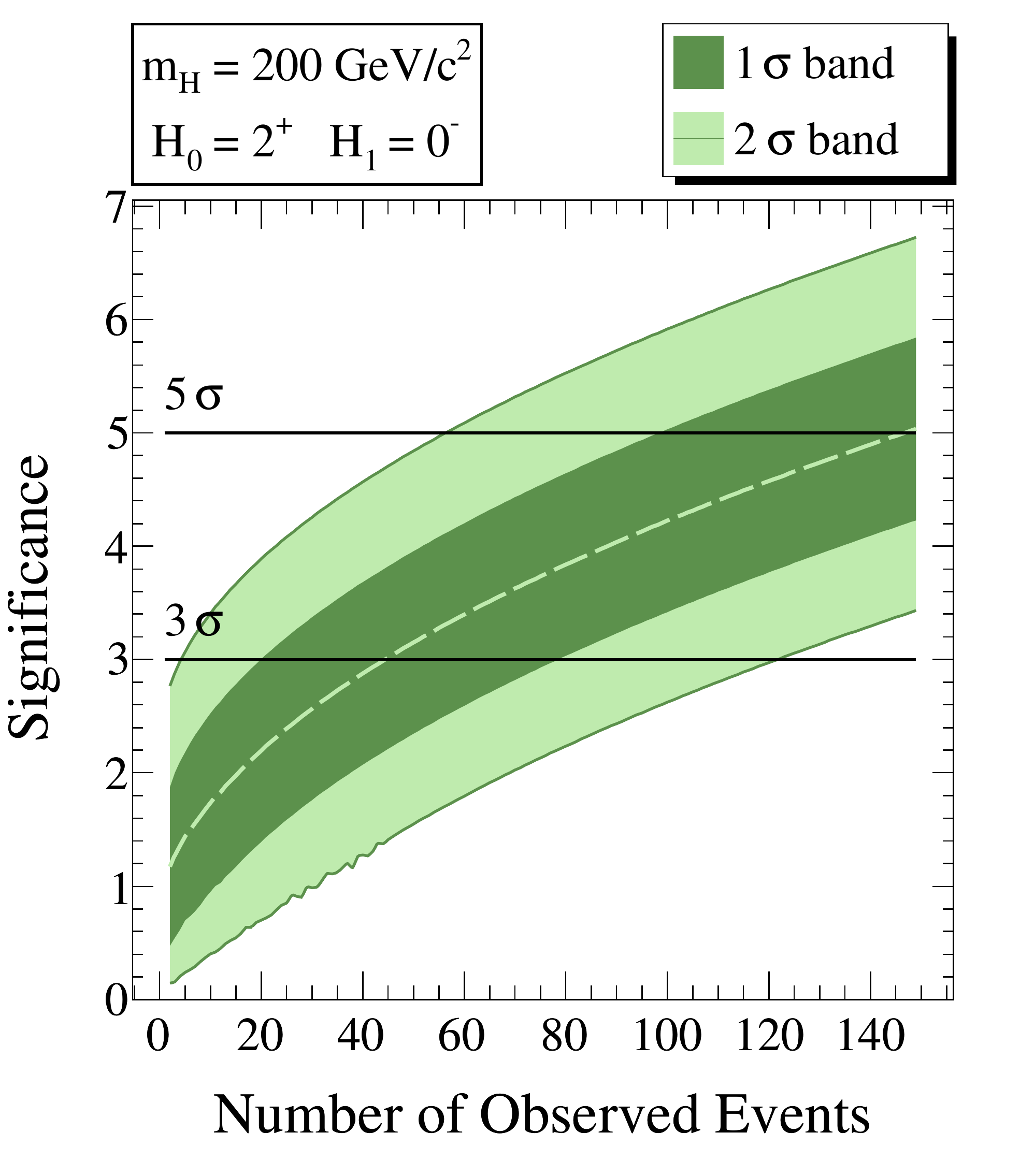}
\includegraphics[width=0.210\textwidth]{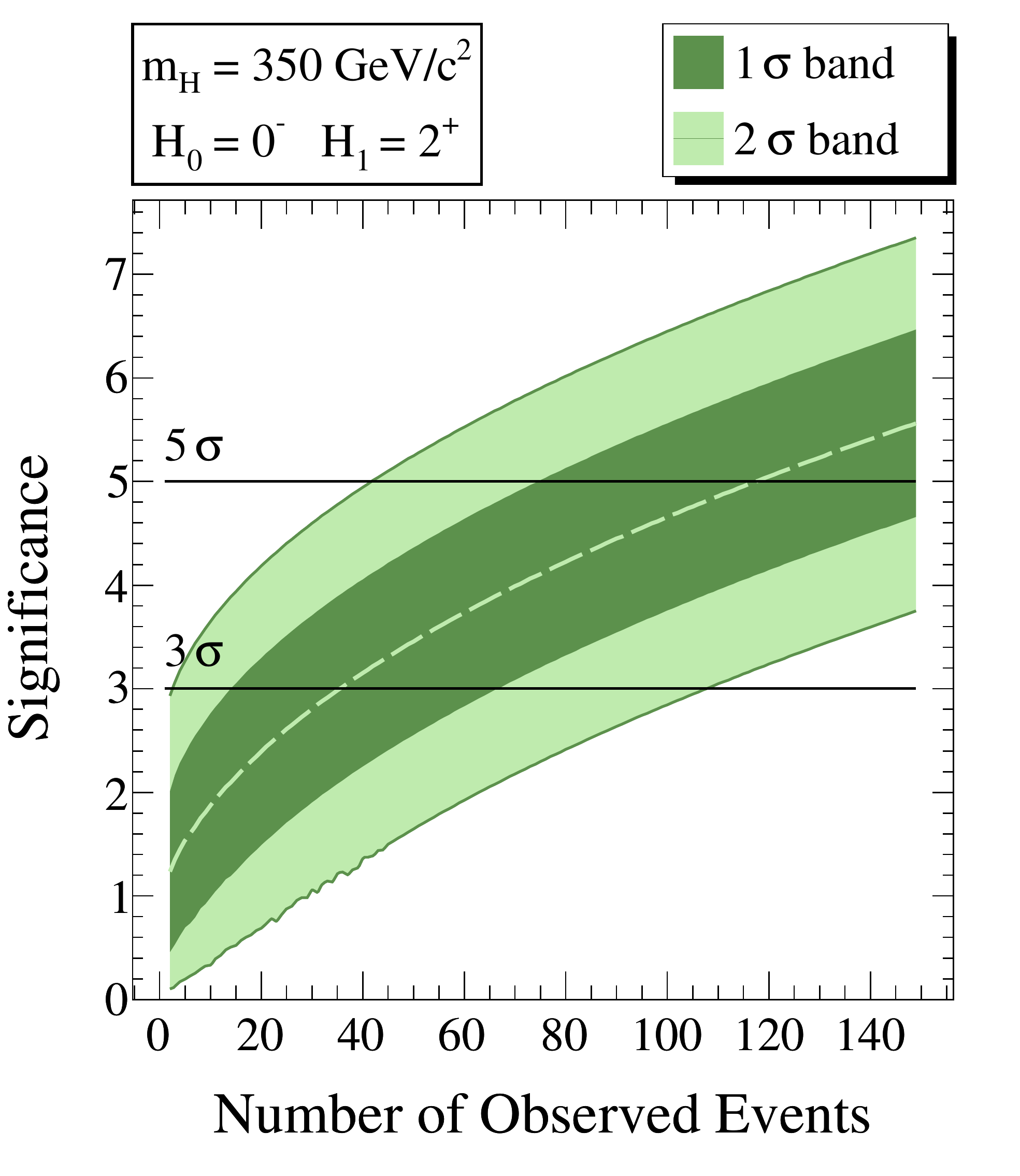}
\includegraphics[width=0.210\textwidth]{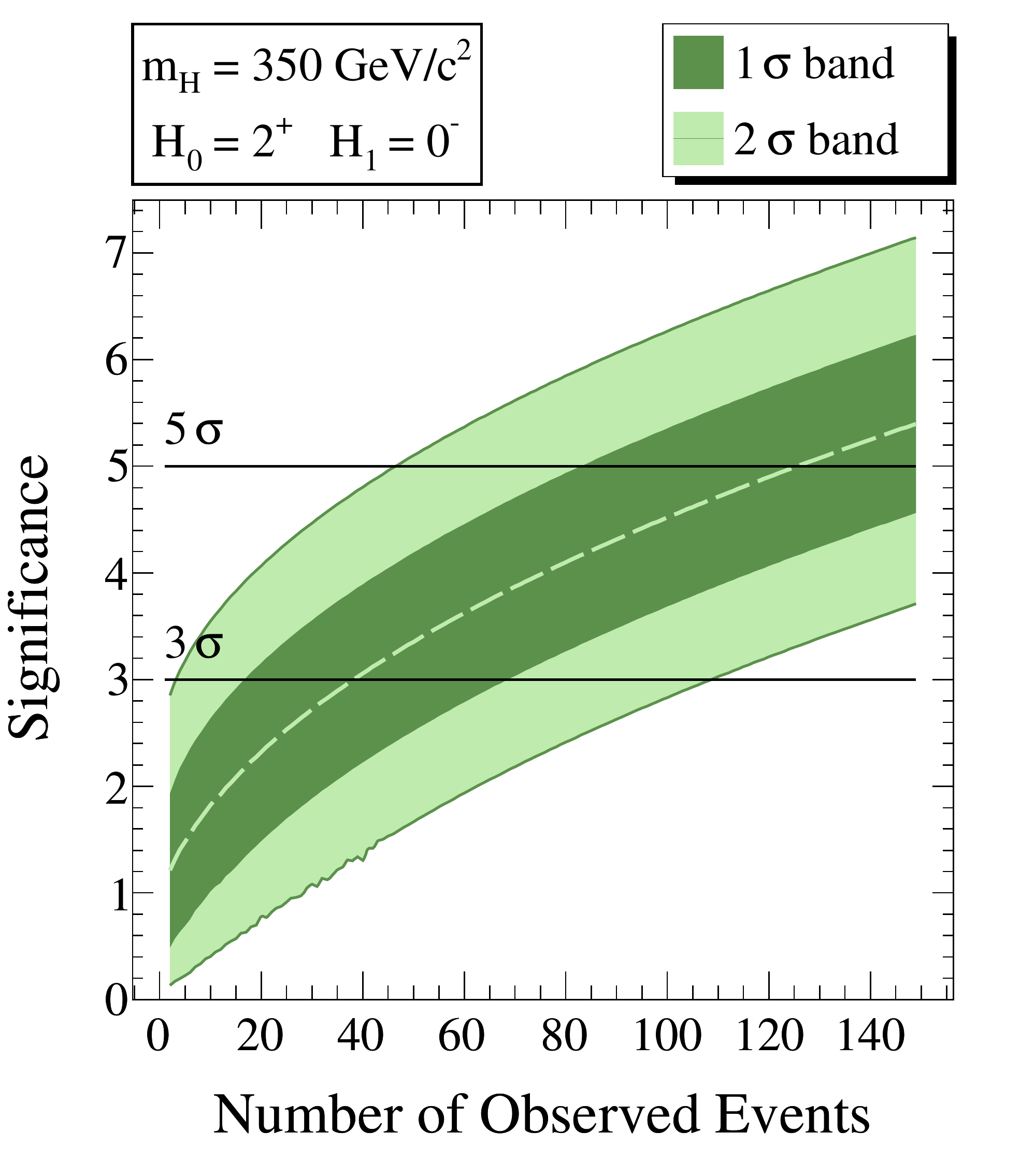}
\caption{Significance for rejecting $0^{-}$ in favor of
  $2^{+}$, assuming $0^{-}$ is true (left) or vice-versa ($0^-\!\leftrightarrow\! 2^+$, right)
  for $m_H$$=$$200$ and 350 GeV/c$^{2}$ (top, bottom).    \label{fig:COMP_PS_v_KK}}
\end{center}
\end{figure}

\begin{figure}[htbp]
\begin{center}
\includegraphics[width=0.210\textwidth]{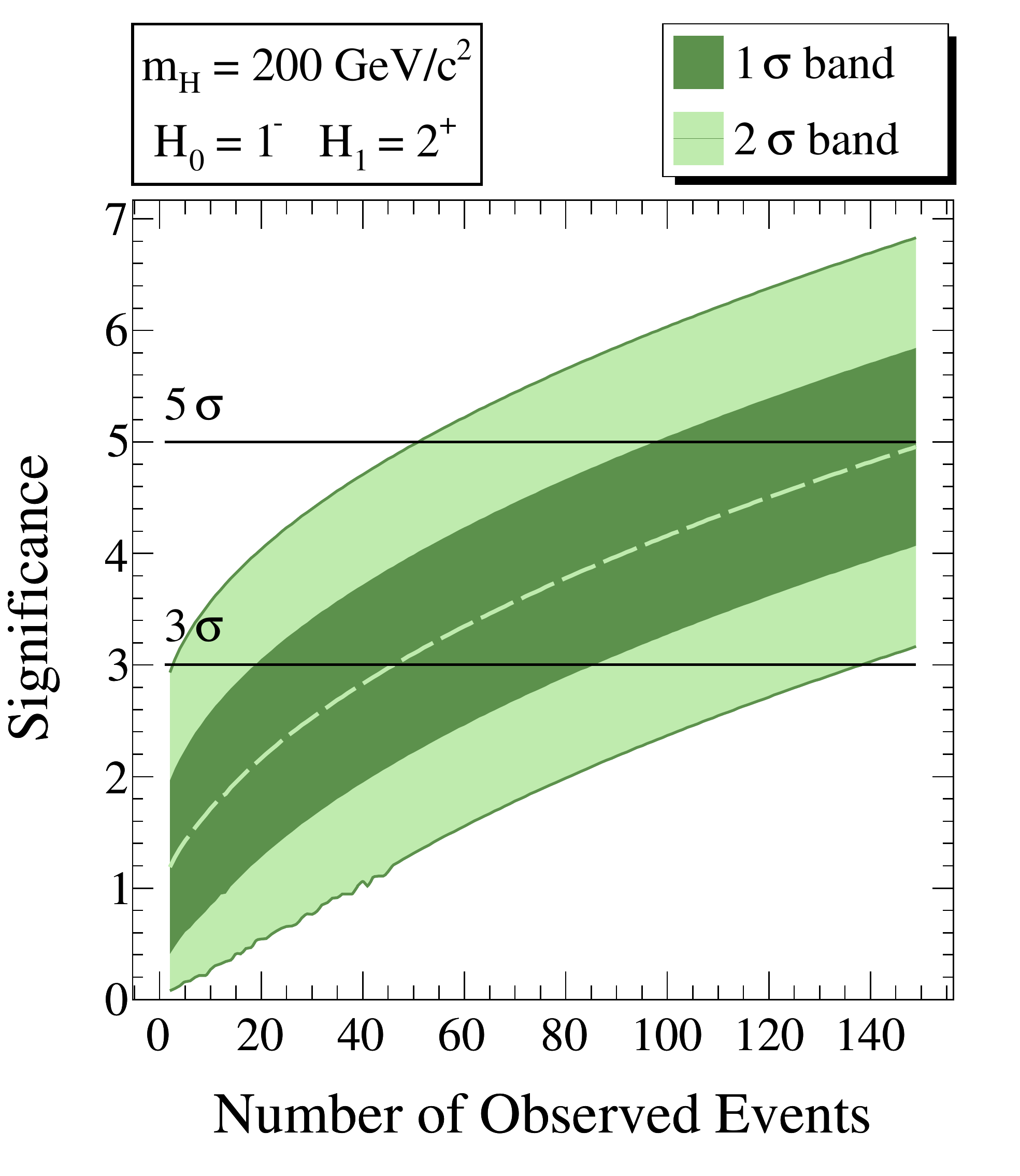}
\includegraphics[width=0.210\textwidth]{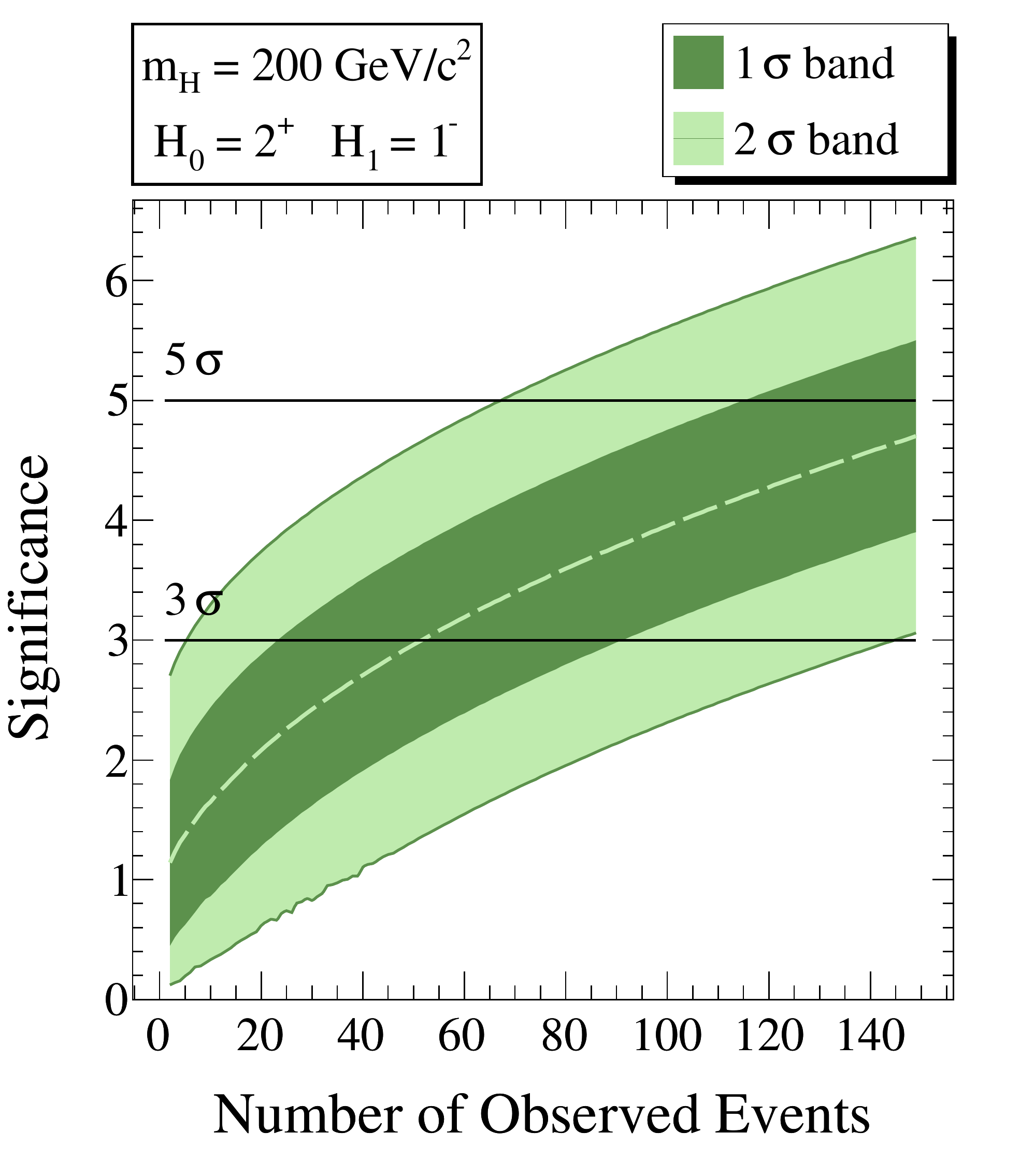}
\includegraphics[width=0.210\textwidth]{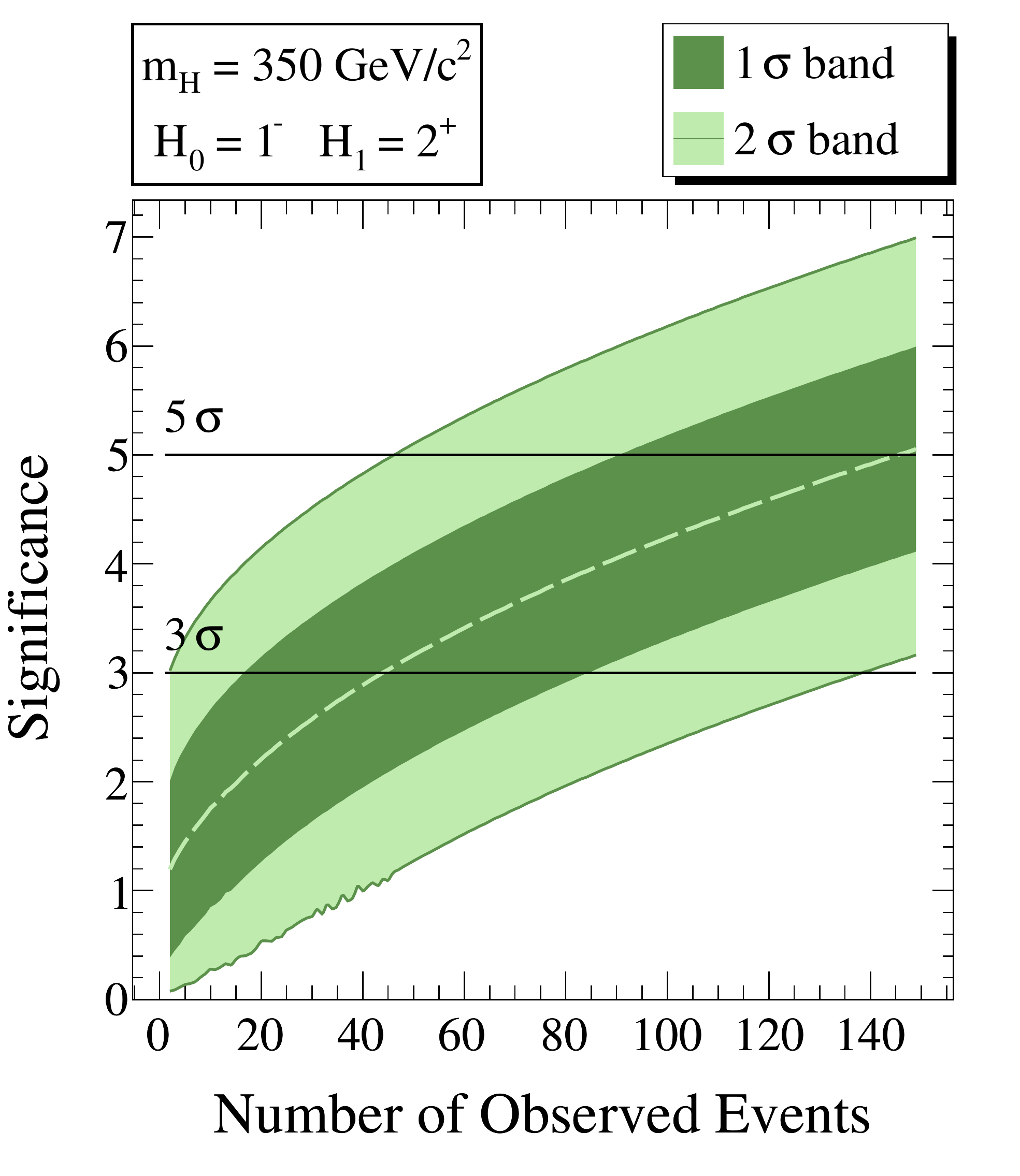}
\includegraphics*[width=0.210\textwidth]{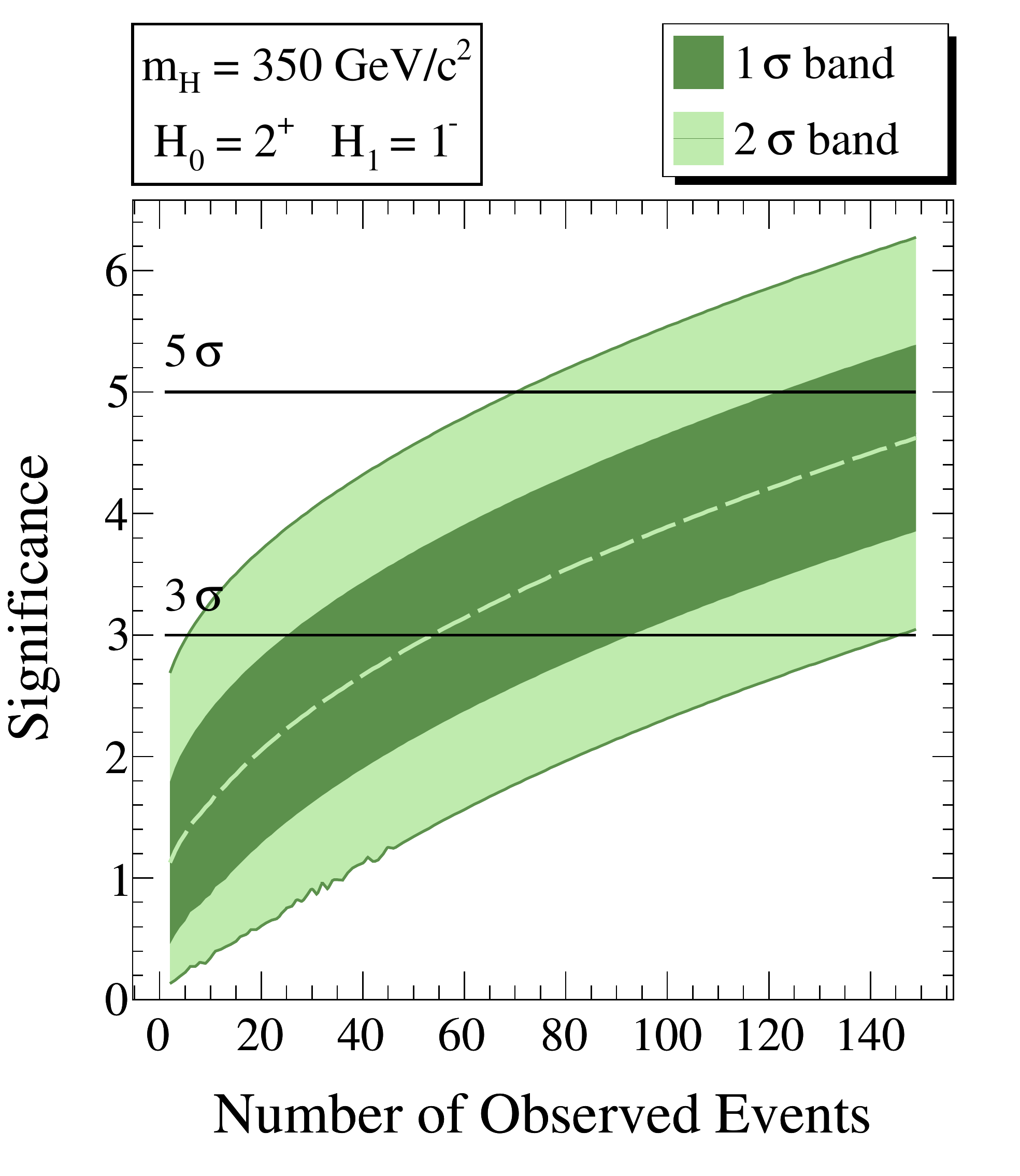}
\caption{Significance for rejecting $1^{-}$ in favor of
  $2^{+}$, assuming $2^{+}$ is true (left) or vice-versa ($1^-\!\leftrightarrow\! 2^+$, right)
  for $m_H$$=$$200$ \ and 350 GeV/c$^{2}$ (top, bottom).  \label{fig:COMP_PV_v_KK}}
\end{center}
\end{figure}

\begin{figure}[htbp]
\begin{center}
\includegraphics[width=0.210\textwidth]{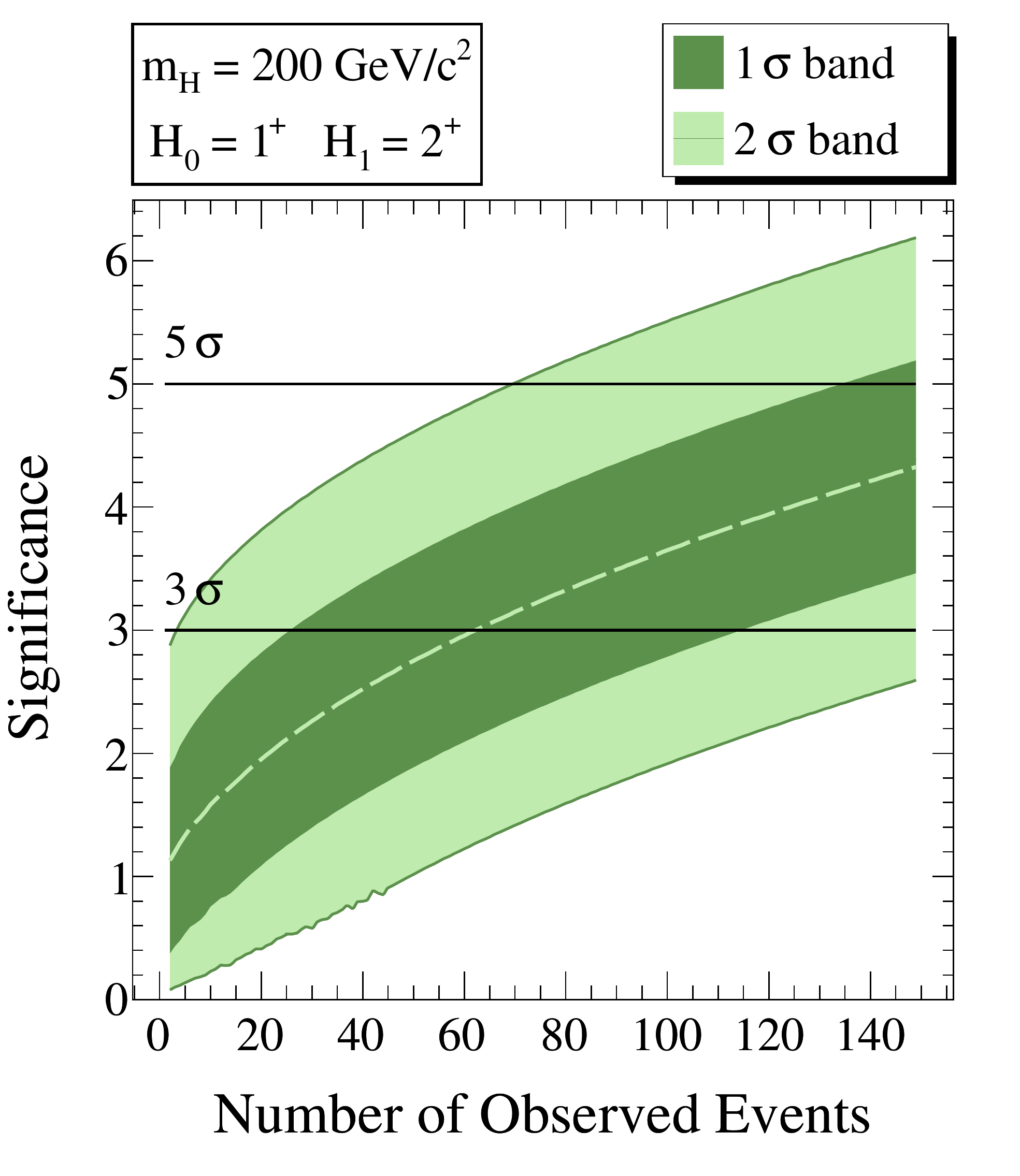}
\includegraphics[width=0.210\textwidth]{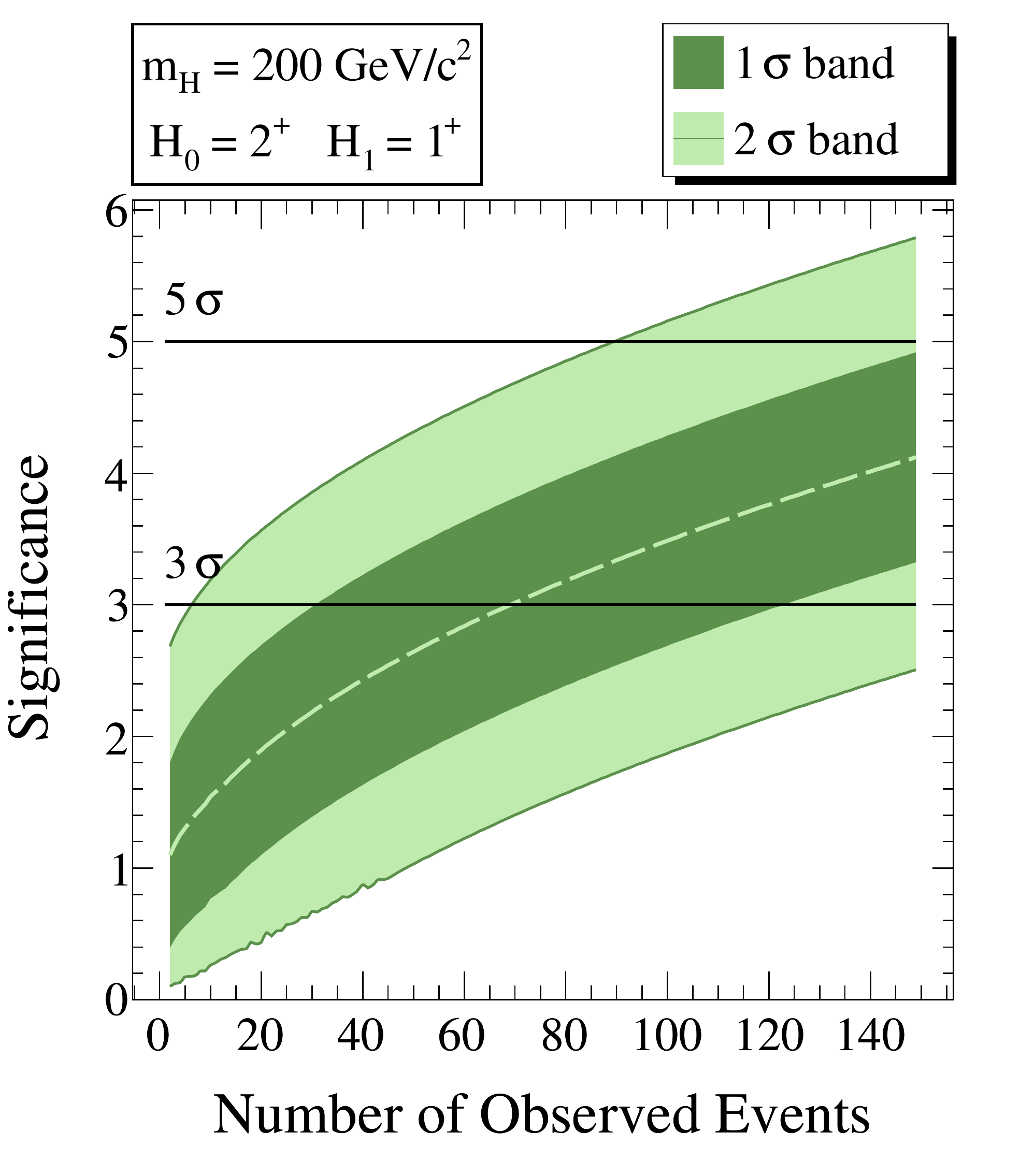}
\includegraphics[width=0.210\textwidth]{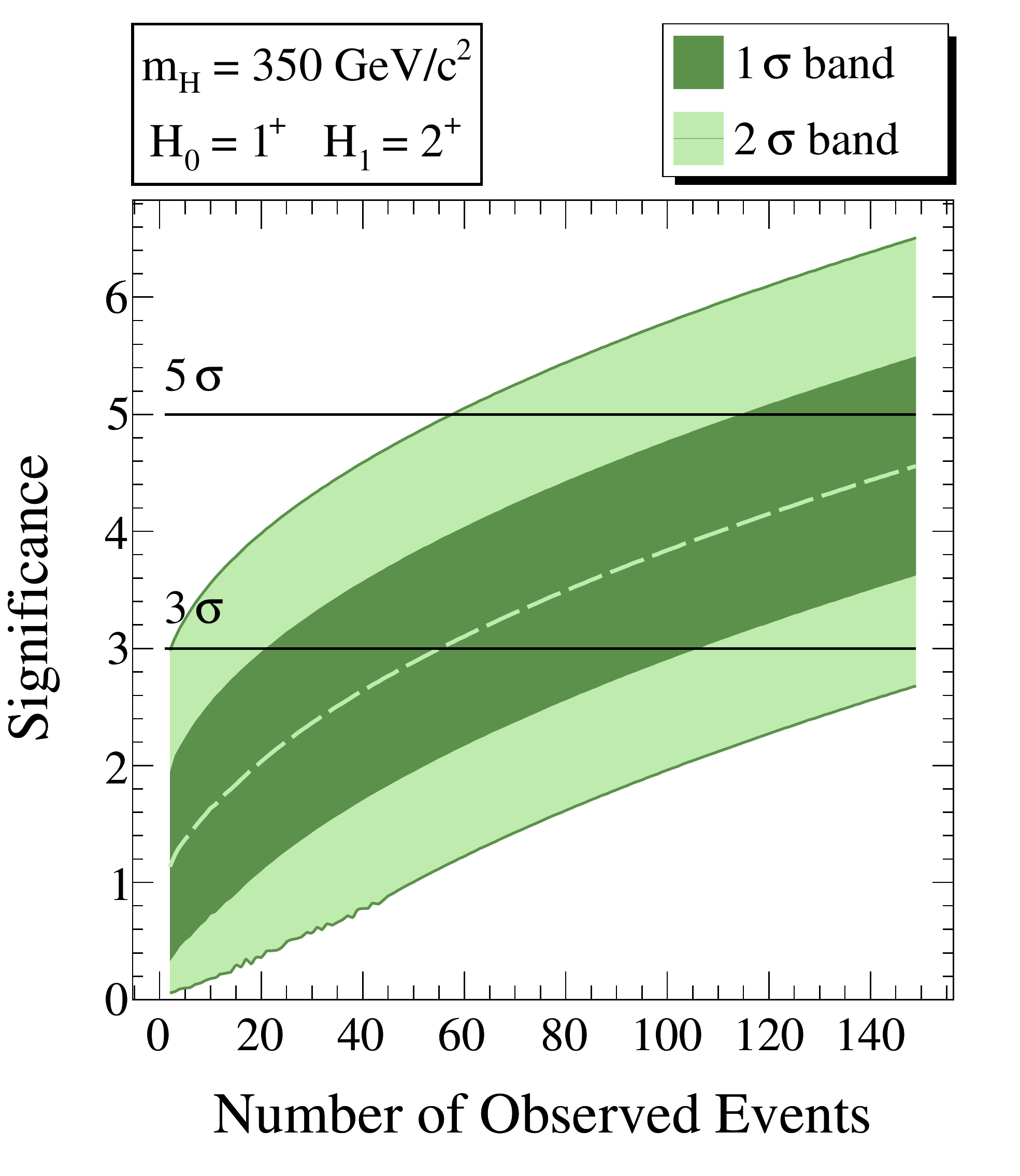}
\includegraphics[width=0.210\textwidth]{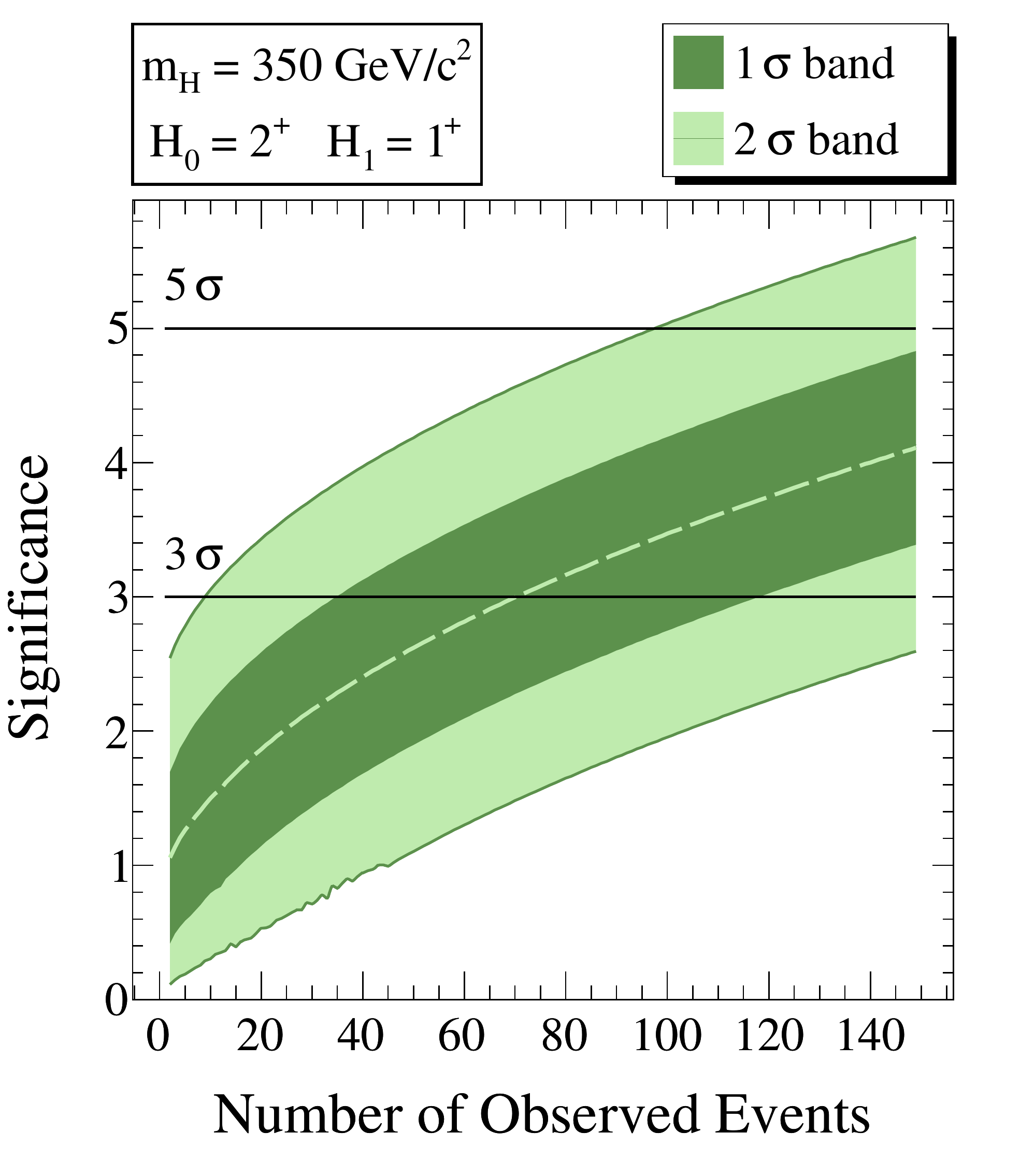}
\caption{Significance for rejecting $1^{+}$ in favor of $2^{+}$,
  assumed to be correct (left) or vice-versa ($1^+\!\leftrightarrow\!
  2^+$, right) for $m_H$$=$$200$ and 350 GeV/c$^{2}$ (top,
  bottom).  \label{fig:COMP_PA_v_KK}}
\end{center}
\end{figure}

The potential to distinguish between $0^{-}$ and $2^{+}$ resonances is
shown in Fig.~\ref{fig:COMP_PS_v_KK} for $m_H$$=$$200$ and 350
GeV/c$^{2}$. If both of the $J$$=$$0$ cases are excluded in favor of
$J$$=$$1$ or $J$$=$$2$, one needs to discriminate between the latter. Relative
to $J$$=$$0$ case, the two pure $J$$=$$1$ resonances have the most similar
{\it pdfs}, as we saw in Sec.~\ref{sec:SM_v_1} while comparing them to
the $0^{+}$ case. The comparison to the $J$$=$$2$ case reflects the same
limitation, as shown in Fig.~\ref{fig:COMP_PV_v_KK}
(\ref{fig:COMP_PA_v_KK}) for $1^{-}$ vs.~$2^{+}$ ($1^{+}$
vs.~$2^{+}$).
%%%%%%%%%%%%%%%%%%%%%%%%%%%%%%%%%%%%%%%%%%%%%%%%%%%%%%%%%%%%%%%%%%%
%%%%%%%%%%%%%%%%%%%%%%%%%%%%%%%%%%%%%%%%%%%%%%%%%%%%%%%%%%%%%%%%%%%
\begin{figure}[htbp]
\begin{center}
\includegraphics[width=0.238\textwidth]{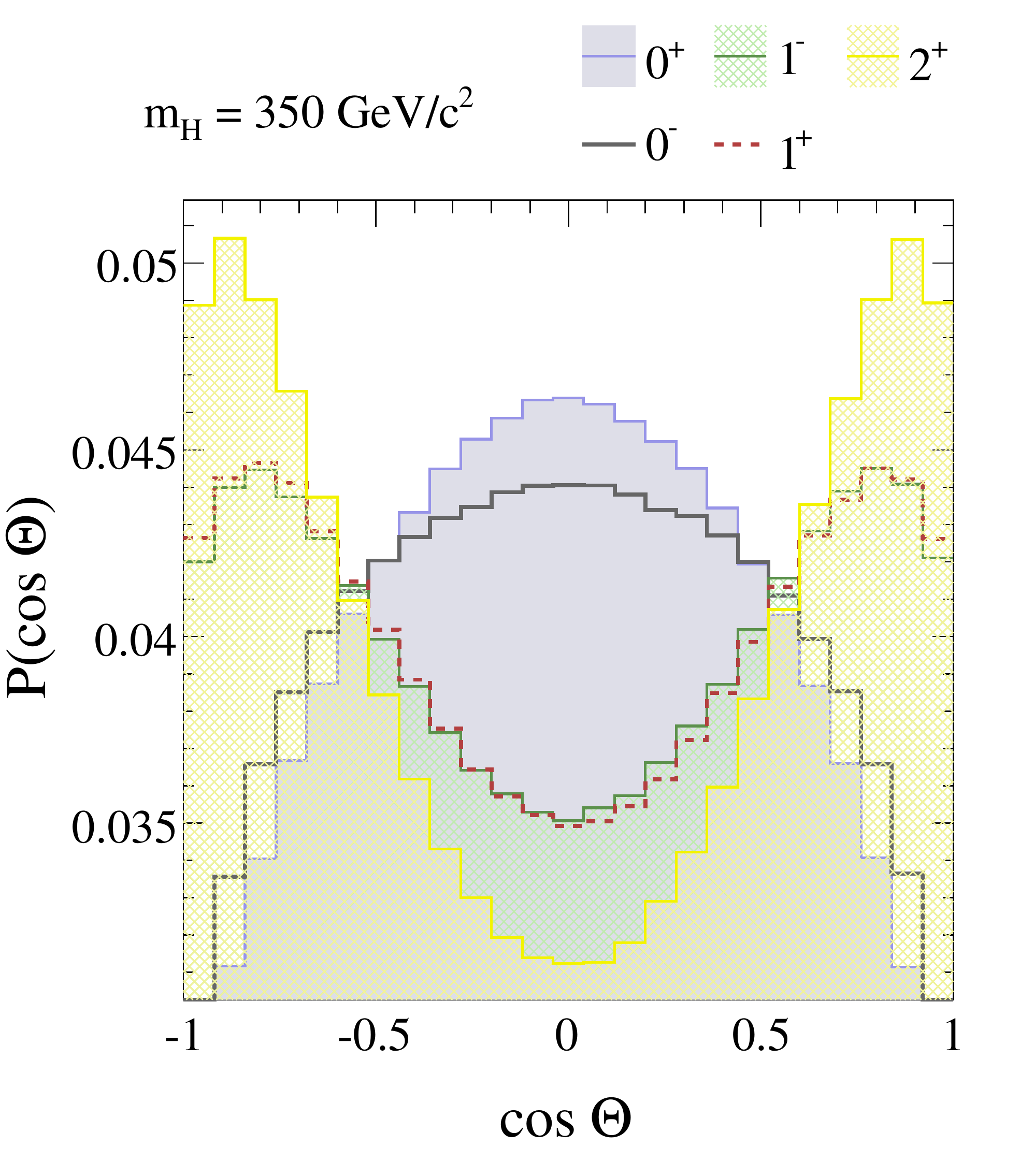}
\includegraphics[width=0.238\textwidth]{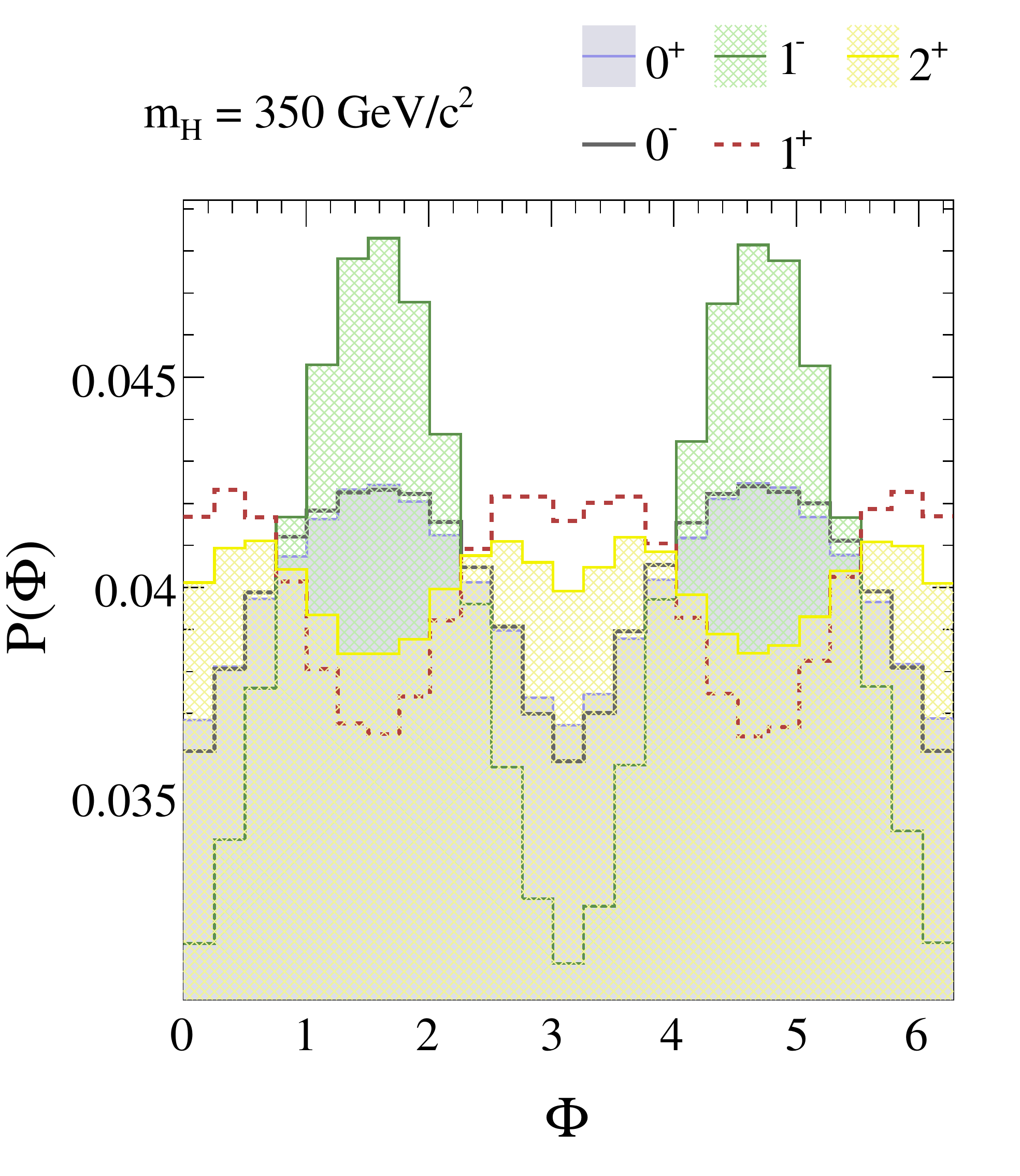}
\includegraphics[width=0.238\textwidth]{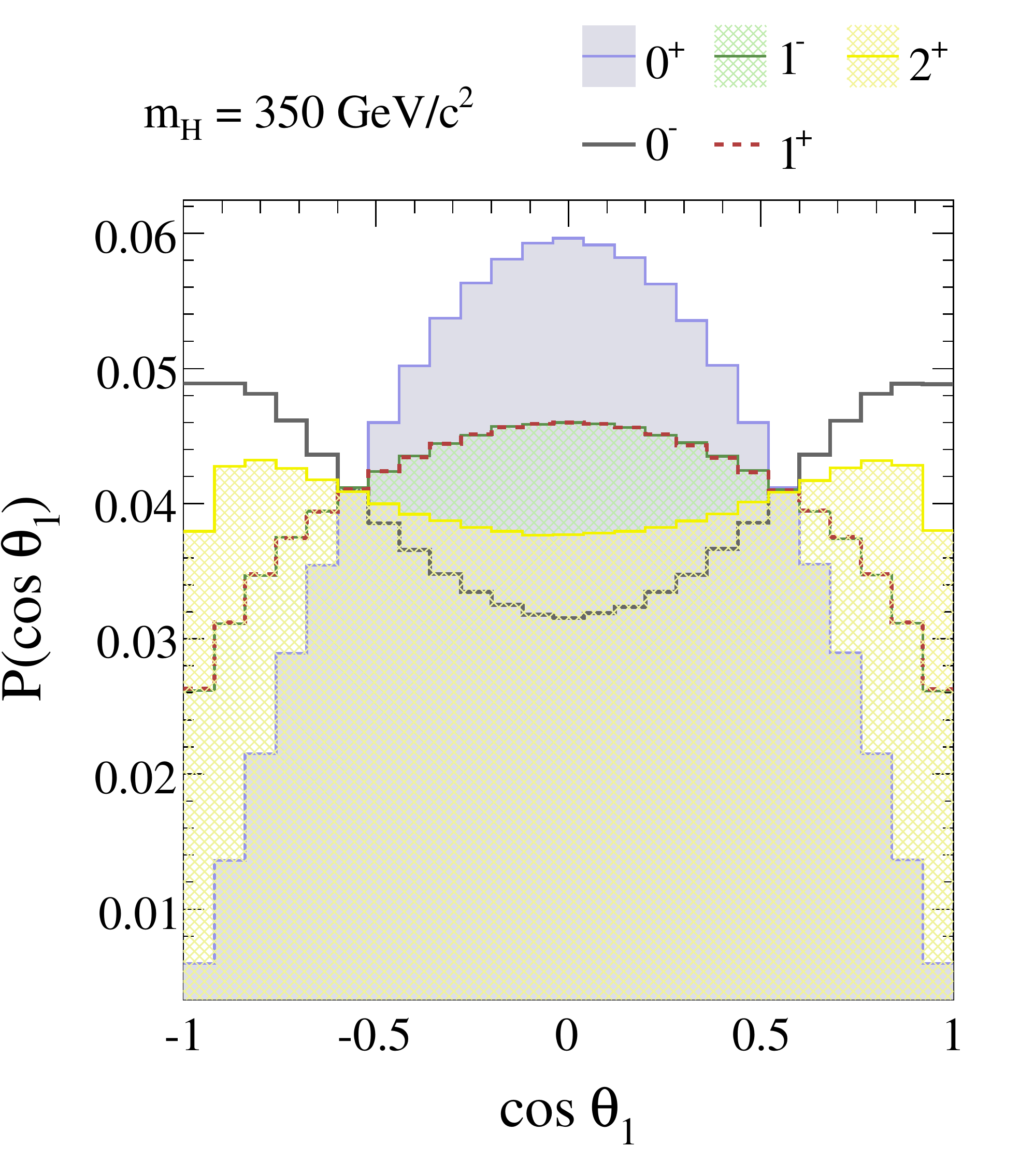}
\includegraphics[width=0.238\textwidth]{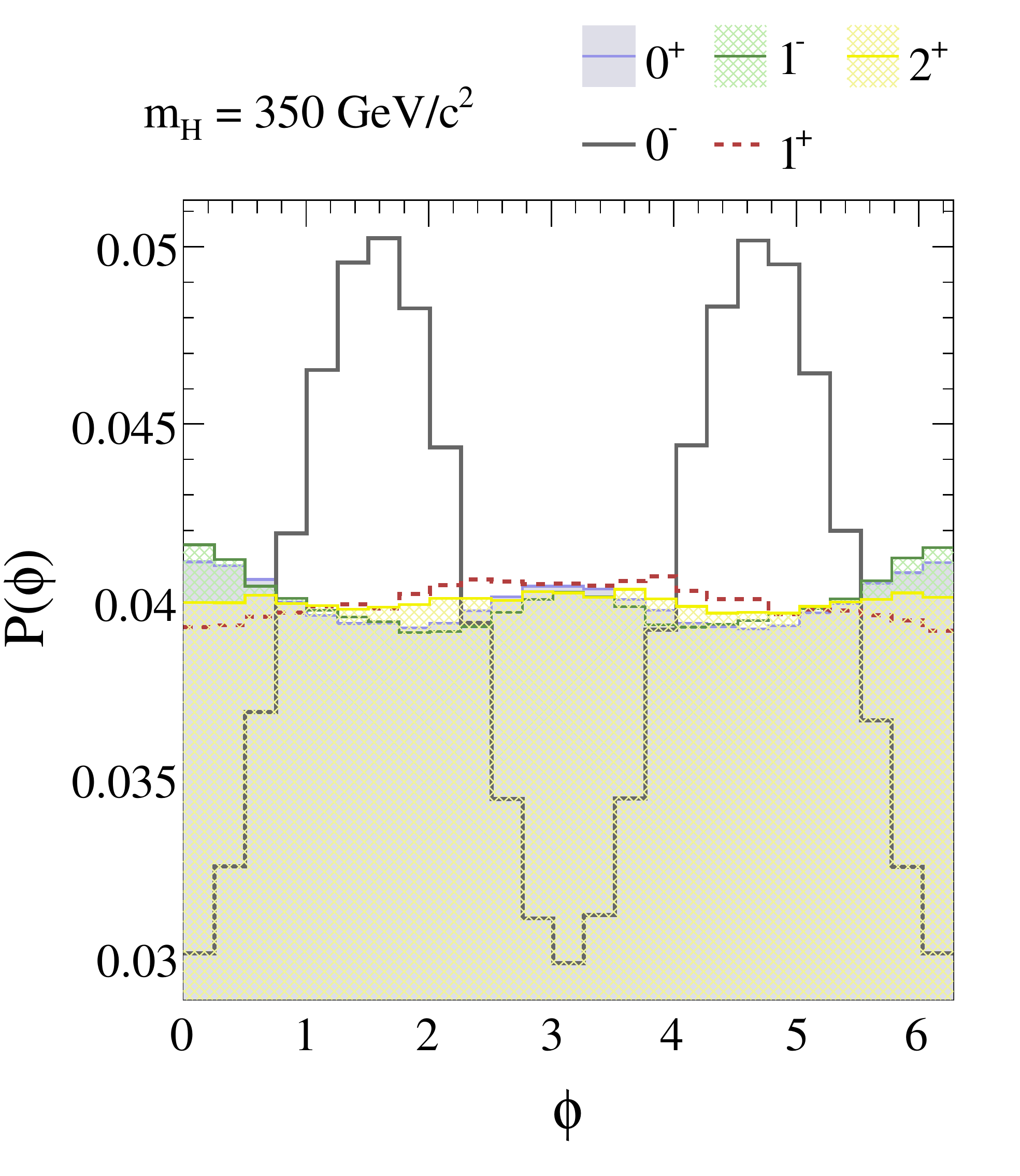}
\caption{Distributions of cos$\,\Theta$ (top left), $\Phi$ (top
  right), cos$\,\theta_{1}$ (bottom left) and $\phi$ (bottom right) for
  all the pure $J^{PC}$ choices we study, for $m_{H}$$=$$350$
  GeV/c$^{2}$. All distributions are normalized to a unit integral.
  \label{fig:KIN_350_all}}
\end{center}
\end{figure}
%%%%%%%%%%%%%%%%%%%%%%%%%%%%%%%%%%%%%%%%%%%%%%%%%%%%%%%%%%%%%%%%%%%
%%%%%%%%%%%%%%%%%%%%%%%%%%%%%%%%%%%%%%%%%%%%%%%%%%%%%%%%%%%%%%%%%%%
%%%%%%%%%%%%%%%%%%%%%%%%%%%%%%%%%%%%%%%%%%%%%%%%%%%%%%%%%%%%%%%%%%%
%%%%%%%%%%%%%%%%%%%%%%%%%%%%%%%%%%%%%%%%%%%%%%%%%%%%%%%%%%%%%%%%%%%

\begin{figure}[htbp]
\begin{center}
\includegraphics[width=0.210\textwidth]{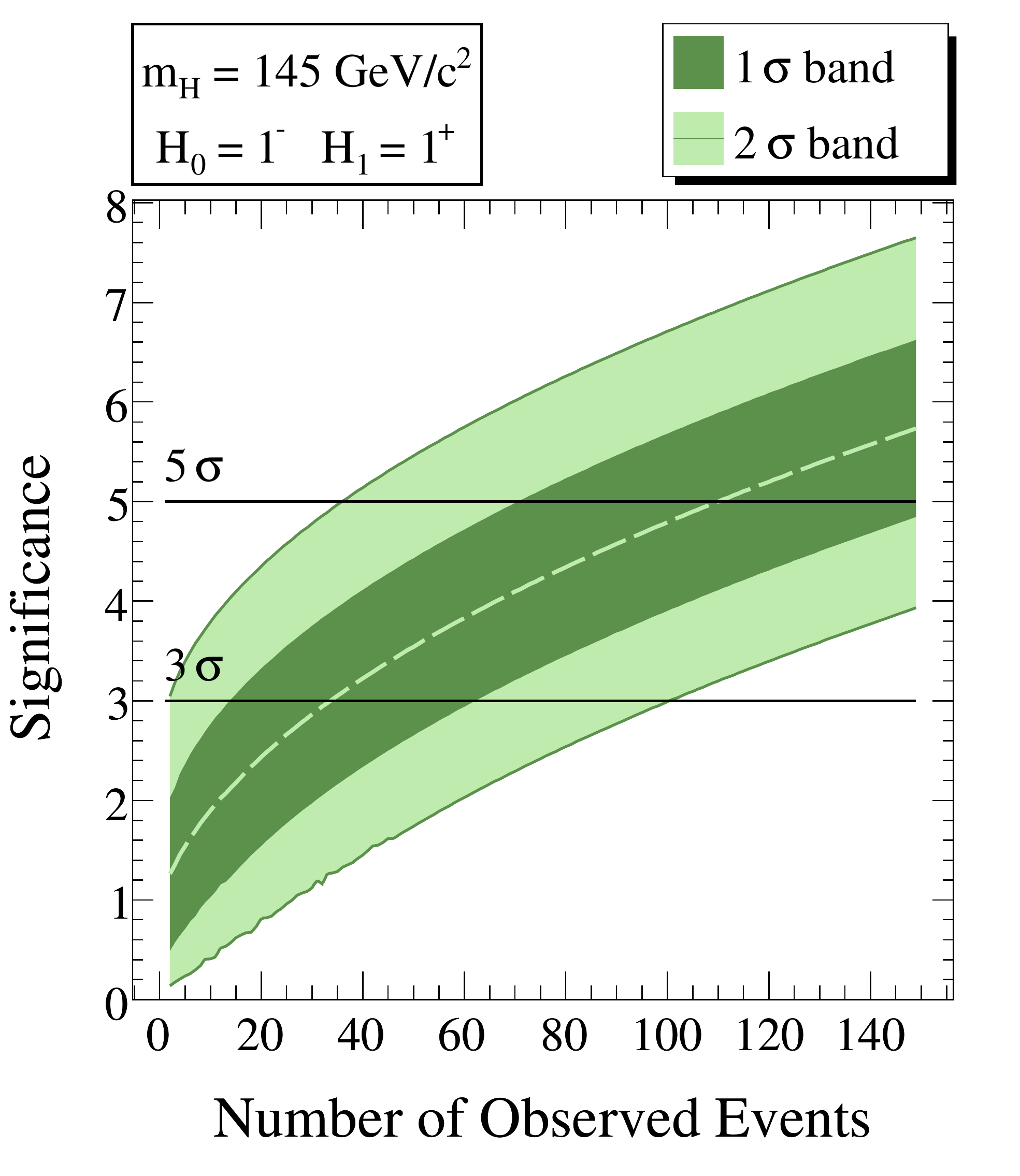}
\includegraphics[width=0.210\textwidth]{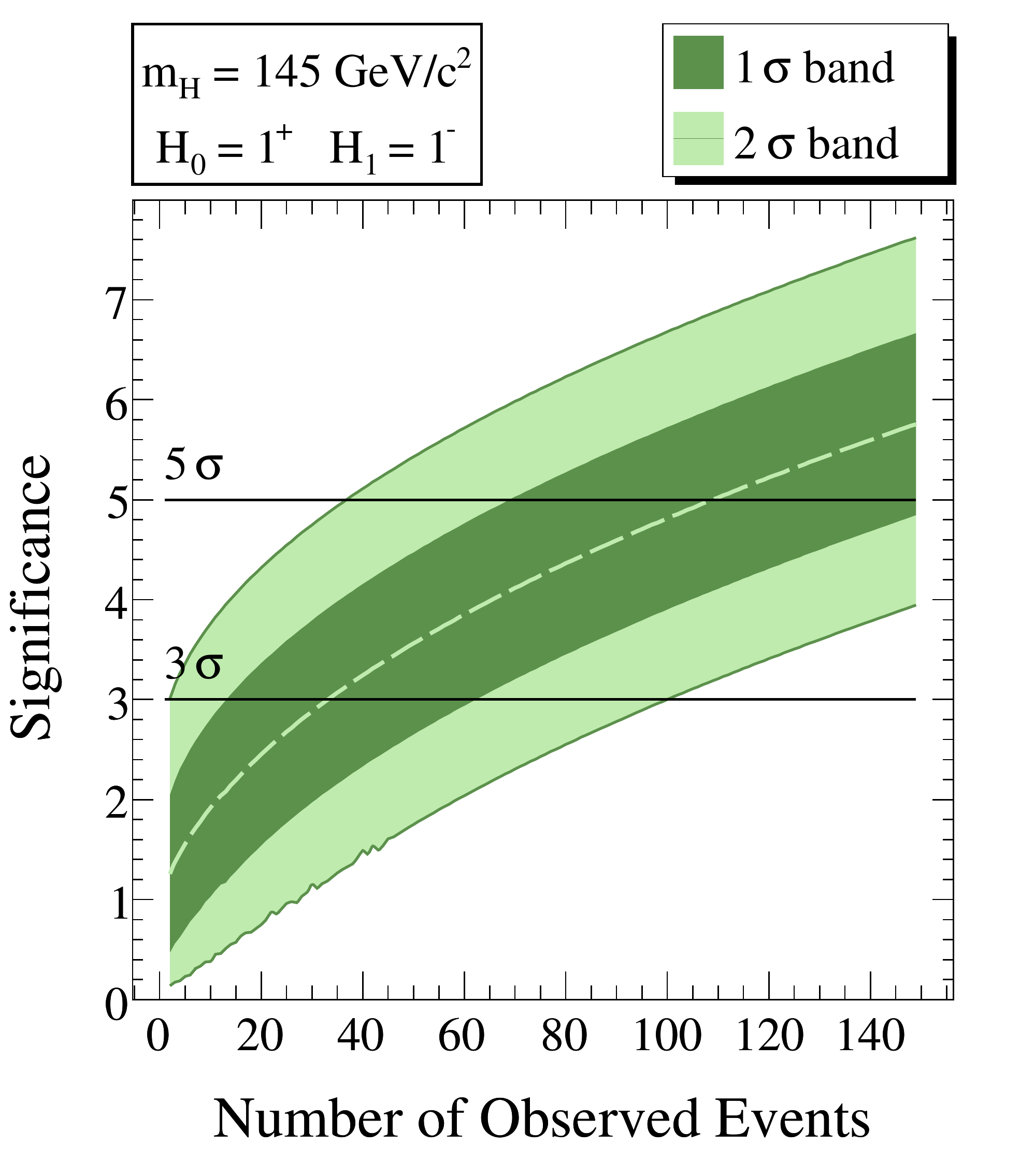}
\includegraphics[width=0.210\textwidth]{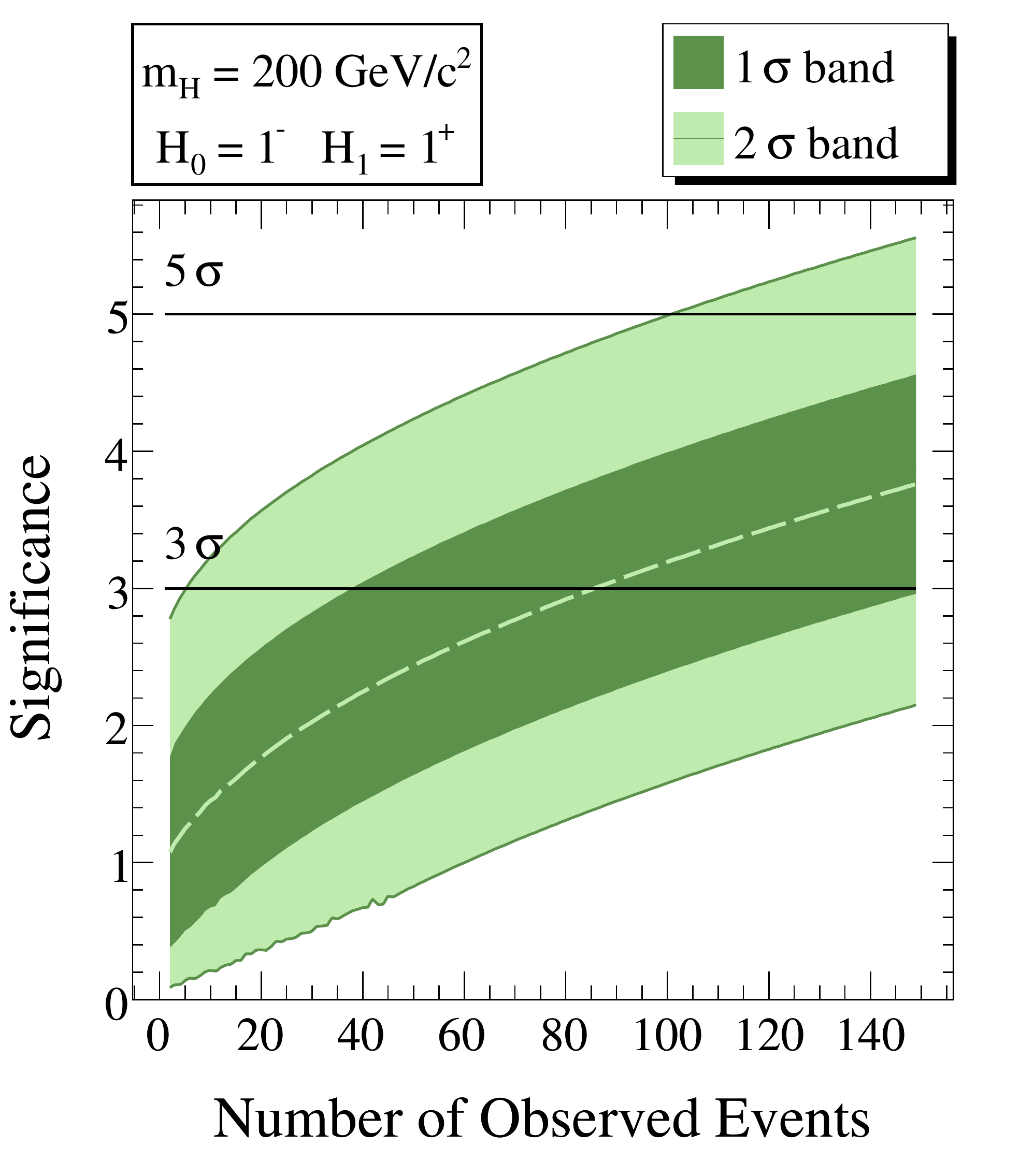}
\includegraphics[width=0.210\textwidth]{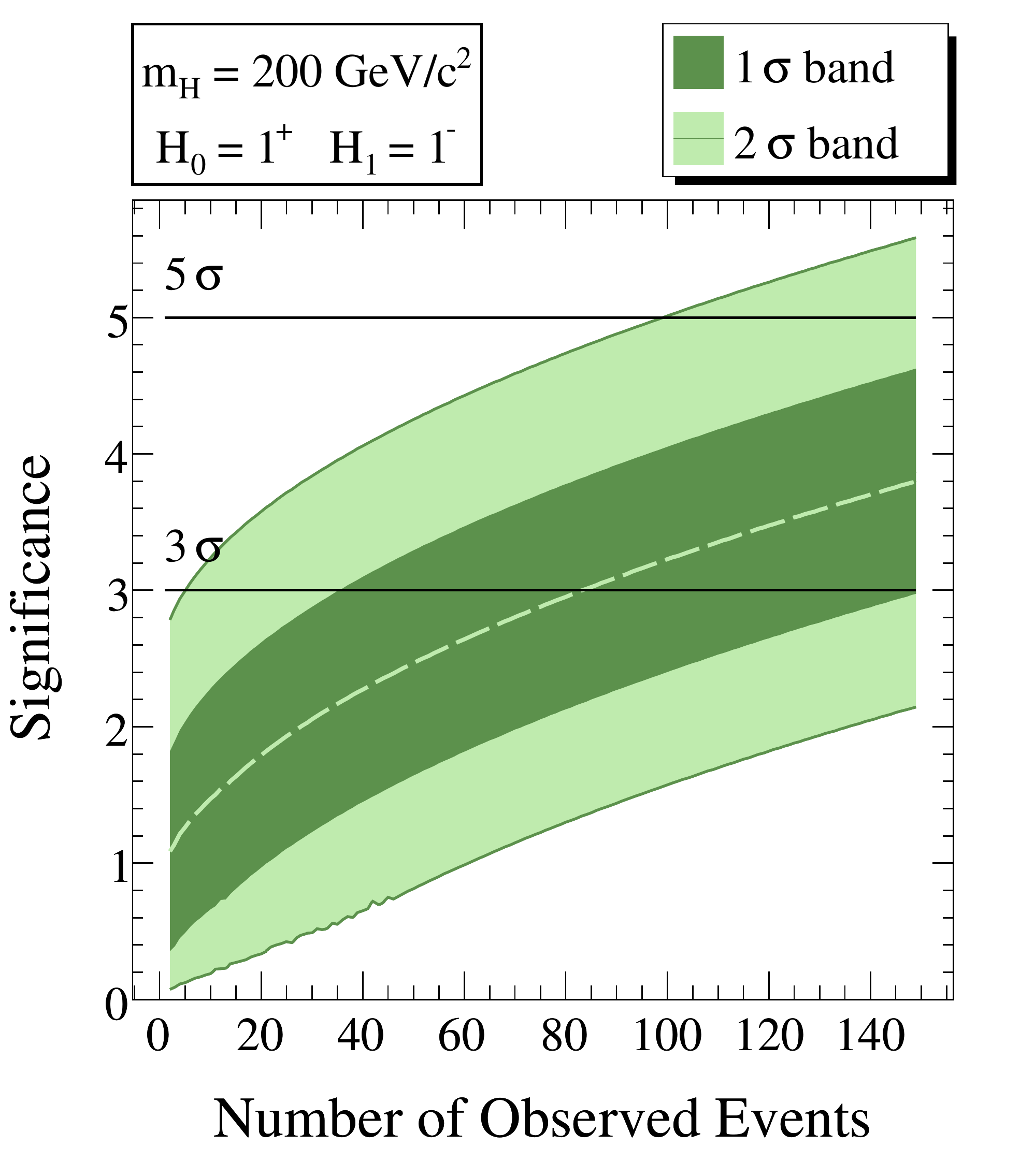}
\includegraphics[width=0.210\textwidth]{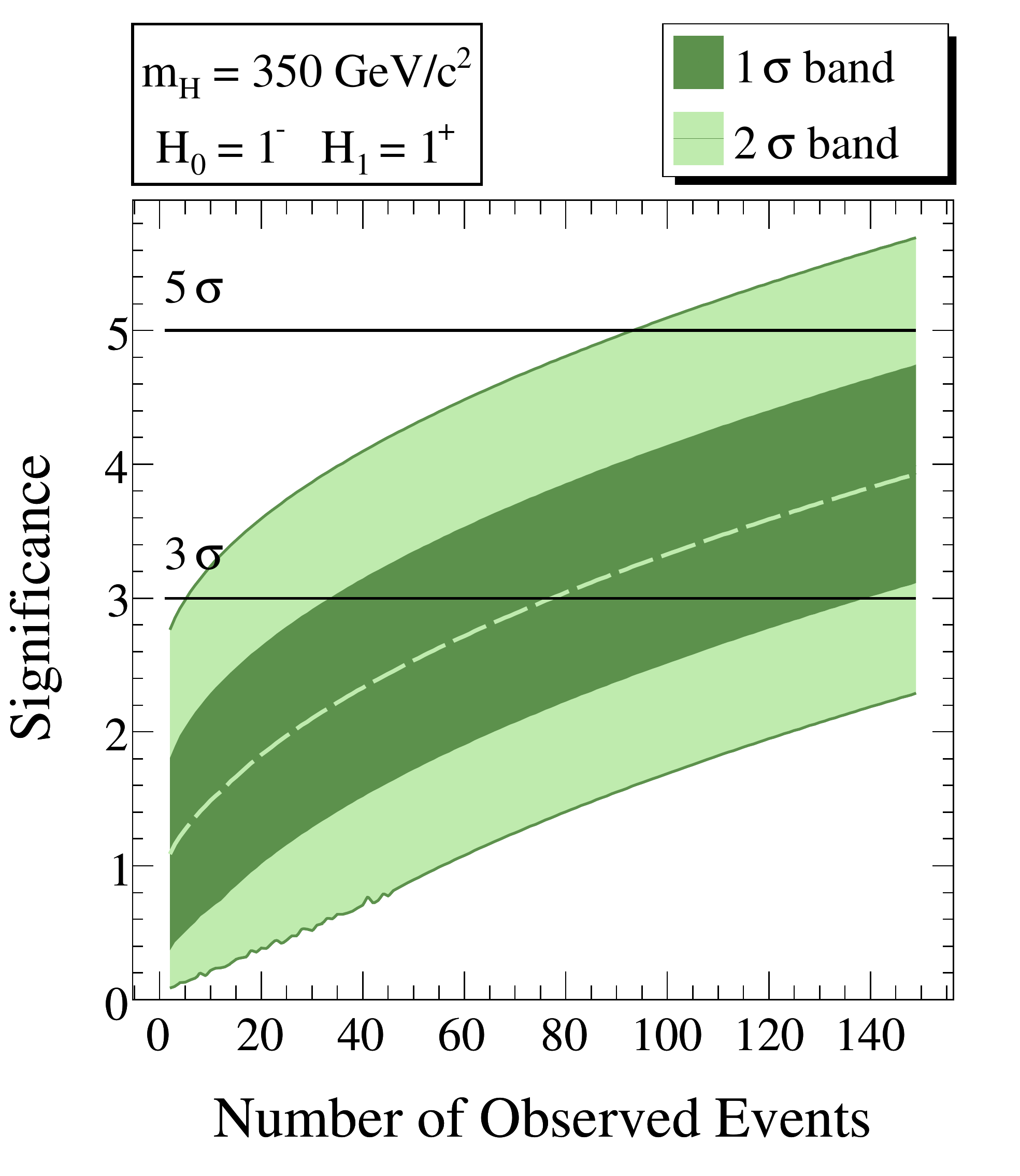}
\includegraphics[width=0.210\textwidth]{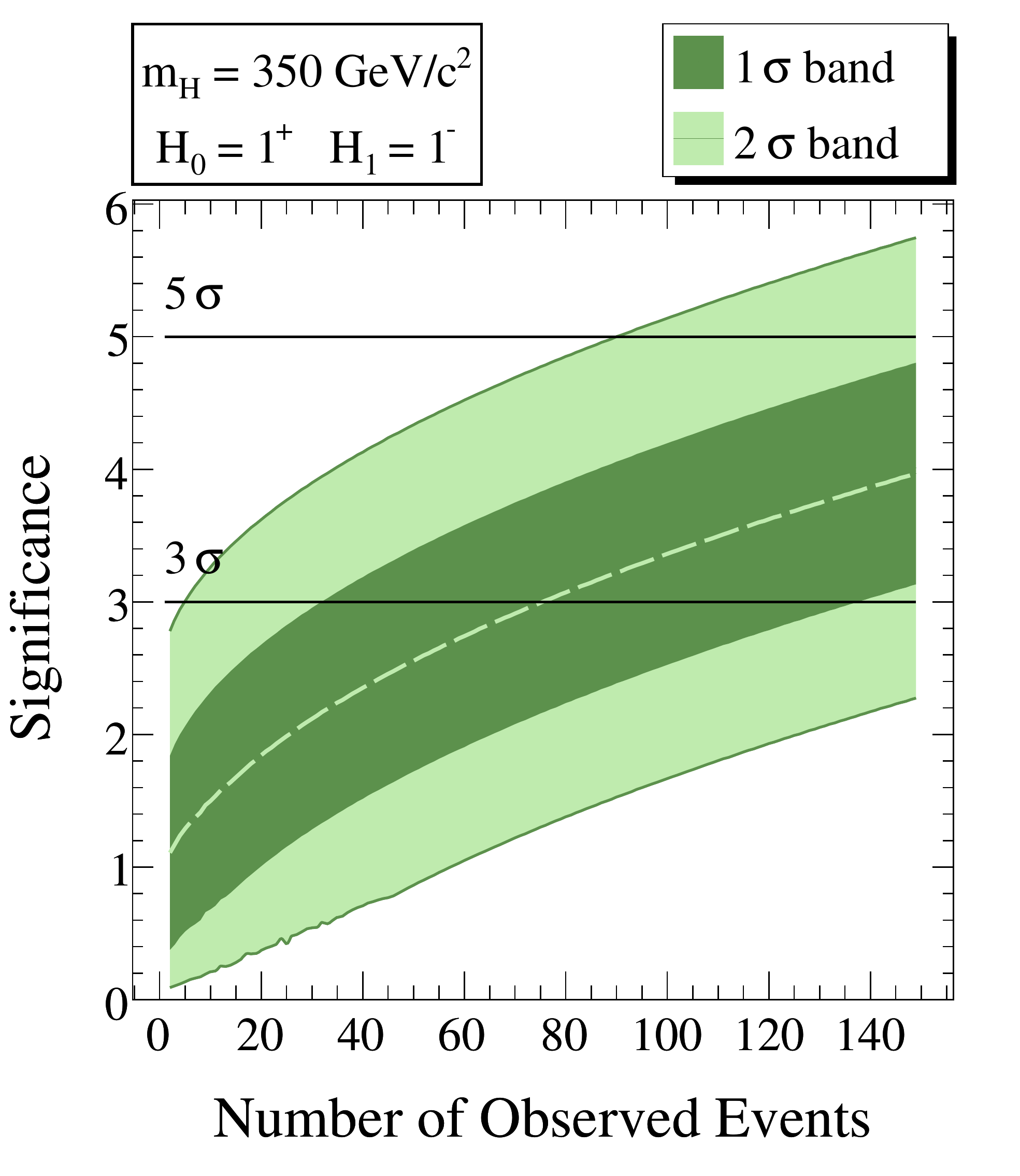}
\caption{Significance for rejecting $1^{-}$ in favor of
  $1^{+}$, assuming $1^{+}$ is true (left) or vice-versa ($1^-\!\leftrightarrow\! 1^+$, right)
  for $m_H$$=$$145$, 200 and 350 GeV/c$^{2}$ (top, middle and bottom).
   \label{fig:COMP_PV_v_PA}}
\end{center}
\end{figure}

The hardest differentiation is between $1^{-}$ and $1^{+}$.  Figures
~\ref{fig:KIN_145},~\ref{fig:KIN_200_all}, and \ref{fig:KIN_350_all},
show that the one-dimensional cos$\,\Theta$, cos$\,\theta_{1}$,
cos$\,\theta_{2}$, and $M_{Z^{*}}$ {\it pdfs} are similar. While the
$\Phi$ and $\phi$ {\it pdfs} provide some discrimination, the phase
space acceptance tends to sculpt the $\Phi$ distributions (and $\phi$
distribution through correlations) in ways that render the two cases
very similar.  The significance for distinguishing between
the two $J$$=$$1$ cases is shown in Fig.~\ref{fig:COMP_PV_v_PA}.  We
conclude that the discriminating potential is weakest for $1^+$
vs.~$1^-$, for all $m_H$. We revisit this result in
Sec.~\ref{sec:PARAM} in the context of measuring mixing parameters in
a general $J$$=$$1$ Lagrangian.
%%%%%%%%%%%%%%%%%%%%%%%%%%%%%%%%%%%%%%%%%%%%%%%%%%%%%%%%%%%%%%%%%%%
%%%%%%%%%%%%%%%%%%%%%%%%%%%%%%%%%%%%%%%%%%%%%%%%%%%%%%%%%%%%%%%%%%%

%%%%%%%%%%%%%%%%%%%%%%%%%%%%%%%%%%%%%%%%%%%%
%% mixed scalar states
%%%%%%%%%%%%%%%%%%%%%%%%%%%%%%%%%%%%%%%%%%%%

\subsection{$\mathbf{0^+}$ vs. mixed scalar states \label{sec:spinzero}}
Consider the vertex Feynman rules of Eq.~(\ref{generalscalar}) for
the most general Lorentz-covariant coupling ${\cal L}_{\mu\alpha}$ of
a spinless object to a $Z$ pair.  Rather than studying the general
case, for which any of the quantities $X$ to $Q$ can be nonzero, we
investigate three cases, each with only two non-vanishing types of
coupling, resulting in one free mixing ``angle'' and an overall
normalization (which we ignore):
\\
\begin{itemize}
\item {$X \ne 0,~P \ne 0$:} A scalar whose $ZZ$ coupling violates
  $CP$, described in terms of an angle $\xi_{XP}$ as:
\begin{equation}
{\cal L}_{\mu\alpha} \propto {\rm cos}(\xi_{XP}) \,g_{\mu\alpha} + {\rm sin}( \xi_{XP})\, \epsilon_{\mu\alpha}{p_{1}p_{2}}/{M_{Z}^{2}}
\nonumber
\end{equation}
\item { $X \ne 0,~Q \ne 0$:} A scalar whose $ZZ$ coupling violates
  $C$, described in terms of an angle as:
\begin{equation}
{\cal L}_{\mu\alpha} \propto {\rm cos}(\xi_{XQ})\, g_{\mu\alpha} + i\, {\rm sin}( \xi_{XQ})\, \epsilon_{\mu\alpha}{p_{1}p_{2}}/{M_{Z}^{2}}
\nonumber
\label{eq:XQdef}
\end{equation}

\item { $X \ne 0,~Y \ne 0$:} A composite $0^+$, parameterized in terms
  of an angle as:
\begin{equation}
 {\cal L}_{\mu\alpha} \propto {\rm cos}(\xi_{XY}) \,g_{\mu\alpha} - {\rm sin}( \xi_{XY})\,{k_{\alpha}k_{\mu}}/{M_{Z}^{2}}
 \nonumber
 \end{equation}
\end{itemize}

As a function of $N_S$ we estimate the significance with which one 
can determine:
\begin{itemize}
\item (a) What range of values of the angles can be excluded in favor
  of a pure $0^+$ for a SM-like resonance;
\item (b) Whether a pure $0^{+}$ can be excluded in favor of a
  non-trivial mixture when the resonance corresponds to one of the
  three mixed cases discussed above.
\end{itemize}

We consider first the example of a $CP$-violating $HZZ$ coupling with
$m_H$$=$$350$ GeV/c$^{2}$.

To address (a) we construct a series of simple hypothesis tests of the
type we considered earlier for distinguishing between pure $J^{PC}$
states. Specifically, for a given number of observed signal events at
a fixed value of $m_H$, we perform a NePe test between two simple
hypotheses: that the resonance is $0^{+}$ (denoted hypothesis
$\mathbb{H}_{1}$) or that the resonance is $J$$=$$0$ with $\xi_{XP}$ {\it
  fixed to a specific nonzero value} (denoted hypothesis
$\mathbb{H}_{0}$). The test statistic we use is $\log
[{\mathcal{L}}^{XP}(\xi_{XP})/{\mathcal{L}}({0^{+}})]$, where
${\mathcal{L}}(0^{+})$ and ${ \mathcal{L}}^{XP}(\xi_{XP})$ denote the
likelihoods for a set of events agreeing with the hypotheses
$\mathbb{H}_{1}$ and $\mathbb{H}_{0}$, respectively.  The test cannot
be performed for $\xi_{XP}$$=$$0$, since in this case the $\mathbb{H}_{0}$
$CP$-violating hypothesis we want to test reduces to the alternative
$\mathbb{H}_{1}$ hypothesis (the CP-conserving SM Higgs).

\begin{figure}[htbp]
\begin{center}
\includegraphics[width=0.38\textwidth]{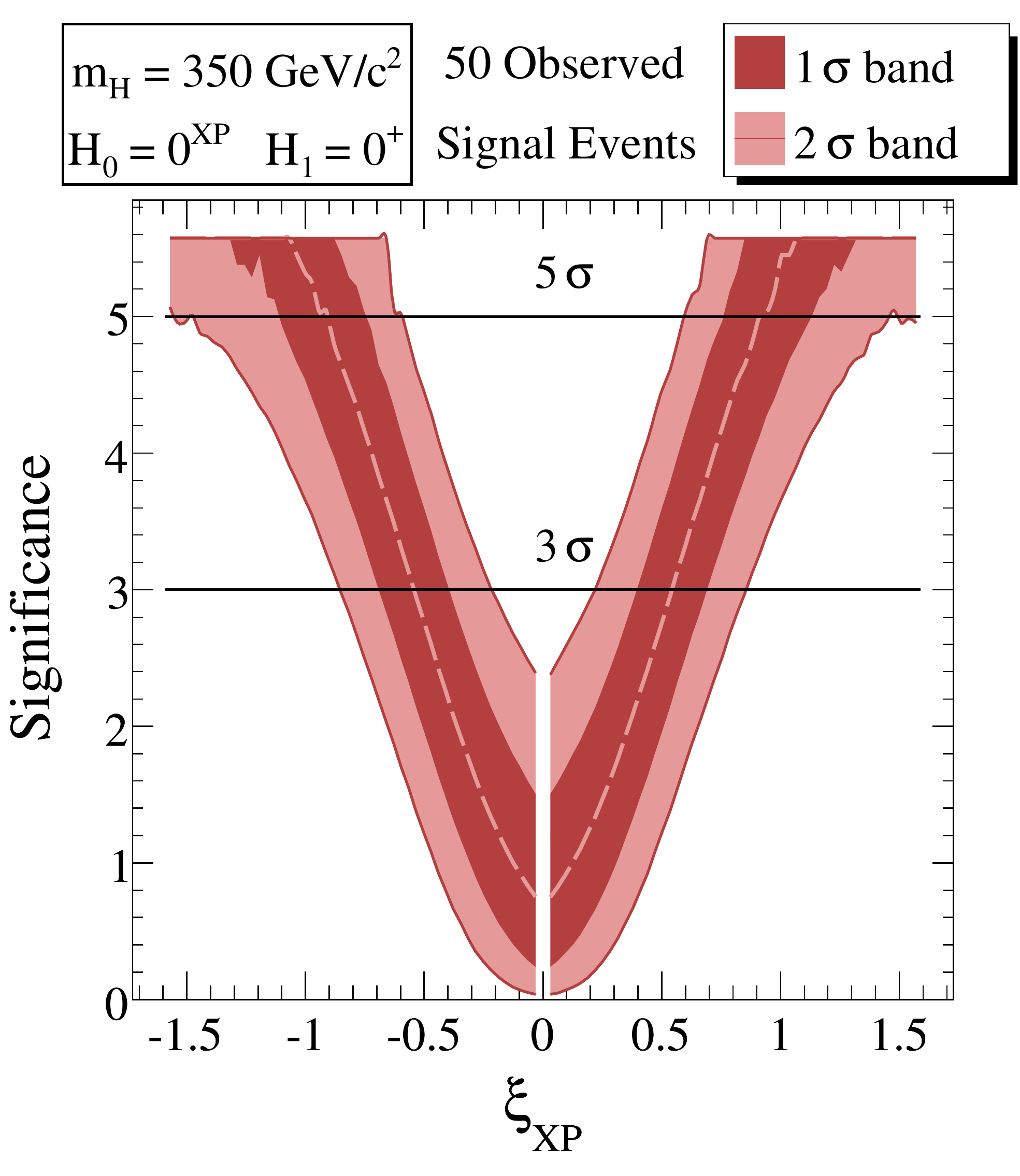}
\caption{Significance for excluding values of $\xi_{XP}$ in the $CP$-violating $J$$=$$0$ hypothesis
  in favor of the $0^{+}$ one, assumed to be 
  correct, for $m_H$$=$$350 $ GeV/c$^{2}$ and $N_S$$=$$50$.
  The dashed line corresponds to the median of the
  significance. The 1 and 2$\,\sigma$ bands correspond to 68\% and
  95\% confidence intervals  centered on the median value.
  \label{fig:COMP1D_H0}}
\end{center}
\end{figure}

The result of this test is the significance with which hypothesis
$\mathbb{H}_{0}$ can be rejected in favor of the hypothesis
$\mathbb{H}_{1}$, or similarly, the significance with which a
particular value of $\xi_{XP}$ can be excluded in favor of the
$0^{+}$ hypothesis. This test is then repeated with different fixed
values of $\xi_{XP}$, i.e.~different NePe tests with different
hypotheses $\mathbb{H}_{0}$. The results for a large ensemble of such
tests are shown in Fig.~\ref{fig:COMP1D_H0}. Here, $\mathbb{H}_{0}=
0^{XP}$ denotes the simple $J$$=$$0$ $CP$-violating hypothesis with
$\xi_{XP}$ fixed at values chosen on the $x$-axis.

In this example we see that, for $N_S$$=$$50$, the significance for
excluding a $CP$-violating coupling exceeds 3$\,\sigma$ for
$|\xi_{XP}| > 0.5$ and 5$\,\sigma$ for $|\xi_{XP}| > 0.9$.

In addressing (b) we cannot construct a simple NePe test between
$0^{+}$ and a fixed-$\xi_{XP}$ hypothesis. Instead, we treat
$\xi_{XP}$ as a nuisance parameter and choose a value, $\hat\xi_{XP}$,
that maximizes the $CP$-violating likelihood for the given set of
observed events. Specifically, we fix $\xi_{XP}$ at a particular value
(the ``true'' value) to generate events and perform NePe tests
comparing $\xi_{XP}$$=$$0$ (denoted hypothesis $\mathbb{H}_{0}$) and
$\xi_{XP} =\hat\xi_{XP}$ ($\mathbb{H}_{1}$). This test is repeated for
many different values of the fixed ``input'' $\xi_{XP}$.

\begin{figure}[htbp]
\begin{center}
\includegraphics[width=0.38\textwidth]{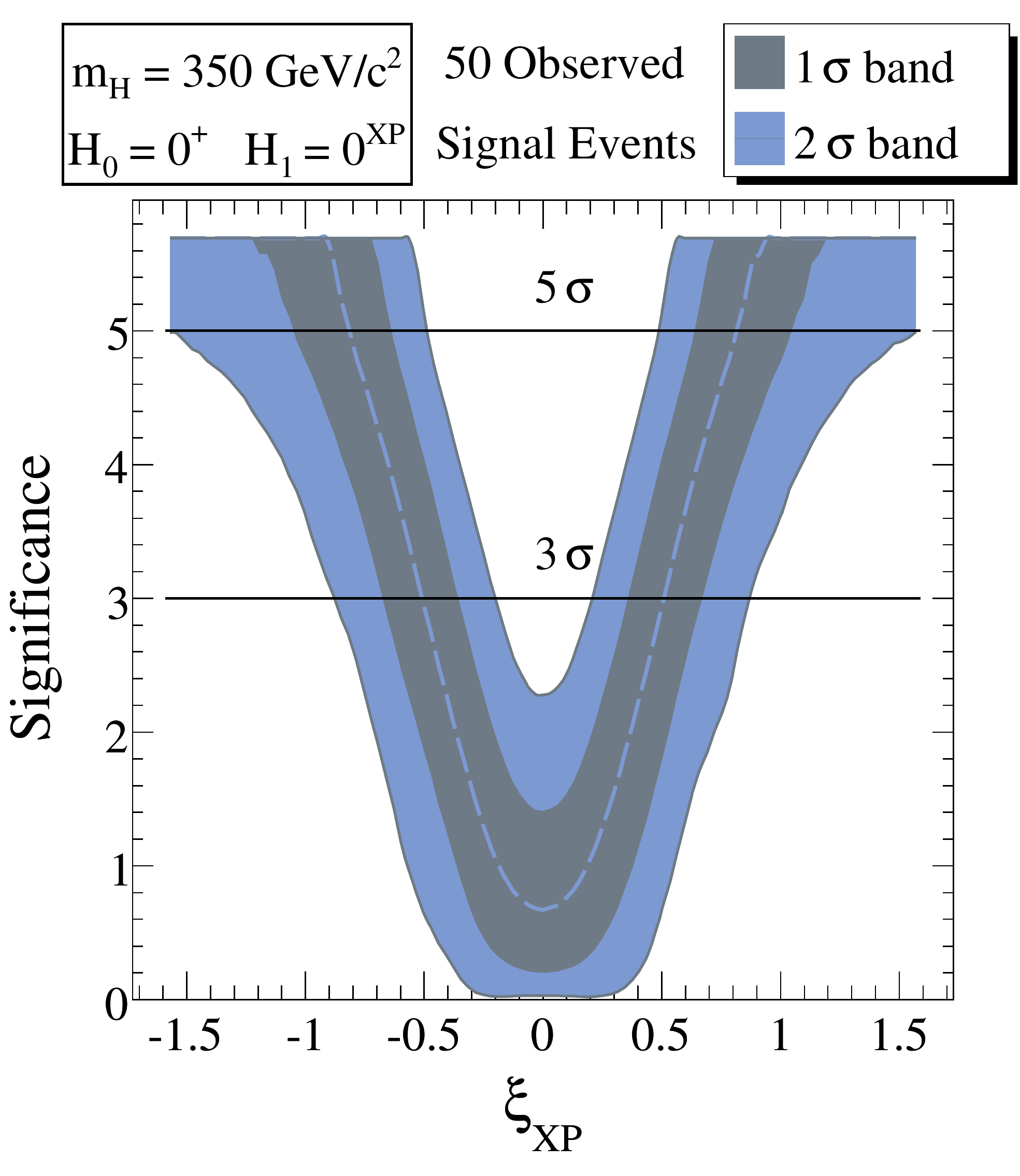}
\caption{The  significance for
  excluding a pure $0^+$ in favor of a $CP$-violating $HZZ$ coupling ($\xi_{XP}\neq 0$),
  assuming the latter to be correct, with $\xi_{XP}$ given by
  its $x$-axis values. Example for $N_S$$=$$50$, $m_H$$=$$350$ GeV/c$^2$.
  Dashed line and bands as in Fig.~\ref{fig:COMP1D_H0}.
  \label{fig:COMP1D_H1}}
\end{center}
\end{figure}

\begin{figure}[htbp]
\begin{center}
\includegraphics[width=0.38\textwidth]{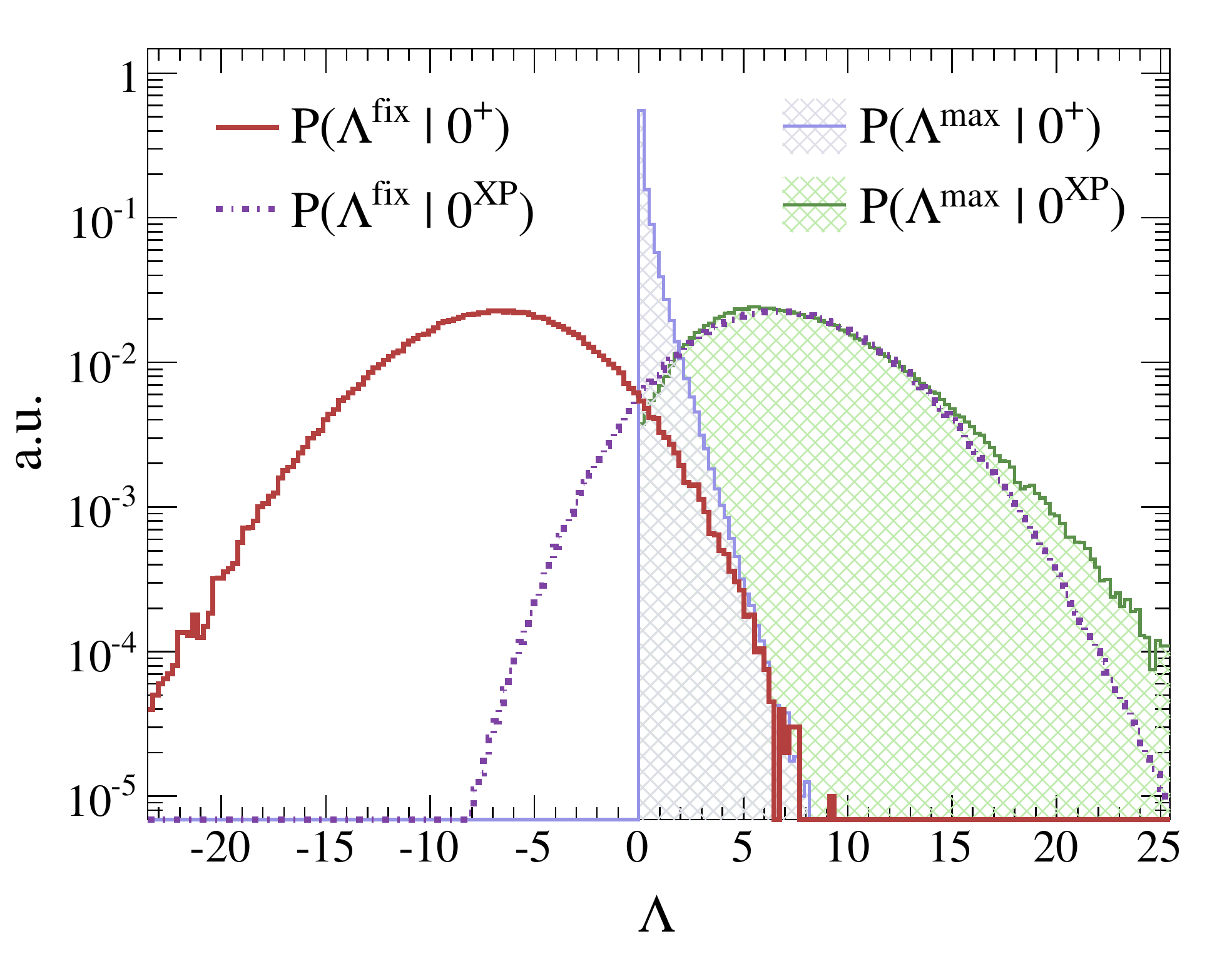}
\includegraphics[width=0.38\textwidth]{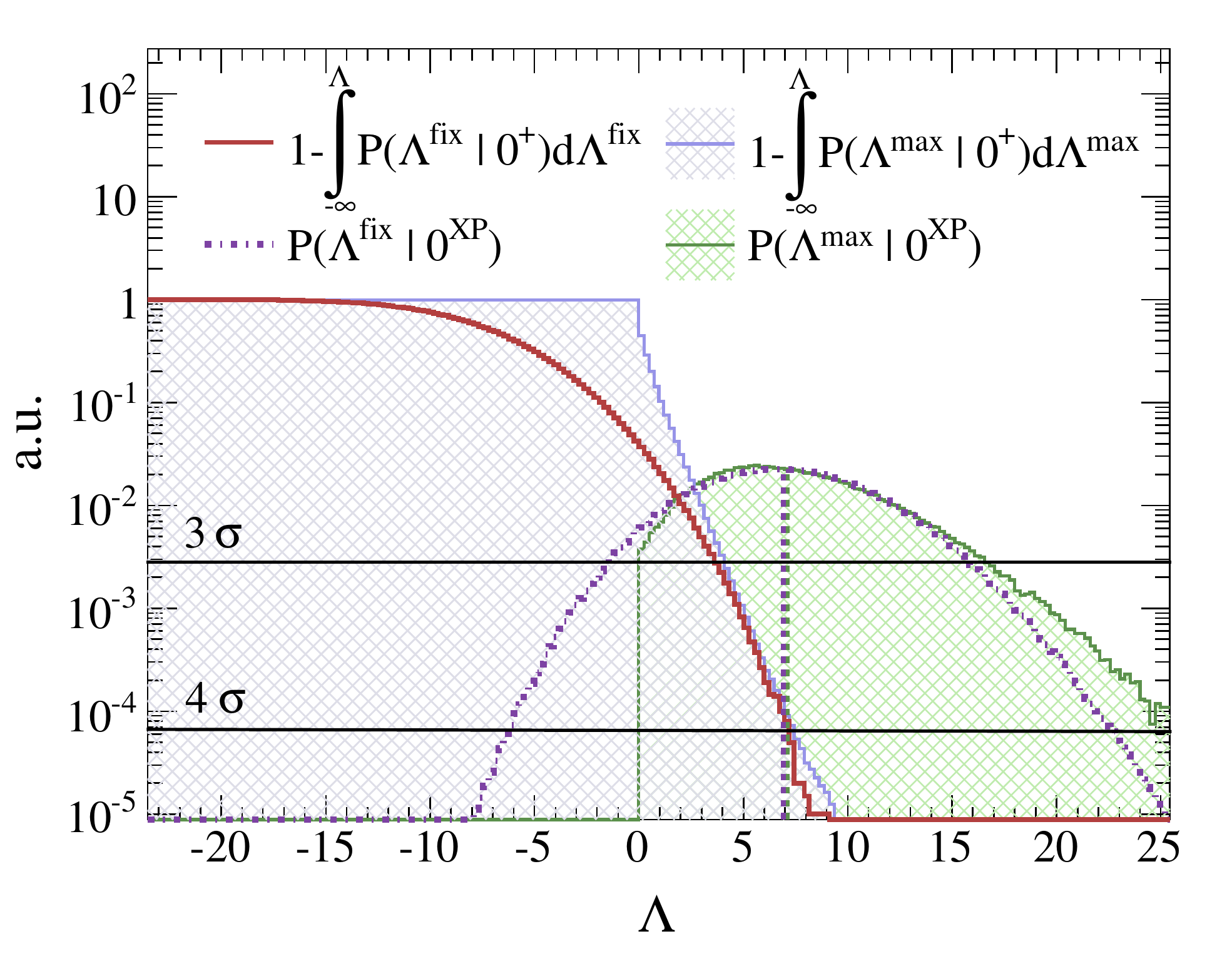}
\caption{Distributions of the two statistics $\Lambda$, defined in the
  text, for $m_H$$=$$350$ GeV/c$^{2}$ and $N_S$$=$$50$. The hypotheses are
  $\mathbb{H}_{0} $$=$$ 0^{+}$, and $\mathbb{H}_{1} $$=$$ 0^{XP}$ with the
  $CP$-phase $\xi_{XP}$ fixed at $\pi/5$. (Top) Probability
  distributions ${\rm P}(\Lambda | \mathbb{H})$. (Bottom) The same
  with the $0^+$ results traded for 1 minus their cumulative values.
  The two nearly indistinguishable vertical dotted lines correspond to
  the median values of the ${\rm P}(\Lambda | \mathbb{H}_1)$
  distributions.
  \label{fig:STAT_mix}}
\end{center}
\end{figure}

An example of results from an ensemble of these tests is shown in
Fig.~\ref{fig:COMP1D_H1}. Because of the addition of a nuisance
parameter, the figure's interpretation is not simply related to the
interpretation of Fig.~\ref{fig:COMP1D_H0}, which answered question
(a). What Fig.~\ref{fig:COMP1D_H1} shows is the expected significance
with which one can exclude the SM hypothesis in favor of the
$CP$-violating hypothesis with  $\xi_{XP} $$=$$\hat\xi_{XP}$, as a function of the
true value of $\xi_{XP}$ (given on the $x$-axis).  No a priori
knowledge of the actual value of $\xi_{XP}$ is required to perform
this test.
From Figs.~\ref{fig:COMP1D_H0} and \ref{fig:COMP1D_H1} we observe that
the expected significances are symmetric around $\xi_{XP}$$=$$0$. This
is due to the {\it pdfs} of the ``pure $0^+$'' and ``pure $0^-$''
terms being even under $\xi_{XP}\to - \xi_{XP}$, while the
$\tilde{T}$-odd interference term vanishes under the integration of
cos$\,\theta_1$, cos$\,\theta_2$ or $\phi$. We shall see that there
are exceptions to this trivial statement. Comparing these two figures
we observe a remarkable similarity of the significances of the two
tests. Since two different statistics are used, this is somewhat of a
coincidence. To explain it, consider the example with 
$\xi_{XP} $$=$$ \pi/5$, which corresponds to vertical slices of
Figs.~\ref{fig:COMP1D_H0} and \ref{fig:COMP1D_H1}.  We denote the two
different test statistics $\Lambda^{\rm fix}$$=$$\log [{ {\mathcal
    L}}^{XP}(\xi_{XP})/{\mathcal L}({0^{+}})]$, with $\xi_{XP}$ fixed
at its true value, corresponding to a simple hypothesis test and
$\Lambda^{\rm max}$$=$$\log [{\rm max}\,{\mathcal
  L}^{XP}(\hat\xi_{XP})/{\mathcal L}({0^{+}})]$, {\it profiled} to
the value $\hat\xi_{XP}$ at which it peaks. The distributions of
$\Lambda^{\rm fix}$ and $\Lambda^{\rm max}$ are shown in
Fig.~\ref{fig:STAT_mix}.

In the top figure the bell-shaped curves $P(\Lambda^{\rm fix} |
0^{+})$ and $P(\Lambda^{\rm fix} | 0^{XP})$ are characteristic of a
simple hypothesis test. The distributions of $\Lambda^{\rm max}$ have
a sharp cut-off at $\Lambda^{\rm max}$$=$$0$, since the $0^{+}$ model is
a member of the $0^{XP}$ family with $\xi_{XP}$$=$$0$, and ${\rm max}\,{
  L}^{XP}(\hat\xi_{XP})/{ L}({0^{+}})\ge 1$, which are also features
characteristic of this type of test.

The reason for two very different hypothesis tests to end up in the
similar-looking results of Figs.~\ref{fig:COMP1D_H0} and
\ref{fig:COMP1D_H1} is that the statistically-significant features of
the different-looking $P(\Lambda)$ distributions shown in
Fig.~\ref{fig:STAT_mix} are actually very similar. $P(\Lambda^{\rm
  fix} | 0^{XP})$ and $P(\Lambda^{\rm max} | 0^{XP})$ differ, but the
distributions of $\xi_{XP}$ close to the maxima are localized around
the true input value, their median values and 68\% and 95\% confidence
intervals are nearly identical (try to tell apart the two vertical
dotted lines in the lower half of Fig.~\ref{fig:STAT_mix}, at
$\Lambda\sim 7$). Also, the tails of one-minus-cumulative
distributions for $P(\Lambda^{\rm fix} | 0^{+})$ and $P(\Lambda^{\rm
  max} | 0^{+})$ coalesce for $p$-values exceeding 2$\,\sigma$
significance, despite large differences in the distributions
themselves.

\begin{figure}[htbp]
\begin{center}
\includegraphics[width=0.238\textwidth]{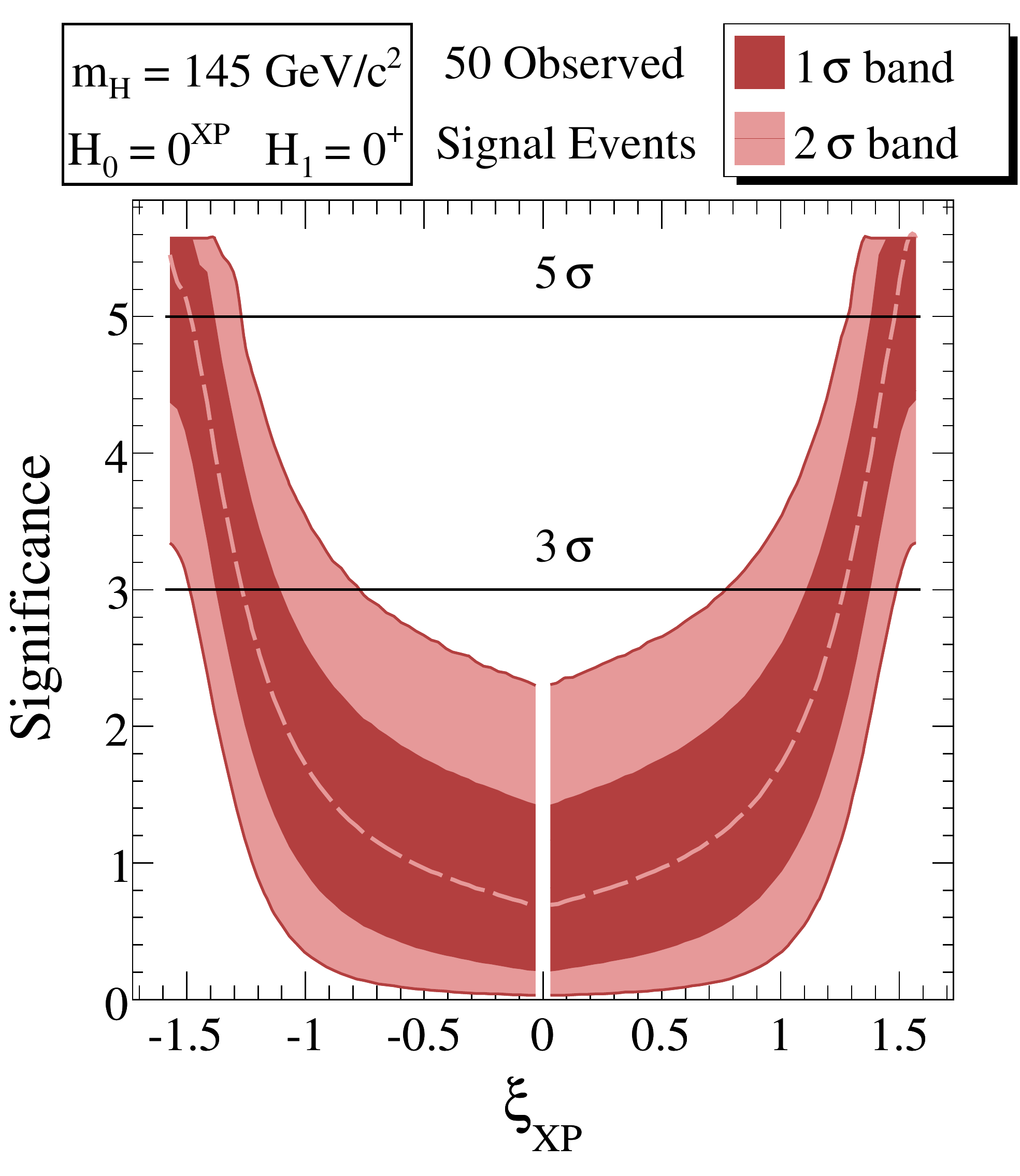}
\includegraphics[width=0.238\textwidth]{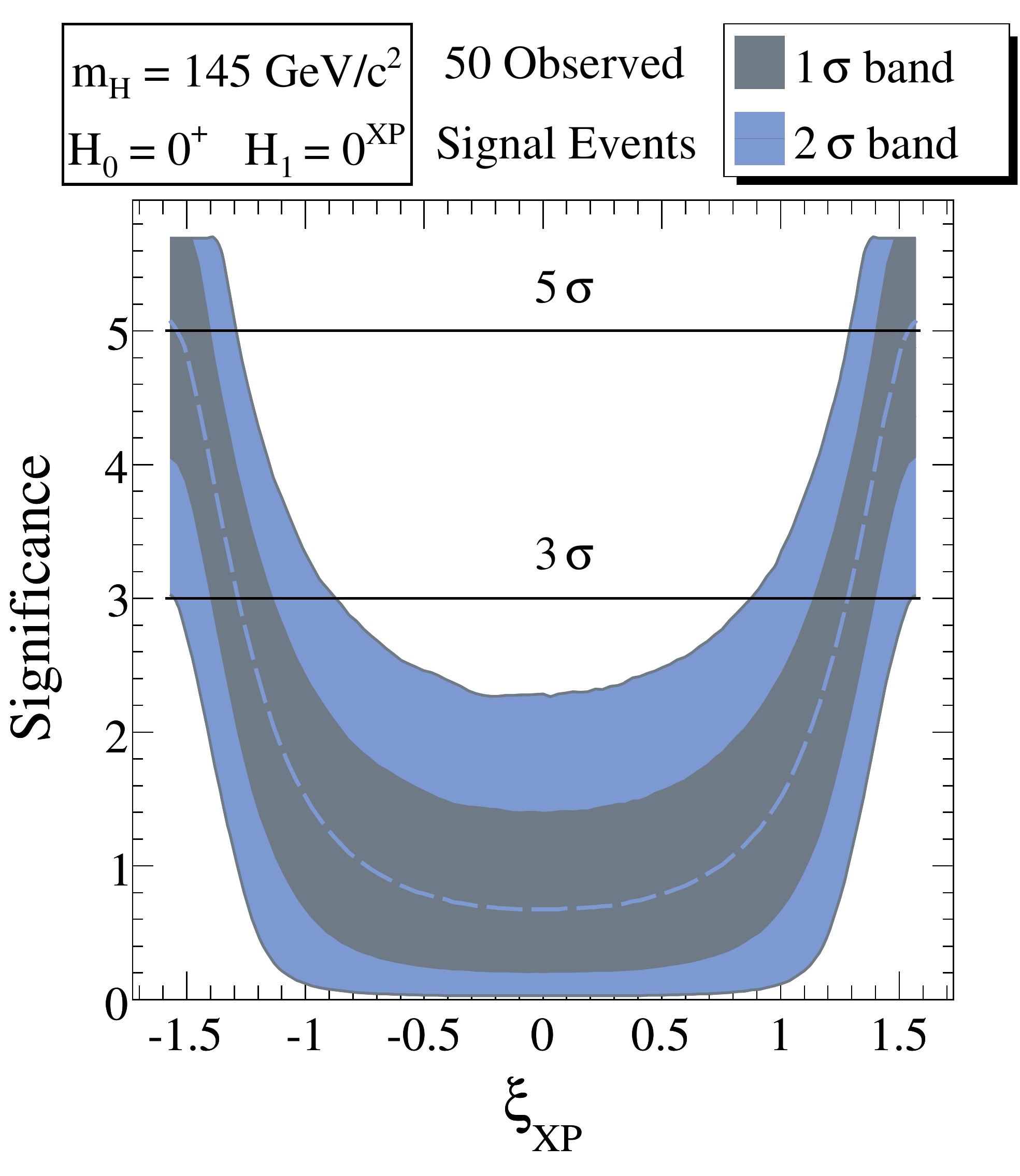}
\includegraphics[width=0.238\textwidth]{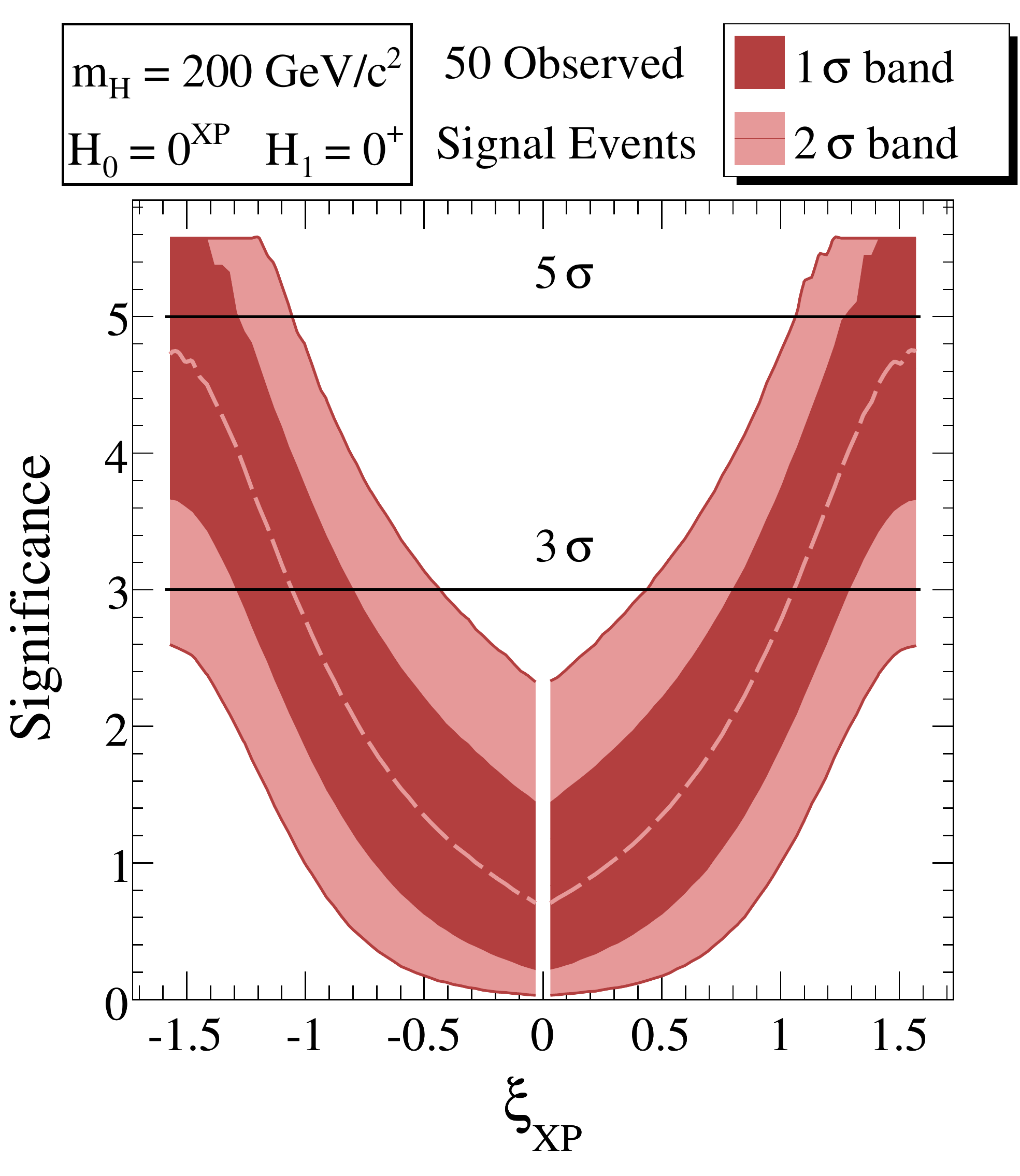}
\includegraphics[width=0.238\textwidth]{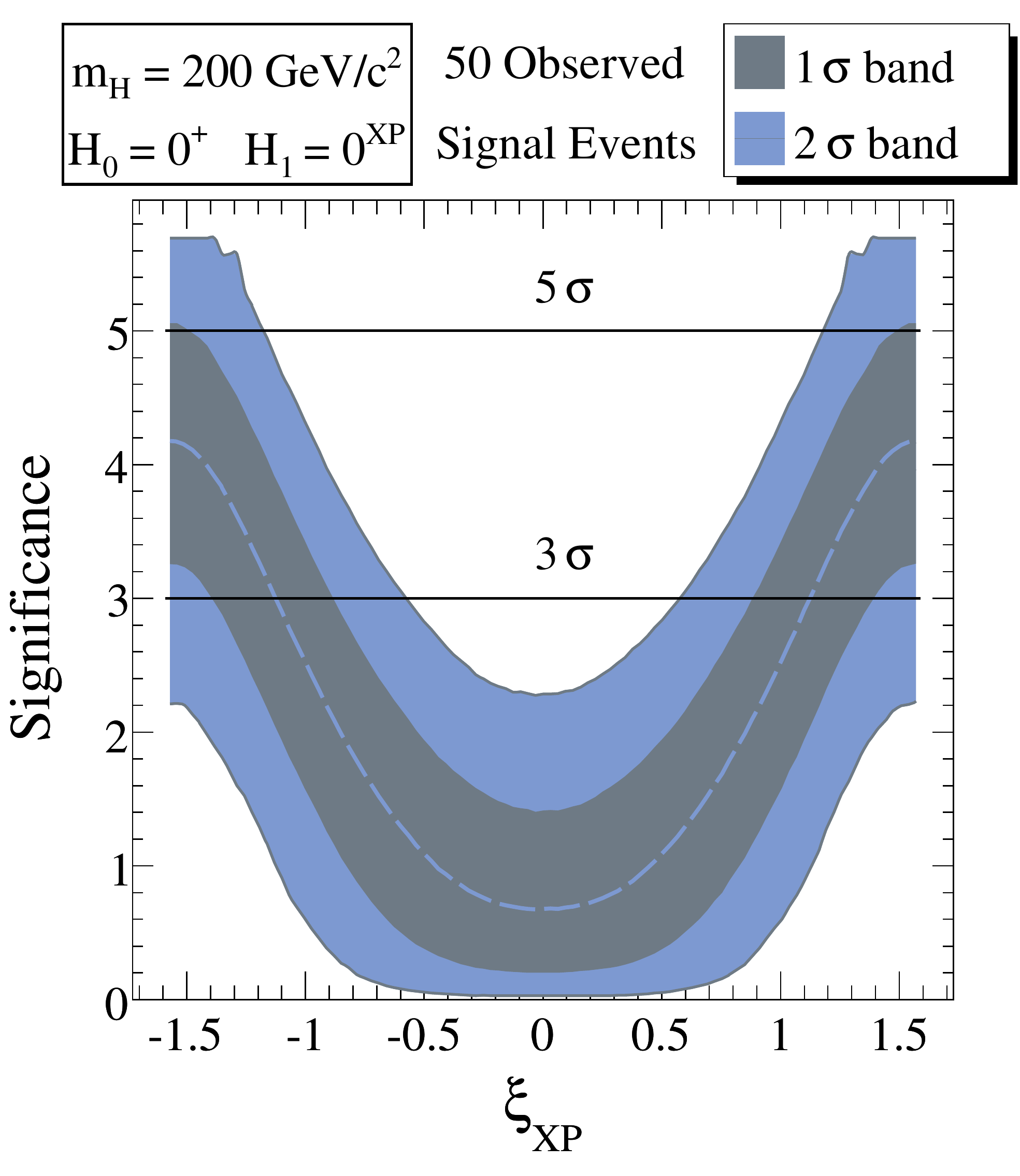}
\caption{Left: Significance for the exclusion of values of a
  $CP$-violating $\xi_{XP}\neq 0$ in favor of $0^+$ ($\xi_{XP}$$=$$0$),
  assumed to be correct.  Right: Significance for excluding a
  pure $0^+$ in favor of $\xi_{XP}\neq 0$, assumed correct with
  $\xi_{XP}$ given by its $x$-axis values. Results for $m_H$$=$$ 145$, 200
  GeV/c$^{2}$ (top, bottom) and $N_S$$=$$50$.
  \label{fig:COMP1D_XP}}
\end{center}
\end{figure}

In Fig.~\ref{fig:COMP1D_XP} we show the results for the distinction
between pure $0^{+}$ and $CP$-violating $J$$=$$0$ hypotheses for $m_H$$=$$145$
and 200 GeV/c$^{2}$. For $m_H$$=$$145$ GeV/c$^{2}$, the ``flat'' behavior
around $\xi_{XP}$$=$$0$ is due to the coupling strength of the $0^{+}$
part relative to $0^{-}$, an order of magnitude larger for $m_H$$=$$145$
GeV/c$^2$ and closer to unity for the higher $m_H$ values. The
corresponding results at $m_H$$=$$350$ GeV/c$^2$ are those of
Figs.~\ref{fig:COMP1D_H0} and \ref{fig:COMP1D_H1}.

The next mixed $J$$=$$0$ case that we consider is that of a $C$-violating scalar, 
with mixing angle $\xi_{XQ}$. This scenario is very
similar to that of the $CP$-violating scalar: only the interference
term between the $0^{+}$ and $0^{-}$ amplitudes is different ($C$-odd, instead of $T$-odd).
\begin{figure}[htbp]
\begin{center}
\includegraphics[width=0.238\textwidth]{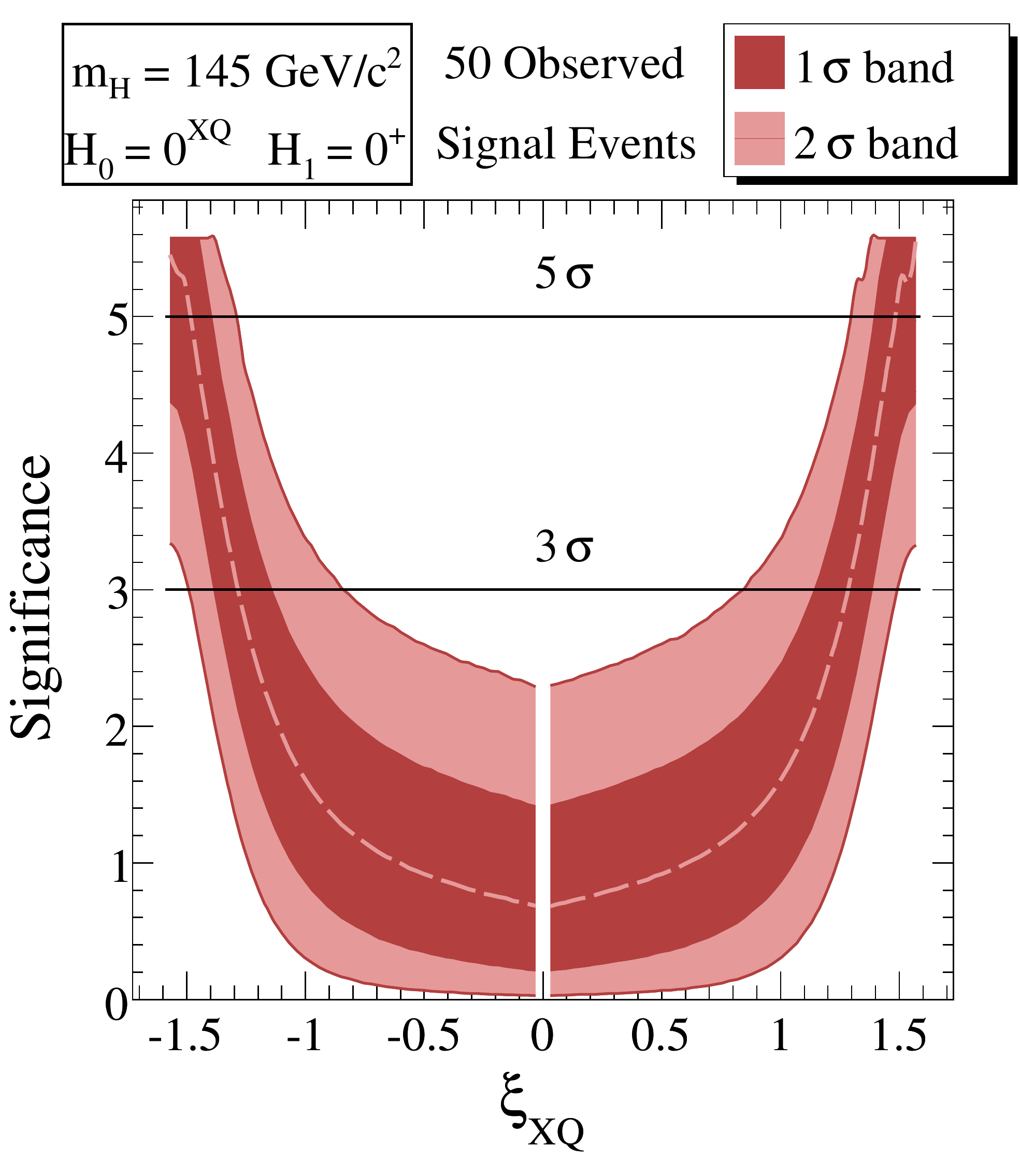}
\includegraphics[width=0.238\textwidth]{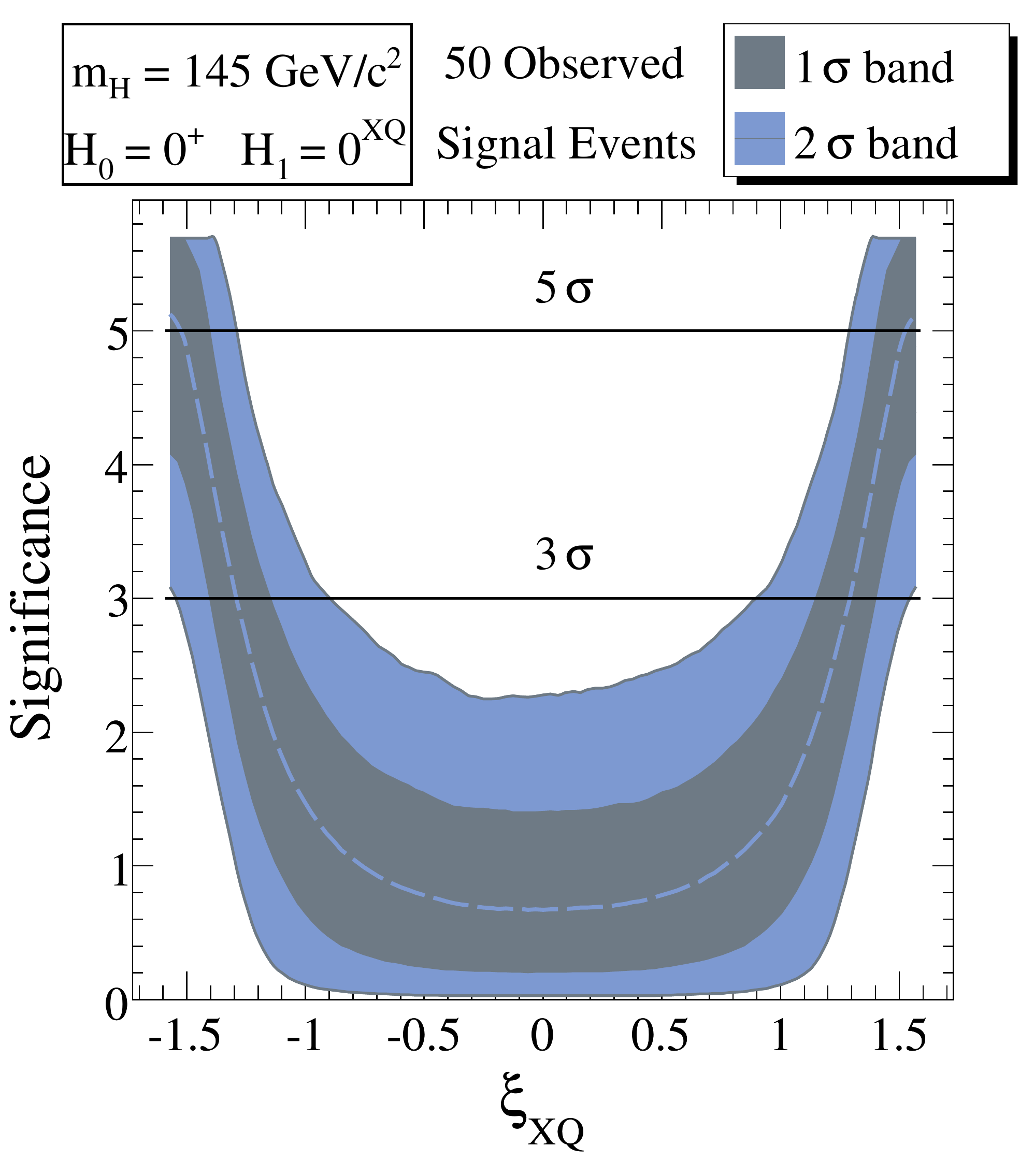}
\includegraphics[width=0.238\textwidth]{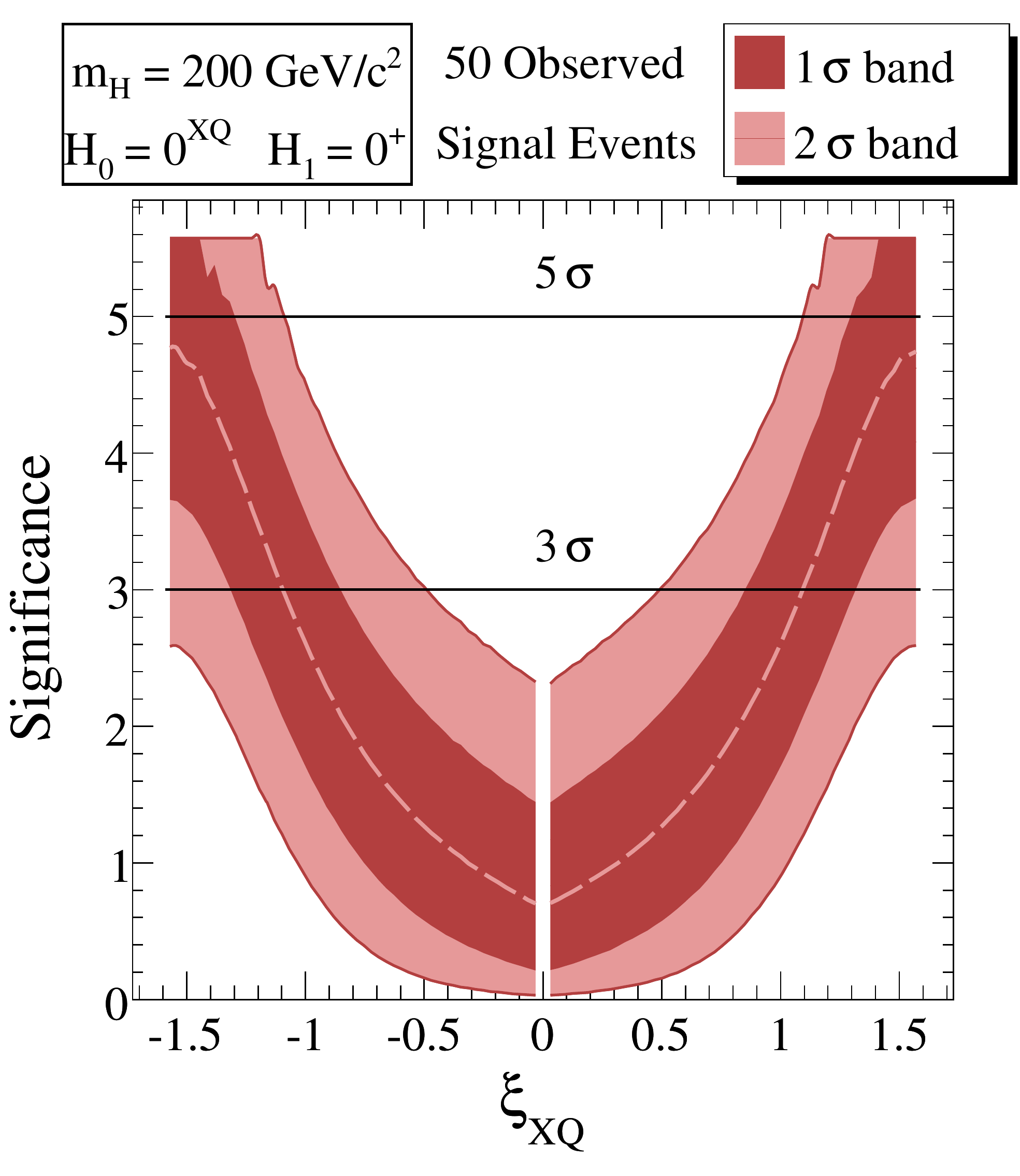}
\includegraphics[width=0.238\textwidth]{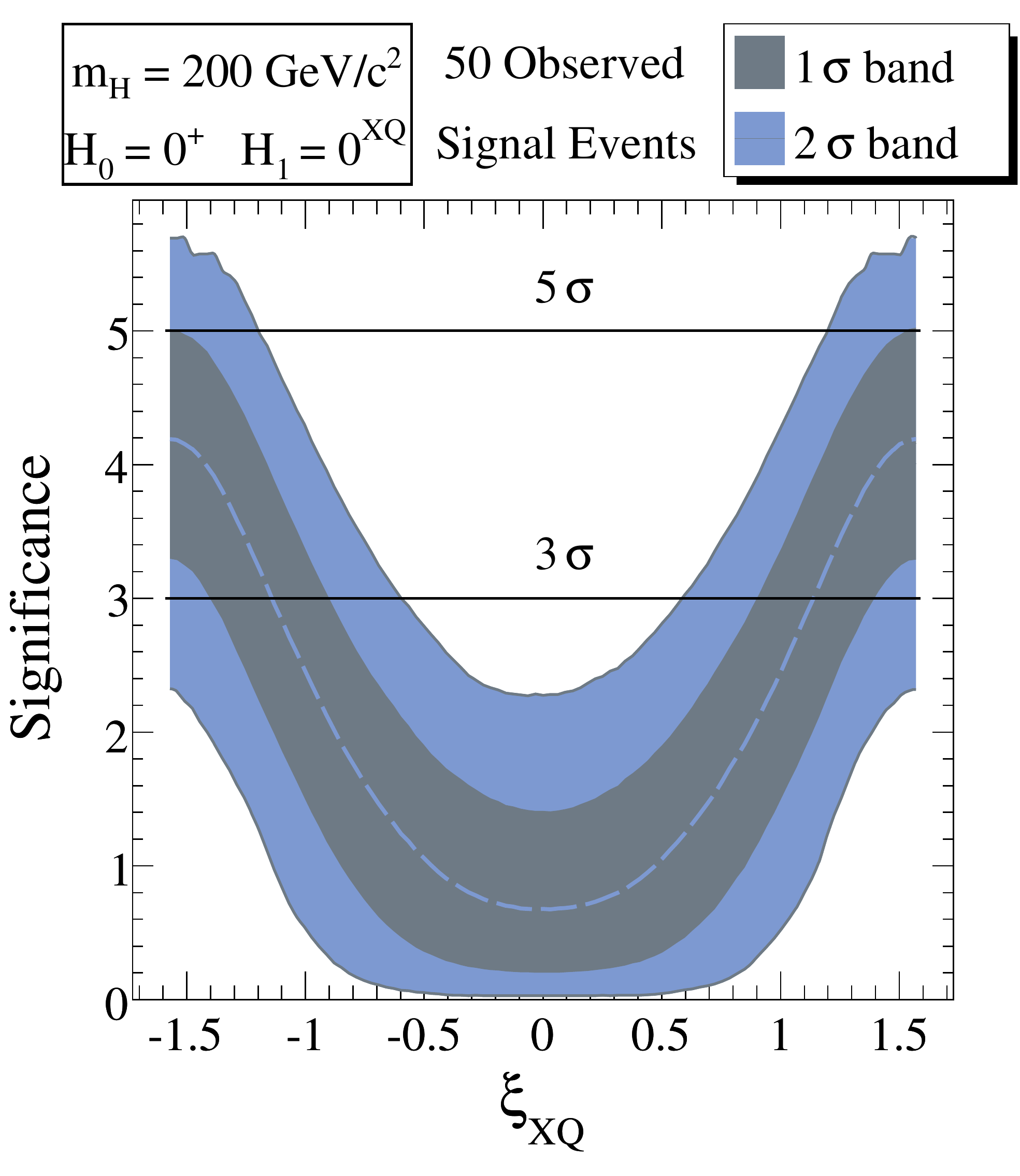}
\includegraphics[width=0.238\textwidth]{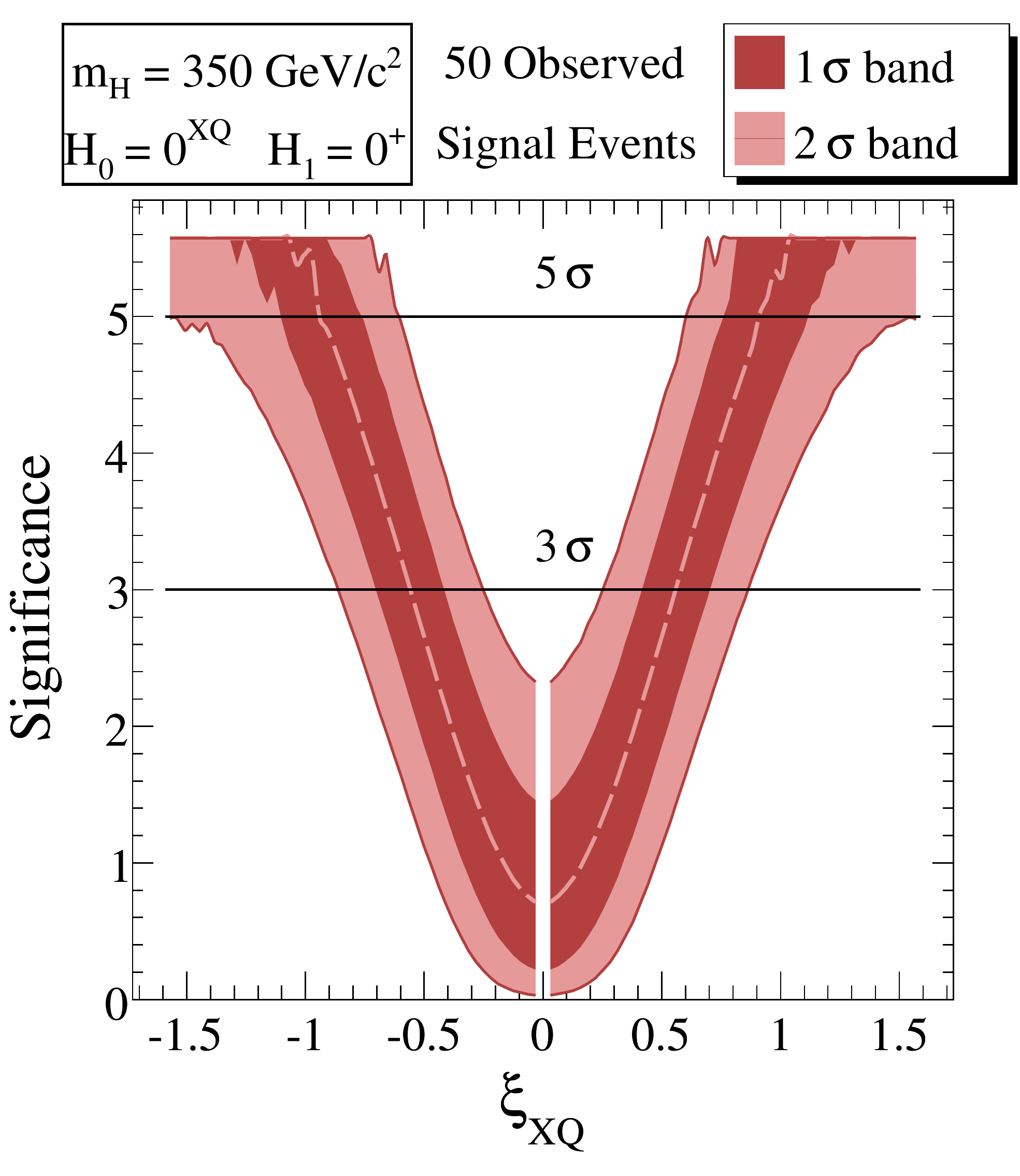}
\includegraphics[width=0.238\textwidth]{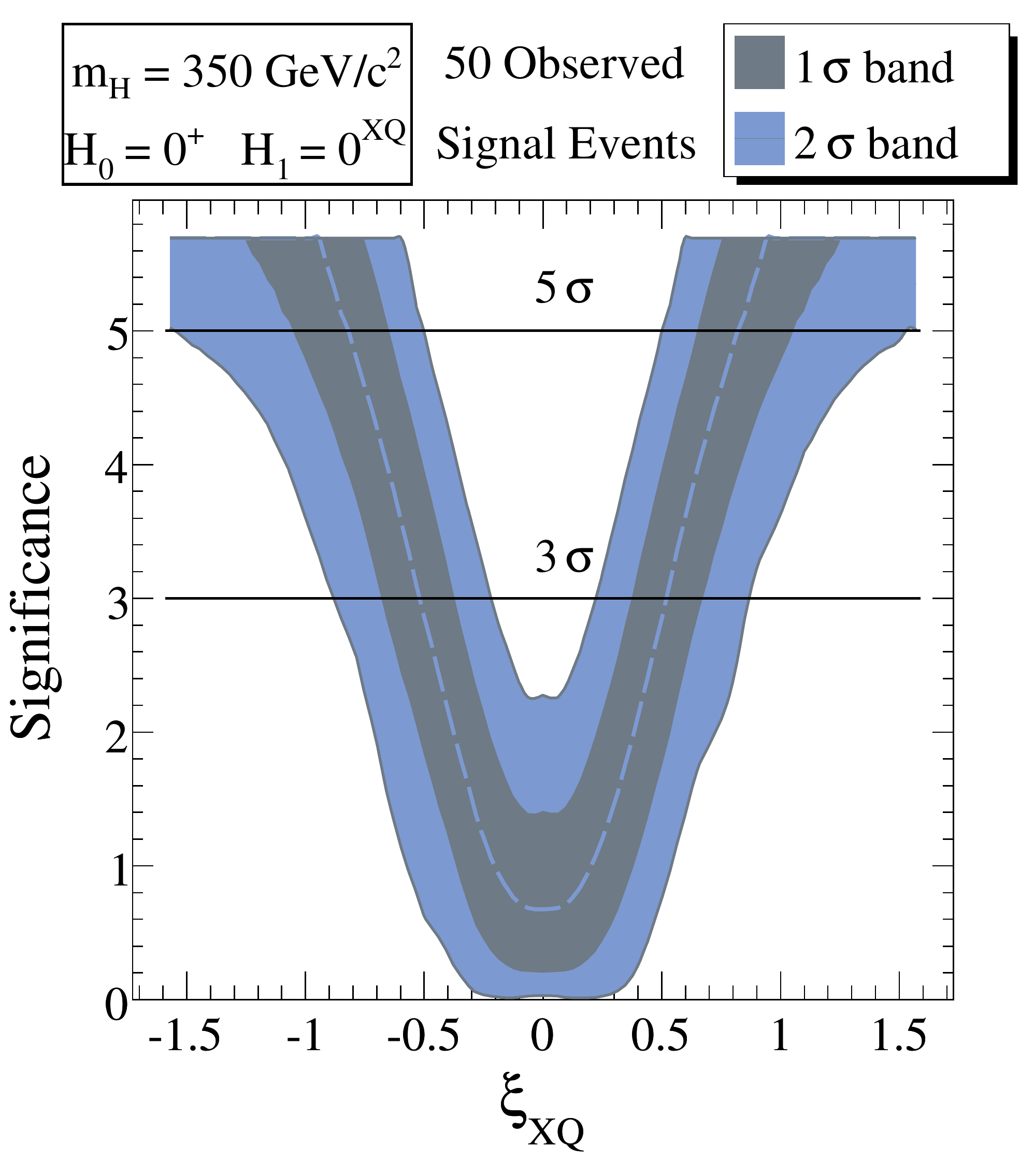}
\caption{Left: Significance for excluding values of a
  $C$-violating $\xi_{XQ}\neq 0$ in favor of $0^+$ ($\xi_{XQ}$$=$$0$),
  assumed to be correct. Right: Significance for excluding a
  pure $0^{+}$ in favor of $\xi_{XQ}\neq 0$, assumed correct for the
  $\xi_{XQ}$-values on the $x$-axis. Hypothesis tests are for
  $m_H$$=$$145$, 200 and 350 GeV/c$^{2}$ (top, middle and bottom), for
  $N_S$$=$$50$.
  \label{fig:COMP1D_XQ}}
\end{center}
\end{figure}

The expected results of hypothesis tests distinguishing between a
$C$-violating scalar and a $0^{+}$ state are shown in
Fig.~\ref{fig:COMP1D_XQ}. Comparing this figure with
Figs.~\ref{fig:COMP1D_H0},~\ref{fig:COMP1D_H1} and
\ref{fig:COMP1D_XP}, we observe identical behavior in all the
results. This shows that the relative strength between the $0^{+}$ and
$0^{-}$ parts of the matrix element squared, rather than the nature of
the interference term, is the most relevant factor in resolving the
values of $\xi_{XP}$ and $\xi_{XQ}$.

\begin{figure}[htbp]
\begin{center}
\includegraphics[width=0.238\textwidth]{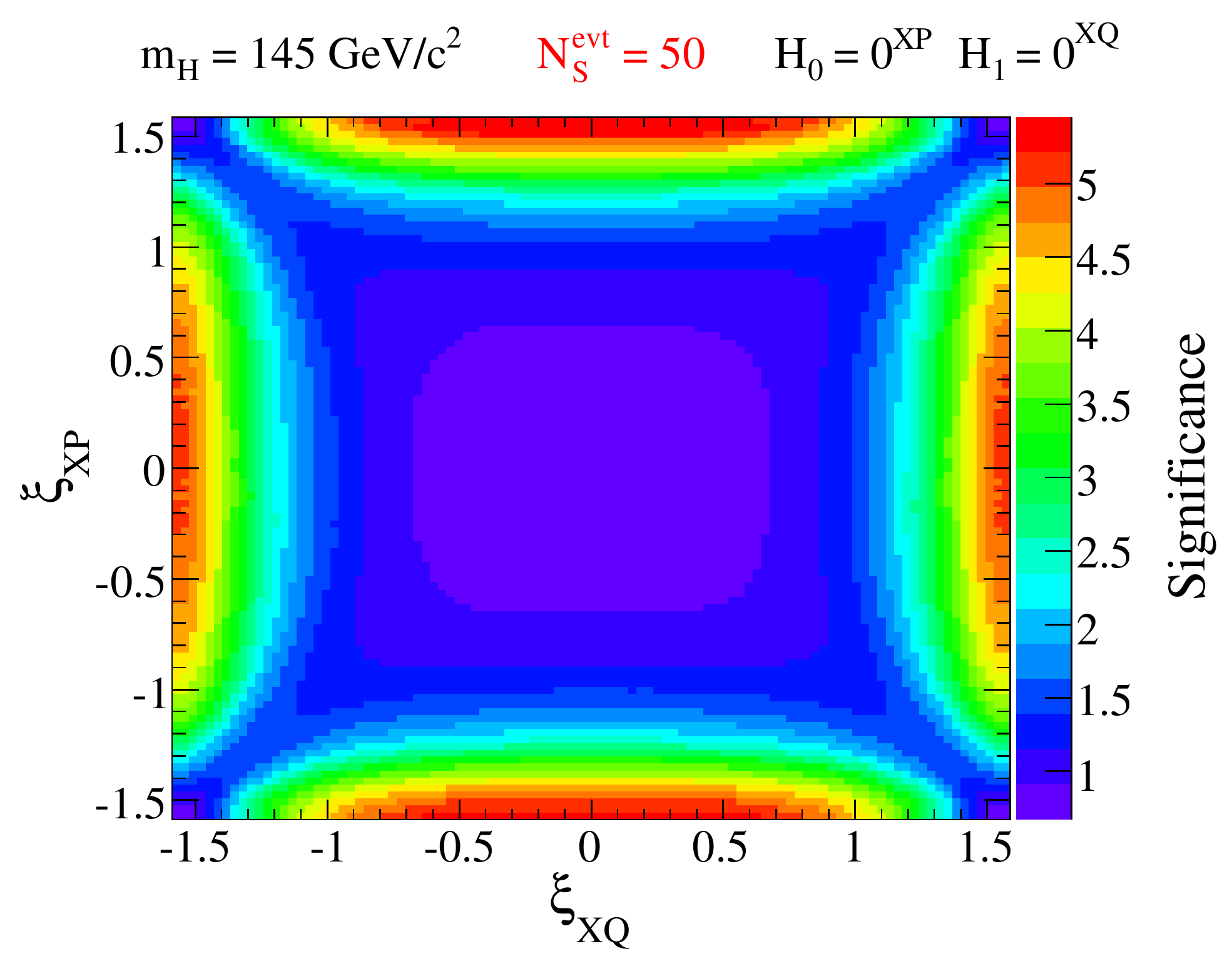}
\includegraphics[width=0.238\textwidth]{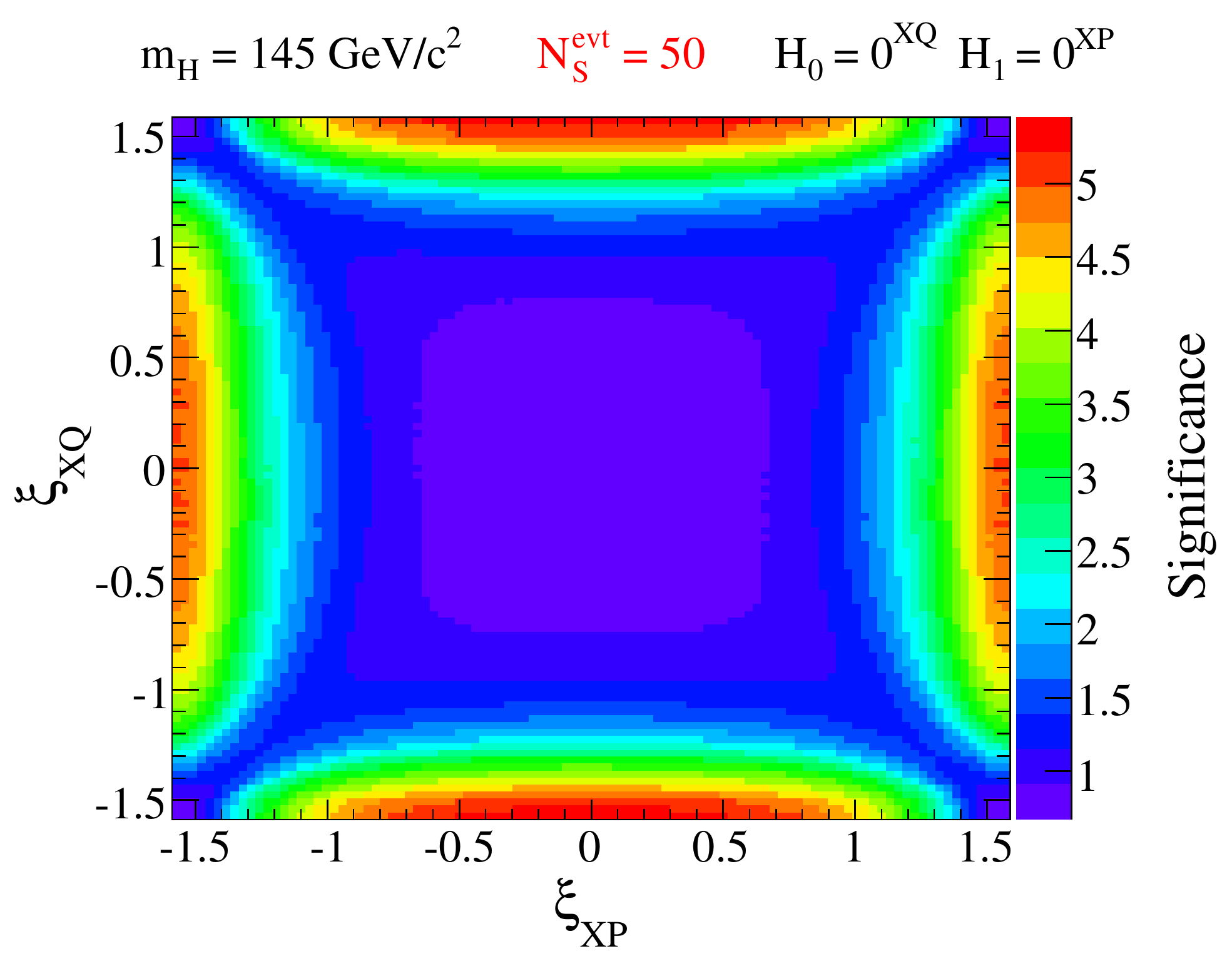}
\includegraphics[width=0.238\textwidth]{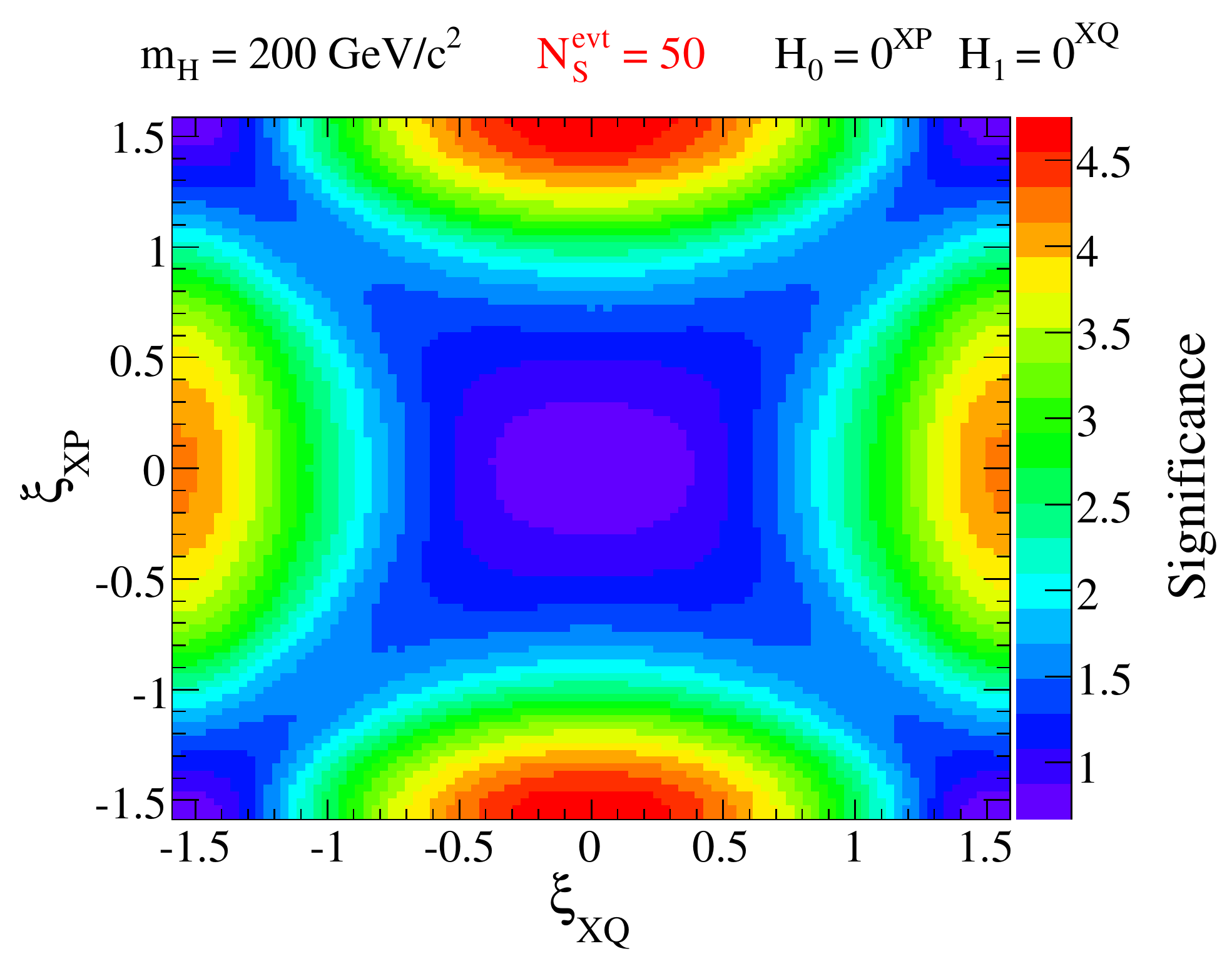}
\includegraphics[width=0.238\textwidth]{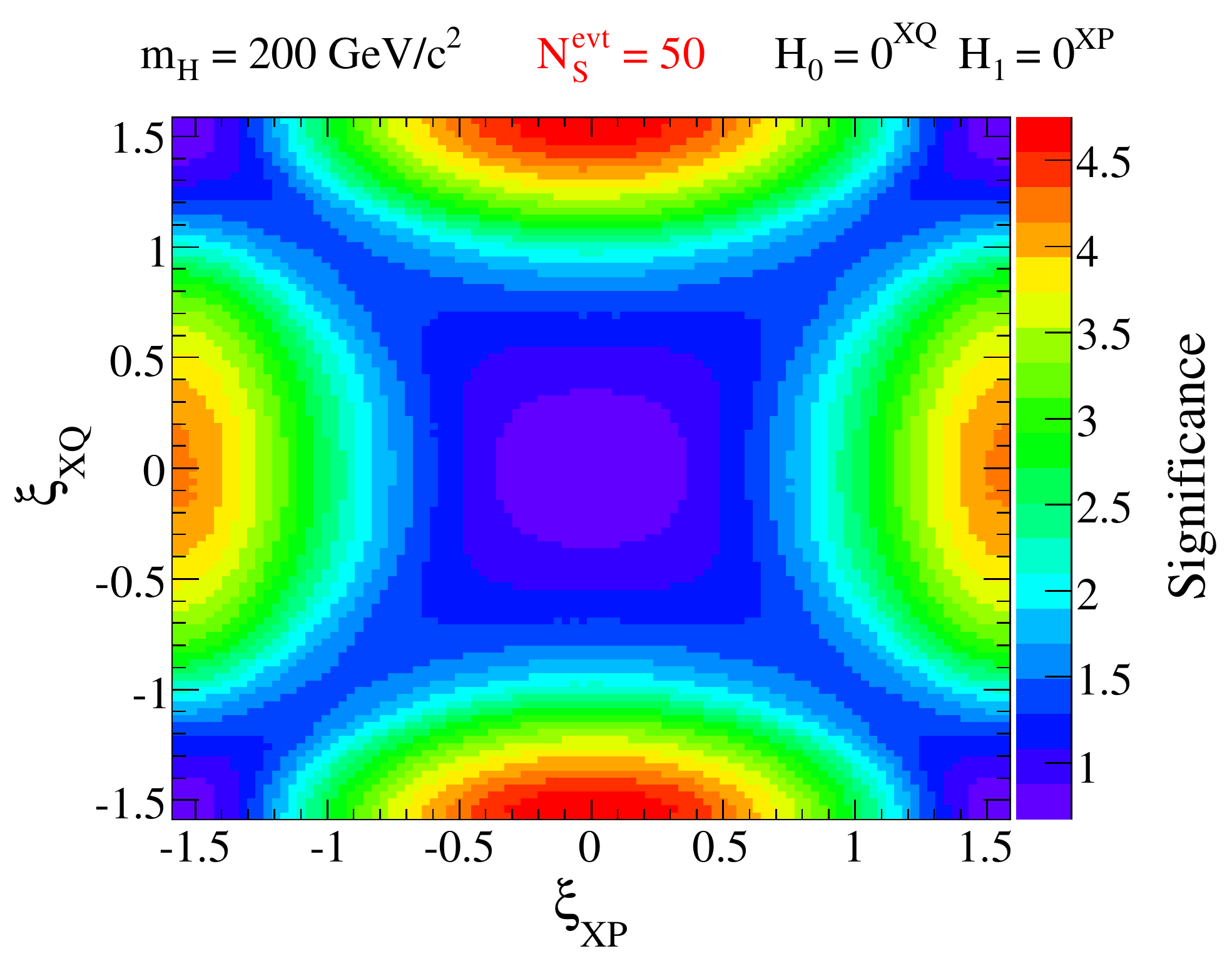}
\includegraphics[width=0.238\textwidth]{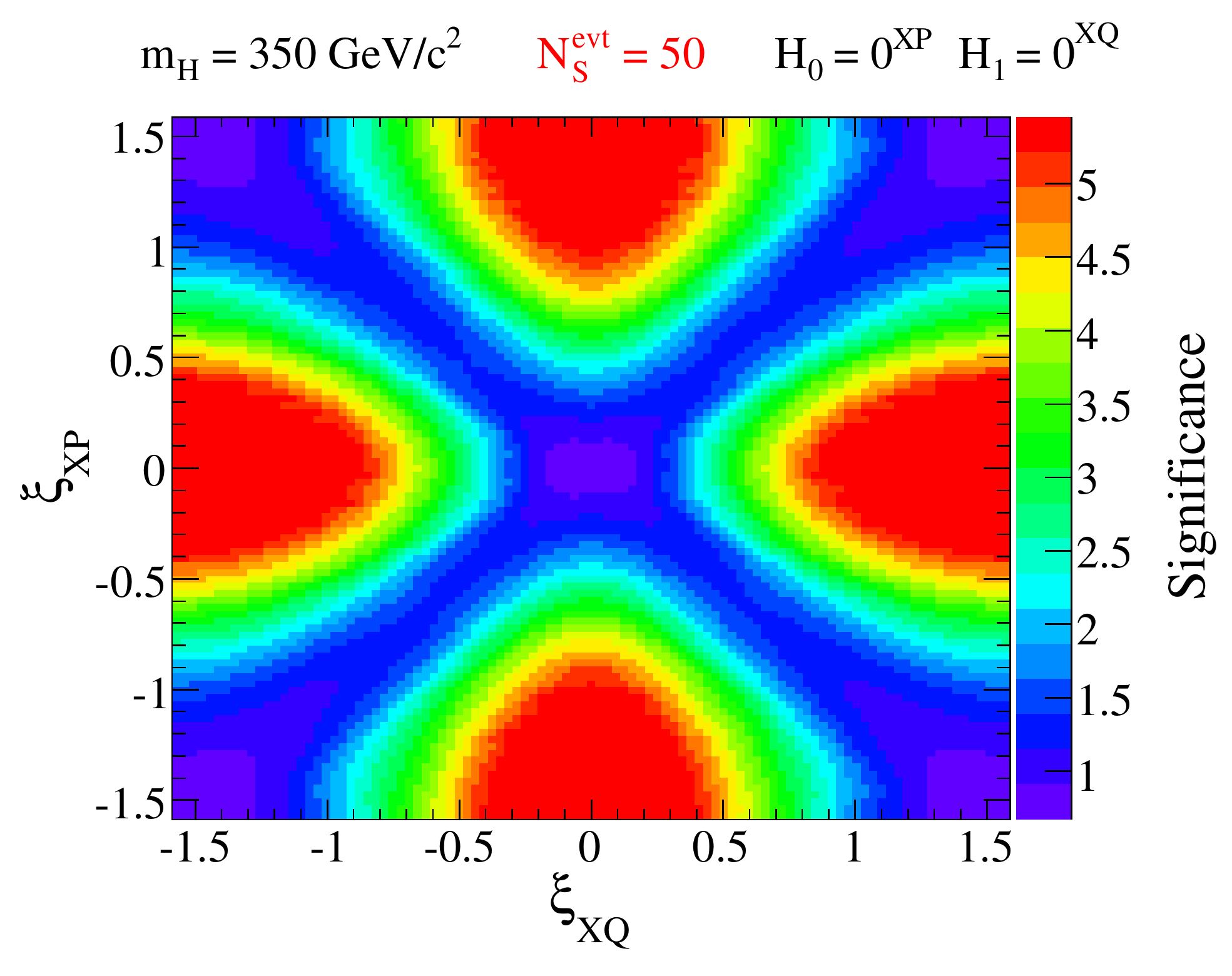}
\includegraphics[width=0.238\textwidth]{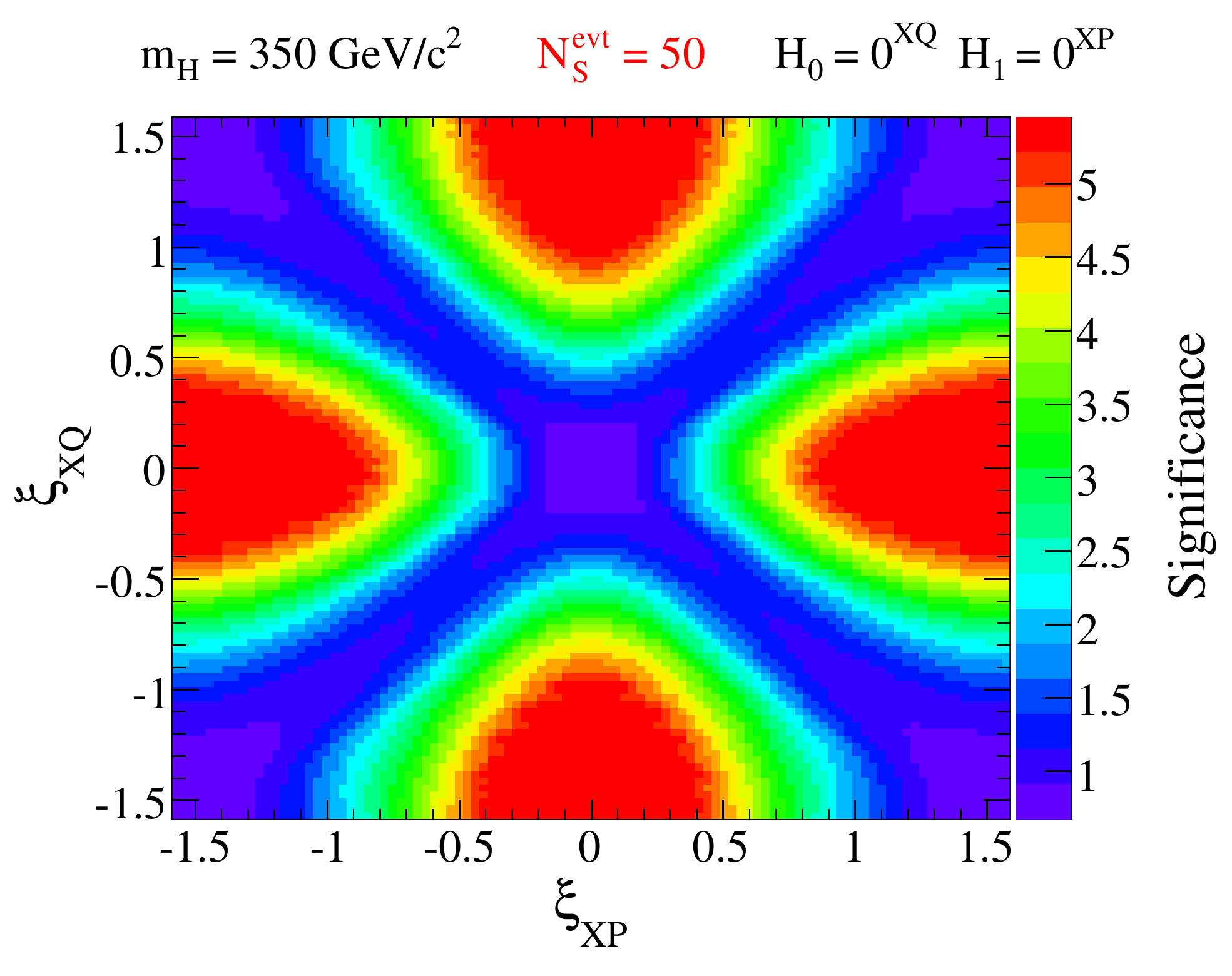}
\caption{The median of the significance (coloured $z$-``axis'') for
  excluding values of $\xi_{\mathbb{H}_{0}}$ ($y$-axis) in favor of
  the $\xi_{\mathbb{H}_{1}} \ne 0$ hypothesis assuming as correct the
  values $\xi_{\mathbb{H}_{1}}$ of the $x$-axis. The tests are
  performed for $\mathbb{H}_{1}$$=$$0^{XP}$, $\mathbb{H}_{0}$$=$$0^{XQ}$
  (left) and $\mathbb{H}_{1}$$=$$0^{XQ}$, $\mathbb{H}_{0}$$=$$0^{XP}$
  (right); $m_H$$=$$145$, 200 and 350 GeV/c$^{2}$ (top, middle and
  bottom), for $N_S$$=$$50$.
\label{fig:COMP2D_XP_v_XQ}}
\end{center}
\end{figure}

If a pure $0^{+}$ hypothesis is rejected in favor of both $\xi_{XP}
\ne 0$ and $\xi_{XQ} \ne 0$, the next question would be whether it is
possible to distinguish between these two cases. To address this
question, we perform a series of hypothesis tests similar to the one
described to answer type (b) questions. Specifically, we first assume
a given $CP$-violating $\xi_{XP}\ne 0$ as ``true''. We then assess the
expected significance with which particular values of $\xi_{XQ}$ can
be excluded in favor of the true hypothesis. Hence, for each fixed
value of $\xi_{XP}$ we perform a test against the $C$-violating case
using a fixed $\xi_{XQ}$. The test statistic is $\Lambda= \log [{\rm
  max}\,{ L}^{XP}(\hat\xi_{XP})/{ L}(\xi_{XQ})]$, where the $0^{XQ}$
hypothesis is simple (fixed $\xi_{XQ}$) and $L(\xi_{XP})$ is {\it
  profiled} ``experiment-by-experiment''. The test is repeated over a
matrix of values for $\xi_{XP}$ and $\xi_{XQ}$. Next, we switch the
roles of the hypotheses to assess the significance for excluding given
values of $\xi_{XP}$ in favor of $\xi_{XQ} \ne 0$. The results are
shown in Fig.~\ref{fig:COMP2D_XP_v_XQ}. The color-coded $z$-``axis''
is the median of the significance for ruling out the hypothesis
$\mathbb{H}_{0}$ with the value of $\xi_{\mathbb{H}_{0}}$ given on the
$y$-axis in favor of the $\mathbb{H}_{1}$ hypothesis with
$\xi_{\mathbb{H}_{1}}\ne 0$, assumed to be correct for
$\xi_{\mathbb{H}_{1}}$-values chosen on the $x$-axis.

\begin{figure}[htbp]
\begin{center}
\includegraphics[width=0.38\textwidth]{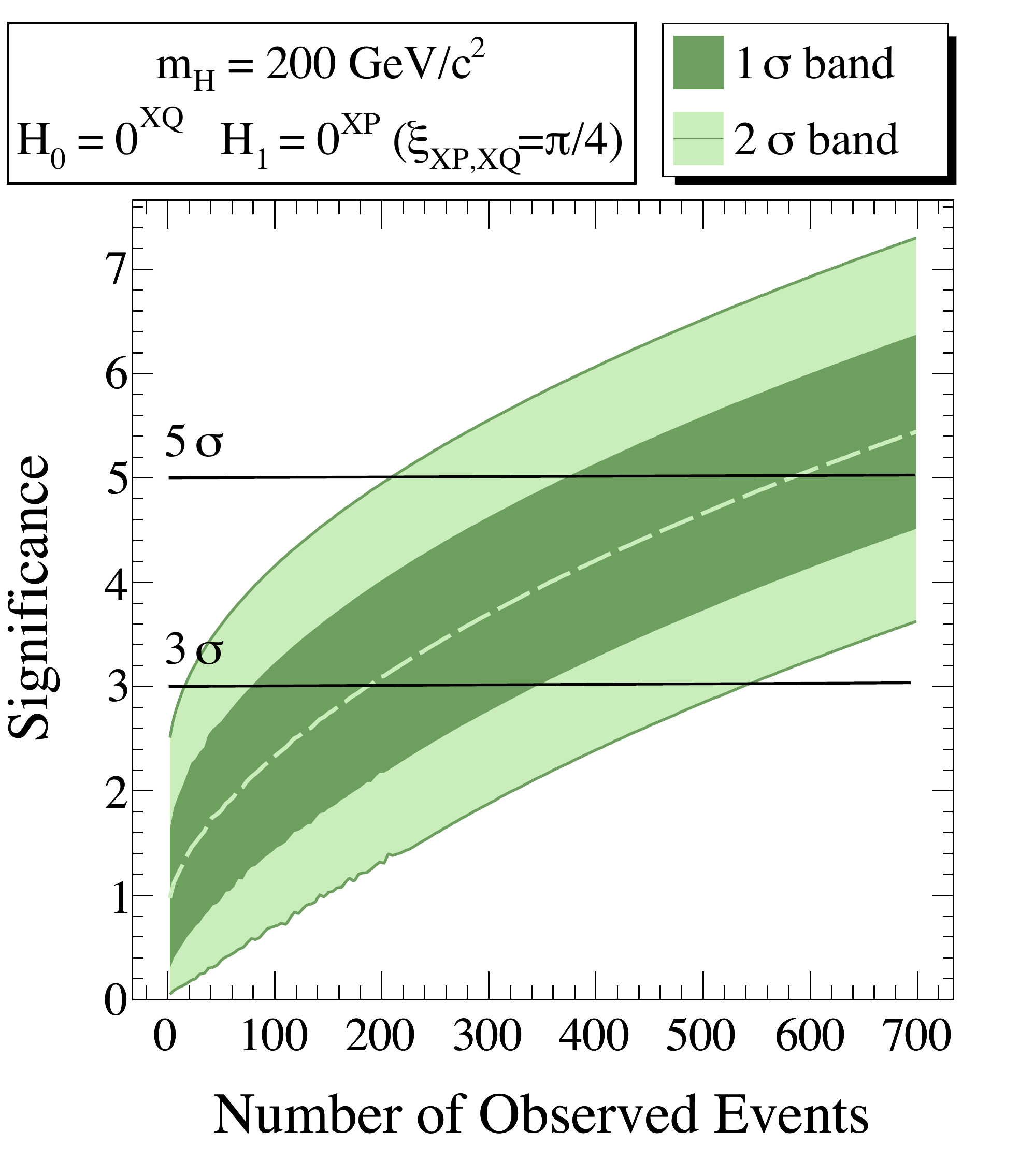}
\caption{The  significance for
  excluding the $C$-violating $J$$=$$0$ hypothesis in favor of a $CP$-violating case,
  assuming the latter to be correct, with $\xi_{XP,XQ}$$=$$\pi/4$. Example for $m_H$$=$$200$ GeV/c$^2$.
  \label{fig:COMP_CP_v_C}}
\end{center}
\end{figure}

The similarities between the $C$- and $CP$- mixed scalars are
reflected in the $y\!\leftrightarrow\! x$ symmetries of
Figs.~\ref{fig:COMP2D_XP_v_XQ}. Moreover, switching the roles of the
two hypotheses (comparing the figures on the left with those on the
right) one only sees small changes. Still, the fact that the diagonals
($|\xi_{XP}|=|\xi_{XQ}|$) are not all at the same significance shows
that the tests are sensitive to the differences between the $\tilde{T}$- and
$C$-odd interference terms, but it would require an order of magnitude
larger $N_S$ to draw $5\,\sigma$-level conclusions over most of the
$(\xi_{XP},\xi_{XQ})$ plane. For example, we show in Fig.~\ref{fig:COMP_CP_v_C} the  significance with which one can distinguish between the two cases, as a function of the number of observed events, for $\xi_{XY,XQ}$$=$$\pi/4$ and $m_{H}$$=$$200$ GeV/$c^{2}$. The ambiguity between $\xi_{XP}^{meas}$,
$-\xi_{XP}^{meas}$, $\xi_{XQ}$$=$$\xi_{XP}^{meas}$ and
$\xi_{XQ}=-\xi_{XP}^{meas}$ would be very hard to lift.

\begin{figure}[htbp]
\begin{center}
\includegraphics[width=0.38\textwidth]{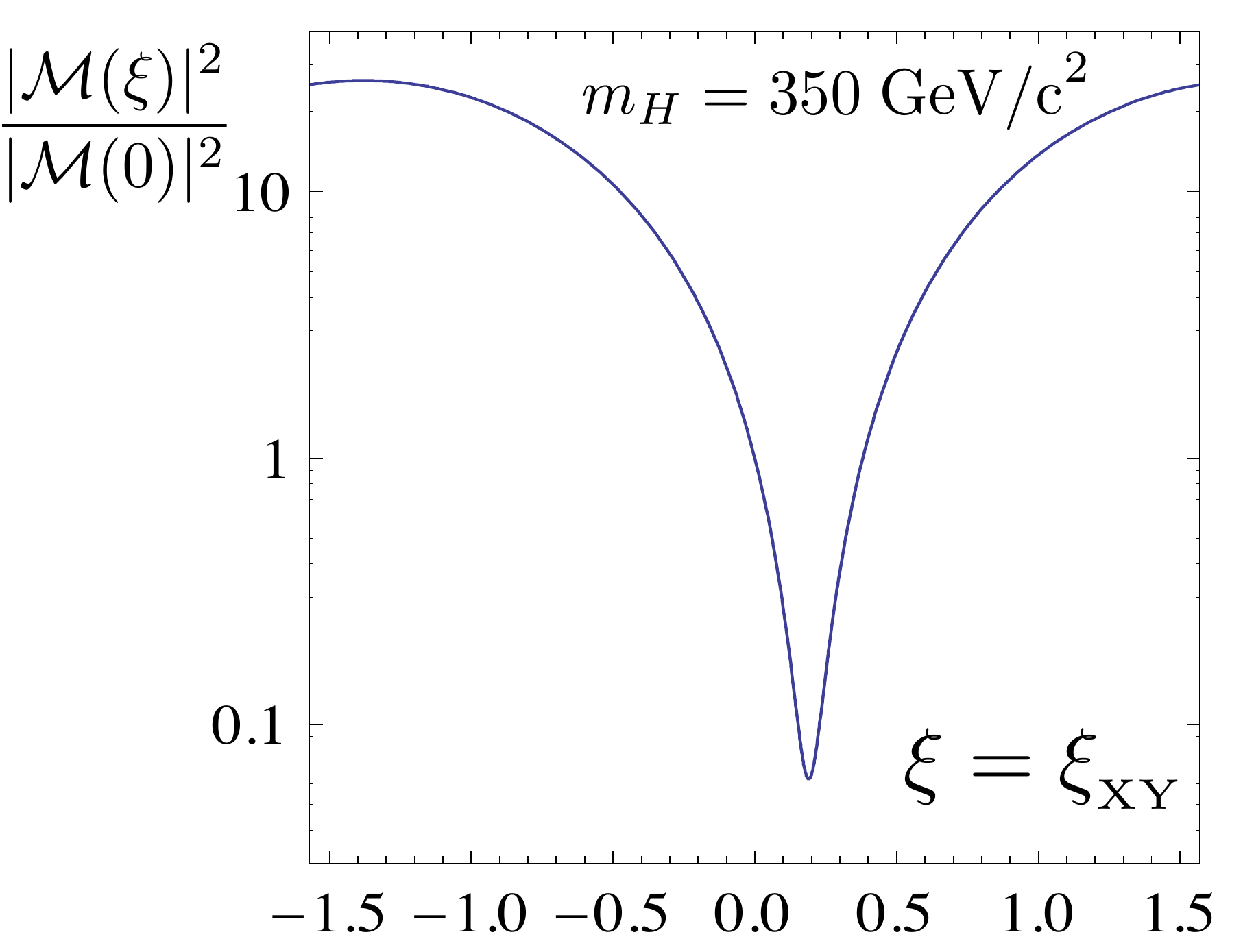}
\caption{The fully angularly-integrated matrix element squared
 for a ``composite'' $0^+$, showing a strong
destructive interference at a given $\xi_{XY}$. The result, shown here for $m_H$$=$$350$ GeV/c$^2$,
is normalized to $\xi_{XY}$$=$$0$. 
\label{fig:DerivativeZero350}}
\end{center}
\end{figure}

The last $J$$=$$0$ mixed case that we consider has unique features; this
is the ``composite Higgs'' in which a term $\propto k_\mu k_\nu$ is
present in the $HZZ$ coupling.  This case is different from the
previous ones in that a composite scalar has well defined
$J^{PC}$$=$$0^{++}$, regardless of the value of the angle $\xi_{XY}$
characterizing the mixing between its pointlike and derivative
couplings.  As a consequence, the angular integrals of their
interference term do not vanish, and there is no symmetry around
$\xi_{XY}$$=$$0$. All the terms in the {\it pdf} having the same discrete
symmetries and similar angular dependences; there happen to be large
cancellations in the {\it pdf} for a `critical' $m_H$-dependent value
of $\xi_{XY}$, as in the example shown in
Fig.~\ref{fig:DerivativeZero350} for the fully angular-integrated
result.

The appearance of an order of magnitude enhancement of the squared
matrix element in Fig.~\ref{fig:DerivativeZero350} for $O(1)$ values
of $\xi_{XY}$ can be regarded as an artifact of our choosing a rather
low mass scale ($M_Z$) in the definition of the dimensionless coupling
$Y$ in Eq.~(\ref{generalscalar}); if e.g.~we instead chose the
compositeness scale at $m_H$$=$$350$ GeV/c$^2$, this enhancement would
be much smaller.  Nevertheless the possible enhancement from a nonzero
$Y$ coupling, and the possible suppression from $XY$ interference,
signifies an interesting scenario: it is possible to discover an HLL
that is in fact a $0^{++}$ resonance, and is produced by exactly the
same $pp$ production processes as a SM Higgs, but for which the cross
section times branching fraction to $ZZ$ is several times higher {\it
  or} several times lower than Standard Model expectation.

\begin{figure}[htbp]
\begin{center}
\includegraphics[width=0.238\textwidth]{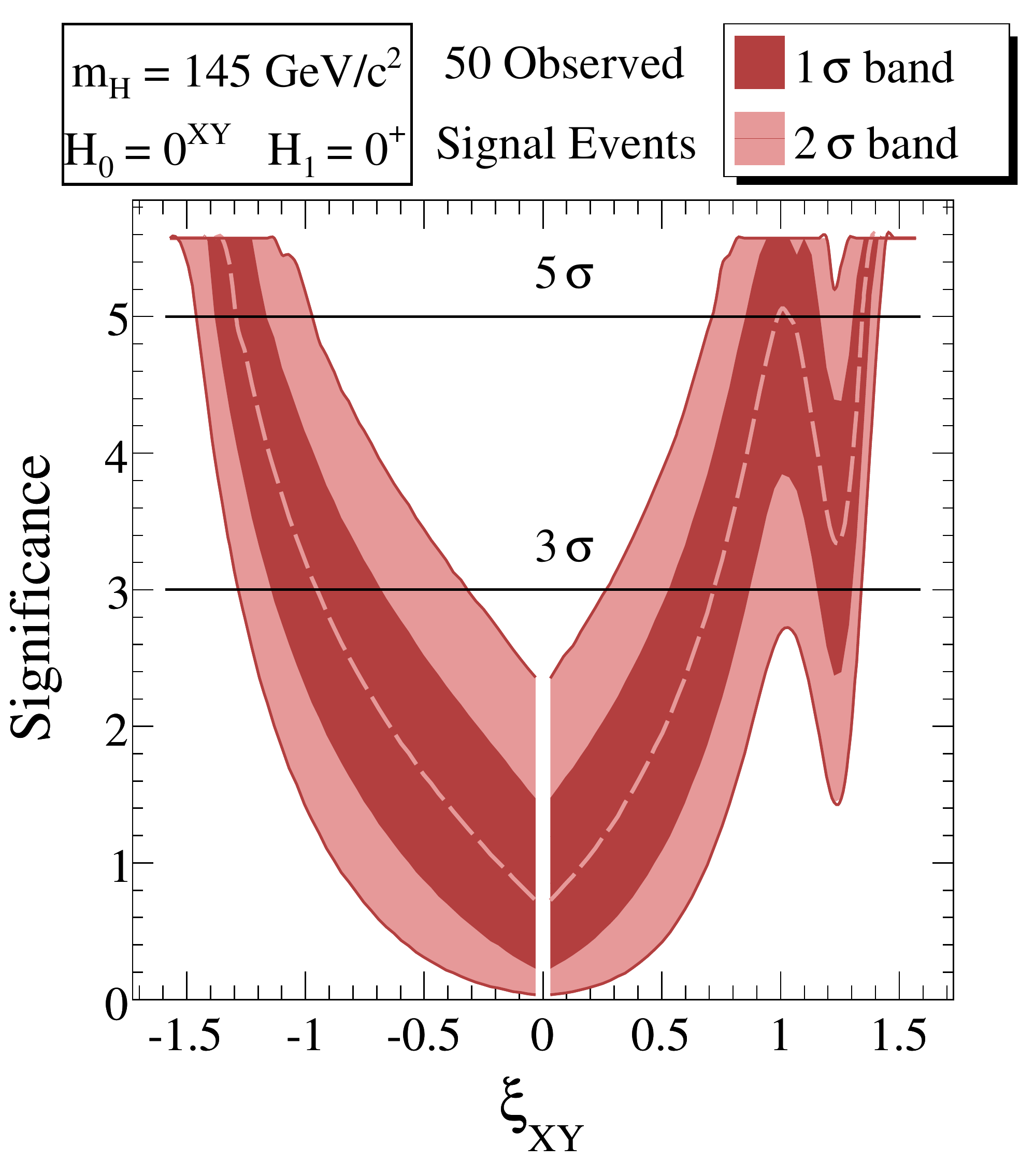}
\includegraphics[width=0.238\textwidth]{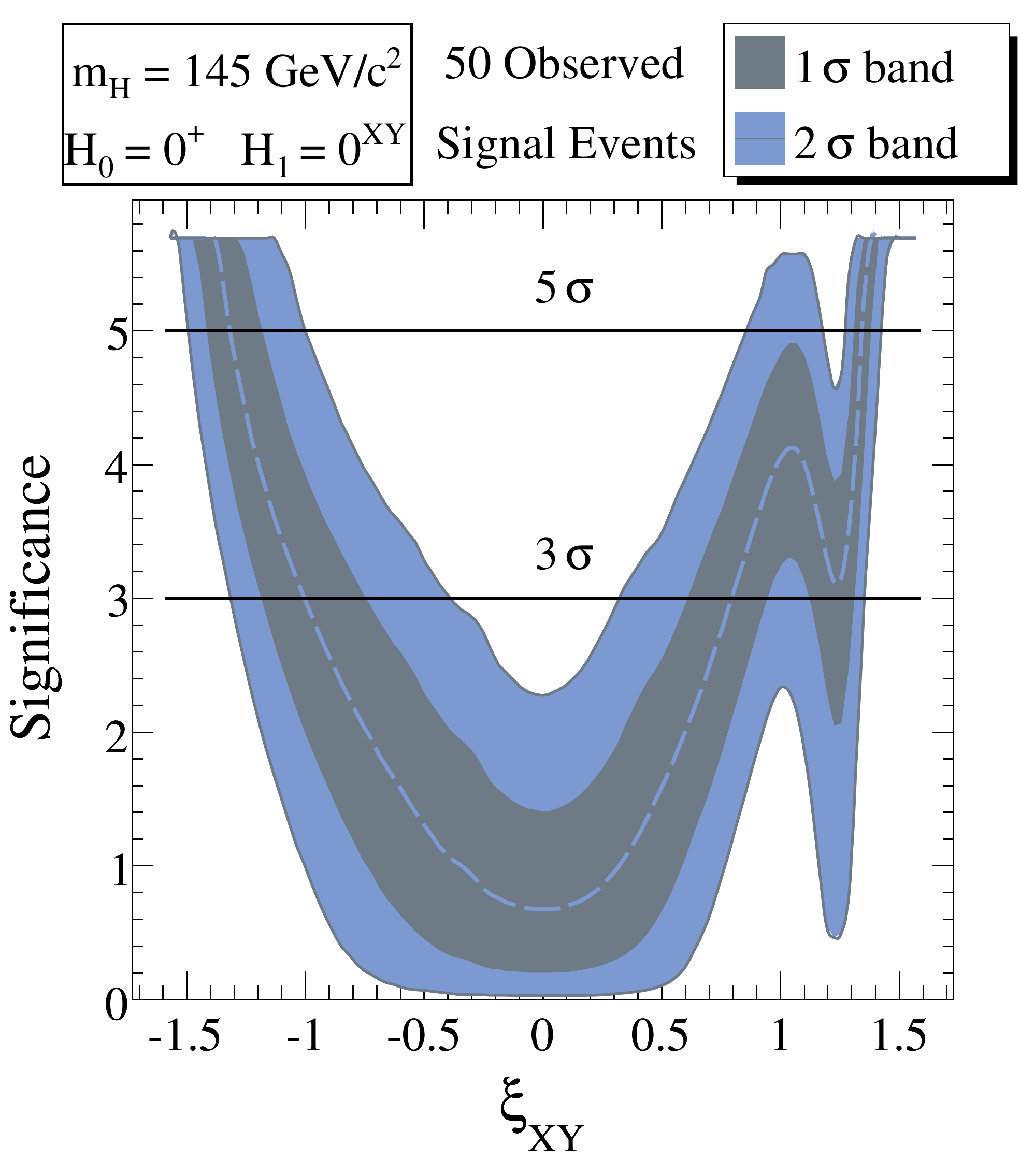}
\includegraphics[width=0.238\textwidth]{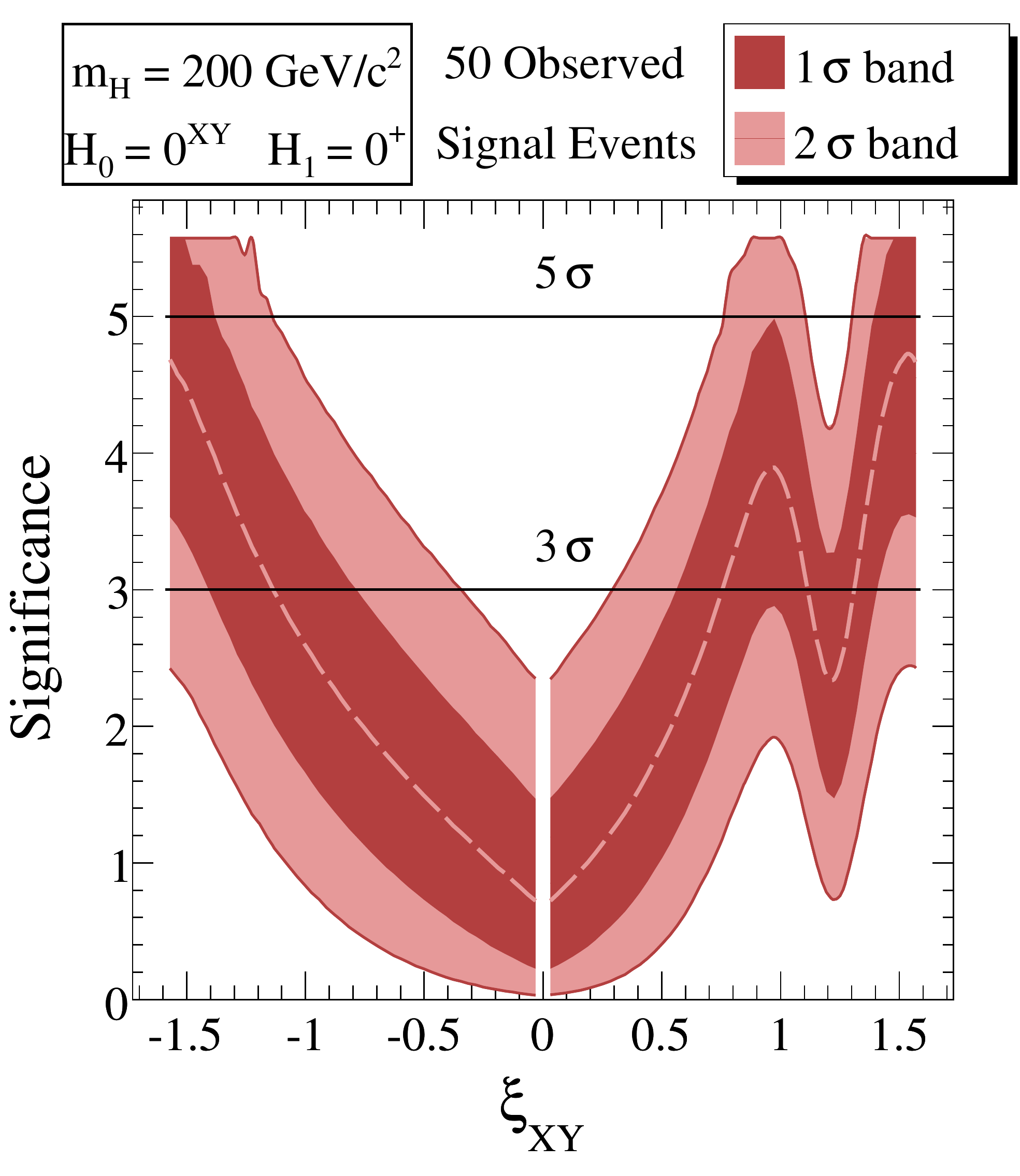}
\includegraphics[width=0.238\textwidth]{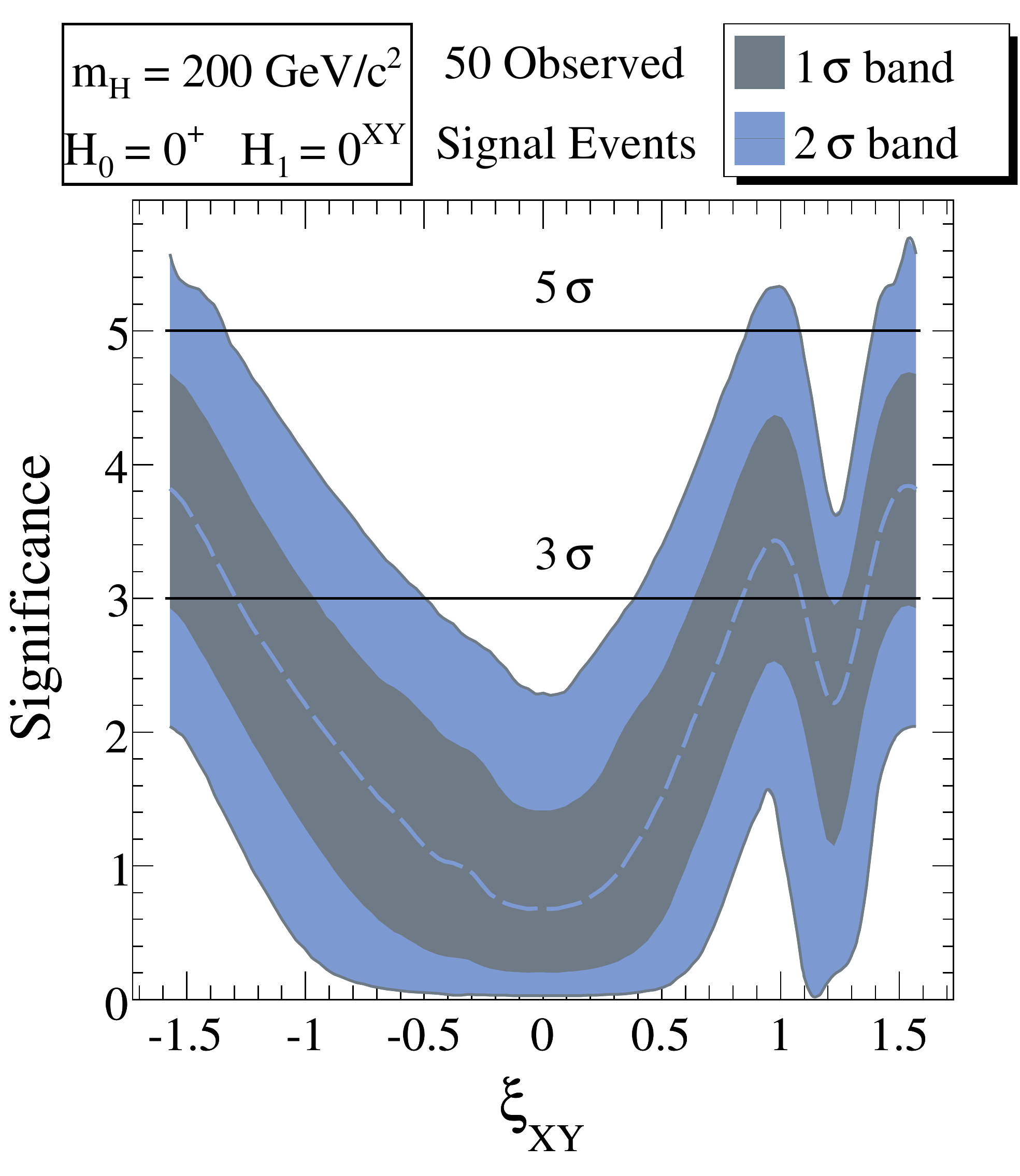}
\includegraphics[width=0.238\textwidth]{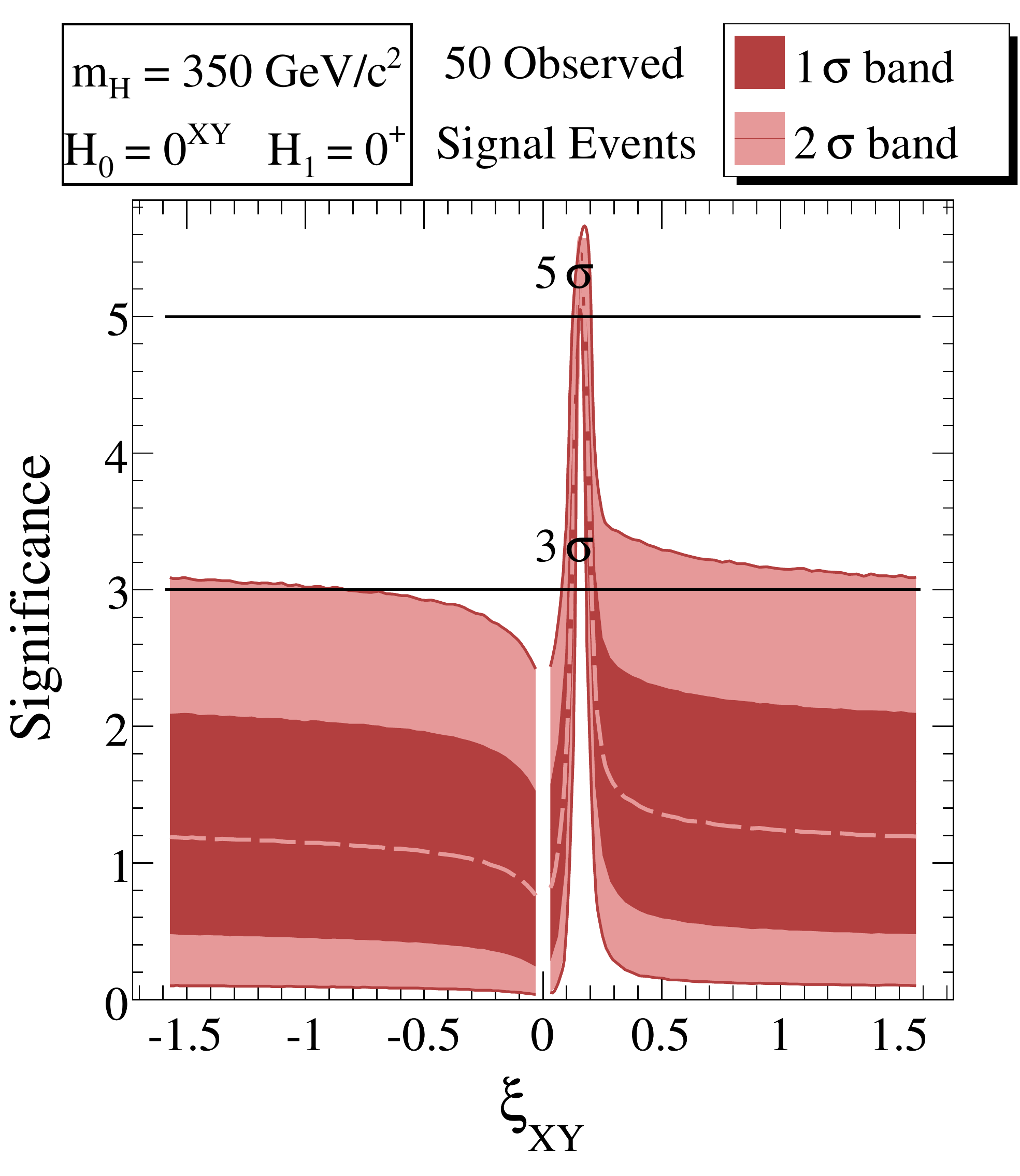}
\includegraphics[width=0.238\textwidth]{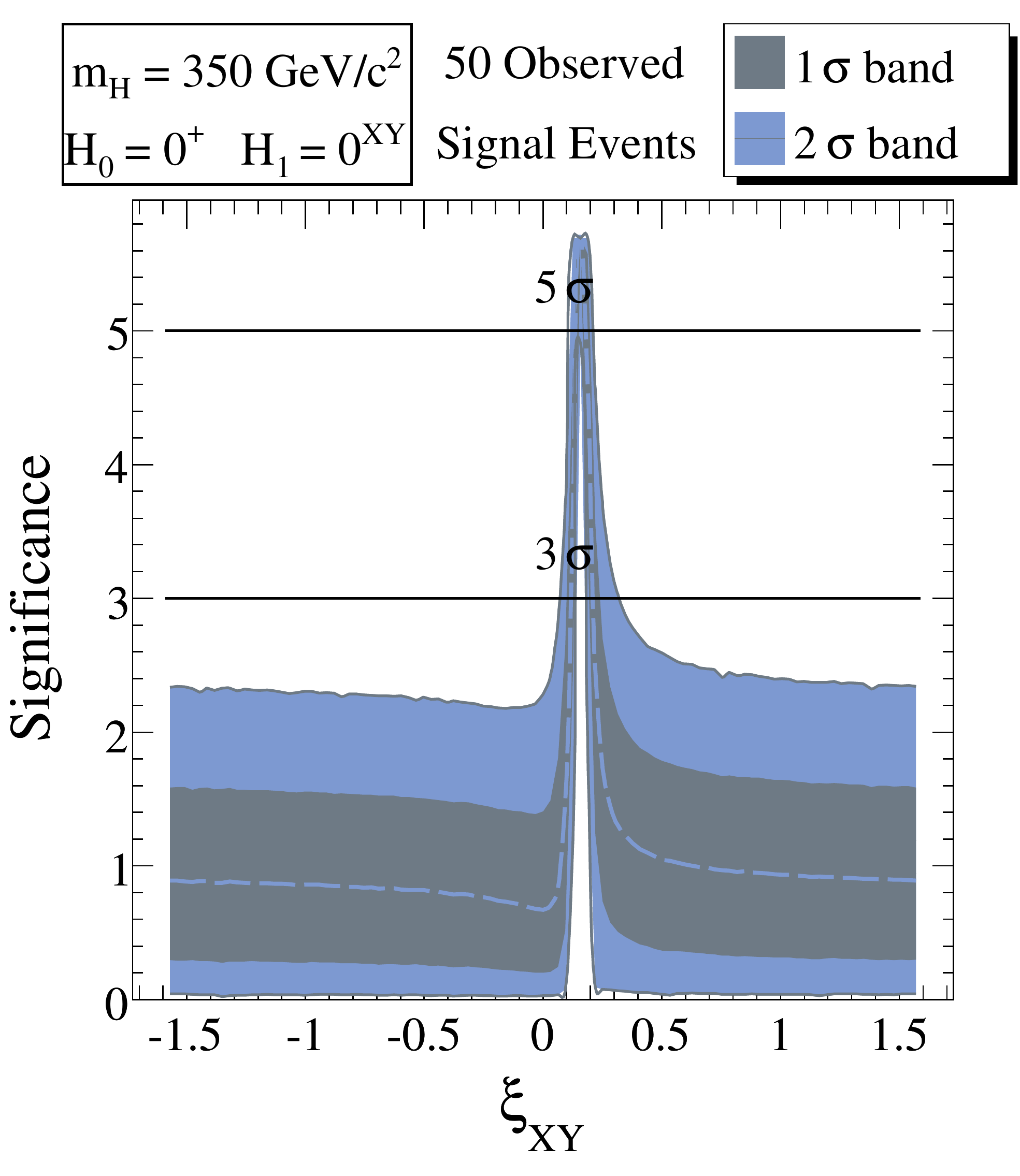}
\caption{Left:  significance for
  excluding values of $\xi_{XY}$ 
  in favor of a pointlike $0^+$ ($\xi_{XY}$$=$$0$), assumed to be correct. 
  Right:  significance for
  excluding a pointlike $0^+$ in favor of a ``composite" one ($\xi_{XY}\ne 0$), assumed correct for the $\xi_{XY}$ values on the $x$-axis, 
  for $m_H$$=$$145$, 200 and 350 GeV/c$^{2}$ (top, middle and bottom) and $N_S$$=$$50$.
  \label{fig:COMP1D_XY}}
\end{center}
\end{figure}

We evaluate the  significance with which one can distinguish
between a pointlike and a composite $0^{+}$ using the same
hypothesis-test approach described earlier for the $CP$-violating
scalar case. The results are shown in Fig.~\ref{fig:COMP1D_XY}.  We
observe a non-trivial behavior of the significance values at and
around the critical $\xi_{XY}$. Interestingly, the qualitative nature
of these cancellations also changes with mass. For $m_H$$=$$145$ GeV/c$^2$
and $m_H$$=$$200$ GeV/c$^2$, the composite scalar with $\xi_{XY}$ near the
critical point is $0^+$-like, relative to nearby values of
$\xi_{XY}$. For $m_H$$=$$350$ GeV/c$^{2}$, it is very difficult to
distinguish between the composite and elementary hypotheses, except if
$\xi_{XY}$ is close to critical. Near this critical value the
significance is greatly improved, because after the large
cancellations the angular distributions of the pure $0^+$ and the
mixed case no longer resemble each other.

\begin{figure}[htbp]
\begin{center}
\includegraphics[width=0.3\textwidth]{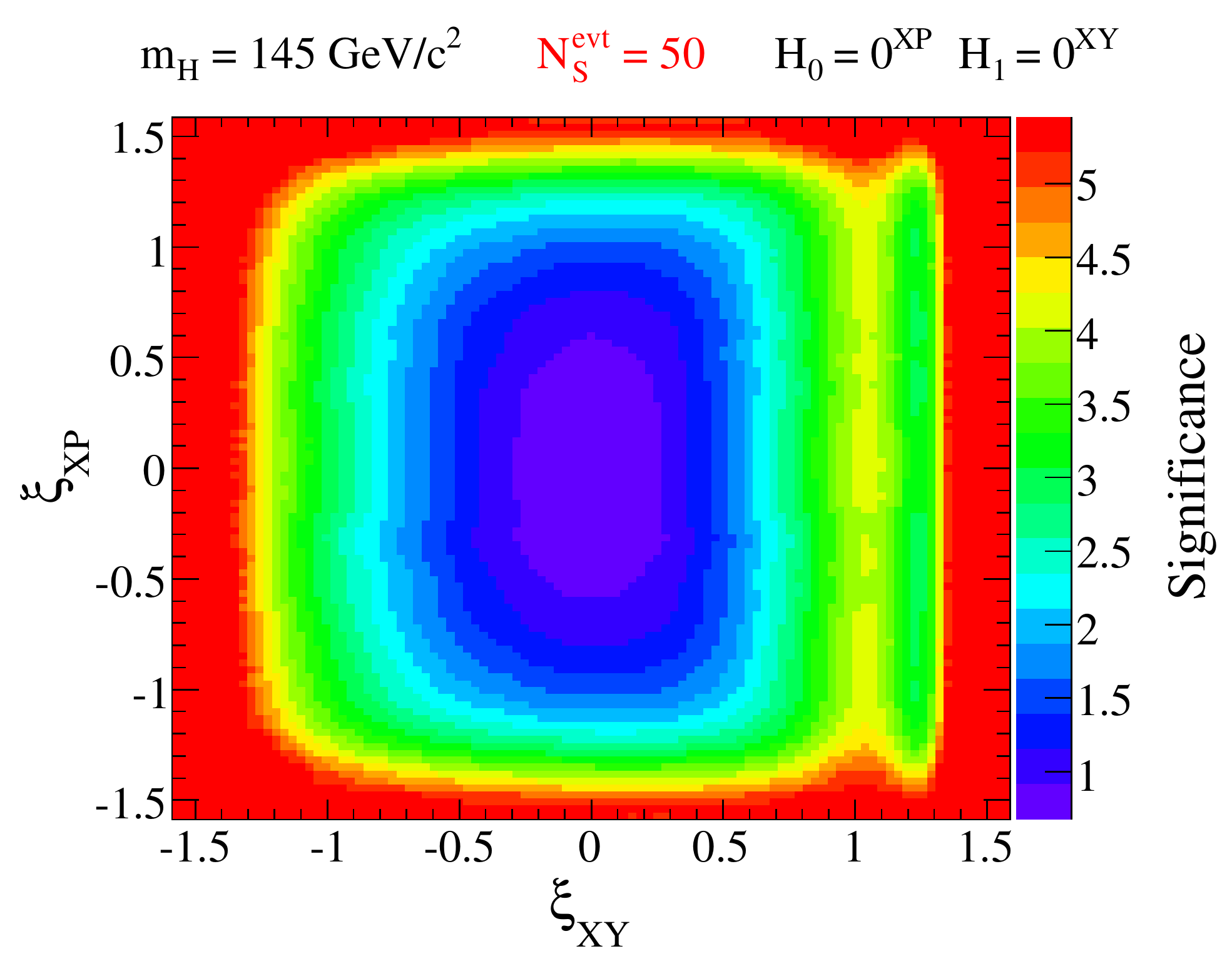}
\includegraphics[width=0.3\textwidth]{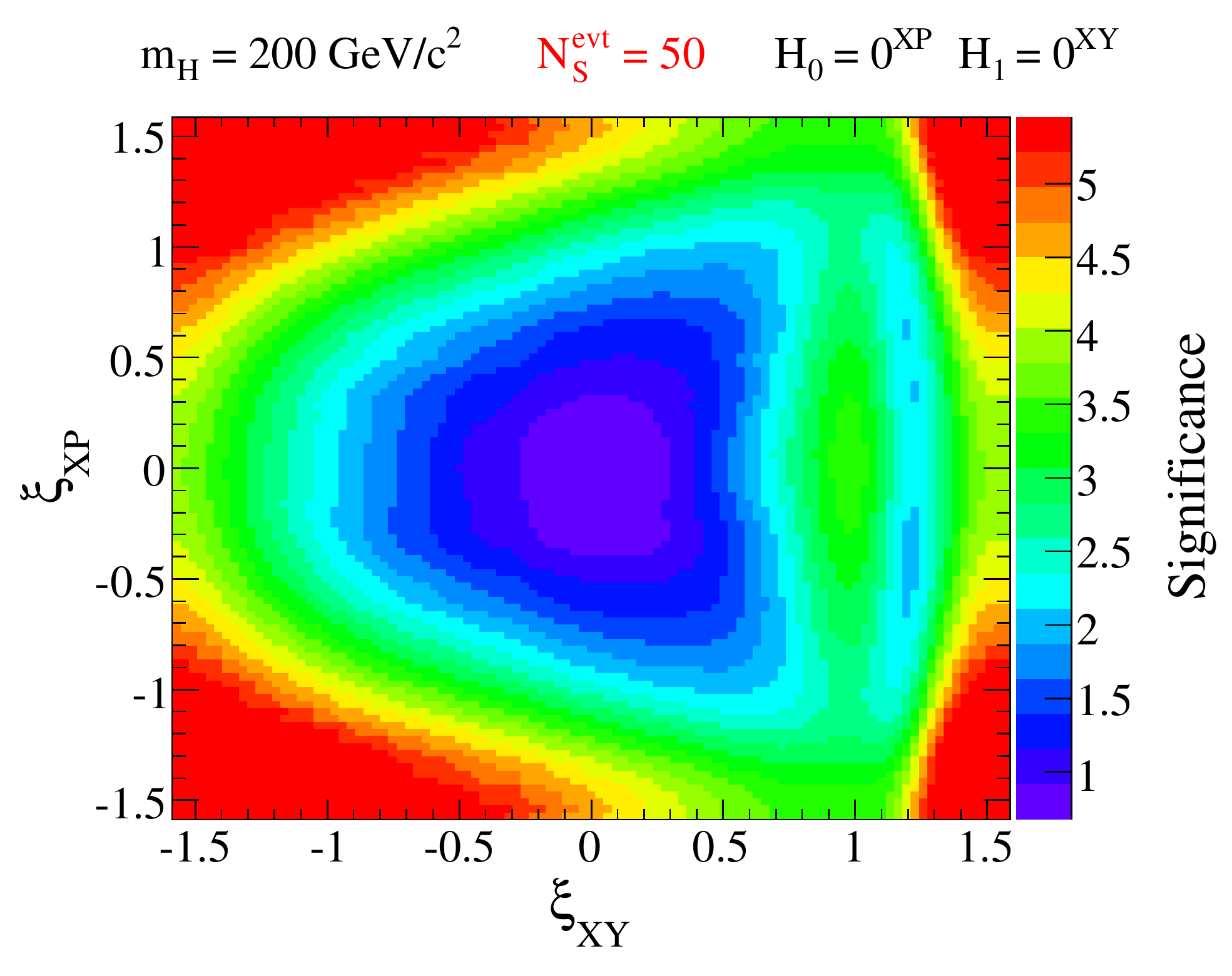}
\includegraphics[width=0.3\textwidth]{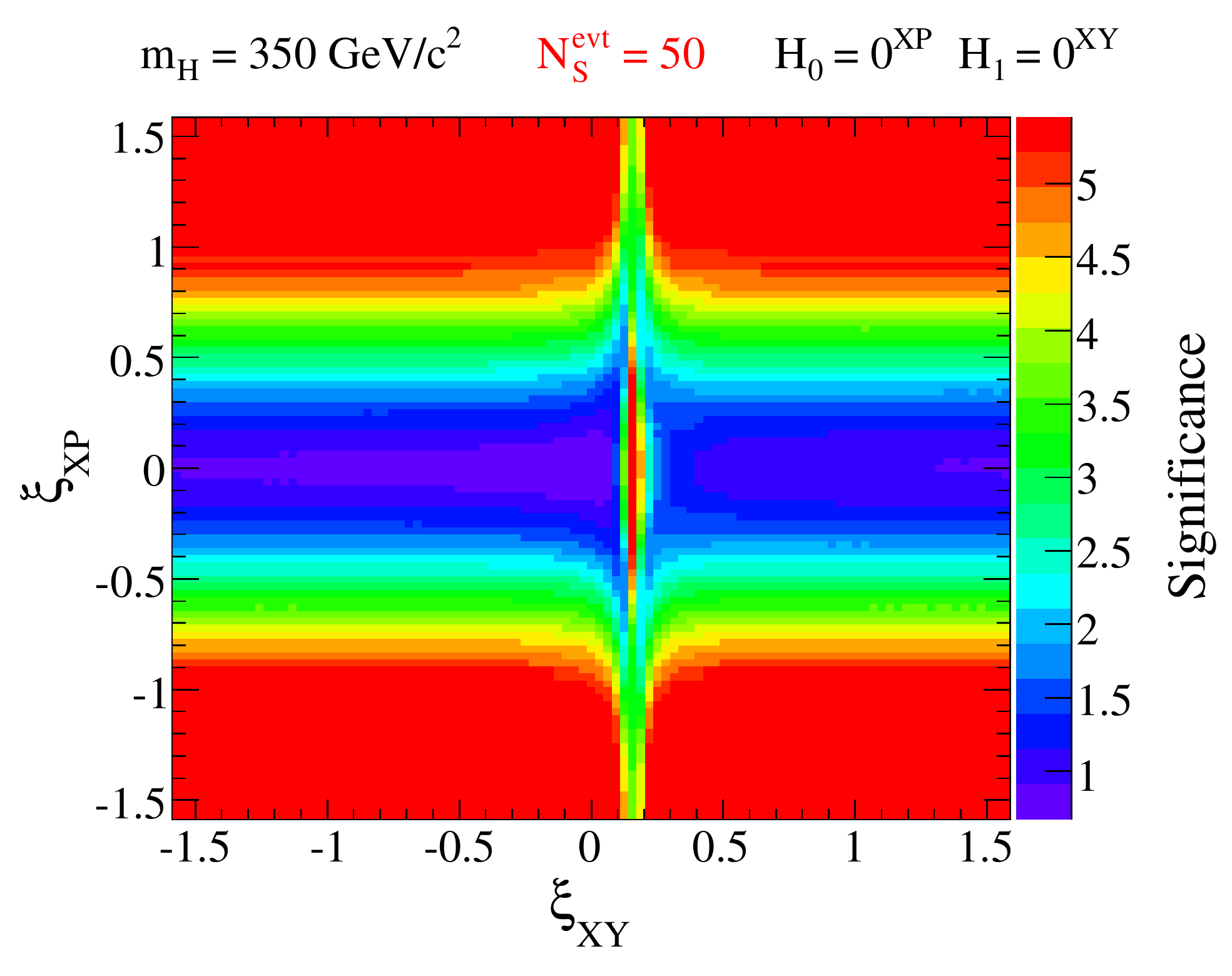}
\caption{The median  of the significance (colored-labeled
  $z$-``axis'') for excluding values of $\xi_{XP}$ ($y$-axis) in
  favor of the composite scalar assuming it to be correct with the
  $\xi_{XY}$ values of the $x$-axis, for $m_H$$=$$145$, 200 and 350
  GeV/c$^{2}$ (top, middle and bottom) and $N_S$$=$$50$.
  \label{fig:COMP2D_XY}}
\end{center}
\end{figure}

As we discussed for the $C$- and $CP$-violating cases, an additional
question is whether one can distinguish a composite scalar from other
mixed scalars. We find that, compared to the composite case, the two
other mixed cases are nearly identical. The results for the distinction
between the $CP$-violating and composite cases are shown in
Fig.~\ref{fig:COMP2D_XY}. For large values of $\xi_{XY}$ and
$\xi_{XP}$, it is possible to distinguish between the two hypotheses
at a large significance with a mere $N_S$$=$$50$. For $m_H$$=$$350$
GeV/c$^{2}$, the composite scalar is very similar to the pointlike
$0^{+}$ --and cannot be distinguished from it-- except if $\xi_{XY}$
is near its critical point.

Replacing the $CP$-violating scalar with the $C$-violating one yields
results nearly identical to the ones in Fig.~\ref{fig:COMP2D_XY}.

%%%%%%%%%%%%%%%%%%%%%%%%%%%%%%%%%%%%%%%
%% general spin 1 case
%%%%%%%%%%%%%%%%%%%%%%%%%%%%%%%%%%%%%%%
\begin{boldmath}
\subsection{$0^+$ vs. general $J$$=$$1$ 
\label{sec:spinone}}
\end{boldmath}

In Sec.~\ref{sec:SM_v_1} we discussed the prospects for distinguishing
a $0^{+}$ from the two pure $J^{PC}$ spin-one objects, vector and
axial-vector. Here, we address a more general question: how well can
one distinguish between $0^{+}$ and the general family of $J$$=$$1$
states?

The most general vertex describing the coupling of a $J$$=$$1$ particle a
$Z$ pair can be parametrized, for non-vanishing $X$, $P$, and $Q$, as:
\bea
&&\hspace*{-25pt}
{\cal L}^{\rho\mu\alpha} 
\hspace*{-3pt}\propto {\rm cos}\,\xi\,(g^{\rho\mu}p_{1}^{\alpha}\hspace*{-2pt}+\hspace*{-2pt}g^{\rho\alpha}p_{2}^{\mu})
+ e^{i\delta}{\rm sin}\,\xi \,\epsilon^{\rho\mu\alpha}(p_{1}\hspace*{-2pt}-\hspace*{-2pt}p_{2}),
\eea
in terms of two mixing angles $\xi$ and $\delta$.

The mixing between the pure vector and axial couplings is described by
$\xi$, while $\delta$ parametrizes the mixing between the $CP$- and
$C$-violating parts of the interference term in the matrix element
squared. In order to quantify the significance at which one can
distinguish between the $0^{+}$ hypothesis and the general $J$$=$$1$ case,
we consider two different types of tests, which answer two similar
questions.

\begin{figure}[tbp]
\begin{center}
\includegraphics[width=0.38\textwidth]{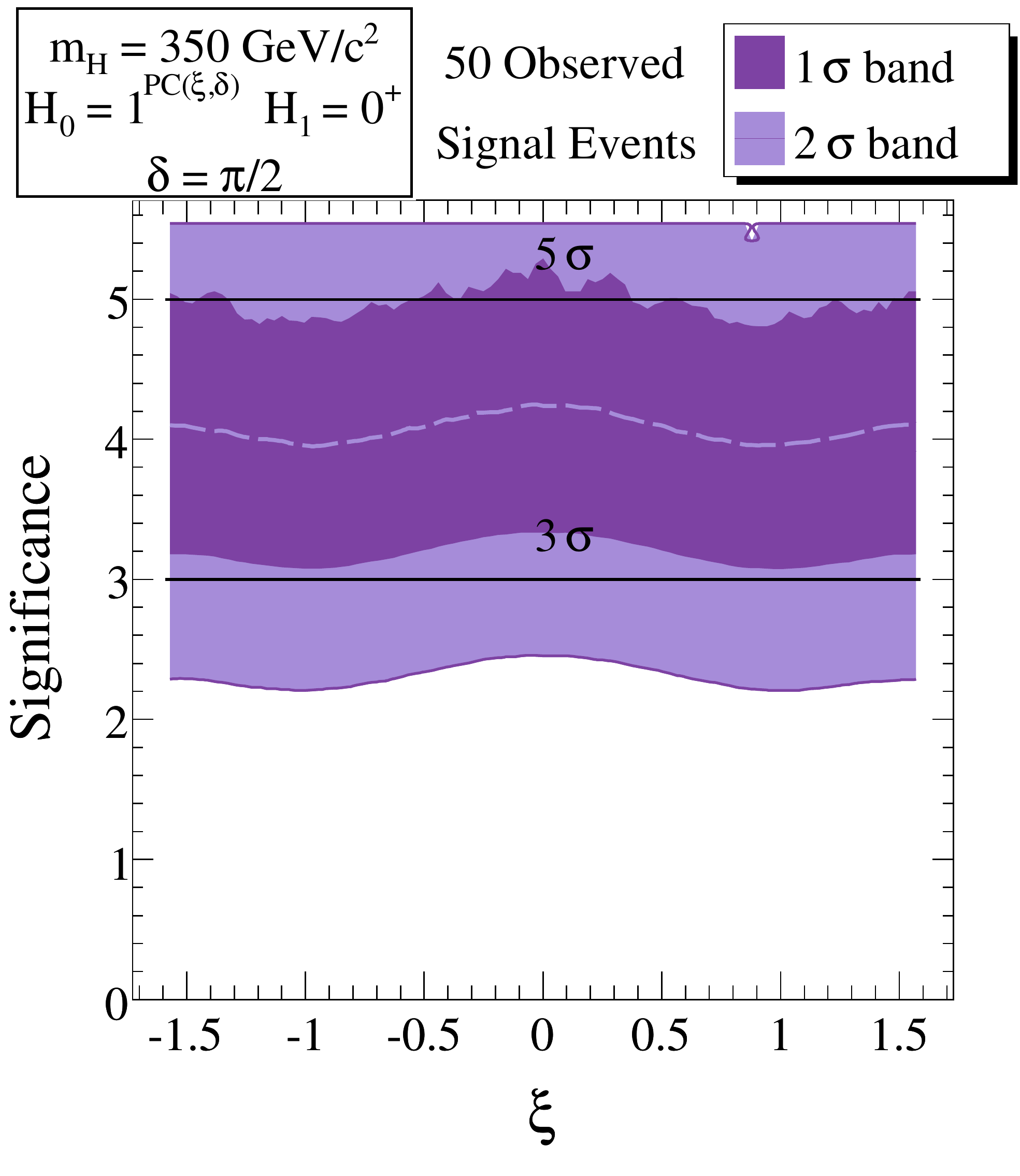}
\caption{Significance for excluding values of $\xi$, for
  $\delta$$=$$\pi/2$, in the general $J$$=$$1$ hypothesis [dubbed $\rm
  1^{PC(\xi\delta)}$] in favor of the $0^{+}$ one, assumed to be
  correct.  Results for $m_H$$=$$350$ GeV/c$^{2}$ and $N_S$$=$$50$.  The
  dashed line is the median of the significance. The 1 and 2 $\,\sigma$ bands
  correspond to 68\% and 95\% median-centered confidence intervals.
  \label{fig:SPINONE1D_H0}}
\end{center}
\end{figure}

Assuming a $0^{+}$ resonance to be the correct choice, we determine
the significance with which can we exclude values of $\xi$ and
$\delta$ for a $J$$=$$1$ hypothesis. We perform a series of simple
hypothesis tests, for each set of fixed values $\xi$ and $\delta$,
between the two hypotheses: the test statistic is
$\Lambda=\log[{\mathcal L}(0^{+})/{\mathcal L}(\xi,\delta)]$.  The
results, as a function of $\xi$ for $\delta$$=$$\pi/2$ and $m_H$$=$$350$
GeV/c$^{2}$, are shown in Fig.~\ref{fig:SPINONE1D_H0}.  The points
$\xi$$=$$0$ and $|\xi|$$=$$\pi/2$ correspond to the pure vector and pure
axial-vector limits, respectively, and are consistent with
Figs.~\ref{fig:COMP_SM_v_PV} and \ref{fig:COMP_SM_v_PA} on these pure
cases.

\begin{figure}[htbp]
\begin{center}
  \includegraphics[width=0.38\textwidth]{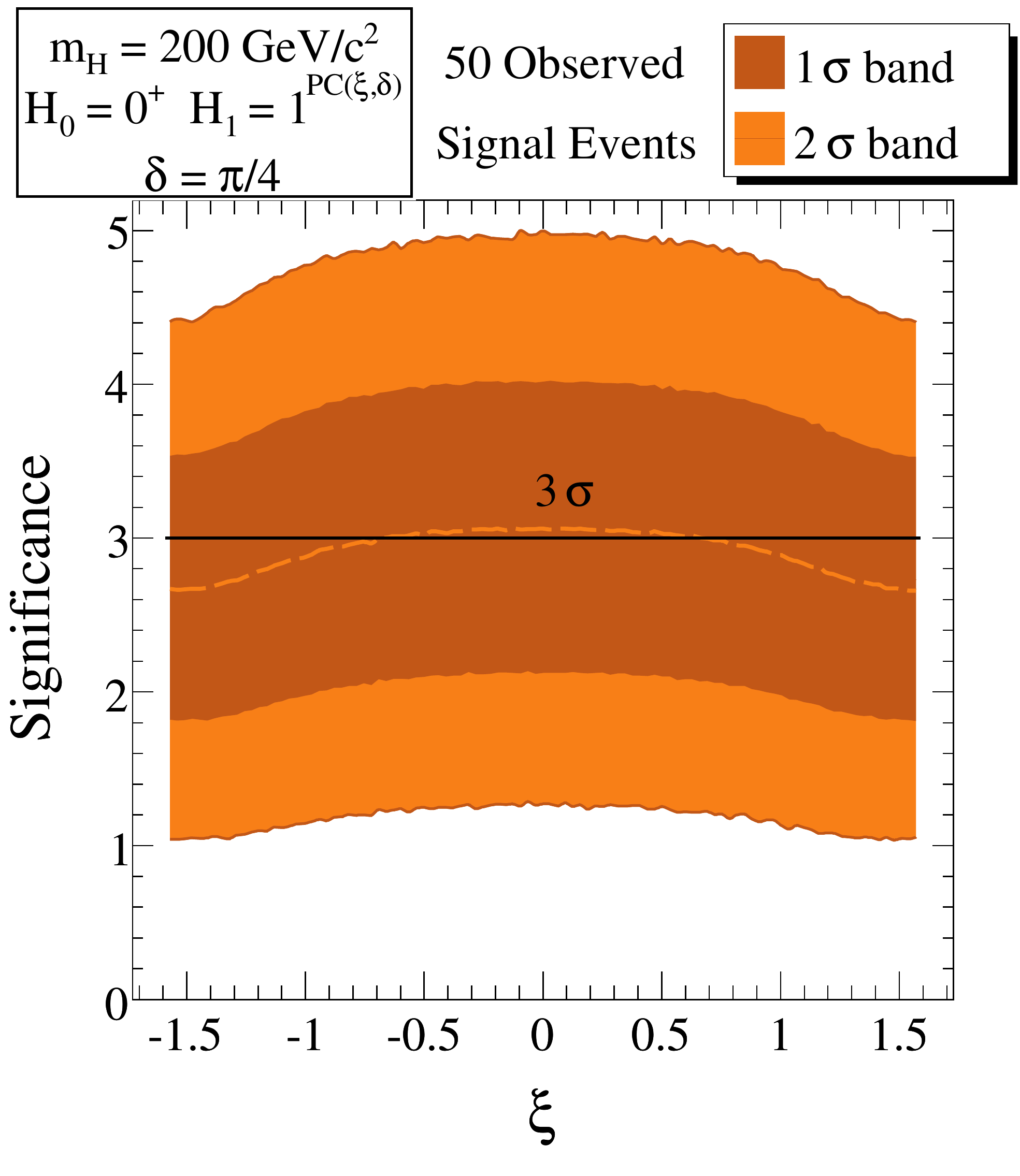}
  \caption{Significance for excluding the $0^{+}$ hypothesis
    in favor of the general $J$$=$$1$ case [dubbed $\rm
    1^{PC(\xi\delta)}$], assumed correct for $\xi$ as in the $x$-axis
    and $\delta$$=$$\pi/4$.  Results for $m_H$$=$$200$ GeV/c$^2$ and $N_S$$=$$50$.
    The dashed line and bands are as in Fig.~\ref{fig:SPINONE1D_H0}.
    \label{fig:SPINONE1D_H1}}
\end{center}
\end{figure}

Assuming a $J$$=$$1$ resonance with given $\xi$ and $\delta$ to be the
correct choice, we determine the significance with which we can
exclude the $0^{+}$ case in favor of $J$$=$$1$. We have to treat
$\xi$ and $\delta$ as nuisance parameters, since we are considering
the {\it general} $J$$=$$1$ case. The statistic is $\log[\max{\mathcal
  L}(\hat{\xi},\hat{\delta})/{\mathcal L}(0^{+})]$. The results, as
functions of $\xi$ for $\delta$$=$$\pi/4$ and $m_H$$=$$200$ GeV/c$^{2}$, are
given in Fig.~\ref{fig:SPINONE1D_H1}, which shows that one can
potentially exclude the $0^{+}$ hypothesis without knowing the actual
values of $\xi$ and $\delta$. Prospects for measuring these angles are
discussed in Sec.~\ref{sec:PARAM}.

\begin{figure}[htbp]
\begin{center}
\includegraphics[width=0.238\textwidth]{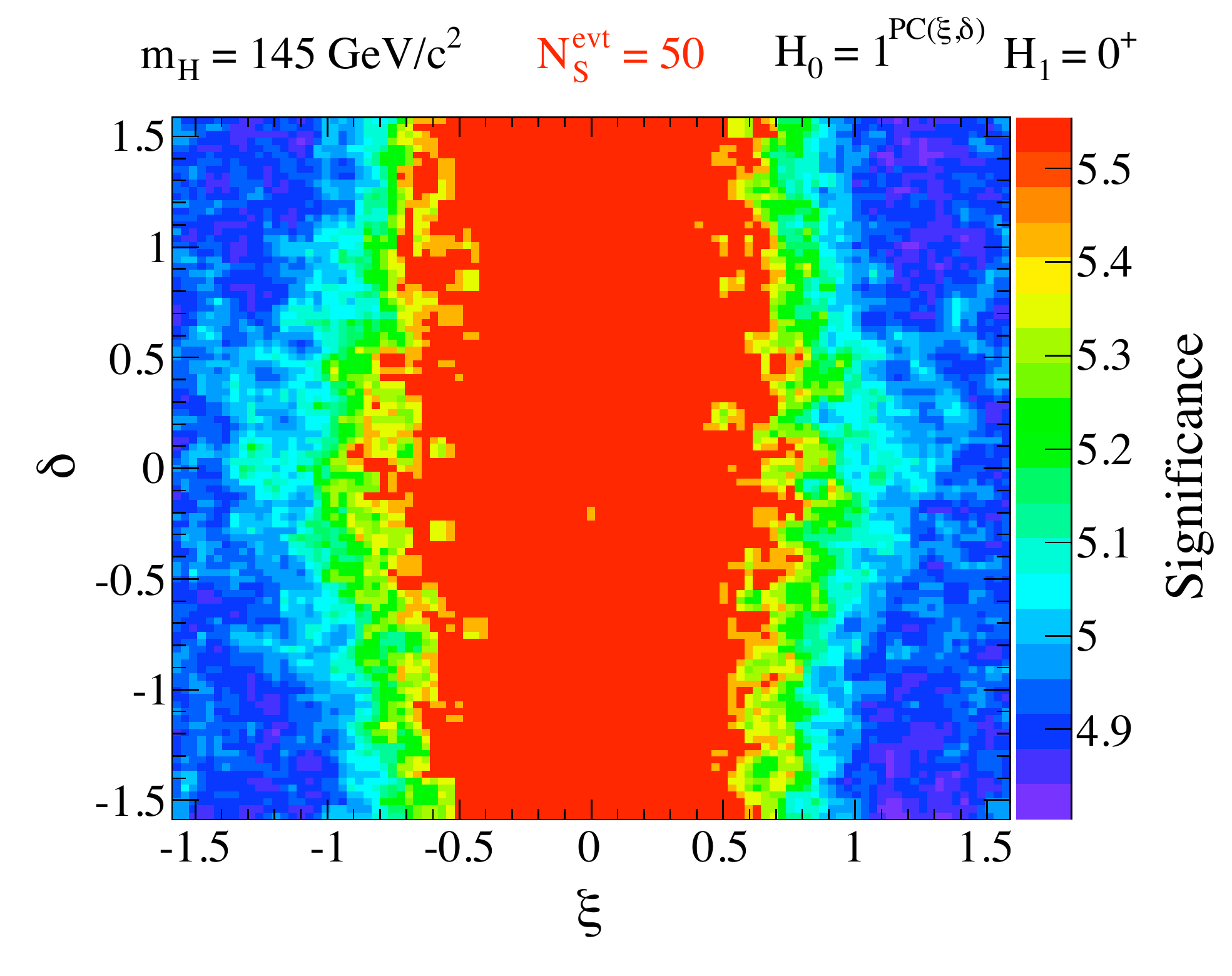}
\includegraphics[width=0.238\textwidth]{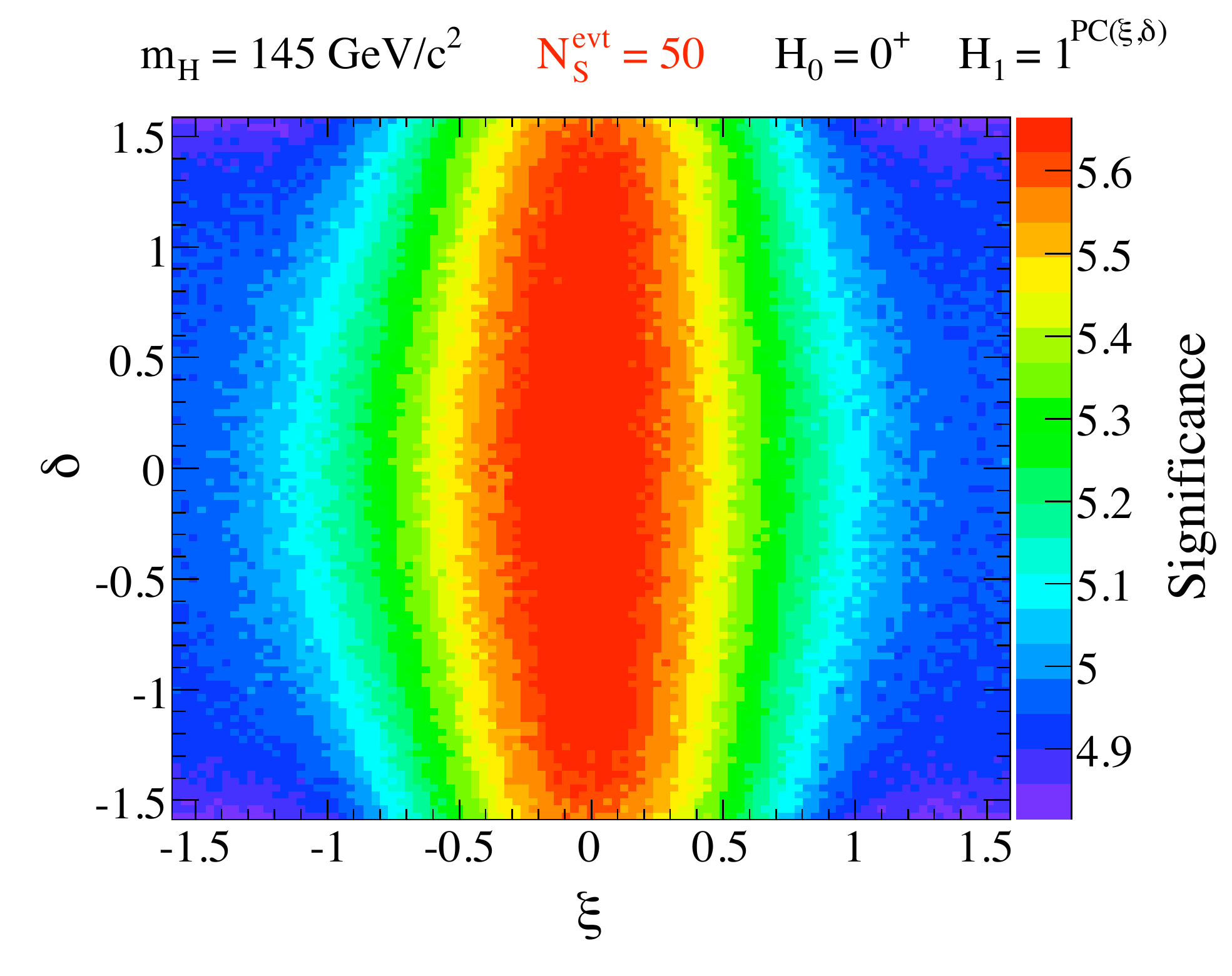}
\includegraphics[width=0.238\textwidth]{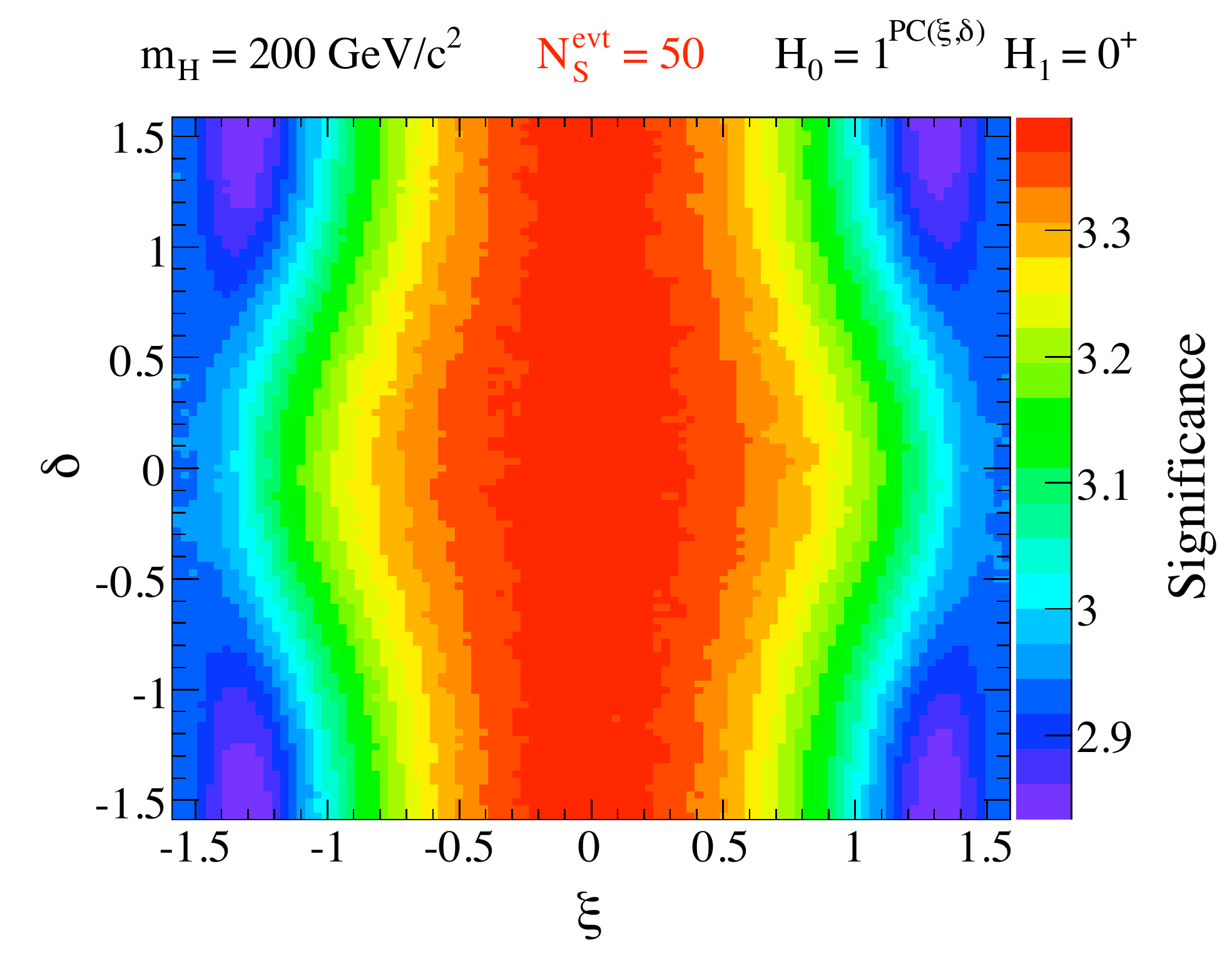}
\includegraphics[width=0.238\textwidth]{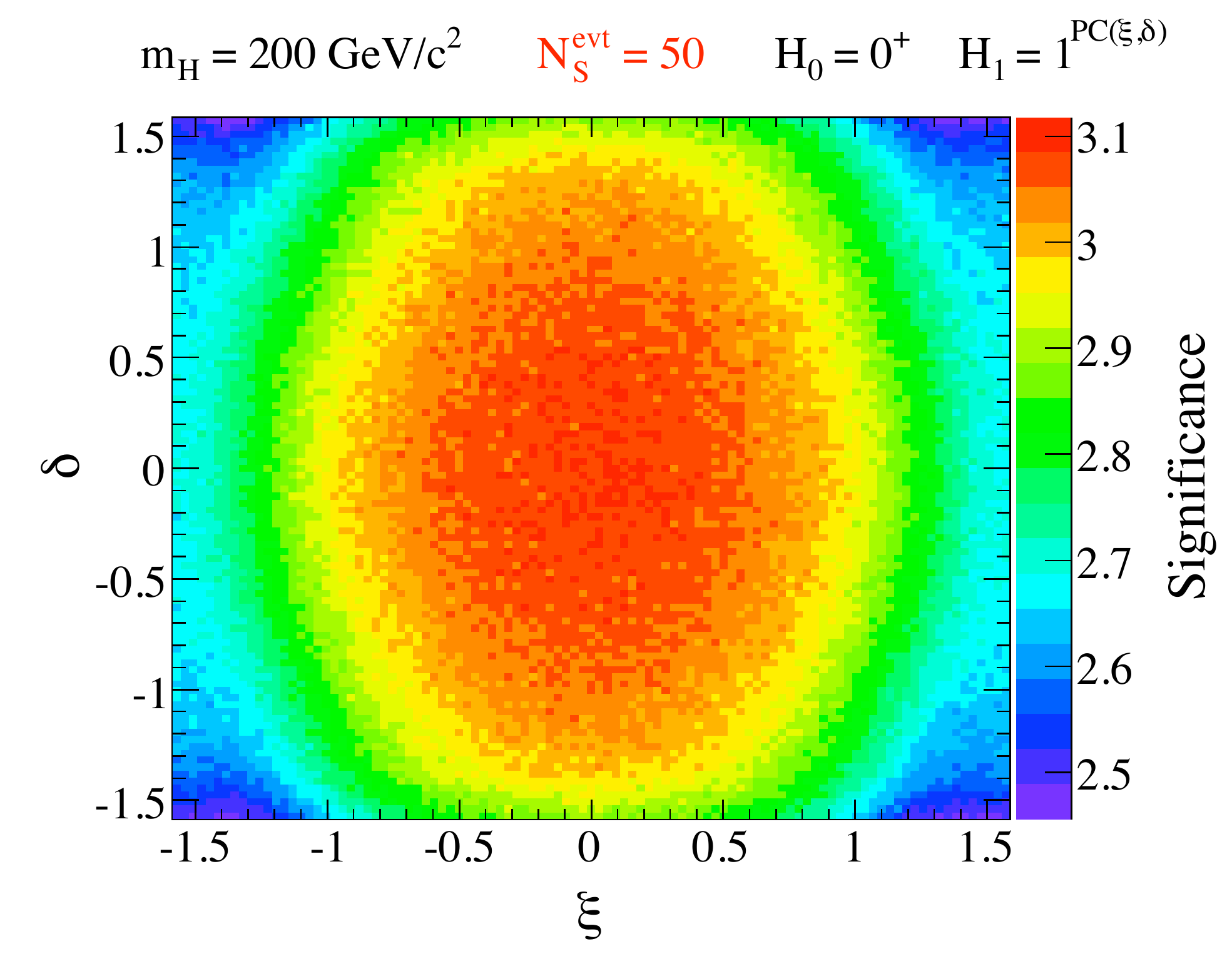}
\includegraphics[width=0.238\textwidth]{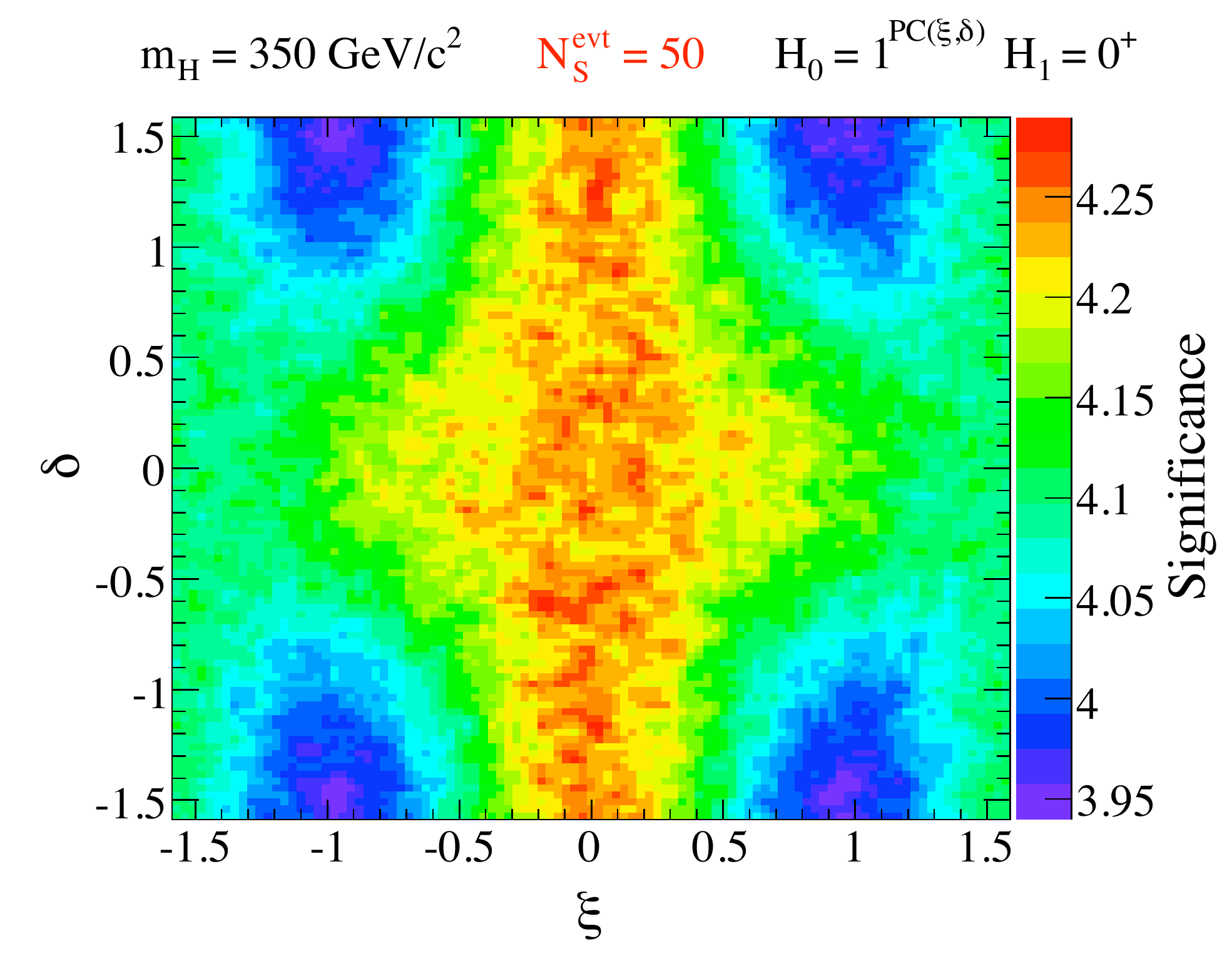}
\includegraphics[width=0.238\textwidth]{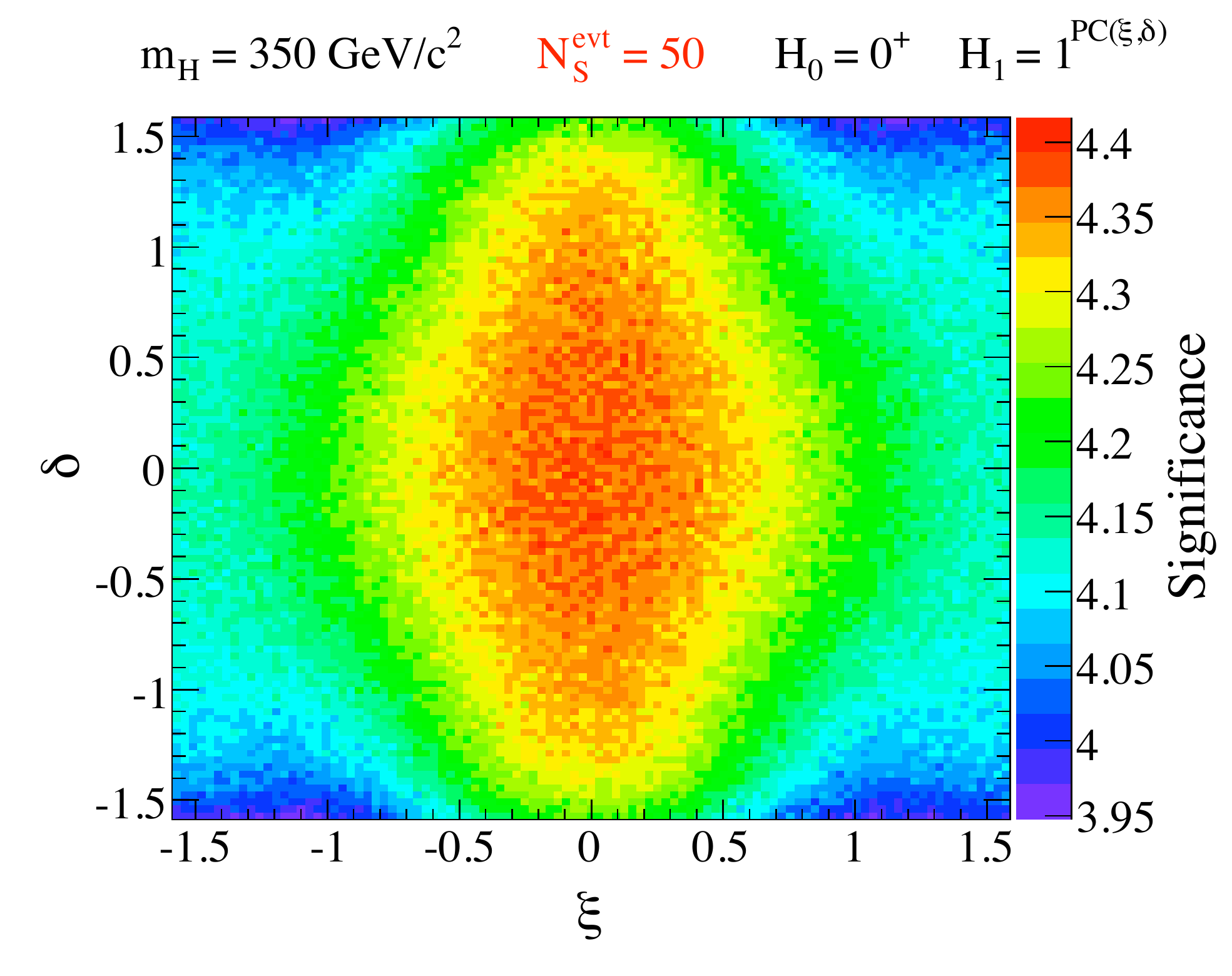}
\caption{Left: Median of the significance (coloured $z$-``axis")
  for excluding values of $\xi$\ and $\delta$ corresponding to a $J$$=$$1$
  hypothesis [dubbed $\rm 1^{PC(\xi\delta)}$] in favor of $0^{+}$, if
  the latter is correct.  Right: vice-versa, with values of $\xi$ and
  $\delta$ indicated on the axes. Results for $m_H$$=$$145$, 200 and 350
  GeV/c$^{2}$ (top, middle and bottom), for $N_S$$=$$50$.
  \label{fig:SPINONE2D}}
\end{center}
\end{figure}

In Fig~\ref{fig:SPINONE2D} we show the significance for the
distinction between the $0^{+}$ and the general $J$$=$$1$ cases, as a
function of $\xi$ and $\delta$, for $m_H$$=$$145$, 200, and 350
GeV/c$^{2}$.  Notice that the significance levels colour-coded as a
$z$-``axis" range over a small interval. This means that the entire
$J$$=$$1$ family is almost ``equally dissimilar'' to $0^{+}$.  In general,
one's ability to exclude $J$$=$$1$ relative to $0^{+}$ is greater than its
opposite, due to the required treatment of $\xi$ and $\delta$ as
nuisance parameters, although the differences are relatively small in
magnitude and in $\xi$- and $\delta$-dependence.

The fact that the significance plane as a function of $\xi$ and
$\delta$ is relatively flat means that, with some $m_H$-dependent
amount of observed events, one shall be able to unambiguously exclude
the general $J$$=$$1$ hypothesis in favor of the $0^{+}$ case (assuming
it to be correct) or vice-versa, regardless of the values of $\xi$ and
$\delta$. Using the pure $J^{PC}$ hypothesis test as a guide, we
conclude that the median expectation for differentiating between
$0^{+}$ and $J$$=$$1$ should exceed 5$\,\sigma$ with $N_S\sim (60,200,85)$
events for $m_H$$=$$(145,200,350)$ GeV/c$^{2}$, respectively.

Additionally, based on our results concerning the distinction between
$0^{-}$ and the two pure $J$$=$$1$ states, and the results on the mixed
$J$$=$$0$ hypotheses, we conclude that it is equally easy, or even easier,
to distinguish between $J$$=$$1$ and a $J$$=$$0$ state other than
$0^{+}$. Hence, with the numbers of events listed above, it is likely
that one will be able to unambiguously exclude the $J$$=$$1$ family of
hypotheses in favor of a general $J$$=$$0$ hypothesis, or vice-versa, if
the resonance is either one or the other.

%%%%%%%%%%%%%%%%%%%%%%%%%%%%%%%%%%%%%%
%% Parameter estimation
%%%%%%%%%%%%%%%%%%%%%%%%%%%%%%%%%%%%%%
\vspace*{2mm}
\subsection{Parameter estimation in mixed $\mathbf{J=0}$ and $\mathbf{J=1}$ 
cases \label{sec:PARAM}}

Were one to find out from real data and the hypothesis tests discussed
in the previous section that a mixed $J$$=$$0$ or $J$$=$$1$ state is the
preferred description, the next item in the context of this analysis would be the
measurement of its mixing parameters (in a larger context one would
include at this stage the measurement of decay branching ratios).

We have seen in Secs.~\ref{sec:spinzero} and \ref{sec:spinone} that
our hypothesis tests can demonstrate --if correct-- and with
computable significance, that a standard $0^{+}$ particle is
disfavored relative to a mixed scalar or vector with unspecified
$HZZ$ coupling ratios (or mixing angles). In these tests, the angles
were treated as nuisance parameters. Their measurement proceeds along
the same line --the preferred value is simply that which maximizes the
likelihood-- but the treatment of confidence intervals need be
different.

More specifically, each mixed hypothesis family is characterized by
mixing angles  $\vec{\xi}$. For each ``experiment'', $N$ events are
simulated, each one characterized by a vector $\vec{x}_{e} =
\{\vec{\omega},\vec{\Omega},M_{Z^{*}}\}|_e$. The likelihood for a
particular family of hypotheses is ${\cal L}(\vec{\xi}) =
\prod_{e=1}^{N}P_{e}(\vec{x}_{e},\vec{\xi})$. The measured values
of the mixing angles, $\vec{\xi}_{\rm meas}$, are chosen to be
those that maximize the likelihood.
%, $\max {\cal L}(\hat{\vec{\rho}})$.  

To assign confidence intervals to these measurements we use a fully
frequentist approach. An ensemble of ``experiments'' is performed with
fixed input values $\vec{\xi}$$=$$ \vec{\xi}_{\rm input}$. For each
experiment, the measured values of $\vec{\xi}$ are taken from the
maximization of the likelihood. This procedure is repeated for a fine
matrix of input values, covering the allowed parameter space. From the
probability distribution functions $P(\vec{\xi}_{\rm meas} |
\vec{\xi}_{\rm input})$, estimated using this ensemble of
experiments, the Feldman-Cousins unified approach~\cite{FC} is used to
choose which elements of probability are included in confidence
intervals.

\begin{figure}[tb!]
\begin{center}
\includegraphics[width=0.32\textwidth]{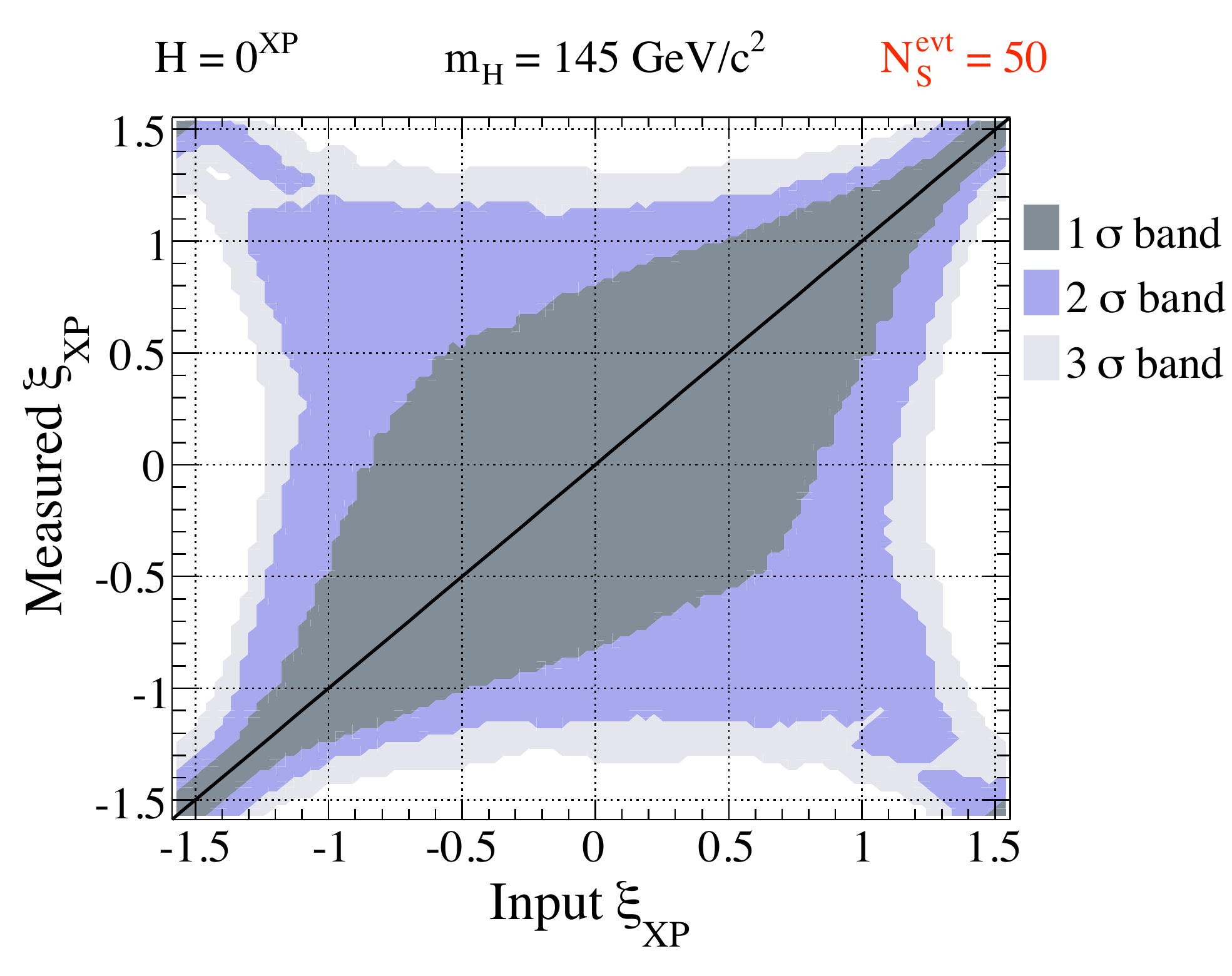}
\includegraphics[width=0.32\textwidth]{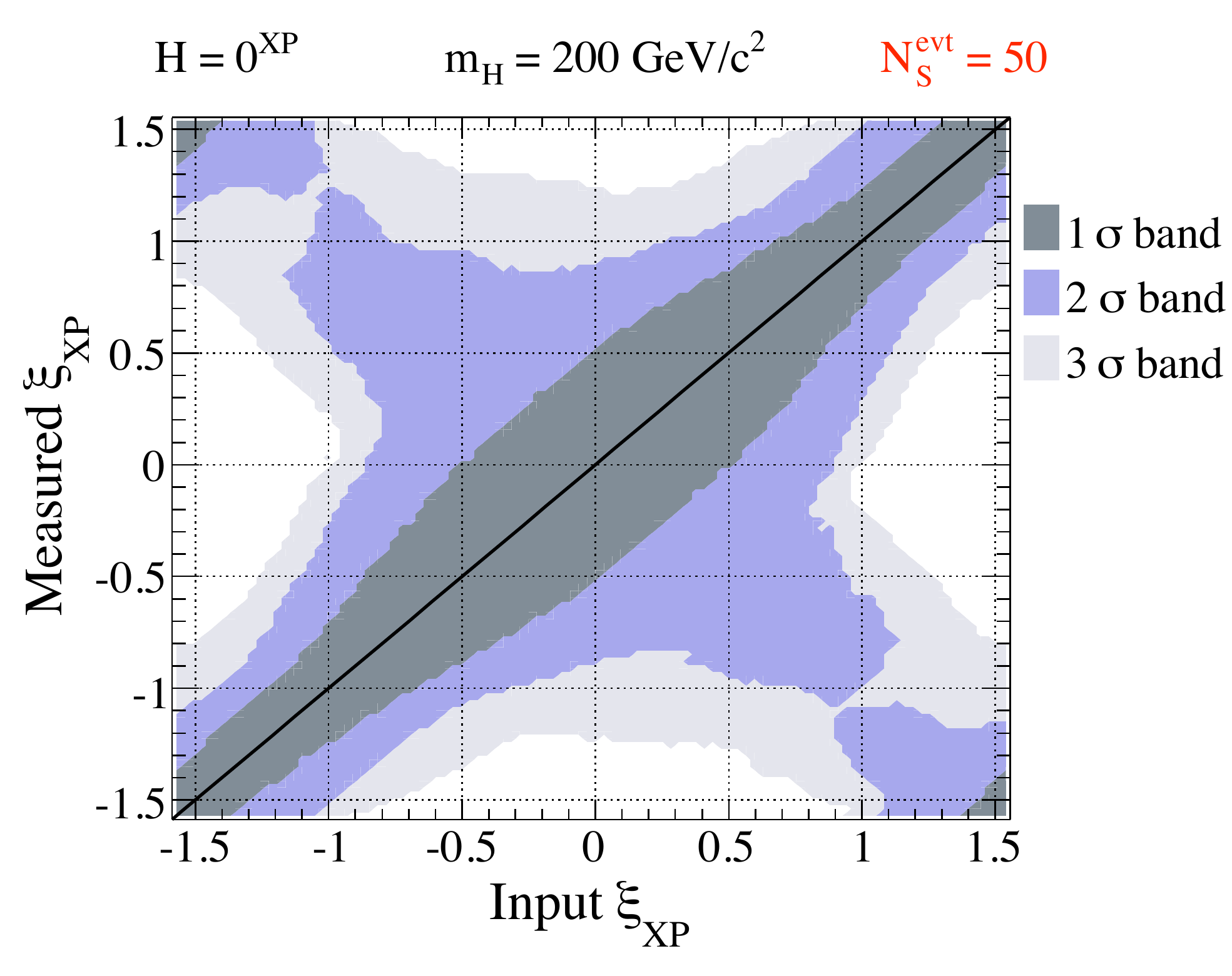}
\includegraphics[width=0.32\textwidth]{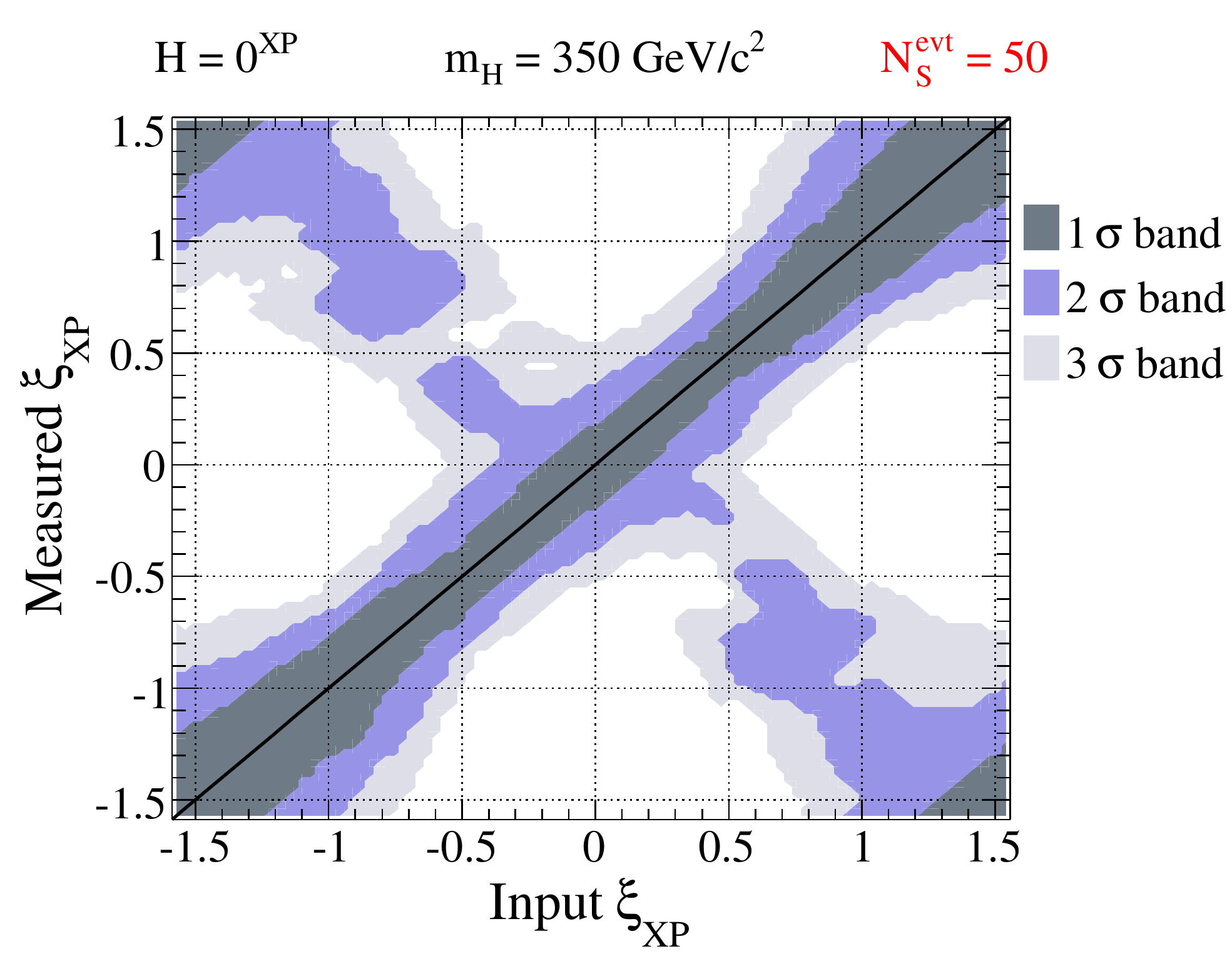}
\caption{Confidence intervals for measured values of $\xi_{XP}$ for a
  $CP$-violating $J$$=$$0$ resonance, for $m_H$$=$$145$, 200 and 350
  GeV/c$^{2}$ (top, middle and bottom), all for $N_S$$=$$50$.  For
  measured values of $\xi_{XP}$ on the y-axis, confidence intervals
  should be read horizontally, see text.
  \label{fig:PARAM_XP}}
\end{center}
\end{figure}

As an example, consider the $CP$-violating scalar case, discussed in
Sec.~\ref{sec:spinzero}. The confidence intervals for measured values
of $\xi_{XP}$ (the mixing parameter that characterizes this
hypothesis) are shown in Fig.~\ref{fig:PARAM_XP} for different values
of $m_H$. The way to interpret these figures is as follows: For a
particular set of data --one experiment, which in this case includes
$N_S$$=$$50$ observed events-- an input value of $\xi_{XP}$ (to be read on
the $x$-axis) results in a measured value to be read (with its error
bands) on the $y$ axis.  The confidence intervals are obtained by
drawing a horizontal line passing through the measured $\xi_{XP}$. The
overlap of this line with the $n\,\sigma$ bands dictates which values
of ``input $\xi_{XP}$" should be included in the $n\,\sigma$
confidence intervals. For example, for $m_H$$=$$200$ GeV/c$^{2}$ (middle
of Fig.~\ref{fig:PARAM_XP}) we see that, if $\xi_{XP}^{\rm meas}$$=$$0$,
the 3$\,\sigma$ confidence interval is approximately $\xi_{XP} \in
[-1,1]$.

The $1\,\sigma$ bands in Fig.~\ref{fig:PARAM_XP} are centered on the
diagonal $\xi_{XP}^{\rm meas}$$=$$\xi_{XP}^{\rm input}$, implying that
there is no significant bias in the measurement. In addition to this,
the $2\,\sigma$ and $3\,\sigma$ bands also cover most of the diagonal
$\xi_{XP}^{\rm meas} = -\xi_{XP}^{\rm input}$. This confirms our
observation from Sec.~\ref{sec:spinone} that our ability to pin down
this parameter comes predominantly from measuring the relative
strengths of the $0^{+}$ and $0^{-}$ parts of the {\it pdf} rather
than the nature ($\tilde{T}$-odd) of its interference term. An
increased number of observed events is needed to fully resolve this
sign ambiguity.

In Fig.~\ref{fig:PARAM_XP} we see that for $m_H$$=$$145$ GeV/c$^{2}$
(but not for $m_H$$=$$200$ GeV/c$^{2}$) the size of the confidence
intervals for $\xi_{XP}$ decreases with increasing $|\xi_{XP}|$. This
is due to the effective coupling strengths of the $0^{+}$ and $0^{-}$
parts of the {\it pdf} differing by a factor of $\sim 10$ at 
$m_H $$=$$145$ GeV/c$^{2}$ but not at the other masses. Hence, at the lowest
mass, only at $\tan^{2}(\xi_{XP}) \sim 10$ does the {\it pdf} exhibit
$0^{+}$- and $0^{-}$-like behaviours of similar magnitude.

\begin{figure}[tb!]
\begin{center}
\includegraphics[width=0.32\textwidth]{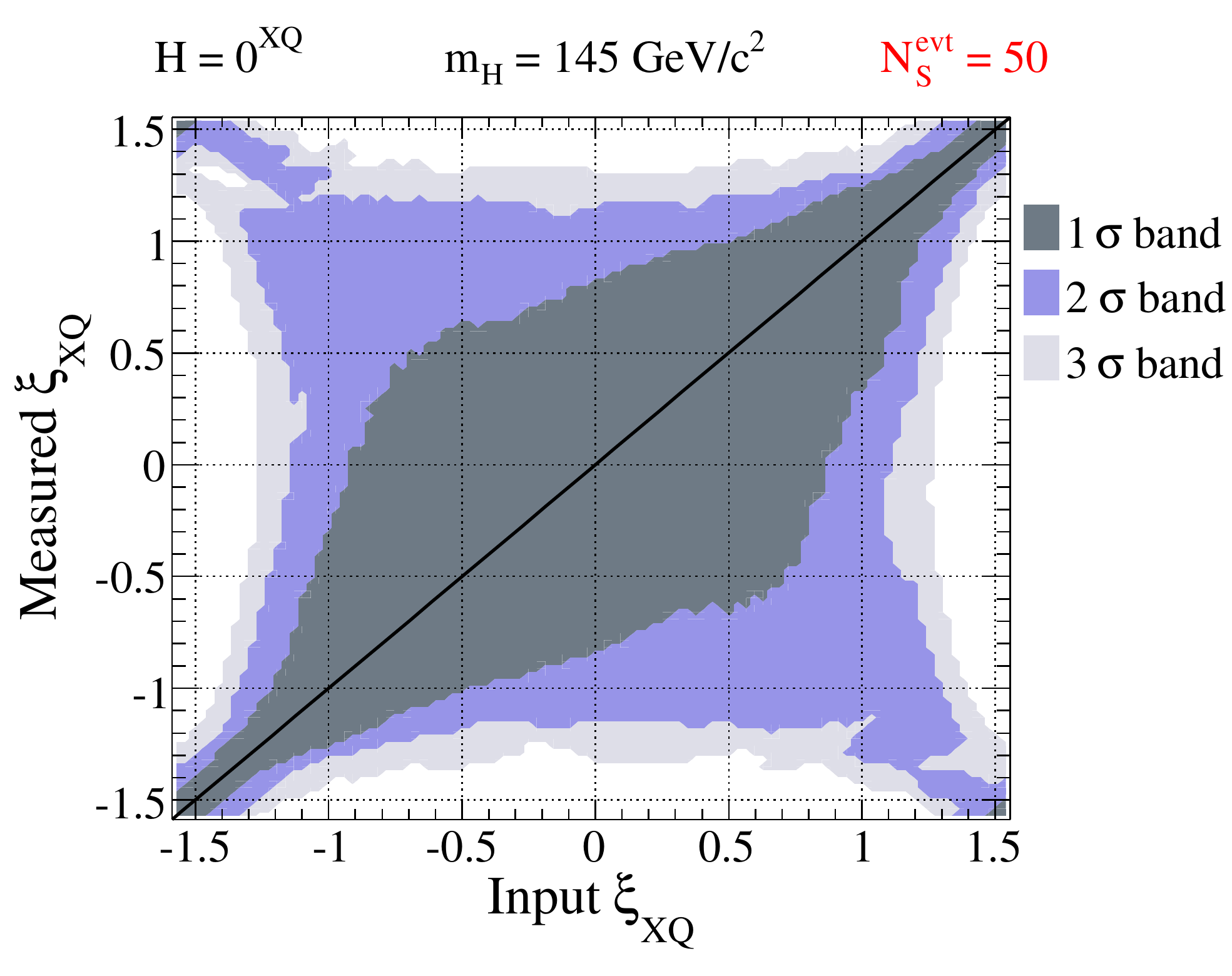}
\includegraphics[width=0.32\textwidth]{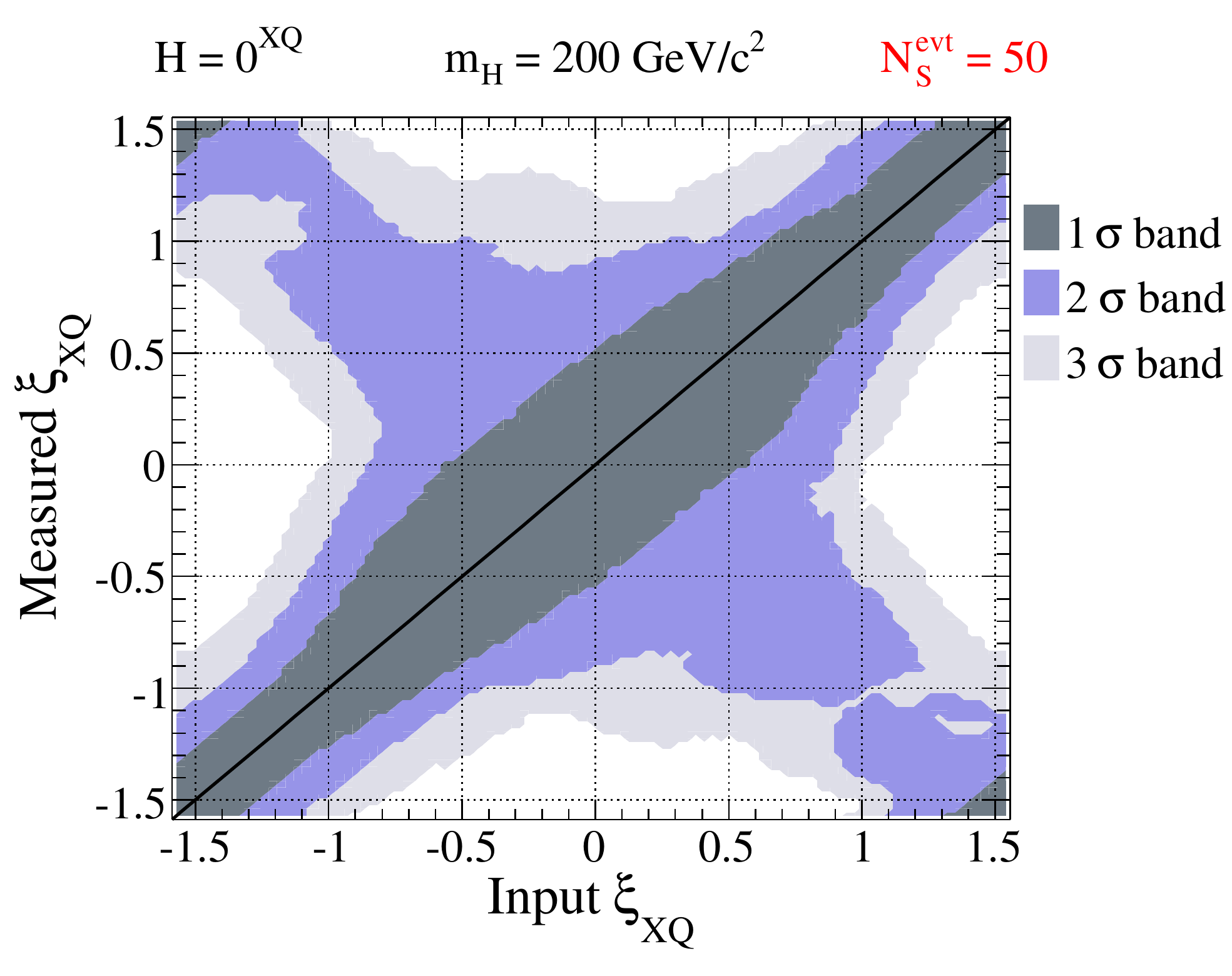}
\includegraphics[width=0.32\textwidth]{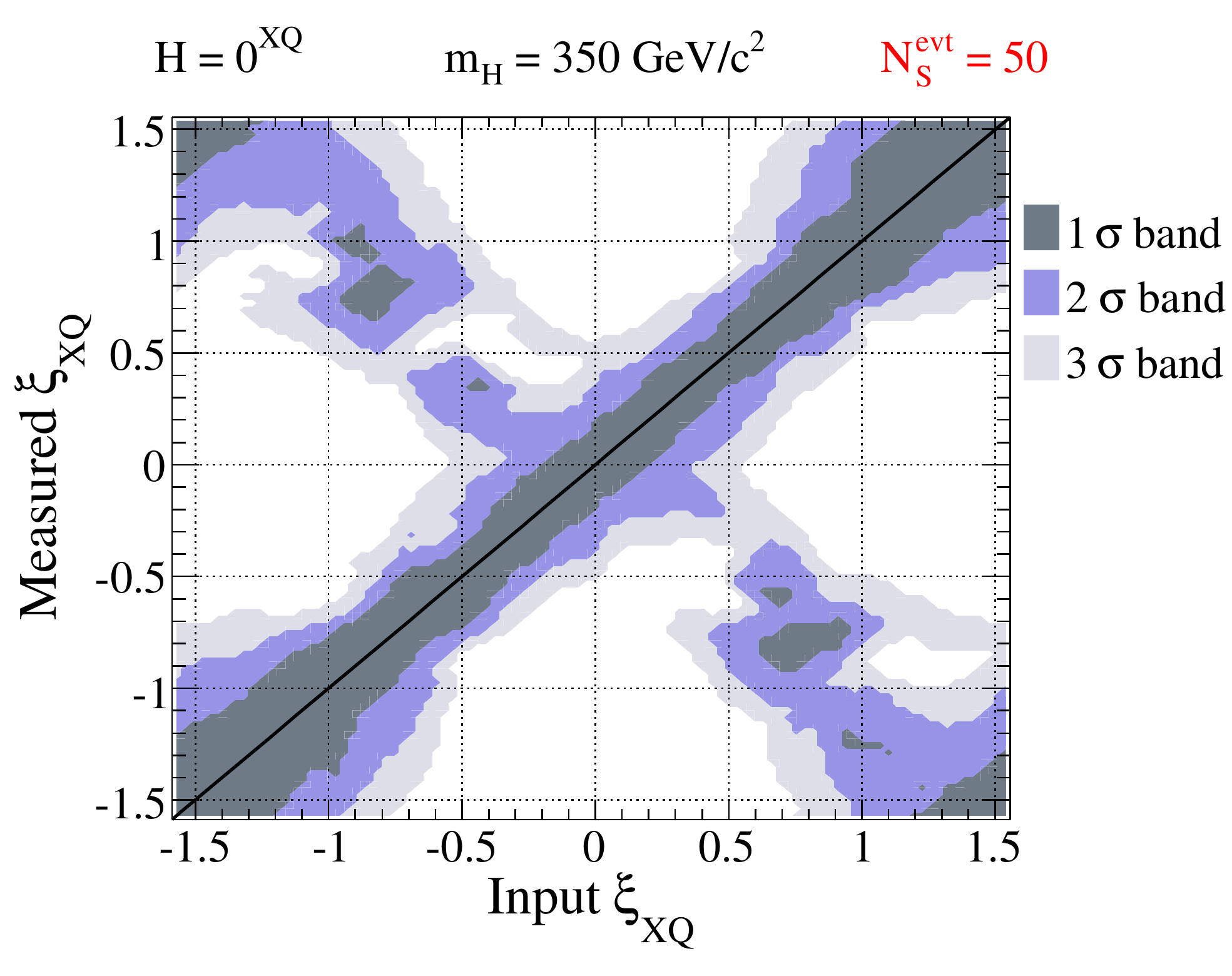}
\caption{Confidence intervals for measured values of $\xi_{XQ}$ for a
  $C$-violating $J$$=$$0$ resonance for $m_H$$=$$145$, 200 and 350 GeV/c$^{2}$
  (top, middle and bottom), all for $N_S$$=$$50$.  For measured values of
  $\xi_{XQ}$ on the y-axis, confidence intervals should be read
  horizontally, see text.
  \label{fig:PARAM_XQ}}
\end{center}
\end{figure}

Confidence intervals for measurements of the parameter $\xi_{XQ}$ for
a scalar with $C$-violating $HLL$ couplings are shown in
Fig.~\ref{fig:PARAM_XQ}; These are nearly identical to those in
Fig.~\ref{fig:PARAM_XP}, reflecting the difficulty of discriminating
the $\xi_{XP}\neq 0$ and $\xi_{XQ}\neq 0$ hypotheses, as discussed in
Sec.~\ref{sec:spinzero}.  For the $C$-odd case, the sign ambiguity of
$\xi_{XQ}^{\rm meas}$ is slightly worse than for the $\tilde{T}$-odd
one as demonstrated by the $1\,\sigma$ confidence bands appearing on
the $\xi_{XQ}^{\rm meas} = -\xi_{XQ}^{\rm input}$ diagonal for 
$m_H $$=$$350$ GeV/c$^{2}$. This is also expected, since the $C$-odd
interference term is proportional to the relatively small number $\eta
\approx 0.15$, see Eq.~(\ref{0Codd}).
\begin{figure}[tbp]
\begin{center}
\includegraphics[width=0.32\textwidth]{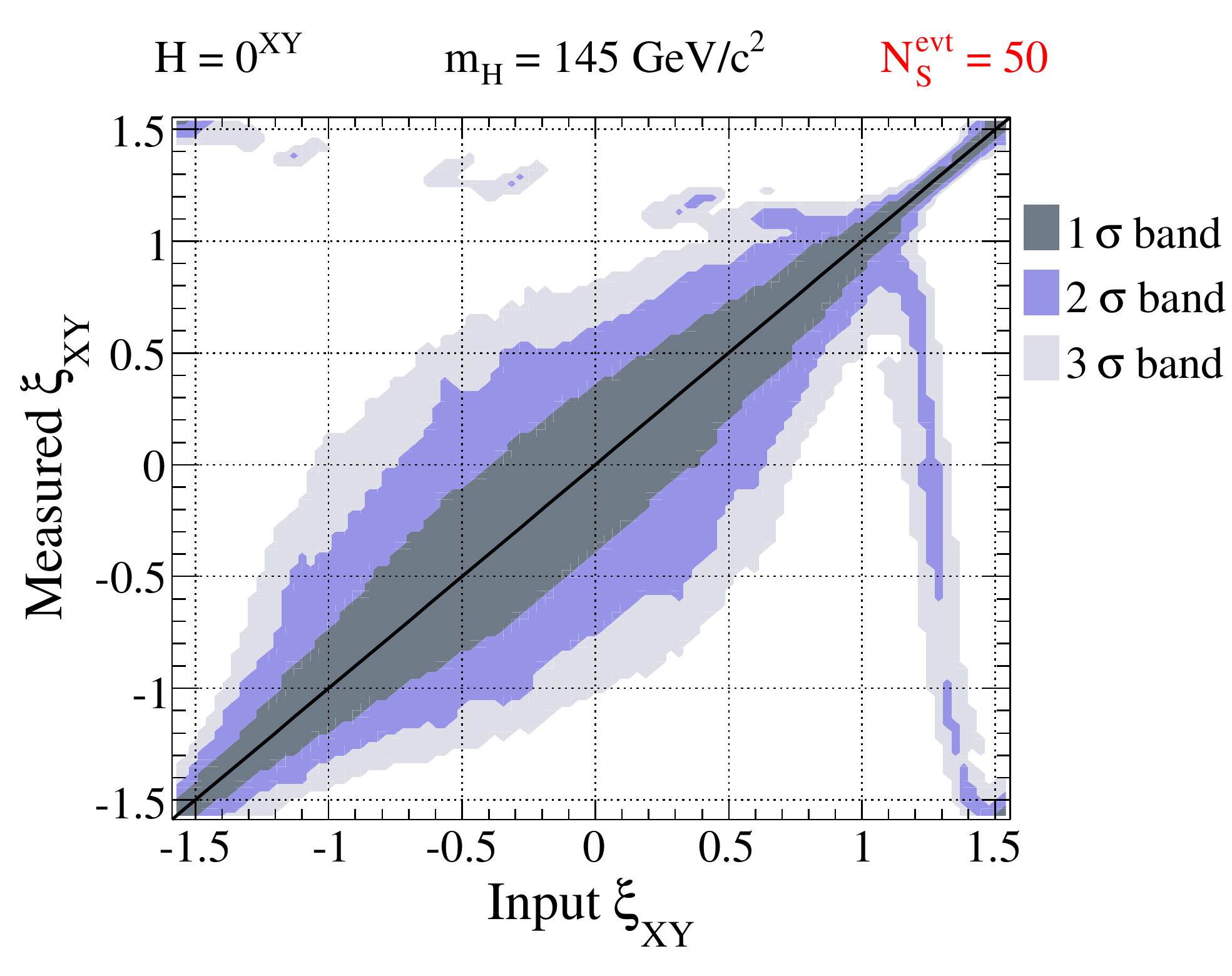}
\includegraphics[width=0.32\textwidth]{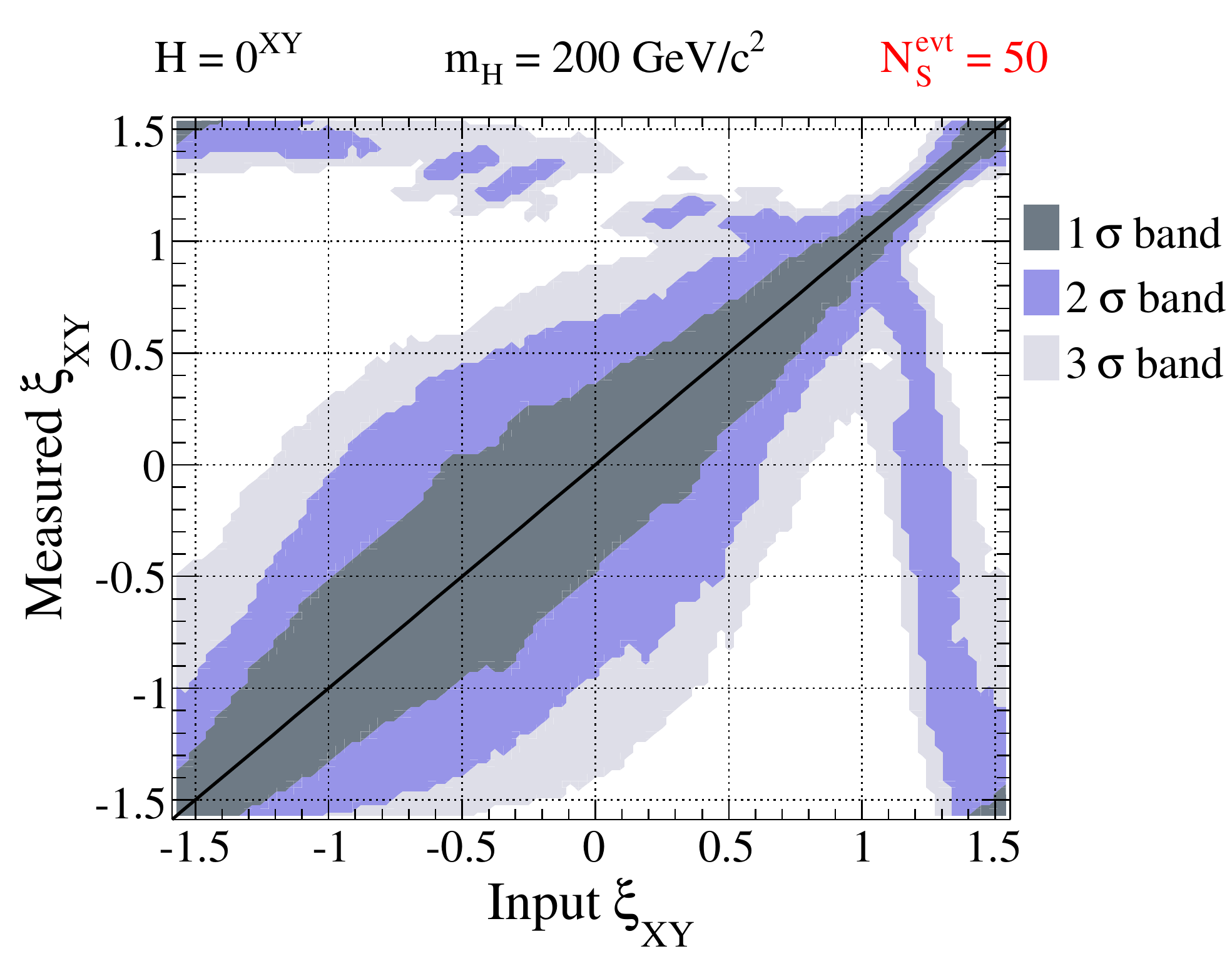}
\includegraphics[width=0.32\textwidth]{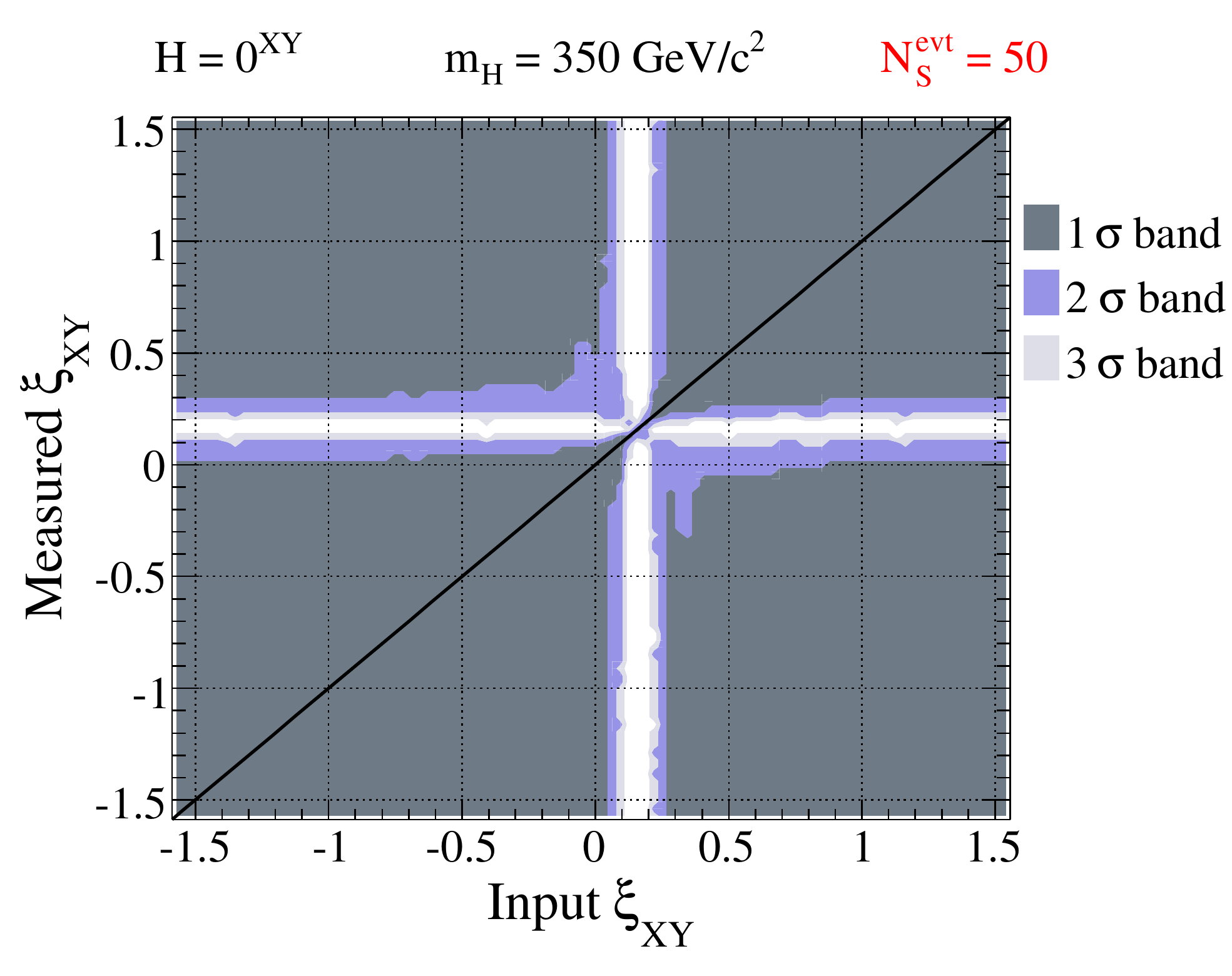}
\caption{Confidence intervals for measured values of $\xi_{XY}$ for a
  ``composite'' $J$$=$$0$ resonance, for $m_H$$=$$145$, 200 and 350
  GeV/c$^{2}$ (top, middle and bottom), all for $N_S$$=$$50$. For measured
  values of $\xi_{XY}$ on the y-axis, confidence intervals should be
  read horizontally.
  \label{fig:PARAM_XY}}
\end{center}
\end{figure}
One's ability to distinguish between $J$$=$$0$ $C$- and $\tilde{T}$-odd
admixtures relies on the resolution of the interference terms. With a
factor of 10 more statistics ($N_S\sim 500$), one would be able to
resolve the sign ambiguity in $\xi_{XP}$ and $\xi_{XQ}$ and to
distinguish between the two cases.

The confidence intervals associated with measurements of $\xi_{XY}$
for a composite scalar are shown in Fig.~\ref{fig:PARAM_XY}. We
observe that, for $m_H$$=$$145$ and 200 GeV/c$^{2}$, the $1\,\sigma$
intervals are centered on the diagonal $\xi_{XY}^{meas} =
\xi_{XY}^{input}$.  There are no bands along $\xi_{XY}^{meas} =
-\xi_{XY}^{input}$, since the interference term is of a different
nature than that of the discrete-symmetry violating cases. The
extensions of the 2 and $3\,\sigma$ bands along almost horizontal and
vertical lines around $\xi_{XY} \sim 1.3$ result from large
cancellations in the {\it pdf}, discussed in Sec.~\ref{sec:spinzero}.

The figure for $m_H$$=$350 GeV/c$^{2}$ is hard to decipher.  With a
magnifier one sees that at the critical value of $\xi_{XY}$ the
confidence intervals are tiny. Everywhere else, the intervals
essentially include all possible values {\it except} the critical
one. This is tantamount to saying that at this mass we cannot tell, on
the basis of our analysis, a composite from a pointlike scalar {\it
  unless} is has a particular value of $\xi_{XY}$, a fact made clearer
by Fig.~\ref{fig:COMP1D_XY}.

The other mixed case we study is that of a general $J$$=$$1$ resonance,
parameterized by angles $\xi$ and $\delta$ as described in
Sec.~\ref{sec:spinone}. We saw in Sec.~\ref{sec:other}, that most
difficult distinction is the one between the two pure $J^{PC}$
spin-one resonances, indicating that these two cases are very
similar. This is what we find again when exploring the potential for
measuring $\xi$ and $\delta$.

\begin{figure}[htbp]
\begin{center}
\includegraphics[width=0.34\textwidth]{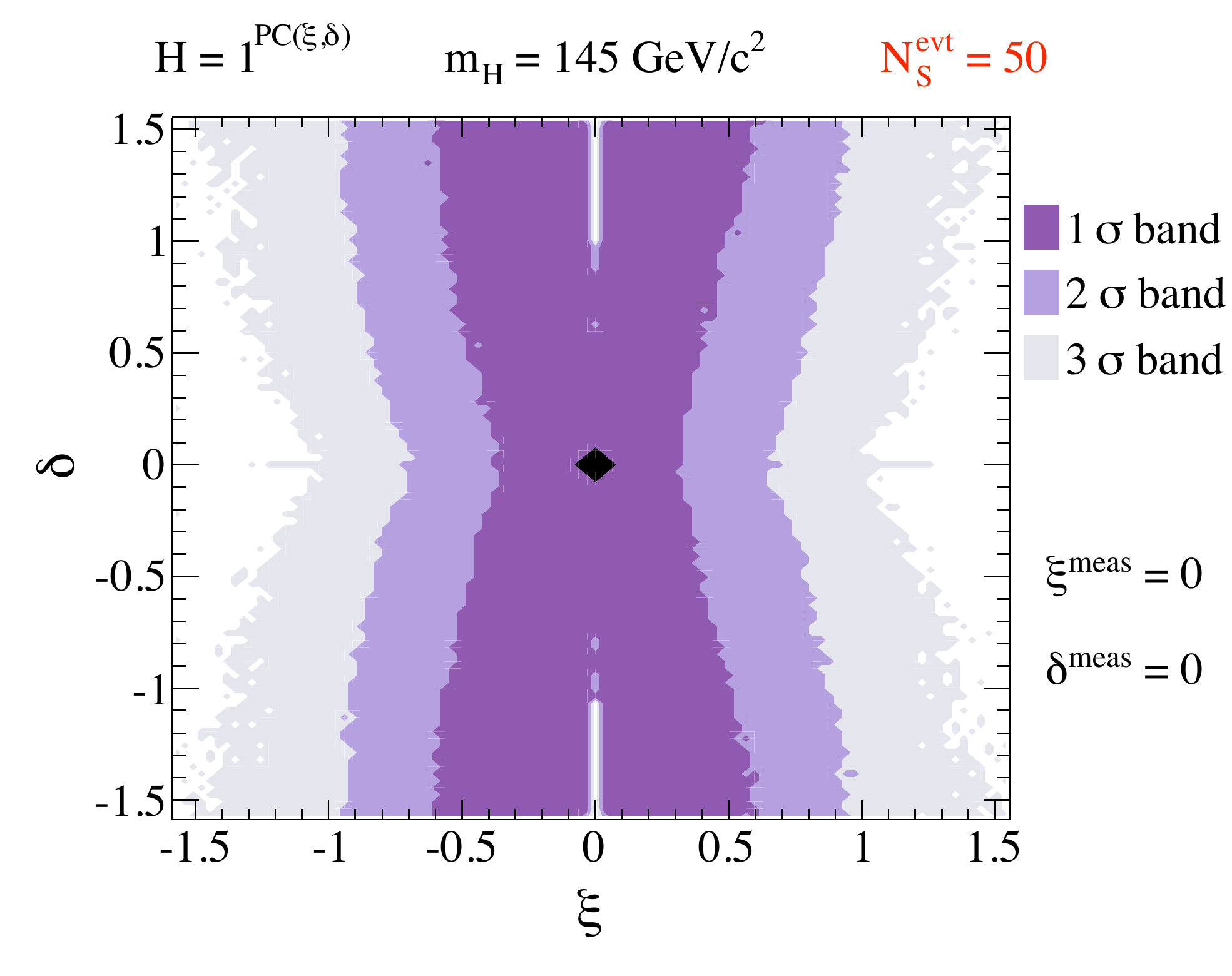}
\includegraphics[width=0.34\textwidth]{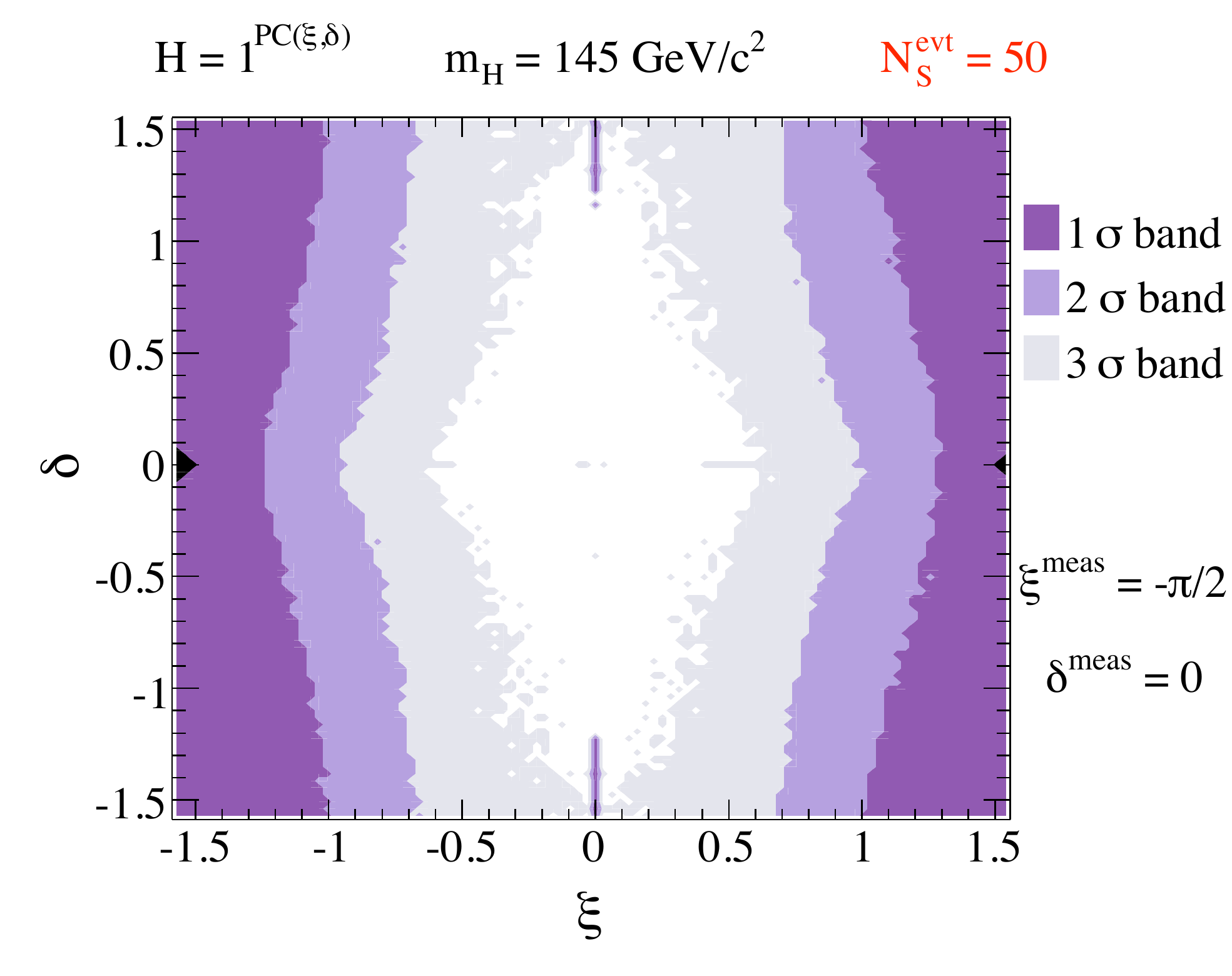}
\includegraphics[width=0.34\textwidth]{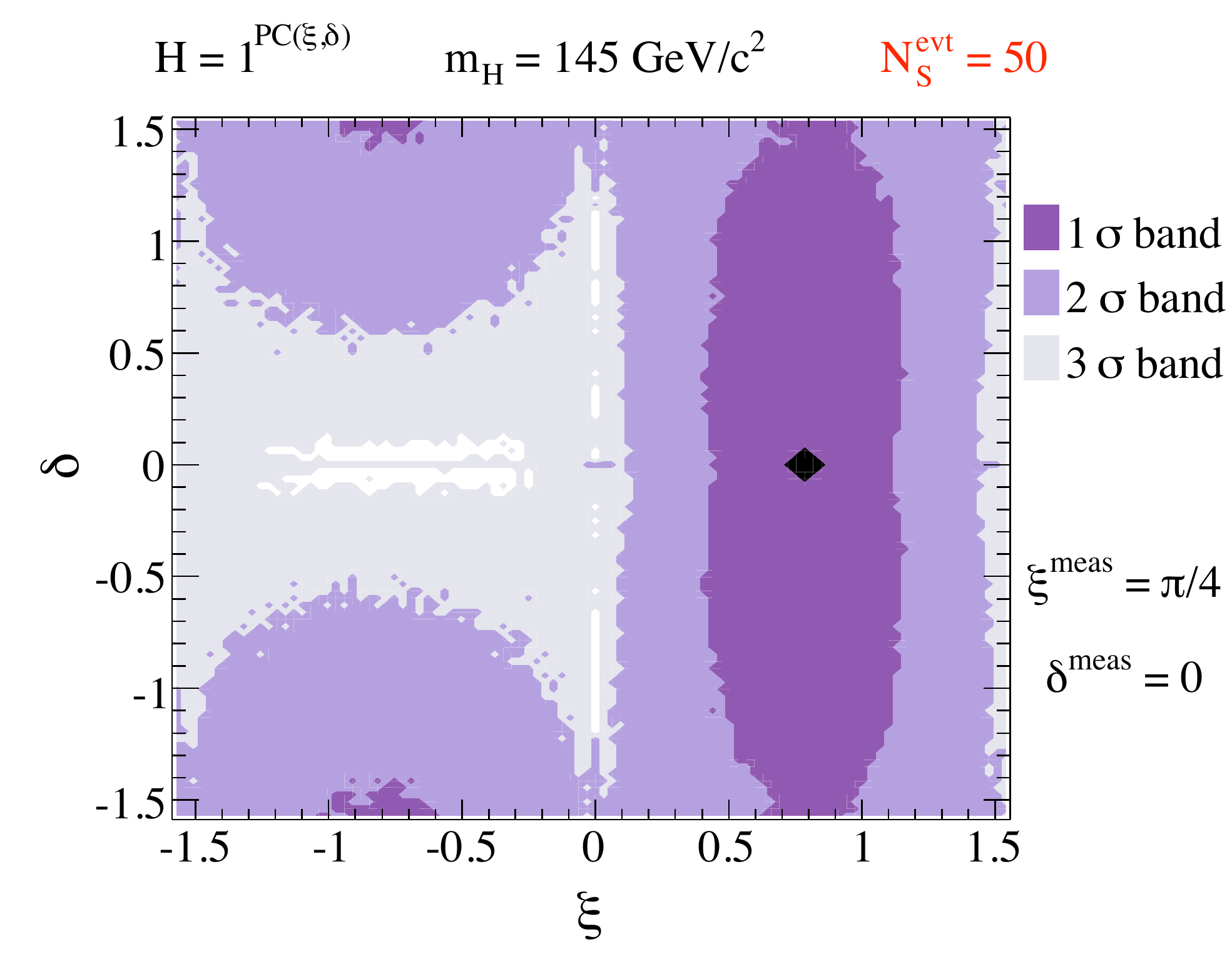}
\includegraphics[width=0.34\textwidth]{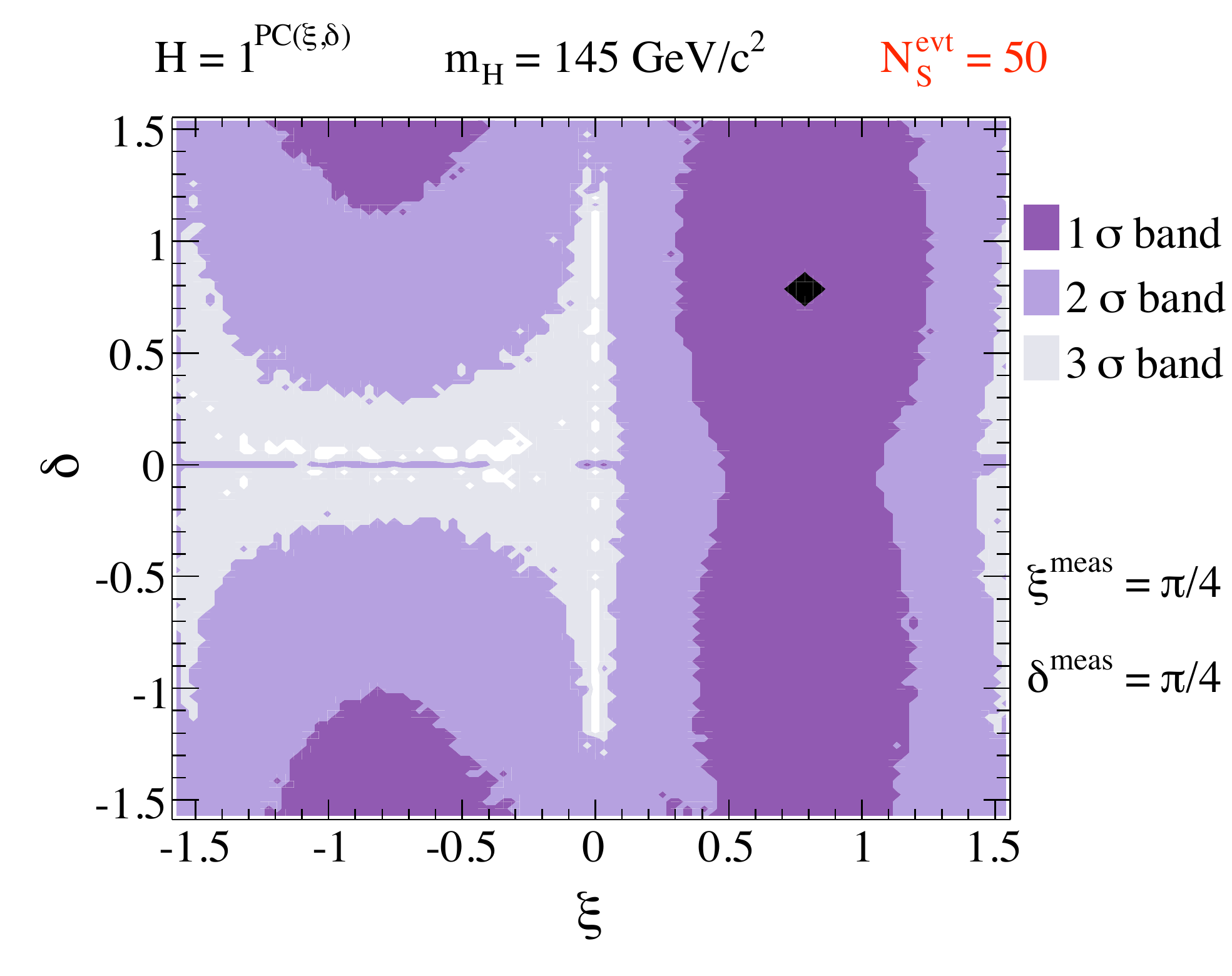}
\caption{Confidence intervals for measured values of $\xi$ and
  $\delta$ for a $J$$=$$1$ resonance with a mass 145 GeV/c$^{2}$
  and $N_S$$=$$50$ events. The input
  values, indicated by diamonds, are reported alongside the figures.
  \label{fig:SPINONEPARAM_145}}
\end{center}
\end{figure}

In Figs.~\ref{fig:SPINONEPARAM_145} we show as an example the
confidence intervals for measurements of $\xi$ and $\delta$ at 
$m_H $$=$$145$ GeV/c$^{2}$
%and 350 GeV/c$^{2}$. 
The ability to resolve the value of the $P$-mixing angle $\xi$ is
modest. The measurement of the $CP$-mixing angle $\delta$ is still
harder. Specifically, we see a large sign ambiguity in the measured
$\delta$, indicating that, with $N_S\sim 50$, it is difficult to
resolve the nature of the interference term, as was the case for
$J$$=$$0$.

Overall, we find that a precise measurement of $\xi$ and $\delta$ for
a $J$$=$$1$ resonance is very difficult. The conclusion of this section
and Sec.~\ref{sec:spinone} is that, if a new $J$$=$$1$ boson is found, a
modest number of events will suffice to exclude $J$$=$0, 2 alternatives
with high significance. Before many more events are gathered, and with
only the tools we have studied, it is hard to make precise statements
about the nature of a $J$$=$$1$ resonance, other than its spin.
%%%%%%%%%%%%%

%               CONCLUSIONS.TEX

%%%%%%%%%%%%%%
%%%%%%%%%%%%%%%%%%%%%%%%%%%%%%%%%%%%%%%%%%%%%%%%%%%%%

\section{Conclusions, caveats, and outlook}
\label{sec:LHCpotential}

\begin{figure}[htbp]
\begin{center}
\includegraphics[width=0.35\textwidth]{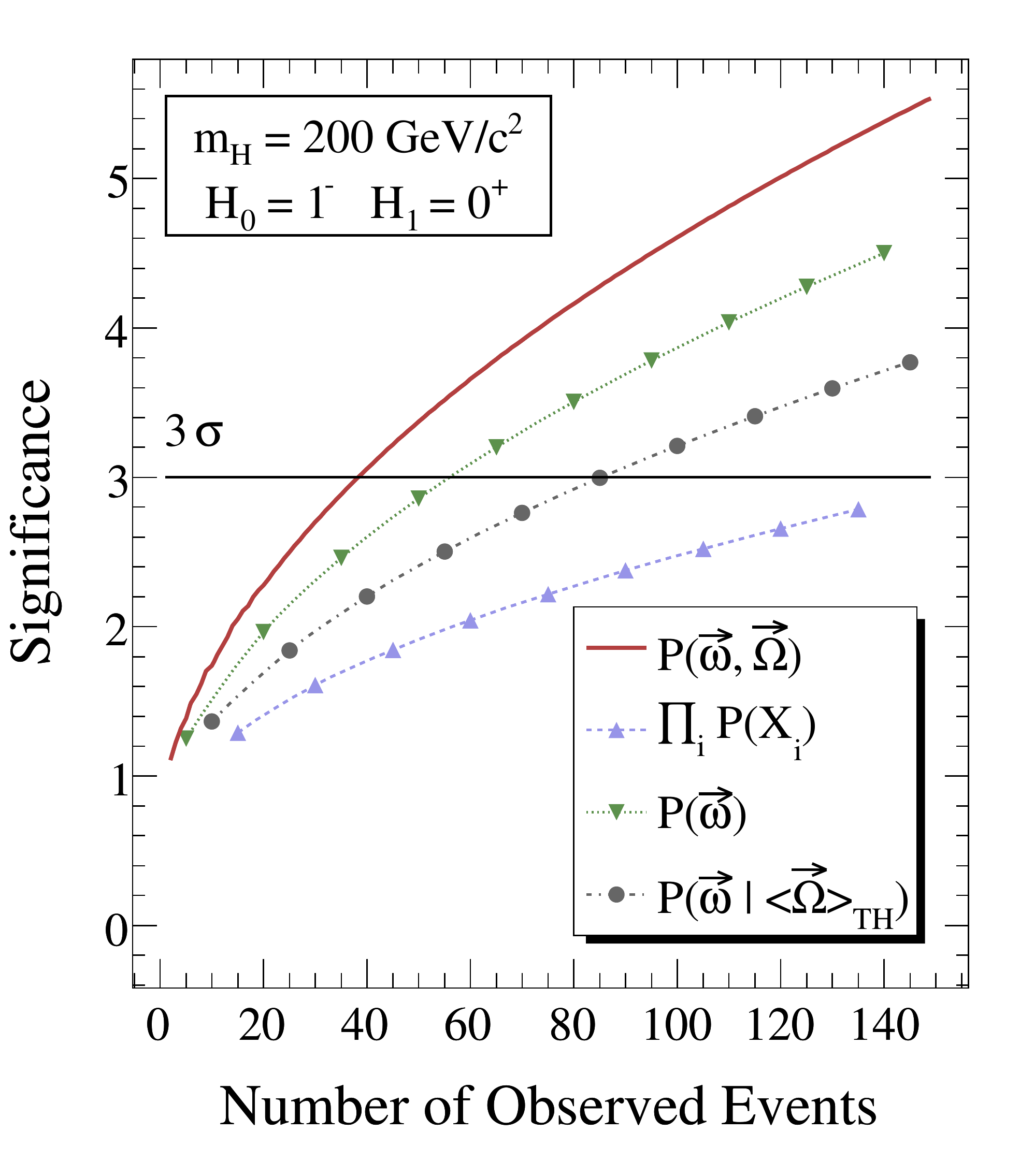}
\caption{Median  significance for rejecting $1^{-}$
in favor of $0^{+}$, for different likelihood constructions used in the
log likelihood ratio test statistic. ${H}_{0}$ is always considered the true
hypothesis. 
\label{fig:prop}}
\end{center}
\end{figure}

It is no surprise that using all of the decay information in a data
sample provides better discrimination of the identity of a new heavy
resonance than examining a single angular distribution or
asymmetry. Nevertheless, one might be tempted, given a small data set
constituting an initial discovery, to settle for a stripped-down
analysis. Our study quantifies the cost, in units of integrated LHC
luminosity, of pursuing such sub-optimal analysis strategies, as
illustrated in Fig.~\ref{fig:prop} for the benchmark $m_H$$=$$200$
GeV/c$^2$.

In this figure we compare the discrimination between the $0^{+}$ and
$1^{-}$ hypotheses for likelihood definitions that exploit different
sets of variables, with the notation that $P(a_{1}, \cdots ,a_{N})$
denotes N-dimensional {\it pdfs} in the correlated variables
$\{a_{1},\cdots,a_{N}\}$. Here $\prod_{i} P(X_{i})$ is constructed
from one-dimensional {\it pdfs} for all variables, ignoring
(erroneously) their correlations. $P(\vec{\omega}\, |
\langle\vec{\Omega}\rangle_{\rm TH})$ are {\it pdfs} including the
variables $\vec{\omega}$ and their correlations, but with the
hypothesis $1^{-}$ represented by a {\it pdf} in which the variables
$\vec{\Omega}$$=$$\{\Phi, {\rm cos}\,\Theta\}$ have been integrated out.
  
The likelihood $P(\vec{\omega}\, | \langle\vec{\Omega}\rangle_{\rm
  TH})$ performs badly even relative to $P(\vec{\omega})$, which uses
fewer angular variables. The two differ only in that the first
construction implicitly assumes a uniform $4\,\pi$ coverage of the
observed leptons (an assumption customary in the literature) as if the
muon $p_{T}$ and $\eta$ analysis requirements did not depend on the
$\vec\Omega$ angular variables. The differing results arise from the
strong correlation between the variables $\Phi$ and $\phi$ in the
$J$$=$$1$ {\it pdfs}, such that phase space acceptance sculpting of the
$\Phi$ distribution alters the $\phi$ distribution, as
discussed in Sec.\ref{sec:ana} and \ref{sec:SM_v_1}.

\begin{figure}[htbp]
\begin{center}
\includegraphics[width=0.42\textwidth]{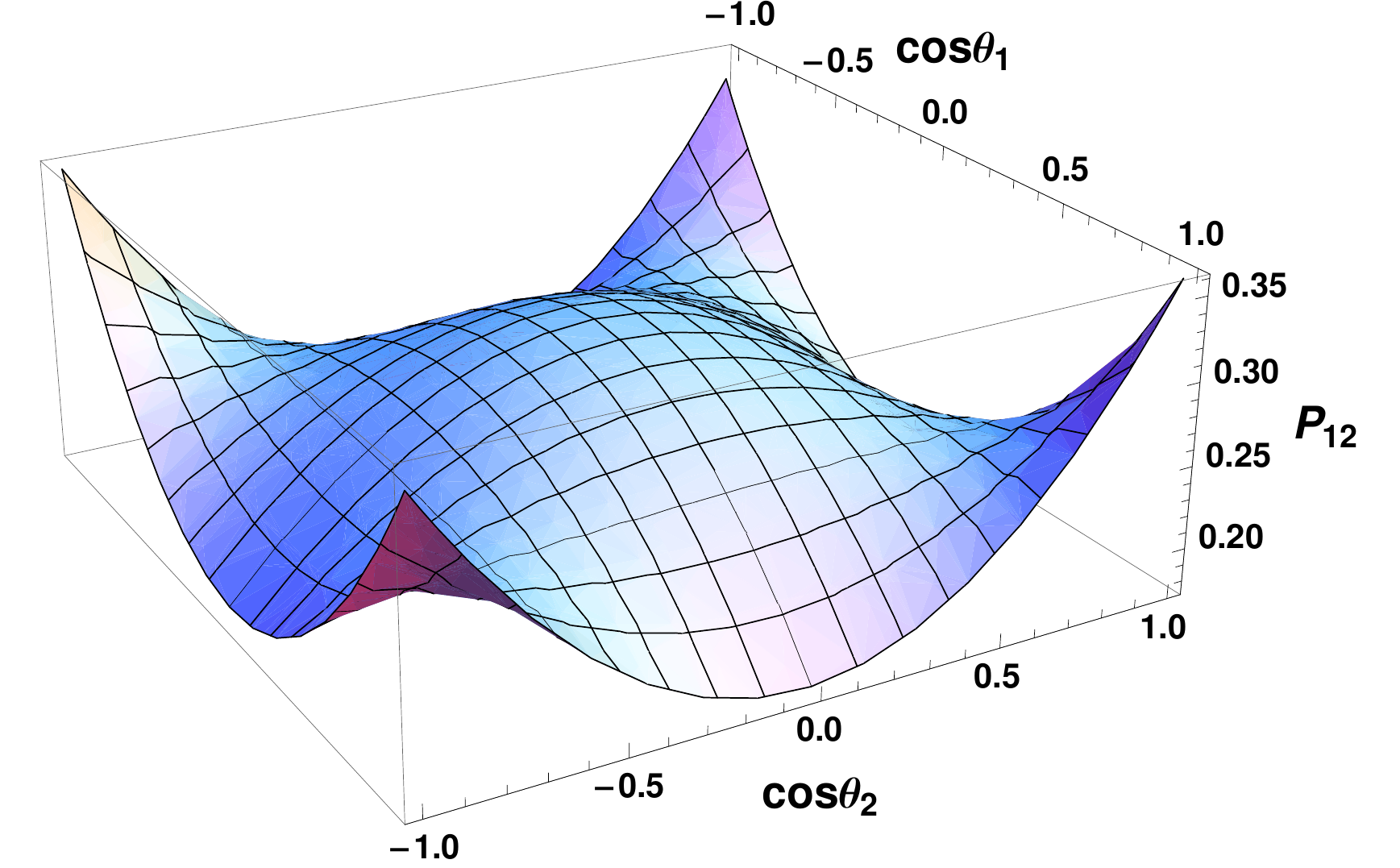}\\
\includegraphics[width=0.42\textwidth]{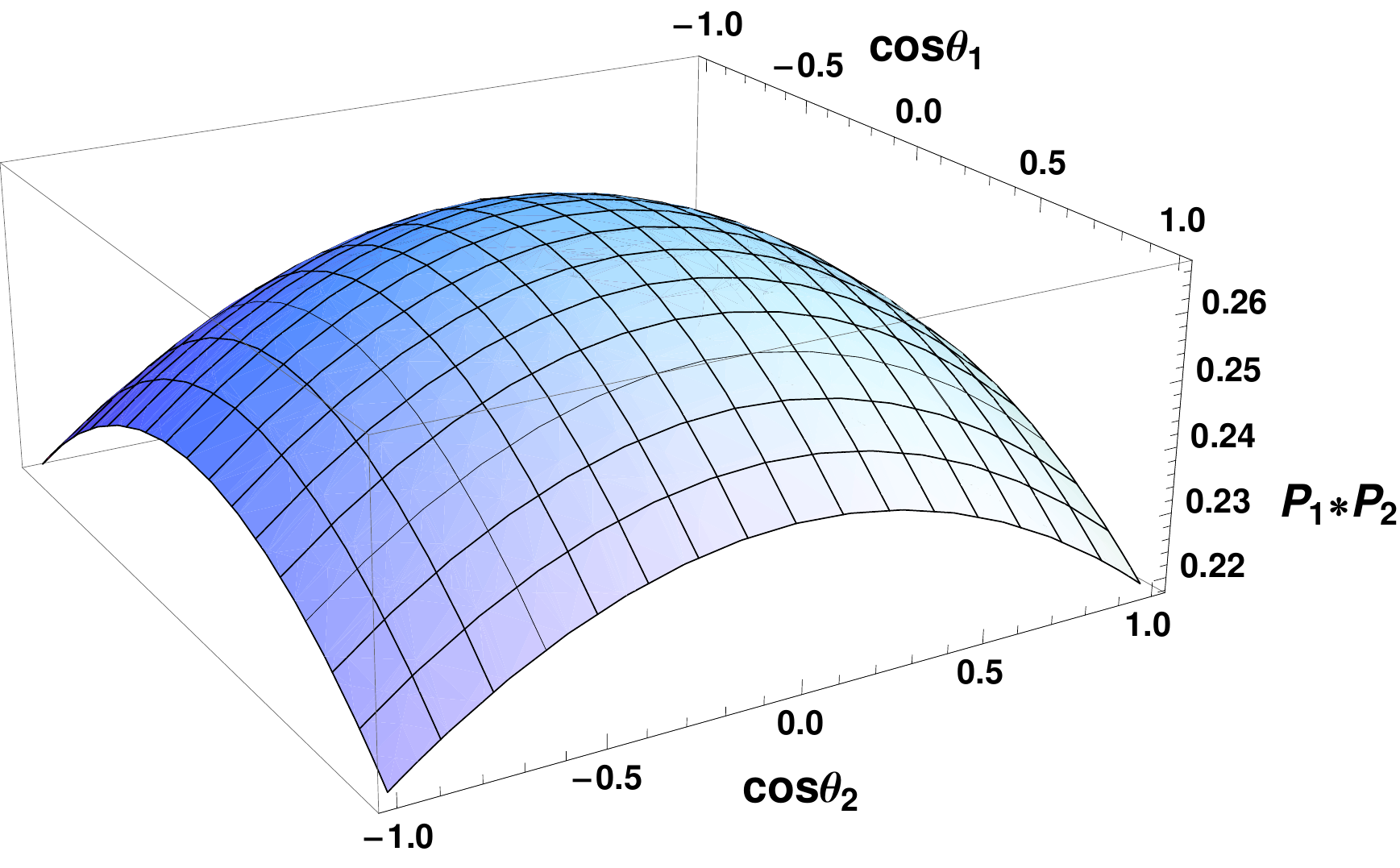}
\caption{The normalized theoretical {\it pdfs} in the variables cos$\,\theta_1$ and cos$\,\theta_2$
(integrated over $\phi$) for $J$$=$$0^+$ and $m_H$$=$$200$ GeV/c$^2$. Top: the 2D {\it pdf}
$P_{12}$$=$$P[{\rm cos}\,\theta_1,\,{\rm cos}\theta_2]$. Bottom: the product $P_1*P_2$$=$$P[{\rm cos}\,\theta_1]\times P[{\rm cos}\,\theta_2]$
of the 1D {\it pdfs}. 
\label{fig:2Dplot}}
\end{center}
\end{figure}

Additionally we find that treating the correlated angular variables as
uncorrelated, as in the $\prod_{i} P(X_{i})$ example of Figure
~\ref{fig:prop}, not only degrades the discrimination significance but
also produces a real chance of falsely labeling the quantum numbers of
the new resonance.
Assume for example the SM, with $m_H$$=$$200$
GeV/c$^2$.  Let the data be fit to either a fully
correlated {\it pdf} or an uncorrelated one. The projections
of the corresponding theoretical {\it pdfs}, involving only the variables 
cos$\,\theta_1$ and cos$\,\theta_2$, are illustrated in Fig.~\ref{fig:2Dplot}. 
On the top (bottom) of the figure we show $P[{\rm cos}\,\theta_1,\,{\rm cos}\,\theta_2]$
($P[{\rm cos}\,\theta_1]$$\times$$P[{\rm cos}\,\theta_2]$). With limited statics -- insufficient
to distinguish between the correlated and uncorrelated distributions --
the correct conclusion will be reached: the data are compatible with the SM.
But, as the statistics are increased, the data will significantly deviate from
the $P[{\rm cos}\,\theta_1]\times P[{\rm cos}\,\theta_2]$ distribution, and a false
rejection of the SM hypothesis would become increasingly supported.
%%%%%%%%%%%%%%%%%%%%%%%%%%%%%%%%%%%%%%%%%%%%%%%%%%%%%%%%%%%%%%%%%%%
%%%%%%%%%%%%%%%%%%%%%%%%%%%%%%%%%%%%%%%%%%%%%%%%%%%%%%%%%%%%%%%%%%%
%%%%%%%%%%%%
%%%%%%%%%%%%%%%%%%%%%%%%%%%%%%%%%%%%%%%%%%%%%%%%%%%%%%%%%%%%%%%%%%%

Overall, we have demonstrated that small signal samples in the $ZZ \to
4\ell$ or $ZZ^* \to 4\ell$ decay channels, as might be available at
the moment of discovery, could be sufficient to characterize a
putative Higgs particle. Below we summarize these results in more
detail.

\subsection{Summary of pure case discrimination}
Amongst the many comparisons considered in our analysis, the ones
between simple hypotheses are the most readily summarized. This we do
in Tables I,II for $m_H$$=$$145$ GeV/$c^{2}$ for all pure-case comparisons
between $J$$=$0, 1 parent particles, and in Tables ~III,IV (V,VI) for
$m_H$$=$$200$ (350) GeV/$c^{2}$, for all pure-case comparisons between
$J$$=$0, 1, 2 parent particles.

\begin{table}[ht!]
\centering
\begin{tabular}{|c||
 c| c| c| c|}
\hline
$\mathbb{H}_{0} \Downarrow \mathbb{H}_{1}\Rightarrow$ & $0^{+}$ & $0^{-}$ & $1^{-}$ & $1^{+}$ \\
\hline
\hline
 $0^{+}$  & -- & 17 & 12 & 16 \\ 
%\hline
 $0^{-}$  & 14 & -- & 11 & 17 \\ 
%\hline
 $1^{-}$  & 11 & 11 & -- & 35 \\ 
%\hline
 $1^{+}$  & 17 & 18 & 34 & -- \\ 
\hline
\end{tabular}
\label{table:1453sigma}
\caption{Minimum number of observed events such that the median significance for rejecting $\mathbb{H}_{0}$ in favor of the hypothesis $\mathbb{H}_{1}$ (assuming $\mathbb{H}_{1}$ is
right) exceeds $3\,\sigma$ with $m_H$$=$$145$ GeV/c$^{2}$. }
\end{table}

\begin{table}[ht!]
\centering
\begin{tabular}{|c||
 c| c| c| c|}
\hline
$\mathbb{H}_{0} \Downarrow \mathbb{H}_{1}\Rightarrow$ & $0^{+}$ & $0^{-}$ & $1^{-}$ & $1^{+}$ \\
\hline
\hline
 $0^{+}$  & -- & 52  & 37 & 50 \\ 
%\hline
 $0^{-}$  & 44 & -- & 34 & 54  \\ 
%\hline
 $1^{-}$  & 33 & 32  & -- & 112  \\ 
%\hline
 $1^{+}$  & 54  & 55  & 109 & -- \\ 
\hline
\end{tabular}
\label{table:1455sigma}
\caption{Same as Table I, but requiring that the median significance exceeds $5\,\sigma$.}
\end{table}

\begin{table}[ht!]
\centering
\begin{tabular}{|c||
 c| c| c| c| c| }
\hline
$\mathbb{H}_{0} \Downarrow \mathbb{H}_{1}\Rightarrow$ & $0^{+}$ & $0^{-}$ & $1^{-}$ & $1^{+}$ & $2^{+}$ \\
\hline
\hline
 $0^{+}$  & -- & 24 & 45 & 62 & 86  \\ 
 $0^{-}$  & 19 & -- & 19 & 19 & 38  \\ 
 $1^{-}$  & 40 & 18 & -- & 90 & 48  \\ 
 $1^{+}$  & 56 & 19 & 85 & -- & 66  \\ 
 $2^{+}$  & 86 & 45 & 54 & 70 & --  \\ 
\hline
\end{tabular}
\label{tab:200_3sigma}
\caption{Minimum number of observed events such that the median significance for rejecting $\mathbb{H}_{0}$ in favor of the hypothesis $\mathbb{H}_{1}$ (assuming $\mathbb{H}_{1}$ is right) exceeds $3\,\sigma$ with $m_H$$=$$200$ GeV/c$^{2}$. }
\end{table}

\begin{table}[htdp]
\centering
\begin{tabular}{|c||
 c| c| c| c| c|}
\hline
$\mathbb{H}_{0} \Downarrow \mathbb{H}_{1}\Rightarrow$ & $0^{+}$ & $0^{-}$ & $1^{-}$ & $1^{+}$ & $2^{+}$  \\
\hline
\hline
 $0^{+}$  & -- & 76 & 146 & 203 & 287  \\ 
%\hline
 $0^{-}$  & 59 & -- & 60 & 61& 123  \\ 
%\hline
 $1^{-}$  & 130 & 57  & -- & 297 & 156  \\ 
%\hline
 $1^{+}$  & 182 & 58 & 278 & -- & 217  \\ 
%\hline
 $2^{+}$  & 287  & 146 & 178 & 230 & --   \\ 
\hline
\end{tabular}
\label{tab:200_5sigma}
\caption{Same as Table III, but requiring that the median significance exceeds $5\,\sigma$.}
\end{table}

\begin{table}[ht!]
\centering
\begin{tabular}{|c||
 c| c| c| c| c|}
\hline
$\mathbb{H}_{0} \Downarrow \mathbb{H}_{1}\Rightarrow$ & $0^{+}$ & $0^{-}$ & $1^{-}$ & $1^{+}$ & $2^{+}$ \\
\hline
\hline
 $0^{+}$  & -- & 8 & 21 & 24 & 11 \\ 
%\hline
 $0^{-}$  & 9 & -- & 22 & 22 & 36 \\ 
%\hline
 $1^{-}$  & 24 & 22 & -- & 81 & 46 \\ 
%\hline
 $1^{+}$  & 26 & 22 & 80 & -- & 56  \\ 
%\hline
 $2^{+}$  & 15 & 39 & 55 & 73 & --  \\ 
\hline
\end{tabular}
\label{tab:350_3sigma}
\caption{Minimum number of observed events such that the median significance for rejecting $\mathbb{H}_{0}$ in favor of the hypothesis $\mathbb{H}_{1}$ (assuming $\mathbb{H}_{1}$ is right) exceeds $3\,\sigma$ with $m_H$$=$$350$ GeV/c$^{2}$. }
\end{table}

\begin{table}[hb!]
\centering
\begin{tabular}{|c||
 c| c| c| c| c|}
\hline
$\mathbb{H}_{0} \Downarrow \mathbb{H}_{1}\Rightarrow$ & $0^{+}$ & $0^{-}$ & $1^{-}$ & $1^{+}$ & $2^{+}$  \\
\hline
\hline
 $0^{+}$  & -- & 25 & 67 & 77  & 35  \\ 
%\hline
 $0^{-}$  & 26 & -- & 68 & 68  & 118  \\ 
%\hline
 $1^{-}$  & 76 & 68 & -- & 268  & 149 \\ 
%\hline
 $1^{+}$  & 83  & 68  & 263  & -- & 184 \\ 
%\hline
 $2^{+}$  & 46 & 127 & 181 & 240 & --  \\ 
\hline
\end{tabular}
\label{tab:350_5sigma}
\caption{Same as Table V, but requiring that the median significance exceeds $5\,\sigma$.}
\end{table}

Overall, the discrimination power of the hypothesis tests is very
impressive.  The $m_H$$=$$200$ GeV/c$^2$ benchmark example is the one requiring
the largest statistics to reach a given discrimination at a given
level of confidence. Compared with the $m_H$$=$$350$ GeV/c$^2$ case, this is
because various coefficients of the angular dependences vanish at the
$m_H$$=$$2\,M_Z$ threshold.  The $m_H$$=$$145$ GeV/c$^2$ example fares better than
the 200 GeV/c$^2$ one for the same reason, amplified by the extra lever-arm
supplied by a non-trivial $M_{Z^*}$ distribution.

The tables also show that the discriminating power between two given
hypotheses is approximately symmetric under the interchange of `right'
and `wrong'. Telling $1^+$ from $1^-$ is always difficult but not
impossible, a fact of relevance for a $Z'$ look-alike analysis.  The
level of significance does not obey a na\"ive $N(\sigma)\propto
\sqrt{N_S}$ law. However we find by inspection that an approximation
of the form $N(\sigma)=a+b\, \sqrt{N_S}$ works well, allowing one to
extrapolate to larger numbers of events than presented here.

Other lessons from the tables are case-by-case specific, reflecting
the mass-dependent quantum-mechanical entanglement between the decay
variables.  Some examples are: distinguishing the `natural-parity'
$J$$=$$0^+$ and $1^-$ hypotheses for $m_H$$=$$145$ GeV/c$^2$ requires only a dozen
signal events for $3\,\sigma$ discrimination.  For 200 GeV/c$^2$,
discriminating $0^+$ from $0^-$ is relatively easy, but distinguishing
$0^+$ from $2^+$ is difficult. For 350 GeV/c$^2$, contrariwise, $2^+$ is
relatively easy to disentangle from $0^+$, but not from $0^-$.

\subsection{Summary of mixed cases, $\mathbf{ CP}$ and compositeness
  discrimination}

We find that direct sensitivity to $CP$ odd, parity odd $XP$
interference effects, or to $CP$ odd, parity even $XQ$ interference
effects, will require signal samples about an order of magnitude
larger than considered here.  We have also observed that with much
smaller statistics it may be possible to conclude that a mix of $X$
and $P$ (or $X$ and $Q$) couplings is favored over just the pure $X$
(i.e.~$0^+$) or pure $P$ (i.e.~$0^-$) couplings alone.  Such a
conclusion would be tantamount to demonstrating $CP$ violation in the
Higgs sector. However this scenario relies on large $CP$ violation,
and even in this favorable case one cannot tell an $X$ and $P$ mixture
from an $X$ and $Q$ mixture without more data than what is required to
establish discovery.

In the case of a composite Higgs, it may be conceivable that the Higgs
is as `soft' as a pion, in the sense of having an inverse radius and a
mass of comparable magnitude. In this scenario we have seen that the
angular distributions associated to the $X$ and $Y$ couplings are
similar after integrating over the decay angles. As a result there can
be strong destructive interference between these contributions.  For
our lighter mass benchmarks we find good discrimination of pure $0^+$
from the mixed composites. For the heavier $m_H$$=$$350$ GeV/c$^2$
example, discrimination based on decay angles is poor unless the
strong interference effects are present; here we also observed that
substantial enhancement or suppression of the HLL$\to ZZ$ branching
fraction can provide another important discriminator.

For mixed cases, one could worry that certain combinations of exotic
couplings might let an HLL successfully masquerade as a $0^+$ Higgs,
even when all the pure case exotics are excluded. For spin 1 HLLs
we have shown that this does not happen.  In fact we find that when we
have an SM Higgs, the entire family of mixed coupling spin 1 HLLs can
be excluded at approximately the same expected level of significance
as for the pure $1^-$ or $1^+$ cases. An even stronger result is that
the general spin 0 hypothesis can be conclusively discriminated from
the general spin 1 hypothesis, at or close to the moment of discovery.

%%%%%%%%%%%%%

\subsection{Analysis limitations}

In our analysis we focused on decay information, 
exploiting an approximate factorization
between observables related to Higgs (or HLL) production and
observables related to decay. The factorization is only
approximate because of phase space acceptance effects
and, in the case of spin $>$$0$ HLLs, correlations between
the initial and final state particles. In a real data analysis
one would want to include production information, which
in turn would require a detailed knowledge of radiative
corrections, PDFs, and full detector simulation for the HLLs.
Such an analysis is beyond the scope of this paper. 
Within our narrower scope we have incorporated
as much as we could all the issues that make a conceptual difference 
in the strategy. When we have made approximations or have neglected
certain effects, it is because the detailed inclusion of these effects would 
not have a qualitative impact on our results.

The QCD corrections to the signal predictions for $d^2\sigma/dp_T\,d\eta$
are large, as is well-studied for the SM Higgs (see, for example, 
\cite{Djouadi:2005gi},\cite{Ahrens:2008nc,deFlorian:2009hc} and references therein.)
The impact on the total cross sections is not relevant to our analysis,
but the corrections to the $(p_T,\eta)$ distributions will modify the
phase space acceptance effects on the distributions of the final-state
leptons. For the SM Higgs we have included these corrections at NLO,
and a recent study shows that the effects of NNLO corrections on the
final-state lepton distributions are not dramatic \cite{Grazzini:2008tf}.
We have not
included the differences between the phase space acceptance effects
for the SM Higgs and those for the various HLLs, but we performed a
comparison at LO to see that these differences are small
compared to the acceptance effects themselves.

There are electroweak radiative corrections that directly
involve the final-state leptons. For the SM Higgs these
corrections have been computed and 
studied in detail \cite{Bredenstein:2006rh};
the corrections are of the order of 5 to 10\% and
cause a mild distortion of the angular distributions.
These effects should be included in a complete analysis,
but they do not introduce anything
conceptually new to the methodology proposed in this study,
and their inclusion involves details of the experimental
treatment of the vertex and subsequent radiations
by electrons and muons.

We only
considered the dominant $ZZ$ background, and only at LO.
It would be useful to include a more comprehensive treatment of
the SM backgrounds to the golden channel and to
use the full signal-to-background discrimination, e.g.~by
adding the Z mass distribution to the likelihood definition.
A complete treatment of the backgrounds would require full
detector simulation.

Our treatment of couplings and HLLs was not exhaustive, since we have
ignored gauge invariant operators with dimension greater than $6$,
have only examined one case of spin 2 HLL, and have not even
mentioned the possibility of HLLs with spins higher than $2$.  At some
point Occam's razor obviates the need for such comparisons:
``Raffiniert ist der Herr Gott, aber boshaft ist Er nicht'', to quote
a known author \cite{Einstein}.

The likelihood analyses pursued here are very computing intensive,
since $5\,\sigma$ discrimination implies simulating sufficiently many
pseudo-experiments to fill out what amounts to the $5\,\sigma$
tails in multidimensional likelihood distributions, where they are typically
highly non-Gaussian.  The analysis
presented here used more than $10^{14}$ pseudo-experiments in
total.

%%%%%%%%%

\subsection{Outlook}

We have seen that by exploiting the full decay information in the
golden channel we should be able to say a lot about the identity of a
putative Higgs resonance around the moment of discovery.  Our results
also show that asymptotically, utilizing the full physics run of the
LHC, it should be possible to explore very detailed properties of such
a resonance.

It has not escaped our attention that there are many processes other
than the $ZZ$ decays of a heavy resonance whose characterization may
benefit from an analysis of the kind that we have performed here.

\subsection*{Note added:} While this manuscript was in preparation we
received the preprint \cite{Gao:2010qx}, reporting on an analysis
similar to what we have presented here.
%%%%%%%%%%%%%%%%%%%%%  
  
\subsection*{Acknowledgments}
We dedicate this work to the memory of our colleagues Andrew Lange and
Juan Antonio Rubio.  The authors are grateful to Andrew Cohen, Bel\'en
Gavela, Keith Ellis, Shelly Glashow, Ken Lane, Ken Lee, Michelangelo Mangano, Chiara
Mariotti, Guido Martinelli, Sezen Sekmen, Riccardo Rattazzi, Raman
Sundrum, Steven Weinberg, Jan Winter and Mark Wise for useful discussions.  JL
acknowledges the hospitality of the CERN Theory Department and support
from the Aspen Center for Physics.  Fermilab is operated by the Fermi
Research Alliance LLC under contract DE-AC02-07CH11359 with the
U.S. Dept. of Energy. CR and MS are supported in part by the
U.S. Dept. of Energy under contact DE-FG02-92-ER40701.

%%%%%%%%%%%%%%%%%%%%%%%

\appendix

\section{$\mathbf{SU(2)_{_L}\times U(1)_Y}$ gauge-invariant couplings\label{GIL}}

To write Lagrangians generating the couplings of Sec.~\ref{generalcouplings} and respecting
the electroweak gauge symmetry one must specify the electroweak
charges of the Higgs look-alikes. Consider the example of HLLs that are ``neutral", 
i.e.~are weak singlets and have zero hypercharge. For the scalar case, in a conventional notation
for isovector and isoscalar gauge fields, the lowest-dimensionality Lagrangian density is:
\begin{eqnarray}
&&\hspace*{-20pt}
L={1\over \Lambda}\, H\,(A_1\, \vec W_{\mu\alpha}\, \vec W^{\mu\alpha}
+A_2 \, B_{\mu\alpha} B^{\mu\alpha})
\nonumber\\
&&
+{1\over \Lambda}\, H\,i\,\epsilon^{\mu\alpha\sigma\tau}
(A_3\, \vec  W_{\mu\alpha}\vec W_{\sigma\tau}
+A_4 \, B_{\mu\alpha} B_{\sigma\tau})\, ,
\label{eq:lagscalar}
\end{eqnarray}
with $A_i$ arbitrary constants and $\Lambda$ a mass parameter. This object
generates, amongst others, the couplings of Eq.~(\ref{generalscalar}). The ``true" 
dimensionality of the operators in Eq.~(\ref{generalscalar})
is that of the ones appearing in Eq.~(\ref{eq:lagscalar}), that is,
dimension five.

The form of Eq.~(\ref{eq:lagscalar}) results in a coupling 
$H Z_{\mu\alpha}\, Z^{\mu\alpha}\to 2\,p_1\cdot p_2\, g_{\mu\alpha}-2\, k_\mu k_\alpha$,
establishing a relation between $X$ and $Y+i\,Z$ in Eq.~(\ref{generalscalar}).
We do not impose it, for it is not general even at tree level.
Consider, for instance, a model with a 
conventionally-charged but otherwise non-standard HLL, dubbed $\Phi$ before
the spontaneous symmetry breaking. Call $V_{\mu\nu}$ any of the field tensors
in Eq.~(\ref{eq:lagscalar}). The operators in this Lagrangian could be ``descendants" 
of dimension 6 operators of the form $\Phi^\dag \Phi\,V^2$, 
with $\Phi\to H+v$, see e.g.~\cite{Plehn:2001nj}.
In such a case there would be a standard-like $g_{\mu\nu}$
coupling plus the one induced by the higher-dimensional operators.

For the case of a spin-1 neutral HLL,  $H_\rho$,
the lowest-dimension gauge-invariant Lagrangian generating the couplings of
Eq.~(\ref{generalvector}) is built of operators of dimension 6:
\begin{eqnarray}
&&\hspace*{-20pt}
\Lambda^2 L=
(\partial^\mu\,H^\alpha+\partial^\alpha\,H^\mu)\,
(A_1 \,\vec W_{\mu}^\lambda\, \vec W_{\alpha\lambda}
+A_2\, B_{\mu}^\lambda B_{\alpha\lambda})
\nonumber\\
&&\hspace*{-20pt}
+{{\epsilon^{\mu\nu\alpha\rho}}}
[A_3(\vec W_{\mu}^{\lambda}\overleftrightarrow{D}_\alpha\vec W_{\nu\lambda})H_\rho
+A_4 (B_{\mu}^{\lambda}\overleftrightarrow{\partial}_\alpha B_{\nu\lambda})H_\rho],
\label{lagvector}
\end{eqnarray}
where $D_\alpha$ is the covariant derivative and
$(M \overleftrightarrow{D}_\alpha N)\equiv M D_\alpha N - (D_\alpha\,M) N$.

For a canonical-dimension spin-2 neutral HLL,  $H_{\mu\nu}$,
the lowest-dimension gauge-invariant Lagrangian has couplings
of dimension 5:

\begin{eqnarray}
&&\hspace*{-20pt}
L={1\over \Lambda}\, H_{\mu\nu}\,(A_1\, \vec W^{\mu}_{\alpha}\, \vec W^{\nu\alpha}
+A_2 \, B^{\mu}_{\alpha} B^{\nu\alpha})
\nonumber\\
&&\hspace{-15pt}
+{1\over \Lambda}\, H^{\nu\rho}\,i\,\epsilon_{\mu\nu\alpha\beta}
(A_3\, \vec  W^{\mu\alpha}\vec W^{\rho\beta}
+A_4 \, B^{\mu\alpha} B^{\rho\beta})\, .
\label{lagtensor}
\end{eqnarray}

The consideration of gauge-invariant constructions for HLLs with
non-trivial electroweak charges would take us well beyond
the scope of this paper.

%%%%%%%%%%%%

\section{Phase space for $\mathbf{ZZ^*}$}

In the case in which one of the two $Z$ bosons is off-shell, the dependence on its mass
($M_{Z^*}$, either $m_1$ or $m_2$) is an extra handle in determining the shapes of signal
and backgrounds.
Let $p_{\rm cms}\equiv |\vec p\,[Z]|=m_1\,\gamma_1\,\beta_1=m_2\,\gamma_2\,\beta_2$
be the momentum of one or the other $Z$ in the $H$ center-of-mass system:
\begin{eqnarray}
&&\hspace*{-25pt}
p_{\rm cms}={1\over 2\,m_H}\,\Theta[m_H-(M_Z+M_{Z^*})]\times
\nonumber\\
&&\hspace*{-15pt}
\sqrt{m_H^2-(M_Z-M_{Z^*})^2}
\,\sqrt{m_H^2-(M_Z+M_{Z^*})^2}
\,.
\label{pcms}
\end{eqnarray}

Let ${\cal M}$ be the matrix element
for the process. The expectation for 
the rate of events, including the dependence on $M_{Z^*}$, is:
    \begin{eqnarray}
&&\hspace*{-25pt}
{dN \over d{\rm cos}\,\theta _1\,d{\rm cos}\,\theta _2\,d\phi\,d{\rm cos}\,\Theta\,d\Phi\,dM_{Z^*}}
\nonumber\\
&&\hspace*{0pt}
\propto  |{\cal M} |^2\;
 {M_{Z^*}\;p_{\rm cms}\over \left(M_{Z^*}^2-M_Z^2\right)^2+M_{Z^*}^2\,\Gamma_Z^2},
 \label{M*}
    \end{eqnarray}
    with $|{\cal M} |^2$ an explicit function of $c_1,\,c_2,\,\phi,\,\Theta,\,\Phi$
    and $M_{Z^*}$ for each specific case to be discussed.

%%%%%%%%%%%%%%%%%%%%%%

\section{General results for spin 0 coupled to $\mathbf{ZZ^*}$}
\label{generalformulae}

In Section \ref{sec:couplings} we have already written the angular distributions 
$d\Gamma[0^+]$ and $d\Gamma[0^-]$ for the pure scalar and pseudoscalar cases,
see Eqs.~(\ref{eq:standang}), (\ref{purepseudoscalarP}). We also
discussed the $T$-odd and $C$-odd
interferences between the standard coupling --proportional to $X$ in Eq.~(\ref{generalscalar})--
and the $P$ and $Q$ terms of the same equation. Thus we defined
$d\Gamma[0,{\rm Todd}]$ and $d\Gamma[0,{\rm Codd}]$ in Eqs.~(\ref{0Todd}), (\ref{0Codd}).
Similarly we discussed
the complete result for the `composite' case with $X\neq 0$ and $Y\neq 0$,
defining $d\Gamma_{XY}$ and $d\Gamma_{YY}$ in 
Eqs.~(\ref{interference}), (\ref{eq:YY}).
This allows us to gather the results corresponding to the most general
deviations from the SM Higgs couplings:
\begin{eqnarray}
{d\Gamma [0]}&&=X^2\,\,d\Gamma [0^+]+(P^2+Q^2)\,\,d\Gamma [0^-]
\nonumber\\
&&+X\,P\,d\Gamma [0,\,{\rm Todd}]+X\,Q\,d\Gamma [0,\,{\rm Codd}]
\nonumber\\
&&+X\,Y\,d\Gamma_{XY}
+(Y^2+Z^2)\,d\Gamma_{YY} \; .
\label{partscalar1}
\end{eqnarray}

To obtain the complete spin 0 result one must add to
Eq.~(\ref{partscalar1}) the interferences between the non-standard
terms themselves:
\begin{eqnarray}
\Delta{d\Gamma [0]}&=&XZ\,d\Gamma_{XZ}+YP\,d\Gamma_{YP}
\nonumber\\
&+&YQ\,d\Gamma_{YQ}+
ZP\,d\Gamma_{ZP}+ZQ\,d\Gamma_{ZQ} \, ,
\label{partscalar2}
\end{eqnarray}
where
\begin{equation}
d\Gamma_{XZ}=2 \,\eta\,m_1^3\, m_2^3\, m_H^2\,\gamma_b^2 \, (c_1 + c_2) \,  s\, s_1\, s_2 \,,
\label{XZscalar}
\end{equation}
\begin{equation}
d\Gamma_{YP}=d\Gamma_{ZQ}=-2 \,m_1^4\, m_2^4 \gamma_b^3 \,s\, s_1 s_2  (c_1 c_2 + \eta^2) \, ,
\label{YPscalar}
\end{equation}
\begin{equation}
d\Gamma_{YQ}=-d\Gamma_{ZP}=2\,\eta\,m_1^4\, m_2^4\, \gamma_b^3\,c \,(c_1 + c_2)  s_1 s_2 \;.
\label{YQscalar}
\end{equation}

%%%%%%%%%%%%%%%%%%%%%%

\section{General results for spin 1 coupled to $\mathbf{ZZ^*}$}
\label{app:genspinone}

We produce a spin 1 HLL from annihilation of $q\bar{q}$ with
quark helicity $\tau/2$, $\tau$$=$$\pm 1$.  To an
excellent approximation the coupling of the HLL to light quarks must
conserve helicity, so the antiquark has helicity $-\tau/2$.  
Then the HLL decays to $ZZ$ (or $ZZ^*$), with $Z_2\to \mu^-\mu^+$
with muon helicity $\sigma_2/2$ and $Z_1\to e^-e^+$ with electron
helicity $\sigma_1/2$.  

The fully differential cross section is a sum over $\tau$, $\sigma_1$,
$\sigma_2$ of the squared absolute values of the helicity
amplitudes. In addition the (unmeasured) helicities 
$\lambda_1$, $\lambda_2$ of $Z_1$, $Z_2$ are summed over 0, $\pm
1$, before squaring.

We use the following notation to denote the helicity-conserving coupling
of a $Z$ boson to a massless fermion of helicity $\sigma /2$,
$\sigma$$=$$\pm 1$:
\begin{eqnarray}
g_\sigma = \frac{1}{2}(c_v - \sigma c_a) \; .
\end{eqnarray}
Similarly, we denote the  helicity conserving coupling
of a vector boson HLL to a massless fermion of helicity $\tau /2$,
$\tau = \pm 1$:
\begin{eqnarray}
g_\tau = \frac{1}{2}(g_v - \tau g_a) \; .
\end{eqnarray}
In the full matrix element squared, the dependence on
these vector-fermion-fermion couplings is
\begin{eqnarray}
&&\hspace*{-15pt}
\frac{1}{64}\Bigl[ 
\left( c_v^2 + c_a^2 \right)^2 \left( g_v^2 + g_a^2 \right)
\nonumber\\
&&\hspace*{-10pt}
-2 c_v c_a \left( c_v^2 + c_a^2 \right) \left( g_v^2 + g_a^2 \right) \,(\sigma_1 + \sigma_2)
\nonumber\\
&&\hspace*{-10pt}
-2 \left( c_v^2 + c_a^2 \right)^2 g_v g_a \;\tau
+4 c_v^2 c_a^2 \left( g_v^2 + g_a^2 \right) \;\sigma_1\sigma_2
\nonumber\\
&&\hspace*{-10pt}
+4 c_v c_a \left( c_v^2 + c_a^2 \right) g_v g_a \,(\sigma_1\tau + \sigma_2\tau)
\nonumber\\
&&\hspace*{-10pt}
-8 c_v^2c_a^2 g_v g_a \;\sigma_1\sigma_2\tau
\Bigr] \; ,
\end{eqnarray}
from which we derive the shorthand notation
\begin{eqnarray}
\label{eqn:ourtotalcouplings}
g_1 &\equiv & \left( c_v^2 + c_a^2 \right)^2 \left( g_v^2 + g_a^2 \right) \nonumber\\
g_{\sigma} &\equiv & -4 c_v c_a \left( c_v^2 + c_a^2 \right) \left( g_v^2 + g_a^2 \right)\nonumber\\
g_{\tau} &\equiv & -2 \left( c_v^2 + c_a^2 \right)^2 g_v g_a\nonumber\\
g_{\sigma\sigma} &\equiv & 4 c_v^2 c_a^2 \left( g_v^2 + g_a^2 \right)\nonumber\\
g_{\sigma\tau} &\equiv & 8 c_v c_a \left( c_v^2 + c_a^2 \right) g_v g_a\nonumber\\
g_{\sigma\sigma\tau} &\equiv & -8 c_v^2c_a^2 g_v g_a \; .
\end{eqnarray}

We allow both $Z$ bosons to be off-shell, with invariant masses $m_1$
and $m_2$. Some useful mass combinations are
\begin{eqnarray}
  &&\hspace*{0pt}  m_d^2 \equiv m_1^2-m_2^2   \; , \\
  &&\hspace*{0pt}  M_1^2 \equiv m_H^2 - 3m_1^2 -m_2^2  \;, \quad
  M_2^2 \equiv m_H^2 - m_1^2 -3m_2^2  \; , \nonumber\\
  &&\hspace*{0pt}  M_3^2 \equiv m_H^2 - 2(m_1^2 + m_2^2)  \;, \quad
  M_4^2 \equiv m_H^2 - (m_1^2 + m_2^2)  \; .\nonumber
\end{eqnarray}

One of the advantages of using helicity amplitudes is that we can keep
track of which contributions come from the longitudinal polarization
of the HLL rather than the transverse polarizations.  We use the
notation $\ell^2$, $\ell_0^2$ to flag the parts of the squared matrix
element that come from the transverse, longitudinal polarizations of
the HLL, and $\ell\ell_0$ to flag contributions from the interference.

We define $\Theta$ to be the polar angle of the incoming quark with
respect to the $z$-axis defined by $Z_2$ in the HLL rest frame.  This
raises a problem since at a $pp$ collider we cannot distinguish the
quark direction from the anitquark direction in a $q\bar{q}$-initiated
process. A solution is to
symmetrize the cross section between the case where $\Theta$ is the
polar angle of the quark direction and the case where $\Theta$ is the
polar angle of the antiquark. In the coupling notation defined in
(\ref{eqn:ourtotalcouplings}), this symmetrization has the the same
effect as setting $g_{\tau}$, $g_{\sigma\tau}$, and
$g_{\sigma\sigma\tau}$ to zero.

The standard convention in the literature for the three azimuthal
angles is somewhat peculiar. The coordinate axes are chosen such that
the outgoing muon moves along the $y$-axis in the rest frame of the
HLL (or equivalently of $Z_2$). Thus the azimuthal angle of the muon
is $\pi /2$, while the azimuthal angle of the outgoing electron is
denoted $\phi -\pi /2$. We denote the azimuthal angle of the incoming
quark by $\Phi$. This choice of conventions leads to rather awkward
expressions for the angular distributions. A better choice is to align
the axes such that the quark azimuthal angle $\Phi = 0$. The remaining
azimuthal dependence is then denoted by $\varphi_1$ and $\varphi_2$,
such that the substitutions $\varphi_1 \to \Phi + \phi$, $\varphi_2
\to \Phi$ regain the previous convention. We will employ this notation
in this appendix, which makes the formulae more symmetrical.

%\subsection*{General formulae}

After the quark-antiquark symmetrization described above, the $XX$
part of the full matrix element squared is given by
\begin{widetext}
\begin{eqnarray}
&&\hspace*{-25pt}
4 m_1^2 m_2^2 X^2 \gamma_b^2\, 
\Bigl[
g_1 S^2 s_1^2 s_2^2\, 
\bigl(
2\ell_0^2 m_d^4 -\ell^2 m_H^2 
\bigr[
m_1^2  \cos{2\varphi_1}
+m_2^2  \cos{2\varphi_2}
\bigr]
\bigr)
\\
&&
\hspace*{-30pt}
+g_1 \ell^2 m_H^2(1+ C^2)\bigl[
2m_2^2 s_1^2+2m_1^2 s_2^2-(m_1^2+m_2^2)s_1^2 s_2^2
\bigr] + 4 \ell\ell_0 g_1 m_H m_d^2\, C\, S\, 
\bigl[
m_1 c_1 s_1 s_2^2 \s{\varphi_1} 
-m_2 c_2 s_2 s_1^2  \s{\varphi_2}
\bigr]
\nonumber\\
&&
\hspace*{20pt}
-2\ell^2 m_H^2 m_1 m_2 s_1 s_2\, 
\bigl(
(1+C^2)  
(g_1 c_1 c_2 -g_{\sigma\sigma})\cos{\varphi_1-\varphi_2}
+S^2  (g_1 c_1 c_2 +g_{\sigma\sigma})\cos{\varphi_1+\varphi_2}
\bigr)
\Bigr] \,.
\nonumber
\end{eqnarray}
%\end{widetext}

The $PP$ part is given by
%\begin{widetext}
\begin{eqnarray}
&&\hspace*{-25pt}
P^2 \Bigl[
\ell^2 g_1 m_H^2 S^2 s_1^2 s_2^2 \,
\bigl[
M_2^4 m_1^2\cos{2\varphi_1}+M_1^4 m_2^2 \cos{2\varphi_2}
\bigr]
\\
&&\hspace*{0pt}
+8 \ell_0^2  m_1^2 m_2^2 m_d^4 S^2\,
\bigl[
g_1\,(c_1^2+c_2^2+s_1^2 s_2^2\sin{\varphi_1-\varphi_2}^2)
+2g_{\sigma\sigma}c_1 c_2
\bigr]
\nonumber\\
&&\hspace*{0pt}
+(1+C^2)\ell^2 g_1 m_H^2
\bigl[
2M_1^4m_2^2s_1^2
+2M_2^4m_1^2s_2^2
%\nonumber \\
%&&\hspace*{50pt}
-(M_2^4 m_1^2+M_1^4 m_2^2)s_1^2 s_2^2
\bigr]
\nonumber\\
&&\hspace*{0pt}
-8 \ell\ell_0  m_H m_d^2m_1 m_2 C\, S\,
\bigl[
M_2^2 m_1 s_2\,  
\bigl(
g_1 c_2 s_1^2\s{\varphi_1}\cos{\varphi_1-\varphi_2}
%\nonumber \\
%&&\hspace*{20pt}
+c_1(g_1 c_1 c_2+g_{\sigma\sigma}) \s{\varphi_2}
\bigr)
\nonumber\\
&&\hspace*{0pt}
- M_1^2 m_2  s_1 \,
\bigl(
g_1 c_1 s_2^2\s{\varphi_2}\cos{\varphi_1-\varphi_2}
%\nonumber\\
%&&\hspace*{45pt}
+c_2(g_1 c_1 c_2+g_{\sigma\sigma})\s{\varphi_1}
\bigr)
\bigr]
\nonumber\\
&&\hspace*{0pt}
+2\ell^2 m_H^2 M_1^2 M_2^2 m_1 m_2 s_1 s_2 
\bigl[
(1+C^2) (g_1 c_1 c_2 
-g_{\sigma\sigma}) 
\cos{\varphi_1-\varphi_2}
%\nonumber\\
%&&\hspace*{85pt}
-S^2  (g_1 c_1 c_2 
+g_{\sigma\sigma})\cos{\varphi_1+\varphi_2}
\bigr]
\Bigr] \,.
\nonumber
\end{eqnarray}
%\end{widetext}
The $XP$ and $XQ$ interference parts are given by
%\begin{widetext}
\begin{eqnarray}
&&\hspace*{-50pt}
4 m_1 m_2 X P \, \gamma_b 
\Bigl[
\ell^2 g_1 m_H^2 S^2 s_1^2 s_2^2 
(M_1^2 m_2^2 \sin{2 \varphi_2}
%\nonumber\\
%&&\hspace*{50pt}
- M_2^2 m_1^2\sin{2 \varphi_1})\\
&&\hspace*{-10pt}
+2\ell\ell_0 g_1 m_H m_d^2 C\,S\, 
\bigl[
m_2 s_1^2 c_2 s_2 (2 m_1^2 \s{\varphi_1} 
\sin{\varphi_1-\varphi_2}-M_1^2 \c{\varphi_2})
\nonumber\\
&&\hspace*{60pt}
- m_1 s_2^2 c_1 s_1  (2 m_2^2 \s{\varphi_2} \sin{\varphi_1-\varphi_2}
+M_2^2 \c{\varphi_1})
\bigr]
\nonumber\\
&&\hspace*{-10pt}
-2 m_1 m_2 s_1 s_2 
\bigl[
(1+C^2) \ell^2 m_H^2 M_3^2  
(g_1 c_1 c_2 -g_{\sigma\sigma})
\sin{\varphi_1-\varphi_2}
\nonumber\\
&&\hspace*{40pt}
+m_d^2 s^2 (g_1 c_1 c_2 +g_{\sigma\sigma})
(\ell^2 m_H^2 \sin{\varphi_1+\varphi_2}
+2 \ell_0^2 m_d^2 \sin{\varphi_1-\varphi_2})
\bigr]
\nonumber\\
&&\hspace*{-10pt}
-4 \ell\ell_0 m_H m_1 m_2 m_d^2\, C\, S\,
\bigl[ 
m_2 s_1\,  (g_1 c_1+g_{\sigma\sigma}c_2)\c{\varphi_1}
+m_1 s_2\, ( g_1 c_2+g_{\sigma\sigma} c_1 )\c{\varphi_2} 
\bigr]
\Bigr] \,,
\nonumber
\end{eqnarray}
%\end{widetext}

%The $XQ$ interference part is given by

%\begin{widetext}
\begin{eqnarray}
&&\hspace*{-30pt}
4 m_1 m_2 X Q\, \gamma_b\, 
\Bigl[
\ell\ell_0 g_{\sigma}m_H m_d^2 C\, S\, 
\Bigl(
m_2 s_1^2 s_2 \,(2 m_1^2 \cos{\varphi_1-\varphi_2} 
\s{\varphi_1}
%\\
%&&\hspace*{50pt}
- M_1^2 \s{\varphi_2})
\\
&&\hspace*{125pt}
- m_1 s_2^2s_1\,(2 m_2^2 \cos{\varphi_1-\varphi_2} \s{\varphi_2}-
%\nonumber \\
%&&\hspace*{125pt}
M_2^2 \s{\varphi_1})
\Bigr)
\nonumber\\
&&\hspace*{-10pt}
+\ell^2 g_{\sigma}m_H^2 (1+c^2)(M_1^2 m_2^2 s_1^2  c_2+M_2^2 m_1^2s_2^2 c_1) 
%\nonumber\\
%&&\hspace*{-10pt}
+m_1 m_2 s_1 s_2\, 
\bigl[
(1+C^2)  \ell^2 g_{\sigma} m_H^2 m_d^2 (c_1-c_2)\cos{\varphi_1-\varphi_2} 
\nonumber\\
&&\hspace*{150pt}
- g_{\sigma} s^2(c_1+c_2) 
(\ell^2 m_H^2 M_3^2 \cos{\varphi_1+\varphi_2}+2 \ell_0^2 m_d^4 \cos{\varphi_1-\varphi_2})
\bigr]
\nonumber\\
&&\hspace*{20pt}
+2 \ell\ell_0 g_{\sigma}m_H m_d^2m_1 m_2  C\, S\, 
\bigl(
1+c_1 c_2 )(m_2 s_1 \s{\varphi_1} -m_1 s_2 \s{\varphi_2}
\bigr)
\Bigr] \, .
\nonumber
\end{eqnarray}
%\end{widetext}
Without the quark-antiquark symmetrization, one adds:

%\begin{widetext}
\begin{eqnarray}
&&\hspace*{-20pt}
8 m_H m_1^2 m_2^2 X^2 \gamma_b^2 g_{\sigma\tau}
\Bigl[
\ell^2 m_H C\,(m_2^2 c_2 s_1^2-m_1^2 c_1 s_2^2
- m_1 m_2(c_1-c_2) s_1 s_2  \cos{\varphi_1-\varphi_2})
\\
&&\hspace*{130pt}
-\ell\ell_0 m_d^2 S\, 
(m_2 s_1^2 s_2 \s{\varphi_2}+m_1 s_2^2 s_1 \s{\varphi_1})
\Bigr]
\nonumber\\
&&\hspace*{-20pt}
+2m_H P^2 g_{\sigma\tau}
\Bigl[
\ell^2 m_H C\,
\Bigl( 
M_1^4m_2^2c_2 s_1^2-M_2^4m_1^2c_1 s_2^2
+ M_1^2 M_2^2 m_1 m_2  (c_1-c_2)s_1 s_2  \cos{\varphi_1-\varphi_2}
\Bigr)
\nonumber\\
&&\hspace*{40pt}
+2 \ell\ell_0 m_1 m_2 m_d^2 S\,(( 1+c_1 c_2 )(M_1^2 m_2 s_1 \s{\varphi_1}
+M_2^2 m_1 s_2 \s{\varphi_2})
\nonumber\\
&&\hspace*{120pt}
-\cos{\varphi_1-\varphi_2}
(M_2^2 m_1 s_1^2 s_2 \s{\varphi_1} +M_1^2 m_2  s_2^2 s_1 \s{\varphi_2}))
\Bigr]
\nonumber\\
&&\hspace*{-20pt}
-4 m_H m_1 m_2 X P \, \gamma_b g_{\sigma\tau}
\Bigl[
2 \ell^2 m_H M_3^2m_1 m_2 C\, (c_1-c_2) s_1 s_2 \sin{\varphi_1-\varphi_2}
\nonumber\\
&&\hspace*{-10pt}
+\ell\ell_0 m_d^2 S\, 
\bigl[
m_2 s_1^2 s_2 (M_4^2 \c{\varphi_2}
-2 m_1^2 \s{\varphi_1} \sin{\varphi_1-\varphi_2})
-m_1 s_2^2 s_1(M_4^2 \c{\varphi_1} +2 m_2^2 \s{\varphi_2} 
\sin{\varphi_1-\varphi_2}
\bigr] 
\nonumber\\
&&\hspace*{80pt}
+2  m_1 m_2 (1+c_1 c_2 )( m_2 s_1 \c{\varphi_1}
-m_1 s_2 \c{\varphi_2}))
\Bigr]
\nonumber\\
&&\hspace*{-20pt}
4 m_H m_1 m_2 X Q\, \gamma_b 
\Bigl[
2 \ell^2 g_{\tau}m_H C\,(2M_1^2 m_2^2 s_1^2-2M_2^2 m_1^2s_2^2 
+M_4^2 m_d^2  s_1^2 s_2^2)
\nonumber\\
&&\hspace*{60pt}
-4\ell^2 m_H m_d^2 m_1 m_2 c (g_{\sigma\sigma\tau}
-g_{\tau}c_1 c_2 )s_1 s_2 \cos{\varphi_1-\varphi_2} 
\nonumber\\
&&\hspace*{-30pt}
+2\ell\ell_0 g_{\tau} m_d^2 S\, 
\bigl[
m_2 c_2 s_1^2 s_2 (2 m_1^2 \cos{\varphi_1-\varphi_2} \s{\varphi_1}
-M_4^2 \s{\varphi_2})
+ m_1 c_1 s_2^2 s_1 (2 m_2^2 \cos{\varphi_1-\varphi_2} \sin{\varphi_2}
-M_4^2 \s{\varphi_1})
\bigr]
\nonumber\\
&&\hspace*{80pt}
+4 \ell\ell_0 m_1 m_2 m_d^2 S\, 
(m_2 s_1 (g_{\sigma\sigma\tau} c_2+g_{\tau}c_1)\s{\varphi_1}
+m_1 s_2  ( g_{\sigma\sigma\tau}c_1
+g_{\tau}c_2 )\s{\varphi_2})
\Bigr]\, .
\nonumber
\end{eqnarray}
\end{widetext}

%\subsection*{On-shell case}

In the limit that both $Z$'s are on-shell, $m_1$$=$$m_2$$=$$M_Z$, we
introduce the notation of Buszello et al.: $x=m_H/M_Z$, $y^2=(x^2-4)/4$.
Then we can simplify using $m_d\to 0$,
$M_1$=$M_2$=$M_3 \to 4m_H^2 y^2/x^2$, $M_4\to m_H^2(x^2-2)/x^2$, and
$\gamma_b \to xy$. For the full symmetrized matrix element squared the
result is:
\begin{widetext}
\begin{eqnarray}
&&\hspace*{-30pt}
\frac{4}{x^6}  \ell^2 m_H^8 y^2 
\Bigl[
2(x^2 X^2+(x^2-4)P^2)
\bigl[
g_1(1+C^2)(1-c_1^2 c_2^2)-S^2(g_{\sigma\sigma}
+g_1 c_1 c_2)s_1 s_2 \cos{\varphi_1+\varphi_2}
\bigr]
\nonumber\\
&&\hspace*{18pt}
- (x^2X^2-(x^2-4)P^2) 
\bigl[
g_1 S^2 s_1^2 s_2^2 (\cos{2 \varphi_1}
+\cos{2\varphi_2})-2 (1+C^2)(g_{\sigma\sigma}
-g_1 c_1 c_2) s_1 s_2 \cos{\varphi_1-\varphi_2}
\bigr]
\nonumber\\
&&\hspace*{-30pt}
-4 X P  \,x \,y \,s_1 s_2
\bigl[
g_1 S^2 s_1 s_2 (\sin{2\varphi_1}
-\sin{2\varphi_2})-2(1+C^2)(g_{\sigma\sigma}-g_1 c_1 c_2)  
\sin{\varphi_1-\varphi_2}
\bigr]
\nonumber\\
&&\hspace*{40pt}
+4 X P \, x \,y \,g_{\sigma} 
\bigl[ 
(1+C^2)(c_2 s_1^2+c_1 s_2^2)
-S^2 (c_1+c_2) s_1 s_2  \cos{\varphi_1+\varphi_2}
\bigr]
\Bigr] \;.
\end{eqnarray}
\label{eq:generalSpinOne}
%\end{widetext}
If we simply set $\Theta = 0$,
the above simplifies to:
%\begin{widetext}
\begin{eqnarray}
&&\hspace*{-30pt}
\frac{16}{x^6} \ell^2 m_H^8 y^2 
\Bigl[
g_1(x^2 X^2+(x^2-4)P^2)(1-c_1^2 c_2^2)
+(x^2 X^2-(x^2-4)P^2)(g_{\sigma\sigma}-g_1 c_1 c_2) 
s_1 s_2 \cos{\varphi_1-\varphi_2}
\nonumber\\
&&\hspace*{30pt}
+4 X P \, x\, y\, s_1 s_2 (
g_{\sigma\sigma}-g_1 c_1 c_2)  
\sin{\varphi_1-\varphi_2}
+2 X P \, x\, y\, g_{\sigma} (c_2 s_1^2+c_1 s_2^2)
\Bigr] \; .
\end{eqnarray}

This agrees with the result of Buszello et al.~\cite{Buszello:2002uu}.
\end{widetext}

%%%%%%%%%%%%%%%

%%%%%%%%%%%%%%%%%%%%%%%%%%%%%%%%%%%%%%%%
%%%%%%%%%
%%%%%%%%%%          BIBLIO.TEX
%%%%%%%%%%
%%%%%%%%%%%%%%%%%%%%%%%%%%%%%

%%%%%%%%%%%%
%%%%%%%%%%%%%%%%%

\end{document}